\documentclass[onecolumn,useamsfonts,mfastrosym]{pasj01}
\Received{$\langle$reception date$\rangle$}
\Accepted{$\langle$acception date$\rangle$}
\Published{$\langle$publication date$\rangle$}

 \usepackage{lineno,color,times}
 \SetRunningHead{Astronomical Society of Japan}{Usage of \texttt{pasj00.cls}}
\definecolor{LinkBlue}{RGB}{6,69,173}
\definecolor{DarkBlue}{RGB}{11,0,128}
\definecolor{red}{rgb}{1,0.,0.}
\usepackage{cuted}
\usepackage{numcompress}

\def\be{\begin{equation}}
\def\ee{\end{equation}}
\def\bea{\begin{eqnarray}}
\def\eea{\end{eqnarray}}
\newcommand{\tb}[1]{\textbf{\texttt{#1}}}
\newcommand{\Mie}{\mathcal{M}}
\newcommand{\Sie}{\mathcal{S}}
\newcommand{\il}{~}
\newcommand{\rtb}[1]{\textcolor[rgb]{1.00,0.00,0.00}{\tb{#1}}}
\newcommand{\gtb}[1]{\textcolor[rgb]{0.17,0.72,0.40}{\tb{#1}}}
\newcommand{\ptb}[1]{\textcolor[rgb]{0.77,0.04,0.95}{\tb{#1}}}
\newcommand{\btb}[1]{\textcolor[rgb]{0.00,0.00,1.00}{#1}}
\newcommand{\otb}[1]{\textcolor[rgb]{1.00,0.50,0.25}{\tb{#1}}}
  \newcommand{\Qa}{\mathcal{Q}}
  \newcommand{\Sa}{\mathcal{S}}
\newcounter{num}
\newcommand{\Rnum}[1]{\setcounter{num}{#1} \Roman{num}}
\newcommand{\rnum}[1]{\setcounter{num}{#1} \roman{num}}
\begin{document}

\title{Dragged surfaces. On the  accretion tori in the ergoregion}
\author{D. Pugliese and Z. Stuchl{\'{\i}}k}%
\altaffiltext{}{Research Centre for Theoretical Physics and Astrophysics\\
Institute of Physics,
  Silesian University in Opava,\\
 Bezru\v{c}ovo n\'{a}m\v{e}st\'{i} 13, CZ-74601 Opava, Czech Republic }
%\email{***@***.***.***}


\maketitle

\begin{abstract}
We discuss the  conditions for the existence of extended  matter configurations orbiting in the ergoregion or
close to the outer ergosurface of the Kerr black hole  ("dragged" configurations).   The  corotating  tori under consideration are  perfect fluid configurations  with barotropic equation of state,   orbiting on the equatorial plane of the central Kerr  black hole.
The possibility of magnetized tori with a toroidal magnetic field is also discussed.
Indications on the attractors where dragged tori can be observed are provided  with   the  analysis of the fluid characteristics and geometrical features, relevant for the tori stability and  their  observations.    QPOs emissions from the inner edges of the dragged tori are also discussed.
We argue that the  smaller
  dragged tori could be  subjected to a characteristic instability, effect of the frame-dragging. This possibility  is thoroughly  explored. This can finally lead to the  destruction of the torus (disk exfoliation) which can  combine with accretion and processes  present in the regions very close to the black hole horizon.
  Tori are characterized  according to the central attractor dimensionless spin.
These structures can be observed  orbiting   black holes  with  dimensionless spin $a>0.9897M$.
\end{abstract}

\def\be{\begin{equation}}
\def\ee{\end{equation}}
\def\bea{\begin{eqnarray}}
\def\eea{\end{eqnarray}}
\newcommand{\bt}[1]{\mathbf{\mathtt{#1}}}

\newcommand{\cc}{\mathrm{C}}

\newcommand{\actaa}{Acta Astron.}

\section{Introduction}
Although restricted by the typical assumptions of the   simplified models, geometrically  thick (stationary) disks provide a striking good approximation of  several  aspects of accretion  in    more complex dynamical models, some of these features include indication on   tori location, the disk   elongation on their symmetry plane, the  inner edge of quiescent and accreting disks, the tori   thickness, the  maximum height, and the critical pressure points in the disks.
 In this article  we  explore accreting  toroidal matter orbiting a central Kerr black hole (\textbf{BH}), within  the hydrodynamic model   introduced  and detailed in a series of works 	
\citep{cc,Pac-Wii,Koz-Jar-Abr:1978:ASTRA:,Abramowicz:1996ap,FisM76,Raine,PuMonBe12,pugtot}.%, and then

This is a fully general
relativistic model of an opaque and super-Eddington, pressure supported disk, cooled by advection
based on the Boyer theory of     the equilibrium and rigidity in general relativity \citep{Boy:1965:PCPS:}.
In the Boyer model,  many   features of the disk dynamics and morphology  like the disk thickness, the disk  stretching  on the equatorial plane and   the disk location,  are predominantly   constrained by the geometric properties of spacetime via an effective potential function  regulating the pressure gradients in  the Euler equation.
The effective potential function contains  two  main components:  a dynamical term, related to the orbiting matter   by means of the fluid angular momentum, here  assumed constant along  the disk,  and a  geometrical one related to the properties of the  spacetime background.
 The boundary of any stationary, barotropic, perfect fluid torus is determined  by the
equipotential surface,  which are the
surfaces of constant pressure, defined by the gradient of a
scalar function, i.e. the effective potential. This  property holds  if, for barotropic fluids,
the relativistic frequency $\Omega$ turns to be function of the fluid angular momentum  $\ell$ only or $\Omega=\Omega(\ell)$
(von Zeipel conditions)
\citep{Koz-Jar-Abr:1978:ASTRA:,M.A.Abramowicz,Chakrabarti,Chakrabarti0}.
Accretion onto the attractor is driven through the vicinity of the cusp (corresponding  to the inner edge of
the cusped toroid) in the self crossed (cusped) closed  configurations   from  a violation of the hydrostatic equilibrium, i.e., due to  \emph{Paczy\'nski mechanism}
\citep{Koz-Jar-Abr:1978:ASTRA:,cc,Pac-Wii}. This mechanism has been proved to be also an
important stabilizing mechanism against the thermal and viscous instabilities locally,
and against the so called Papaloizou\&Pringle instability (\textbf{PPI}) globally \citep{Blaes1987,abrafra,Pac-Wii,cc,Koz-Jar-Abr:1978:ASTRA:}.
During the evolution of dynamical processes, the functional form of the angular
momentum and entropy distribution depends on the initial conditions of the system and on
the details of the dissipative processes.

We focus  on tori orbiting in  regions close to the \textbf{BH} static limit and  in the  \textbf{BH} ergoregion, which is  the region between the outer ergosurface and the (outer) event horizon, defining the partially contained and dragged  surfaces.
(The outer ergosurface is also called
 stationary limit surface  or static limit, for the absence of
any  (timelike) static observers inside the region).
A partially contained torus is   centered (location of maximum pressure point) in the  outer region, but a part of the torus  inner region, which is bounded from the below by the inner edge  and from the above by the torus center,  orbits in the ergoregion.

Thick accretion disks are usually  associated with very compact objects like  black holes and super massive \textbf{BHs} (\textbf{SMBHs}). Directly involved   in   the equilibrium phases of the attractors, they are  very likely the base of    the jet formation and dynamics,  the  Active Galactic Nuclei (\textbf{AGN}) and
Gamma Ray Bursts (\textbf{GRBs}) emissions.
In this context the dynamics inside the ergoregion  is extremely  relevant in Astrophysics: accreting matter  can support, for example,  jets of matter or radiation originated inside the ergoregion \citep{Meier,Gariel:2013hwa}, which
 can  empower jets with  powerful Poynting flux,
as  from ergospheric disks  studied  in \citet{Punsly:2007ka}.

  The  processes  occurring inside the  \textbf{BH} ergoregion  are essential
for understanding the central engine mechanism of many  emissions \citep{Meier,Fro-Z}.
 The mechanism, by which energy from compact spinning objects is extracted,
is of  great astrophysical interest:
\textbf{BH}  rotational energy can be extracted through the classical  Penrose process, and indeed the
super-radiance
as seen by asymptotic observers.
Black hole thermodynamics  has been  grounded on the observation that the \textbf{BH} energy can actually  be extracted from the black hole, and  ergoregion is  the background  geometrical engine of many processes of energy extraction   from the Kerr \textbf{BH}.
The ergosurface is also the geometric basic of  super-radiant scattering of incident waves which is the  wave  analog of the Penrose  process,  based on the fact that  the  particles energy, within the ergoregion, as  measured  by an observer at infinity, can also  be negative\footnote{The presence of negative   energy particles is also a distinctive feature of the  ergoregion in weakly rotating naked singularities (\textbf{WNS}) differentiating these from the \textbf{BHs}. (The superirradiance and ergoregion instability is a further characteristic differentiating \textbf{BHs} from  \textbf{NSs}.). The static limit can  act   as a semi-permeable membrane separating the spacetime  region, filled with negative energy particles  from the external one, filled with positive. The \textbf{WNSs} ergoregion  is characterized   by the existence of   zero and  negative energy states  circular orbits
(stable circular geodesics with negative energy)\citep{1980BAICz..31..129S,2012CQGra..29f5002S,2018EPJC...78..879C,2018-ZC}.
 The presence of this special matter  in an  ``antigravity''
sphere,  possibly filled with negative energy matter formed according to the Penrose process,
  and bounded by orbits with zero angular momentum,   could have  an important role in the source evolution\citep{Hawking,Penrose71}. }
(\textbf{BH} super-radiance also appears to
distinguish  bosonic waves from the fermions
in the super-radiant amplification).

The  rotational energy can be extracted from the source, lowering  its angular momentum.
The Penrose  energy extraction, and the super-radiance are classical phenomena  due to the frame-dragging of the spinning spacetime.
Another possibility is the extraction of  energy from a rotating black hole  through the Blandford-Znajek mechanism \citep{BZ,Pei:2016kka,Komissarov:2008yh,Lasota:2013kia,Penrose} and magnetic Penrose process or radiative Penrose process\citep{2021PhRvD.103b4021K,2020ApJ...895...14T}.
 The Hawking radiation instead  arises from  the vacuum fluctuations   in the regions close to the \textbf{BH} horizon, it is  the spontaneous emission of thermal radiation which is created in the vacuum regions surrounding  a \textbf{BH}, and it may eventually  lead to a decrease of the  \textbf{BH} mass \citep{Penrose71,[3],[4],[6]}.

In this article  we characterize   the    orbiting  extended  matter configurations  orbiting  in the ergoregion of   a Kerr \textbf{BH}, introducing the concept of dragged  tori  and partially contended tori made by
  corotating perfect fluids   thick tori.
We discuss possible instabilities by exploring  the effects of  geometry  frame-dragging on
the  disk exfoliation and evaluating  the pressure  gradients which are  the main factor for the tori formation and    formation of  a possible atmosphere of  (free) particles  swarm. A  possible outcome of the instabilities  of the  smallest tori  could be an emission of swarm of   particles and photons, enhanced accretion, or the torus destruction. Dragged and partially contained tori are also subjected to   processes typical  of  the regions very close to the black hole horizon as, for example, the Runaway Instability.
To  evaluate this situation, we  asses
the torus verticality, the dragged torus  geometrical thickness,  the    influence of the dragging frame on the disk thickness, and the location of the tori extreme of pressure  with the respect to the static limit.
We proceed to the exploration of the   tori  characteristics in dependence on  the central attractor dimensionless spin,   providing eventually indications on the attractors where dragged tori can be observed.
Some extreme configurations in the ergoregion are also described, as  proto-jets  which are open cusped  General Relativistic Hydrodynamic (GRHD) configurations, limiting solutions for the thick disks models, and the possibility of  orbiting agglomerates of toroids  composed by an aggregate of multi  toroids orbiting  in the ergoregion.

In the second part of this work, dragged and partially contained tori  are   studied in  relation to  the quasi-periodic oscillations  (\textbf{QPOs}) in emission from the tori  inner edges.
We  relate also the  dragged tori  in the  ergoregion and the maximum extractable rotational  energy  $\xi$ from the \textbf{BH} horizon,   constituting   the state prior the total extraction of the energy from the \textbf{BH}, following the  approach introduced in \citet{Daly0}, which is  quite independent from  the details of the
specific process  of energy extraction, and based on
 the definition of  \textbf{BH} irriducible mass function,  rotational energy and using
 the  \textbf{BH} classical thermodynamical law.
(The energy extraction   through the accretion process, the rotational  energy converted into
radiation corresponds to  the binding energy of the fluid).
 For large part of this analysis we shall consider perfect fluids defined by a barotropic equation of state; however, we also
investigate the possibility that the frame dragging  could   differentiate  different   polytropics.
In this context we provide  an estimation of the   the flux thickness, mass-flux and   enthalpy-flux
   for dragged and partially contained   thick  disks with a fixed polytropic fluid.

\medskip
The   article is organized as follows:
 In Sec.\il(\ref{Sec:model})  we introduce  the thick  accretion   model and  the fluid  effective potential for the  toroidal configurations in a  Kerr spacetime background.
Dragged surfaces and partially contended tori are introduced in Sec.\il(\ref{Sec:deta-hall}). In Sec.\il(\ref{Sec:dv})
the tori verticality is studies  and disk exfoliation is introduced.
Dragged disks thickness is the focus of Sec.\il(\ref{Sec:influ-ergosra}) where the  influence of the dragging frame on the disk thickness is investigated.
In Sec.\il(\ref{Sec:mid-w-t}) we  explore the  process of
tori exfoliation.
{Extreme configurations as orbiting agglomerates of toroids and proto-jets are briefly considered in Sec.\il(\ref{Sec:gir-c2}).
Characteristic frequencies of the oscillations of the  toroids are studied in Sec.\il(\ref{Sec:carac-fre}) and  in Sec.\il(\ref{Sec:qpos}) where we discuss the
origin of the \textbf{QPOs} emission  in relations to the  toroids in  the ergoregion.
Analysis of possible polytropic fluids  for dragged and partially contained tori is in Sec.\il(\ref{Sec:poly-altr}).
Discussion on some aspects of  tori energetics for these tori  is in Sec.\il(\ref{Sec:ener}).
Conclusions are  Sec.\il(\ref{Sec:conclusion}).
In Appendix\il(\ref{Sec:magn}) there are some
notes on stationary  magnetized tori in the ergoregion.
\section{Fluid configuration on the Kerr spacetime}\label{Sec:model}
We consider a perfect fluid orbiting  in  the Kerr
spacetime background,  where the   metric tensor can be
written in Boyer-Lindquist (BL)  coordinates
\( \{t,r,\theta ,\phi \}\)
as follows
\bea&&\label{Eq:metric-1covector}
d s^2=-\alpha^2 d t^2+\frac{A \sigma}{\rho^2} (d \phi- \omega_z d t)^2+\frac{\rho^2}{\Delta}d r^2+\rho^2 d\theta^2,\\&&\nonumber \sigma\equiv\sin^2\theta,\quad A\equiv (r^2+a^2)^2-a^2 \Delta \sigma
\eea
where
$\alpha=\sqrt{(\Delta \rho^2/A)}$  and $\omega_z=2 a M r/A$ are the lapse function and the frequency of the zero angular momentum fiducial observer (\textbf{ZAMOS}) \citep{observers}, whose four velocity is $u^a=(1/\alpha,0,0,\omega_z/\alpha)$, being  orthogonal  to the surface   of constant $t$.
here $M$ is a mass parameter and the specific angular momentum is given as $a=J/M$, where $J$ is the
total angular momentum of the gravitational source and  $\rho^2\equiv r^2+a^2\cos\theta^2$, $\Delta\equiv r^2-2 M r+a^2$. In the following it will be also  convenient to introduce  the quantity
$\sigma \equiv\sin^2\theta$. We will consider  the Kerr black hole (\textbf{BH}) case defined by $a\in ]0,M[ $,  the extreme black hole source $a=M$, and the non-rotating  limiting case $a=0$ of the  Schwarzschild metric.
 The horizons $r_-<r_+$ and the static (or stationary) limit $r_{\epsilon}^+$ are respectively
\bea
r_{\pm}\equiv M\pm\sqrt{M^2-a^2};\quad r_{\epsilon}^{+}\equiv M+\sqrt{M^2- a^2 \cos\theta^2};
%&&\nonumber
\eea
where  $r_+<r_{\epsilon}^+$ on the planes  $\theta\neq0$  and it is $r_{\epsilon}^+=2M$ i on the equatorial plane $\theta=\pi/2$.
In the  {region $r\in]r_+,r_{\epsilon}^{+}[\equiv \Sigma_{\epsilon}^+$} ({\em ergoregion} or outer ergoregion),  there  is  { $g_{tt}>0$} (time component of the metric tensor) and $t$-Boyer-Lindquist coordinate becomes spacelike,
this fact implies that a  { static observer} cannot exist inside
the ergoregion. The radius $r_{\epsilon}^+$ is the ergosurface, or, precisely, outer ergosurface.
%Al
In this work we investigate toroidal  configurations of  perfect fluid orbiting a Kerr \textbf{SMBH}, it will be therefore convenient to consider first the properties of the test particle circular motion.
Since the metric is independent of $\phi$ and $t$, the covariant
component of a particle four--momentum, $p_{\phi}$ and $p_{t}$,  are
conserved along  the   geodesics or\footnote{We adopt the
geometrical  units $c=1=G$ and  the $(-,+,+,+)$ signature, Latin indices run in $\{0,1,2,3\}$.  The   four-velocity  satisfy $u^a u_a=-1$. The radius $r$ has unit of
mass $[M]$, and the angular momentum  units of $[M]^2$, the velocities  $[u^t]=[u^r]=1$
and $[u^{\varphi}]=[u^{\theta}]=[M]^{-1}$ with $[u^{\varphi}/u^{t}]=[M]^{-1}$ and
$[u_{\varphi}/u_{t}]=[M]$. For the seek of convenience, we always consider the
dimensionless  energy and effective potential $[V_{eff}]=1$ and an angular momentum per
unit of mass $[L]/[M]=[M]$.}
\bea&&\label{Eq:after}
{E} \equiv -g_{ab}\xi_{t}^{a} p^{b},\quad L \equiv
g_{ab}\xi_{\phi}^{a}p^{b}\ , \quad \mbox{or}
\quad
{E}=-(g_{tt} u^t +g_{\phi t} u^\phi),\quad L=(g_{\phi \phi}u^\phi +g_{\phi t} u^t)
,\\&&\label{Eq:uf}
\mbox{where}\quad
u^t= \frac{{E} g_{\phi \phi}+g_{\phi t} L}{\mathcal{A}_{\mathrm{T}}},\quad u^\phi=- \frac{{E} g_{\phi t}+g_{tt} L}{\mathcal{A}_{\mathrm{T}}}, \quad
\left(\mathcal{A}_{\mathrm{T}}\equiv g_{t \phi }^2-g_{{tt}}g_{\phi \phi }\right).
\eea
The
 momentum $p^a= \mu u^a$ of the  particle with  mass $\mu$ and four-velocity $u^{a}$
can be normalized so that
$g_{ab}u^{a}u^{b}=-k$, where $k=0,-1,1$ for null, spacelike and timelike
curves, respectively.
Quantities $(E,L)$ are  constants of motion for circular test particles geodesics where  $\xi_{t}=\partial_{t} $  is
the Killing field representing the stationarity of the Kerr geometry and  $\xi_{\phi}=\partial_{\phi} $
is the
axial Killing field; the vector $\xi_{t}$ is   spacelike in the ergoregion.
In general,  we may interpret $E$, for
timelike geodesics, as representing the total energy of the test particle
 coming from radial infinity, as measured  by  a static observer at infinity, and  $L$ as the axial component of the angular momentum  of the particle.
The  { normalization condition}  on the four-velocity
can be solved  for the energy  ${E}$. %  to obtain the  solutions,
 For circular motion  there is $u^r=0$. Considering the motion  on the  fixed equatorial plane $\sigma=1$, no
motion is  in the $\theta$ angular direction and $u^\theta=0$ (the  Kerr metric is  symmetric  under reflection through  the  equatorial plane $\theta=\pi/2$).
Within these assumptions  we obtain  the  effective potential $V_{eff}(a;L,r)\equiv \left. E_{\pm}/\mu\right|_{u^r=0}$:
circular orbits are therefore described by
\be\label{Eq:Kerrorbit}
\dot{r}=0,\quad V_{eff}={E}/\mu,\quad \partial V_{eff}/\partial r=0.
\ee
Some  notable  radii regulate the particle dynamics, namely the \emph{marginal  circular orbit} for timelike particles,  $r_{\gamma}^{\pm}$ which is also a photon orbit,   the \emph{marginal  bounded orbit}  at $r_{mbo}^{\pm}$, and the \emph{marginal  stable circular orbit} at $r_{mso}^{\pm}$ with angular momentum and energy  $(\mp L_{\pm},E_{\pm})$ respectively,  where $(\pm)$ is for  counterrotating or corotating orbits with respect to the attractor.
It is convenient to  introduce also   the  relativistic angular frequency  $\Omega$ and the specific angular momentum $\ell$ as follows
\bea&&\label{Eq:flo-adding}
\Omega \equiv\frac{u^\phi}{u^t }=-\frac{{E} g_{\phi t}+g_{tt} L}{{E} g_{\phi \phi}+g_{\phi t} L}= -\frac{g_{t\phi}+g_{tt} \ell}{g_{\phi\phi}+g_{t\phi} \ell},
\quad\mbox{and}\\&&\nonumber
\ell\equiv\frac{L}{{E}}=-\frac{u_{\phi}}{u_t}=-\frac{u^\phi  g_{\phi\phi}+g_{\phi t} u^t }{g_{tt} u^t +g_{\phi t} u^\phi } =-\frac{g_{t\phi}+g_{\phi\phi} \Omega }{g_{tt}+g_{t\phi} \Omega }.
\eea
}

\begin{figure*}\centering
  \includegraphics[width=7.49cm]{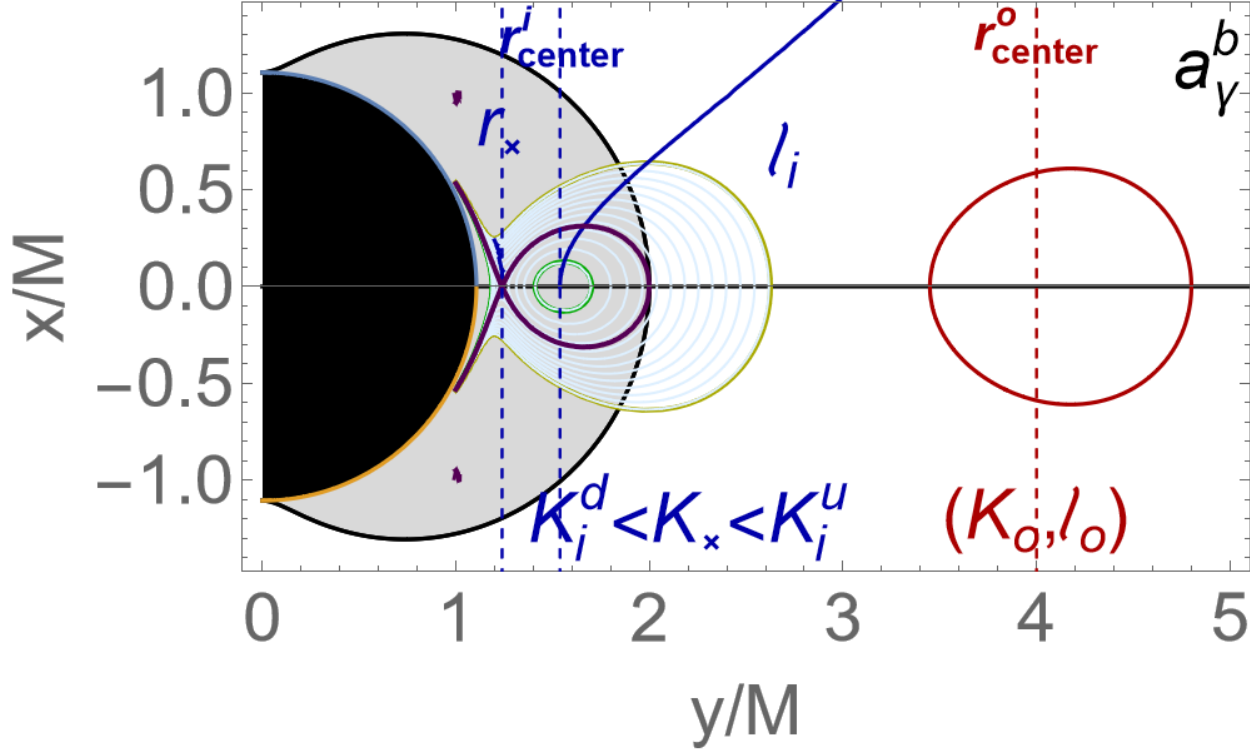}
  \includegraphics[width=7.49cm]{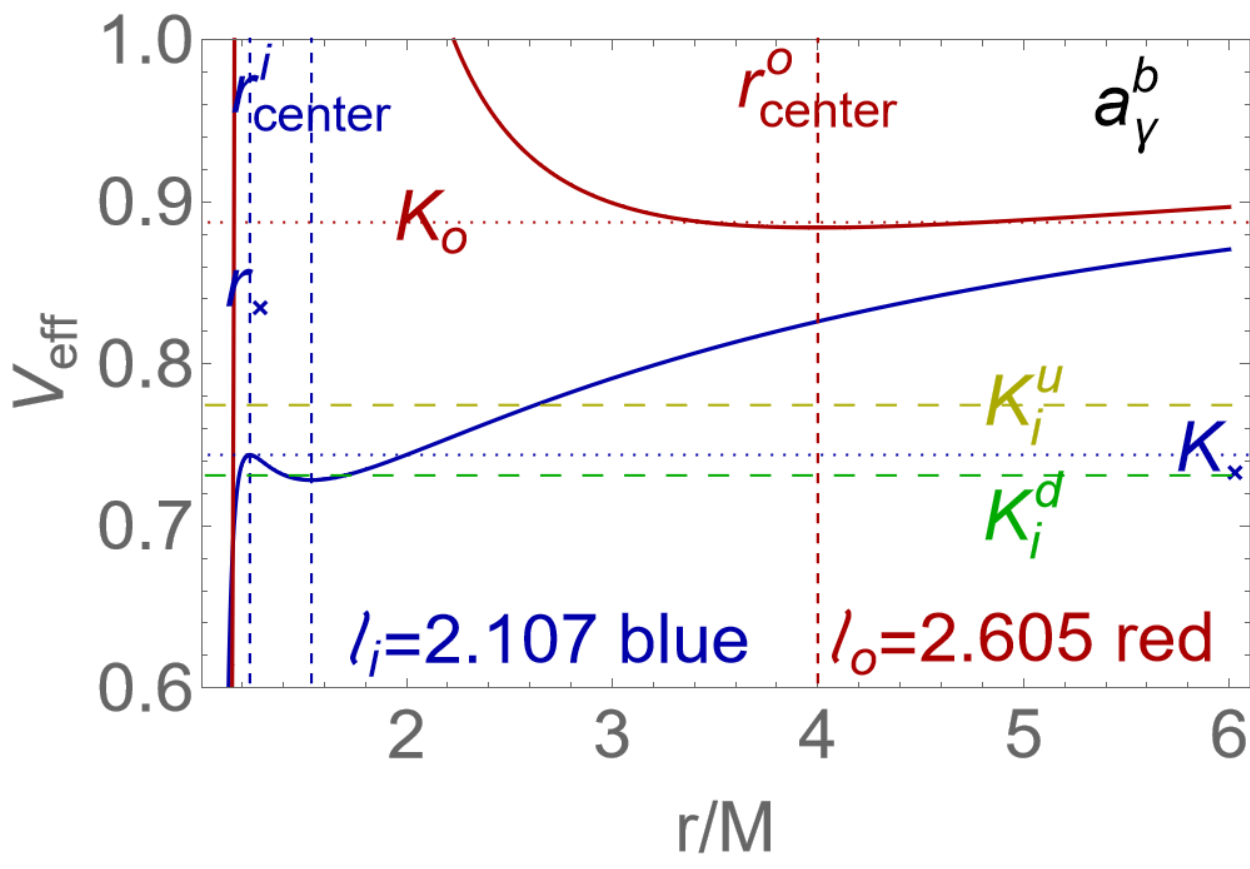}\\
\includegraphics[width=7.49cm]{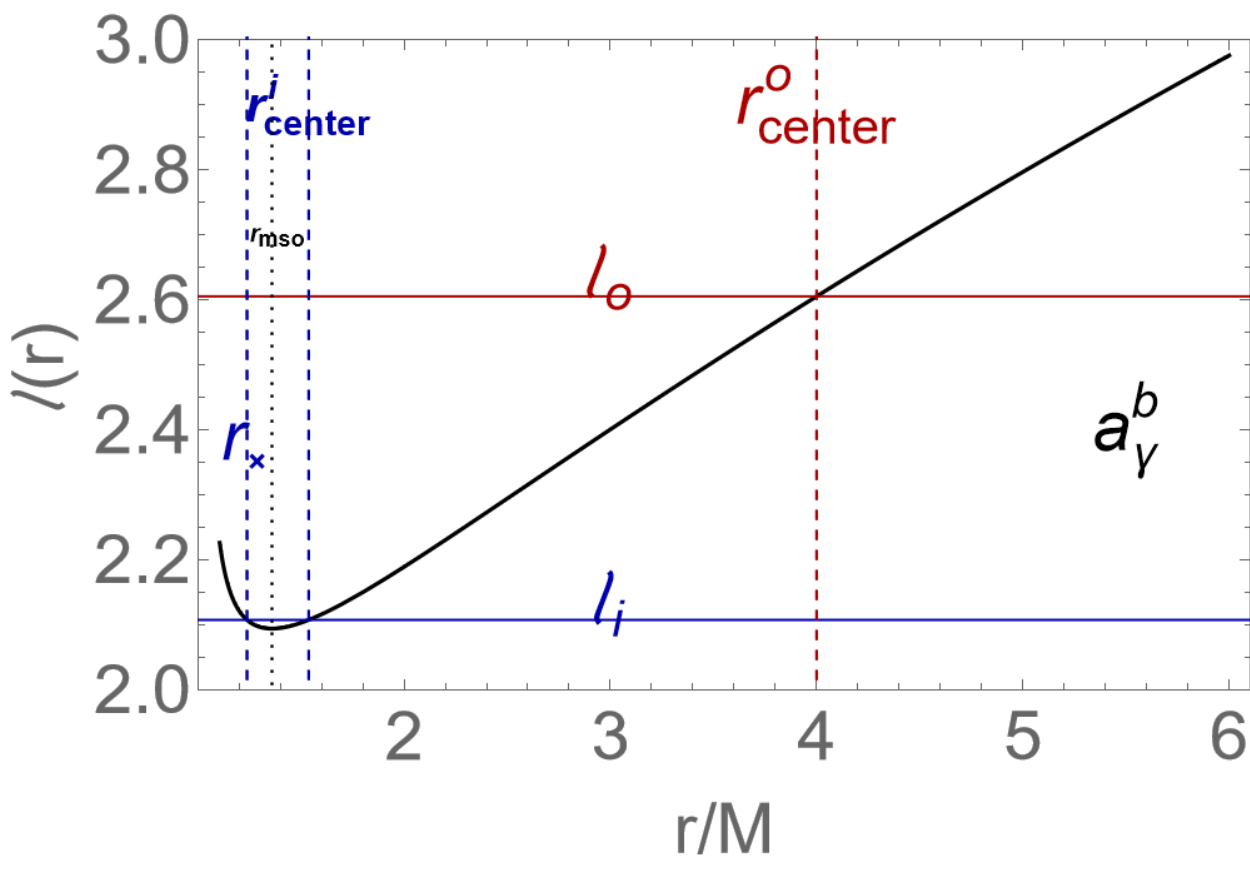}
\includegraphics[width=7.49cm]{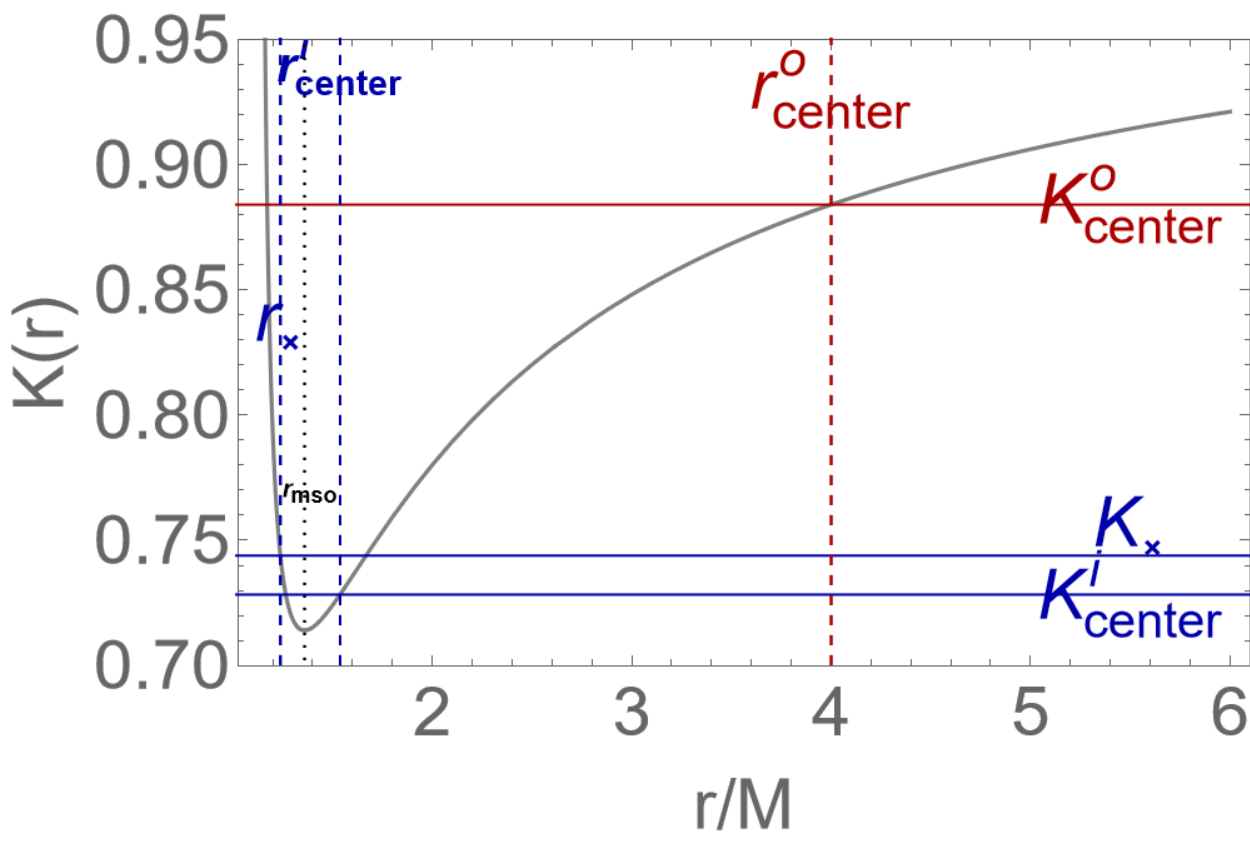}
    \caption{Tori construction as equi-pressure surfaces. Upper left panel shows as set of  corotating  tori (equipressure surfaces-Boyer surfaces). Upper right panel shows the  tori effective potentials $V_{eff}$, defined in  Eq.\il(\ref{Eq:scond-d}) as functions of $r/M$. Bottom left panel shows the leading function $\ell(r)$ (angular momentum distribution) as function of $r/M$. Bottom right panel shows the energy function $K(r)$ as function of $r/M$. $\ell(r)$ and $K(r)$ are defined in   Eq.\il(\ref{Eq:leading-energy}).
   Black hole with spin $a_{\gamma}^b\equiv 0.9943M$ is considered--see also Eq.\il(\ref{Eq:strateg}).
   Black central region is the \textbf{BH}, gray region is the outer ergoregion $\Sigma_{\epsilon}^+\equiv ]r_+,r_{\epsilon}^+[$ ($r_+$ is the \textbf{BH} outer horizon, and $r_{\epsilon}^+$ is the outer ergosurface). The tori and the \textbf{BH} equatorial plane is  at $x=0$ (there is $r=\sqrt{x^2+y^2}$ and $ \sin\theta= {y}/{\sqrt{x^2+y^2}}$). Tori with specific angular momentum $\ell_i$ (inner) and $\ell_o$ (outer-red curve) are plotted. Inner set of tori at $\ell=\ell_i$ are shown for different values of the $K$ parameter in $K \in [K_i^d, K_i^u]$, for $K=K_{\times}$ there is the formation of the torus cusp  $r_{\times}$ and it corresponds to the  maximum point of the  torus effective potential. The torus at $\ell=\ell_o$ has $K=K_o$ parameter.  The centers of maximum density (pressure) $(r_{center}^i, r_{center}^o)$ in the tori  and the cusp $r_{\times}$ are also shown.  Blue curves connect the maximum density point in the disk $r_{center}^i$  (determined by $\ell_i$) with the geometrical  tori maximum determined by the $K$ parameter, and the minimum pressure point $r_{\times}$ with the tori surfaces minima at $r<r_{center}^i$ for $K\in [K_{\times}, K_{i}^u]$-- see also  Figs\il(\ref{Fig:polodefin1}). In the right upper panel the tori effective potentials are shown, red curve for the  outer torus with  $\ell=\ell_o$, and the blue curve for  the inner tori with $\ell=\ell_i$. Radii  $(r_{center}^i, r_{center}^o)$, minima of the effective potential are  shown, the cusp $r_{\times}$, maximum of the effective potential  with $\ell=\ell_i$, is shown.
   Values of the $K$ parameter $K_{\times}$ (at maximum point of the effective potential defining the cusped tori), $K_i^u$ and  $(K_o, K_i^d)$ for closed tori   are shown in correspondence with the  right panel tori. Bottom panels show functions $\ell(r)$ and $K(r)$   in the \textbf{BH} spacetime with spin  $a=a_{\gamma}^b$ as functions of $r/M$  and  the relevant values for the tori at  $\ell=\ell_i$ and $\ell=\ell_o$.
   Left panel: couples $(\ell_o, r_{center}^o)$ (red lines) and $(\ell_i, r_{center}^i)$ (blue lines) are shown. Dotted  black  line is the marginally stable circular orbits $r_{mso}$: at $r>r_{mso}$ there are the tori  centers  (maximum pressure points in the tori), at $r<r_{mso}$ there are  the minima of the pressure (tori cusps).  The radius $r_{\times}$ is also signed.
   Right panel: couples $(K_{center}^o, r_{center}^o)$ (red lines) and $(K_{center}^i, r_{center}^i)$ (blue lines) are shown, where  $(K_{center}^i, K_{center}^o)$ are the $K$ parameter values  the maximum points of pressure in the tori with  $\ell=\ell_i$ and $\ell=\ell_o$ respectively (correspondent to the minimum values of the tori effective potentials). The value $K_{\times}$ at the cusp of the inner torus  having  $\ell=\ell_i$ is also shown. Dotted  black  line is the marginally stable circular orbits $r_{mso}$: at $r>r_{mso}$ there are the tori  centers (maximum pressure point), at $r<r_{mso}$ there are  the minima of the pressure (tori cusps).  The radius $r_{\times}$ (torus cusp) is also showed.}\label{Figs:PlotTRialfull}
    \end{figure*}

We consider a one-specie particle perfect  fluid (simple fluid),  where
\be\label{E:Tm}
T_{a b}=(\varrho +p) u_{a} u_{b}+\  p g_{a b},
\ee
is the fluid energy momentum tensor,  $\varrho$ and $p$ are  the total energy density and
pressure, respectively, as measured by an observer moving with the fluid. For the
symmetries of the problem, we always assume $\partial_t \mathbf{Q}=0$ and
$\partial_{\varphi} \mathbf{Q}=0$, being $\mathbf{Q}$ a generic spacetime tensor
(we can refer to  this assumption as the condition of  ideal hydrodynamics of
equilibrium).
  The timelike flow vector field  $u^a$  denotes now the fluid
four-velocity.
The motion of the fluid  is described by the \emph{continuity  equation} and the \emph{Euler equation} respectively:
\bea\label{E:1a0}
u^a\nabla_a\varrho+(p+\varrho)\nabla^au_a=0\, ,\quad
%\label{Eulerif0}
(p+\varrho)u^a\nabla_au^c+ \ h^{bc}\nabla_b p=0\, ,
\eea
where the projection tensor $h_{ab}=g_{ab}+ u_a u_b$ (from the conservation equation $\nabla_a T^{ab}=0$ along the tensor $(u^a,h_{ab})$ defined by the fluid, where $\nabla_ag_{bc}=0$.).
 We investigate in particular  the case of a fluid circular configuration, defined by the constraint
$u^r=0$. %\btb{(**********)}
 Tori  are centered on the central \textbf{BH}, orbiting on  its equatorial plane which is also  the  tori symmetry plane--see for example Figs\il(\ref{Figs:PlotTRialfull}).  On  the fixed plane $\sigma=1$, no
motion is assumed in the $\theta$ angular direction, which means $u^{\theta}=0$.
We assume moreover a barotropic equation of state $p=p(\varrho)$. The  continuity equation %Eq.\il(\ref)
% %
%
 %
is  identically satisfied as consequence of these conditions.
From the Euler  equation (\ref{E:1a0}) we obtain
\be\label{Eq:scond-d}
\frac{\partial_{\mu}p}{\varrho+p}=-\frac{\partial}{\partial \mu}W+\frac{\Omega \partial_{\mu}\ell}{1-\Omega \ell},\quad \mbox{where}\quad W\equiv\ln V_{eff}(\ell),\quad\mbox{with}\quad \ell\equiv \frac{L}{E}.
\ee
The function $W$ is Paczynski-Wiita  (P-W) potential, function $V_{eff}(r)$ is the  torus (fluid) effective potential \footnote{We can relate the fluid  effective potential  $V_{eff}^{\pm}(a;\ell,r)$ with the test particles circular orbits  potential   $V_{eff}^{\pm}(a;L,r)$   as
  %
%\bea&&\label{Eq:def-lvl}\nonumber
$
u_t^2=V_{eff}^{\pm}(a;\ell,r)^2={\mathcal{A}_{\mathrm{T}}}{\mathcal{A}^{-1}}=-{\mathcal{A}_{Q}^{-1}(g_{tt} u^t +g_{\phi t} u^{\phi})^2}={{E}^2\mathcal{A}_{\mathrm{T}}}{\mathcal{A}_{\mathrm{P}}^{-1}}
$ where
% %
$\mathcal{A}_{Q}^{-1}\equiv g_{tt} (u^t) ^2+2 g_{\phi t} u^tu^\phi +g_{\phi\phi}(u^\phi)^2$ and ${\mathcal{A}_{\mathrm{P}}^{-1}}\equiv {E}^2 g_{\phi\phi}+2 {E} g_{\phi t} L(\ell)+g_{tt}  L(\ell)^2$ and
where $
\mathcal{A}\equiv \ell ^2 g_{tt}+2 \ell  g_{t\phi}+g_{\phi \phi }=\ell^2(\mathcal{L}\cdot\mathcal{L})= \ell^2\left (g_{tt}+
     2\omega_{\ell} g_ {t\phi} +\omega_{\ell}^2 g_ {\phi\phi} \right)$,
here
$\omega_\ell\equiv 1/\ell$,  and $\mathcal{A}$   is related  to the norm of  stationary observers whose frequency is $\omega_{\ell}$ and $\mathcal{A}_{\mathrm{T}}$ is defined in Eq.\il(\ref{Eq:uf}). Light surfaces  have  $\mathcal{A}=\mathcal{L}\cdot \mathcal{L}=0$.
Note, that the effective potential of the fluids and  the particles,  as defined here in the hydrostatic model,  is related to the constant energy  $E$ of the  test particle motion and $u_t$. These  quantities could be negative in the ergoregion. The presence of (geodetical) circular motion with $E=0$ or $E<0$ in the ergoregion is a well known feature of the Kerr naked singularity solutions   which has also zamos, zero angular momentum, $L=0$  observers, which has not equivalent  in the ergoregion of the \textbf{BH} and it is  a distinctive feature of certain naked singularities with "small"  spin--mass ratios. This  characteristic has been  attributed to the repulsive gravity effects of the \textbf{NSs}.
We also note that this feature is capable to  define an effective ergoregion in the field of Reissner-Norstr\"om geometry (electrically charged and static metric). In the case we consider here circular orbit has $E>0$  and therefore effective potential is always positive. (The energy $E$ may be  negative as measured at infinity, locally it is always positive.) \citep{2013CQGra..30g5012S,2012CQGra..29f5002S}  Here we specify that $V_{eff}^2<0$ (not well defined) in some circumstances inside the ergoreigon, it is related to the normalization conditions on the fluid 4-momentum which is stationary (i.e. the associated fluid four momentum has $u^a= u^t+\omega u^\phi $, where $\omega=\Omega(\ell)$ is the relativistic angular velocity which satisfies  to the von Zeipel conditions) therefore well defined in the ergoregion.
There is  $V_{eff} (r)^2 < 0$ for $ r =
 r_ {\epsilon}^+$ and  $a/\in]0, a_{\gamma}$[ ($a_{\gamma}\equiv {M}/ {\sqrt {2}}$) with $r/M\in] 1,
  2[$ with $a\in[\sqrt {2 r(M -
      r)},\sqrt {(r -
        3M)^2 rM}/(2M)[$ which is a photon orbit in the ergoregion.}, reflecting the background Kerr geometry and the centrifugal effects through the fluid specific angular momentum $\ell$. The fluid equilibrium is therefore regulated by the balance of the gravitational and pressure terms versus centrifugal factors
arising due to the fluid rotation and the curvature effects of the Kerr background, encoded in the effective
potential function $V_{eff}$. Here $\mu=\{r,\theta\}$ and, for  large part of this analysis, we  shall consider  mainly   the radial $(r)$ gradients of pressure.
Assuming  the fluid  is   characterized by the specific  angular momentum  $\ell=$constant across the torus (see also \citet{Lei:2008ui,abrafra}),  we consider, from Eq.\il(\ref{Eq:scond-d})   the  critical points of the pressure in the torus as  the extremes of the effective potentials considering  the solutions   $V_{eff}=K=$constant, introducing  therefore the couple of tori parameters $(\ell, K)$.
The toroidal surface are  equipressure surfaces, therefore  equipotential surfaces  with  $V_{eff} = K =$constant, for a given $\ell=$constant
\footnote{This fully general relativistic model  of  pressure supported torus, traces back to the
Boyer theory of the equilibrium and rigidity in general relativity, i.e. the analytic theory of equilibrium configurations of
rotating perfect fluids \citep{Boy:1965:PCPS:,Raine}.
Each toroid is governed by the General Relativity hydrodynamic  condition of equilibrium configurations of rotating perfect fluids applied to toroidal surfaces.These correspond also
to the surfaces of constant density $\varrho$, specific angular momentum $\ell$, and constant relativistic angular frequency
$\Omega$, where $\Omega = \Omega(\ell)$ as a consequence of the von Zeipel theorem \citep{Chakrabarti0,Chakrabarti,Zanotti:2014haa}. In this model the entropy is constant along the flow  and the rotation
law  $\ell =\ell(\Omega)$ is independent of the equation of state \citep{abrafra}. In these tori in fact the functional
form of the angular momentum and entropy distribution during the evolution of dynamical processes, could be considered as
dependent only on the initial conditions of the system see \citet{abrafra}.  In this situation  the  surfaces of constant enthalpy, pressure and density coincide with the  surfaces of constant effective
potential $V_{eff}$, fixed by the integration parameter $ K $ related to the boundary conditions on the integration of  the Euler equation.
 The torus surfaces are surfaces of constant pressure  or $\Sigma_{i}=$constant for $i\in(p,\rho, \ell, \Omega)$, \citep{Raine}, where it is indeed $\Omega=\Omega(\ell)$ and $\Sigma_i=\Sigma_{j}$ for ${i, j}\in(p,\varrho, \ell, \Omega)$.}--\citep{abrafra,pugtot,ringed,PuMonBe12}  and for example \citet{M.A.Abramowicz,Japan,Abramowicz:1996ap,adamek,FisM76,Font:2002bi,Raine,Komissarov:2006nz,Lei:2008ui,Montero:2007tc,EPL,Fi-Ringed,open,dsystem,Multy,long,letter,PS21,Zanotti:2014haa}.

 According to Eq.\il(\ref{Eq:scond-d}), tori centers (maximum pressure points) and cusps (minimum pressure points),  are    the  critical points of  $V_{eff}(r)$  as function of the radius $r$. The maximum  of the hydrostatic pressure corresponds to the minimum of the  effective potential $V_{eff}$, and it is the \emph{torus center}   $r_{center}$.   The instability points of the tori (P-W mechanics), are located at the minima of the pressure, correspondent to the  maximum of $V_{eff}$ as functions of $r$--see Figs\il(\ref{Figs:PlotTRialfull}).

\begin{figure}\centering
  % Requires \usepackage{graphicx}
  \includegraphics[width=5.58cm]{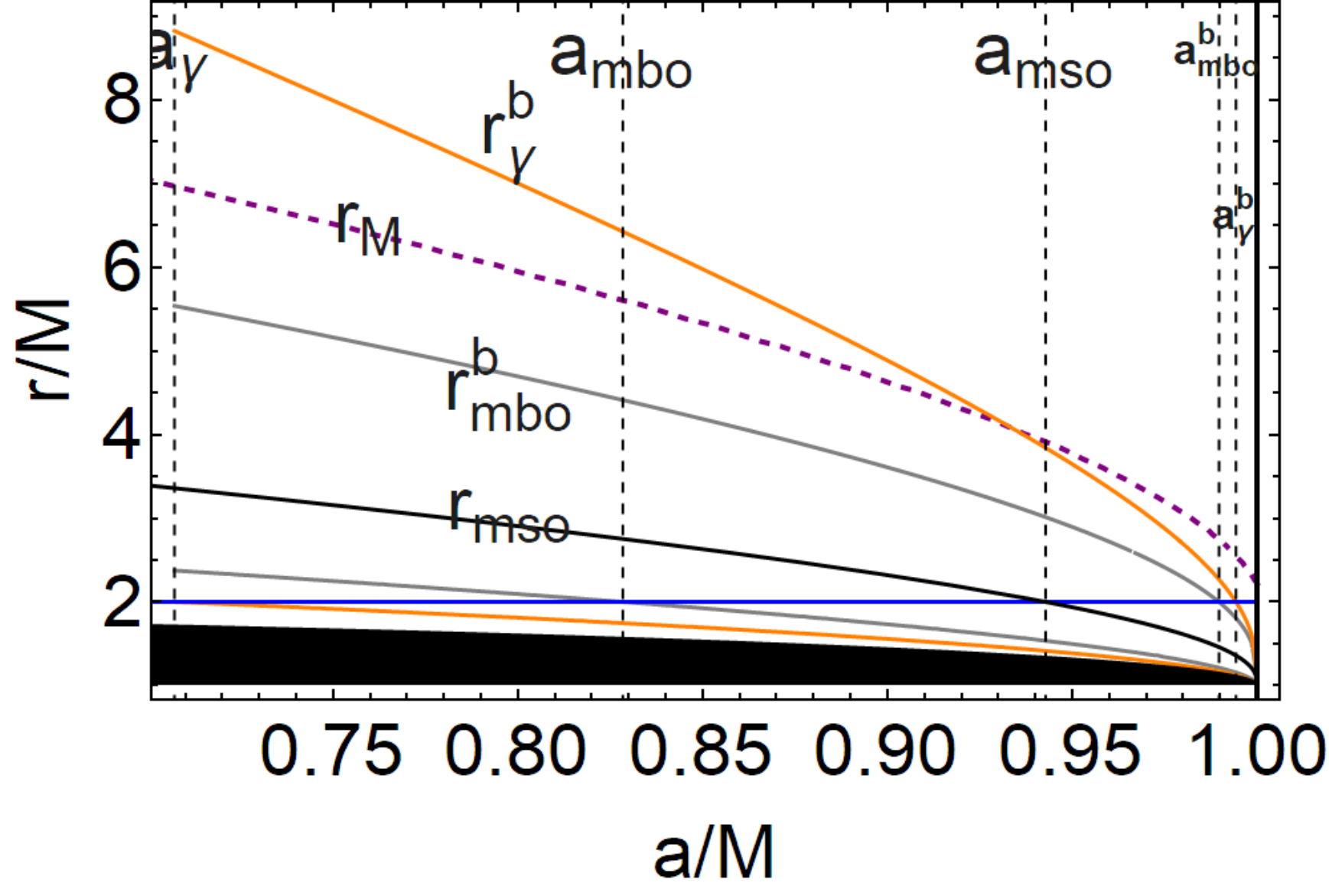}
  \includegraphics[width=5.58cm]{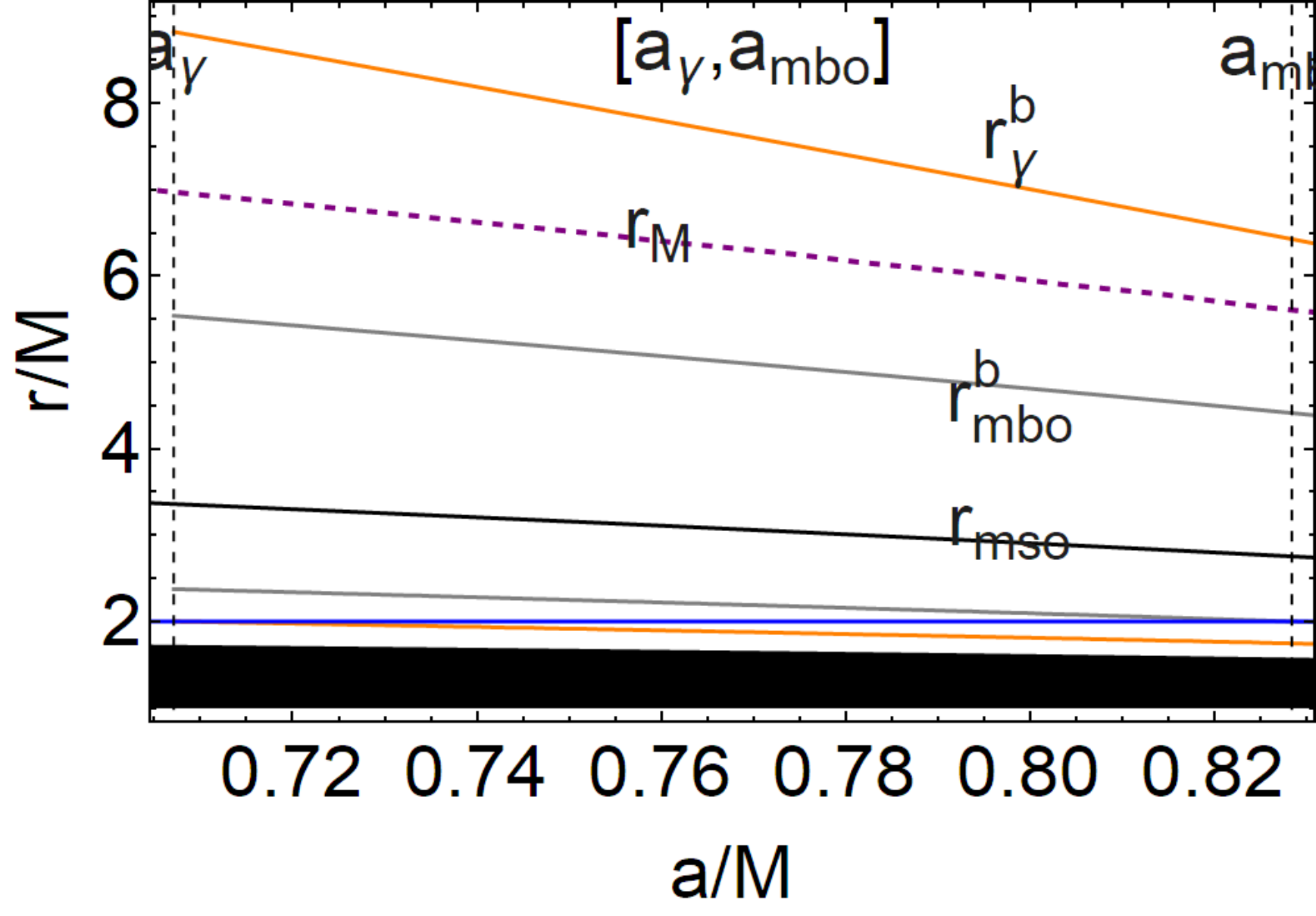}
  \includegraphics[width=5.58cm]{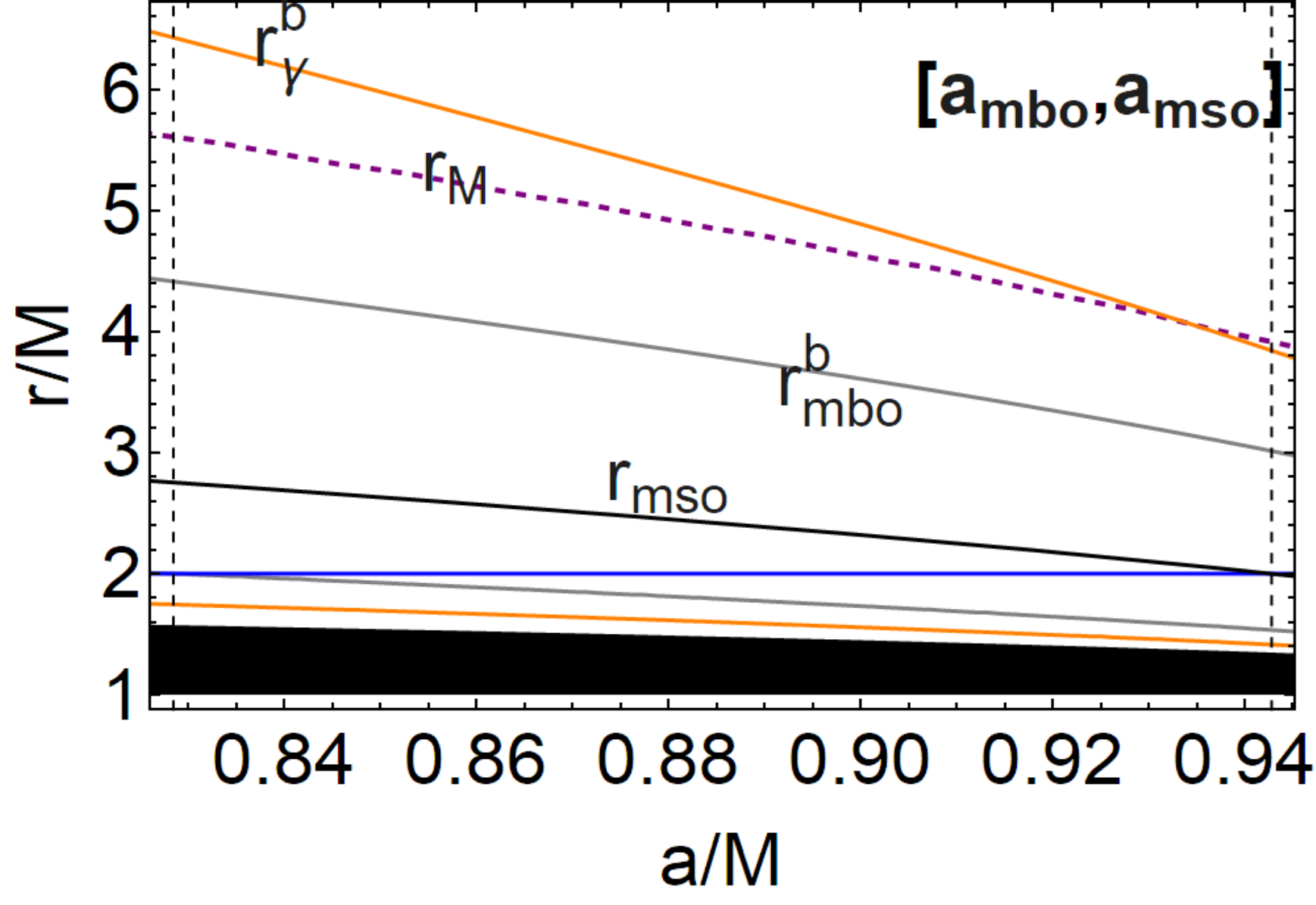}\\
  \includegraphics[width=5.58cm]{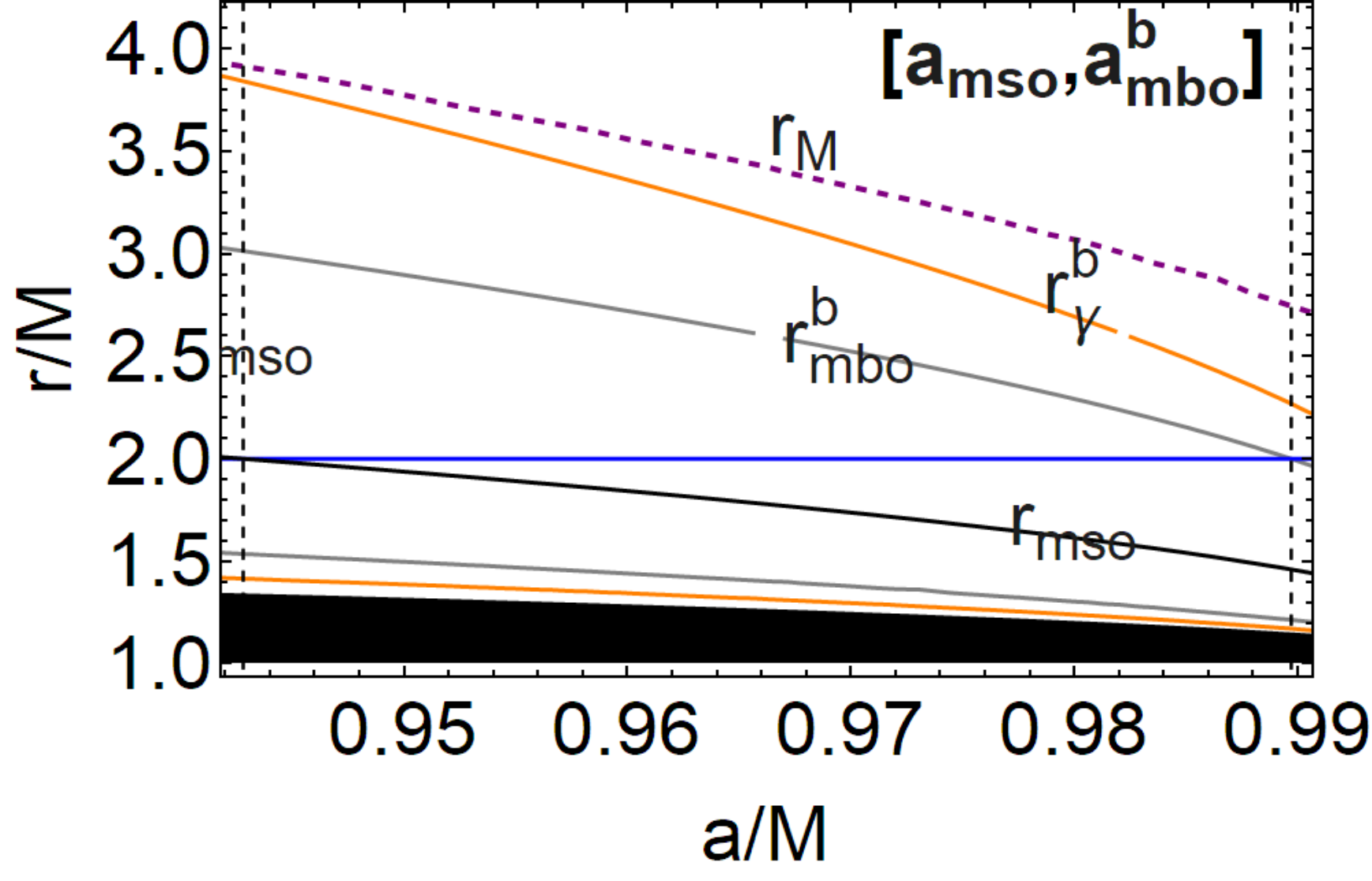}
  \includegraphics[width=5.58cm]{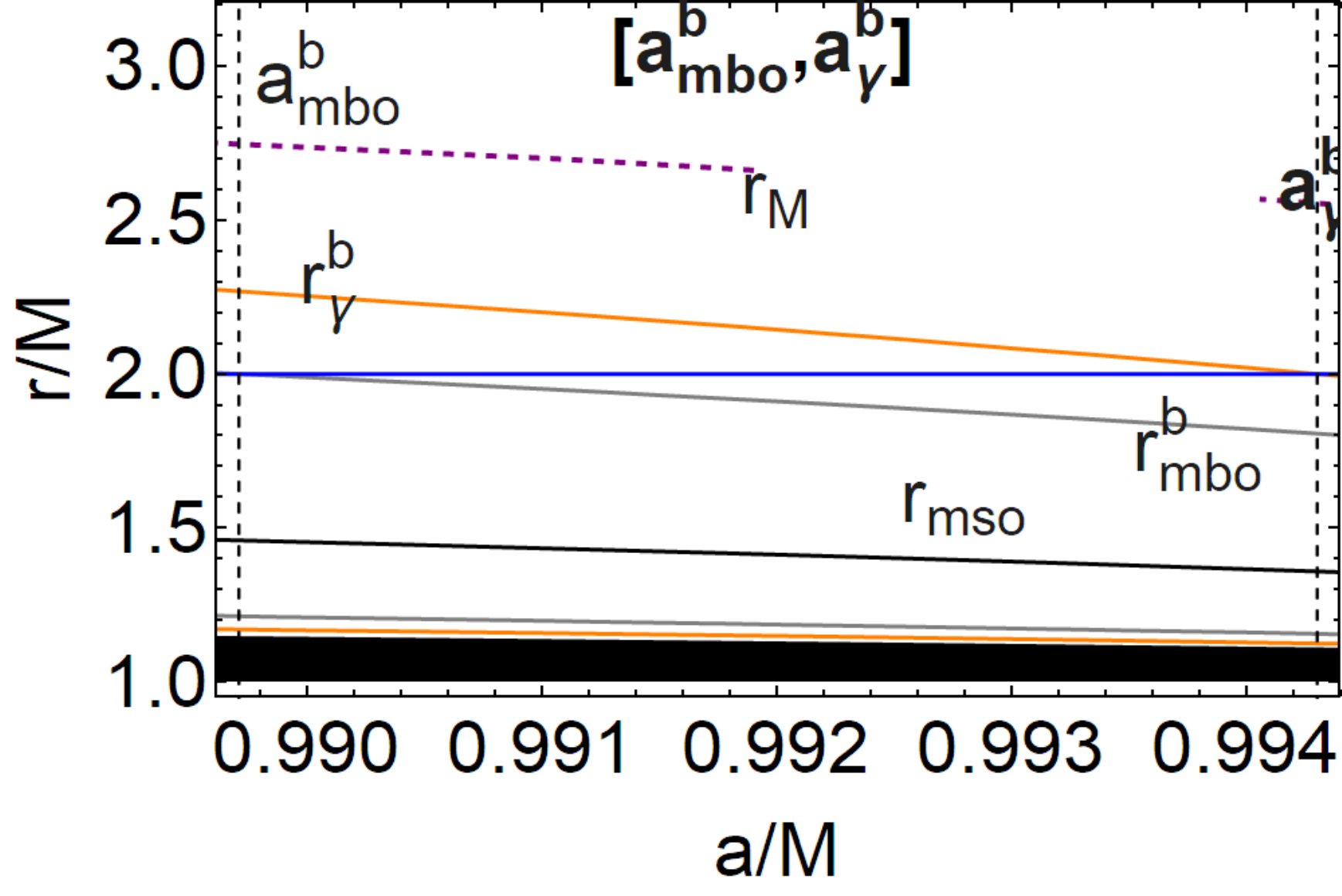}
  \includegraphics[width=5.58cm]{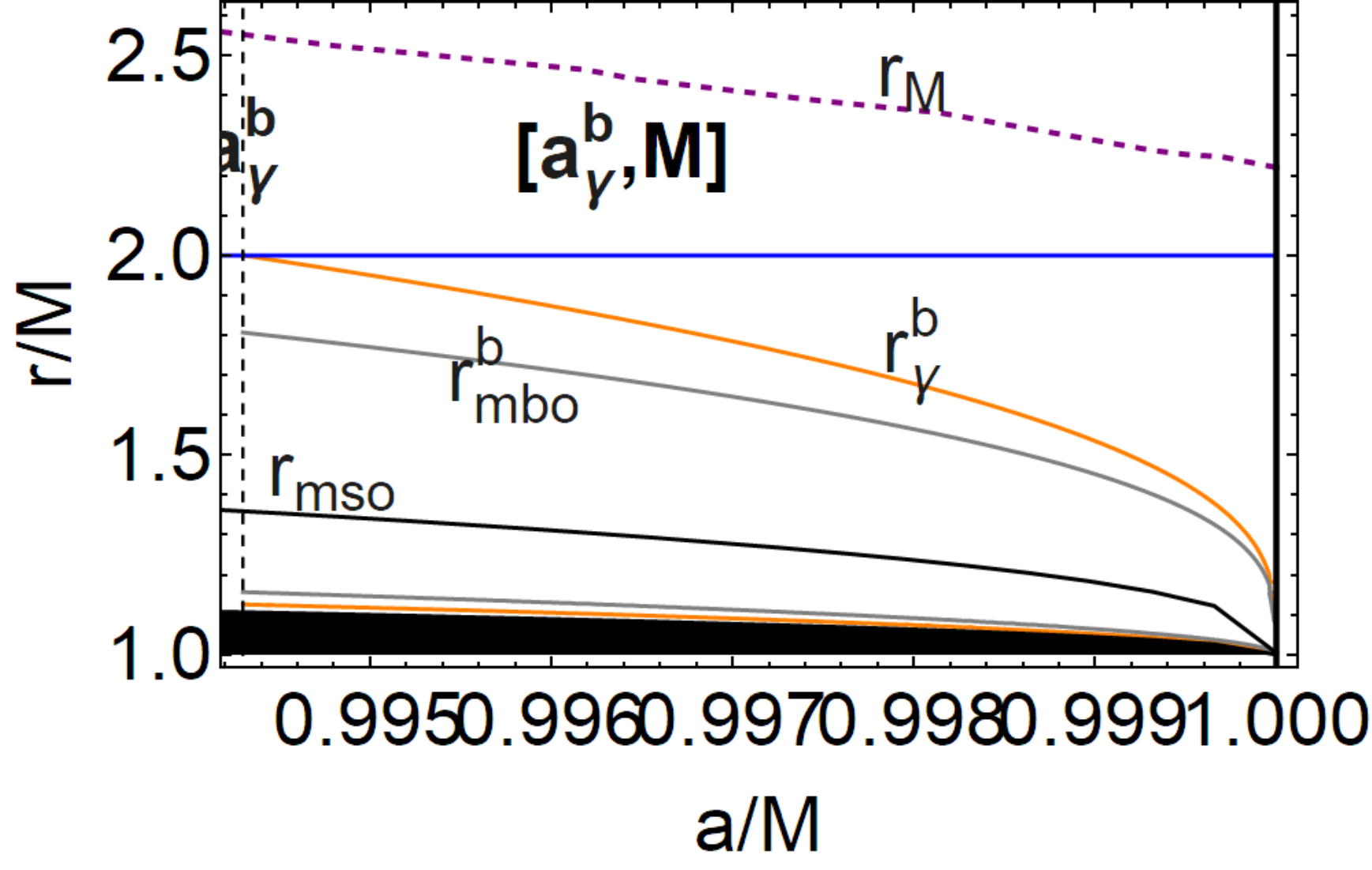}
  \caption{Notable radii of the geodetic structure  are represented as function of the \textbf{BH} dimensionless spin. $r_{\gamma}$ is the photon orbit, $r_{mbo}$ the marginally bounded orbit, $r_{mso}$ marginally stable orbit.   Radius $r_{\Mie}: \partial^2_r \ell(r)=0$ is dashed purple line. Black region is the \textbf{BH} at $r<r_+$ (outer horizon).Center of quiescent $\cc_3$ configuration with specific momentum in $\ell>\ell_{\gamma}$ are in $r>r^b_{\gamma}$. Spins   $\mathbf{A}_{\epsilon}^+\equiv \{a_{\gamma},a_{mbo},a_{mso},a_{mbo}^b,a_{\gamma}^b\}$ of Eqs\il(\ref{Eq:strateg}) are dashed lines, see discussion  in Sec.\il(\ref{Sec:deta-hall}). Upper left panel shows the entire range $a\in [0,M]$, while other panels show the ranges \textbf{Range $\mathbf{A}_1\equiv[a_{\gamma},a_{mbo}]$};
\textbf{Range $\mathbf{A}_2\equiv[a_{mbo},a_{mso}]$};
\textbf{Range $\mathbf{A}_3\equiv[a_{mso},a_{mbo}^b]$};
\textbf{Range $\mathbf{A}_4\equiv [a_{mbo}^b, a_{\gamma}^b]$};
\textbf{Range $\mathbf{A}_5\equiv[a_{\gamma}^b,M]$}. }\label{Fig:PlotVampb1}
\end{figure}

More generally, equation  $\partial_r V_{eff}=0$   can be  solved for the fluid specific angular momentum  obtaining the function $\ell(r)$  "leading function".
We also introduce the  function
$ K(r)$ ("energy function") which, as
function $\ell(r)$, is  also dependent on the \textbf{BH} dimensionless spin,  and are  defined as:
\bea\label{Eq:leading-energy}
\ell^{\pm}(r):\;\partial_r V_{eff}=0,\quad  K^{\pm}(r)\equiv V_{eff}(r,\ell^{\pm}(r)),
\eea
   for  counterrotating $(+)$ and corotating $(-)$ fluids respectively.
   Functions $\ell(r)$ and $K(r)$ are showed in connection with the tori surfaces in Figs\il(\ref{Figs:PlotTRialfull}).
 Functions $(\ell(r),K(r))$  define the   hydrostatic pressure critical points  $r_{crit}$  in the torus, centers of maximum pressure $r_{center}=r_{crit}>r_{mso}^{\pm}$ or cusps $r_{cusp}=r_{crit}<r_{mso}^{\pm}$.
Curves $K^{\pm}(r)$ locate the tori centers and provide information on torus elongation and  density and,  for a  torus accreting onto the central \textbf{BH}, determine the inner,  $r_{inner}$, and outer, $r_{outer}$, torus edges, where the elongation on the equatorial plane is $\lambda\equiv r_{outer}-r_{inner}$--Figs\il(\ref{Figs:PlotTRialfull}) and Figs\il(\ref{Fig:PlotVampb1}).
 The  couple of  constant parameters $(\ell,K)$ uniquely identifies each toroidal  surface and these can be directly reduced to a single parameter   $\ell$,  in presence of a cusp. (The critical points of the pressure are determined by the $\ell$ parameter only, therefore in the case of cusped tori, the cusp fixes uniquely also the $K$-parameter, coincident with the maximum value of the torus effective potential--Figs\il(\ref{Figs:PlotTRialfull})).

More specifically,  solutions of the problem $V_{eff}=K=$constant  lead to  three  classes of configurations  corresponding  to  --{$\cc $}--cross sections of the{ closed}  surfaces (equilibrium quiescent torus); $\cc_{\times}$--cross sections of the {closed cusped}   surfaces (accreting torus)--
{${J_\times}$}--cross sections of the {open cusped}   surfaces, generally associated to proto-jet configurations.

 The
closed, not cusped,   $\cc$  surfaces are  associated to  stationary equilibrium (quiescent)  toroidal configurations.
 For  the  cusped and   closed equipotential surfaces,  $\cc_{\times}$,  the accretion onto the central black hole can occur through the
cusp $r_{\times}$ of the equipotential surface. In this situation  the outflow of matter through the cusp occurs as the torus  surface exceeds
the critical equipotential surface (having a cusp), leading to a mechanical
non-equilibrium process,  i.e. an instability in the balance
of the gravitational and inertial forces and the pressure gradients,  where  matter inflows  into the central  black hole (a violation of the hydrostatic equilibrium known as Paczy\'nski mechanism).  In the case considered here the accreting flow "starts" across a "Lagrange
point" (Roche lobe overflow, due to Paczynski accretion mechanics \citep{Pac-Wii,cc}. This is the
cusp of the orbiting toroidal surface, which is an important aspect of thick disks since its presence also stabilizes the tori
against several instabilities (thermal and viscous local instabilities, and globally against the Papaloizou-Pringle instability-PPI
and it could be possibly connected to QPOs emission.) \citep{abrafra}.

Therefore, in this
accretion model we shortly indicate the  cusp of the  self-crossed closed toroidal surface as  the "inner edge of accreting torus".
Finally, the open equipotential surfaces, which are briefly considered in Sec.\il(\ref{Sec:gir-c2}), have been associated to  the formation of  proto-jets \citep{open,proto-jet,long}.

\begin{figure}\centering
  % Requires \usepackage{graphicx}
    \includegraphics[width=8.4cm]{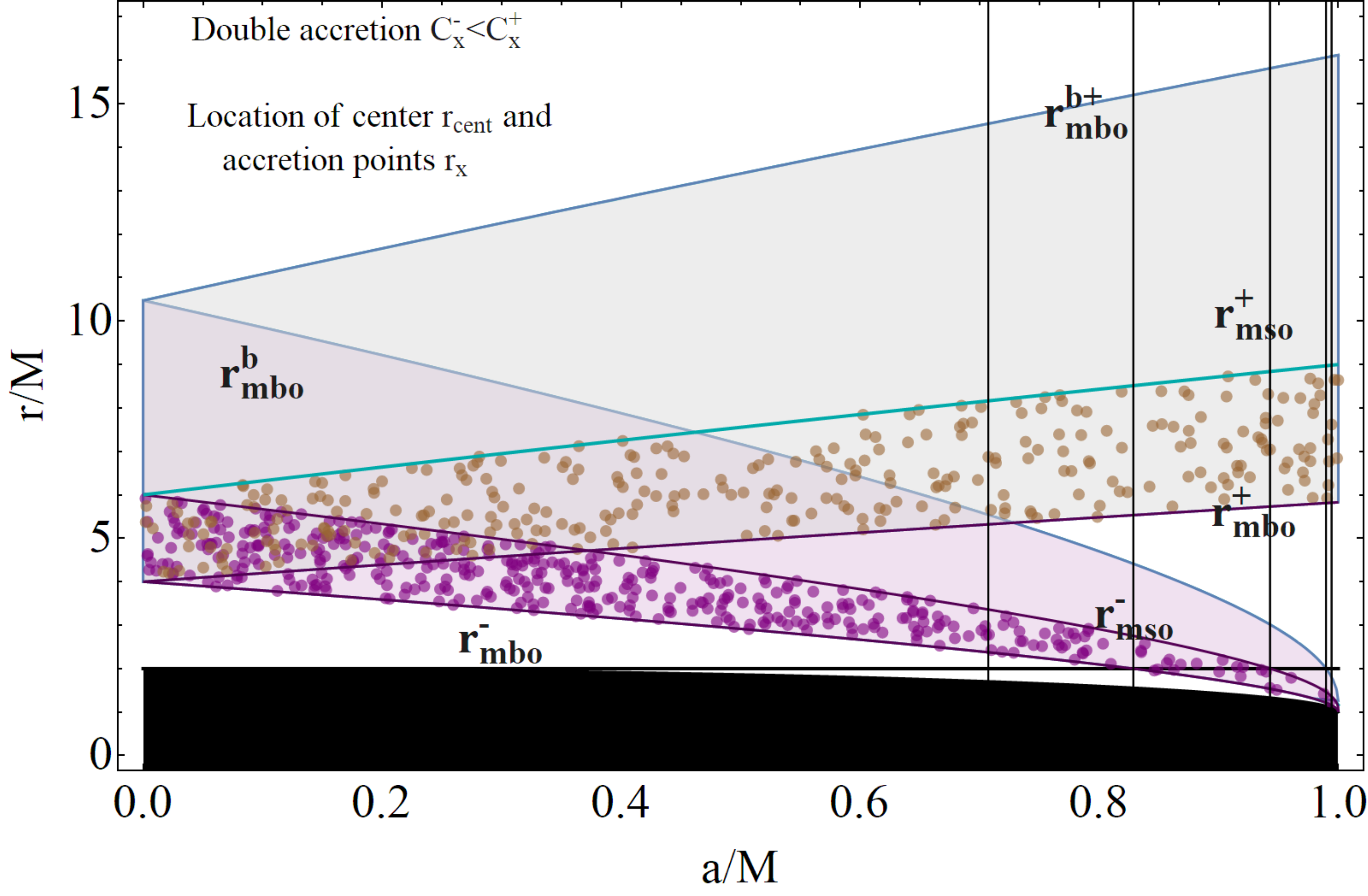}
  \includegraphics[width=8.4cm]{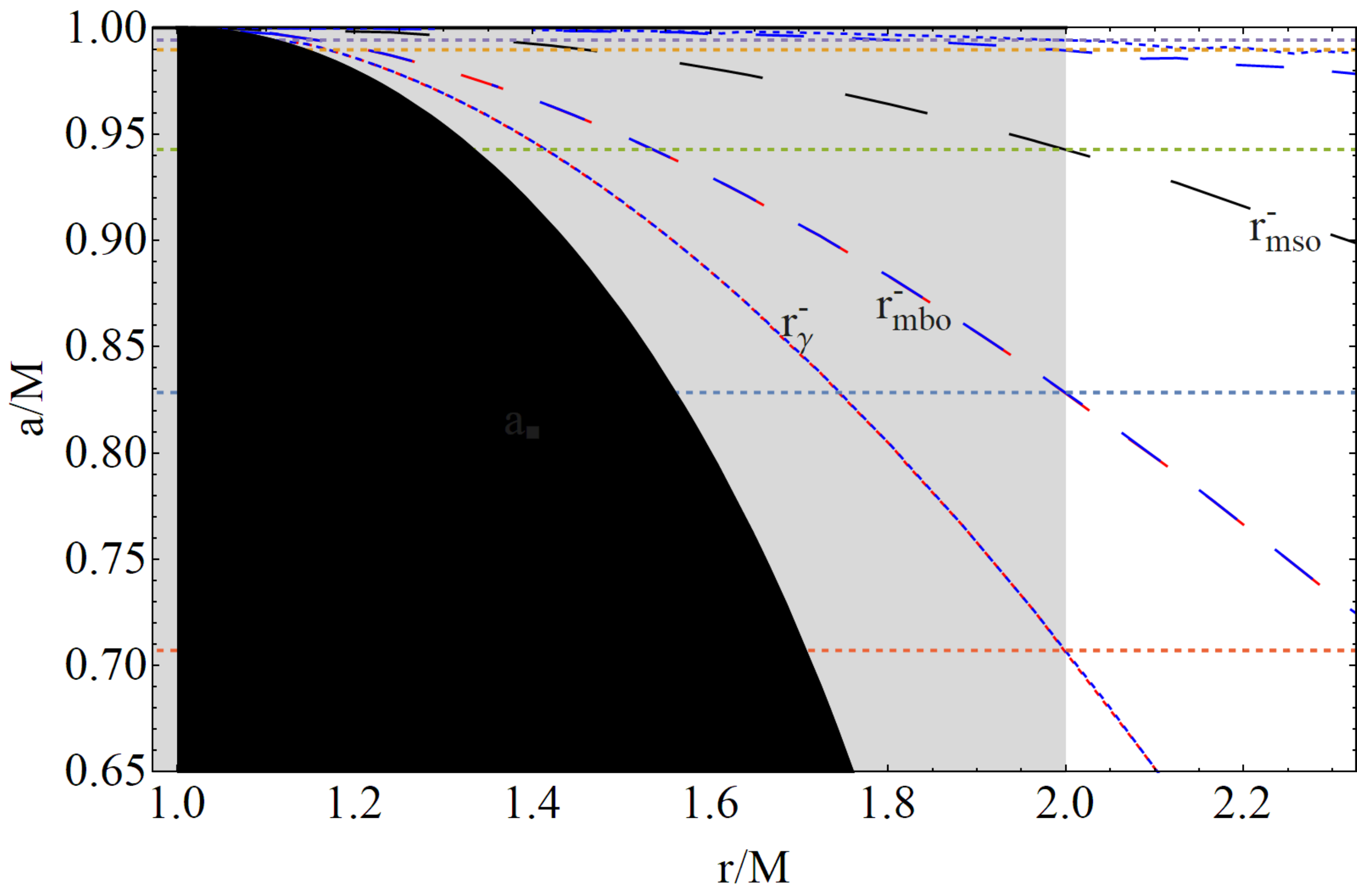}
  \caption{Existence condition for accreting tori an  double accretion (left panel), location of notable radii of the geodetic structure (right panel). Black region is $r<r_+$, $r_+$ is the outer horizon as function of the \textbf{BH} spin $a/M$. Left panel: description of double accretion $\cc_{\times}^-<\cc_{\times}^+$, (inner corotating cusped closed  torus and outer counter-rotating closed cusped torus) location of tori centers $r_{center}^{\pm}$ (dotted strips) and accretion points  $r_{\times}^{\pm}$ (shaded regions) for corotating $(-)$ and counterrotating $(+)$ fluids. $r_{mbo}$ is the marginally bounded orbits  $r_{mbo}^b: \ell(r_{mbo})=\ell(r))$, $r_{\gamma}$ is the photon orbit, $r_{mso}$ is the marginally stable orbit. Black lines are  $\mathbf{A}_{\epsilon}^+\equiv\{a_{mso},a_{mbo},a_{msbo}^b,a_{\gamma},a_{\gamma}^b\}$    defined in Figs\il(\ref{Fig:PlotVampb1}).
  Right panel: gray region is the outer ergosurface $r_{\epsilon}^+=2M$ dotted lines are the spins    $\mathbf{A}_{\epsilon}^+$. }\label{Fig:corso}
\end{figure}

 The minimum  points of the hydrostatic pressure are constrained   by  the  {geodesic structure} of the  Kerr geometry.
Alongside the geodesic structure of the Kerr spacetime,  we include  radii $r_{\mathcal{M}}^{\pm}$  solution  of  $\partial_r^2\ell=0$,
and    radii $r_{mbo}^{b\pm}$ and $r_{\gamma}^{b\pm}$  or more generally $(r_{mbo}^{b\pm}, r_{\gamma}^{b\pm},r_{\mathcal{M}}^{b\pm})$, regulating   the maximum points of the pressure,  and
defined as the solutions of  the  following equations
\bea&&\label{Eq:conveng-defini}
{r}_{{mbo}}^{b\pm}:\;\ell^{\pm}(r_{mbo}^{\pm})=
 \ell^{\pm}({r}_{{mbo}}^{b\pm})\equiv {\ell_{mbo}^{\pm}},
\quad
  r_{\gamma}^{b\pm}: \ell^{\pm}(r_{{\gamma}}^{\pm})=
  \ell^{\pm}(r_{\gamma}^{b\pm})\equiv {\ell_{{\gamma}}^{\pm}}, \quad\mbox{and}
 \\&&\nonumber
r_{\mathcal{M}}^{b\pm}: \ell^{\pm}(r_{\mathcal{M}}^{b\pm})= \ell_{\mathcal{M}}^{\pm}\quad\mbox{where}
 \\ &&\label{Eq:pre-Nob}
r_{\gamma}^{\pm}<r_{mbo}^{\pm}<r_{mso}^{\pm}<
 {r}_{mbo}^{b\pm}<
 r_{\gamma}^{b\pm},
\eea
 respectively for the counterrotating $(+)$ and corotating $(-)$ orbits--Figs\il(\ref{Fig:corso}). %
 Radii $\{r^{\pm}_{mso}, r^{\pm}_{mbo},r^{\pm}_{\gamma}\}$,  $\{ r^{b\pm}_{mbo}, r^{b\pm}_{\gamma}\}$ and $\{r^{\pm}_{\Mie}, r^{b\pm}_{\Mie}\}$ constitute the  \emph{extended} geodetic structure of the spacetime.  (Location of radii
  $\{r^{\pm}_{\Mie}, r^{b\pm}_{\Mie}\}$  with respect to the  radii Eqs\il(\ref{Eq:pre-Nob}) depend on the \textbf{BH} spin).

 We summarize the constraints, defining ranges  of fluids specific angular momentum $(\mathbf{L_1,L_2,L_3})$ defining the closed quiescent, cusped or open configurations and the tori location  (cusp and tori centers) with respect to the central \textbf{BH}:

\begin{description}
\item[--Tori $(\cc_1, \cc_{\times})$]
{For $\ell\in \mathbf{L_1}\equiv[ \ell_{mso},\ell_{mbo}[$ there are quiescent  and cusped tori},
  with   topologies $(\cc_1, \cc_{\times})$;  accretion point  is  $r_{\times}\in]r_{mbo},r_{mso}]$ and center with maximum pressure is  $r_{center}\in]r_{mso},r_{mbo}^b]$;
\item[--Tori $(\cc_2, J_{\times})$]
%\hline
For $\ell\in \mathbf{L_2}\equiv[ \ell_{mbo},\ell_{\gamma}[$ there are  quiescent  tori and proto-jets,
with  topologies    $(\cc_2, J_{\times})$;   unstable point  is on  $r_{j}\in]r_{\gamma},r_{mbo}]$  and  center with maximum pressure $r_{center}\in]r_{mbo}^b,r_{\gamma}^b]$;
\item[--Tori $\cc_3$]
%\hline

For $\ell\in \mathbf{L_3}: \ell \geq\mp\ell_{\gamma}$ there are quiescent  tori
  $\cc_3$  with center $r_{center}>r_{\gamma}^b$, holding respectively for corotating and counterrotating tori,
\end{description}
Let $(K_{center}, K_{\times})$ be function  $K(r)$ evaluated at $r_{center}$  and $r_\times$ respectively. Closed, not cusped configurations $\cc$ are for $K\in [K_{center},K_{*}]$ where $K_*\in \{K_{\times},1\}$ according to the momentum $\ell\in \mathbf{L_1}$ or $\ell\in\{\mathbf{L_2},\mathbf{L_3}\}$ respectively, cusped configurations $\cc_{\times}$ are for $K=K_{\times}$--see Figs\il(\ref{Fig:corso}).
{As we study the configurations inside the ergoregion, we consider mainly corotating fluids,  therefore notation $(-)$ to indicate the corotating tori  will be generally omitted.}
For any quantity $\mathcal{Q}$ evaluated  on  a general radius $r_{\bullet}$, we  adopt notation $\mathcal{Q}_{\bullet}\equiv \mathcal{Q}(r_{\bullet})$
We also  adopt the notation $\Qa_{\times}$ or $\Qa^{\times}$  for any quantity $\Qa$  relative to the cusped torus  $\cc_{\times}$. Where appropriate,  to easy the reading of   complex expressions, we will  use  the  geometric units where $r\rightarrow r/M$  and $a\rightarrow a/M$. %\btb{(**********)}

\section{In the outer ergoregion}
Here we characterize the orbiting  tori in the outer  ergoregion or close to the static limit.
In section (\ref{Sec:deta-hall}) we introduce the concept of dragged surfaces and partially contained surfaces which are analyzed  in Sec.\il(\ref{Sec:dv}) particularly with the  investigation of the
the tori verticality and  introducing the disk exfoliation for dragged tori.
The influence of the dragging frame on the disk thickness is the focus of Sec.\il(\ref{Sec:influ-ergosra}),
while in  Sec.\il(\ref{Sec:mid-w-t}) we deepen the discussion on the
tori exfoliation.
We conclude this section  in Sec.\il(\ref{Sec:gir-c2}),  with  some notes on limiting
 configurations  as the possibility of multi-tori (aggregates of multi-toroids) in the  ergoregion,  the proto-jets and  $\cc_2$ tori.
\subsection{Dragged surfaces}\label{Sec:deta-hall}
We explore    two types of  closed configurations in  $\Sigma_{\epsilon}^+$:
\begin{description}
\item[
 \textbf{{{(1)}}}]The closed quiescent  configurations $\cc$;
\item[
 \textbf{{(2)}}]  The closed  tori $\cc_{\times}$ with cusps $r_{\times}$, having  specific angular momentum $\ell\in \mathbf{L_1}$.
 \end{description}
In Sec.\il(\ref{Sec:gir-c2}) we  consider also  the proto-jets, open toroidal configurations, in the ergoregion.

As discussed in \citet{pugtot}, we can consider an evolutive parameter following the  configurations evolution, from  the  phase of disk  formation, assuming a simplified two-phase accretion tori evolution,  to the  accretion phase. Naturally one can consider the  parameter  $\ell$ evolution  with a negative "time-gradient",   where    $\ell>\ell_{mso}$  decreasing in  region
  $\ell \in [\ell_{mso},\ell_{mbo}] $,  with  the consequent evolution of $K$ to  $K_{\times}$ and  to $K_{*}\in[K_{\times},1[$  with an "over-critical" toroid  where  there is  formation of an  accretion throat,  corresponding to  the  torus  cusp enlargement and  associated to  matter flowing-- Figs\il(\ref{Figs:PlotTRialfull}).
Cusped tori  $\cc_{\times}$ are considered  limiting configurations for the  accretion phases.  The disk  may also be subsequently stabilized (in quiescent tori)  to a sequence of interrupted phases of super-Eddington accretion as often considered in models of \textbf{SMBHs} evolution \citep{apite1,apite2,apite3,Volonteri:2002vz,Li:2012ts,Oka2017,Kawa,Allen:2006mh}, occurring, for example, in composed systems of tori aggregates as  the  \textbf{eRAD} \citep{ringed,dsystem,letter}.  A further  possibility consists in the   formation of an accretion throat, for over critical tori, with  $K>K_{\times}$, and formation of a fat torus  with accretion material very close to the horizon.

 We  focus specifically  on  the following two types of configurations:

\begin{description}
\item[
 \textbf{1. \emph{Dragged surfaces}}]

 Dragged surfaces are closed  quiescent,  $\cc$, or closed cusped $\cc_{\times}$ tori  entirely  contained in the ergoregion. This definition considers the   tori radial dimension,  i.e.,   we assume   the  torus inner, $r_{inner}$, and outer  edge  $r_{outer}$,  be   both located in  the range  $]r_+,r_{\epsilon}^+]$.
 This  condition however  does not imply  the  torus entire   {containment}  in $\Sigma_{\epsilon}^+$ at any plane $\theta\neq\pi/2$. Considering   the disk vertical dimension,  tori  defined by constant fixed $K$ may cross the ergosurface at a plane  $\theta\neq\pi/2$, including the  tori geometrical maximum which is  related to the maximum pressure point.
We focus the analysis of the disks verticality in  $\Sigma_{\epsilon}^+$  considering the  ergosurface crossing at any $\theta$    in Figs\il(\ref{Fig:polodefin1},\ref{Fig:weirplot},\ref{Fig:gatplot8},\ref{Fig:gatplot17},\ref{Fig:gatplot5}) and Sec.\il(\ref{Sec:dv}).

\item[
\textbf{2. \emph{Partially contained configurations}}]

The partially contained tori  are characterized  by the conditions $r_{inner}<r_{\epsilon}^+$ and $r_{outer}>r_{\epsilon}^+$.  This case  has  the following  two sub-cases, which can also be  associated to different phases of the   tori evolution inside  the ergoregion: \textbf{(a)} In the first sub-class of partially contained tori,  the \emph{disk inner part}, region $[r_{inner},r_{center}]$, is fully contained in the ergoregion, which may include also  a section   the \textit{torus outer part}, $]r_{center},r_{outer}]$. \textbf{(b) } In the second sub-class of tori,   only a section of the torus  inner part  is  contained in the ergoregion (that is a region $[r_{inner},r_{*}]$ with $r_*<r_{center}$. This  condition implies  that the projection  on the equatorial plane of the maximum geometrical point $r_{\max}>r_{center}$   is out of the ergoregion.
(In the following we consider, when not otherwise stated, $r_{\max}$ the projection of the geometrical maximum of the surface on the equatorial plane.).
\end{description}
 The   tori stability     in $\Sigma_{\epsilon}^+$ is an  important issue  to be assessed, that can  involve
a series of hypothesis  on the dragged and partially contained surfaces formation. Tori evolution may develop  from a configuration formed outside the ergoregion, then evolving   in  partially contained  and from partially contained  to dragged surfaces, featuring, therefore,    disk formation  in three  stages.   This hypothesis  can  be grounded on the assessment of  appropriate details  of  disk formation, as  growing  from accreting matter from the  embedding,  and  conditions on the accretion process, for example in the case where  the central \textbf{SMBH} is a part of a binary system  having  an accreting companion  star.
A second possibility  consists in   the formation of a torus  in the ergoregion crossing the ergosuface from the outer part of the torus. (Note that  there is $r_{inner}<r_{center}<r_{\max}<r_{outer}$,  therefore  there is a  first phase when   $r_{inner}$  crosses  the ergosurface,  implying  also the torus crossing   the ergosurface  on  planes different from the equatorial plane).
Finally, a third hypothesis  features a disk formed inside the ergoregion,  growing   in the ergoregion  and accreting into the central attractor. The problem of  disk  approaching and crossing the static limit  or the formation of the closed quiescent    torus  across the static limit is also faced by the analysis of the   disk vertical dimension,  rephrasing the problem for a torus  crossing  the ergoregion on different  planes.
Following the torus growing, and   the accreting  disk  stretching    on the equatorial plane, the torus outer edge  may cross the static limit, leading the initial  dragged surface to a subsequent  partially contained configuration.
It may be possible that the dragged configurations could  appear as   seeds, formed in-loco, on  a range centered on the orbit $r_{center}\in ]r_+,r_{\epsilon}^+[$. This situations can be also largely affected by the details of  fluids characteristics  as the equations of state or differentiated in  more complicated frames as the GRMHD models, we consider  this issue briefly in Sec.\il(\ref{Sec:poly-altr}) and Appendix\il(\ref{Sec:magn}). (However as we note from the Euler equation in the GRHD model, the  "pressure-forces"  are entirely regulated  by the  fluid effective potential gradient, and possibly the fluid relativistic angular velocity  and the gradient of the  specific angular momentum for model with $\ell\neq$constant, while it is independent from the details of the  polytropics).

The disks embedding is another relevant  aspect of the dragged configurations,  i.e.,   tori could be formed  following  accretion in a binary system  or  they can result  also as the consequence, for instance, of the Bardeen--Petterson effect on an originally  misaligned torus,  broken due to the  frame dragging and  other factors,   as the fluids viscosity, in an inner  corotating torus  and an outer torus which may also  be counter-rotating \citep{BP,2015MNRAS.448.1526N}.
A second possibility can also  consider   the Bardeen-Petterson  effect  on an original, partially contained,  torus-\citep{2015MNRAS.448.1526N,Martin:2014wja,King:2018mgw,2012ApJ...757L..24N,2012MNRAS.422.2547N,2015MNRAS.448.1526N,
 2006MNRAS.368.1196L,Feiler,King2005}.
Vertical stresses in the disks, and the polar gradient of  pressure, can combine with the  Lense-Thirring  effect.
Consequentially, a relevant issue   of these processes for  the stability of the partially contained configurations, is whether   the inner or outer  part of the disk  is in the ergoregion and,  the case of  crossing of the outer ergosurface on  planes different from the equatorial plane, which  may affect the torus  (global) stability when   stretching towards the horizon during  accretion.
For this reason we consider more in details the  torus inner and outer edges location  with respect to the static limit.
\begin{description}
\item[\textbf{--The inner edge}]
The  inner edge of tori with specific angular momentum  in  $\mathbf{L_1}$  satisfies the relation  $r_{inner}\geq r_{\times}\in[r_{mbo},r_{mso}]$ with  $r_{center}\in [r_{mso},r^b_{mbo}]$.
  For any dragged or partially contained  configuration, we concentrate on  the problem of penetration  of the inner edge of the tori in the outer ergoregion, assuming  firstly $\ell\in [\ell_{mso},\ell_{mbo}]$, and    considering  cusped tori $\cc_{\times}$.  A necessary but not sufficient condition for partially contained or dragged surfaces,  is the occurrence of
  $r_{mbo}\leq 2M$, distinguishing    therefore two classes of \textbf{BH}  attractors, which we consider more in details in Figs\il(\ref{Fig:PlotVampb1}).
  The value  $r_{mbo}=2M$ is  a limiting condition for configurations approaching  the ergoregion,   being  very close to the outer ergosurface.
 Condition  $r_{mso}=2M$   is a  sufficient but not necessary condition for a partially contained torus (a section of the
 torus inner part being contained). It is clear  that the necessary but not sufficient condition for the torus to
be  centered in the ergoregion, i.e.  the maximum pressure point
would be inside the ergoregion, is that  $r_
{mso}\in \Sigma_ {\epsilon}^+$.
A sufficient but not necessary condition for the  torus center to be included in the ergoregion
is   $r_
{mbo}^b\in \Sigma_ {\epsilon}^+$.
The analysis of the limiting situation {$r_{center}=r_{\epsilon}^+$} is faced in   Figs\il(\ref{Fig:principmill}) solving the equivalent problem $\ell_{\epsilon}^+\in ]\ell_{mso},\ell_{mbo}[$ for a given spacetime $a/M$, where $\ell_{\epsilon}^+\equiv \ell(r_{\epsilon}^+)$. %
\begin{figure}\centering
  % Requires \usepackage{graphicx}
  \includegraphics[width=5.5cm]{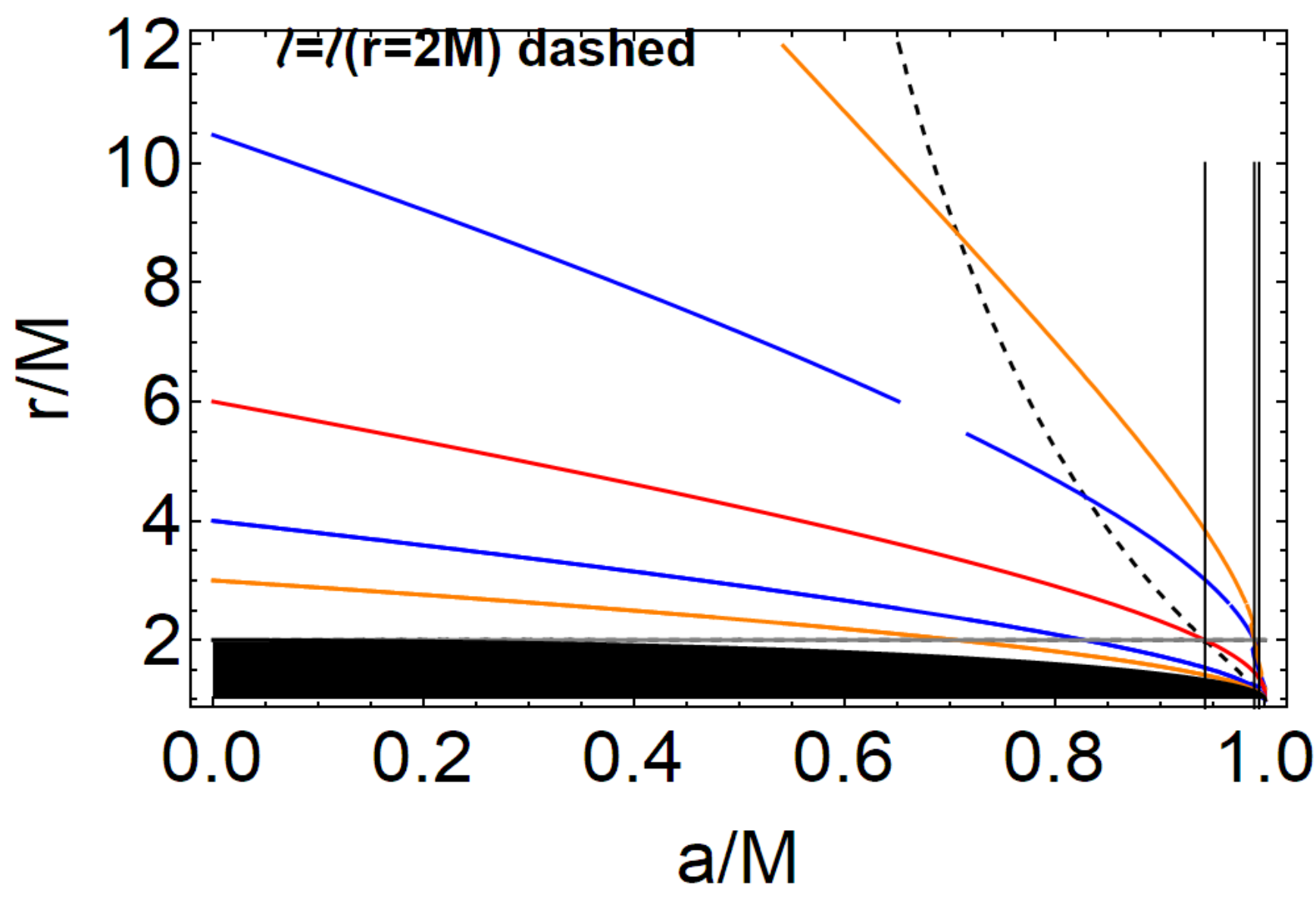}
  \includegraphics[width=5.5cm]{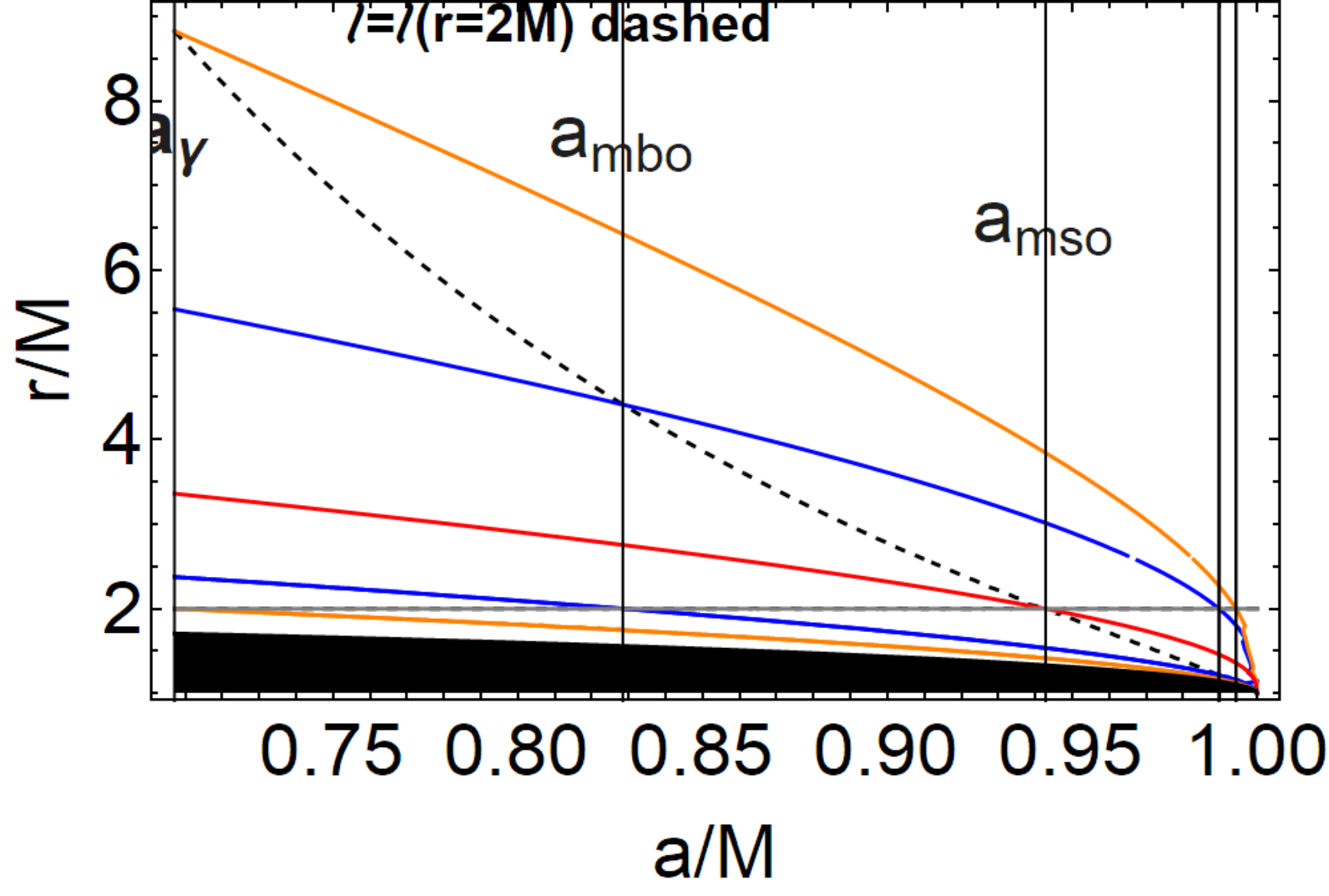}\\
  \includegraphics[width=5.5cm]{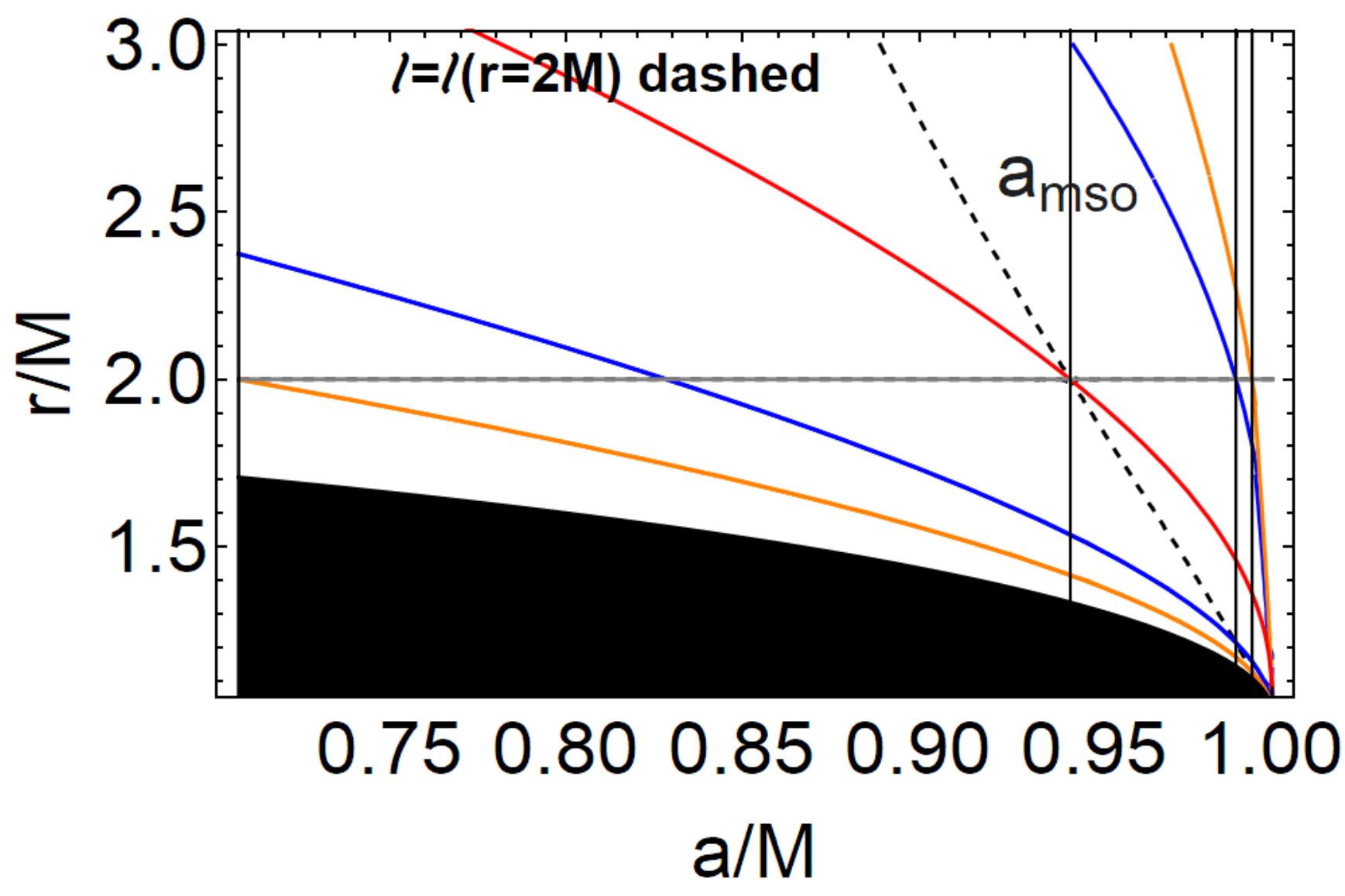}
  \includegraphics[width=5.5cm]{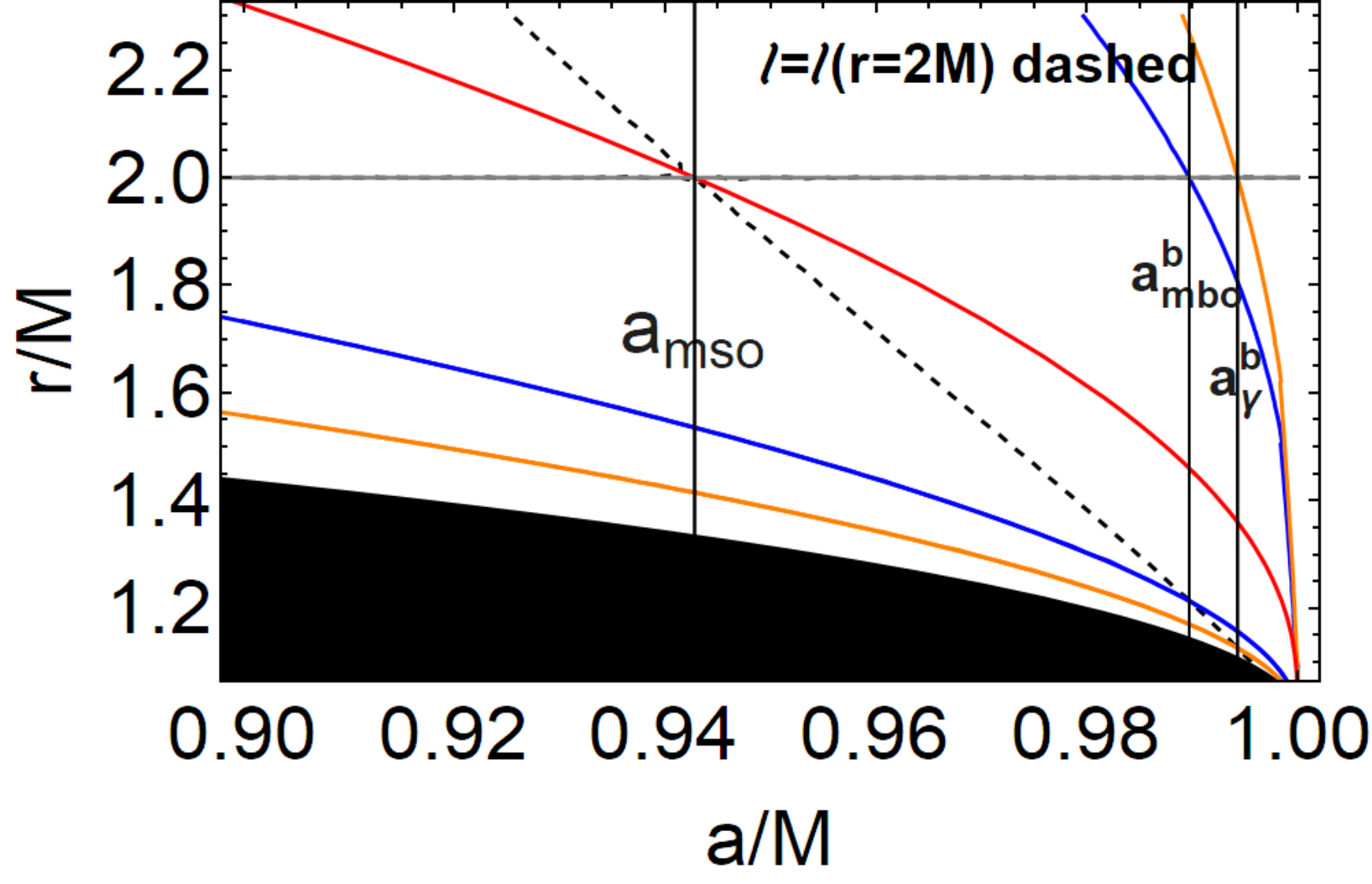}
   \includegraphics[width=5.5cm]{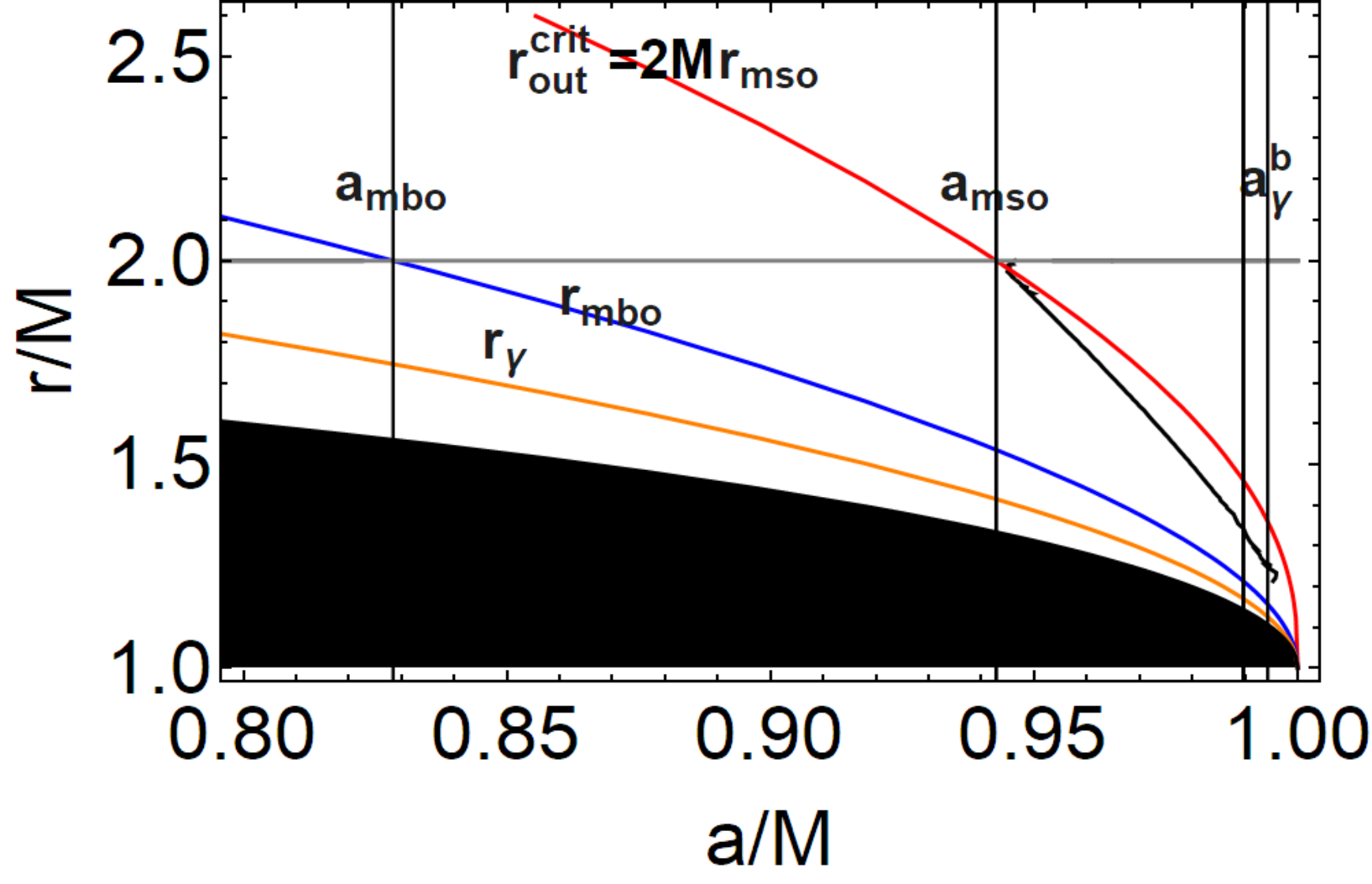}
  \caption{Notable radii of the spacetime geometry. Dashed curve is the radius $r:\ell=\ell(r_{\epsilon}^+)$, radius $r_{\epsilon}^+=2M$ is the outer ergosurface, showing the inner edge of the disk whose center is in $r_{center}=r_{\epsilon}^+>r_{mso}$ or viceversa the center of the disk whose cusp is in $r_{\times}=r_{\epsilon}^+<r_{mso}$. Black curve is the outer edge of the accreting  torus  $r_{outer}^{crit}=2M$. Black region is $r<r_+$ the \textbf{BH} horizon. Marginally bounded orbit  is $r_{mbo}$, blue curve as is radius $r_{mbo}^b: \ell(r_{mbo})=\ell({r})$, red curve is  marginally stable orbit $r_{mso}$, orange curve  is the photon orbit $r_{\gamma}$, where radius  $r_{\gamma}^b>r_{\gamma}: \ell(r_{\gamma})=\ell({r})$ is also represented as orange curve.  Panels show different ranges of the \textbf{BH} dimensionless spin $a/M$.}\label{Fig:principmill}
\end{figure}
(We note that the center of maximum pressure and the minimum pressure point in the torus are determined  by the specific angular momentum only, while the geometrical maximum is  regulated by the parameter $K$ once fixed $\ell$).
The torus cusp $r_{cusp}$ may also be in the near-horizon region for very high spins $a/M$ of the central \textbf{SMBH} (partially contained for $a>a_{\gamma}$ and entirely contained for $a>a_{\gamma}^b$).
(For $r_{cusp}$  we intend the minimum points of the pressure $r_{cusp}\in]r_{\gamma},r_{mso}]$, which can be a cusp of the closed torus $r_{cusp}=r_{\times}\in]r_{mbo},r_{mso}[$ or the cusp of a proto-jet $r_{cusp}=r_j\in]r_{\gamma},r_{mbo}]$). Analysis of the inner edge location with  respect to the static limit is in Figs\il(\ref{Fig:PlotBlakPurp7}) and (\ref{Fig:Plotssot}).
\item[\textbf{--The outer edge}]
If  $r_{mso}<2M$,  the condition $r_{outer}\leq2M$  is possible for  some  $\ell$  and $K$ values. The condition
${r}^b_{mbo}<r_{\epsilon}^+$ is necessary but  not sufficient condition for the outer edge of a $\cc_2$ configuration to be included in the ergoregion.  The inner part of such tori can be in part  $\Sigma_{\epsilon}^+$.  The inner part of a $\cc_1$ configuration must be in the ergoregion. The condition
${r}^b_{\gamma}<r_{\epsilon}^+$ is necessary but  not sufficient condition for the outer edge of a $\cc_3$ configuration to be included in the ergoregion.  The inner part of such tori can be in part in $\Sigma_{\epsilon}^+$. The inner part of a $\cc_2$ configuration must be in the ergoregion. Analysis of the outer  edge location with the respect to the static limit is in Figs\il(\ref{Fig:PlotVampa1},\ref{Fig:spessplhoke1},\ref{Fig:PlotVamp1},\ref{Fig:PlotJurY}).
\end{description}
The maximum elongation of the dragged torus is at most $\lambda\approx2M$.
The surfaces with very large  specific angular momentum $\ell$,   $\ell\geq\ell_{\gamma}^-$, tend to be stable   against the cusp formation, leading to quiescent closed configurations. An interesting question is whether these configurations, stabilized by a large centrifugal component (the cusp absence), are  dragged or partially contained  toroidal surfaces rather then tori   with  lower specific angular momentum  characterized with the cusp formation $\cc_{\times}$ or the quiescent  configurations with $\ell\in\mathbf{L_2}$ \footnote{This issue is indeed relevant for the stability of these configurations as
 the relativistic
Roche lobe overflow at the cusp of  the equipotential surfaces is also  the
 stabilizing mechanism against the thermal and viscous instabilities {locally},
and against the so called Papaloizou--Pringle instability {globally} \citep{Blaes1987}.
(For a discussion on the relation  between  Papaloizou-Pringle (\textbf{PP})  global incompressible modes  in the tori, the  Papaloizou-Pringle Instability
(\textbf{PPI}),
a global, hydrodynamic, non-axi-symmetric instability   and the  Magneto-Rotational Instability (MRI)  modes see  \citet{Fi-Ringed,Bugli}).}.
These configurations can also grow very huge, with the center of maximum pressure in the torus at  $r>{r}^b_{\gamma}\gtrsim9M$  for large spin $a\gtrsim a_{\gamma}$.

\begin{figure}\centering
  % Requires \usepackage{graphicx}
  \includegraphics[width=8cm]{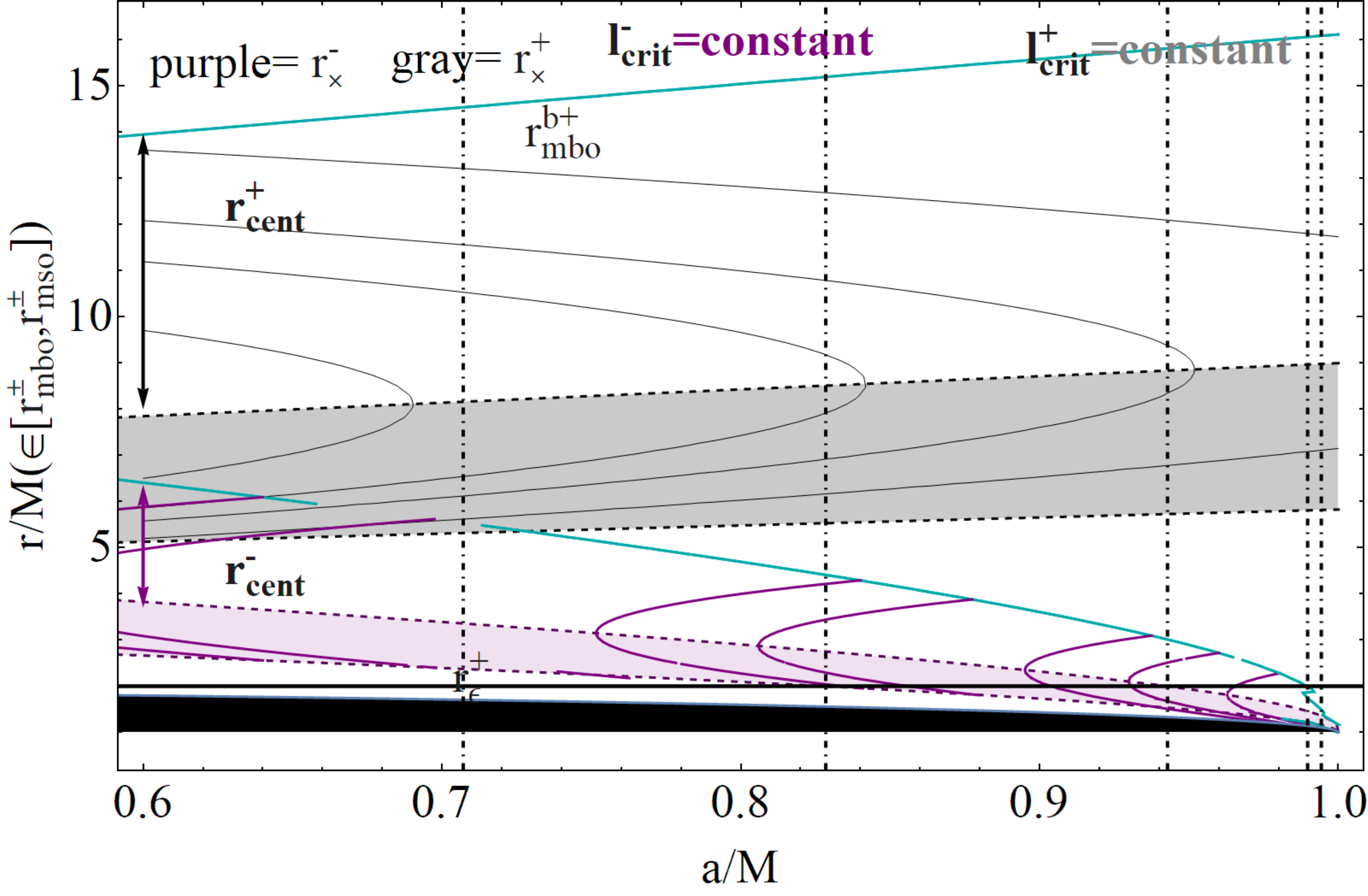}
  \includegraphics[width=8cm]{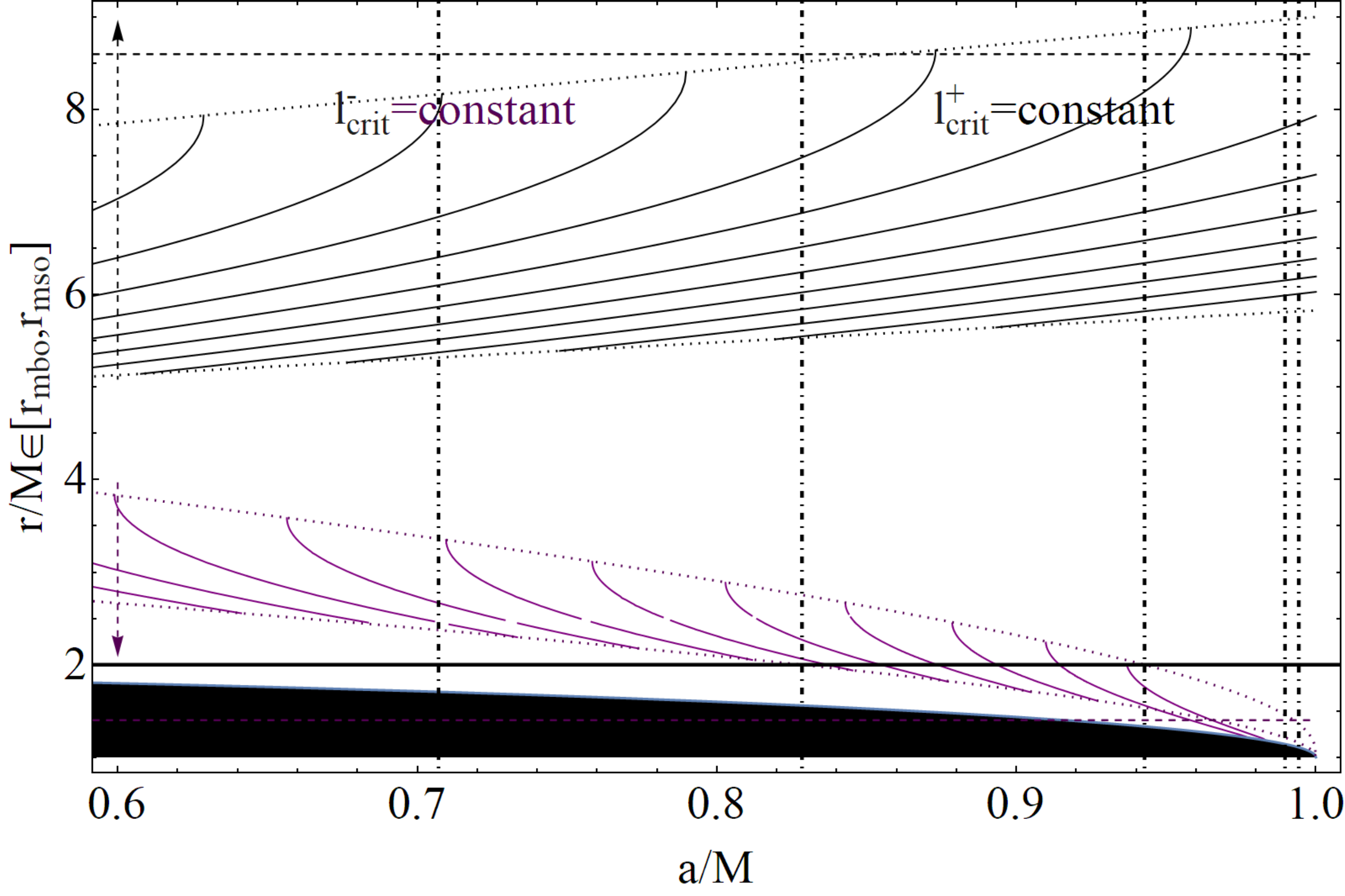}
    \includegraphics[width=8cm]{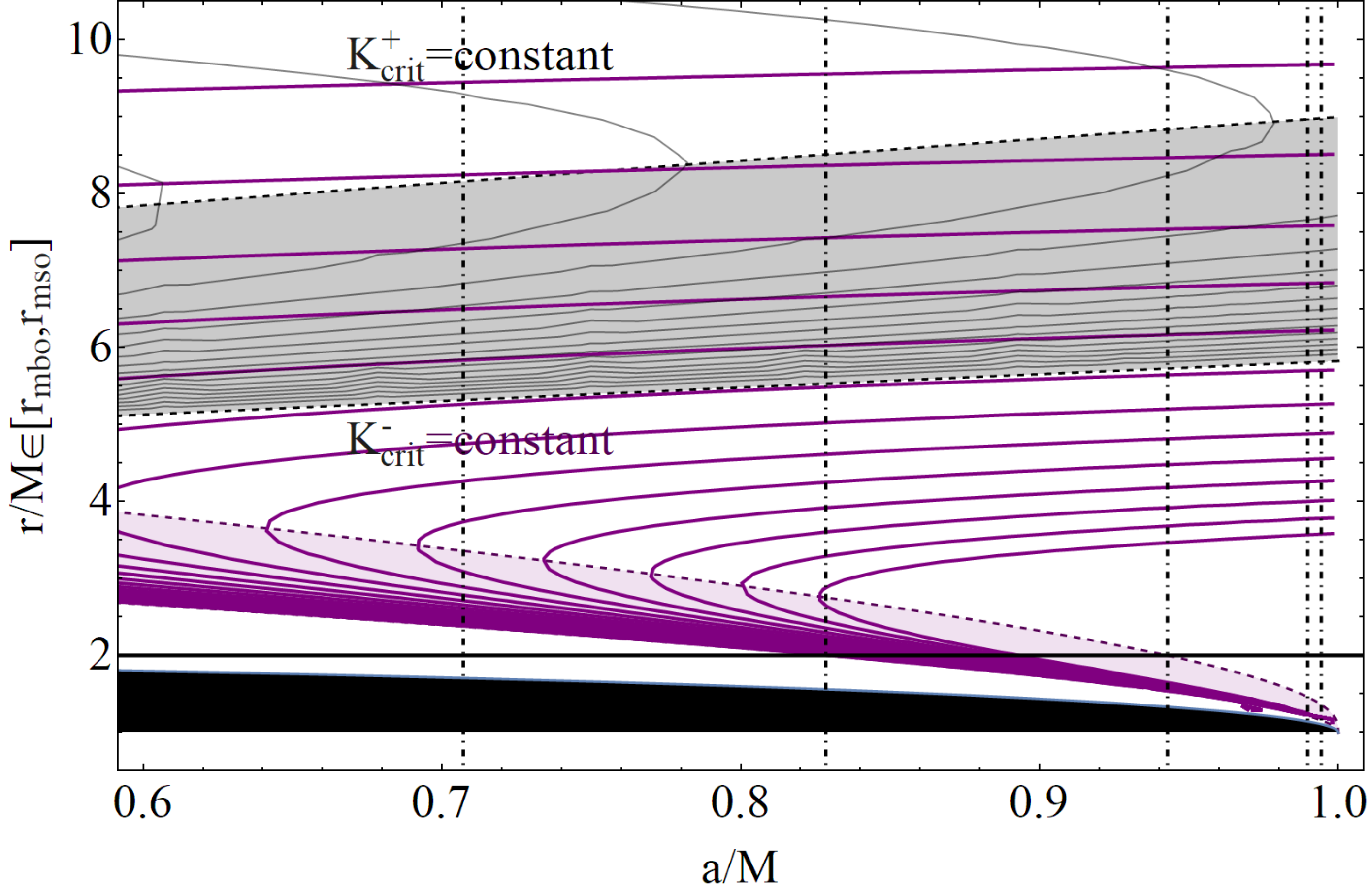}
  \includegraphics[width=8cm]{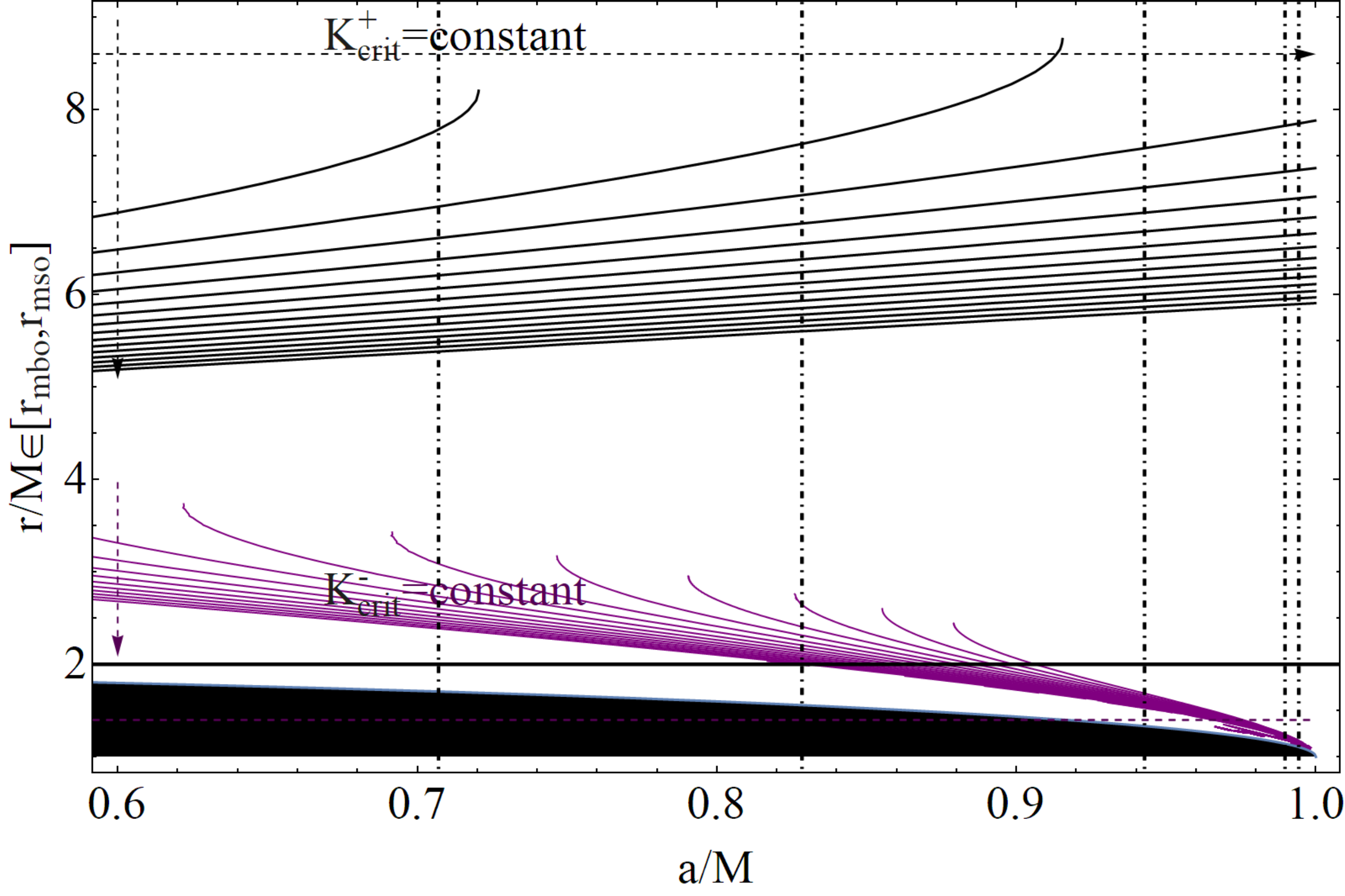}
  \caption{  \emph{Upper  panels}:
Fluid specific angular momentum $\ell^{\pm}=$constant in the plane
$r\in[r^{\pm}_{mbo},r^{\pm}_{mso}]$ and $a\in[0,M]$,
for corotating (purple curves) and counterrotating  fluids (black curves).
\emph{Bottom  panels}: Curves $K_{crit}^{\pm}=$constant  for $a\in[0, 0.9998M]$  and $r\in[r^{\pm}_{mbo},r^{\pm}_{mso}]$ for corotating (purple curves) and counterrotating (black curves) fluids.
 Arrows  in the plots follow  increasing values of the plotted functions. Spins  $\mathbf{A}_{\epsilon}^+\equiv\{a_{mbo},a_{mbo}^b,a_{\gamma},a_{\gamma}^b,a_{mso}\}$  of  Figs\il(\ref{Fig:polodefin1}) are also plotted.  Black region is the \textbf{BH} at $r<r_+$ where $r_+$ is the outer horizon, the outer ergosphere at $r_{\epsilon}^+=2M$  is also plotted. Right panel focuses on the location of the critical points.}\label{Fig:slowtoga}
\end{figure}

In Figs\il\ref{Fig:corso} we show the regions of  tori parameters for the existence of the closed tori, featuring also  the possibility of double cusped configurations for \textbf{BHs}  with  high-spin, while in Figs\il(\ref{Fig:slowtoga})  curves $\ell=$constant and $K=$constant are shown, fixing  the location of the critical  points  of the pressure, $r_{center}$ and $r_{cusp}=(r_{\times},r_{j})$. We investigate the distance between the two tori in the aggregates of  toroidal solutions, with one  dragged  surface with an  outer  corotating torus,   considering the distance between the critical points an the centers.
We use the leading function   $\ell(r)$, distribution of the tori specific   angular momentum. We also evaluated the  energy function,  related to   tori densities and  tori energetics.
The  function $K(r)$ is related  to  an independent tori parameter $K$ which  regulates the torus elongation $\lambda$ on {its} symmetry plane and  the emergence of hydro-dynamical instability. It is also associated  to the torus density,  the torus thickness and other characteristics related to the tori energetics, as cusp luminosity and accretion rates. The  function $K(r)$ gives the critical values $K_{crit}=\{K_{cusp},K_{center}\}$ of the tori with different specific angular momentum.
The study of $\ell(r)$ and $K(r)$ functions is also important to set constrains for the tori  collision.
Function $K(r)$ provides the distribution of the effective  potential values corresponding to the points of  maximum and minimum  of the  density (and of the HD pressure), for different spacetimes and angular momentum,
The  "energy function",  defined as $K(r)\equiv V_{eff}(\ell(r),r)$,   is therefore a function of $r$  with parameter $a/M$, and it  parameterizes each torus with equal angular momentum  $\ell$ with  the  $K$-parameter governing  the   torus center and  eventually $K_{cusp}$  at its cusp. Function $r_p(r): K(r_p)=K(r)$ identifies  a pair of tori  ($T_1$, $T_2$) with different angular momentum $\ell$, with  $K_{center}(T_1)=K_{crit}(T_2)=K({r})$  where $r_p={r}=$constant, and   $r_{crit}$  is  $r_{\times}$ where $K\in K_{\max}\in]K_{mso},1[$, or $r_j$ where $K\geq1$--Figs\il(\ref{Fig:PlotVampb1}).
For fixed  $\ell=$constant, the solution of $\ell=\ell(r)$ gives a torus center  $r_{center}$,  the maximum density and HD pressure points;  tori with different $\ell\in [\ell_{mso},\ell_{\gamma}]$  can have also equal minimum  pressure  (and  density)-points (different inner edges, only one torus cusp). In general, tori  characterized by  same value of specific angular momentum $\ell$  have not same  geometrical  maximum, which  depends  on  the $K$ parameter. The  $K$ parameter sets  the disk "verticality", defining the  solution of  {$\partial_y V_{eff}(x,y)=0$}, where $V_{eff}(x,y)$ is the fluid effective potential on any plane $\sigma$  in Cartesian coordinates (there is ($x=r\cos\theta, y=r\sin\theta)$, on the equatorial plane there is $y\equiv r$).

\medskip
In   Figs\il(\ref{Fig:PlotVampb1}),
we explore five classes of the   Kerr geometries   defined  according to the circular geodesics related to the ergosphere boundary that give rise to the characteristics values of the \textbf{BH} spin $\mathbf{A}_{\epsilon}^+\equiv \{a_{\gamma},a_{mbo},a_{mso},a_{mbo}^b,a_{\gamma}^b\}$.  The  geodetic radii used in the definition and related spins are given as
\bea\nonumber
&&\frac{a_ {\gamma}}{M}\equiv\frac{1}{\sqrt {2}} = 0.7071: r_{\gamma}=r_{\epsilon}^+,\quad \frac{a_{mbo}}{M}\equiv2\left (\sqrt {2} -1\right)\approx 0.828: r_{mbo}=r_{\epsilon}^+,\\&&\nonumber  \frac{a_{mso}}{M}\equiv\frac{2 \sqrt{2}}{3}\approx 0.9428: r_{mso}=r_{\epsilon}^+
\eea
regulating  the location of the points of the  minimum of pressure in the torus with respect to the static limit, and
\bea\label{Eq:strateg}
&&\frac{a_{mbo}^b}{M}\equiv0.9897: r^b_{mbo}=r_{\epsilon}^+,
\quad \frac{a_{\gamma}^b}{M}\equiv0.9943M: r^b_{\gamma}=
r_{\epsilon}^+,
\eea
regulating  the  location of the centers of maximum pressure with respect to the static limit. Spins $\mathbf{A}_{\epsilon}^+$ are
represented in Figs\il(\ref{Fig:principmill}).
In terms of the fluid  specific angular momentum, spins of the set $\mathbf{A}_{\epsilon}^+$ can also be defined as  $a_{mso}: \ell_{\epsilon}^+=\ell_{mso}$,  where  $
(a_ {mbo}, a_{mbo}^b)$   satisfy   $\ell_ {\epsilon}^ += \ell_ {mbo}$, and  for the  spins $
(a_ {\gamma},a_ {\gamma}^b) $ there is $ \ell_ {\epsilon}^ += \ell_ {\gamma}$-- Figs\il(\ref{Fig:principmill}) and (\ref{Fig:Plotsoorr}).
 Some configurations orbiting in the geometries defined in $\mathbf{A}_{\epsilon}^+$ are shown in  Figs\il(\ref{Fig:polodefin1}).

We study  the conditions for the tori have outer edge $r_{outer}$   in $\Sigma_{\epsilon}^+$, in  the different  geometries of the set  $\mathbf{A}_{\epsilon}^+$   in Figs\il(\ref{Fig:PlotVampa1}).
A similar analysis for  the outer edge of   cusped surfaces  $\cc_{\times}$  is in  Figs\il(\ref{Fig:spessplhoke1}), showing the conditions in terms of  torus specific angular momentum, and   in Figs\il(\ref{Fig:PlotVamp1})  where the conditions are expressed as functions of the disks inner edge $r_{inner}$ location.
The crossing of outer edge of the disk with the outer ergosurface is  considered in Figs\il(\ref{Fig:principmill}).

Below we consider the dragged and partially contained tori in the  five classes of geometries delimited by spins $\mathbf{A}_{\epsilon}^+$:

\begin{description}
\item[
\textbf{--Range $\mathbf{A}_1\equiv[a_{\gamma},a_{mbo}]\in \mathbf{A}_{\epsilon}^+$}]

In the  geometries of this range, with  increasing dimensionless  spin $a/M$,  the radius
$r_{\times}$, cusp of the $\cc_{\times}$  torus  with momentum $\ell\in\mathbf{ L_1}$, approaches  the static limit. An inner Roche lobe   of orbiting  matter appears close  to the horizon. The closed configurations with  $\ell\in\mathbf{L_3}$, have  center  in $r>r^b_{\gamma}$ which can also be,  for small \textbf{BH} spins, very far from the outer ergosurface. These  tori can be very huge. The inner edges of $\cc_1$ tori are out of the ergoregion, but approaching the outer ergosurface  at distance that can be less then $\approx 0.5 M$. The inner edge of  the quiescent  $\cc_2$  torus may be included in the ergoregion, the cusp of the proto-jet can be in the ergoregion, but the inner part of the disk ($[r_{inner},r_{center}]$)  is  partially contained in the ergoregion as the center is at $r>r^b_{mbo}$.
\item[
\textbf{--Range $\mathbf{A}_2\equiv[a_{mbo},a_{mso}]\in \mathbf{A}_{\epsilon}^+$}]

 The center of the
$\cc_1$ configurations  is  out the ergoregion. The cusp $r_{\times}$ of the cusped closed torus $\cc_{\times}$  is  inside the ergoreigon, therefore  these tori must be  partially contained  with their inner part. Quiescent $\cc_1$ tori  can be partially contained  in $\Sigma_{\epsilon}^+$   and they cannot evolve into  dragged tori.
The quiescent $\cc_2$ and  $\cc_3$ tori have center out of the ergoregion. The inner edge of $\cc_2$ can be contained in the ergoregion and the cusp of proto-jet must be inside the ergoregion.
These tori can be  huge, with elongation of the  disk inner part of     $
\approx 4M$ for $\cc_2$ to  $\approx 6M$.
\item[
\textbf{--Range $\mathbf{A}_3\equiv[a_{mso},a_{mbo}^b]\in \mathbf{A}_{\epsilon}^+$}]

 In these  \textbf{BH} spacetimes, the center of configurations  $\cc_{\times}$ and $\cc_1$ can  be in the ergoregion, while  the  center of $\cc_2$ and  $\cc_3$ tori  cannot  be in the ergoregion. The cusp $r_{\times}$ of $\cc_{\times}$ tori must be in $\Sigma_{\epsilon}^+$, therefore the inner part of the cusped surface must be partially contained in the ergoregion. Configurations $\cc_1$ can be dragged.

\item[
\textbf{--Range $\mathbf{A}_4\equiv [a_{mbo}^b, a_{\gamma}^b]\in \mathbf{A}_{\epsilon}^+$}]

In the geometries of this class, the center  of $\cc_1$   configurations must be contained inside the ergoregion.
 Center  of $\cc_2$ configurations approaches the static limit.

\item[
\textbf{--Range $\mathbf{A}_5\equiv[a_{\gamma}^b,M]\in \mathbf{A}_{\epsilon}^+$}]

The center of  $\cc_1$ and $\cc_2$ tori  must be in the ergoregion, while the  $\cc_3$  tori centers  can be in the ergoregion. Both $\cc_2$ and $\cc_3$ can be dragged and must be partially contained.
%Note that this analysis  consider the crossing of tori of the outer ergosurface on the equatorial plane.
%
\begin{figure}\centering
  % Requires \usepackage{graphicx}
  \includegraphics[width=7cm]{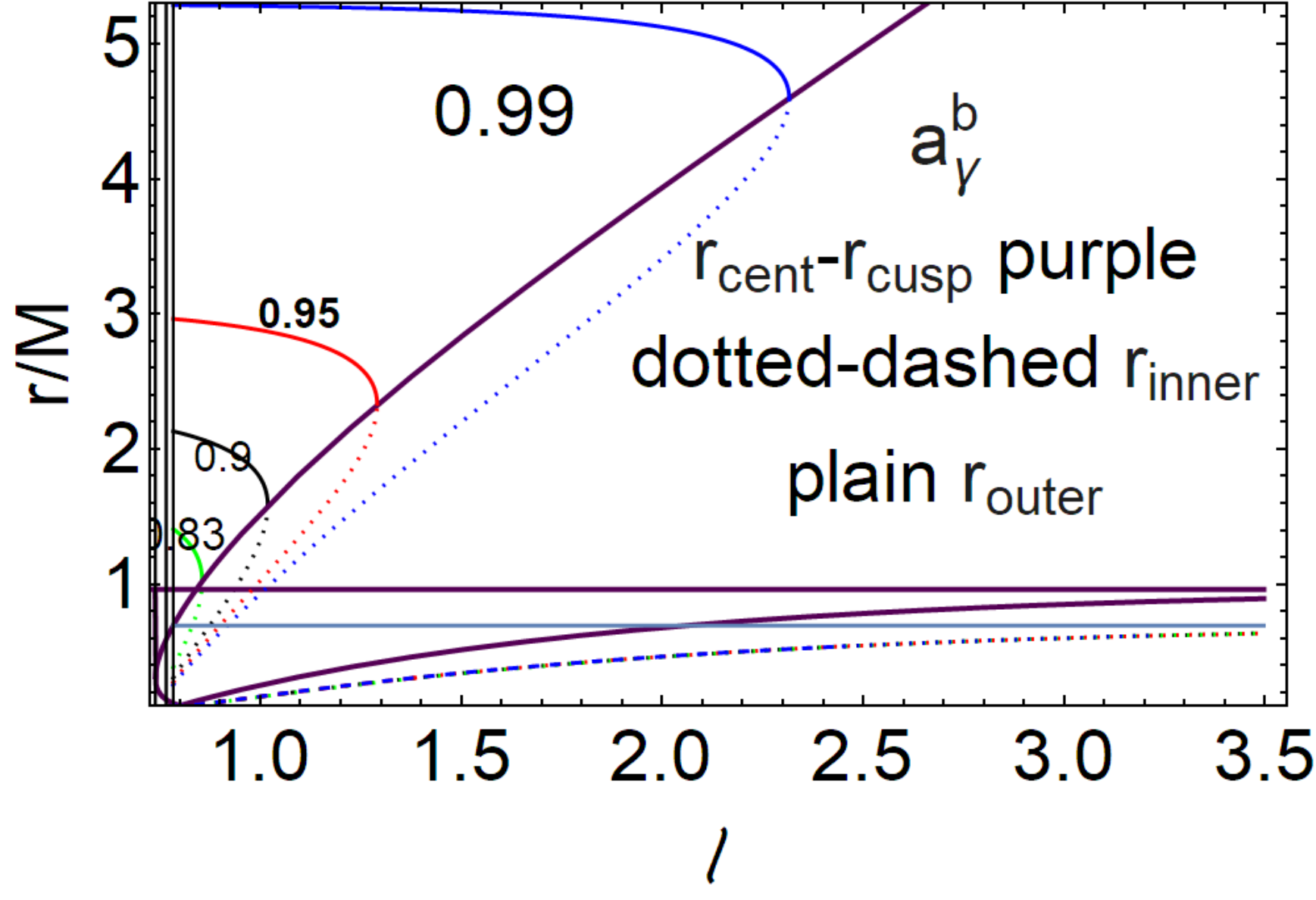}
  \includegraphics[width=7cm]{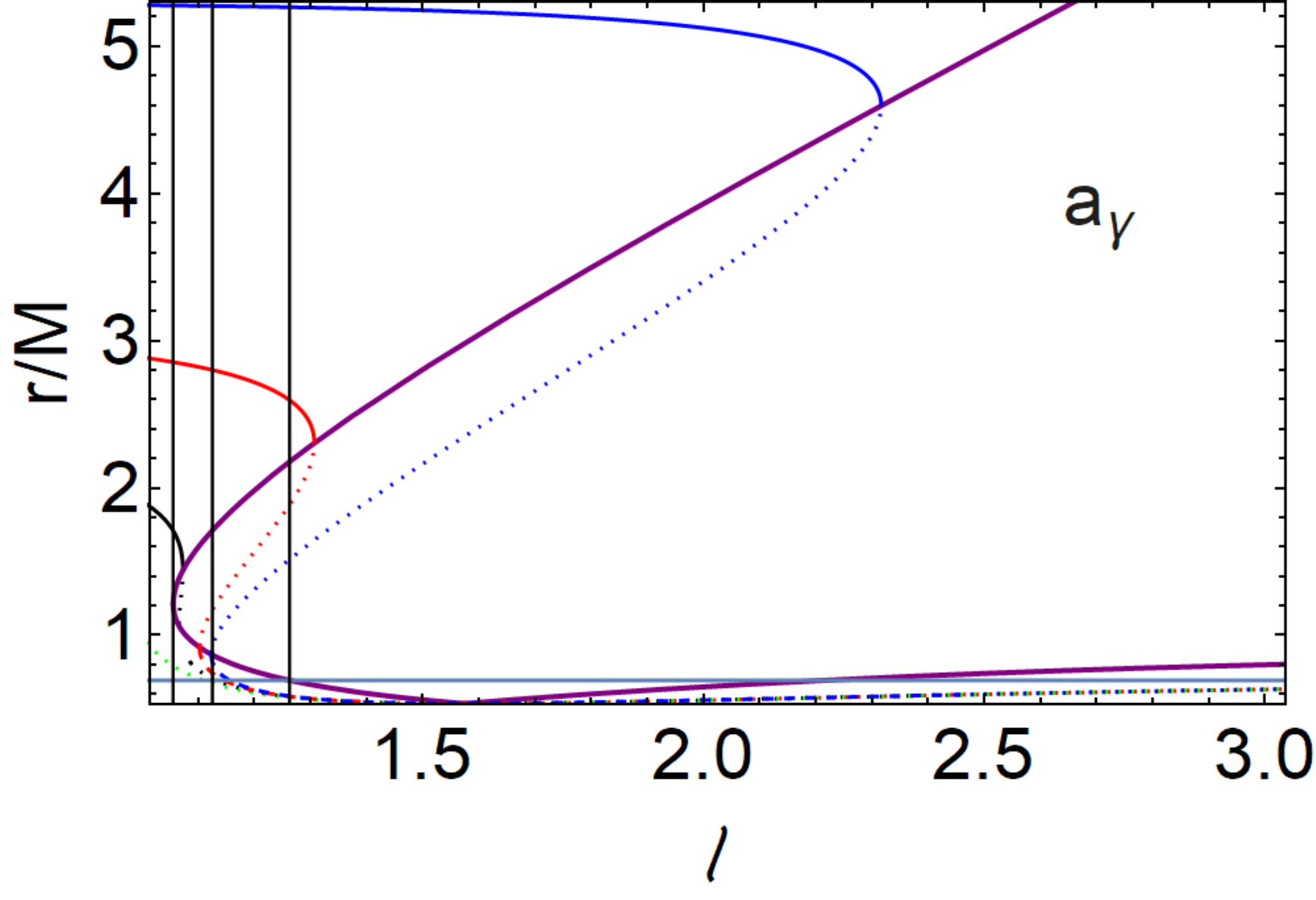}\\
  \includegraphics[width=7cm]{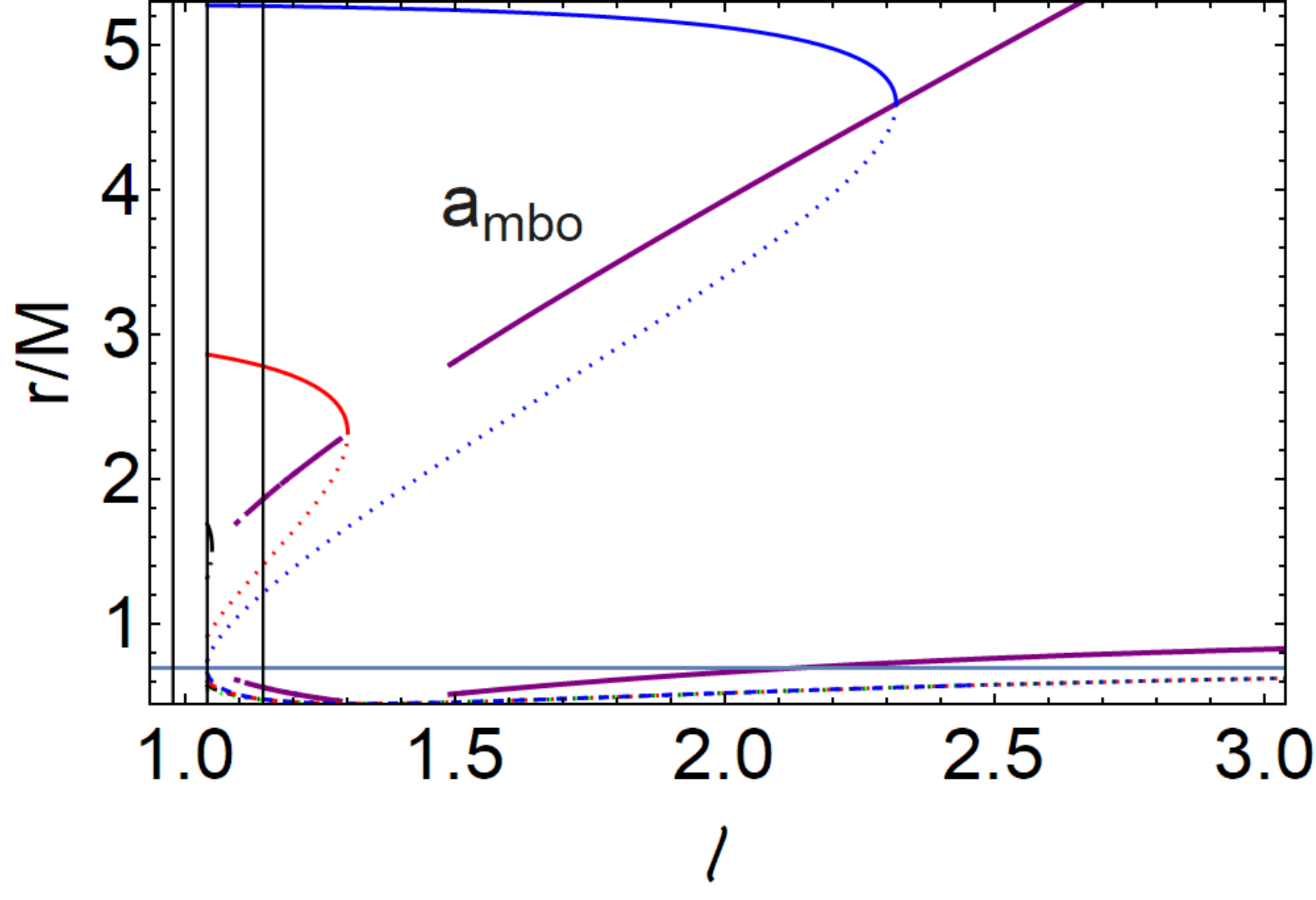}
  \includegraphics[width=7cm]{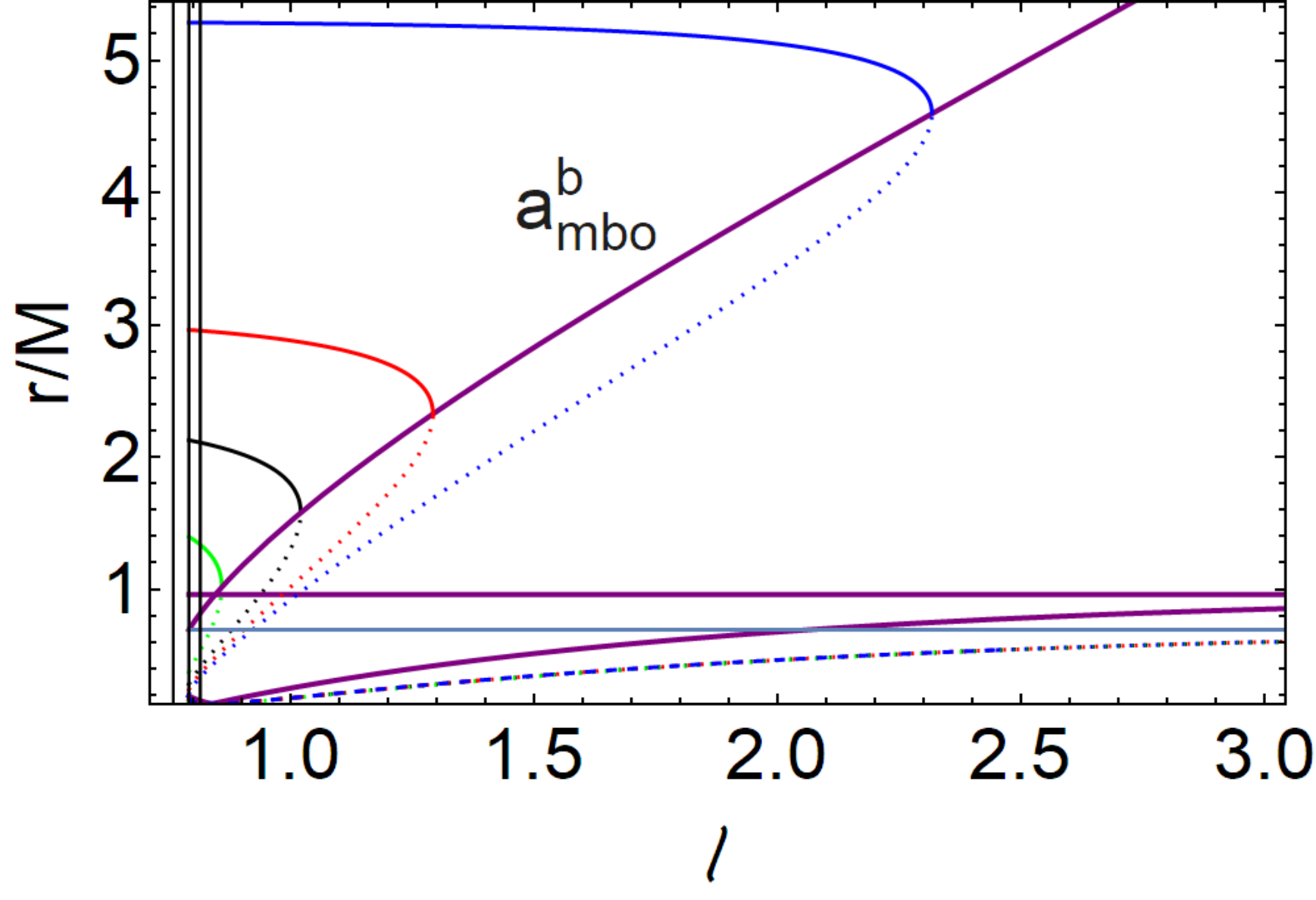}
  \caption{Tori edges and pressure extremes are shown as functions of the fluid specific angular momentum $\ell$. Purple line is the couple of radii $r_{center}>r_{\times}$ center of maximum pressure point and minimum pressure point in the disk respectively. Number  sets the values of $K$ parameter of the disks. $r_{inner}$ is the inner edge of the disk $r_{outer}$ is the outer edge of the disk. $\ell$ is the fluid specific angular momentum. Spins   $\mathbf{A}_{\epsilon}^+\equiv\{a_{mso},a_{mbo},a_{msbo}^b,a_{\gamma},a_{\gamma}^b\}$ are   defined in Figs\il(\ref{Fig:PlotVampb1}).  }\label{Fig:PlotVampa1}
\end{figure}
Location of the inner edge, center and outer edge of the disk is shown in Figs\il(\ref{Fig:PlotVampa1}) for different spins.
Maximum elongation and location of the inner edge, outer edge and center of the disks  are represented in Figs\il(\ref{Fig:spessplhoke1}).

\begin{figure}\centering
  % Requires \usepackage{graphicx}
  \includegraphics[width=5.6cm]{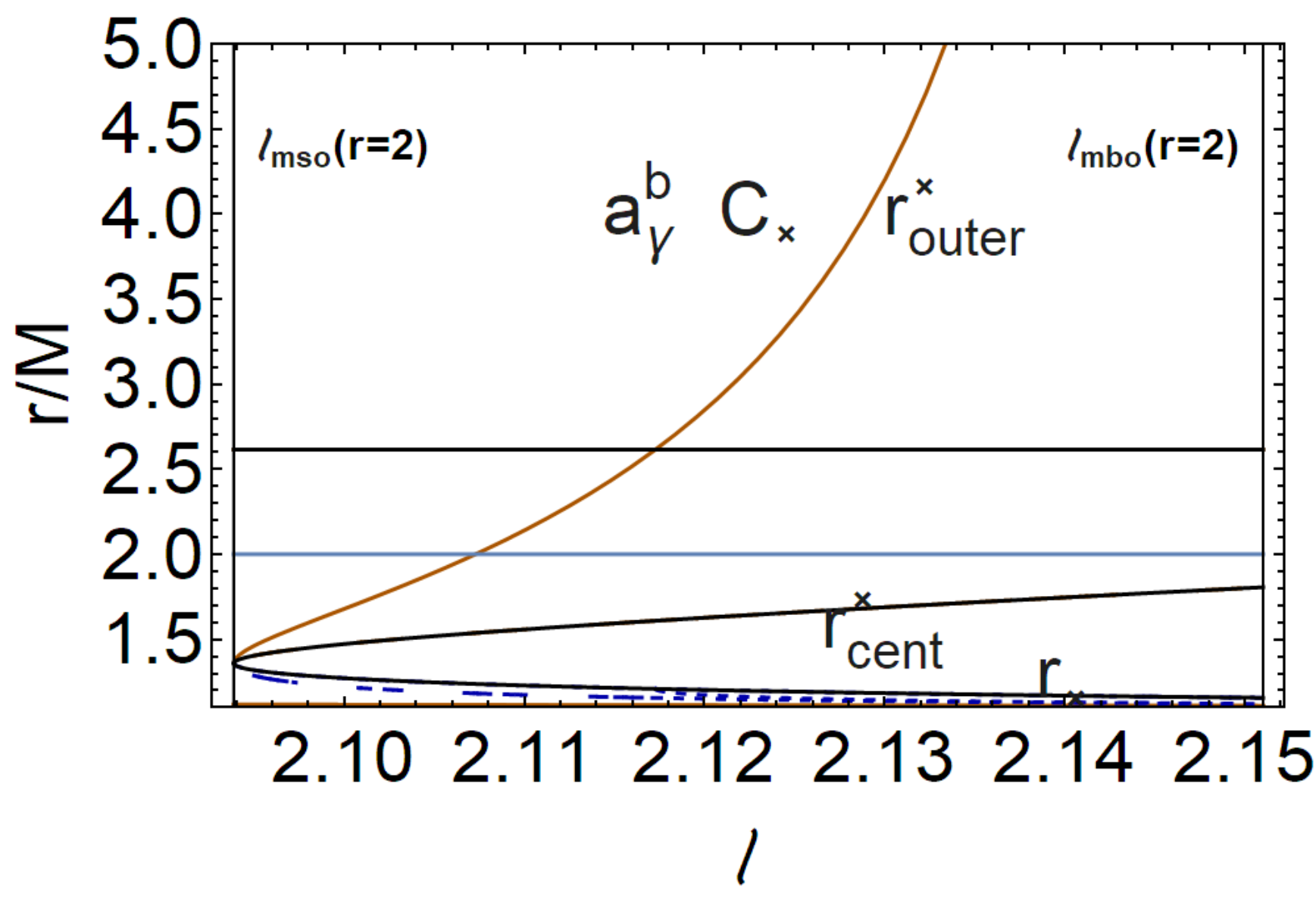}
  \includegraphics[width=5.6cm]{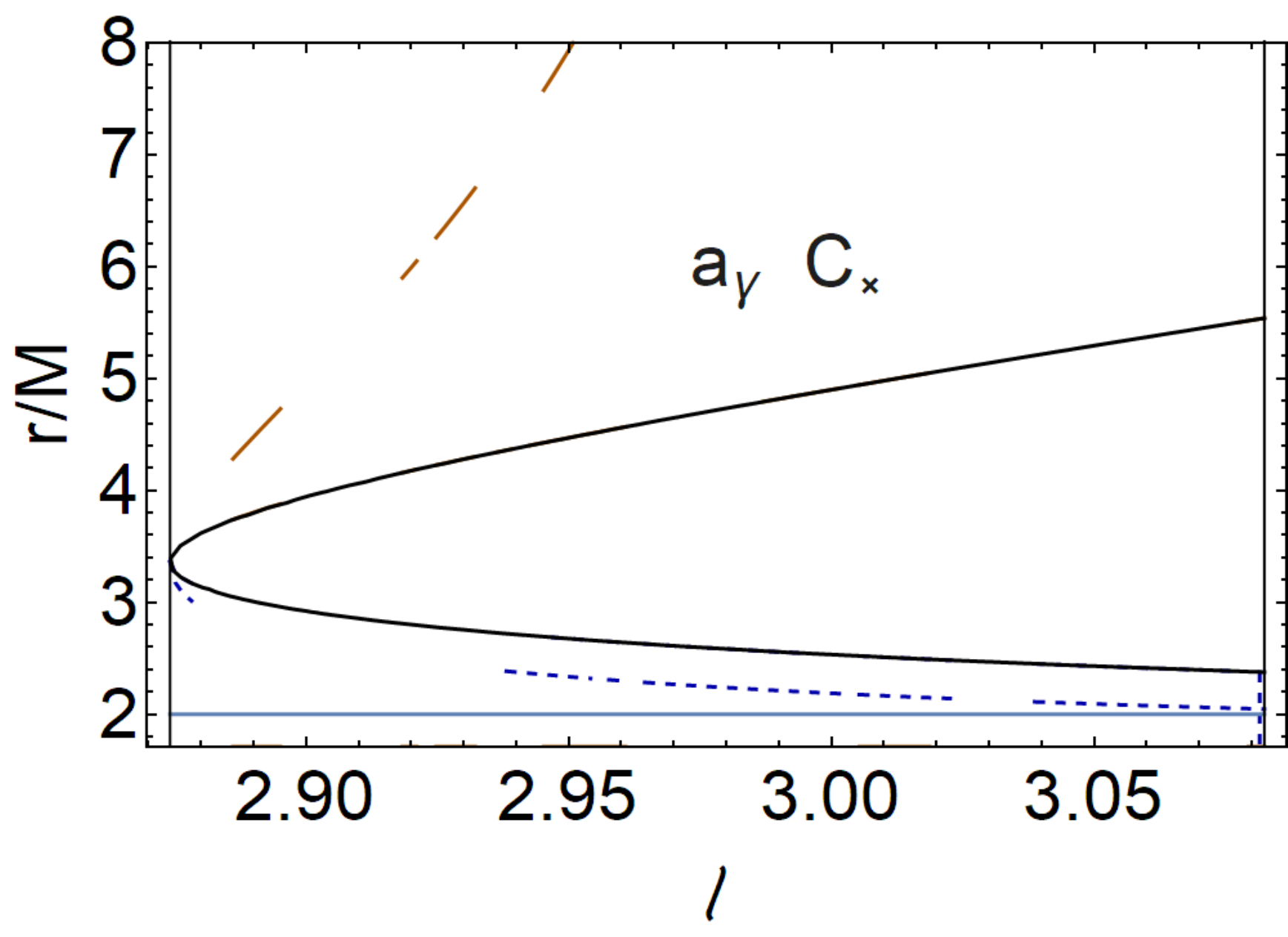}
  \includegraphics[width=5.6cm]{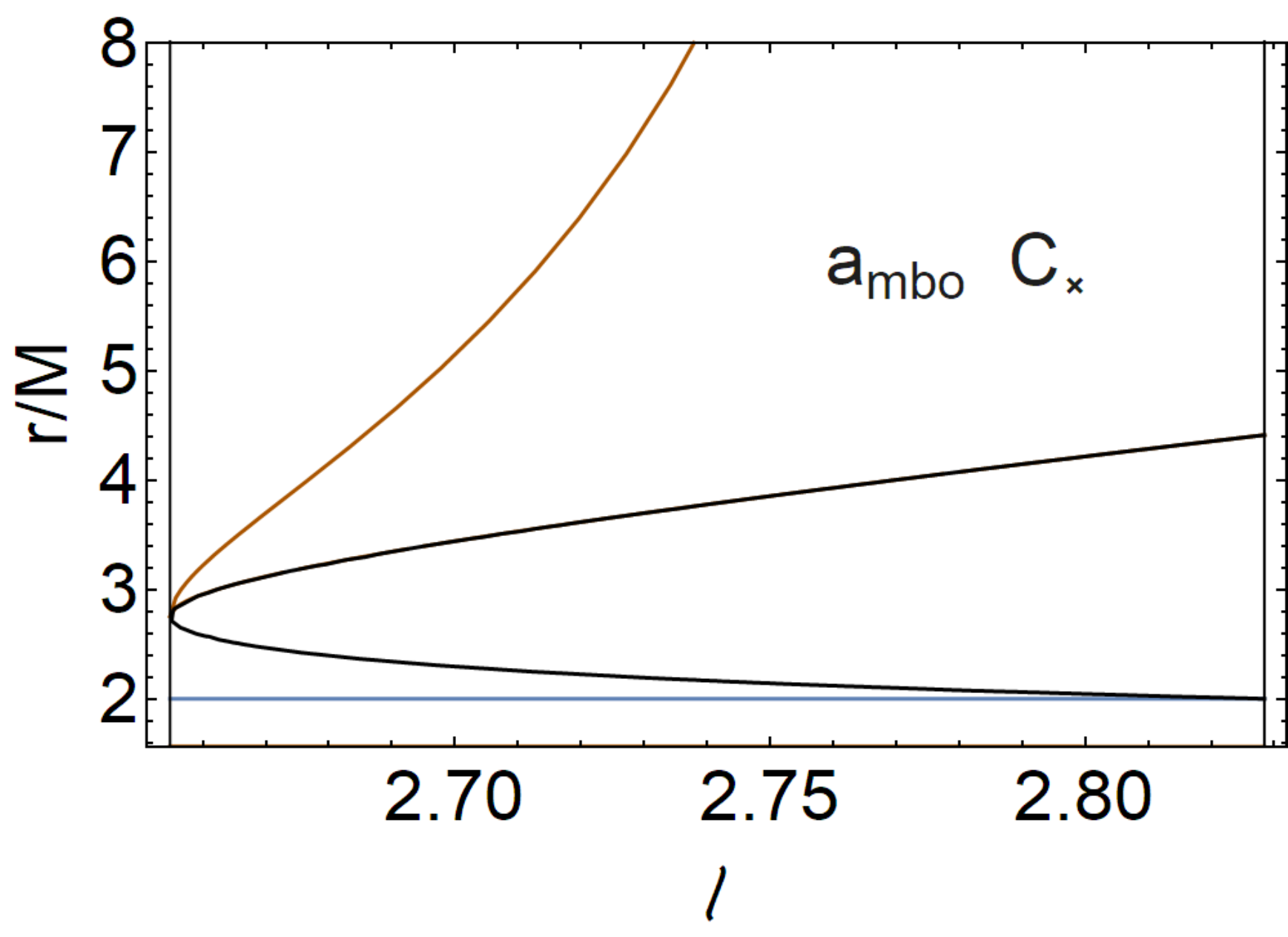}
  \includegraphics[width=5.6cm]{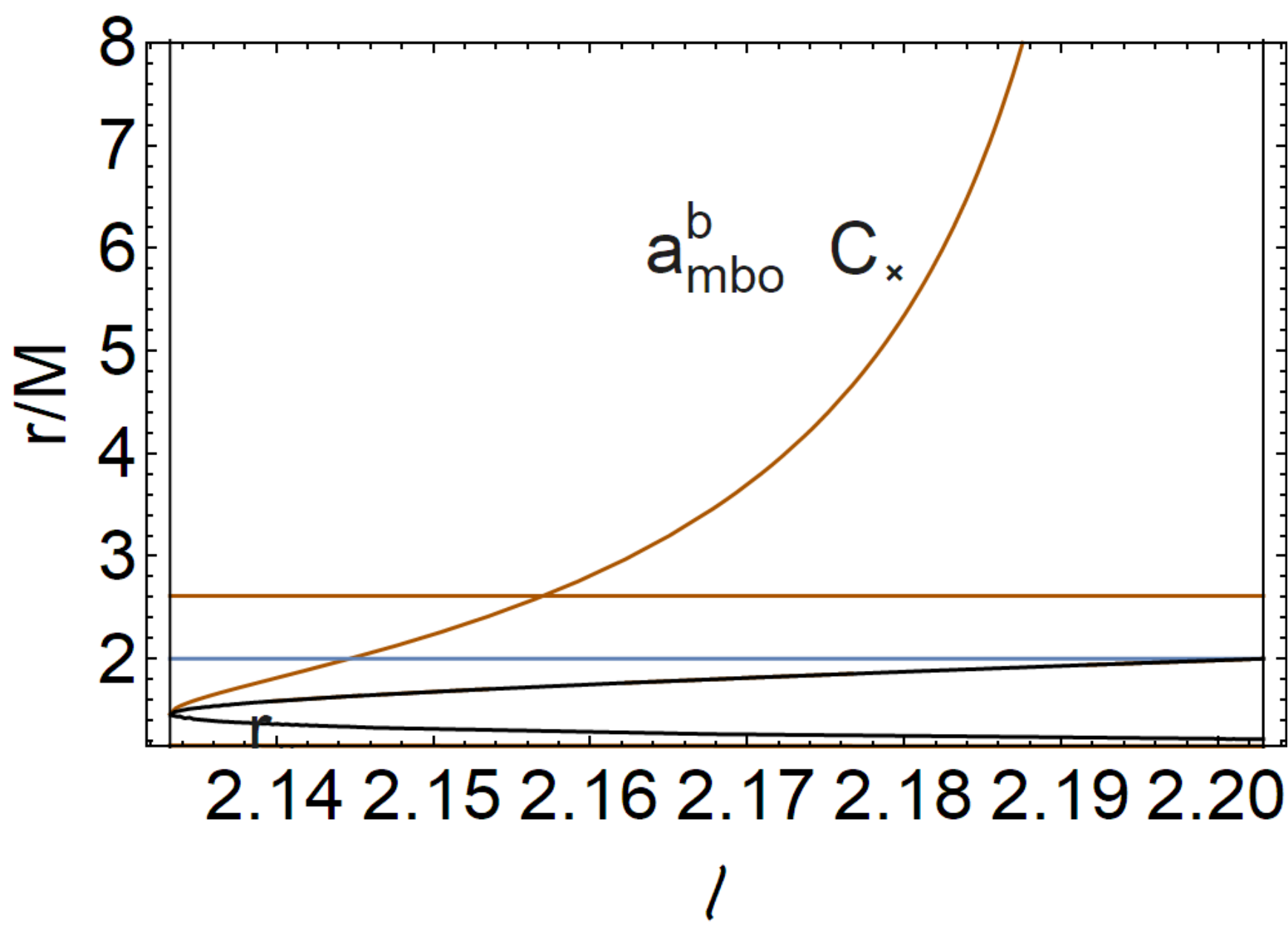}
    \includegraphics[width=5.6cm]{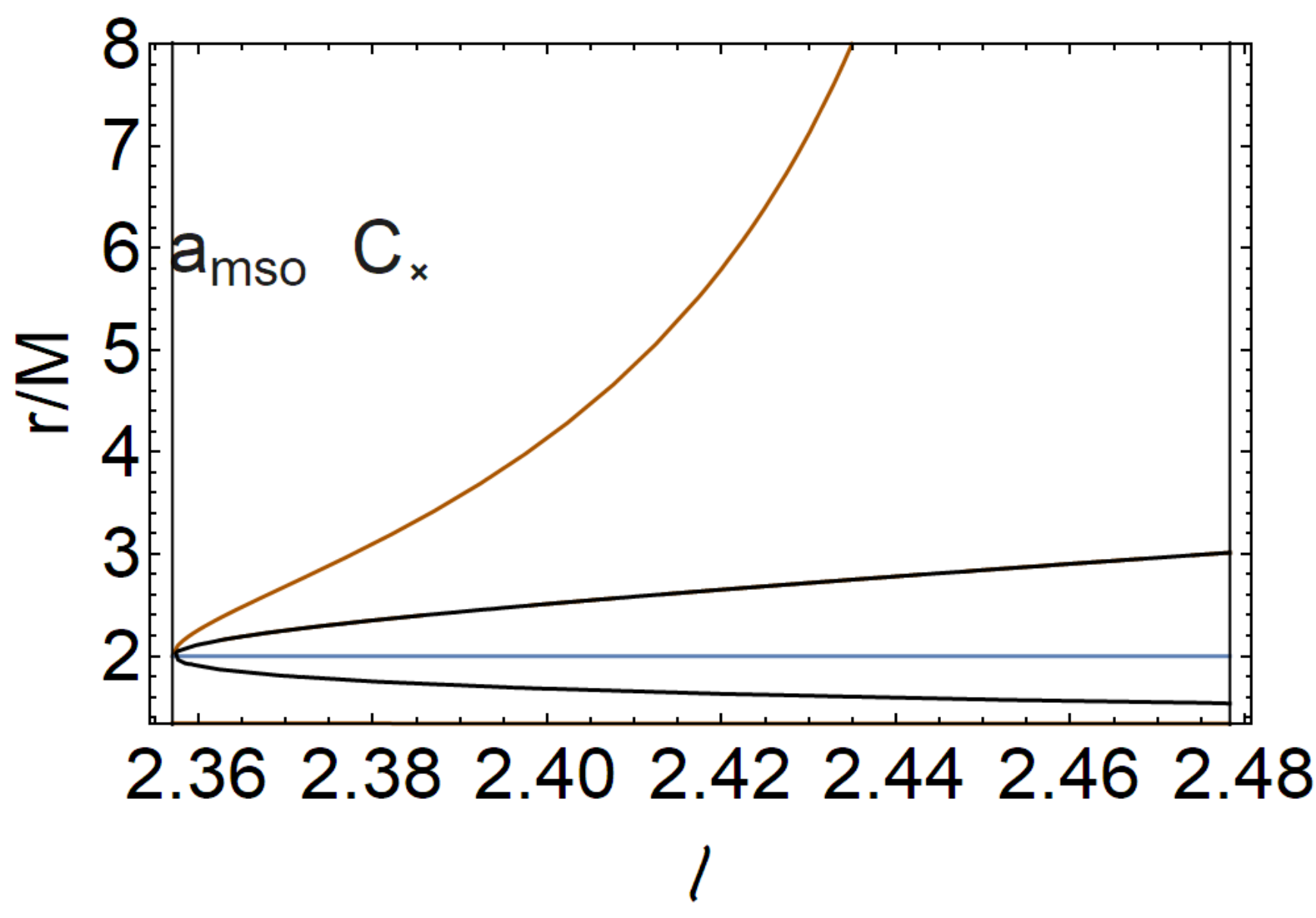}
  \caption{Plot of the outer edge $r_{outer}^{\times}$, inner edge and the center $r_{center}$ of the cusped tori $\cc_{\times}$ as functions of the fluid  specific angular momentum, for spins   $\mathbf{A}_{\epsilon}^+\equiv\{a_{mso},a_{mbo},a_{msbo}^b,a_{\gamma},a_{\gamma}^b\}$    defined in Figs\il(\ref{Fig:PlotVampb1}). $\ell\in[\ell_{mso},\ell_{mbo}]$ is the fluid specific angular momentum, $mso$ is for marginally stable orbit, $mbo$ is for marginally bounded orbit, the outer ergosurface on the equatorial plane is $r_{\epsilon}^+=2M$ (gray line).  }\label{Fig:spessplhoke1}
\end{figure}
\begin{figure}\centering
  % Requires \usepackage{graphicx}
  \includegraphics[width=5.6cm]{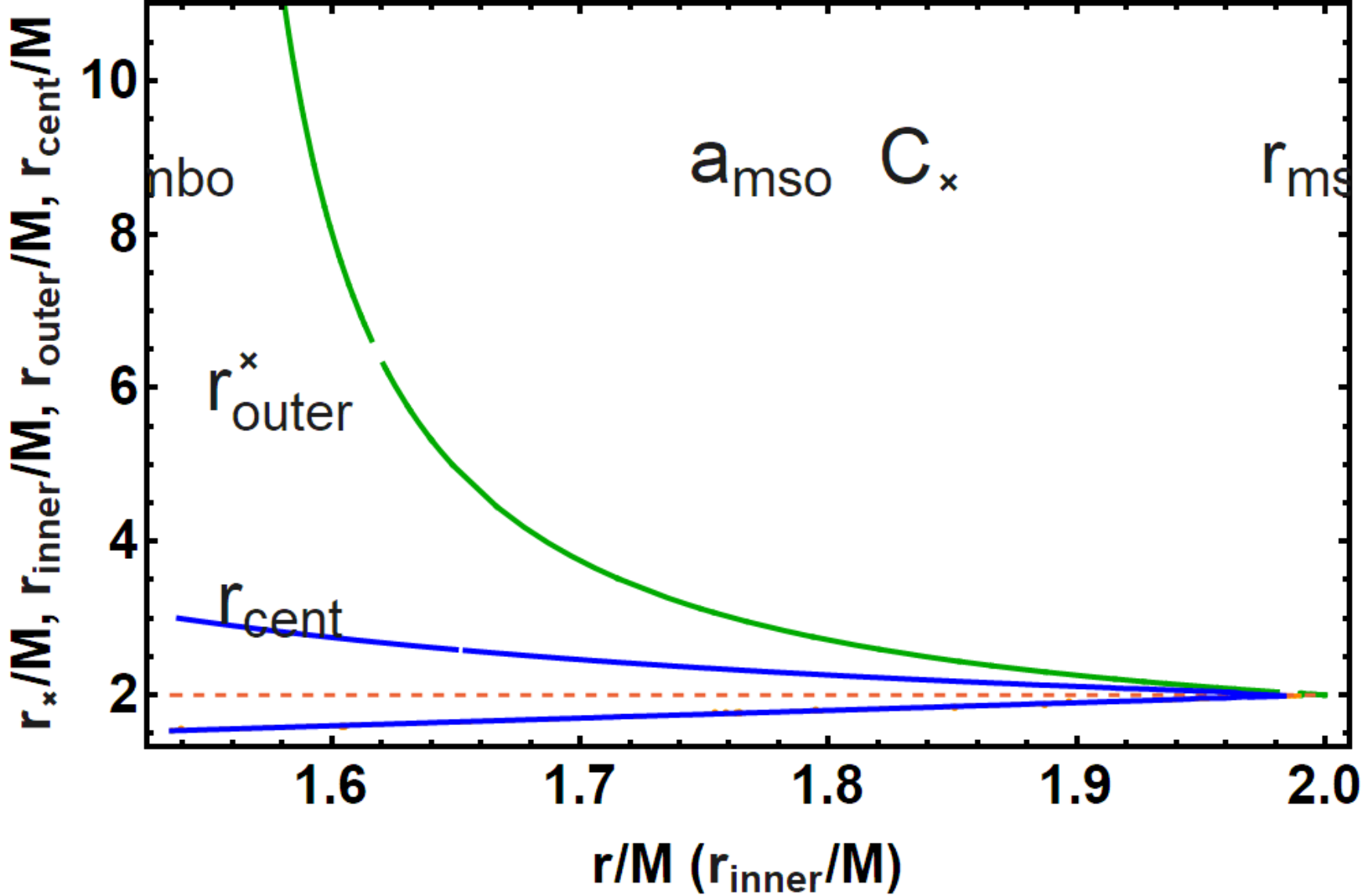}
  \includegraphics[width=5.6cm]{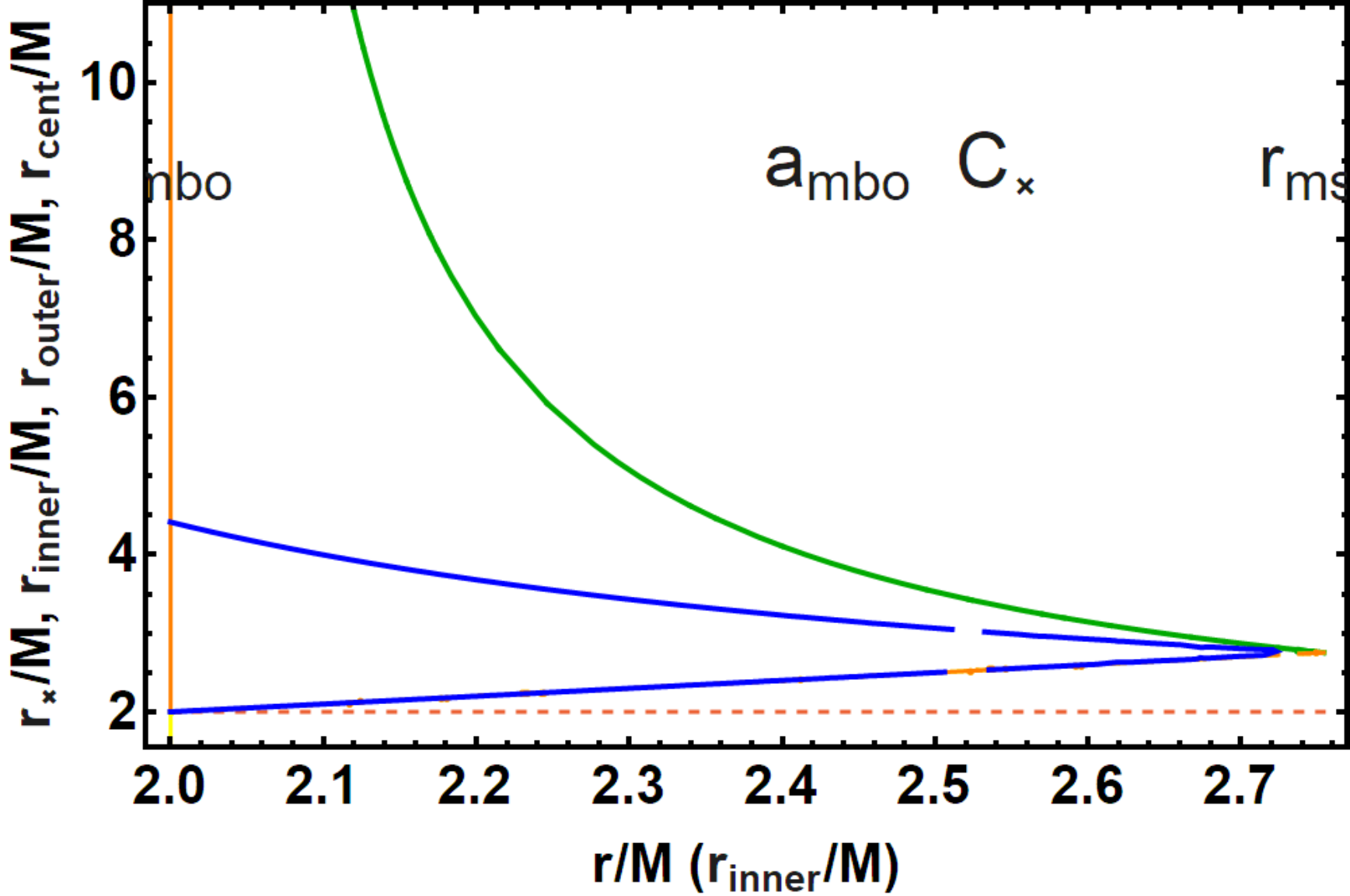}
  \includegraphics[width=5.6cm]{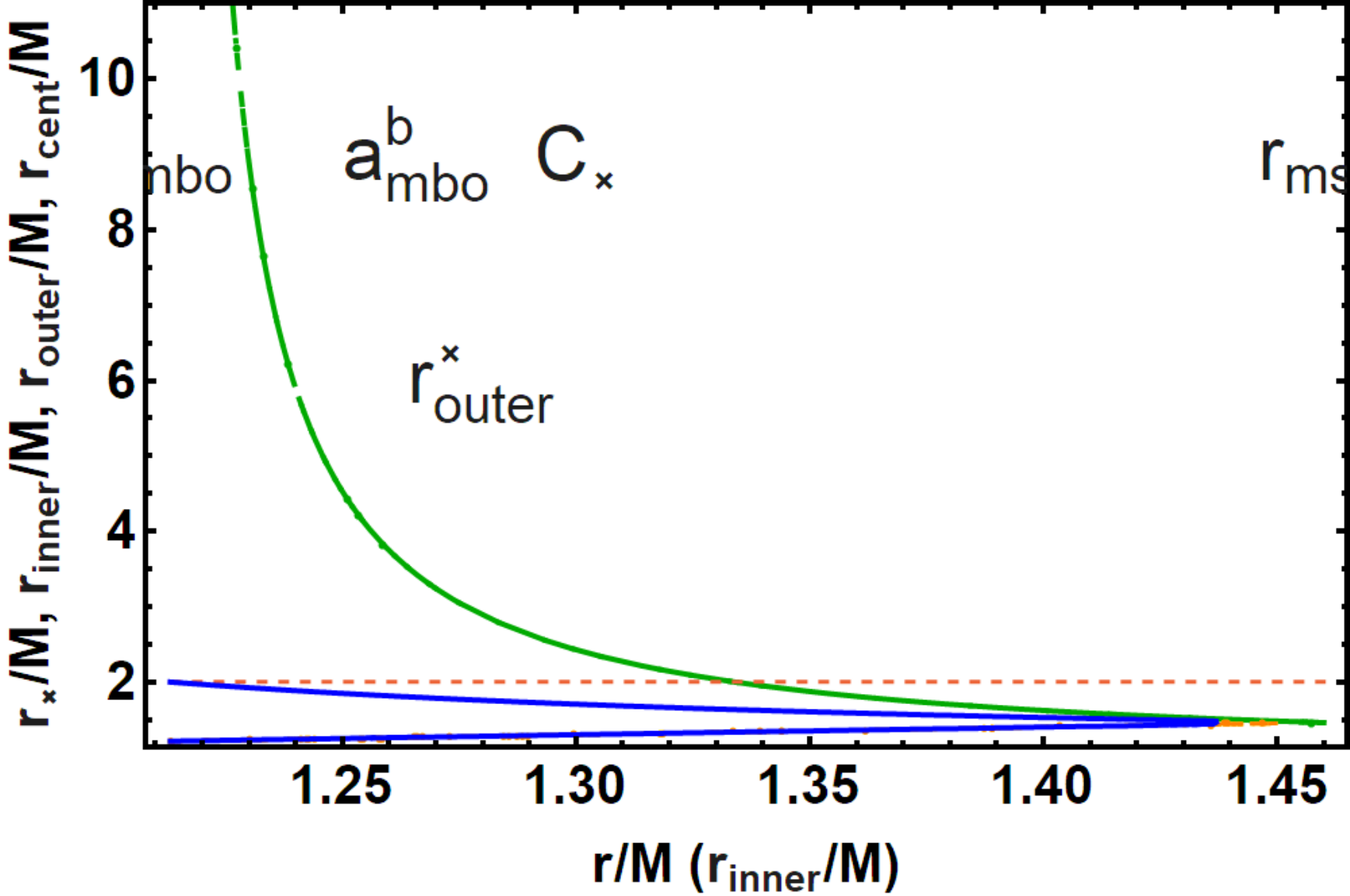}
  \includegraphics[width=5.6cm]{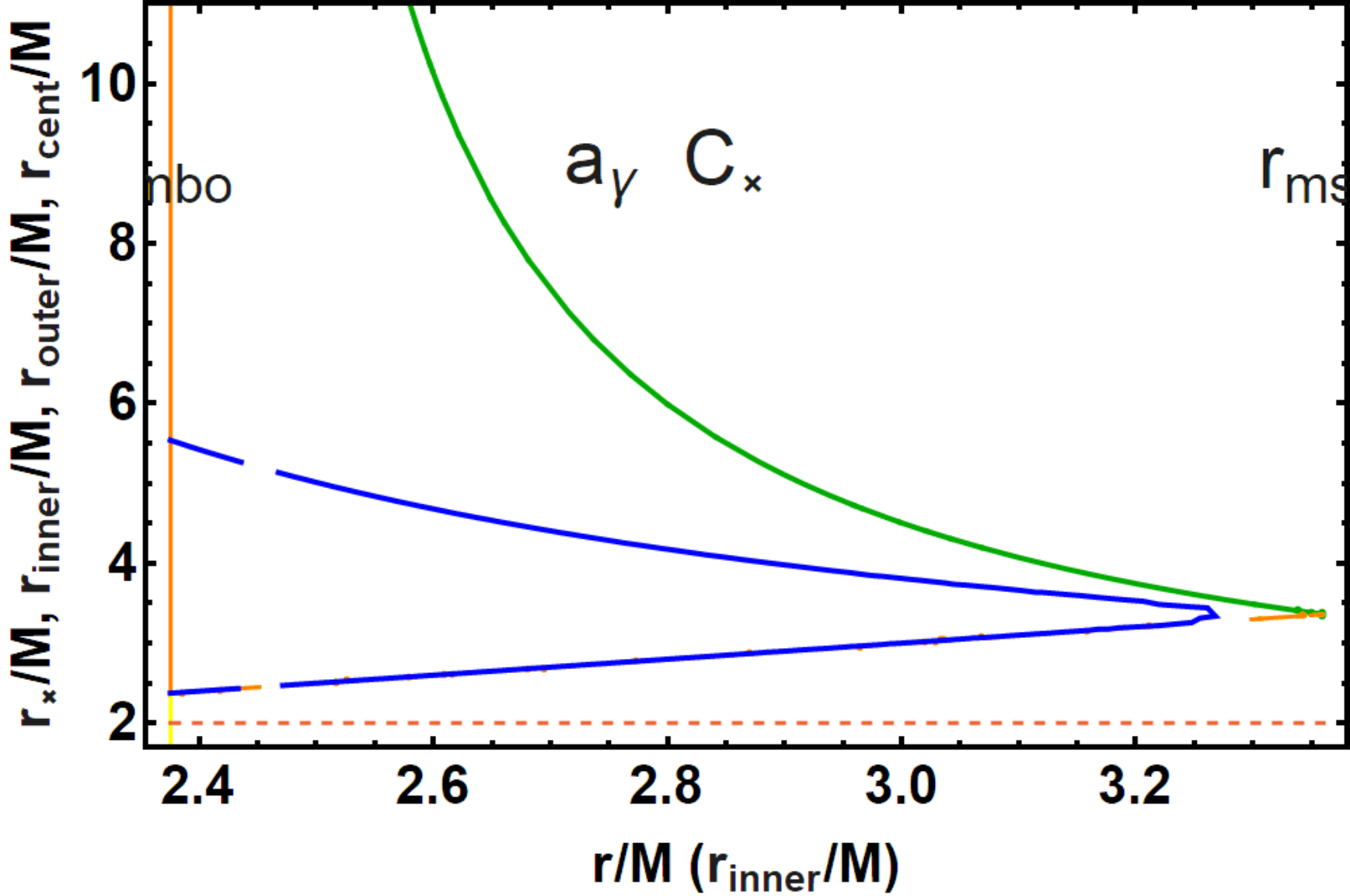}
  \includegraphics[width=5.6cm]{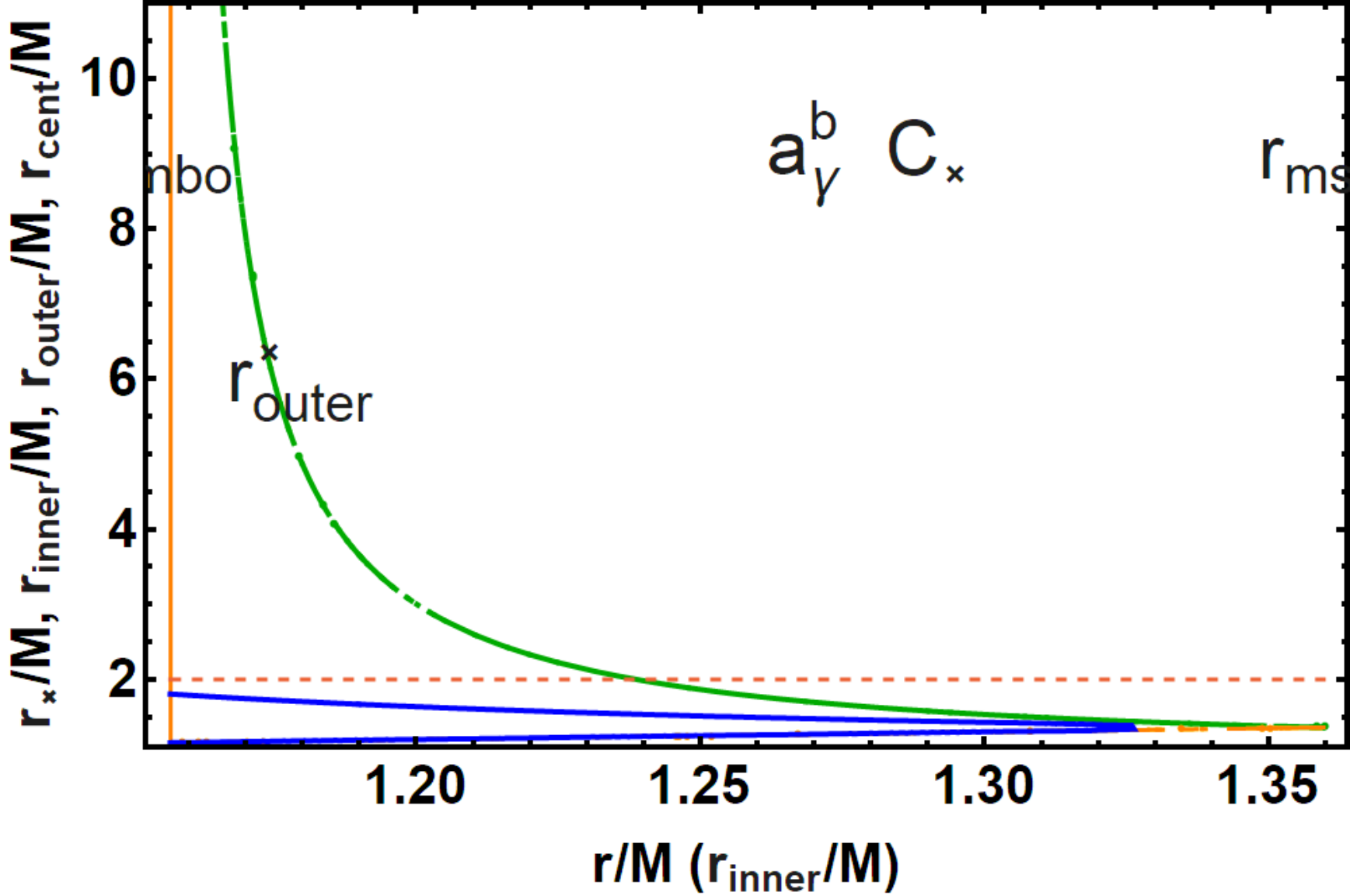}
  \caption{Configurations $\cc_{\times}$ for closed cusped equipressure surfaces where the fluid specific angular momentum is at $\ell\in[\ell_{mso},\ell_{mbo}]$, for selected spins $a/M$, the center $r_{center}$ of the configurations (maximum point of pressure and density), inner edge and cusp $r_{\times}$ (minimum of pressure) and outer edge $r_{outer}$ of the closed surface are shown where $r_{\times}<r_{center}<r_{outer}$. The outer ergosurface $r_{\epsilon}^+=2M$  on the equatorial plane is dashed line. Radii are shown as function of the inner edge $r_{inner}=r_{\times}$ in the ranges $r_{\times}\in[r_{mbo},r_{mso}]$, spins $\mathbf{A}_{\epsilon}^+\equiv\{a_{mso},a_{mbo},a_{mso}^b,a_{\gamma},a_{\gamma}^b\}$ are represented  according to the analysis of Figs\il(\ref{Fig:PlotVampb1}). The plots show the regions where the center $r_{center}$, the cusp $r_{\times}$ and the outer edge $r_{outer}^{\times}$ of the cusped configuration cross the ergosurface and are contained in the ergoregion, consequently it shows also the maximum elongation of the disk on the equatorial plane. The related tori analysis  is in Figs\il(\ref{Fig:polodefin1}).}\label{Fig:PlotVamp1}
\end{figure}
Results of equations $r: \ell(r)=\ell_{\epsilon}^+$ are shown  in Figs\il(\ref{Fig:principmill}) and Figs\il(\ref{Fig:Plotsoorr}),
where there is
the study of the  outer edge  approaching  the  outer ergosurface.
\begin{figure}\centering
  % Requires \usepackage{graphicx}
   \includegraphics[width=8cm]{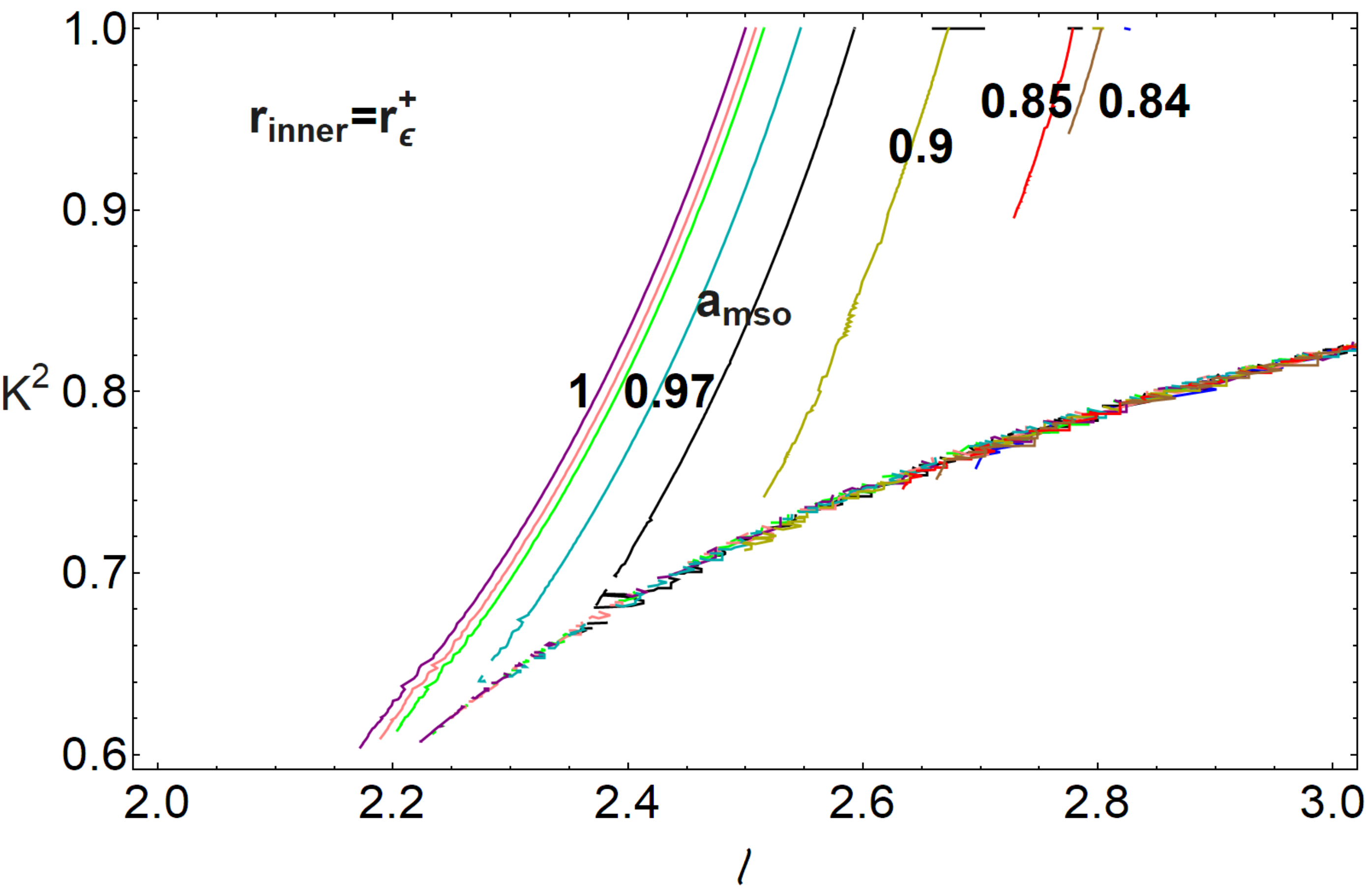}
  \includegraphics[width=8cm]{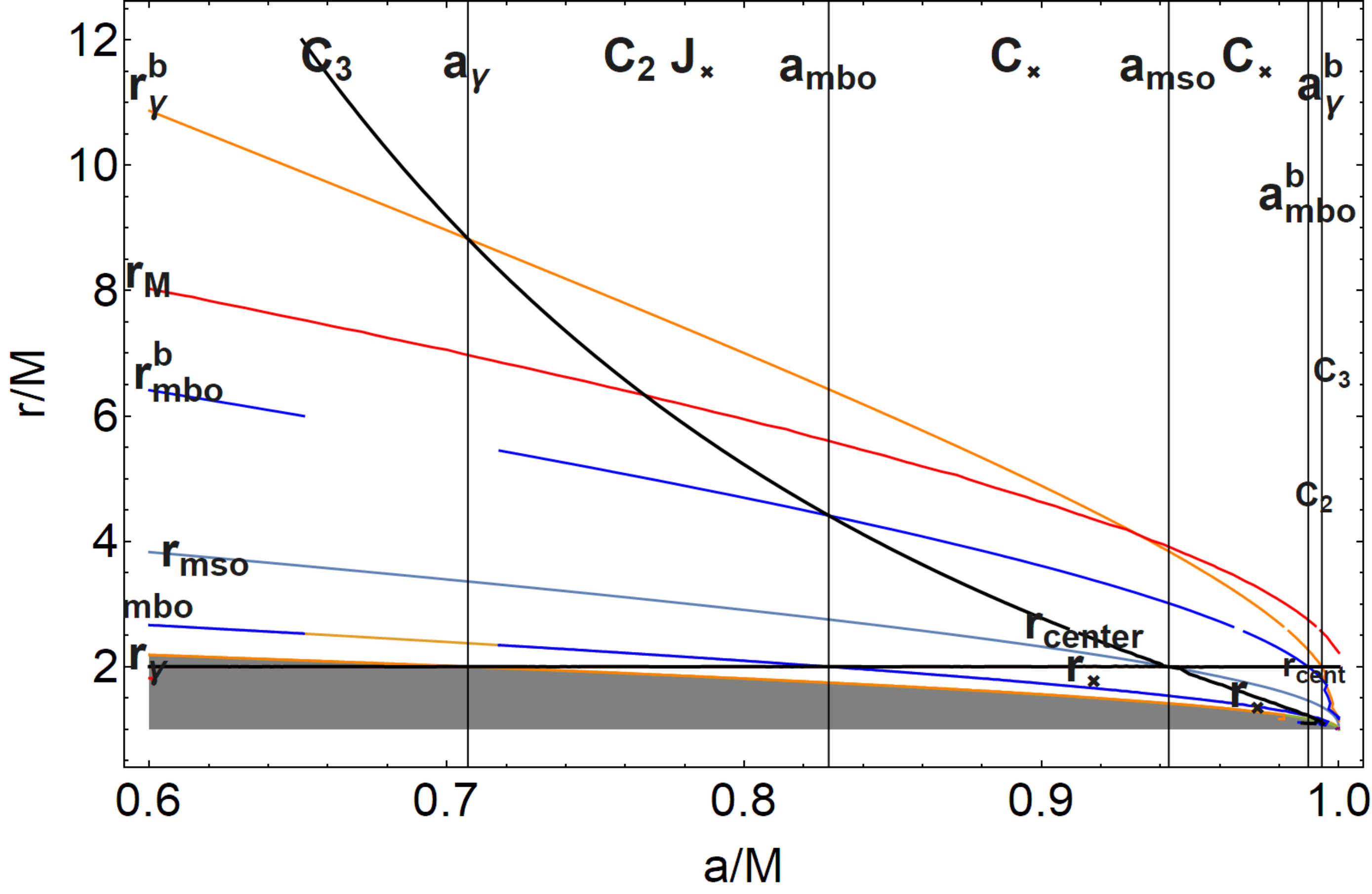}
  \includegraphics[width=8cm]{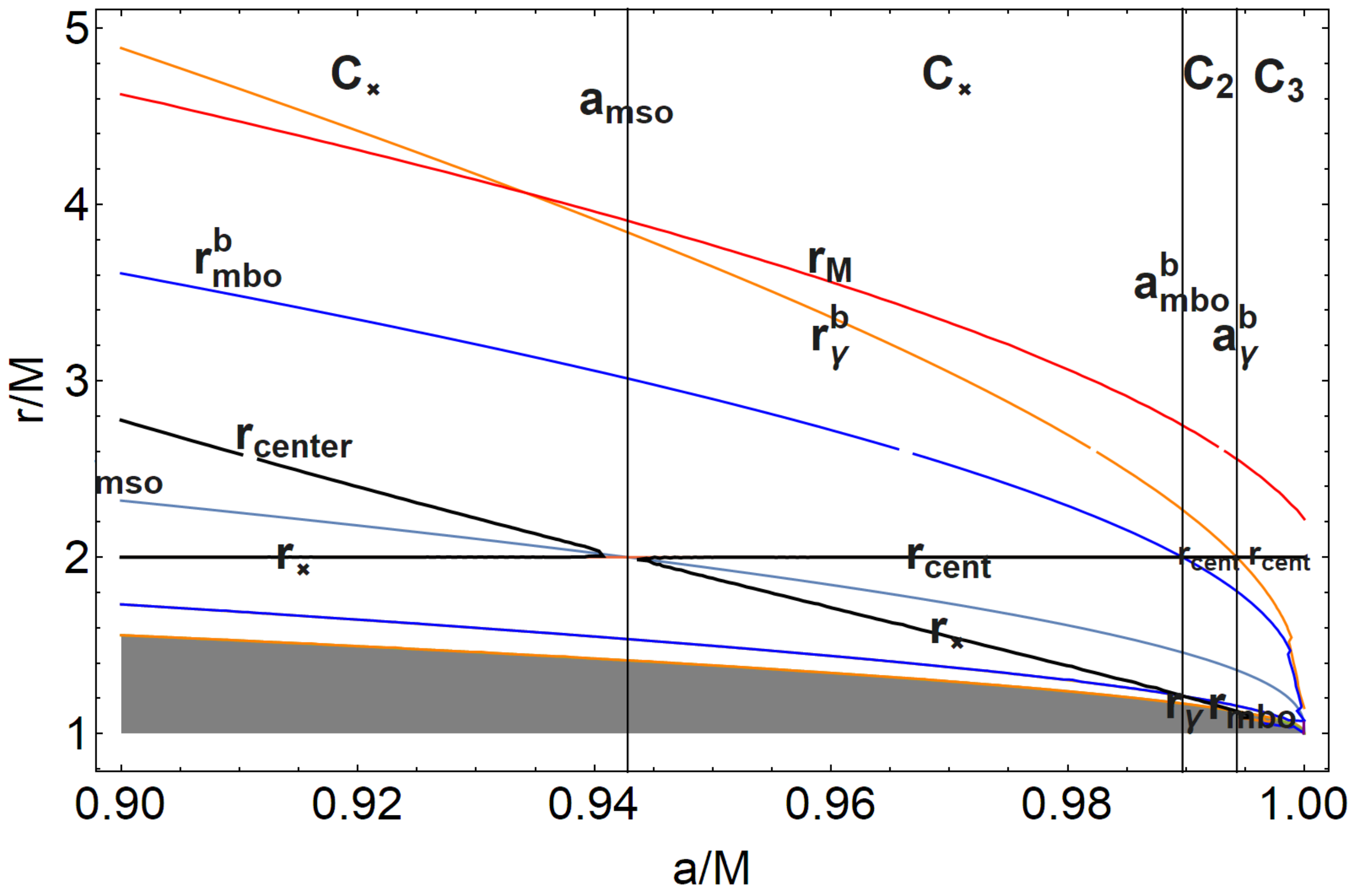}
  \includegraphics[width=8cm]{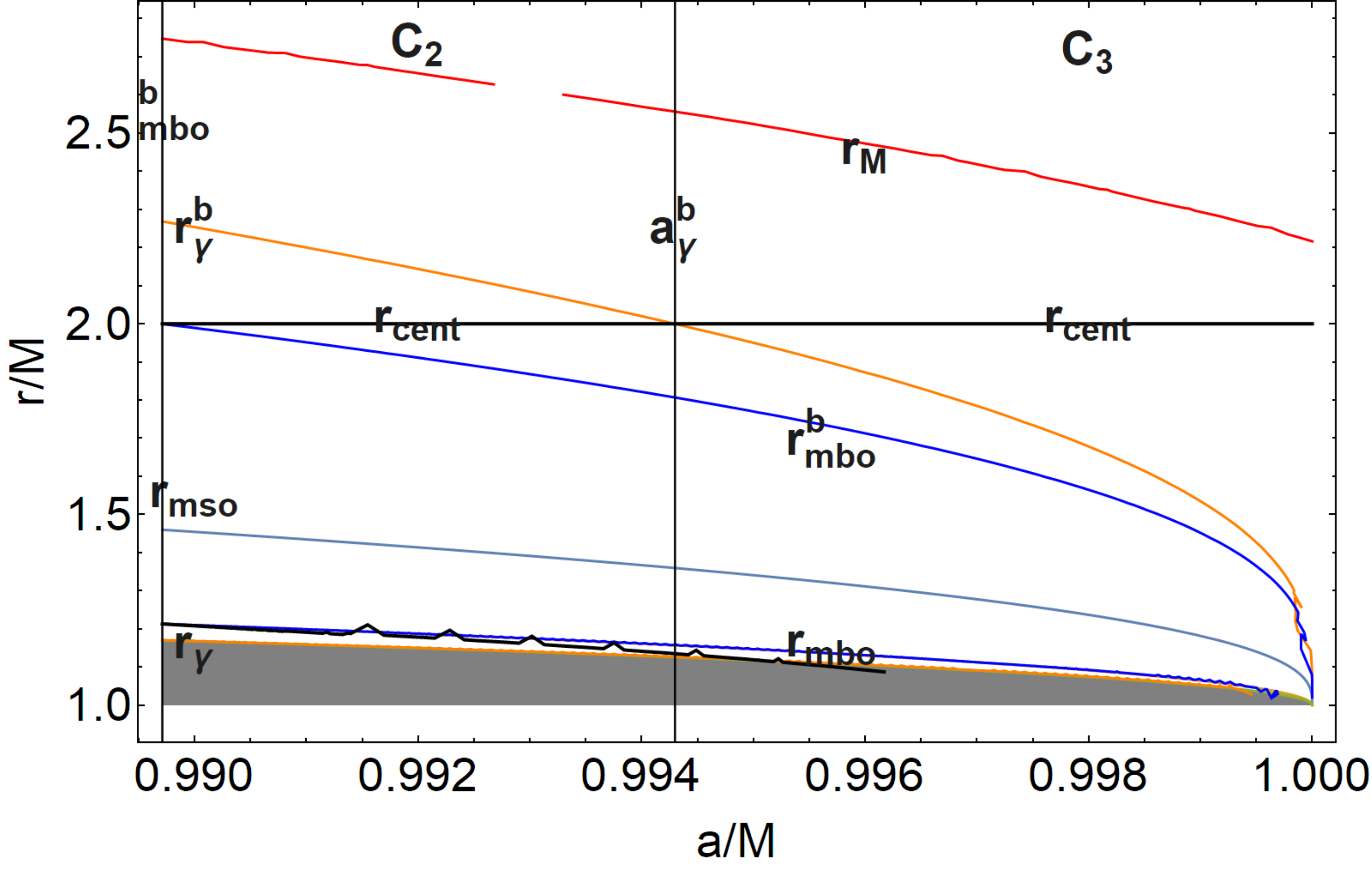}
  \caption{Analysis of tori center and inner edge  with respect to the stationary limit. Left upper panel: Curves show  the parameters $K^2>K_{mso}^2$  and $\ell>\ell_{mso}$ ranges  for  the (quiescent and cusped) tori  inner edge be  coincident with the outer ergosurface $r_{inner}=r_{\times}$,  for different \textbf{BH} dimensionless spin $a/M$
signed  along the curves. Right upper panel, and bottom panels: solutions $r:\ell=\ell(r_{\epsilon}^+)$ (black curves) where   $r_{\epsilon}^+=2M$ is the outer ergosurface. For cusped surfaces, the curves locate  the  cusp of the configuration  whose center is in $r_{center}=r_{\epsilon}^+>r_{mso}$ or, viceversa, curves locate the  center of the torus  whose cusp is in $r_{\times}=r_{\epsilon}^+<r_{mso}$,  according to the different ranges of $r/M$ and \textbf{BH} spins $a/M$.  Location of  center $r_{center}$ and cusp $r_{\times}$ is signed on the black curve. Gray region is $r<r_\gamma$, the  radius $r_{\gamma}$ is  the marginally circular orbit and photon orbit where  $r_{\gamma}^b>r_{\gamma}: \ell(r_{\gamma})=\ell({r})$ (orange curves). Marginally bounded orbit   $r_{mbo}$  and  radius $r_{mbo}^b: \ell(r_{mbo})=\ell({r})$ are the blue curves. Light-blue  curve is  the marginally stable orbit $r_{mso}$.
 Radius $r_M: \partial^2_r \ell=0$ is the red curve.
  Spins $\mathbf{A}_{\epsilon}^+\equiv\{a_{mbo},a_{mbo}^b,a_{\gamma},a_{\gamma}^b,a_{mso}\}$ are represented.
We show the regions where  there can be   quiescent tori with specific angular momentum $\ell\in\mathbf{L_3}$  (i.e. tori $\cc_3$),  quiescent tori   $\cc_2$ and proto-jets   $J_{\times}$ (having specific angular momentum  $\ell\in \mathbf{L_2}$)  and cusped tori  $\cc_\times$  having center or cusp located on $r_{\epsilon}^+$ or regulated by  the relation $r:\ell=\ell(r_{\epsilon}^+)$. Panels shows  different  ranges of spins.
}\label{Fig:Plotssot}
\end{figure}
\begin{figure}\centering
 \includegraphics[width=5.6cm]{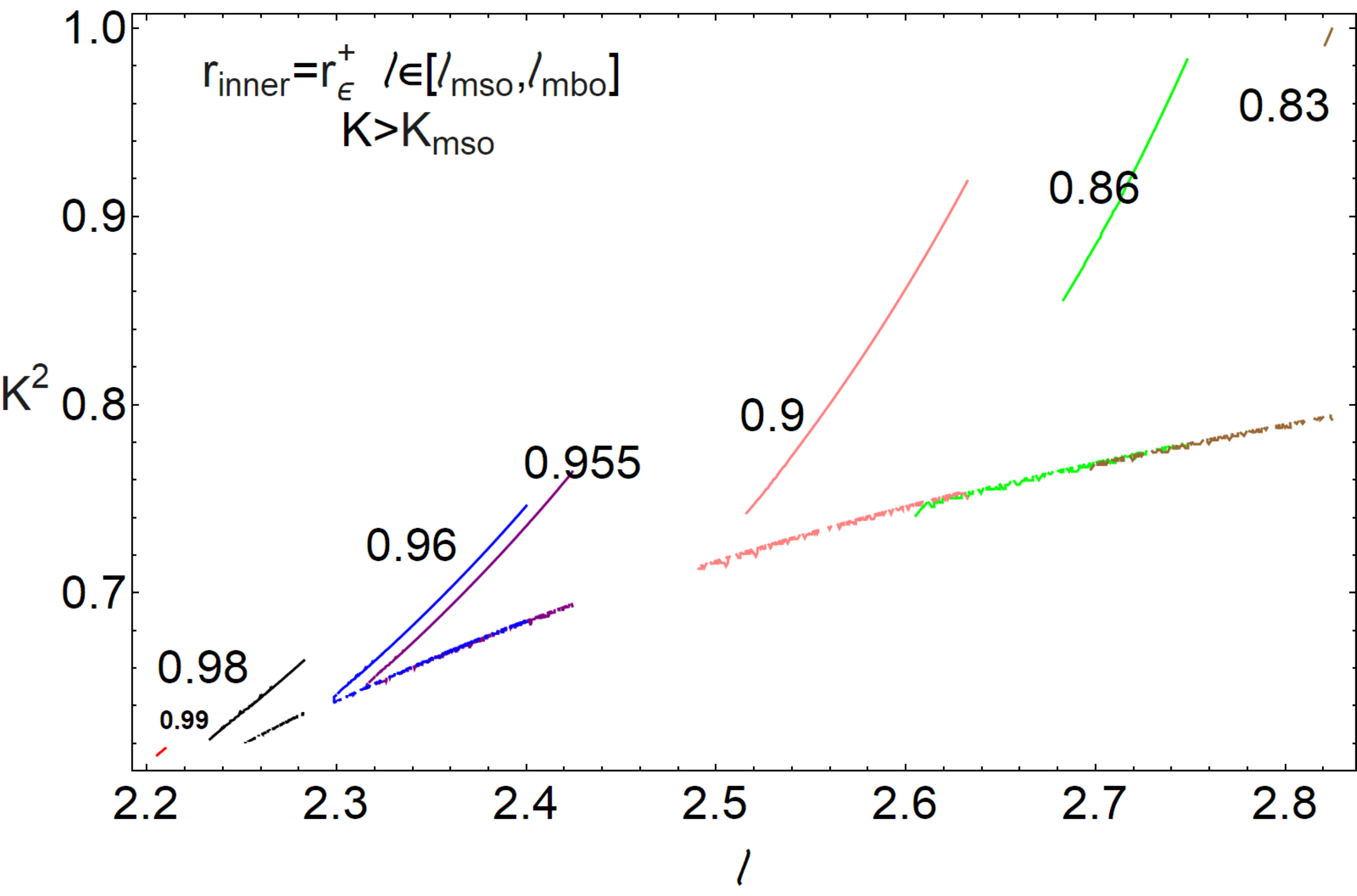}
  \includegraphics[width=5.6cm]{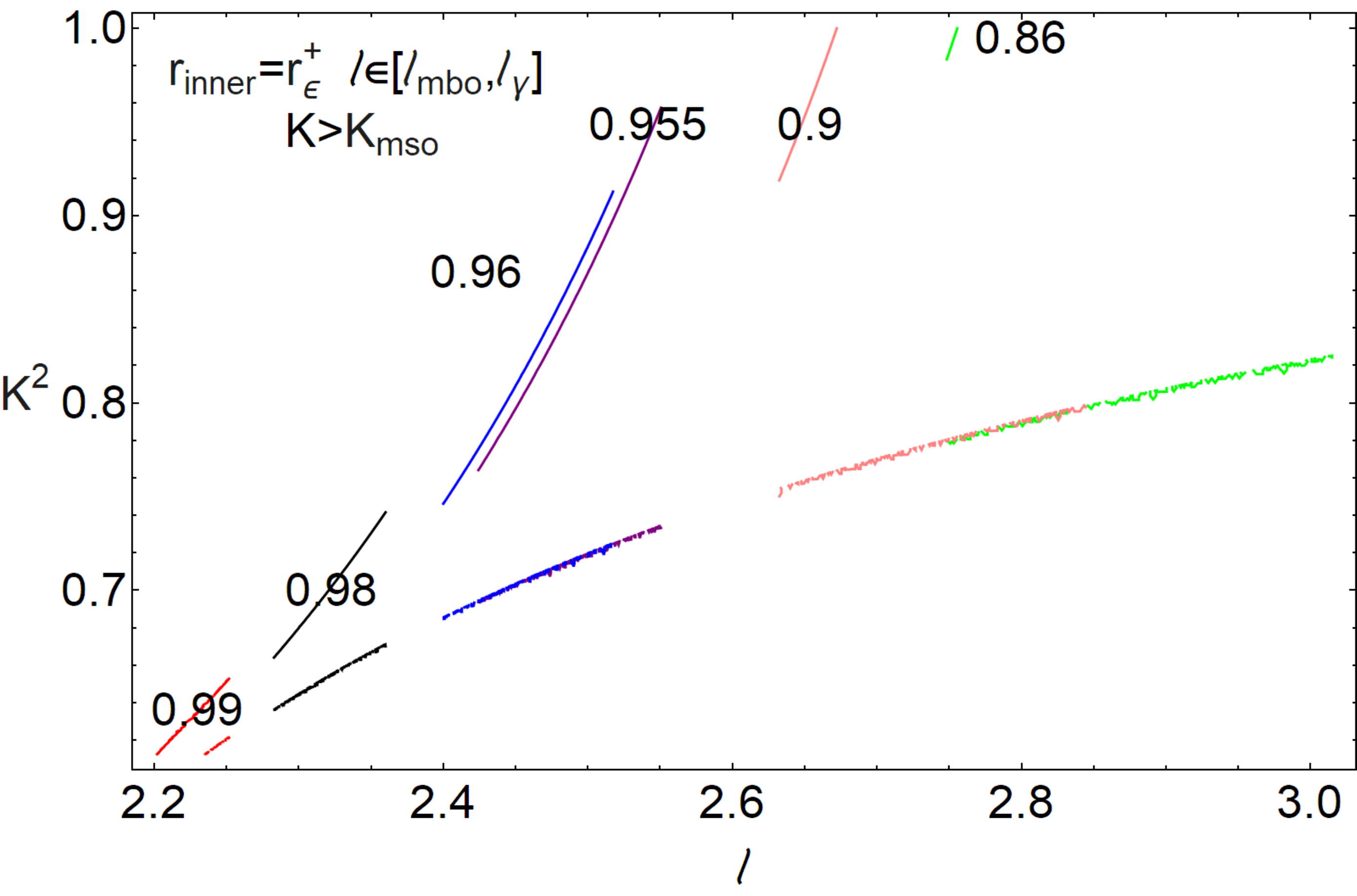}
    \includegraphics[width=5.6cm]{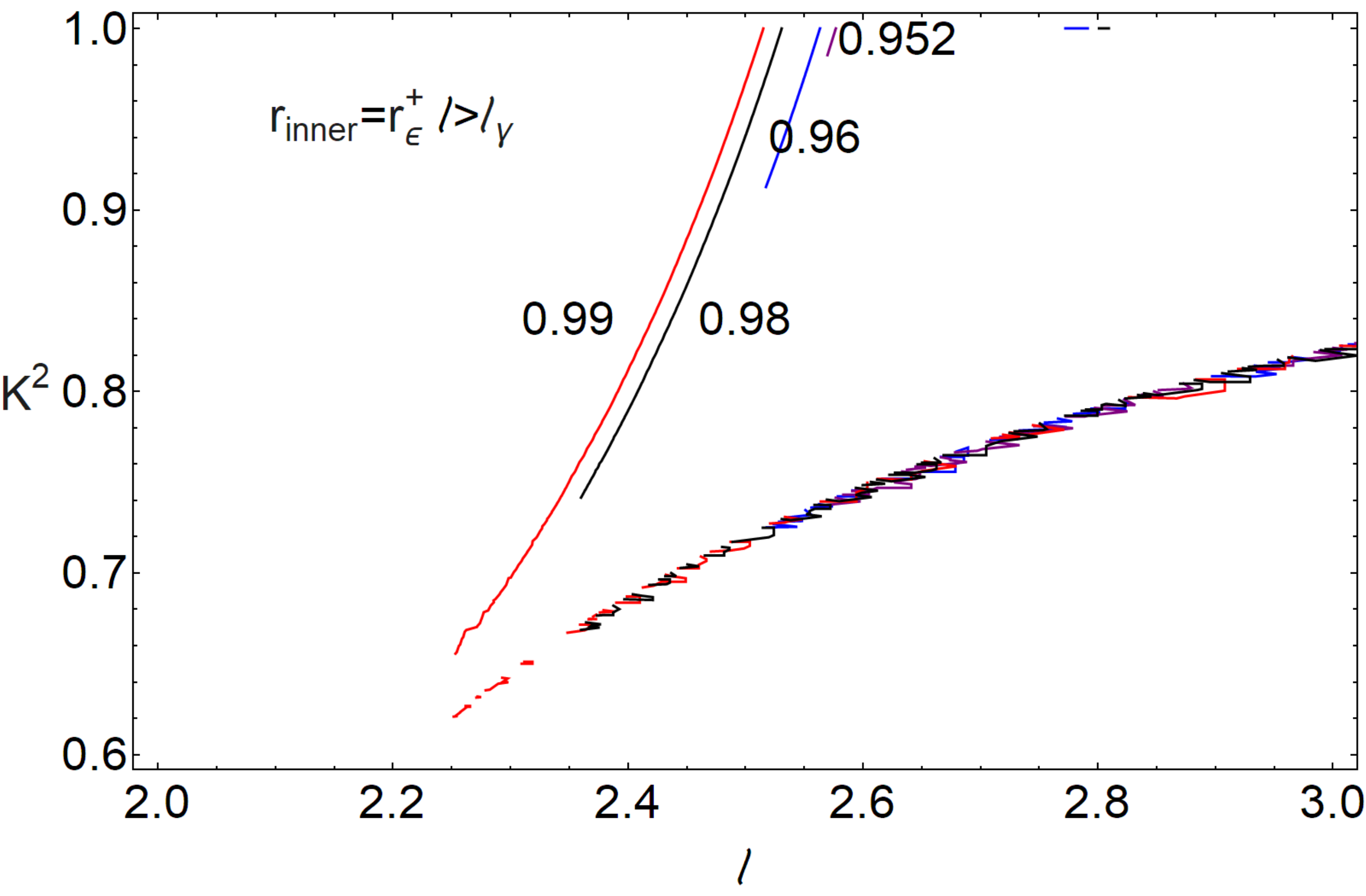}
  \caption{ Solutions $r_{inner}=r_{\epsilon}^+$, inner edge of the torus coincident with the outer ergosurface on the equatorial plane, in terms of the parameter $K^2>K_{mso}^2$ for different \textbf{BH} dimensionless spin $a/M$ signed close to the curve and for specific angular momentum $\ell\in [\ell_{mso},\ell_{mbo}]$ (for  tori $\cc_1$ and $\cc_{\times}$)-left panel;  for $\ell\in [\ell_{mbo}, \ell_\gamma]$ (for  tori $\cc_2$ and  protojets $j_{\times}$)-center  panel and $\ell>\ell_\gamma$  (for  tori $\cc_3$)-right  panel.
}\label{Fig:PlotBlakPurp7}
\end{figure}
\begin{figure}\centering
  % Requires \usepackage{graphicx}
   \includegraphics[width=5.6cm]{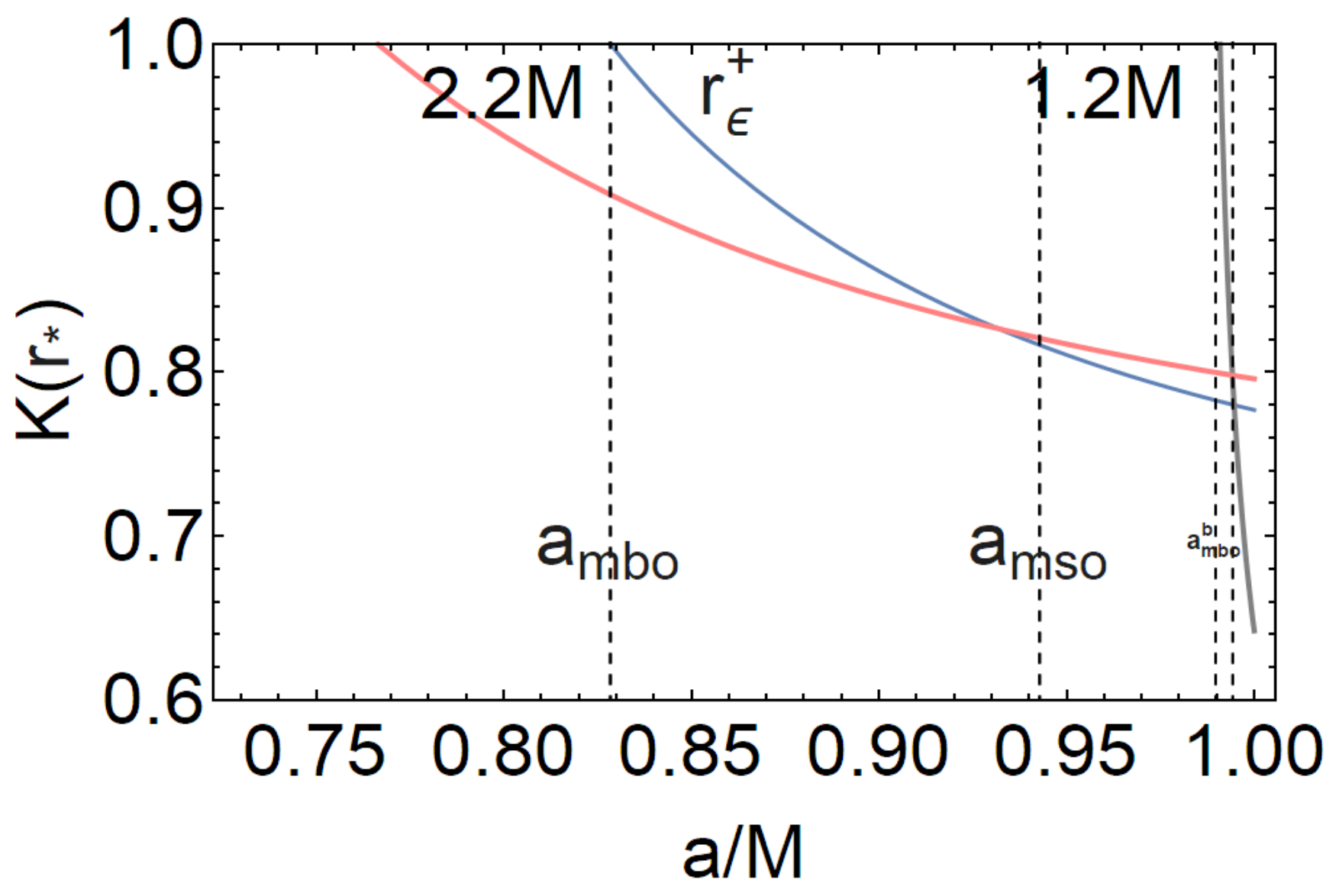}
  \includegraphics[width=5.6cm]{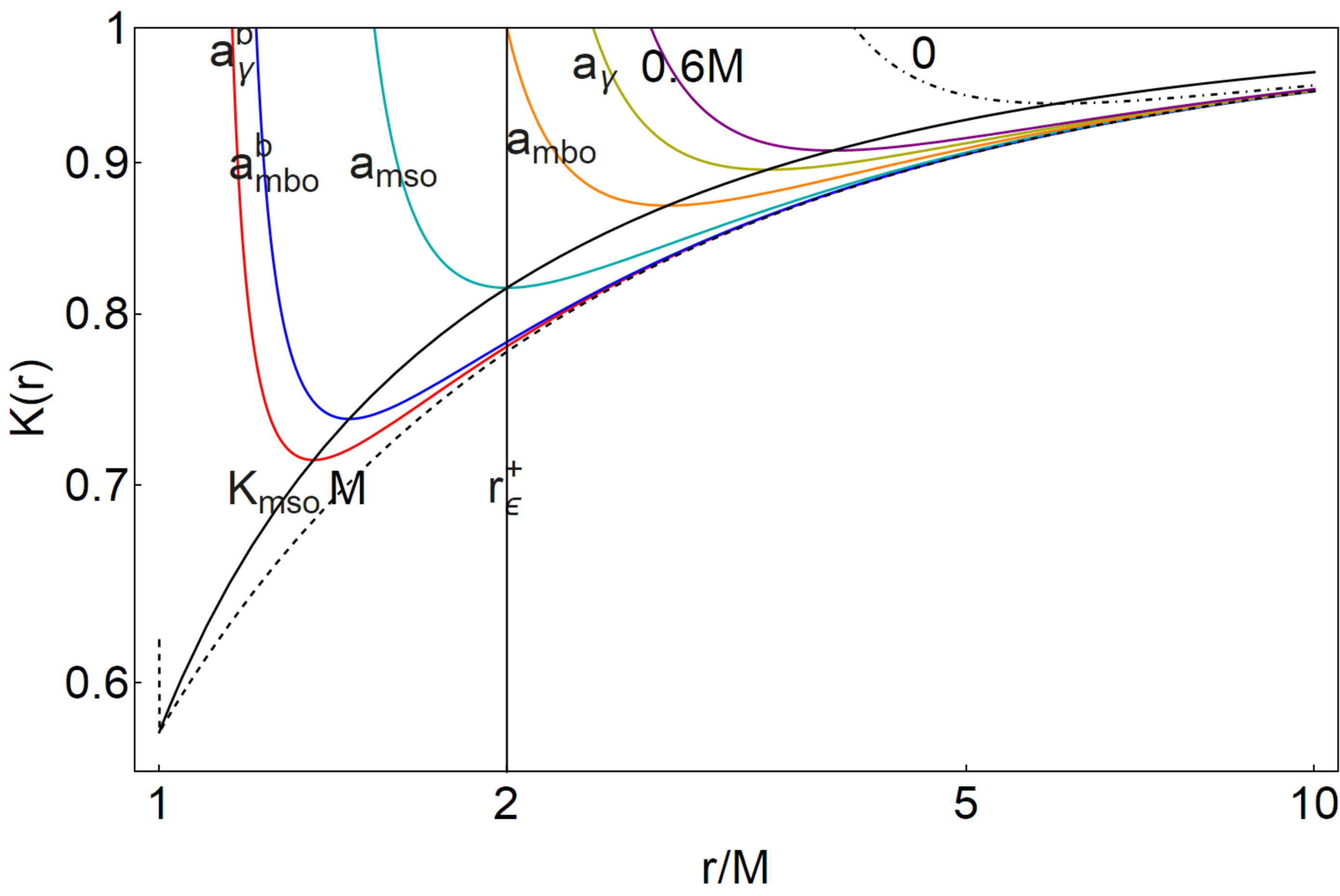}
  \includegraphics[width=5.6cm]{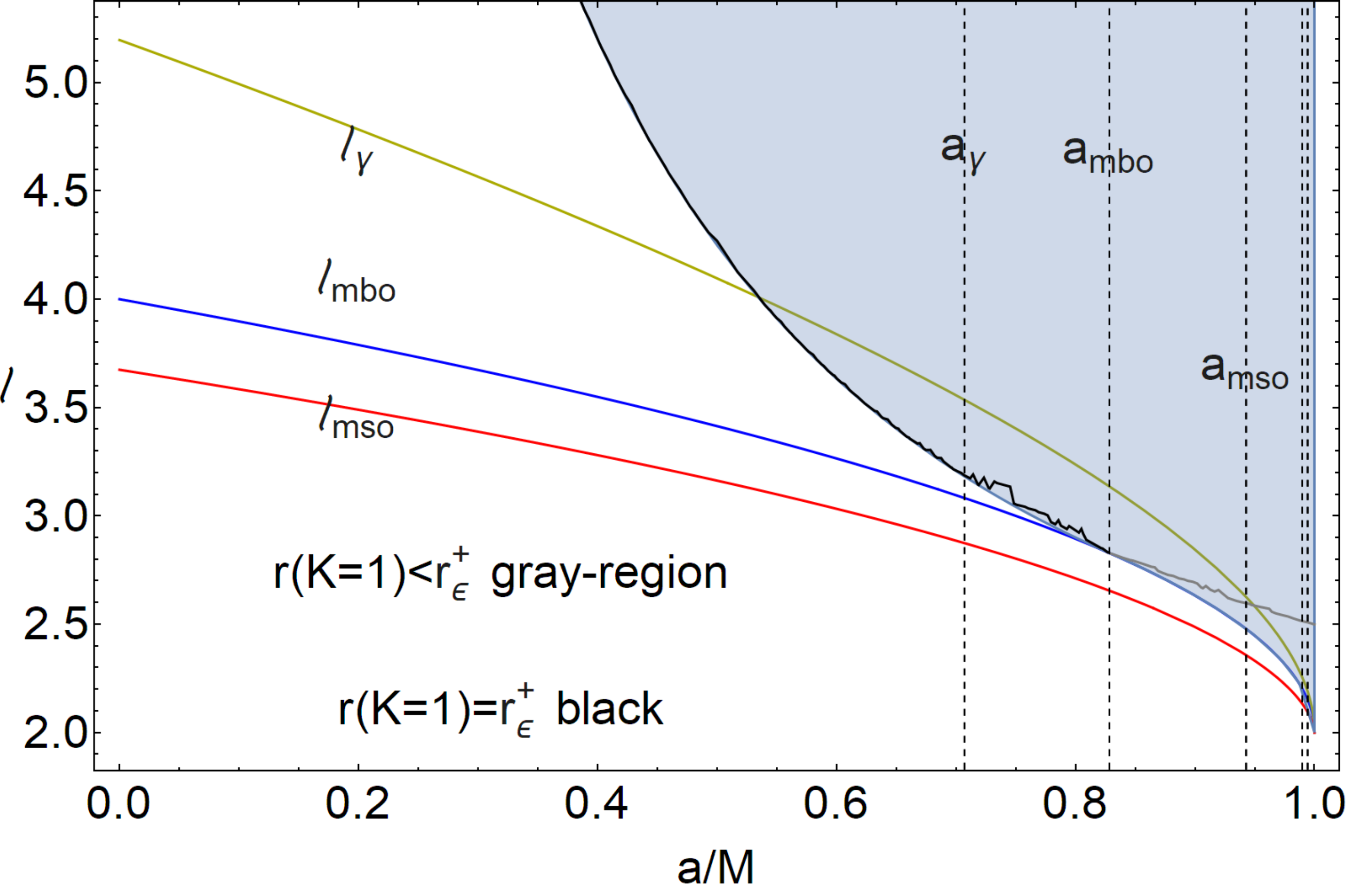}\\
   \includegraphics[width=7cm]{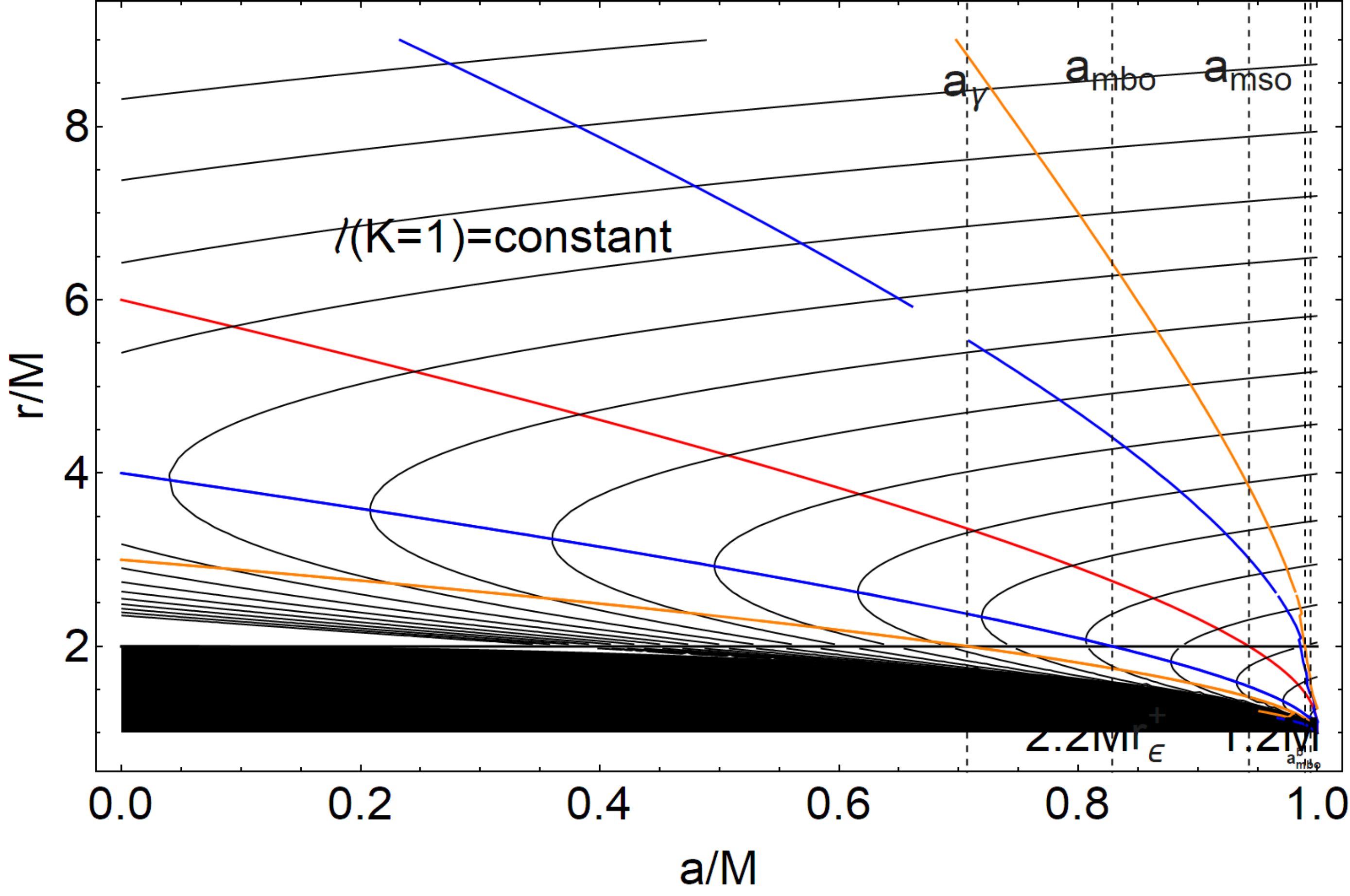}
     \includegraphics[width=7cm]{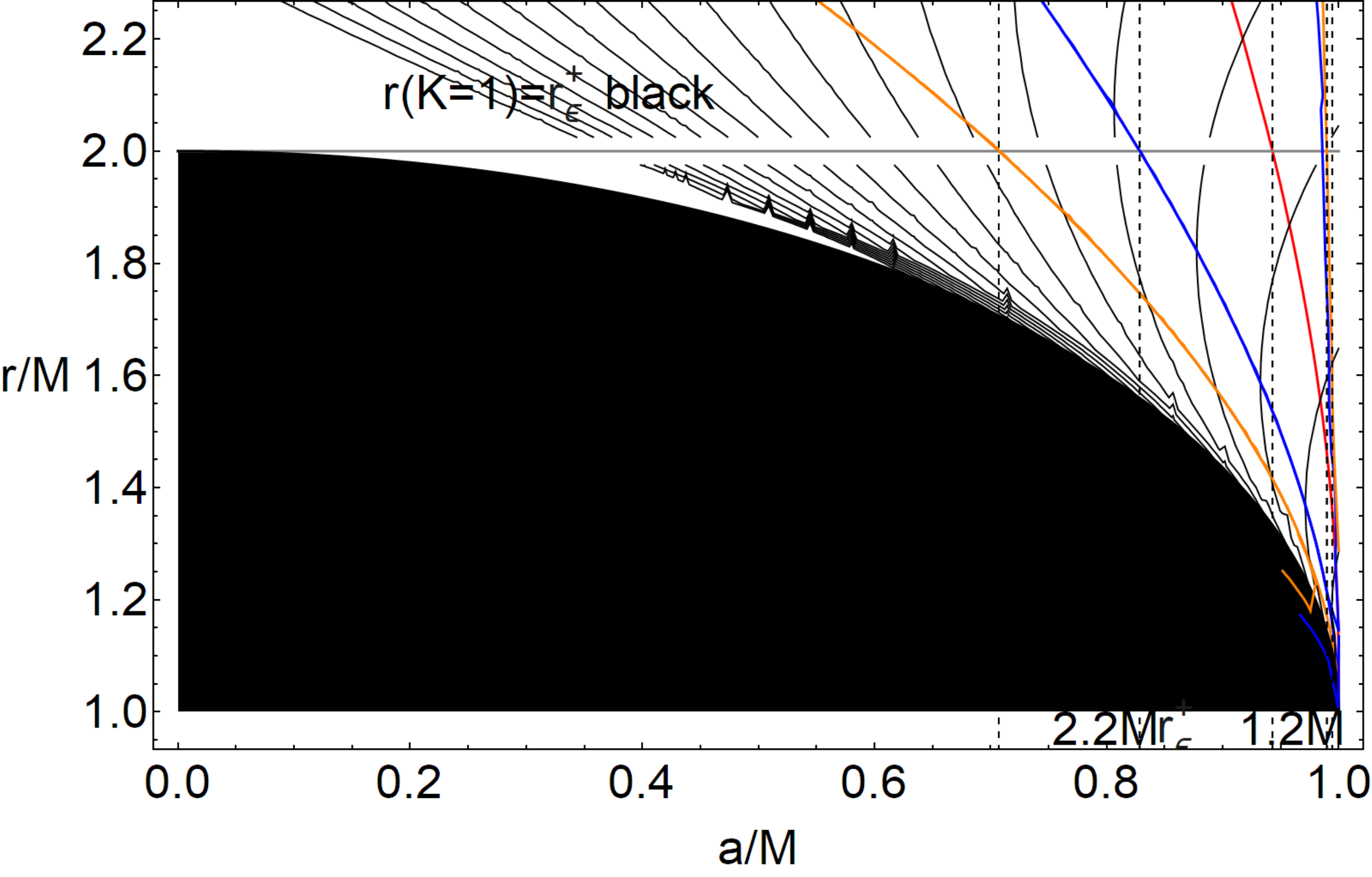}
  \caption{Analysis of solutions $V_{eff}=1$ on the ergosurface, $r_{\epsilon}^+=2M$, on the equatorial plane.
Upper left panel:  solution of
$K(r_*)=1$ for different radii  signed on the curves close to  the outer ergoregion. Dashed lines are the
 spins $a_{mbo}$ and $a_{mso}$.
 Central panel:  plots of $K(r)=V_{eff}(\ell(r),r)$ as function of the radius $r/M$,
 for different spins  of the set $\mathbf{A_{\epsilon}^+}$ signed in  figures.
 Each point for the curves sets  the value of the  $K$ parameter  on the minimum and maximum of the fluid effective potential, maximum and minimum of the pressure in the disk respectively.
 The limiting Schwarzschild spacetime $a=0$ is the dotted-dashed  curve,  the extreme Kerr spacetime
 with spin  $a=M$   is the dashed curve.
 The curve  $K_{mso}(r)\equiv K(r_{mso}(a))$ is shown.
 Left panel:  radius $r(\ell,a): K(r)=1$ is shown, each radius at
$r>r_{mso}$ black curve can be a torus center. Gray region is region
$K(r)<1$. The   fluid specific angular momentum $\ell_{\gamma}=\ell(r_{\gamma}),\ell_{mso}=\ell(r_{mso}),\ell_{mbo}=\ell(r_{mbo})$ are shown
and dashed lines are the spins $\{a_{mso}, a_{\gamma}, a_{mso}\}$.
Below panels: solutions $\ell: V_{eff}=1$, black region is  $r<r_+$, where $r_+$ the outer horizon
Colored curves show $r_{\gamma}<r_{mbo}<r_{mso}<r^b_{\gamma}<r^b_{mbo}$.}\label{Fig:PlotBlakPurp2}
\end{figure}
\begin{figure}\centering
  % Requires \usepackage{graphicx}
   \includegraphics[width=8cm]{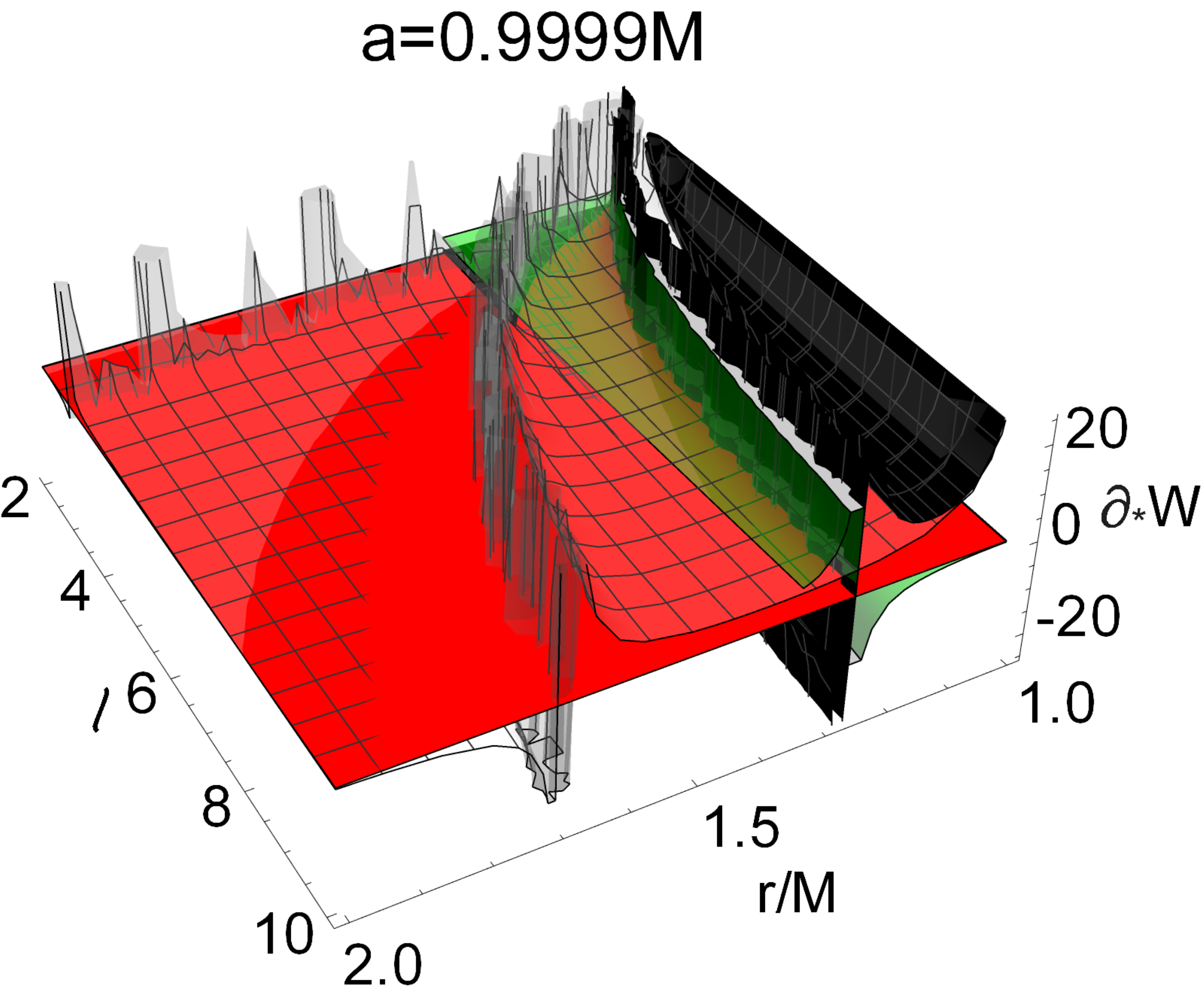}
  \includegraphics[width=8cm]{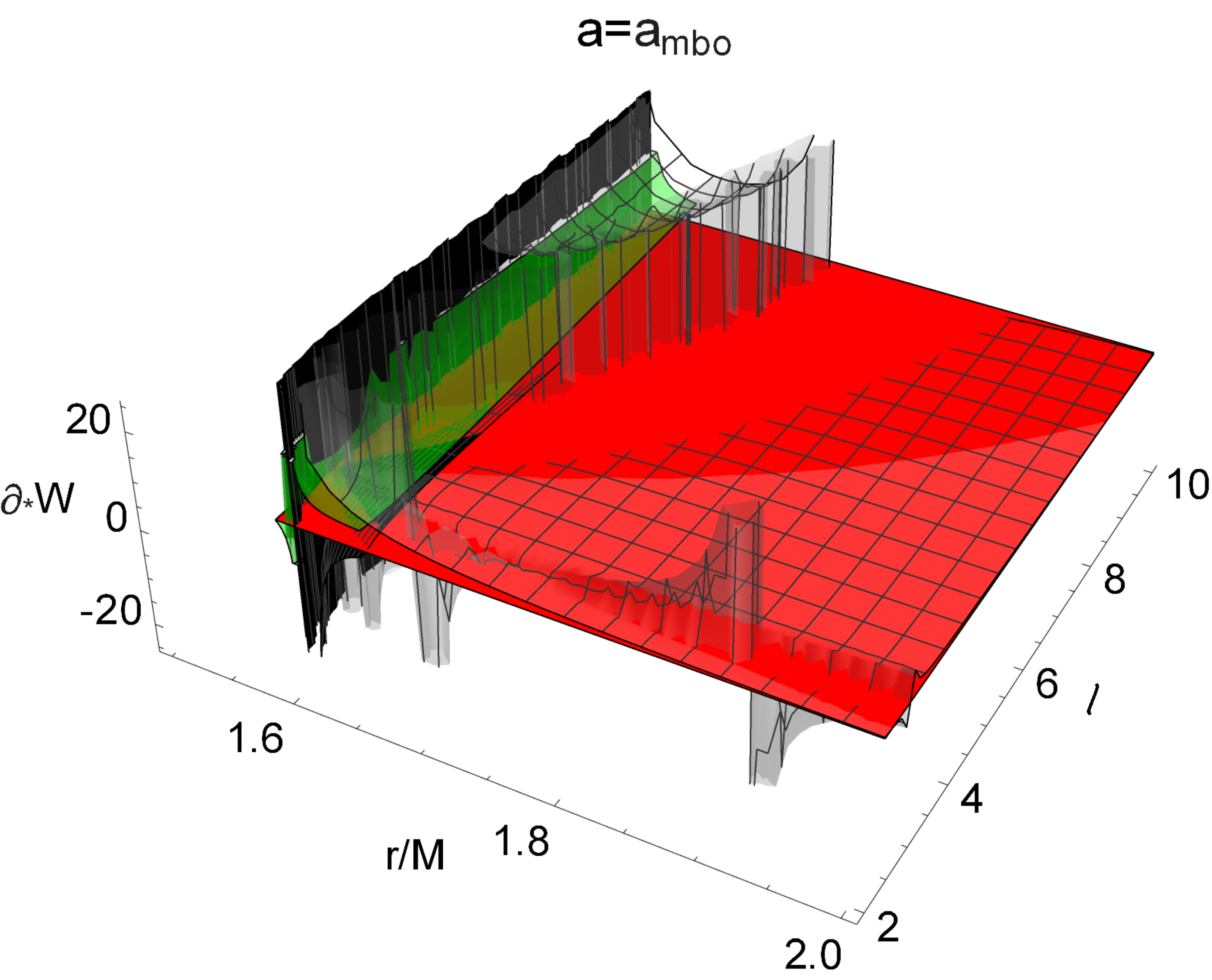}
  \caption{ Analysis of the pressure gradients, in the ergoregion $\Sigma_{\epsilon}^+$  regulated by  Eq.\il(\ref{Eq:scond-d}).
  3D plots show the  function $\partial_* W$ where $W$ is the P-W potential and  $*=r$ (light gray surface for $\theta=\pi/2$, black surface is $\theta=\pi/9$),
   $*=\theta$ (green surface for $\theta=\pi/9$),  for tori orbiting  \textbf{BHs} with spins $a=0.9999M$ and $a=a_{mbo}$.
Plane $\partial_* W=\partial_*p=0$ is shown as red surface. $\ell>\ell_{mso}$ is the fluid specific angular momentum and $r\in [r_+(a),r_\epsilon^+(a,\theta)]=\Sigma_{\epsilon}^+$}\label{Fig:PlotBlakPurp}
\end{figure}
{In Figs\il(\ref{Fig:PlotBlakPurp}) we show  the pressure gradients in the ergoregion $\Sigma_{\epsilon}^+$  according to Eq.\il(\ref{Eq:scond-d}), for different planes and spins.}
\end{description}

\medskip

\textbf{Notes on the  torus inner edge, the static limit and tori topology}

\medskip

\begin{figure}\centering
  % Requires \usepackage{graphicx}
   \includegraphics[width=5.6cm]{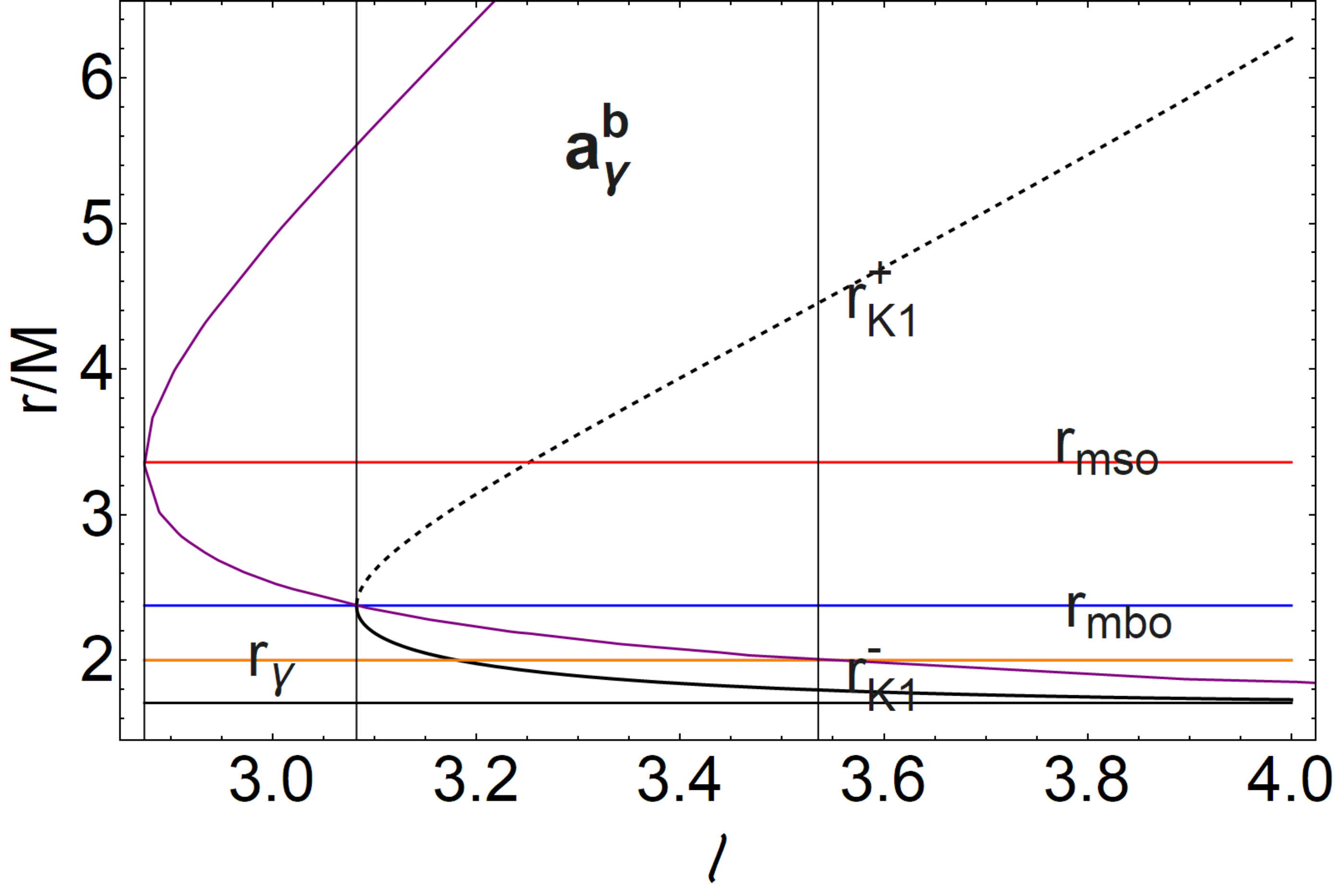}
       \includegraphics[width=5.6cm]{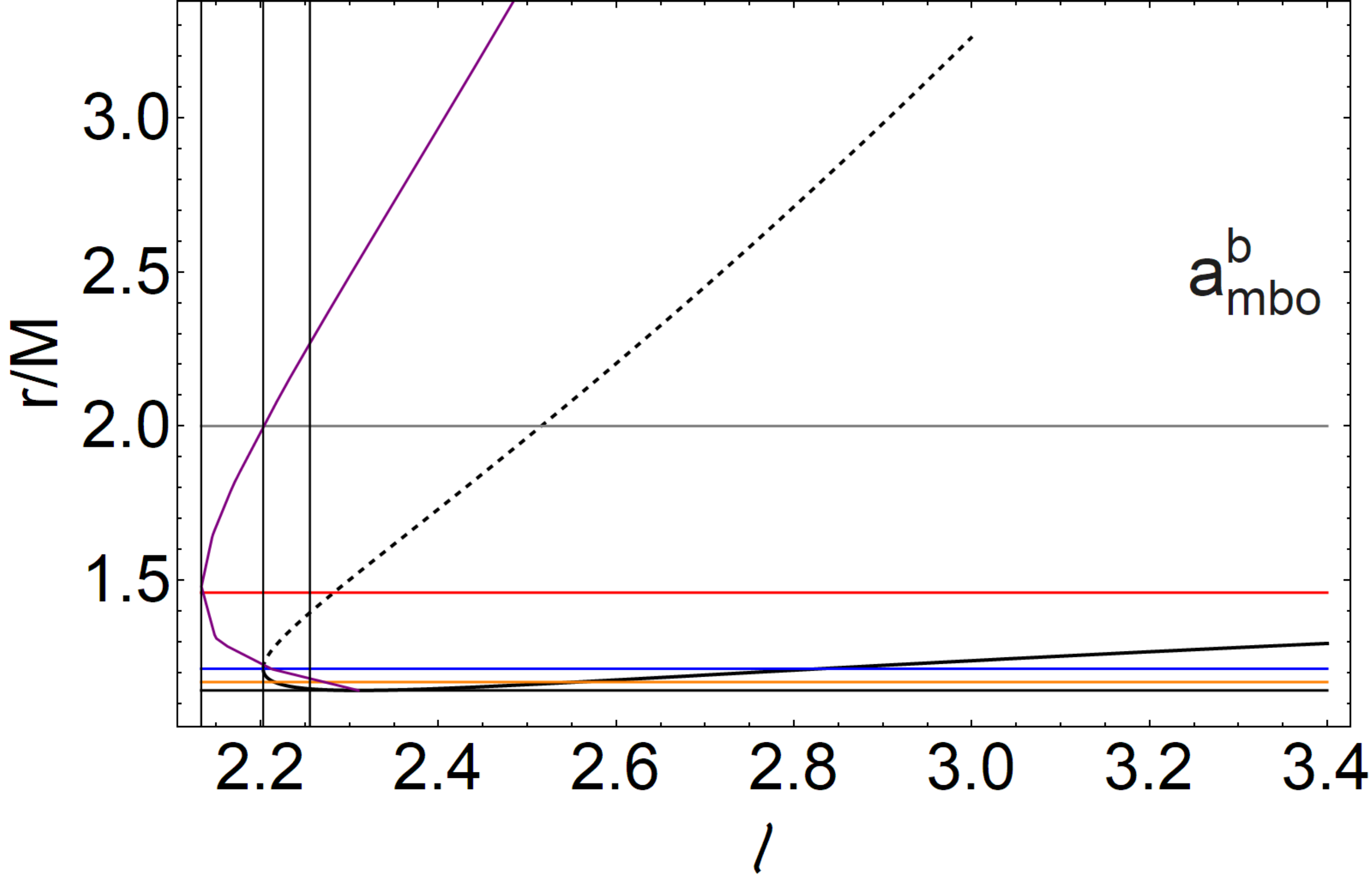}
  \includegraphics[width=5.6cm]{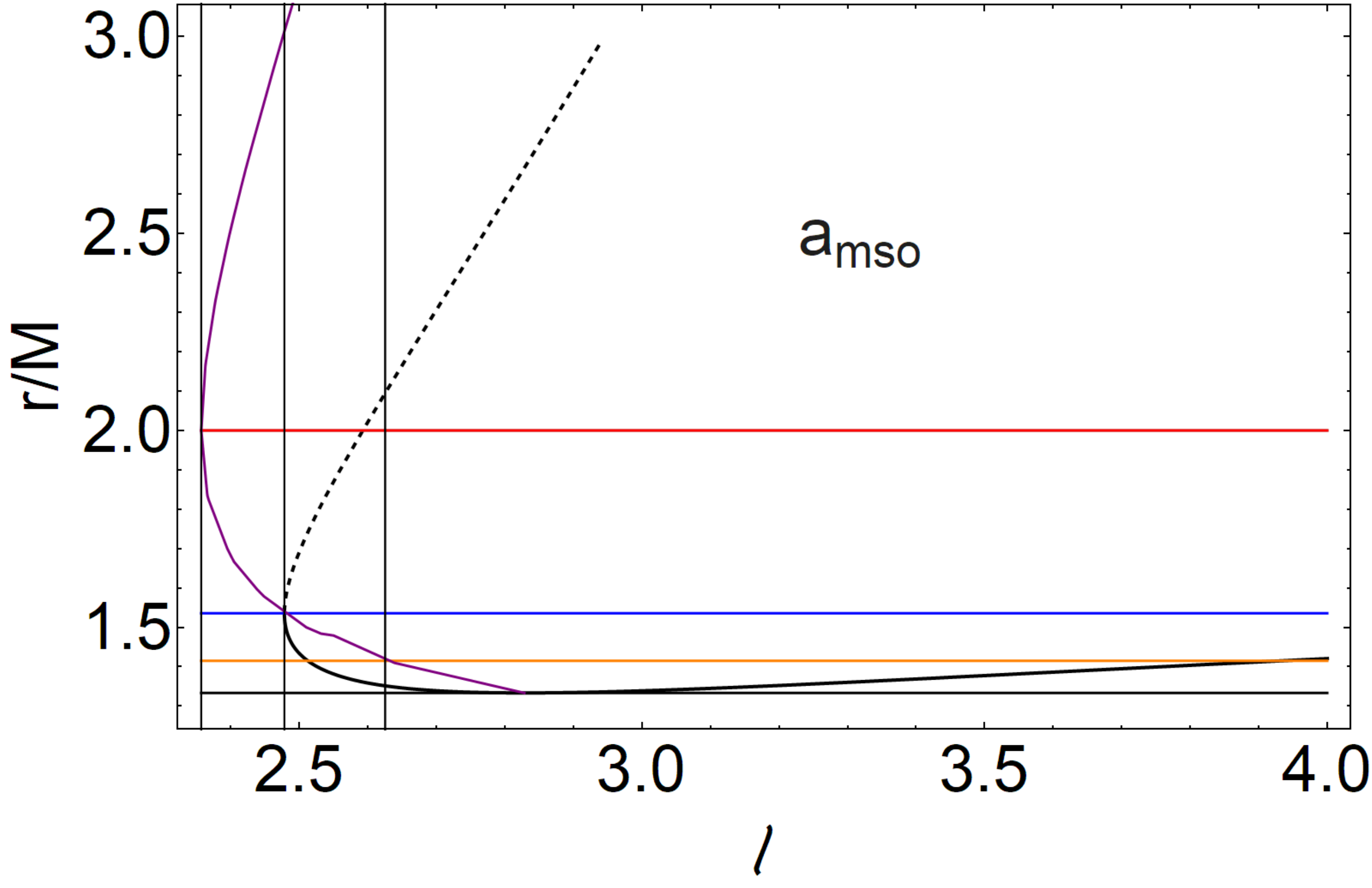}
  \caption{ Plots of radii  $r_{K1}^{\pm}: V_{eff}=1$  of Eq.\il(\ref{Eq:rkpm}) as functions of  the fluids specific angular momentum $\ell$.  Marginally bounded orbit $r_{mbo}$, marginally stable orbit $r_{mso}$ and marginally circular orbit $r_{\gamma}$ with the \textbf{BH} horizon $r_+$ and the outer ergosurface $r_{\epsilon}^+=2M$ on the equatorial plane are  plotted. There is $a_{\gamma}^b>a_{mbo}^b>a_{mso}$, see  Eqs\il(\ref{Eq:strateg}). Purple curve is solution $r:\ell=\ell(r)$ for the critical points (cusps) and center of maximum pressure in the tori. Vertical lines are $\ell_{mso}<\ell_{mbo}<\ell_{\gamma}$.}\label{Fig:titlePlot}
\end{figure}

From Figs\il(\ref{Fig:Plotssot})--right panel,  we see that condition  $r_{inner}=2M$   holds only  for tori  orbiting  \textbf{BHs}  with sufficiently large  spin-mass ratios ($a\geq a_{mbo}$),  where  $\ell$
and  $K$ are progressively smaller with  increasing $a/M$.
 Figs\il(\ref{Fig:Plotssot}) detail  a  complicated  situation where  there are two classes, $\mathbf{\mathbf{\mathbf{(1)}}}$ and $\mathbf{(2)}$, of $\cc_\times$, $\cc_2$ and $\cc_3$ configurations  (defined according to the specific angular momentum classes in Sec.\il(\ref{Sec:model})) such that  $\cc_3\mathbf{\mathbf{(1)}}<\cc_2\mathbf{\mathbf{\mathbf{(1)}}}<\cc_{\times}\mathbf{\mathbf{\mathbf{(1)}}}<\cc_{\times}\mathbf{(2)}<\cc_2\mathbf{(2)}<\cc_3\mathbf{(2)}$, where the ordering  relation expresses the  ordering   relation  between the central \textbf{BH} attractor  dimensionless spins of the geometries  where the configurations  are orbiting.
Therefore, for sufficiently small spins,   $a<a_{\gamma}$, solutions $r:\ell=\ell(r_{\epsilon}^+)$ define    quiescent tori $\cc_3\mathbf{\mathbf{\mathbf{(1)}}}$  which are generally very huge.
Whether the quiescent tori $\cc_3\mathbf{\mathbf{\mathbf{(1)}}}$, orbiting   small spin attractors,   can have inner edges  approaching, or even crossing,  the static limit, remains to be clarified. It is clear that as for the cases of  $\cc_2$ tori, the limiting condition  is  $V_{eff}(r_{\epsilon}^+)=1$ (or more precisely $V_{eff}(r_*)=1$ where $r_*\leq r_{\epsilon}^+$)  whose properties are described  in Figs\il(\ref{Fig:PlotBlakPurp2},\ref{Fig:titlePlot}).

Solutions of the problem $V_{eff}=1$ are
\bea&&\label{Eq:rkpm}
r_{K1}^{\pm}\equiv\frac{1}{4} \left(\ell^2\pm\sqrt{\ell^4-16 \ell^2+32 a \ell-16 a^2}\right), \quad\mbox{alternately} \quad
a_{K1}\equiv\ell-\frac{\sqrt{r \left(\ell^2-2 r\right)}}{\sqrt{2}}\\&&\nonumber \mbox{and}\quad \ell^{\pm}_{K1}\equiv\frac{\pm\sqrt{2} \sqrt{r \Delta}-2 a}{r-2},
\eea
see Figs\il(\ref{Fig:titlePlot}).
In the case of torus  $\cc_2$, we consider the limiting case of the proto-jet  cusp on the  static limit and, therefore, the quiescent disks could be at a relatively large distance from the static limit.
For the cusped  configurations,   the cusp is  the lower extreme of the inner edge.
However, the disks $\cc_3$ although being  (always) centered more external than the  $\cc_2$ and $\cc_1$ tori, they  might  have the   inner edges  lower than the tori $\cc_2$ (where $r_{inner}\geq r_{j}\geq r_{\gamma}$) and $\cc_1$
(where $r_{inner}\geq r_{\times}\geq r_{mbo}$).
As clear also from the analysis of Figs\il(\ref{Fig:spessplhoke1},\ref{Fig:PlotVamp1},\ref{Fig:Plotssot},\ref{Fig:PlotBlakPurp7}), this is not the case for $\cc_3\mathbf{\mathbf{\mathbf{(1)}}}$ an $\cc_2\mathbf{\mathbf{\mathbf{(1)}}}$ tori.
 For $\cc_2\mathbf{\mathbf{\mathbf{(1)}}}$ tori  \textbf{BH}  with  $a\approx a_{mbo}$ is a limiting attractor  as
there is $r_{mbo}<r_{\epsilon}^+$ for
 $a>a_{mbo}$, where   tori of any topology can have, under proper conditions on $\ell$ and $K$, inner edge or center on $r_{\epsilon}^+$.
Figs\il(\ref{Fig:spessplhoke1},\ref{Fig:PlotVamp1},\ref{Fig:Plotssot},\ref{Fig:PlotBlakPurp7})  provide also an indication on the  inner region elongation, $\lambda_{inner}\equiv r_{center}-r_{inner}$. Tori  can have   large inner  elongation for small  \textbf{BH} spin.

Therefore,  including also range  $\mathbf{A}_0\equiv a<a_{\gamma}$, there is:
\begin{description}
\item[--]
For $\mathbf{A}_0: a<a_{\gamma}$, there are no tori with inner edge on $r_{\epsilon}^+$ as follows from  the analysis of Figs\il(\ref{Fig:PlotVamp1},\ref{Fig:spessplhoke1},\ref{Fig:Plotssot},\ref{Fig:PlotBlakPurp7}).
 Whereas there is $r_{\gamma}>r_{\epsilon}^+$, therefore a proto-jet cusp  can approach the static limit.
 The analysis of tori $\cc_3$ and $\cc_2$  with inner edge close to the static limit (or for $\cc_2$ on the static limit) is regulated by the solution of the problem $V_{eff}(r_{\epsilon})\leq1$ for these tori (note tori $\cc_1$ cannot have inner edge on the static limit).
\item[--]
For    $\mathbf{A}_1:] a_{\gamma}, a_{mbo}[$
there is  $r_{mbo}> r_{\epsilon}^+$. The  cusp of $\cc_1$  tori is always out of the ergoregion, approaching the static limit, where can be however   a proto-jets cusp. Therefore, there  is $r_{\epsilon}^+\in  [r_{\gamma}, r_{mbo}]$, and  to prove that $r_{\epsilon}^+$ can be the inner edge of a  quiescent torus $\cc_2$, we need to asses if $V_{eff}(r_{\epsilon}^+)<1$ for $\ell\in \mathbf{L_2}$ in these geometries,which is done in Figs\il(\ref{Fig:PlotBlakPurp2},\ref{Fig:PlotBlakPurp}) showing that this is not the case.
Tori $\cc_2$ are defined in two sets of configurations with centers in $r_{center}\in[r_{\gamma}^b, r_{\Mie}]$, and  in $r_{center}\in[r_{\Mie}, r_{mbo}^b]$ respectively.
\item[--]
For    $\mathbf{A}_2: [a_{mbo},a_{mso}]$  there are tori  $\cc_1\mathbf{\mathbf{\mathbf{(1)}}}$ whose cusp can be  on the static limit as there is
$r_{\epsilon}^+\in [r_{mbo}, r_{mso}]$; therefore  the  outer ergosurface  can be the inner edge of $\cc_1$, $\cc_2$ and $\cc_3$ torus--see for  details  Figs\il(\ref{Fig:spessplhoke1},\ref{Fig:PlotVamp1},\ref{Fig:PlotJurY}).
\item[--]
For   $\mathbf{A}_3: [a_{mso},a_{mbo}^b]$
 tori $\cc_\times\mathbf{(2)}$ have maximum pressure point $r_{center}=r_{\epsilon}^+$, while the cusp has to be inside the ergoregion.
 In this case  there is $r_{\epsilon}^+\in  [r_{mbo}^b,r_{mso}]$, therefore radius $r_{\epsilon}^+$ can coincide with the center of $\cc_1$ tori, and can be the  inner edge of $\cc_1, \cc_2$ and $ \cc_3$  tori. See  Figs\il(\ref{Fig:spessplhoke1},\ref{Fig:PlotVamp1},\ref{Fig:PlotJurY}) for the analysis of  the torus outer edge coincidence with the static limit.
\item[--]

For  $\mathbf{A}_4: ] a_{mbo}^b, a_{\gamma}^b[$ there are  tori $\cc_2$ centered on $r_{\epsilon}^+$.  The static limit  can be the center and the inner edge of the  $\cc_2$ tori, and the inner edge of $\cc_3$ tori (the analysis of  the outer edge coincidence with the static limit is in  Figs\il(\ref{Fig:spessplhoke1},\ref{Fig:PlotVamp1},\ref{Fig:PlotJurY}).
\item[--]

For  $\mathbf{A}_5: ]a_{\gamma}^b, M]$ the static limit can be   also the inner edge or the center of $\cc_3$  tori (the analysis of the outer edge coincidence with the static limit  has been done in  Figs\il(\ref{Fig:spessplhoke1},\ref{Fig:PlotVamp1},\ref{Fig:PlotJurY}).
\begin{figure}\centering
  % Requires \usepackage{graphicx}
  \includegraphics[width=5.5cm]{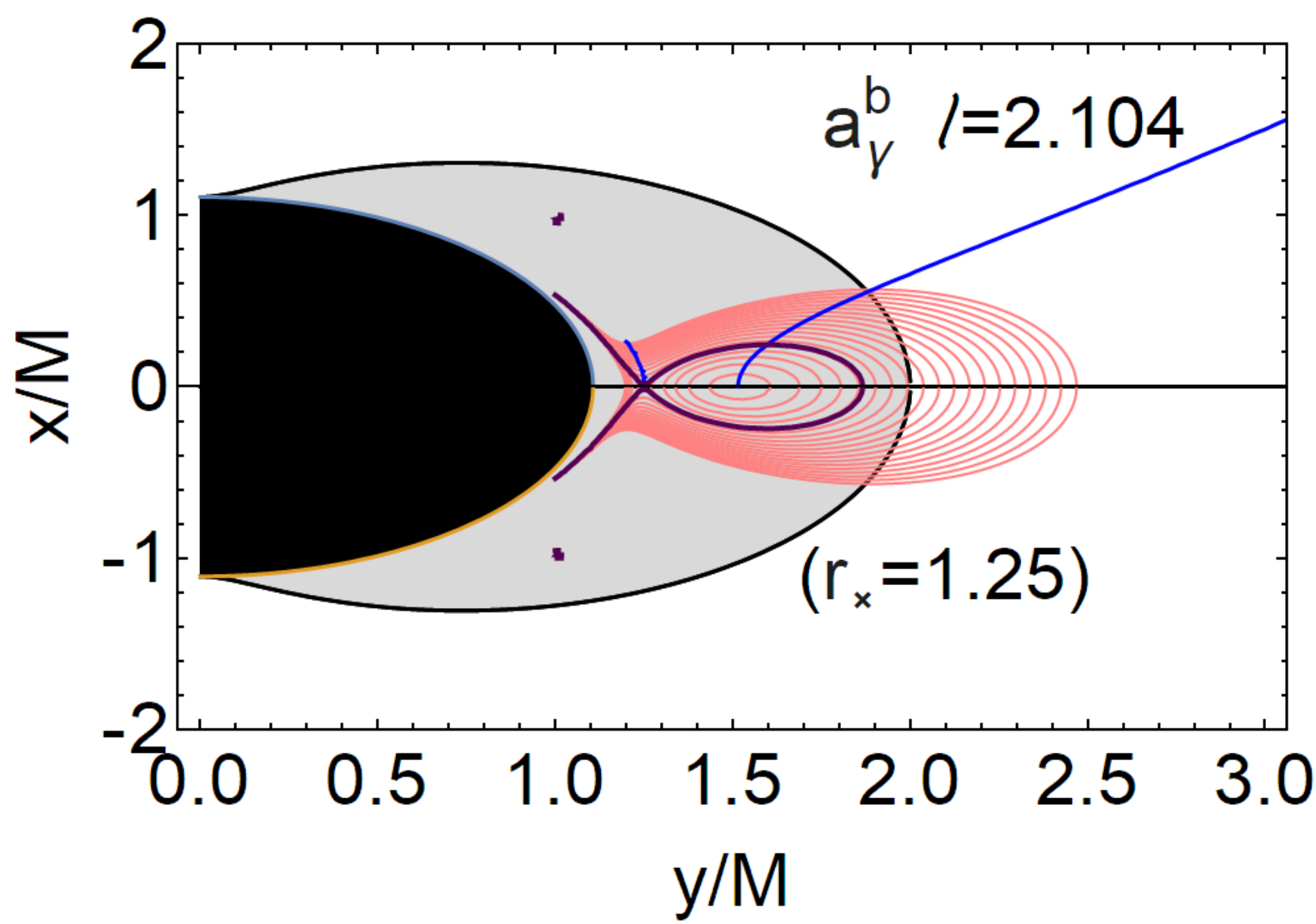}
  \includegraphics[width=5.5cm]{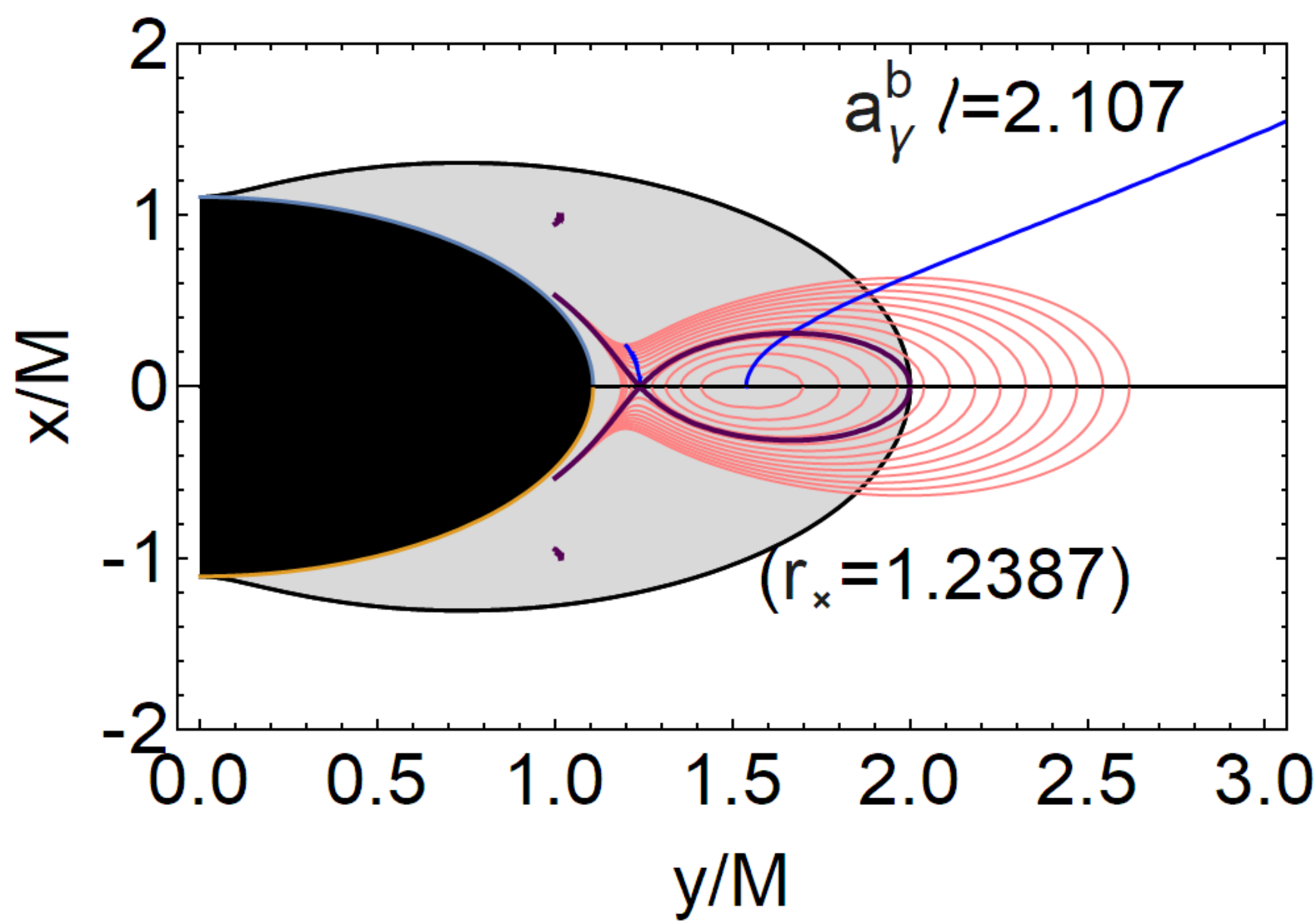}
    \includegraphics[width=5.5cm]{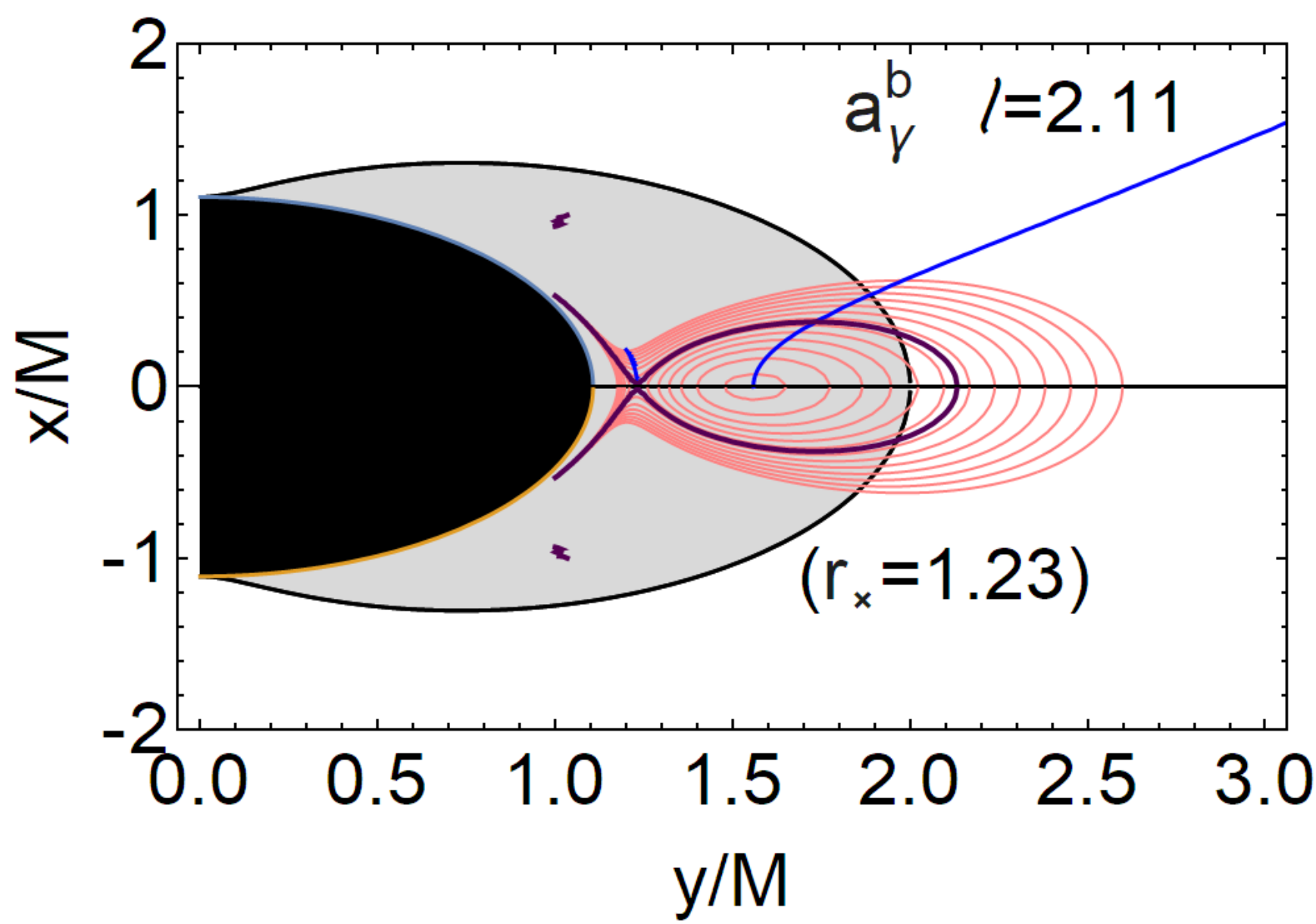}
  \includegraphics[width=5.5cm]{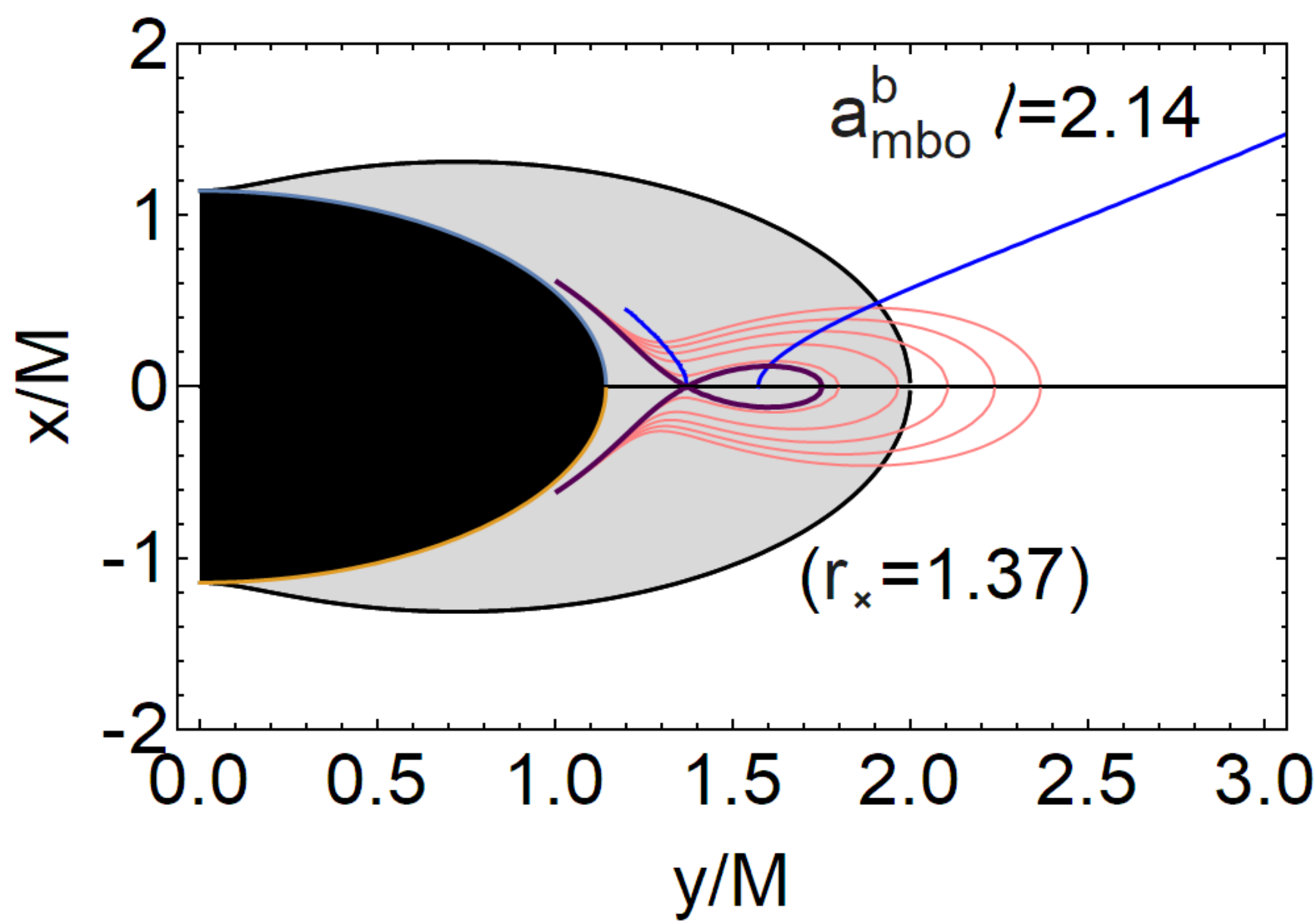}
    \includegraphics[width=5.5cm]{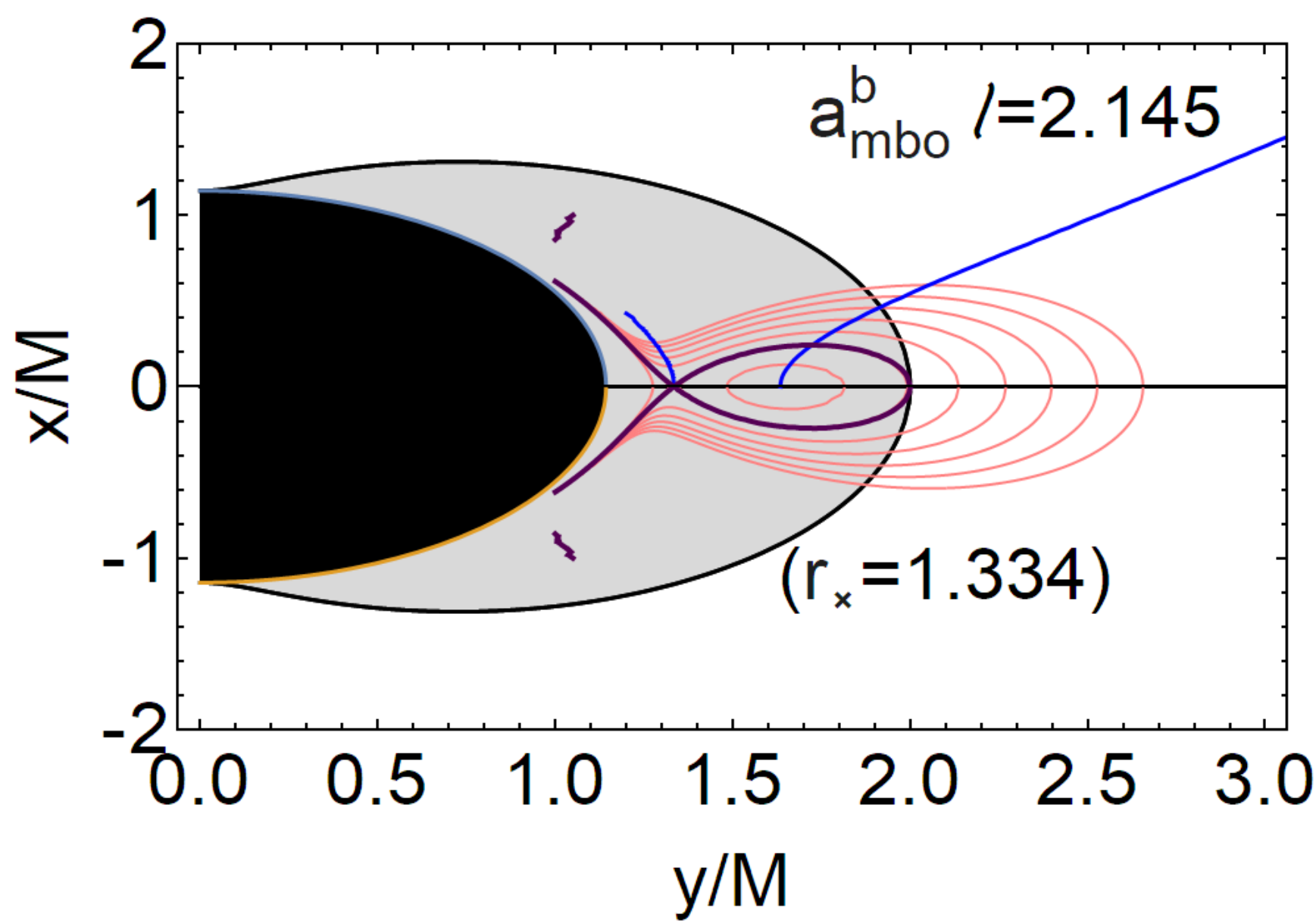}
   \includegraphics[width=5.5cm]{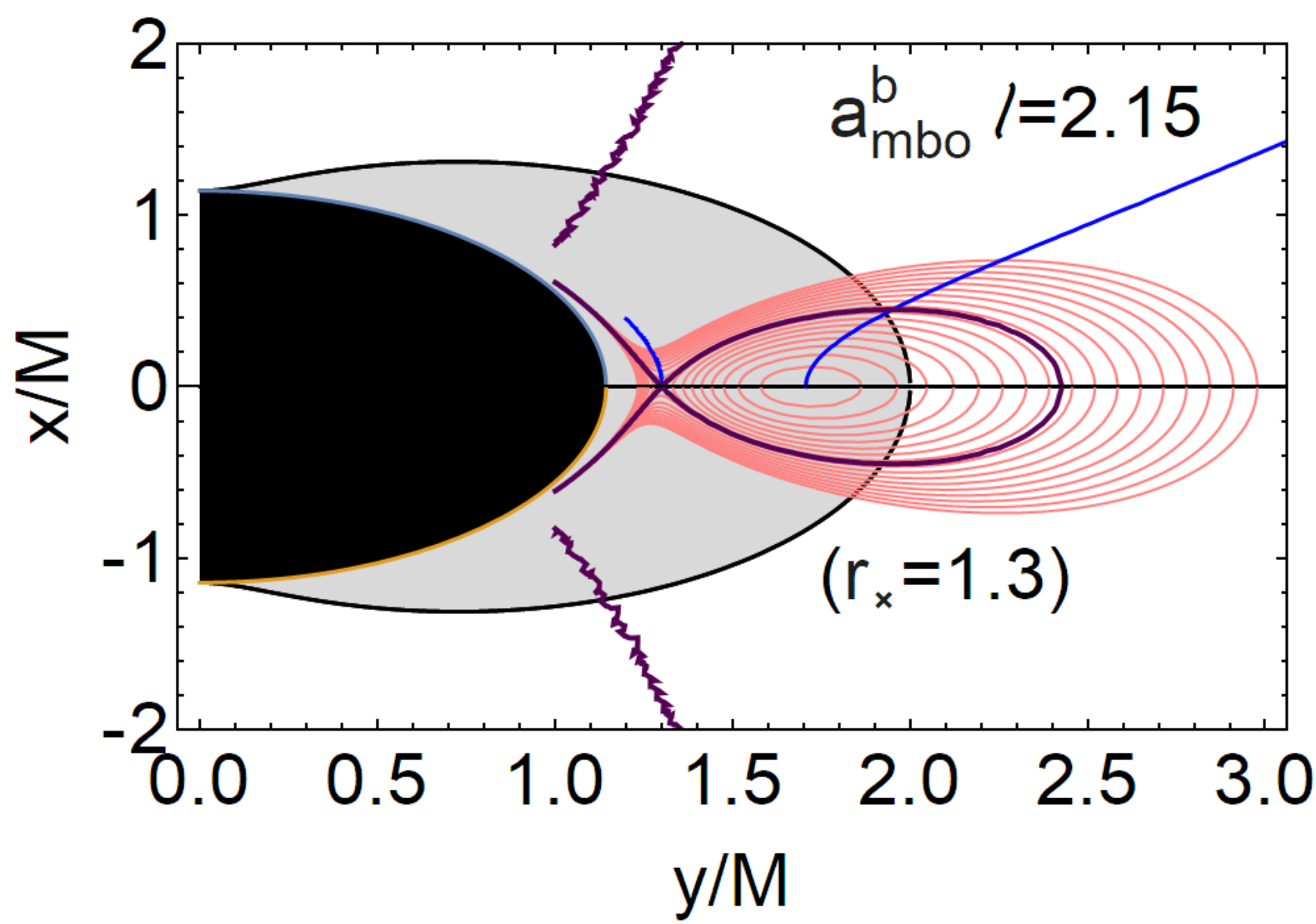}
  \includegraphics[width=5.5cm]{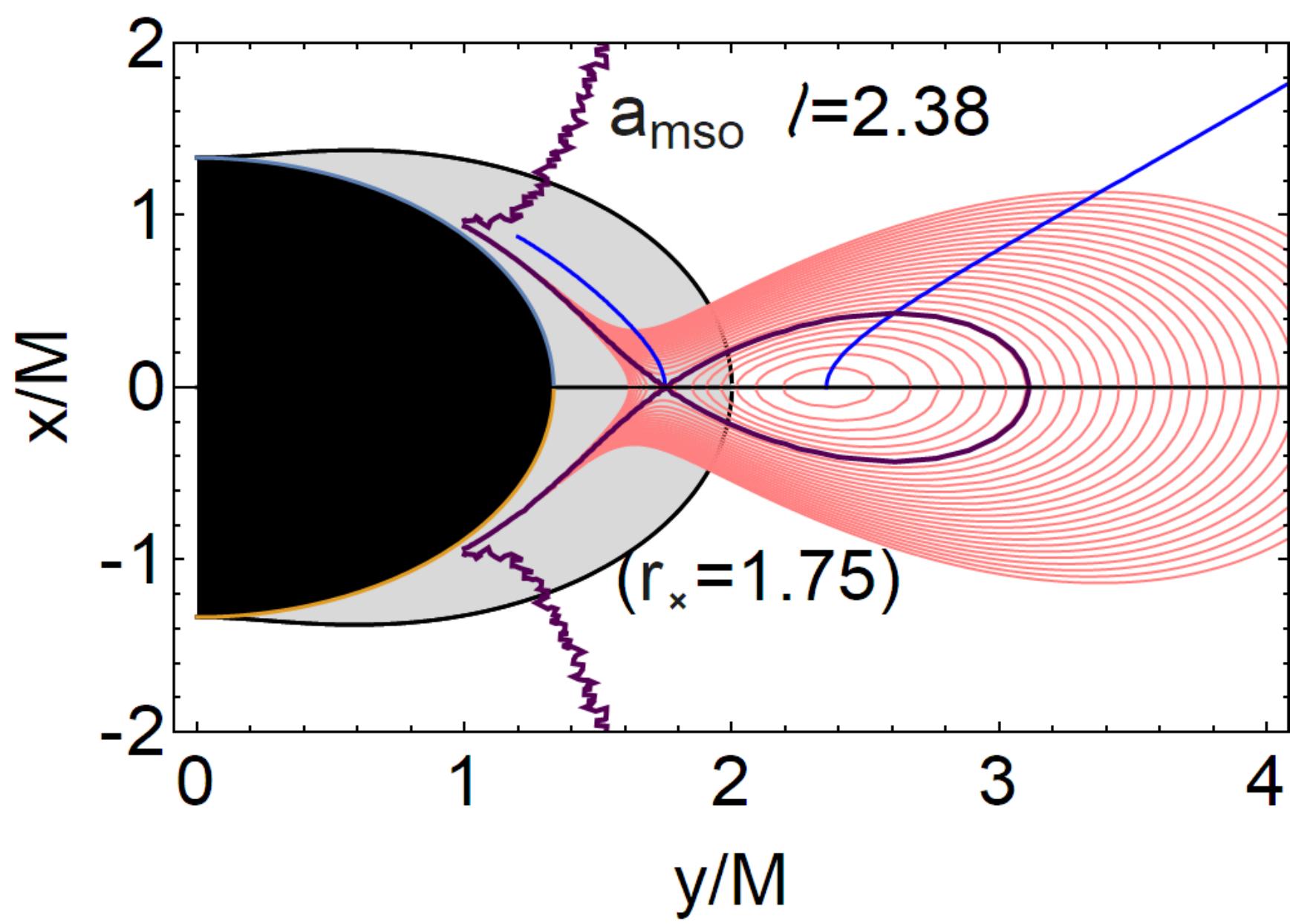}
  \includegraphics[width=5.5cm]{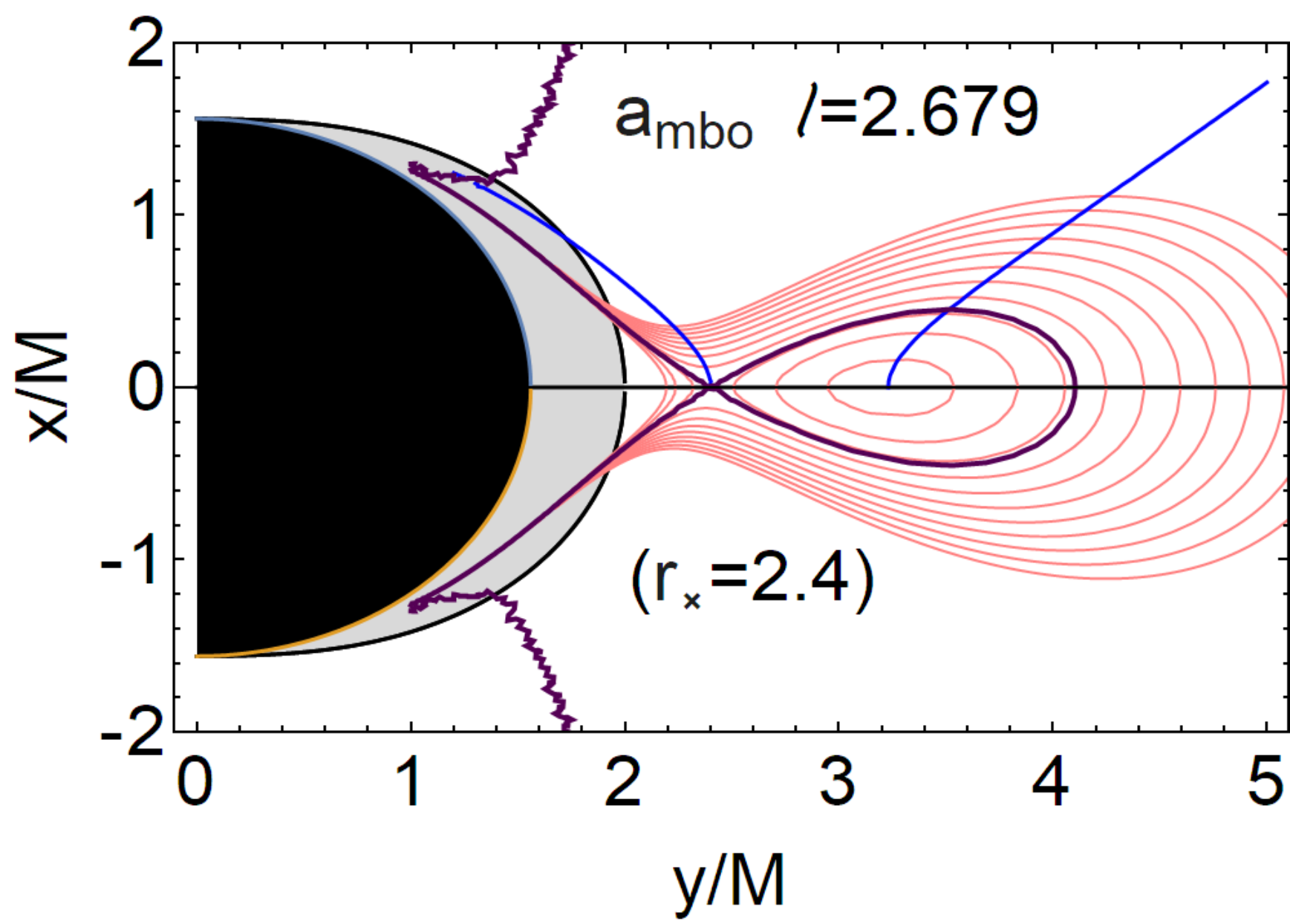}
  \includegraphics[width=5.5cm]{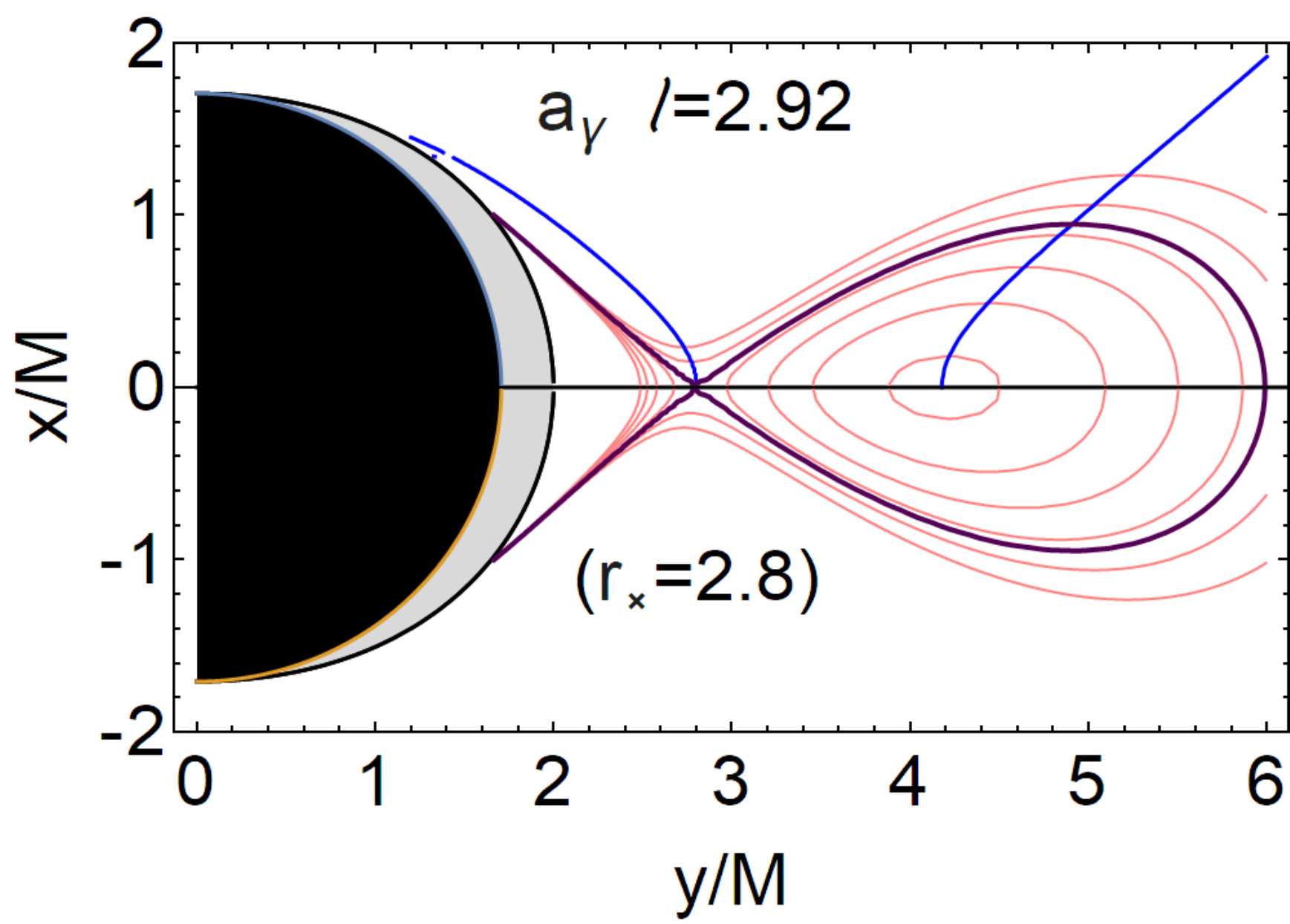}
  \caption{Dragged  tori in the five regions of Figs\il(\ref{Fig:PlotVamp1}) for critical $\cc_{\times}$ and quiescent configurations $\cc$ in the ergoregion or partially contained in the ergoregion or proximate to the outer ergosurface. Spins $\mathbf{A}_{\epsilon}^+\equiv\{a_{mbo},a_{mbo}^b,a_{\gamma},a_{\gamma}^b,a_{mso}\}$ are defined in Eqs\il(\ref{Eq:strateg})  and
Figs\il(\ref{Fig:PlotVampb1}), . Black region is the \textbf{BH} $r<r_+$, gray region is the outer ergosurface, equipotential and equipressure surfaces are shown. $r_{\times}$ is the cusp of the critical configurations $\cc_{\times}$ chosen considering  Figs\il(\ref{Fig:PlotVamp1})  for the purple surface. Blue curves set the extremes of the pressure and density in the disks: the maximum pressure inside the disk from the center of the configurations to the geometrical maximum  for different $K$ surfaces, and the inner minimum point of the pressures from the cusp. $\ell$ is the specific fluid angular momentum.}\label{Fig:polodefin1}
\end{figure}
\begin{figure}\centering
  % Requires \usepackage{graphicx}
  \includegraphics[width=5.5cm]{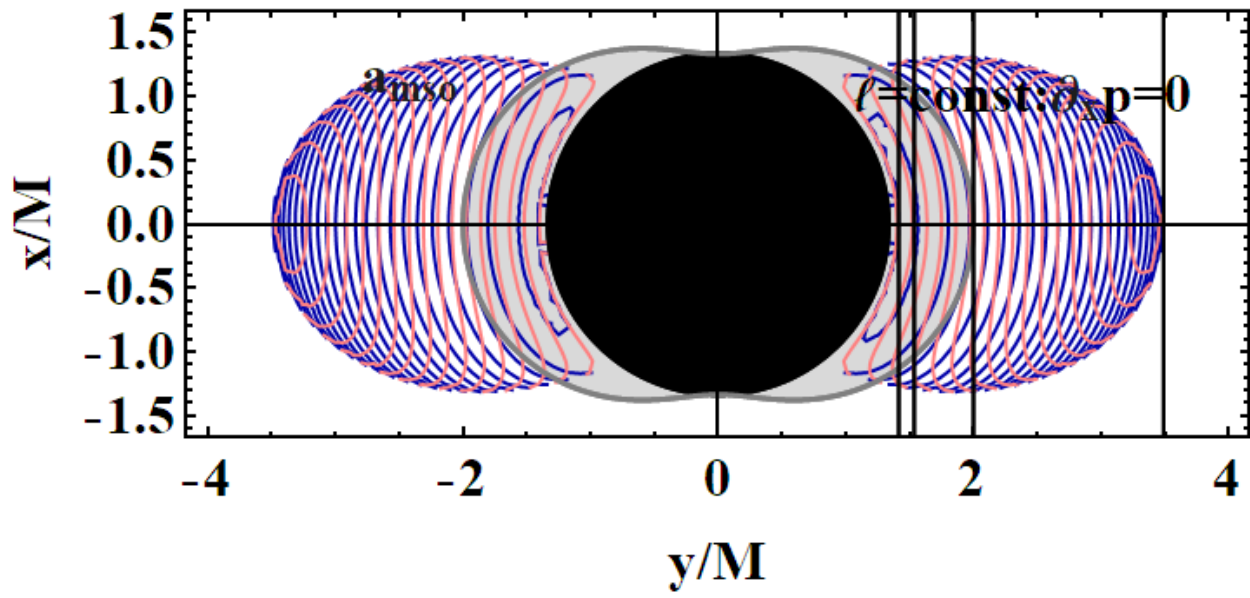}
  \includegraphics[width=5.5cm]{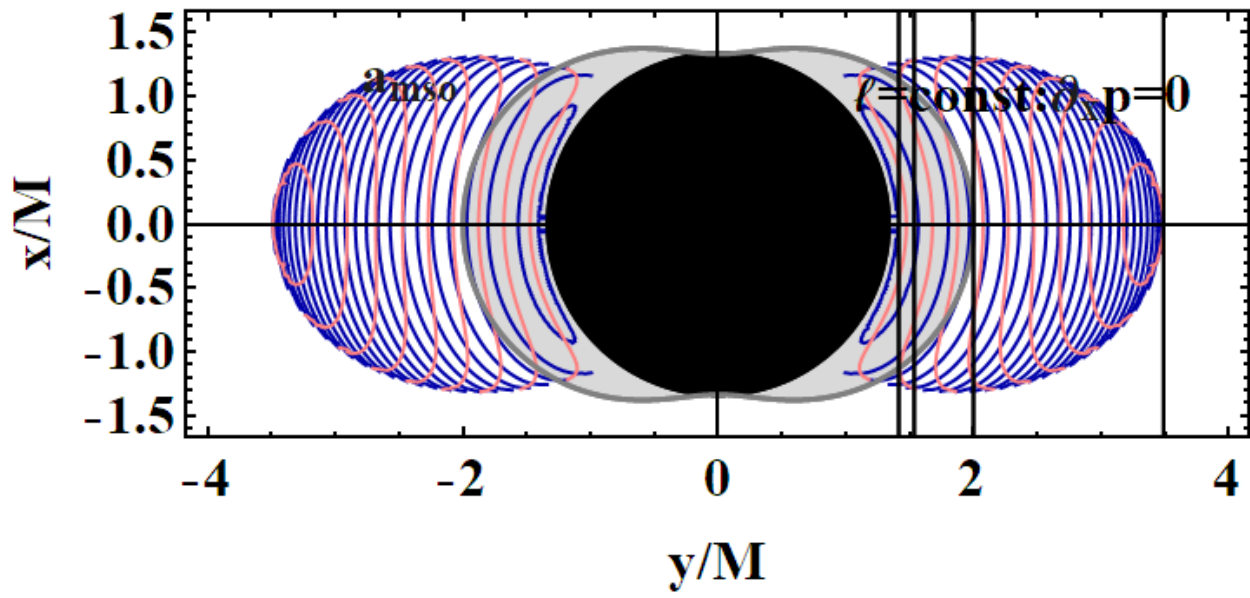}
    \includegraphics[width=5.5cm]{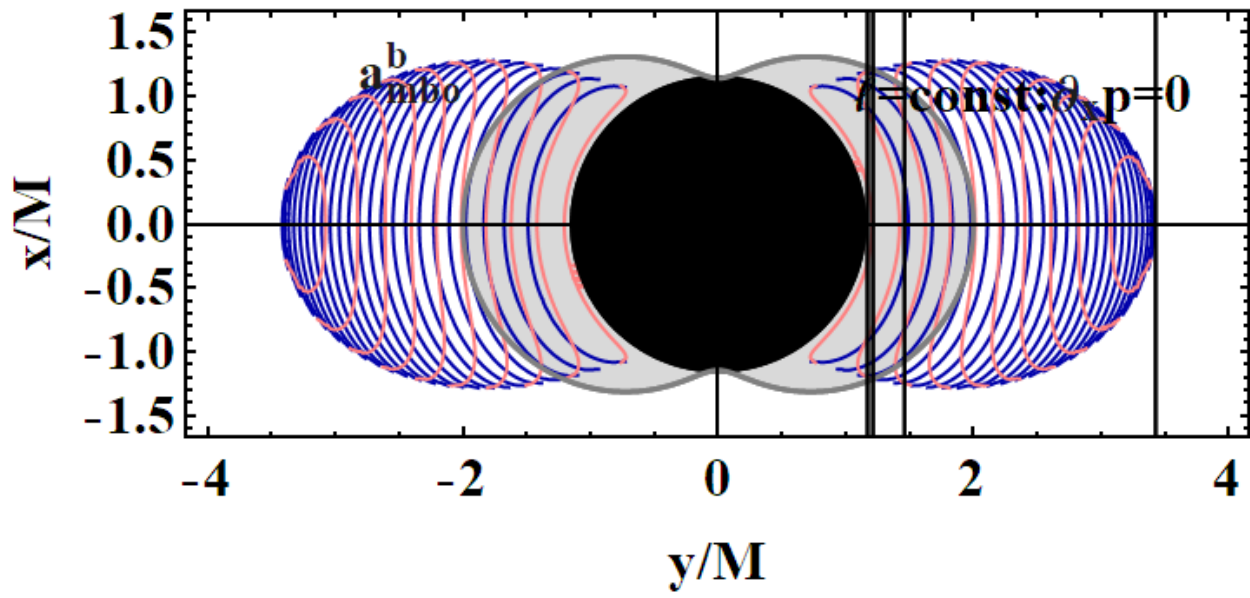}
  \includegraphics[width=5.5cm]{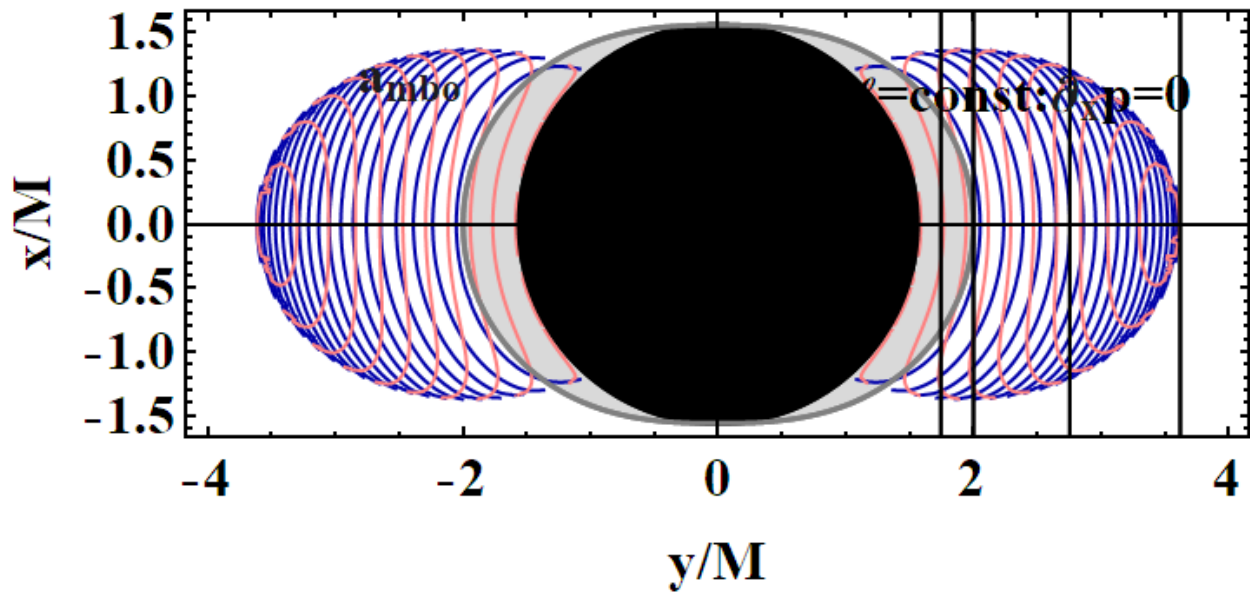}
    \includegraphics[width=5.5cm]{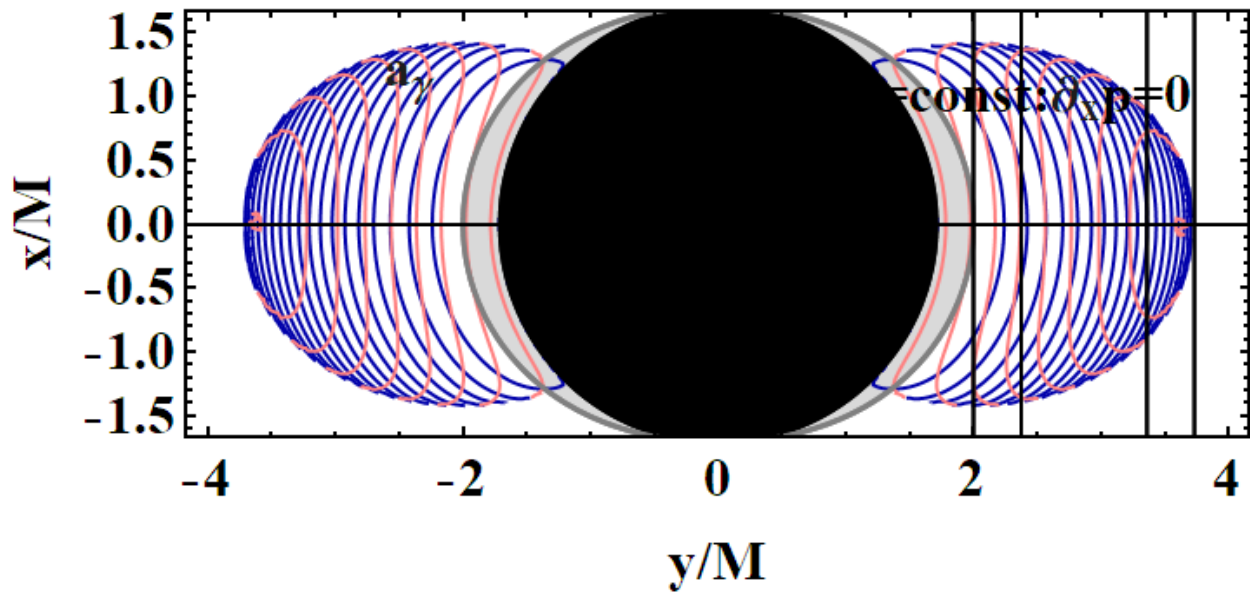}
   \includegraphics[width=5.5cm]{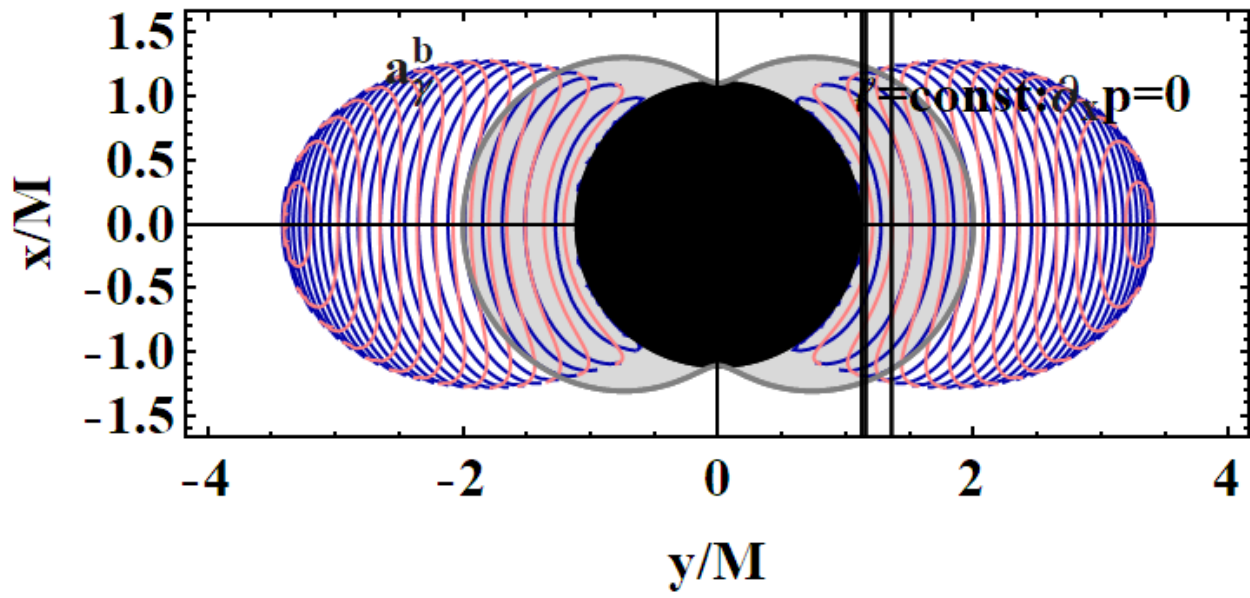}
  \caption{Study of the verticality of the disk, analysis of the vertical pressure gradients inside the disk. Black region is the \textbf{BH}, gray region is the outer ergosurface, $x/M$ and $y/M$ are Cartesian coodinates, different spins $\mathbf{A}_{\epsilon}^+\equiv\{a_{mbo},a_{mbo}^b,a_{\gamma},a_{\gamma}^b,a_{mso}\}$ are represented --Figs\il(\ref{Fig:polodefin1}). Curves are the solutions of $\ell=constant$ where $\ell:\partial_xV_{eff}=0$, providing a description of the vertical pressure gradient in the disks (there is ($x=r\cos\theta, y=r\sin\theta)$, on the equatorial plane there is $y\equiv r$). The range of definition $\ell=$constant is bounded in a region dependent on the spin $a$ only. The boundary conditions are analyzed in Figs\il(\ref{Fig:collag}). }\label{Fig:weirplot}
\end{figure}
\end{description}
\subsubsection{The tori verticality and disk exfoliation}\label{Sec:dv}
Relevant aspects of the toroidal  geometrically  thick disks centered on the   \textbf{BH} equatorial plane, are regulated  by  the dependence of the  tori     characteristics   on the radial  dimension, as the radial pressure gradients, which ultimately  fixes    the  tori  rotational law $\ell(r)$. For the tori orbiting in the ergoregion it is necessary to explore     the vertical pressure gradients inside the disk. The disk verticality is  in fact a significant   factor for  the accretion disks also  in presence of the magnetic fields, and in many models of tori oscillation, as for the analysis of \textbf{QPOs} from accretion disks, which we face  here in Sec.\il(\ref{Sec:qpos}).
The disk verticality in this model is regulated by the  pressure  $x$-gradient and the curve of the extremes of pressure,   including  the points  $r_{\times}$,  the disks center $r_{center}$, which is a  maximum pressure point,   and the minimum in the accretion  flow at  radius  $r_*<r_{\times}$ (for the  overcritical, $\cc_1$, configurations $K>K_{\times}$) and the disk  geometrical maximum-- Figs\il(\ref{Fig:polodefin1}). The  curve relates the geometrical maxima and  the extremes  of the HD pressure,  solution of $\ell:\partial_y V_{eff}(x,y)=0$ (there is  $\partial_y V_{eff}=\partial_r V_{eff}$). This curve  depends on  the \textbf{BH} dimensionless spin $a/M$ only.

The  curves, solution  of  equation $\partial_x V_{eff}=0$, are  bounded in a range analyzed in  Figs\il(\ref{Fig:collag})  where we show that the zeros and  the maximal extensions of the  range  of definition for the solutions $\ell=$constant for the vertical gradient of the pressure are limited in a bounded region of the plane $x-y$.
More precisely, in  Figs\il(\ref{Fig:weirplot})  we  note how the vertical pressure  gradient   is limited by a surface $y(x)$  dependent on the spin only.  The extension of this range   is largely unchanged by the \textbf{BH} spin as clarified  in Figs\il(\ref{Fig:collag}).

The boundary curve is
 \bea\label{Eq:quesalyl}
y_\ell\equiv\sqrt{2(2M^2- a^2)}+2M\quad \mbox{or, equivalently, }\quad a_\ell\equiv\sqrt{y\left(2M -\frac{y}{2}\right)},
\eea
represented in Figs\il(\ref{Fig:collag}), where the maximum  $y_{\ell}\leq 4M$ is shown.
It is evident that increasing the  \textbf{BH} spin, the curves $\ell=$constant are confined in the ergoregion,  the density of curves in the region increases approaching the horizon. For the limiting case of a static   Schwarzschild  \textbf{BH},  the boundary radius  corresponds to  $y_{\ell}=r_{mbo}=4M$ (while the specific angular momentum $\ell$ and the energy function $K$ on this curve decreases with the spin, which can be seen analyzing the function  $\ell(r)$  and  the function $K(r)$).

 To ensure the disk is entirely contained in $\Sigma_{\epsilon}^+$,  at  a general plane $\theta$ we need to  consider the condition
$\partial_y V_{eff}(x,y))=0$. In fact, the necessary but not sufficiently condition to be satisfied for   the outer part of the disk  to be in the ergoregion (and also for the inclusion of the torus geometrical maximum)  is   $r_{center}\in \Sigma_{\epsilon}^+$, while this is  a sufficient condition for the inner part of the disk to be included in the ergoregion.
 It is clear from Figs\il(\ref{Fig:weirplot}) that
 the curve of  geometrical maxima and minima depends on   $y$ ($r$ on the equatorial plane) but also on  the parameter  $K>K_{center}$ ($x_{\max}(y)=x_{\max}(K)$). There are  two branches of the curve, one  from $r\leq r_{\times}$ and  one $r\geq r_{center}$.  %\btb{Formula}
The  torus approaches the \textbf{BH} poles rotational axes, when the  torus geometrical maximum crosses the  outer ergosurface.
\begin{figure}\centering
  % Requires \usepackage{graphicx}
  \includegraphics[width=8cm]{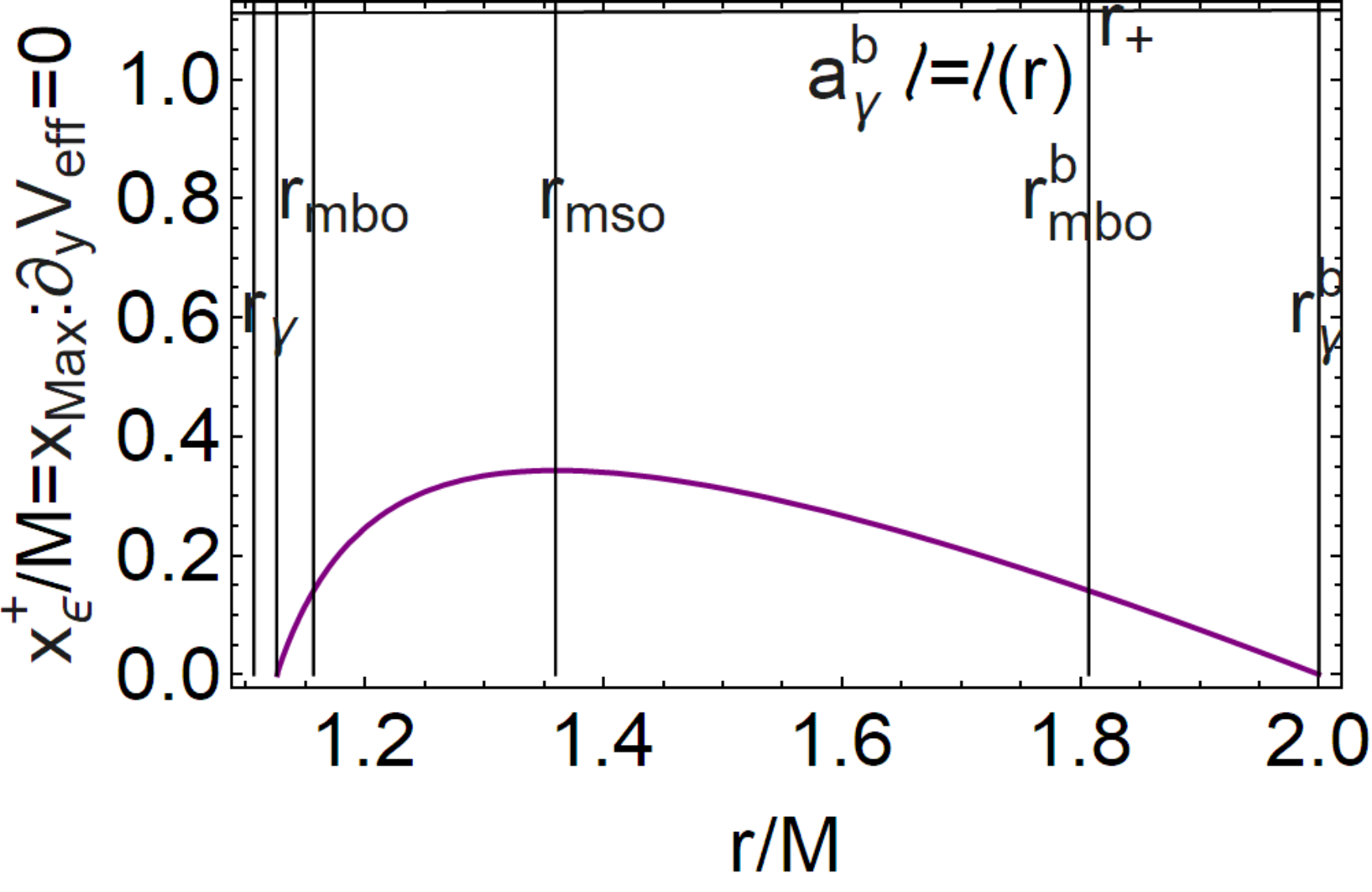}
  \includegraphics[width=8cm]{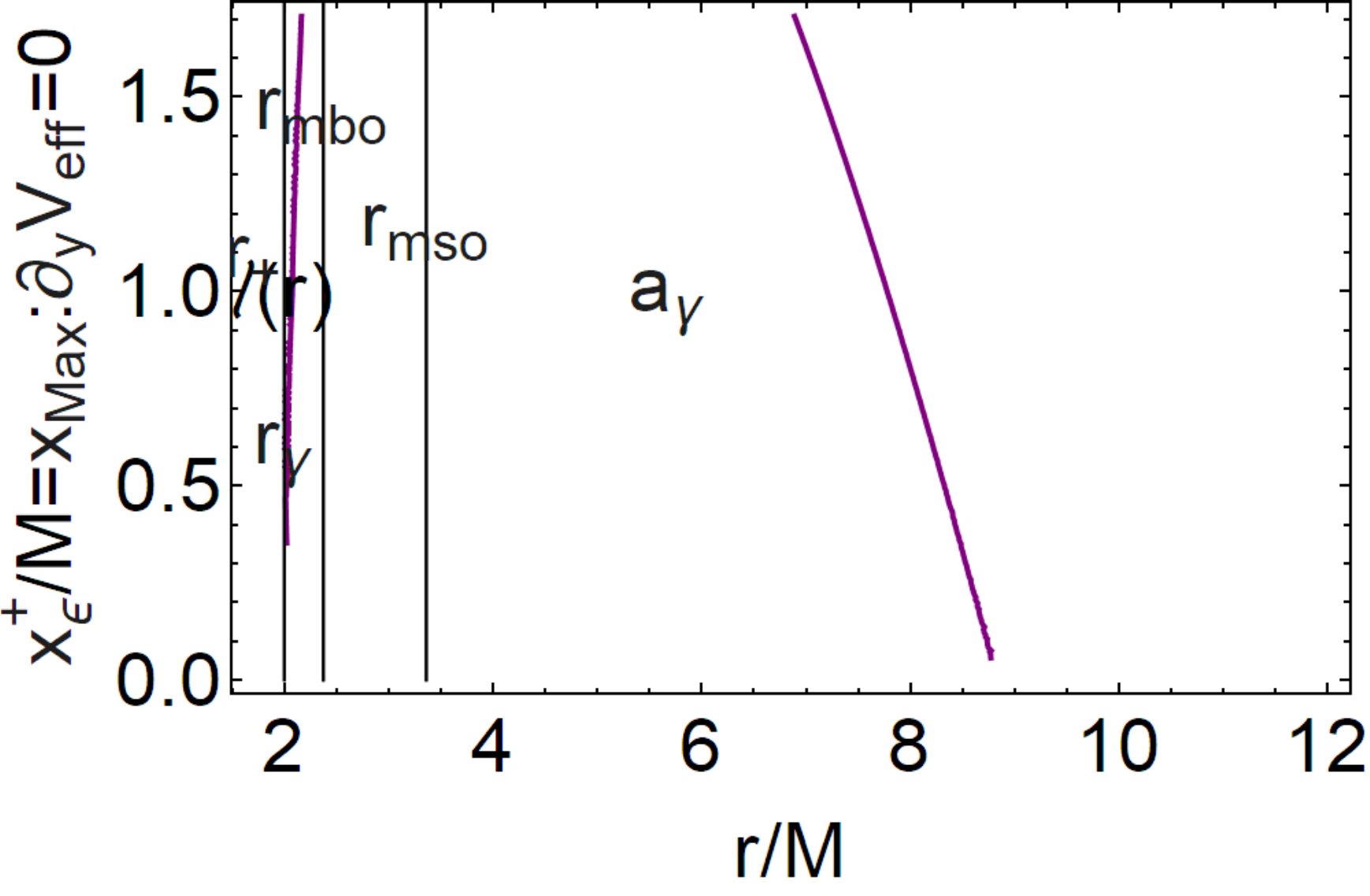}
     \includegraphics[width=8cm]{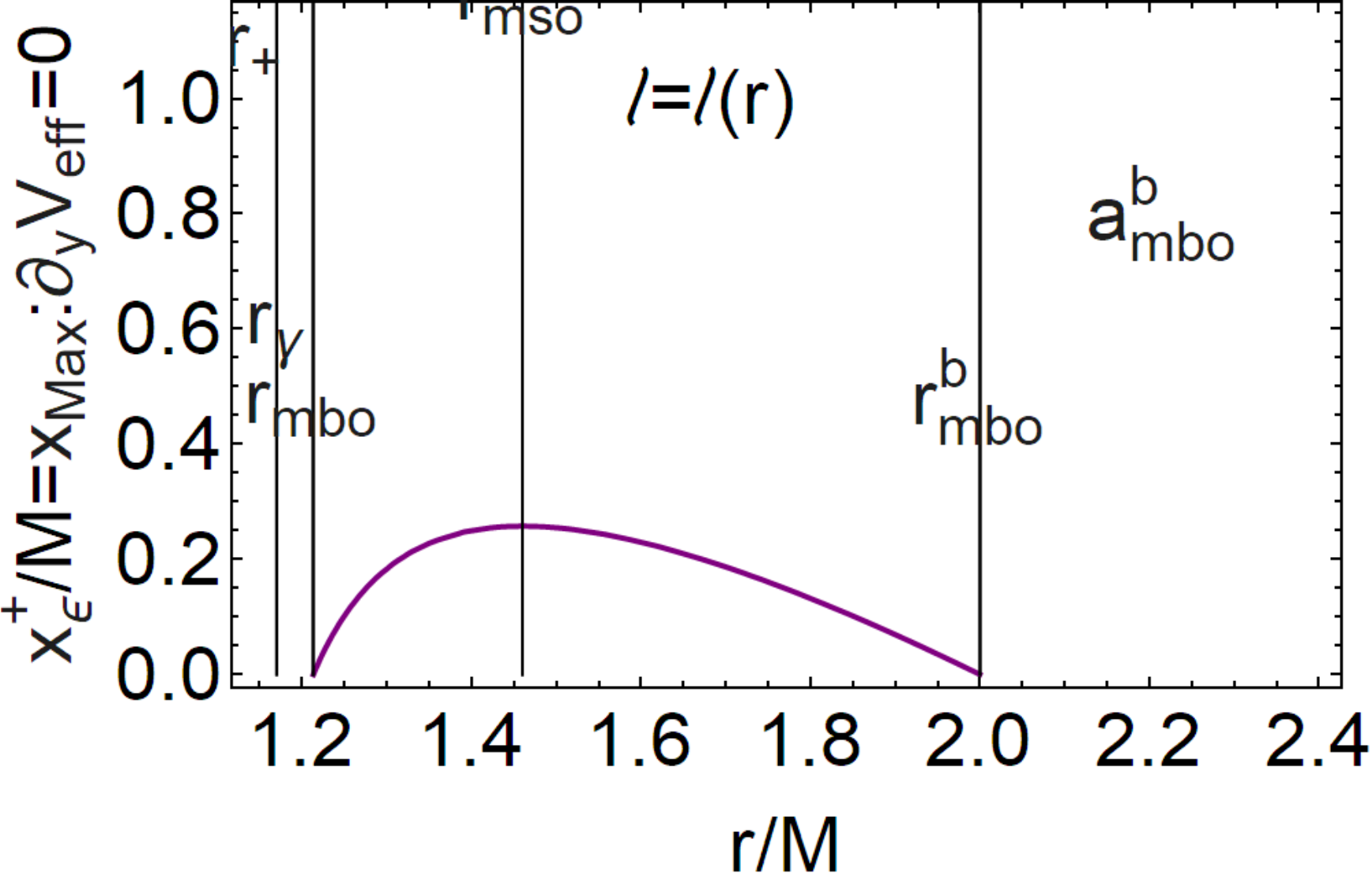}
    \includegraphics[width=8cm]{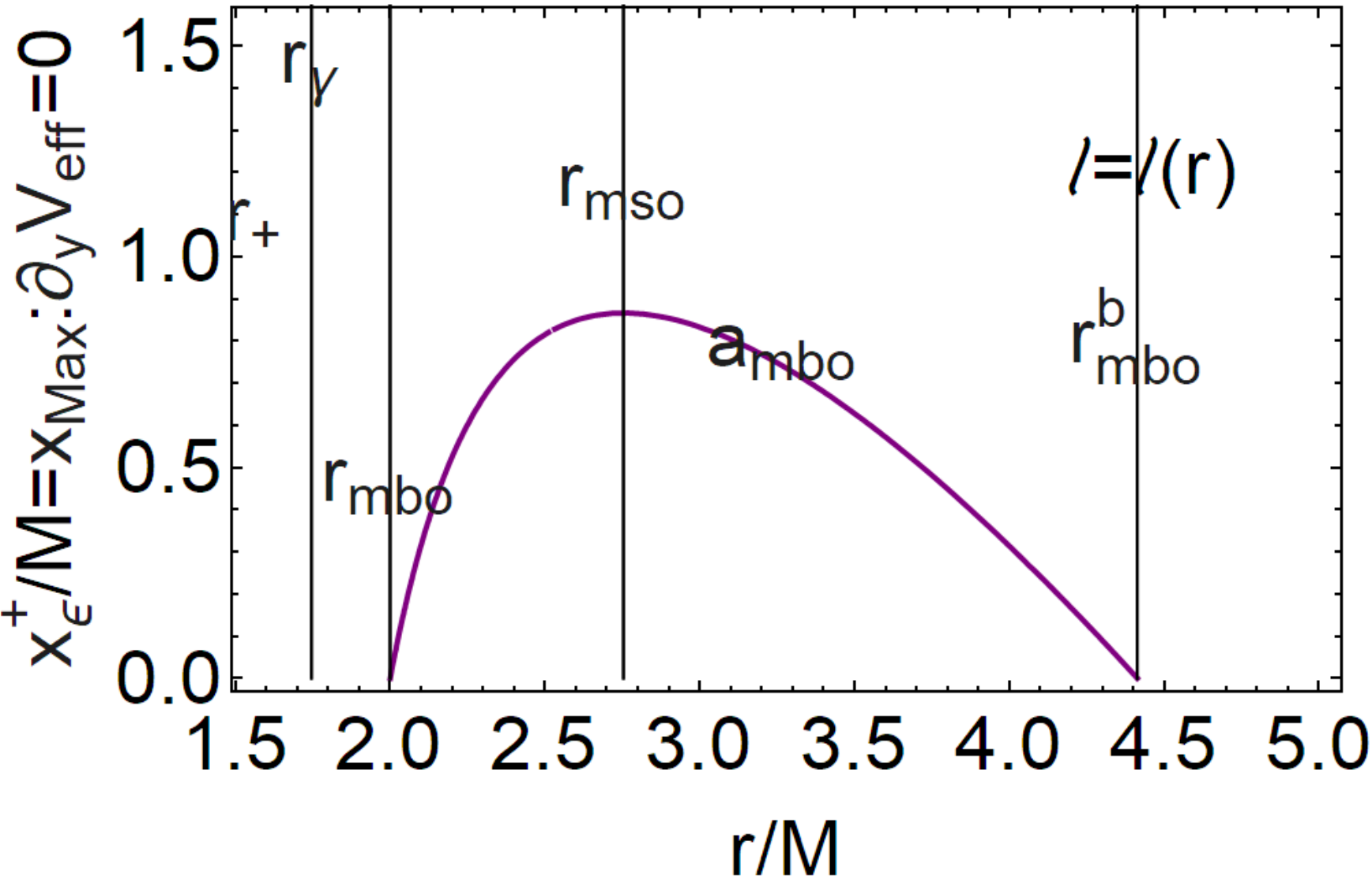}
    \caption{Study of the tori crossing the  outer  ergosurface. Plot of solutions  $x/M$ (vertical axis, there is ($x=r\cos\theta, y=r\sin\theta)$, on the equatorial plane there is $y\equiv r$)  crossing of the line of geometrical maxima $\partial_y V_{eff}=0$ on the ergosuface as function of  $r>r_{+}$ (outer Killing horizon) $\ell$ is the fluid specific angular momentum.  We consider the critical configurations therefore  we evaluated $V_{eff}$ on $\ell(r)$ setting centers or cusps. $r_{mbo}$ is  the  marginally bounded orbit, $r_{mso}$ is  the   marginally stable  orbit, $r_{\gamma}$ is the corotating photon orbit.   Different spins $\mathbf{A}_{\epsilon}^+=\{a_{mbo},a_{mbo}^b,a_{\gamma},a_{\gamma}^b,a_{mso}\}$ are represented as  in Figs\il(\ref{Fig:polodefin1}). Therefore  $r<r_{mso}$ sets a cusp, while  $r>r_{mso}$ is a center.  Center of (quiescent) tori with $\ell>\ell_{\gamma}$ is at $r>r^b_{\gamma}$, center of (accreting) tori with $\ell\in[\ell_{mso},\ell_{mbo}]$ with $r_{center}\in[r_{mso},r_{mbo}^b]$, while tori with $\ell\in[\ell_{mbo},\ell_{\gamma}]$ associated to quiescent tori and protojects have center in $r_{center}\in[r^b_{mbo},r_{\gamma}^b]$. }\label{Fig:gatplot8}
    \end{figure}
Figs\il(\ref{Fig:gatplot8})  show the results of the analysis of the  conditions for  tori crossing the  outer  ergosurface considering the line of geometrical maxima. Radius $r_{mso}$ is  a limiting radius, being a maximum of the curve, as one value of the fluid specific angular momentum $\ell$   corresponds to  two radii $r_1<r_2: \ell(r_1)=\ell(r_2)$. Solutions  $x/M$ (vertical axis)  crossing  of geometrical maximum line  with the  ergosuface as function of  $r>r_{+}$ (outer  horizon)  are plotted (as solution of $\partial_y V_{eff}=0$). The analysis refers to critical (cusped) configurations. The smaller  (larger)    $x_{\epsilon}^+$ (the intersection of the toroidal surface with the static limit on $\theta\neq\pi/2$)  is and  the closest to the equatorial plane (rotational axis) of the \textbf{BH} is the crossing.
\begin{figure}\centering
     \includegraphics[width=5.5cm]{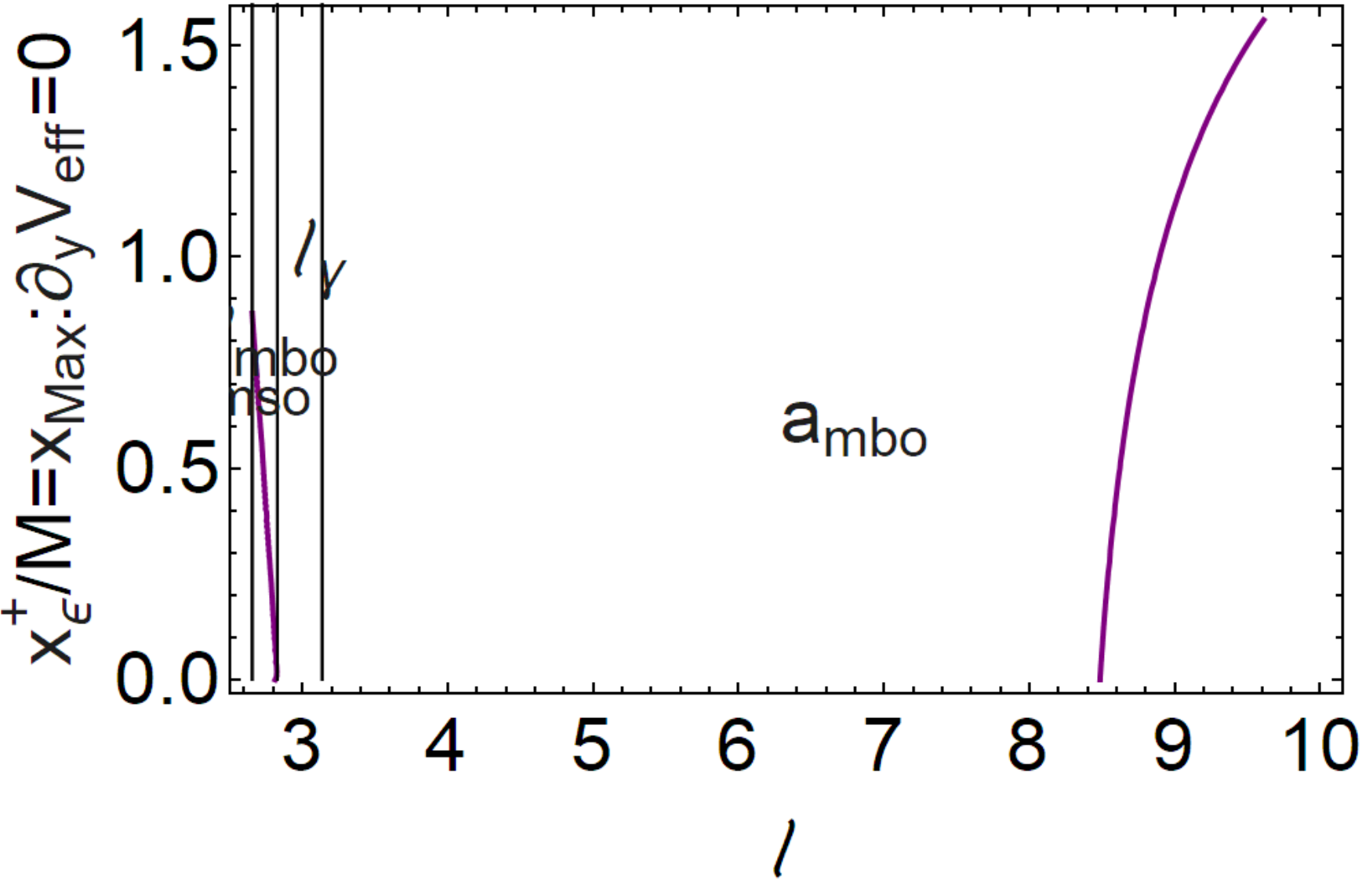}
  \includegraphics[width=5.5cm]{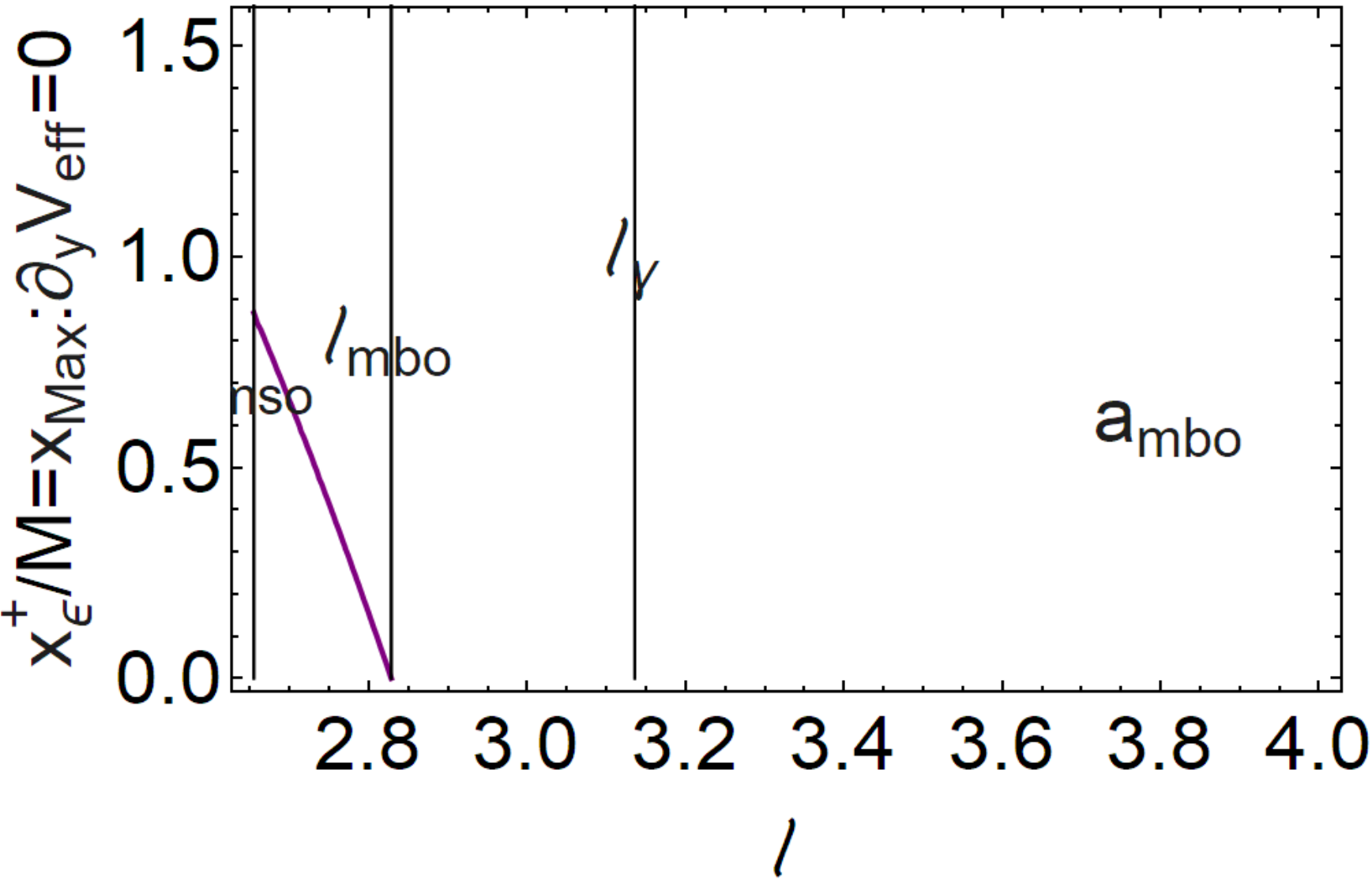}
    \includegraphics[width=5.5cm]{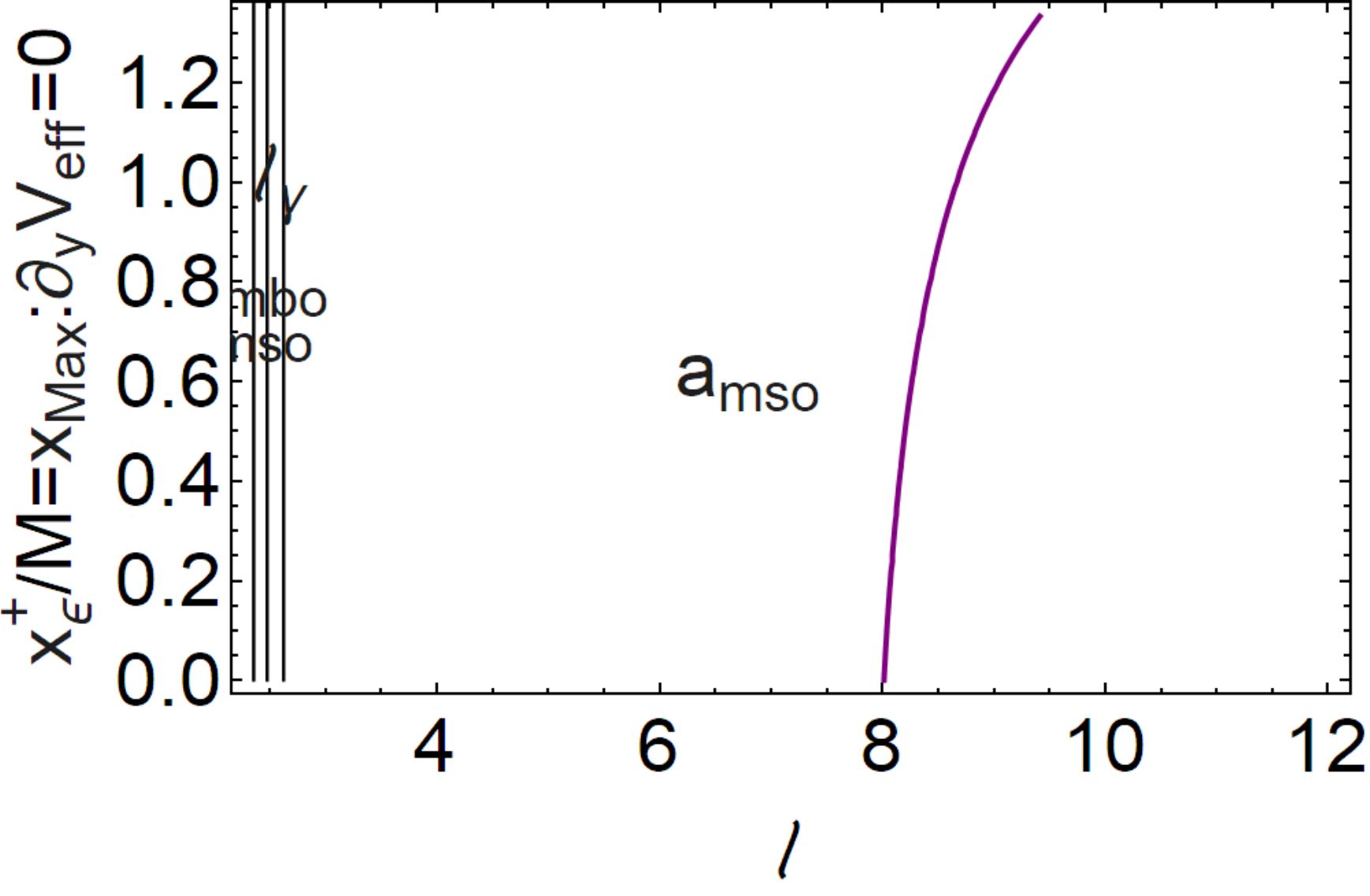}
     \includegraphics[width=5.5cm]{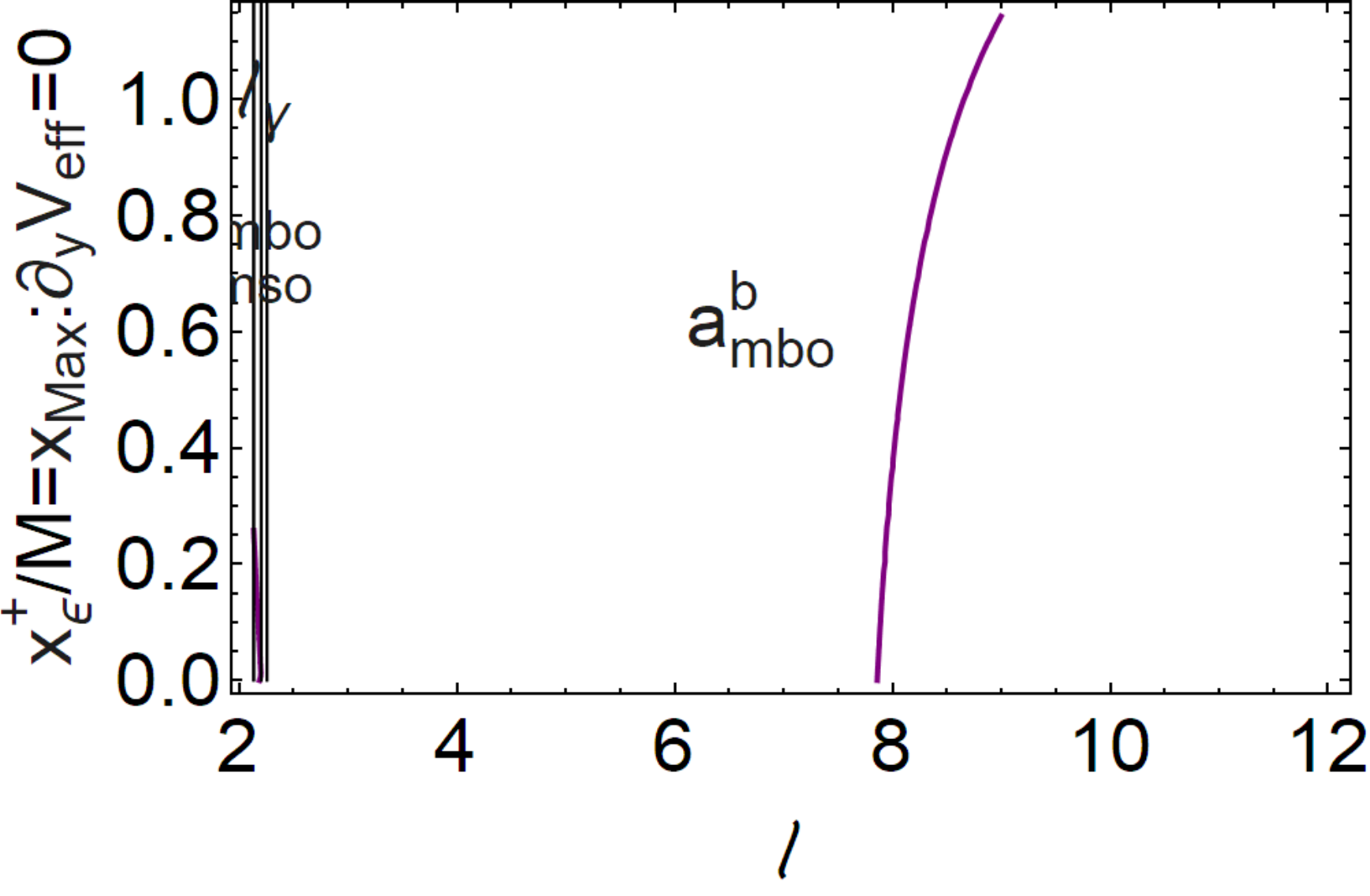}
  \includegraphics[width=5.5cm]{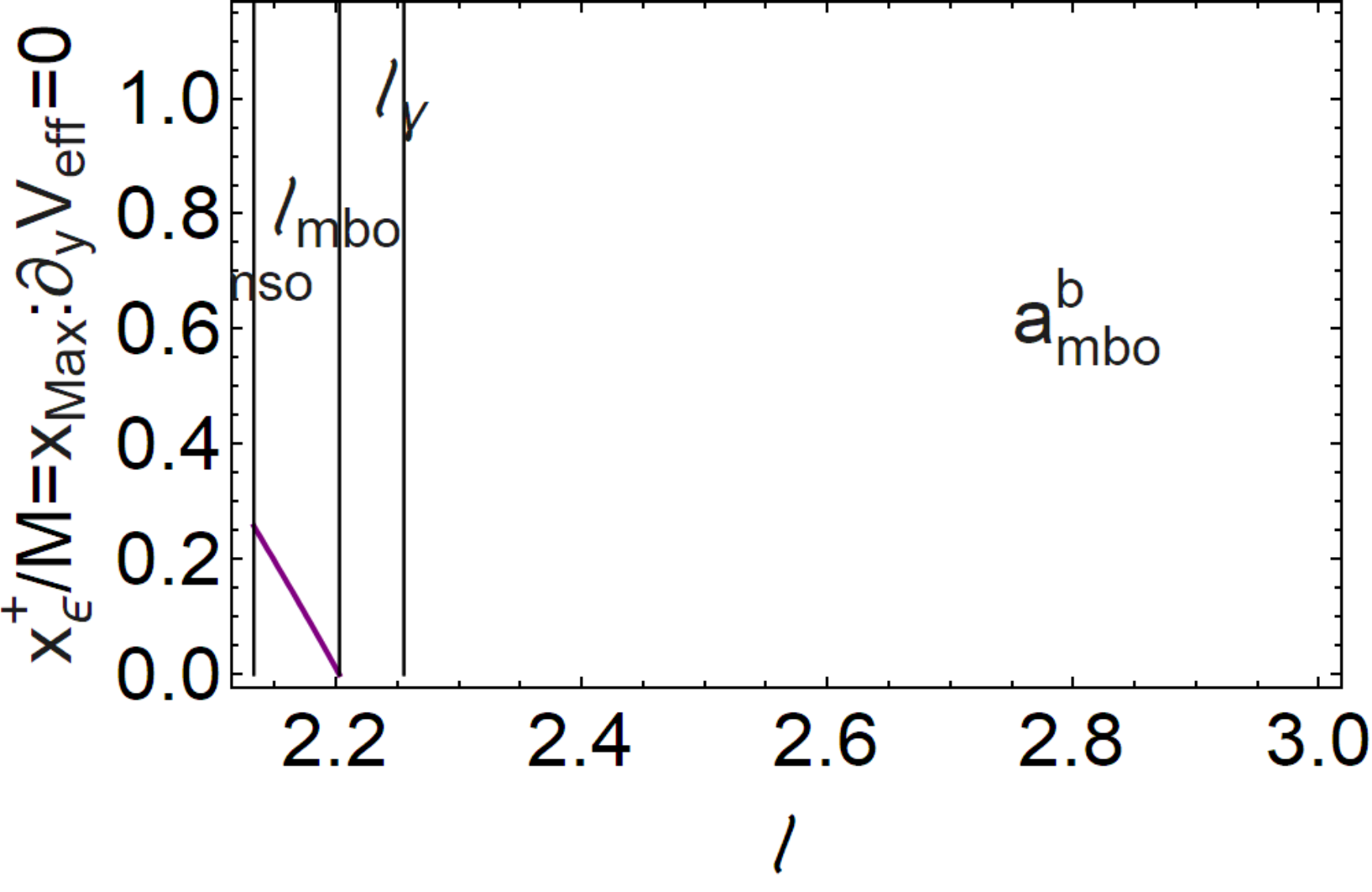}
        \includegraphics[width=5.5cm]{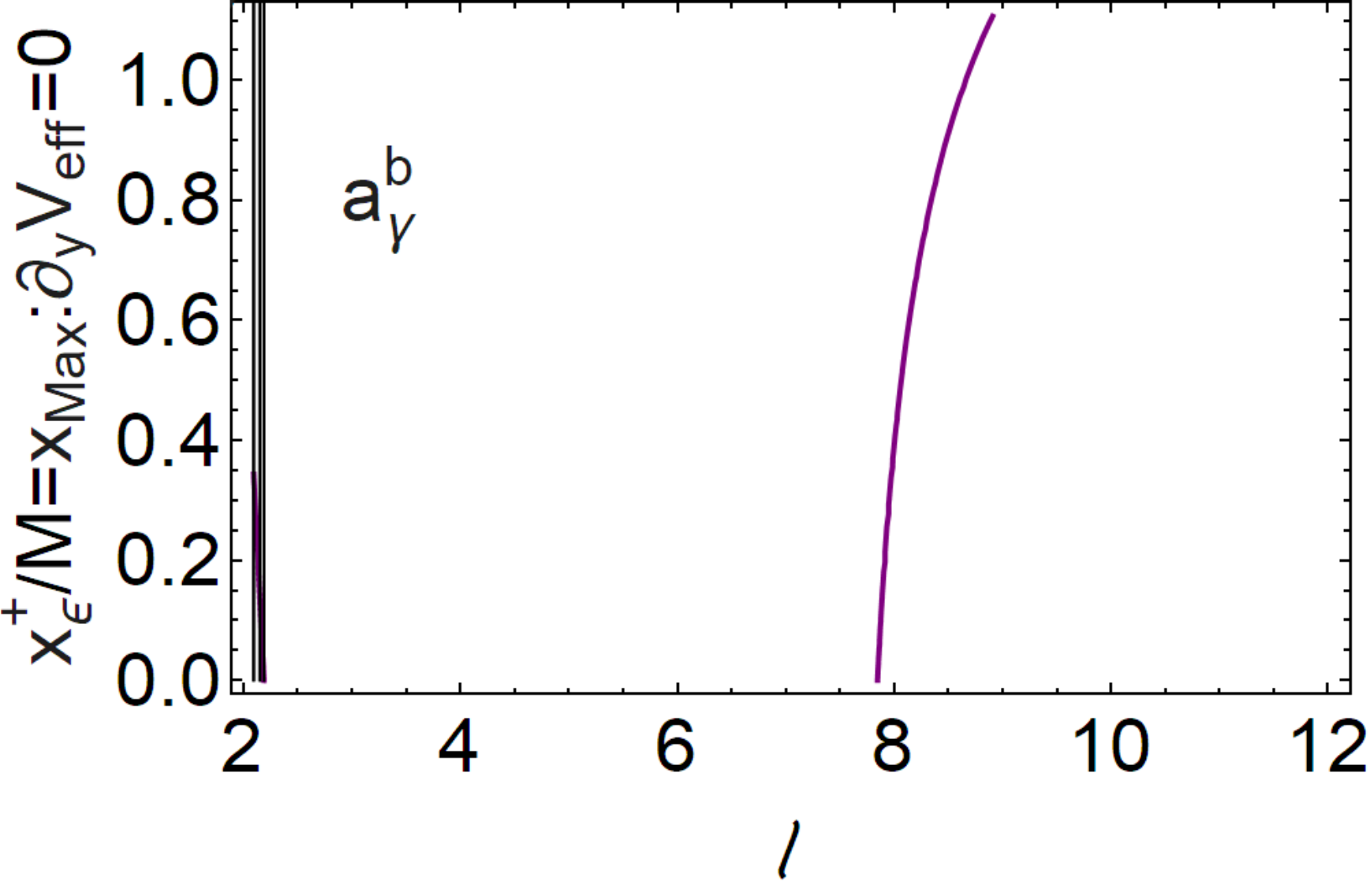}
                \includegraphics[width=5.5cm]{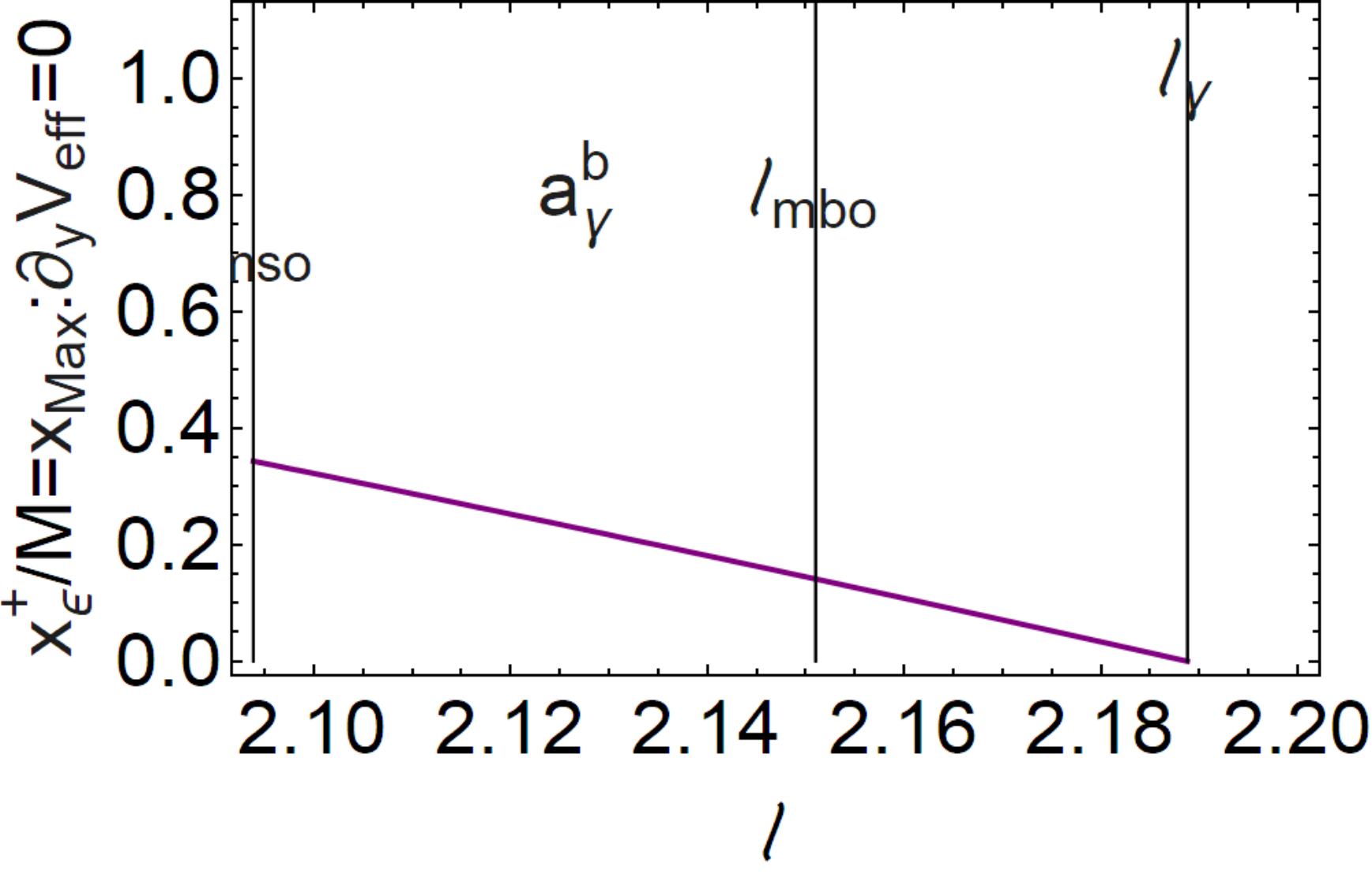}
    \includegraphics[width=5.5cm]{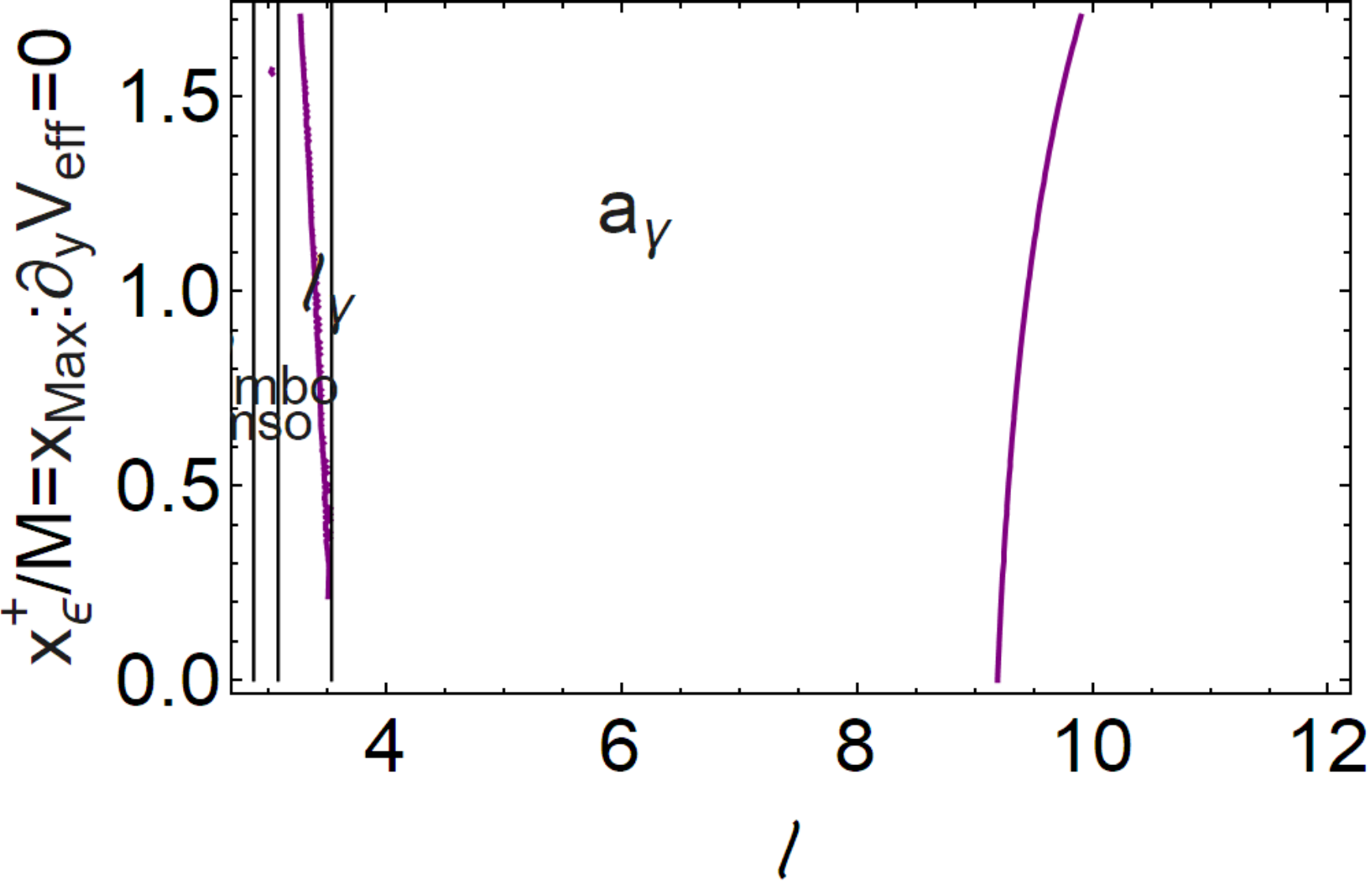}
     \includegraphics[width=5.5cm]{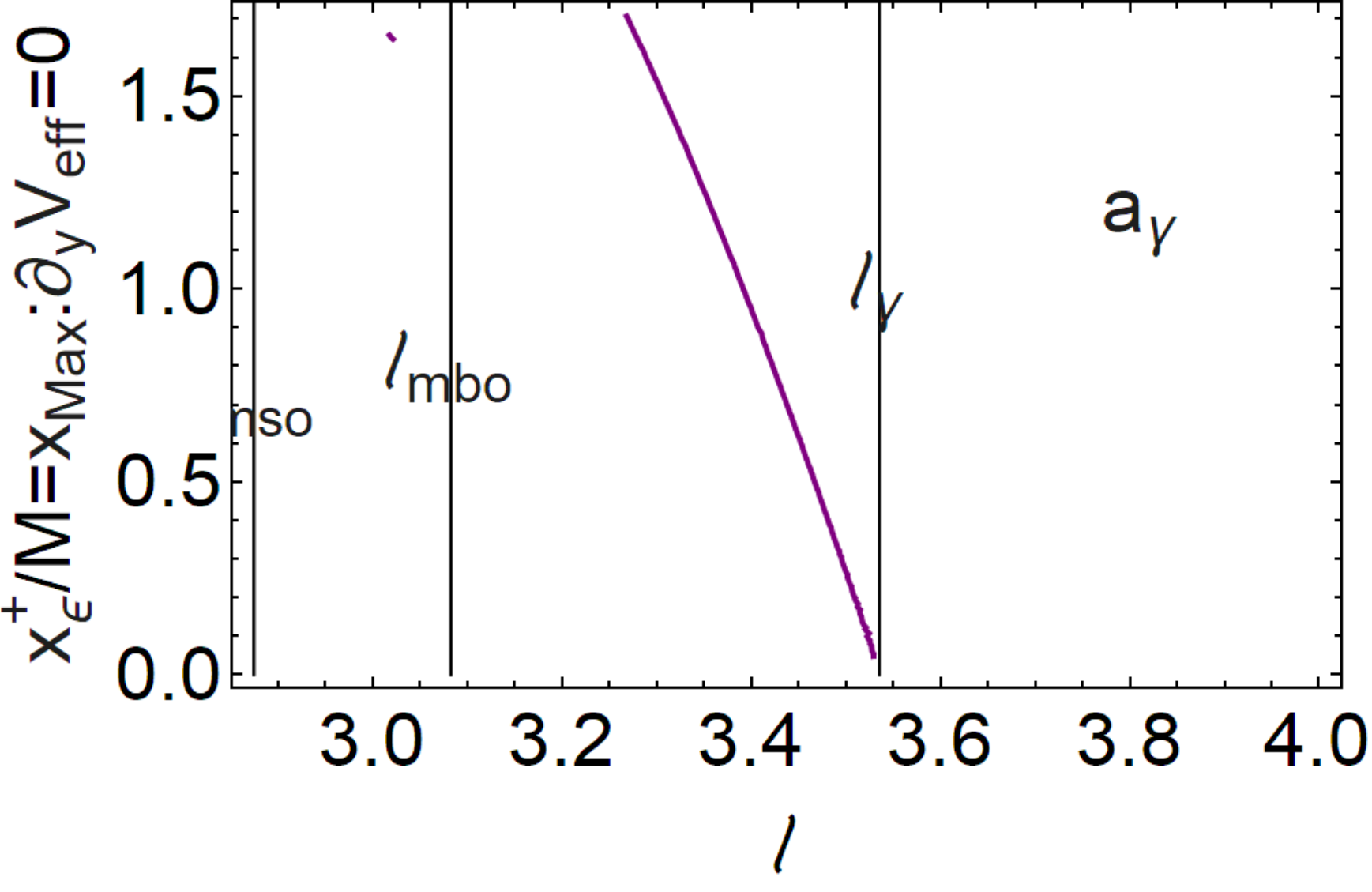}
  \caption{Study of the vertical penetration of the outer  ergosurface of  the disk. Plot of solutions  $x/M$ (vertical axis, (here is ($x=r\cos\theta, y=r\sin\theta)$, on the equatorial plane there is $y\equiv r$)  crossing of the line of geometrical maxima $\partial_y V_{eff}=0$ on the ergosuface for $\ell>\ell_{mso}$ where $\ell$ is the specific angular momentum of the fluid. mbo is for marginally bounded orbit, mso is for  marginally stable  orbit, $r_{\gamma}$ is the corotating photon orbit.  $r_+$ is the outer horizon.  Different spins $\mathbf{A}_{\epsilon}^+\equiv\{a_{mbo},a_{mbo}^b,a_{\gamma},a_{\gamma}^b,a_{mso}\}$ are represented as  in Figs\il(\ref{Fig:polodefin1}). Momenta $\ell_{mso}$, $\ell_{mbo}$ and $\ell_{\gamma}$ are black lines. For $a_{mbo}$, $a_{mbo}^b$ and $a_{\gamma}$, a zoom in the region $\ell<\ell_{\gamma}$ is shown. Cusped tori are for $\ell\in [\ell_{mso},\ell_{mbo}]$.}\label{Fig:gatplot17}
\end{figure}

\begin{figure}\centering
  \includegraphics[width=8cm]{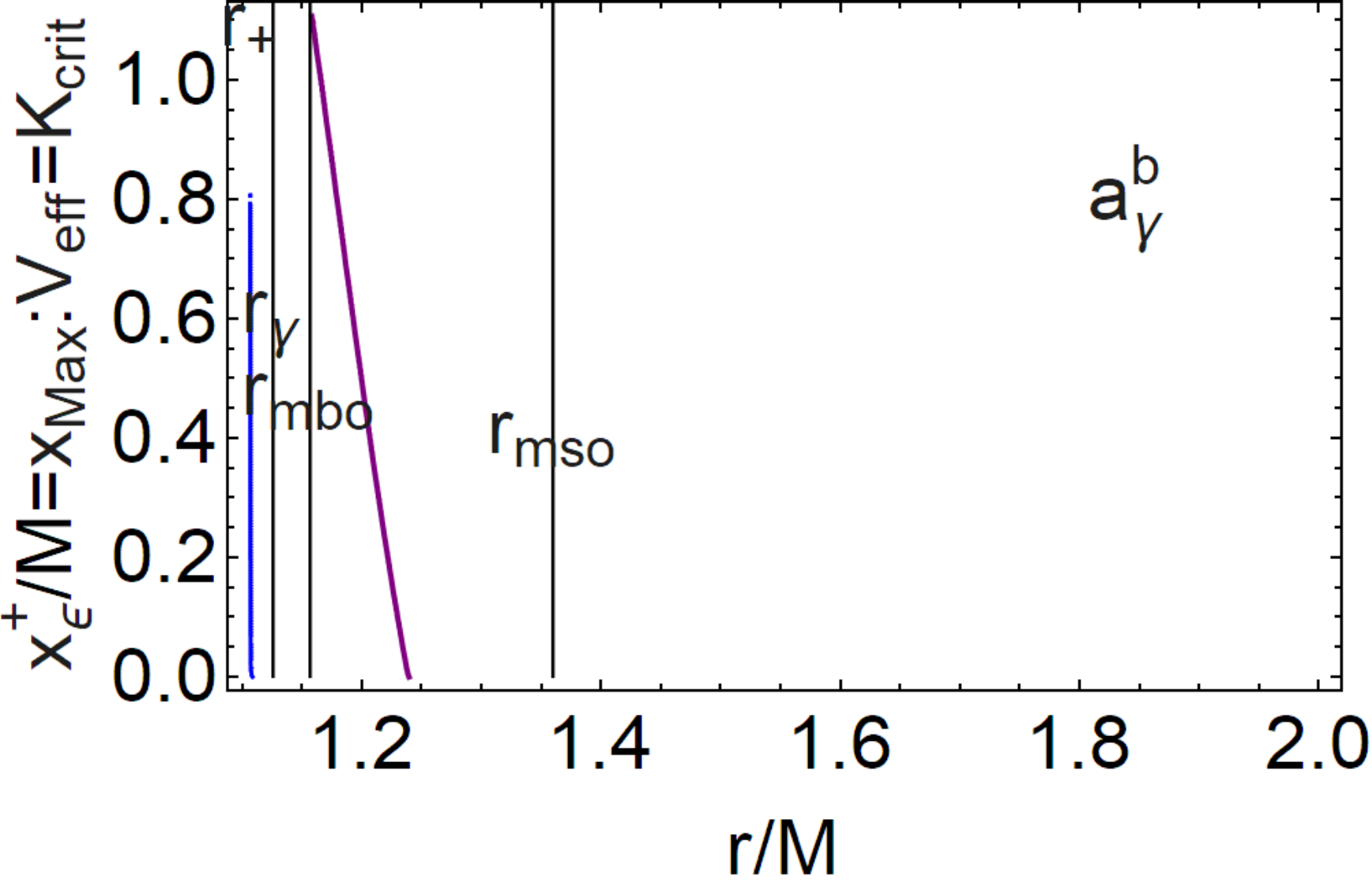}
     \includegraphics[width=8cm]{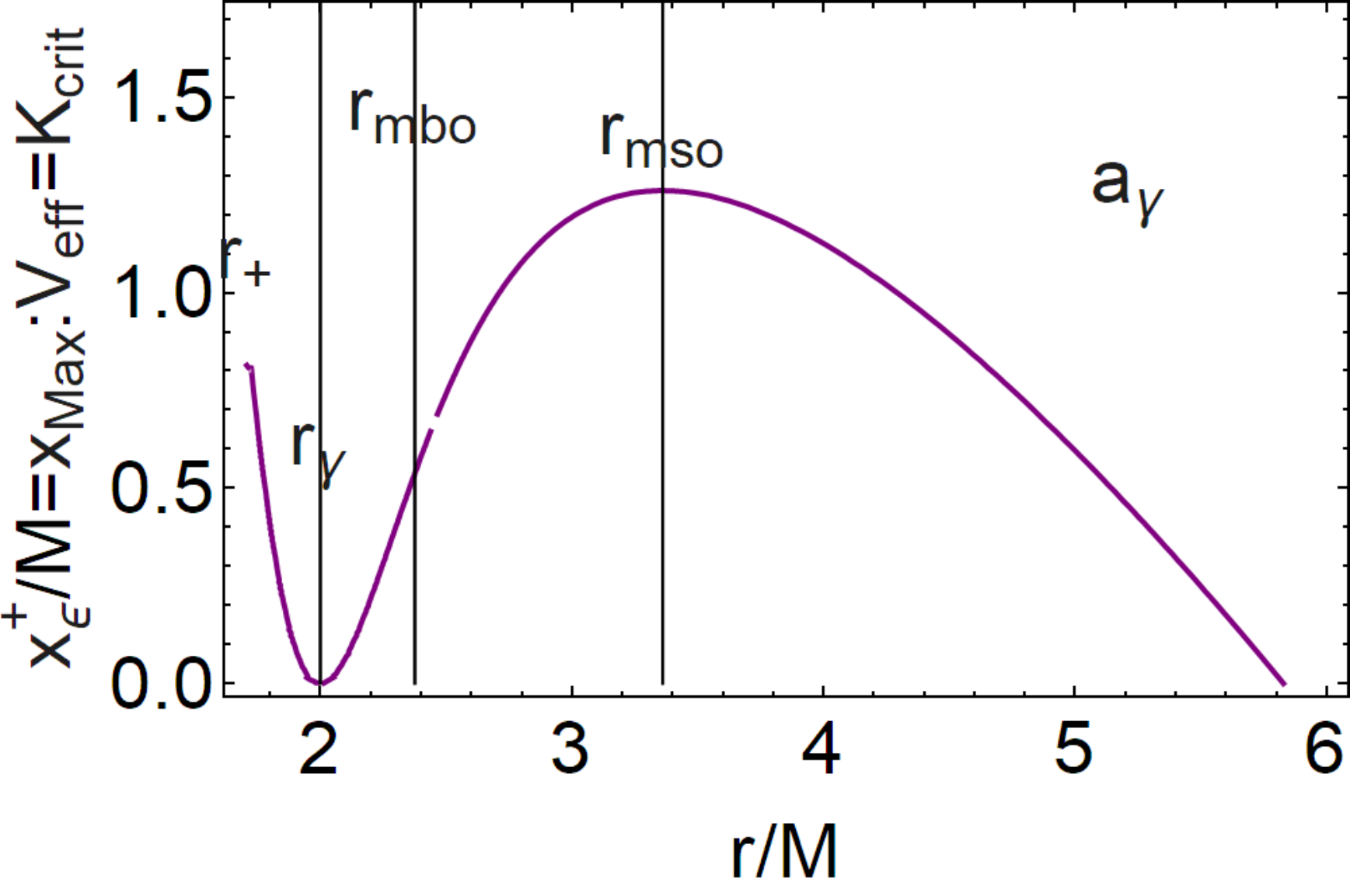}
  \includegraphics[width=8cm]{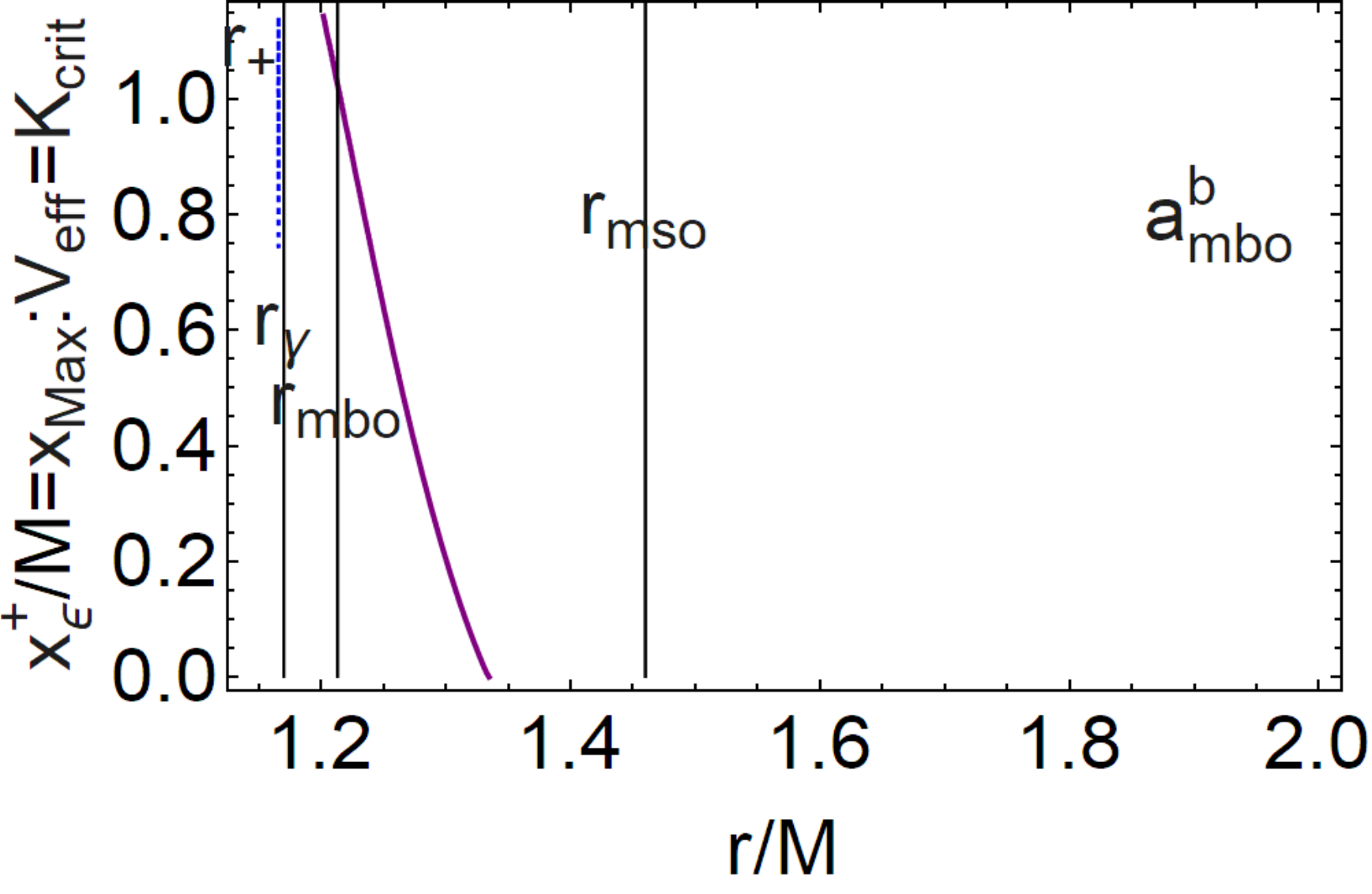}
    \includegraphics[width=8cm]{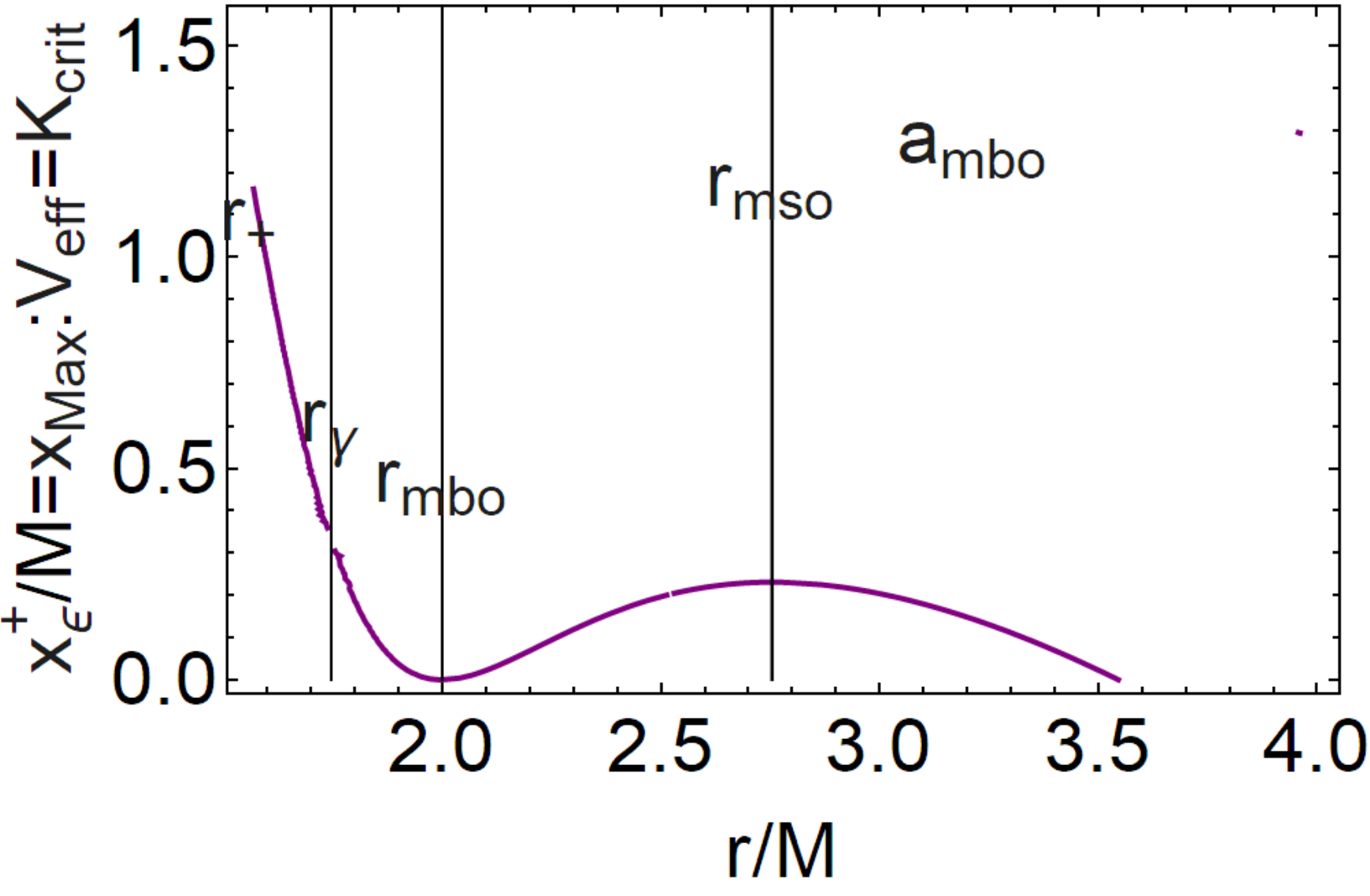}
    \caption{Study of the tori crossing the  outer  ergosurface.  Different spins $\mathbf{A}_{\epsilon}^+\equiv\{a_{mbo},a_{mbo}^b,a_{\gamma},a_{\gamma}^b,a_{mso}\}$ are represented as  in Figs\il(\ref{Fig:polodefin1}). Plot of solutions  $x/M=x_{\epsilon}^+/M\in[0,r_+]$ the outer ergosurface (vertical axis, there is ($x=r\cos\theta, y=r\sin\theta)$, on the equatorial plane there is $y\equiv r$)  crossing  the torus surface $V_{eff}=K_{\times}$ on the ergosurface as function of  $r>r_{+}$ (outer Killing horizon). $\ell$ is the fluid specific angular momentum.  We consider the critical configurations,   evaluating $V_{eff}$ on $\ell(r)$, setting centers or cusps $V_{eff}(\ell(r))=K_{\times}(r)$. $r_{mbo}$ is  the  marginally bounded orbit, $r_{mso}$ is  the   marginally stable  orbit, $r_{\gamma}$ is the corotating photon orbit.   Therefore for $r<r_{mso}$ sets a cusp, while for $r>r_{mso}$ is a center.  Center of (quiescent) tori with $\ell>\ell_{\gamma}$ is at $r>r^b_{\gamma}$, center of (accreting) tori with $\ell\in[\ell_{mso},\ell_{mbo}]$ with $r_{center}\in[r_{mso},r_{mbo}^b]$, while tori with $\ell\in[\ell_{mbo},\ell_{\gamma}]$ associated to quiescent tori and proto-jets have center in $r_{center}\in[r^b_{mbo},r_{\gamma}^b]$.}\label{Fig:gatplot5}
\end{figure}

We argue that  for   the small surfaces  crossing   the outer ergosurface or in the ergoregion, the  partially contained or dragged tori  might undergo a form of  instability, which could be also  coupled with the usual internal processes, of the disk,  induced by the geometry frame dragging, occurring in  regions of the disks also far  from the center of maximum pressure, leading to a process of disk "exfoliation" ("\textit{peeling}").
We consider  this process more in details below.
In Figs\il(\ref{Fig:gatplot8})  we show the conditions to be satisfied  for the geometrical  maximum of the cusped surfaces intersecting  the outer  ergosurface.
In Figs\il(\ref{Fig:gatplot17})  we approach the analysis of the torus exfoliation studying  the solutions of the problem of the
disks crossing  the outer  ergosurface  on $\theta\neq\pi/2$,  by its geometrical maximum (or the minimum) according to the analysis of  Figs\il(\ref{Fig:weirplot}). We show the solutions  $x/M$ (vertical axis),   crossing of the line of geometrical maxima, defined by $\partial_y V_{eff}=0$, with  the ergosurface for fluid specific angular momentum $\ell>\ell_{mso}$, in the geometries    of the set $\mathbf{A}_{\epsilon}^+=\{a_{mbo},a_{mbo}^b,a_{\gamma},a_{\gamma}^b,a_{mso}\}$. The maximum heigh of the tori  defined in the ergoregion is   rather small. We note, in accordance with the analysis of  Figs\il(\ref{Fig:gatplot8}), (\ref{Fig:gatplot17}), (\ref{Fig:gatplot5}),  that for the geometries with  spins $a=a_{mbo}^b$, $a=a_{mbo}$ and $a=a_{\gamma}^b$ there is a solution  of this problem for cusped tori which have specific  angular momentum with $\ell\in\mathbf{ L_1}$. At larger momentum, $\ell>\ell_{\gamma}$,  there is $r_{center}\geq r^b_{\gamma}$, and the torus can be also  very large. The results of the  analysis of the  torus inner edge are  in  Figs\il(\ref{Fig:spessplhoke1}).  In the restricted range of tori parameters where
cusped tori are,   only  geometries of a particular  range of dimensionless \textbf{BH} spin $a/M$  admit  tori with  center  of maximum pressure $r_{center}$ in the ergoregion, which is  necessary but  not  sufficient condition for the  intersection of the  geometrical maxima with the ergosurface--see Figs\il(\ref{Fig:PlotVamp1}).
%Therefore
An analogue situation  is  in Figs\il(\ref{Fig:gatplot5}) where is shown   the crossing of the critical  toroidal configurations (any point)    with the outer ergosurface.
\begin{figure}\centering
  % Requires \usepackage{graphicx}
  \includegraphics[width=5.5cm]{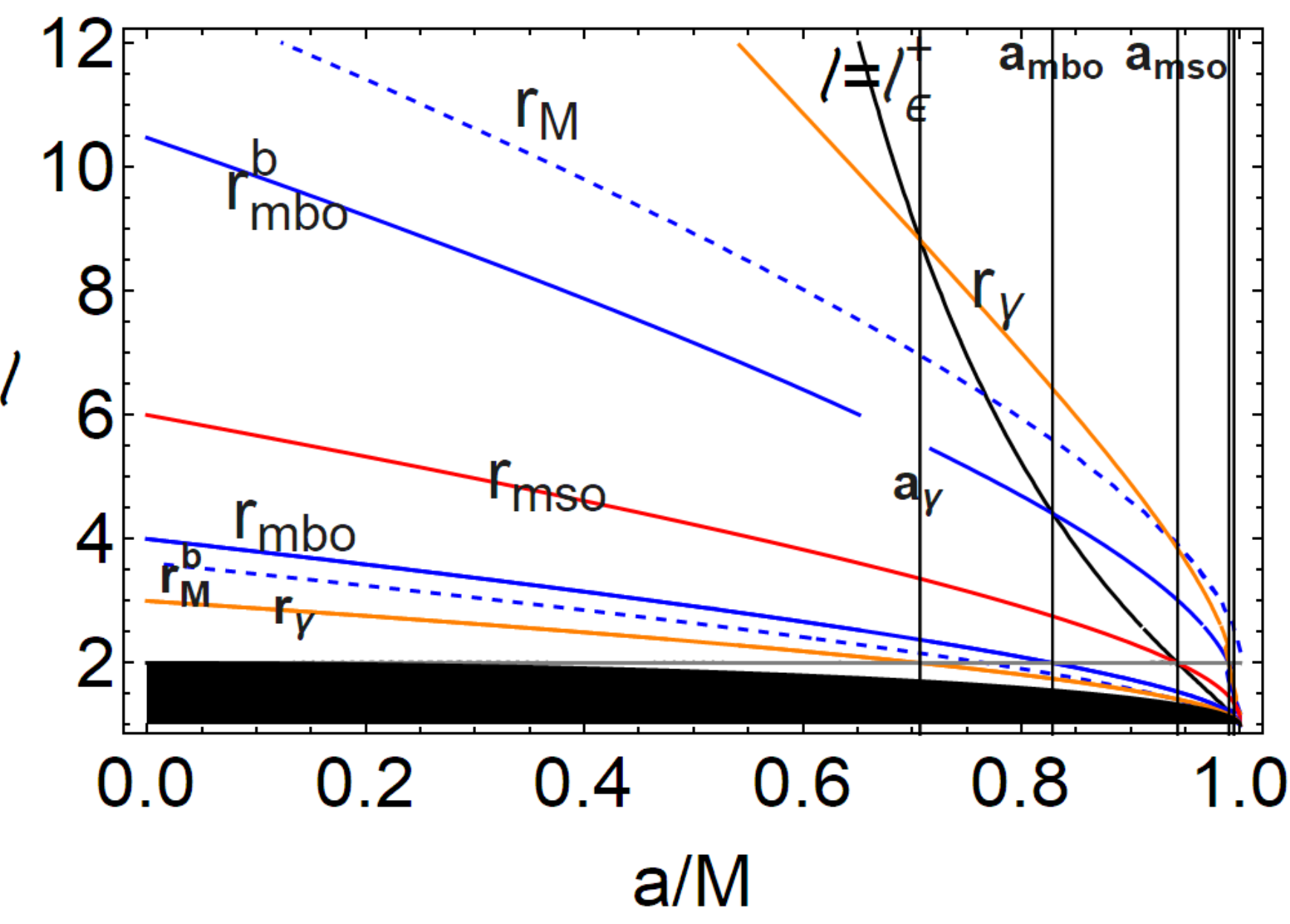}
  \includegraphics[width=5.5cm]{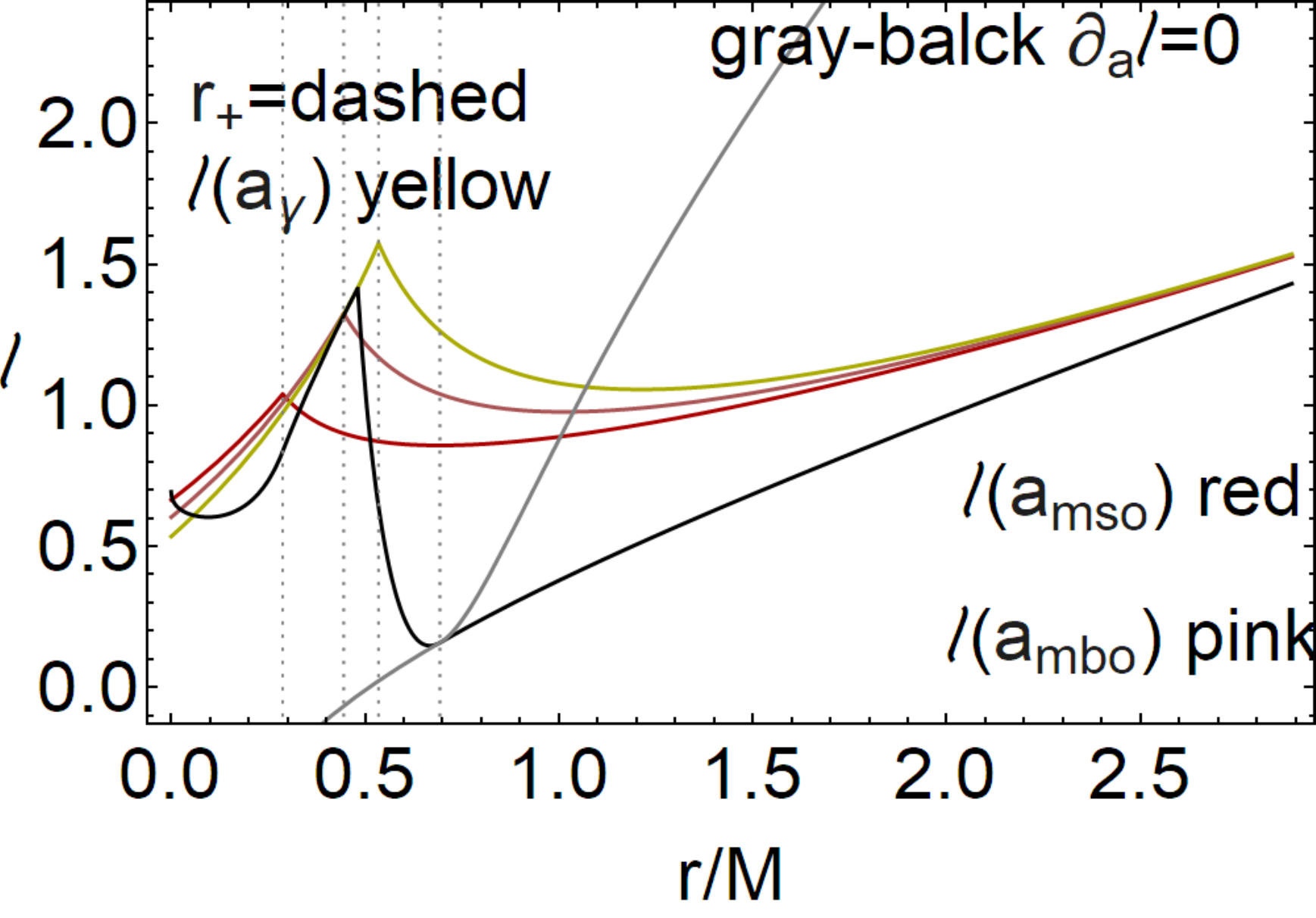}
    \includegraphics[width=5.5cm]{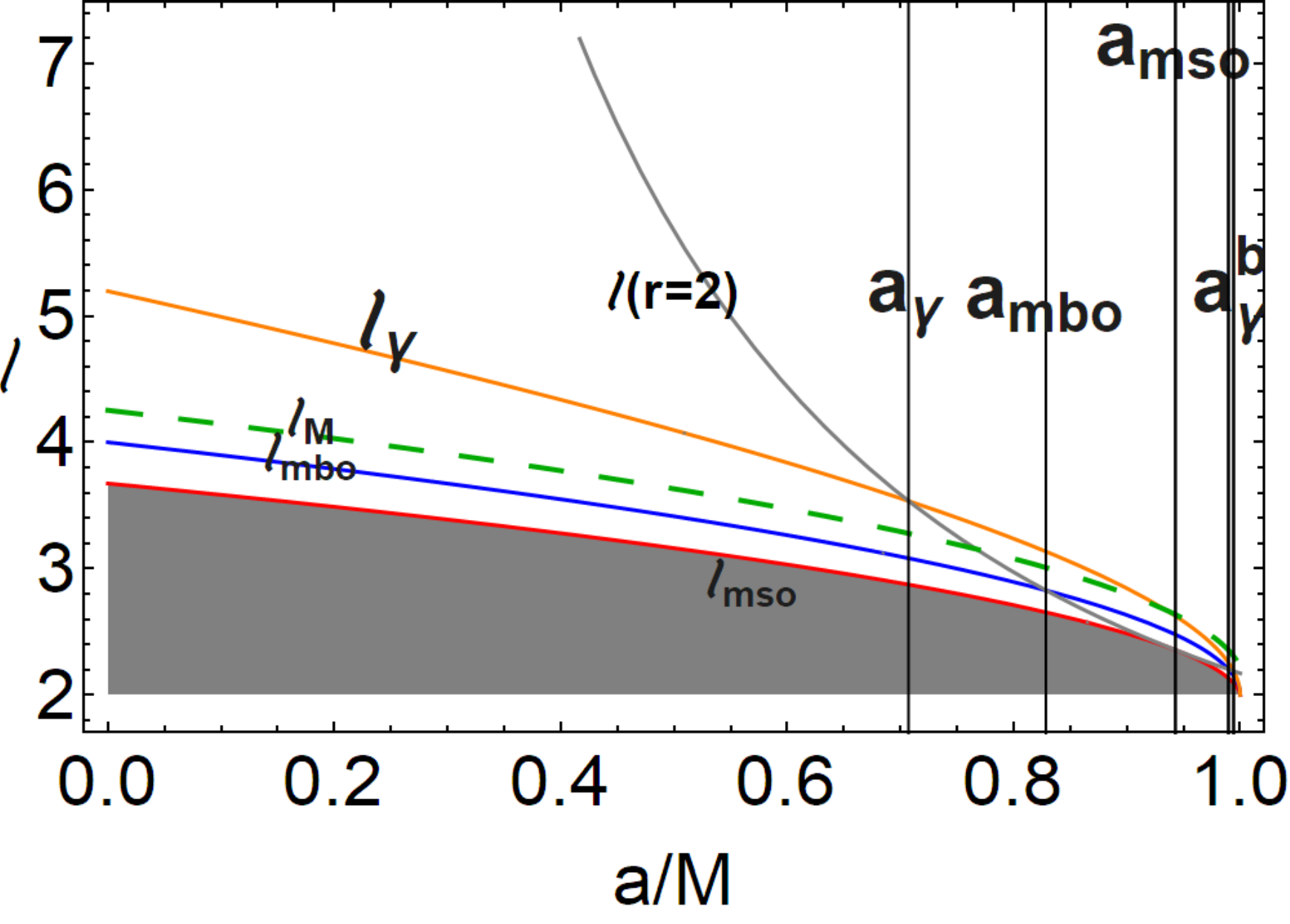}
  \caption{Notable momenta and radii. Black region is the region $r<r_+$, where $r_+$ is the outer horizon. $\ell$ is the specific angular momentum of the fluid. mbo is for marginally bounded orbit, mso is for  marginally stable  orbit, $r_{\gamma}$ is the corotating photon orbit. Gray region is region $\ell<\ell_{mso}$ where no accretion  tori can be defined. Dashed blue line  is the curve $\partial_r^2\ell=0$, the curve $r:\ell=\ell_{\epsilon}^+\equiv \ell(r_{\epsilon}^+)$ is also represented.  Central panel shows specific angular momentum of the fluid $\ell$ for different spins $\mathbf{A}_{\epsilon}^+\equiv\{a_{mbo},a_{mbo}^b,a_{\gamma},a_{\gamma}^b,a_{mso}\}$ are represented as  in Figs\il(\ref{Fig:polodefin1}), and solution of $\partial_a \ell=0$. }\label{Fig:Plotsoorr}
\end{figure}
Notable radii and values of momenta are represented in Figs\il(\ref{Fig:Plotsoorr}).
\begin{figure}\centering
  % Requires \usepackage{graphicx}
  \includegraphics[width=8cm]{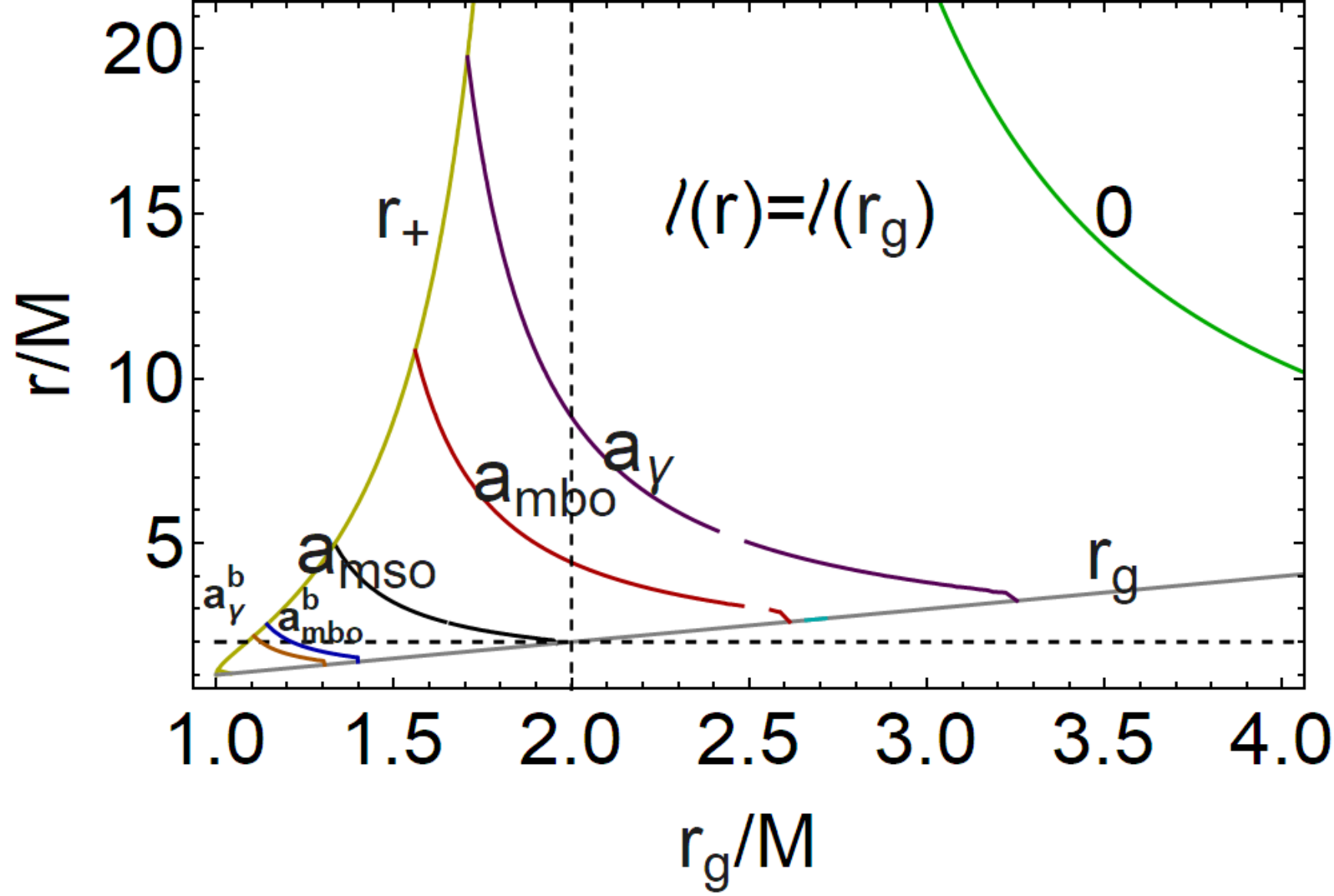}
  \includegraphics[width=8cm]{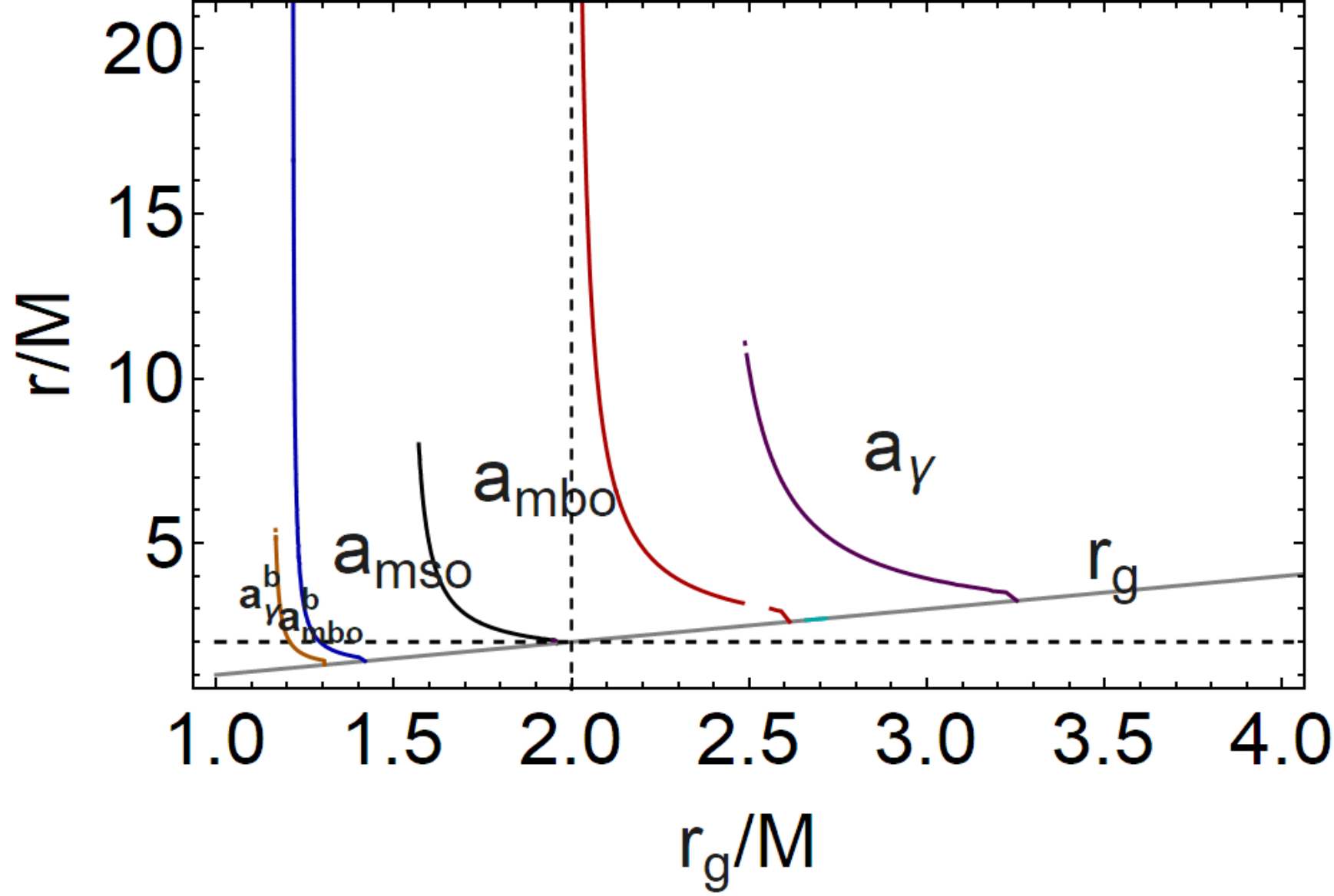}
  \caption{Solution of $r: \ell(r)=\ell(r_g)$ (left panel) and $r:K(r)=K(r_g)$ (right panel) as function of the radius $r_g/M$, for the spins $\mathbf{A}_{\epsilon}^+\equiv\{a_{mbo},a_{mbo}^b,a_{\gamma},a_{\gamma}^b,a_{mso}\}$--see Figs\il(\ref{Fig:polodefin1}).  $r_+$ is the outer \textbf{BH} horizon,  yellow curve is $\ell(a_\pm)$, where $a_{\pm}\in[0,M]$ is the horizon curve in the plane $a-r$. The static Schwarzschild case is the green curve for $a=0.$ Dashed lines are the outer ergosurface $r_{\epsilon}^+=2M$.}\label{Fig:Plotbicilia}
\end{figure}
Figs\il(\ref{Fig:Plotbicilia}) describe  solutions of the problem $f(r)=f(r_g)$, for two radii $r\neq r_g$ and  for $f\in\{\ell(r),K(r)\}$ at different spins   of the set  $\mathbf{A}_{\epsilon}^+$. For $f=\ell$, solutions are the radii $(r,r_g)$, which are  the center $r>r_{mso}$ or the  critical point of the tori as  $r_{\times}$ or $r_j$,  where  $r_{crit}<r_{mso}$. Therefore  the analysis show the {distance} between maximum and minimum pressure  points in the cusped disk, as well as the  location of the points. The curves shown Figs\il(\ref{Fig:Plotbicilia})  are bounded in the region delimited by $r_g$ and the  radius $r:\ell=\ell(a_{\pm})$ where $a_{\pm}\equiv \sqrt{r(2M-r)}$ is the horizon curve in the plane $a-r$.
The solution of the problem for $f=K$ distinguishes  the points $(r,r_g)$ such that $K_{center}(\ell^*)=K_{crit}(\ell^\flat)$ for two tori  with specific angular momenta $\ell^*\neq \ell^\flat$ respectively--see \citet{ringed}.
\subsubsection{Dragged disks thickness: influence of the dragging frame on the disk thickness}\label{Sec:influ-ergosra}
\begin{figure}\centering
  % Requires \usepackage{graphicx}
    \includegraphics[width=5.6cm]{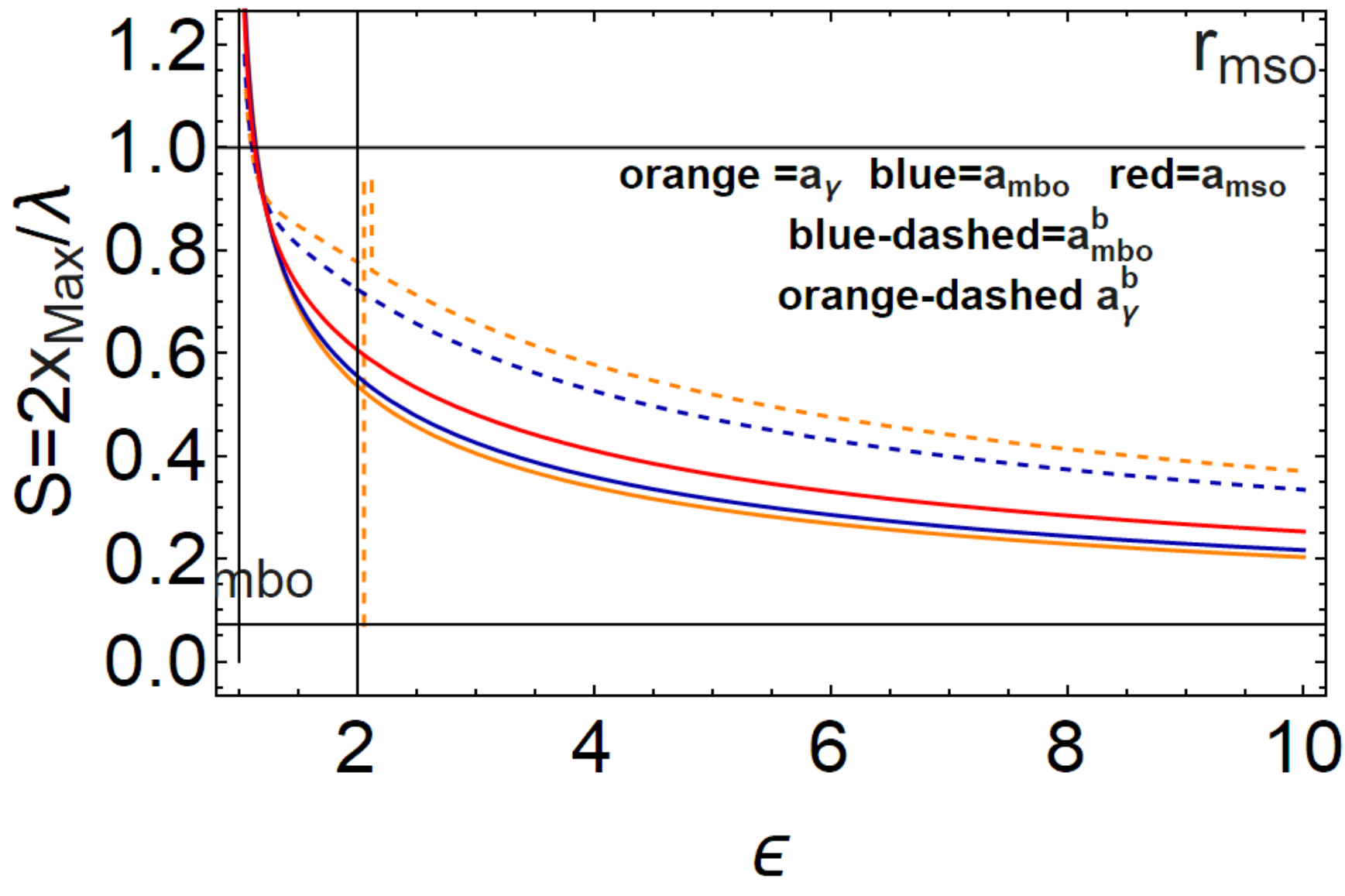}
  \includegraphics[width=5.6cm]{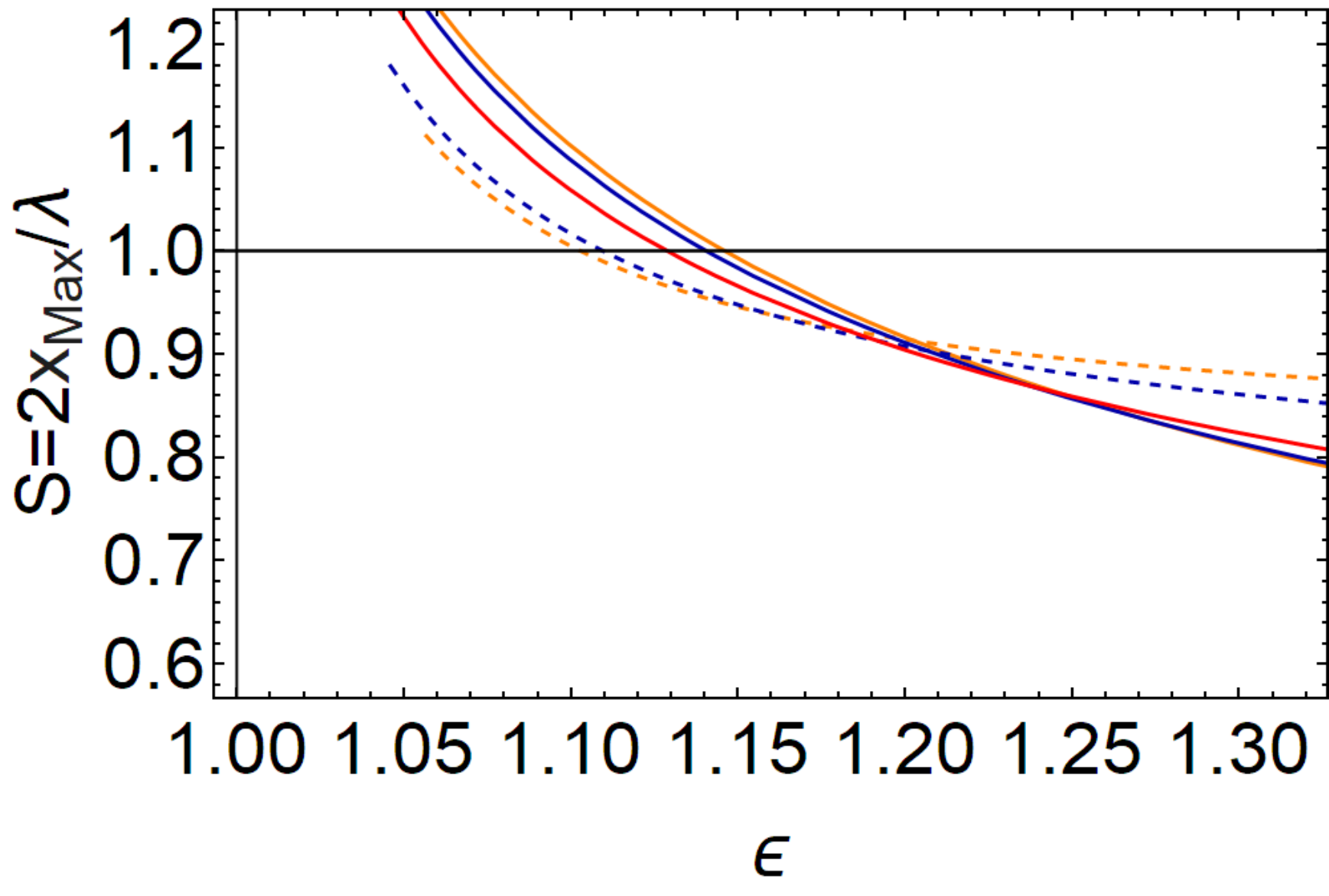}
  \includegraphics[width=5.6cm]{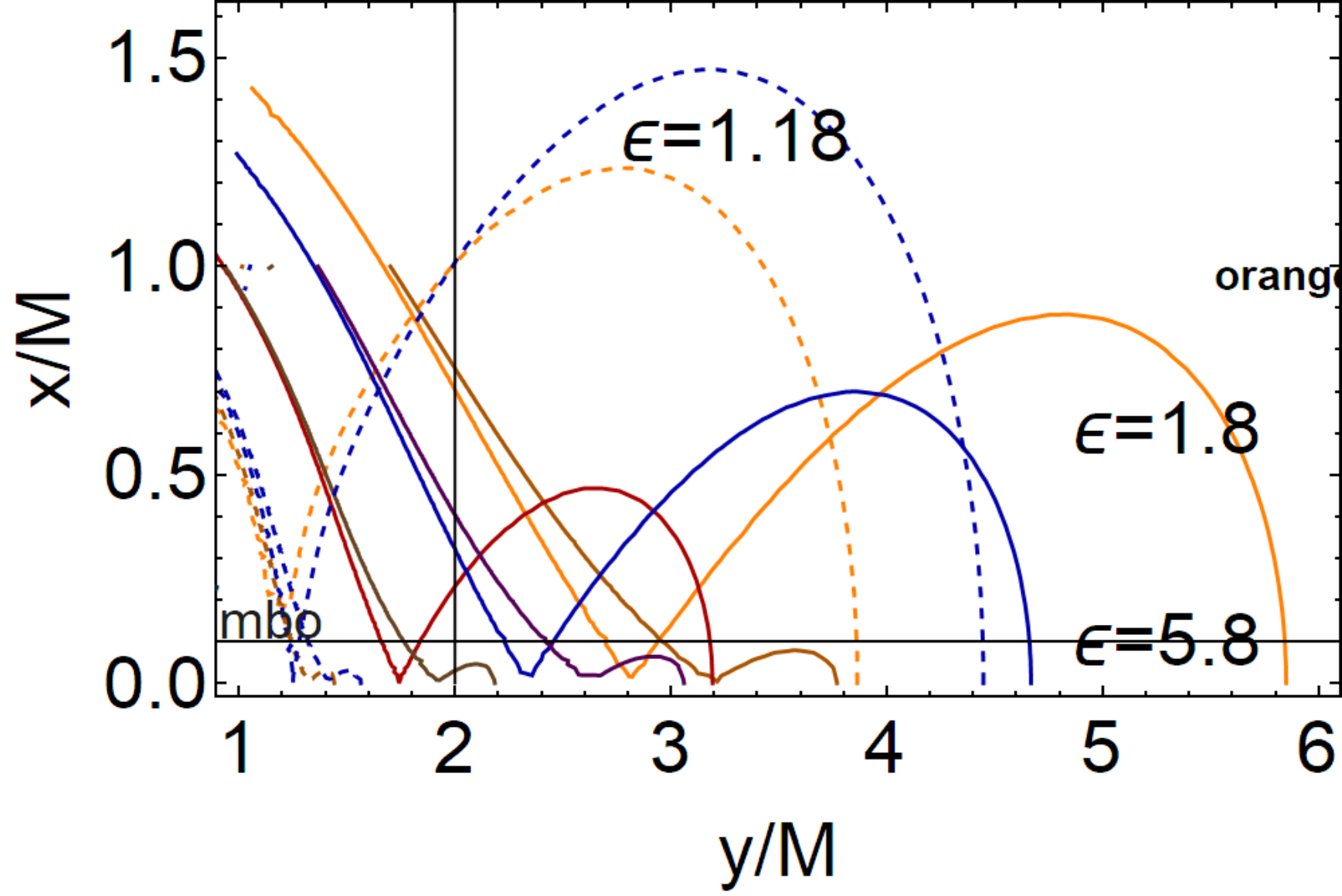}
 \includegraphics[width=5.6cm]{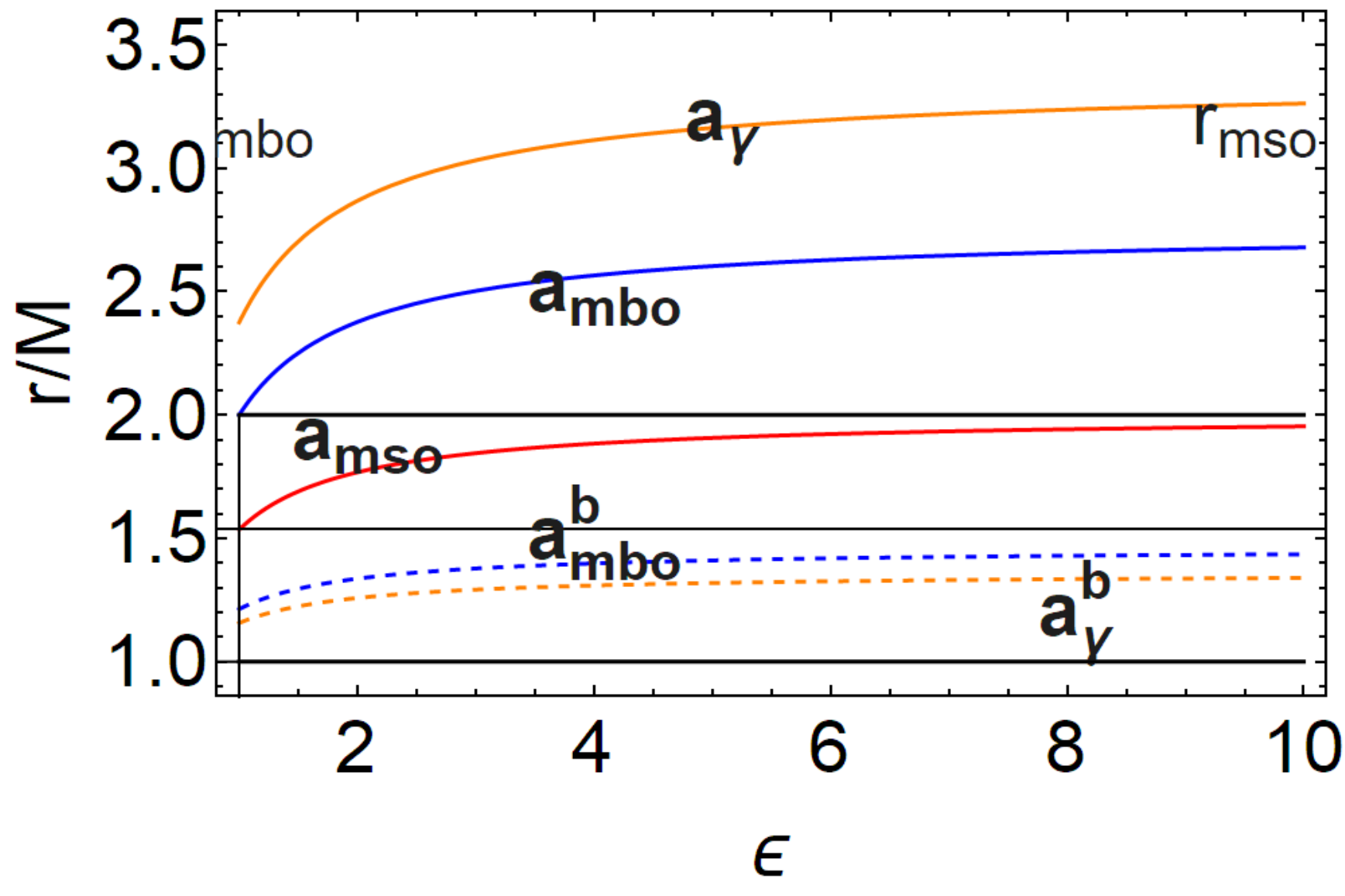}
\includegraphics[width=5.6cm]{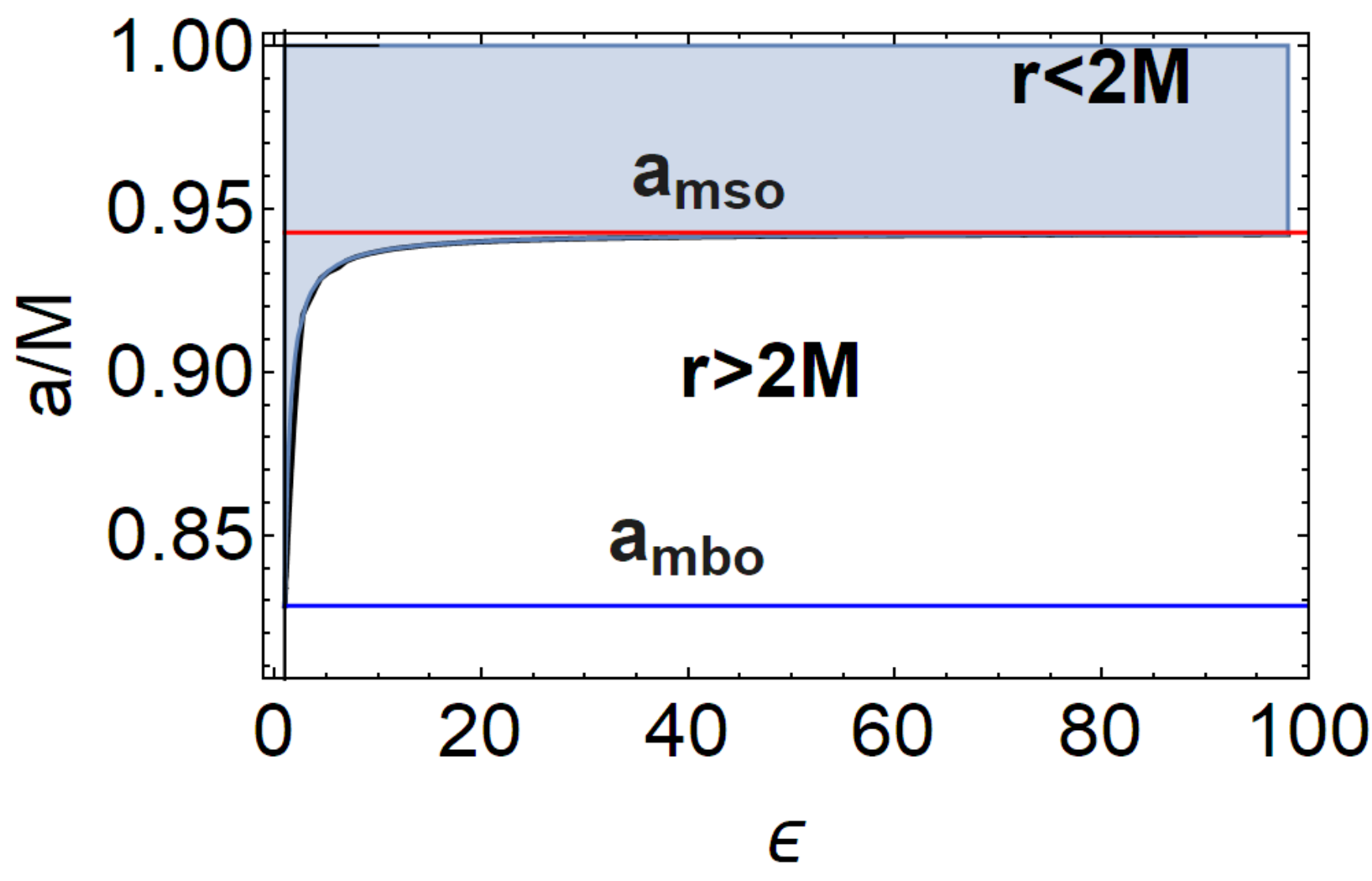}
   \caption{Study of the tori geometrical thickness $\Sie\equiv 2x_{\max}/\lambda$ of the cusped tori in the ergoregion (dragged surfaces) or close to the outer the ergosurface. \textbf{BH} geometries with $\mathbf{A_{\epsilon}^+}\equiv a\in\{a_{mbo},a_{mbo}^b,a_{\gamma},a_{\gamma}^b,a_{mso}\}$ are considered--see also Figs\il(\ref{Fig:polodefin1}).  On the equatorial plane $(x=0)$ the outer ergosurface is $y=r_{\epsilon}^+=2M$. $ x_{\max}$ is the geometrical maximum of the toroidal surface, solution of  $\partial_y V_{eff}=0$ and $V_{eff}=K_{\times}$ (the energy parameter $K$ at the cusp $r_{\times}$) (see also Figs\il(\ref{Fig:polodefin1})); $\lambda\equiv r_{outer}-r_{\times}$ is the torus elongation on the equatorial plane--see Figs\il(\ref{Fig:PlotVamp1}), (\ref{Fig:spessplhoke1}). (There is ($x=r\cos\theta, y=r\sin\theta)$, on the equatorial plane there is $y\equiv r$.). The geometrical thickness is represented as function of $\epsilon$, a parameter regulating the location of the critical radius (cusp) $r_{times}\in[r_{mbo},r_{mso}]: r_{\times}\equiv r_{mso}-(r_{mso}-r_{mbo})/\epsilon$. For $\epsilon=1$ there is $r_{\times}=r_{mbo}$ the limiting case of a proto-jet, for $\epsilon\rightarrow+\infty$ there is $r=r_{mso}$, the limiting case of one ring torus.
   Below panels show the range of existence of the radius $r(\epsilon)$ for selected spacetime in relation to the presence of cusp in the ergoregion $r_{\epsilon}^+=2M$ (outer ergosurface). It is evident how only tori orbing \textbf{BHs} with spins   $a\in\{a_{\gamma}^b,a_{mbo}^b,a_{mso}\}$  of the considered set have $r_{\times}<r_{\epsilon}^+$. Upper panels: thickness as function of $\epsilon$. Central panel shows a zoom in the region $\epsilon\leq1.30$. Right panel: tori for selected spacetimes according to the color range of the first panel. Smallest tori are for $\epsilon=5.8$ middle plain color tori are for $\epsilon=1.8$, larger dashed tori are at $\epsilon=1.18$.}\label{Fig:raisePlot}
\end{figure}

  In   Figs\il(\ref{Fig:weirplot}) and  Figs\il(\ref{Fig:Plotexale})   we show the results of the analysis of the  disk verticality in terms of the polar gradient of the pressure,  considering the lines of extremes  of the HD  pressure,  which  provide also the  surfaces geometric maximum,  obtaining  clear indication of the maximum vertical extension of the torus in $\Sigma_{\epsilon}^+$. The  disk thickness is an  essential parameter   in the study   tori oscillation processes,  in the regulation of the instability processes  and in  the accretion.
The analysis of   Figs\il(\ref{Fig:raisePlot})  clarifies the situation for the geometries of the set $\mathbf{A_{\epsilon}}^+$  (represented   Figs\il(\ref{Fig:polodefin1})).
Figs\il(\ref{Fig:raisePlot}) show  the geometrical thickness $\Sie\equiv 2x_{\max}/\lambda$ of the cusped tori in the ergoregion  for the dragged surfaces and  for tori close to the outer  ergosurface for the   \textbf{BH} geometries with  spins in $a\in \mathbf{A}_{\epsilon}^+$. In the definition of $\Sie$,  $ x_{\max}$ is the geometrical maximum of the cusped  toroidal surface, obtainable from  the  solution of  $\partial_y V_{eff}=0$ and defined by the condition $V_{eff}=K_{\times}$ (the energy parameter $K$ at the cusp $r_{\times}$), which is also studied in Figs\il(\ref{Fig:polodefin1}),  $\lambda\equiv r_{outer}-r_{\times}$ is the torus elongation on the equatorial plane--Figs\il(\ref{Fig:PlotVamp1}),  Figs\il(\ref{Fig:spessplhoke1}). Because we restricted our analysis on the cusped tori, we assumed $r\in [r_{mbo},r_{mso}]$: for  $r_{\times}=r_{mbo}$ there is  the limiting case of a proto-jet, for $r=r_{mso}$ there is  the limiting case of one ring torus located on the marginally stable orbit.
   It is evident how only tori for $a\in\{a_{\gamma}^b,a_{mbo}^b,a_{mso}\}$ of the set $\mathbf{A}_{\epsilon}^+$ have $r_{\times}<r_{\epsilon}^+$.

For larger  $r_{\times}\lesssim r_{mso}$, tori can be also very small, while  for smaller $r_{\times}\gtrsim r_{mbo}$ tori, including  partially contained tori in the ergoregion, are larger.
   For radii $r_{\times}$ very close to the marginally bounded orbit, for any spacetime considered, there is a turning point in the  tori thickness,  i.e.,
   at larger $r$ there is $\Sie_{a_{\gamma}}<\Sie_{a_{mbo}}<\Sie_{a_{mso}}
   <\Sie_{a_{mbo}^b}<\Sie_{a_{\gamma}^b}<1$. As value  $\Sie=1$ is   the limit for geometrically thick disks prevalently tori  under this condition  are small tori.
 However,  for cusp close   to  $r_{mbo}$,  for all the spacetimes there is a region where $\Sie_{a_{\gamma}^b}<\Sie_{a_{mbo}^b}<\Sie_{a_{mso}}<
 \Sie_{a_{mbo}}<\Sie_{a_{\gamma}}<1$  (right range) and   a second region (left range at $r_{\times}$ closer to $r_{mbo}$) where  $1<\Sie_{a_{\gamma}^b}<\Sie_{a_{mbo}^b}<
 \Sie_{a_{mso}}<\Sie_{a_{mbo}}<\Sie_{a_{\gamma}}$.
\subsubsection{Tori exfoliation}\label{Sec:mid-w-t}
The Lense-Thirring effect  of the  Kerr  geometry  is expected to hugely  affect partially contained and  dragged    tori (also if viscosity and radiation enter the play). The small tori can  undergo a process induced by the dragging effects in the ergoregion, at different planes  resulting   in (almost vertical) fractiousness  of  the tori surface,  which can  combine with the  accretion  processes onto the central \textbf{BH} attractor,
    leading to  a swarm of  (initially corotating)  particles and eventually  photons, which can constitute a tori atmosphere.    According to the initial data on the torus, they  can  escape in the outer region, remain trapped in the region $r<r_{inner}$, impacting back to the tori surface and  environment (analogously to a Poynting--Robertson effect), and eventually be  captured by the central \textbf{BH}   (\emph{disk exfoliation}).
  This process, relevant for the  smaller disks, may   also occur in combination with  the  Bardeen\&Petterson   effect in tori with a slight inclination  with respect to the equatorial plane.
  In this regard it is  necessary to reassess the "pressure forces" in the disks orbiting in $\Sigma_{\epsilon}^+$ which,   in the GRHD model with $\ell =$constant, are independent from the details of the polytropic equation of state, and depend only on the fluid effective potential gradients. The gradients of pressure represent the main factor for the  formation and stability  of the dragged  tori  (we also note there are no ZAMOs geodetics  in the \textbf{BH} ergoregion).  On the other hand, the pressure  gradients are also  the main factor to be evaluated for  the  exfoliation and the consequent  formation of  a possible atmosphere of  (free) particles  swarm.
 In this context it can be also important the role of external large scale magnetic field--see for example \citet{2021PhRvD.103b4021K,2016EPJC...76...32S,2016PhRvD..93l4055K}.

\begin{figure}\centering
  % Requires \usepackage{graphicx}
  \includegraphics[width=8cm]{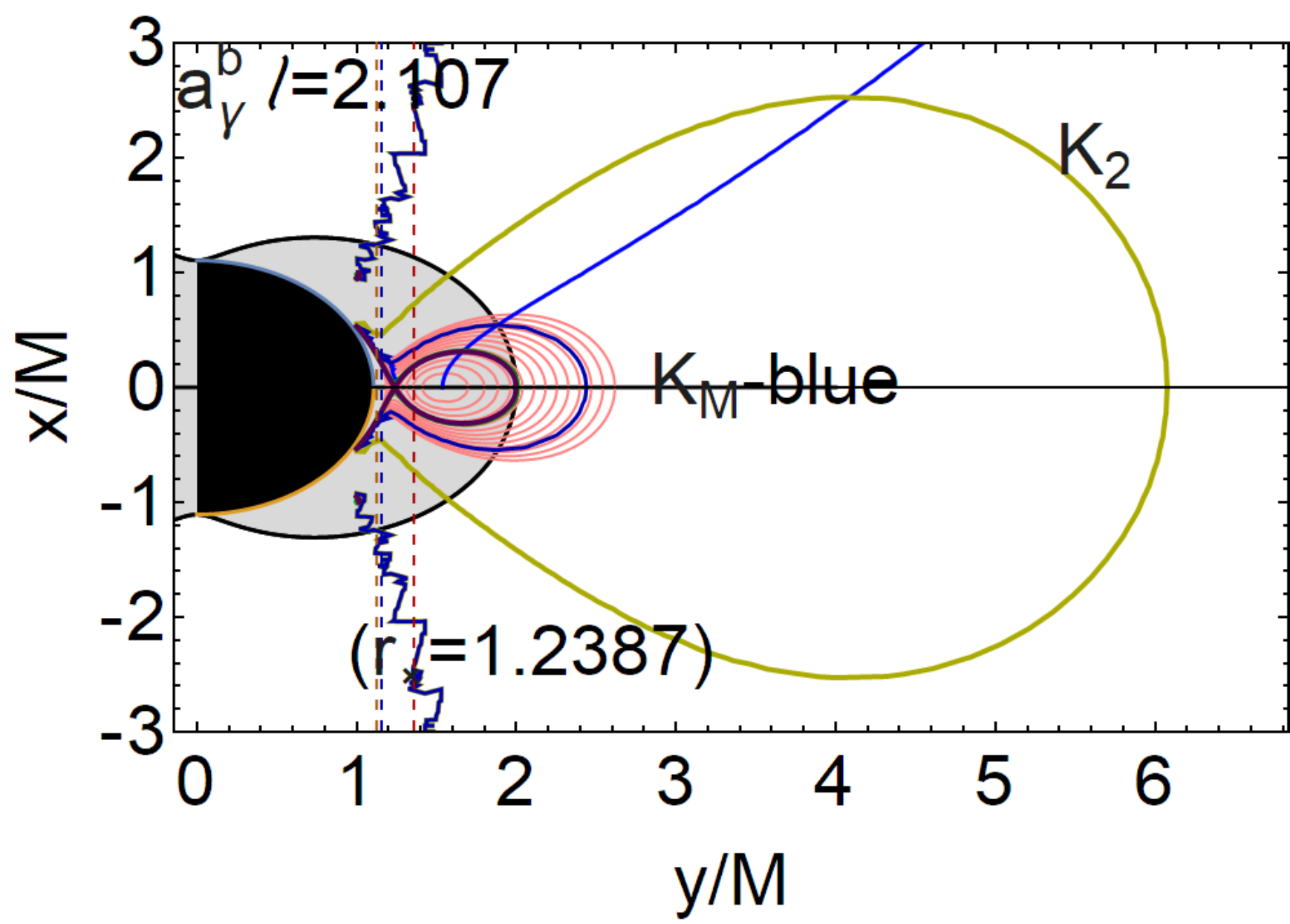}
  \includegraphics[width=8cm]{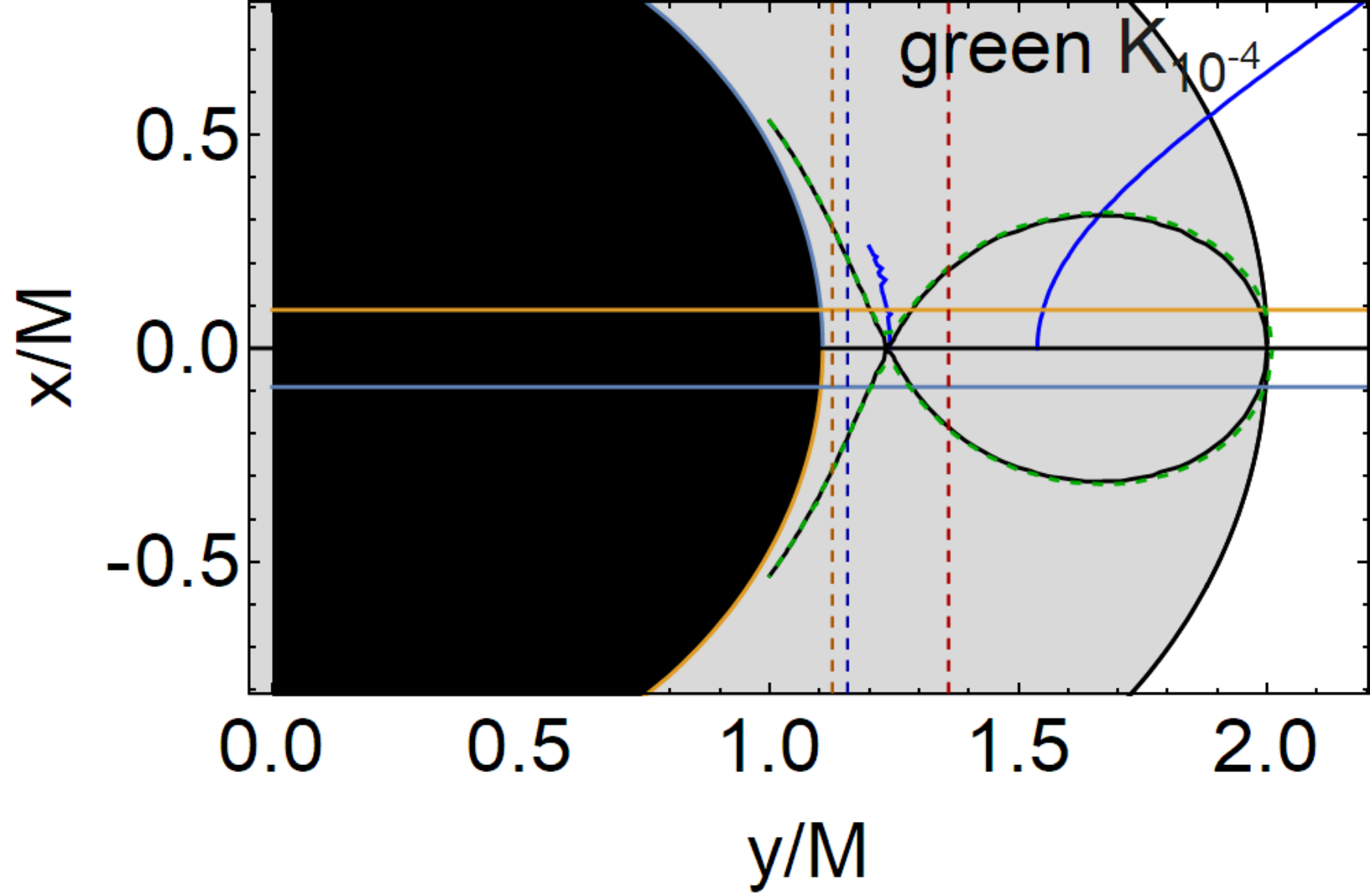}
    \caption{ Equipressure surfaces for GRHD tori are shown at different $K$ (the energy parameter). Black region is the central \textbf{BH} with spin   $a=a_{mbo}^b$ as  in Figs\il(\ref{Fig:polodefin1})--(upper line central panel).  Gray region is the outer ergosuface.  Vertical dashed lines are $r_{mbo}$  the  marginally bounded orbit, $r_{mso}$   the   marginally stable  orbit, $r_{\gamma}$  the corotating photon orbit on the equatorial plane. $\ell$ is the fluid specific angular momentum.  $r_{\times}$ is the torus cusp.
     $K_2\equiv K_{\times}+(1-K_{\times})/2$ and $K_{10^{-4}}\equiv K_{\times}+5\times10^{-4}$ and $K_M= 0.765671$ (blue line torus whose geometrical maximum is on the outer ergosurface)  see also analysis in Figs\il(\ref{Fig:ManyoBoF}). Blue curves set the extremes of the pressure and density in the disks: the maximum pressure inside the disk from the center of the configurations to the geometrical maximum  for different $K$ surfaces, and the inner minimum point of the pressures from the cusp.  The cusped torus has outer edge in $r=2M$ the outer ergosurface one the equatorial plane. Orbits of fluid particles and photons in this system is shown in Figs\il(\ref{Fig:Plotcredcatra})  and Figs\il(\ref{Fig:Plotcredcatr}),(\ref{Fig:zzPlotcredcatr}), (\ref{Fig:zzPlotcredcatr1}) respectively. }\label{Fig:PlotTRial}
    \end{figure}

Here we do not focus on the  particulars of the processes leading to the  formation,  reflection, or emission of particles and photons, but on the evaluation of the effectiveness of this process in the considered hydrodynamic models, giving thus a groundwork for the analysis of more complex systems.
This process may also give  rise   to  effective dissipative  forces  (or inducing  internal turbulence), especially in small tori  with a small  inner elongation  $\lambda=r_{center}-r_{inner}$ (distance between the maximum pressure point and the inner edges), and  crossing   the ergosurface  at some plane $\theta$, as considered  for example in Figs\il(\ref{Fig:polodefin1}),  characteristic of the \textbf{BHs} geometries  with   spins $a>a_{\gamma}^b$.
Clearly,  these  particles and photons  interact  with  the  \textbf{BH} environment, for example in the presence of a magnetosphere, with  surrounding magnetic fields and other accreting materials. However,  it is necessary to understand the characteristic  time scales for the eventual destruction of the disk, which  would  depend on the mass (regulated by the $K$ parameter) and on the  disk and \textbf{BH} specific momentum.
In the context considered here, this phenomenon involves disks  with  $r_{inner}\leq 2M $, which can combine  with   jet and proto-jets emission,  regulated  by the limiting
light surfaces provided by the solutions of {$\mathcal{L}\cdot \mathcal{L}=0$}-- Figs\il(\ref{Fig:Plotexale}).
A particularly interesting  situation would occur when $r_{center}\leq 2M$ as this condition, where the inner part of the disk is dragged and the outer part of the disk is fully contained or partially contained in the ergoregion,  may imply the geometrical  maximum of the disk approaches the outer ergosurface.
Therefore in the process of  torus exfoliation we can distinguish the following two cases:
\begin{description}
\item[\textbf{(I)}] In the first case  there is  $r_{center}\leq r_{\max}\leq r_{\epsilon}^+=2M$, where $r_{\max}$ is the projection on the equatorial plane of the geometrical maximum of the torus.   This condition  is sufficient for partially contained tori with the inner region, $[r_{inner},r_{center}]$, in $\Sigma_{\epsilon}^+$. The first inequality,  $r_{center}< r_{\max}$, is always satisfied, while condition $r_{center}= r_{\max}$ occurs in  the limit of $K \approx K_{center}$.
\item[\textbf{(II)}] In the second case there  is $r_{\max}\geq r_{\epsilon}^+$.
We note that even with $r_{\max}\in \Sigma_{\epsilon}^+$ the geometrical maximum of  the torus can be outside the ergoregion.
 \end{description}
 We  also consider the following three cases for the toroidal configurations:
 \begin{description}
\item[\emph{\textbf{(i)}}] $K>K_{crit}$  correspondent to overcritical tori  with an accretion throat (there is $K\in]K_{crit},1[$);
 \item[\emph{\textbf{(ii)}}]$K=K_{crit}$,  correspondent to a cusped torus;
 \item[\emph{\textbf{(iii)}}] $K<K_{crit}$  associated to a quiescent torus (there is $K\in]K_{center},K_{crit}[$)
  \end{description}
  where $K_{crit}$ is the value of the $K$ parameter at the maximum of the fluid effective potential (minimum of the pressure).
Tori  in these three cases  also have equal constant  fluid specific angular momentum.
Cusped tori
are considered in the Figs\il(\ref{Fig:weirplot})-- first line-center panel. In some cases the toroidal  configuration  crosses the outer ergosurface  on a point  $z_s\in]r_{inner}, z_{\max}[$  or  for external $z_s\in]z_{\max}, r_{outer}[$  {(where $z$ is here  the crossing  that occurs on a point at $\theta\neq0$)} along the toroid surface. These  cases are  characterized by two different scenarios  concerning accreting matter colliding with the embedding tori materials.
The small  dimensions of the tori   (see Figs\il(\ref{Fig:raisePlot}))  allow  to perform a first analysis in the hypothesis of free test particle from the tori surface.
  We shall  assume free  particles with initial conditions dictated by the toroidal  configurations  studied in  Figs\il(\ref{Fig:weirplot}) from   four regions of the configurations crossing the static limit, which are the set $\mathcal{R_{\emph{free}}}=($\emph{\textbf{1.} the inner edge,  \textbf{2.} the accretion throat,  \textbf{3.}  the outer edge,   \textbf{4.}  the geometrical maximum}). In this first analysis we do not consider the details of the different  accretion mechanisms which may also involve plasmas particles and magnetic fields\footnote{For the analysis of  timelike  orbits, integration can be done  equivalently on two  sets of equations.
Considering derivation with respect to proper time $\tau$, the constraint equation  (normalization condition) can be included in the set of four equations and integrated  numerically.
The three  equations for the  particles coordinates can be numerically integrated   solving the constraint
for $t'(\tau)$, according to the initial data. Note we  numerically integrated  the geodesic equations $u^a\nabla_a u^b=0$ assuming zero pressure terms in Eq.\il(\ref{E:1a0}) and $u^a$ the particles four-velocity with normalization condition $u_au^a=-1$.
To fix the initial data of  Table\il(\ref{Table:Particle-models}) from the toroidal models of Figs\il(\ref{Fig:PlotTRial}),  particles inherit    aspects of the torus configuration reflected in the particle motion initial data, that is we express  $(\phi',
t')$ (at the initial time) from the  $(E, L)$ definition of Eq.\il(\ref{Eq:after}) (expressed   in terms of  $t'$ and  $\phi'$),
with   $L = \ell E$, where $\ell$ is from the torus model detailed in  Figs\il(\ref{Fig:PlotTRial}).
As we consider  timelke particles, parameter $E$  coincides with the $K$-parameter of the torus
(this can also be tested by applying the constraint to the initial data, as done  in the case of photon particles, used also  for example for the cases where, for some particles models, there is  $E\neq K$).
  We adapted the constraint to the case of photons, and $(E, L)$ parameters cannot be provided from  the tori  model  as for timelike particles,  depending on  the particular reflection and emission model. We used   $\phi'$ and  $t'$, in the initial data, expressed in terms of the $(E,L)$ parameters and, similarly to the timelike particles cases, with
 $\ell = L/E$  adapted to the torus,
the normalization condition on the initial surface is used to recover
the parameter $E$.}.
Particles models considered here  inherit the initial data  from the    tori construction (values of $\ell$ and $K$ parameters).  Therefore particles leave the tori  from  different points   of the toroidal  surfaces (data set $\mathcal{R_{\emph{free}}}$).  %with  initial data $\Qa_0\equiv \Qa(0)$, where $\Qa_0$ is for any quantity evaluated at initial time $\tau=0$.
From the definition of  $E$ and  $L$,
we find  $\phi'(0)=u^{\phi}(0)$  expressed in terms of $(\ell,E)$  using definition $L=\ell E$,
having
 $\ell$ settled according to the torus model and $E$  for the  timelike particles coincident to parameter   $K$  of the torus.
 Initial data $t'(0)=u'(0)$ is managed according to the   equations set up solving the constrain at the initial time.
 (Note $\ell$ and $K$ are  torus "global" parameter in the sense that are constant on each toroidal surface).
 Particles are initially  circularly orbiting on an orbit  defined by the location $r(0)$ and $\theta(0)$ on the torus surface  (there is no need to define explicitly $\phi(0)$) according with
  $\mathcal{R_{\emph{free}}}$, therefore $\theta'(0)=0$ and $r'(0)=0$.

\begin{figure}\centering
  % Requires \usepackage{graphicx}
    \includegraphics[width=5cm]{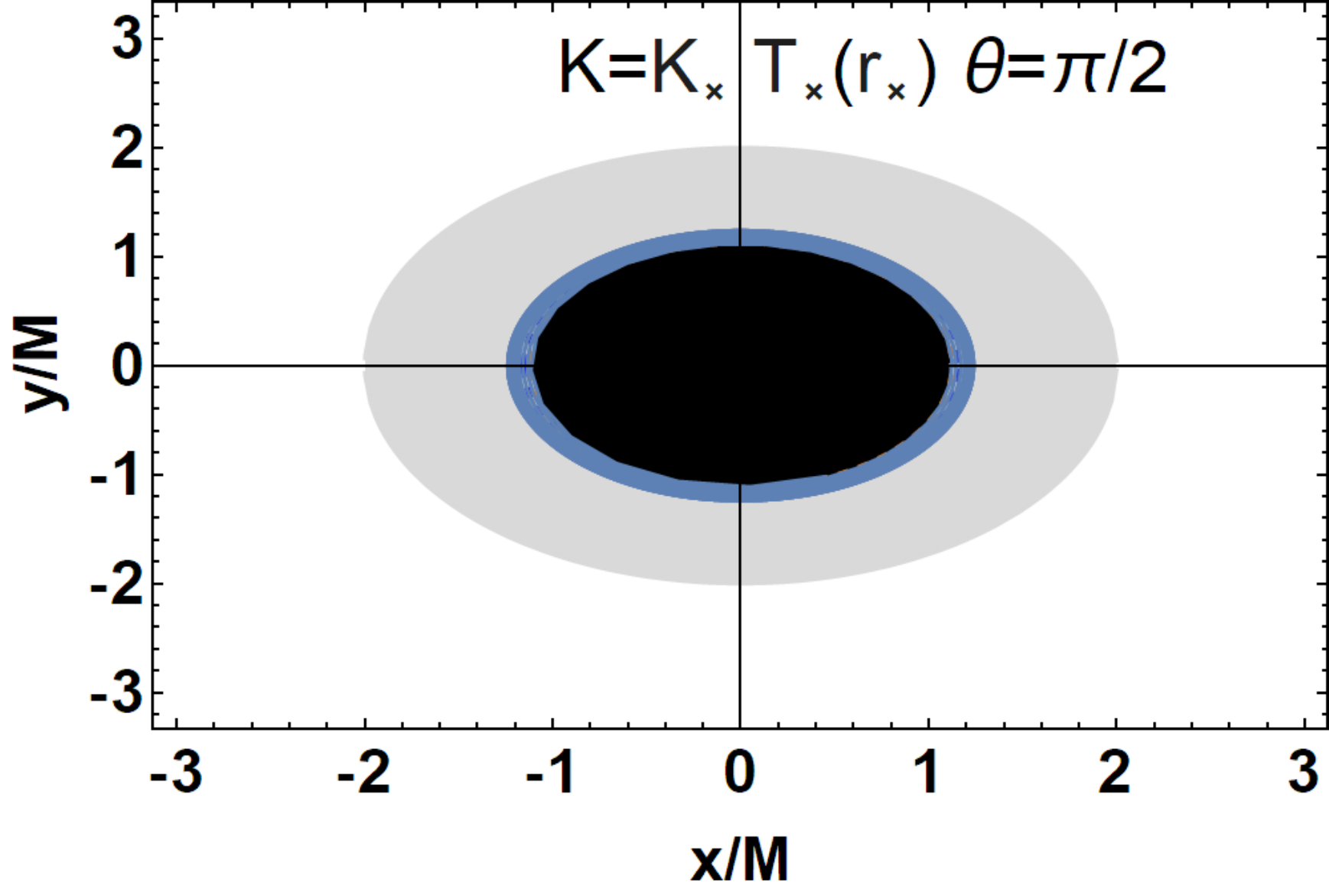}
  \includegraphics[width=5cm]{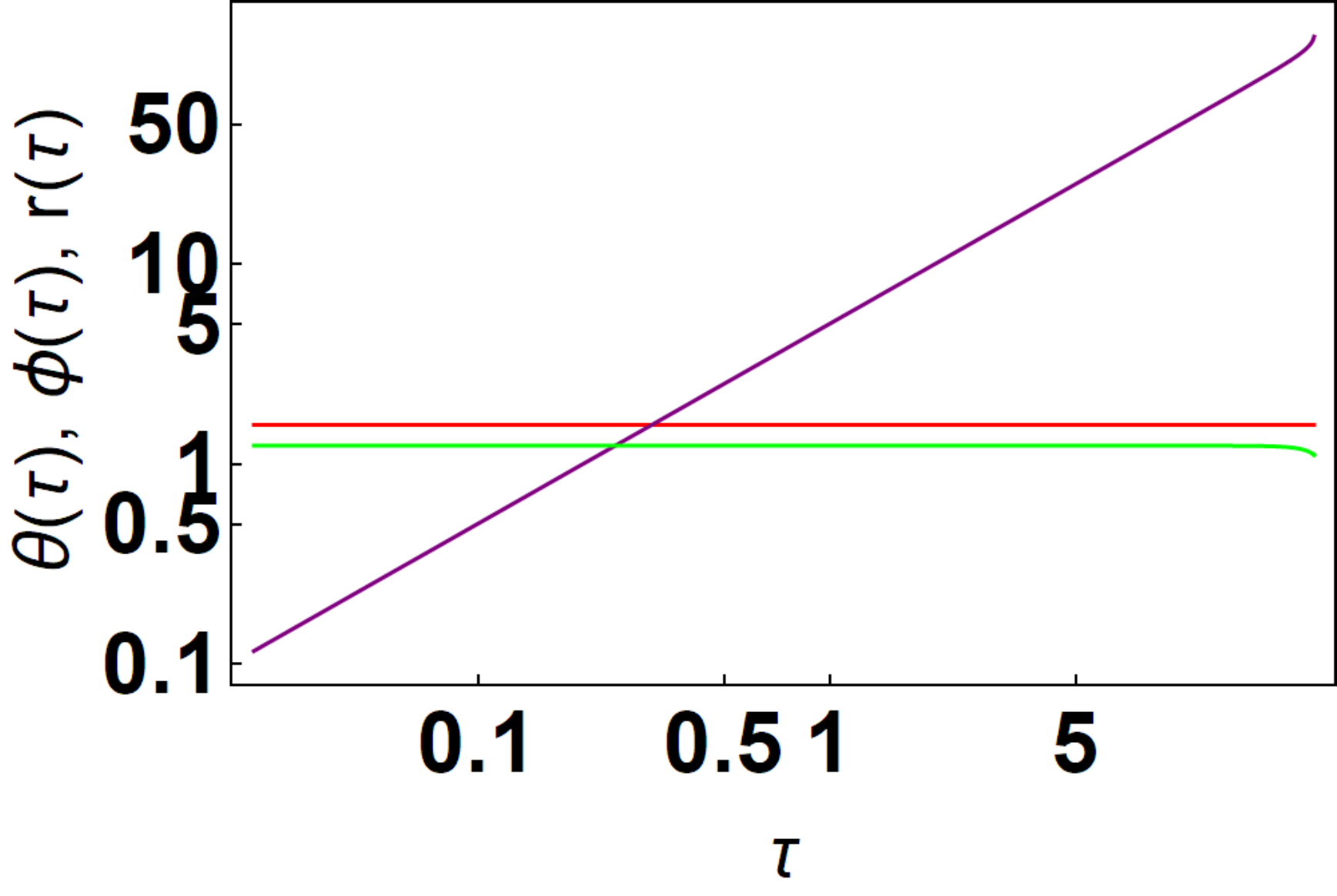}
  \includegraphics[width=5cm]{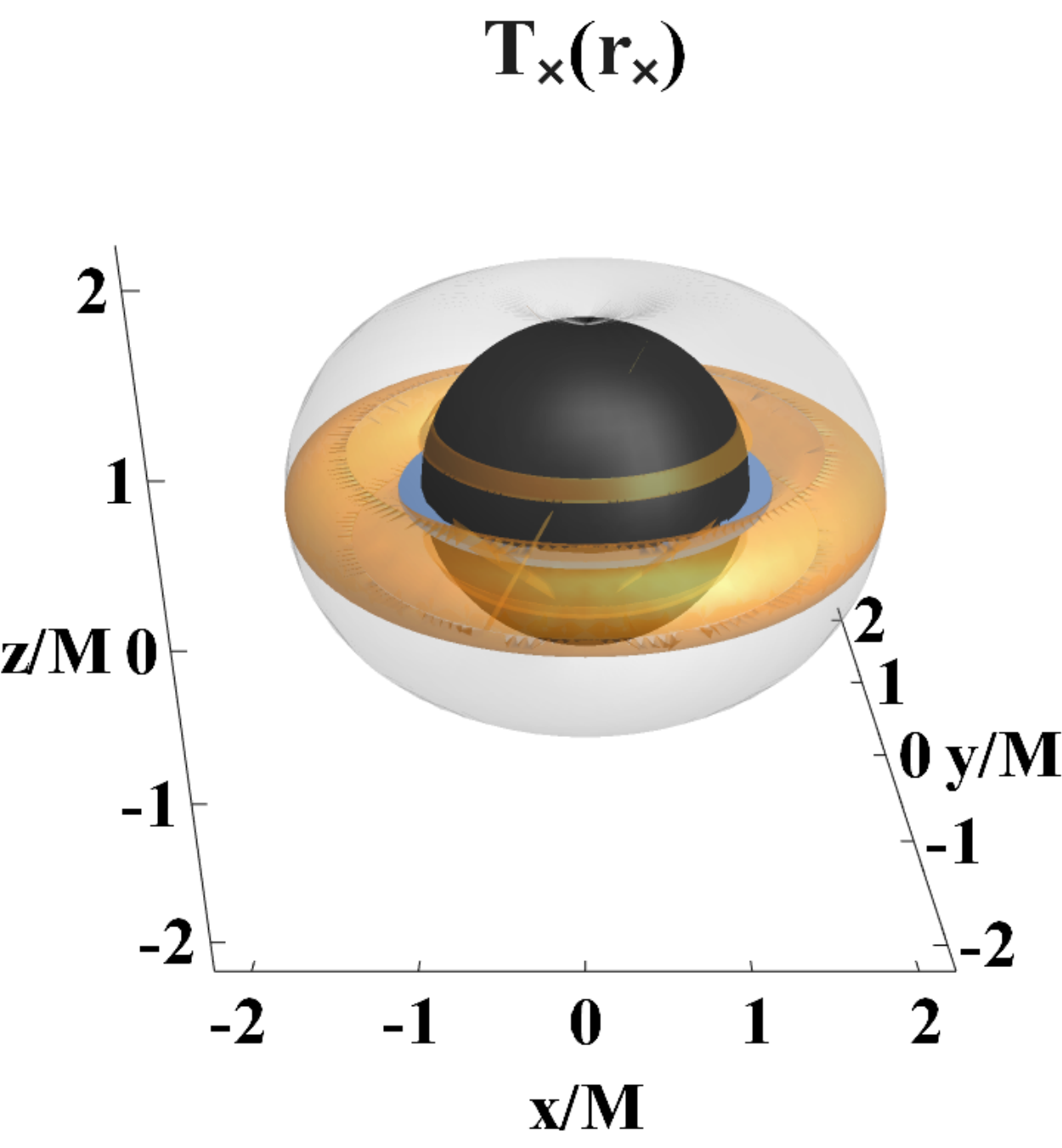}
  \includegraphics[width=5cm]{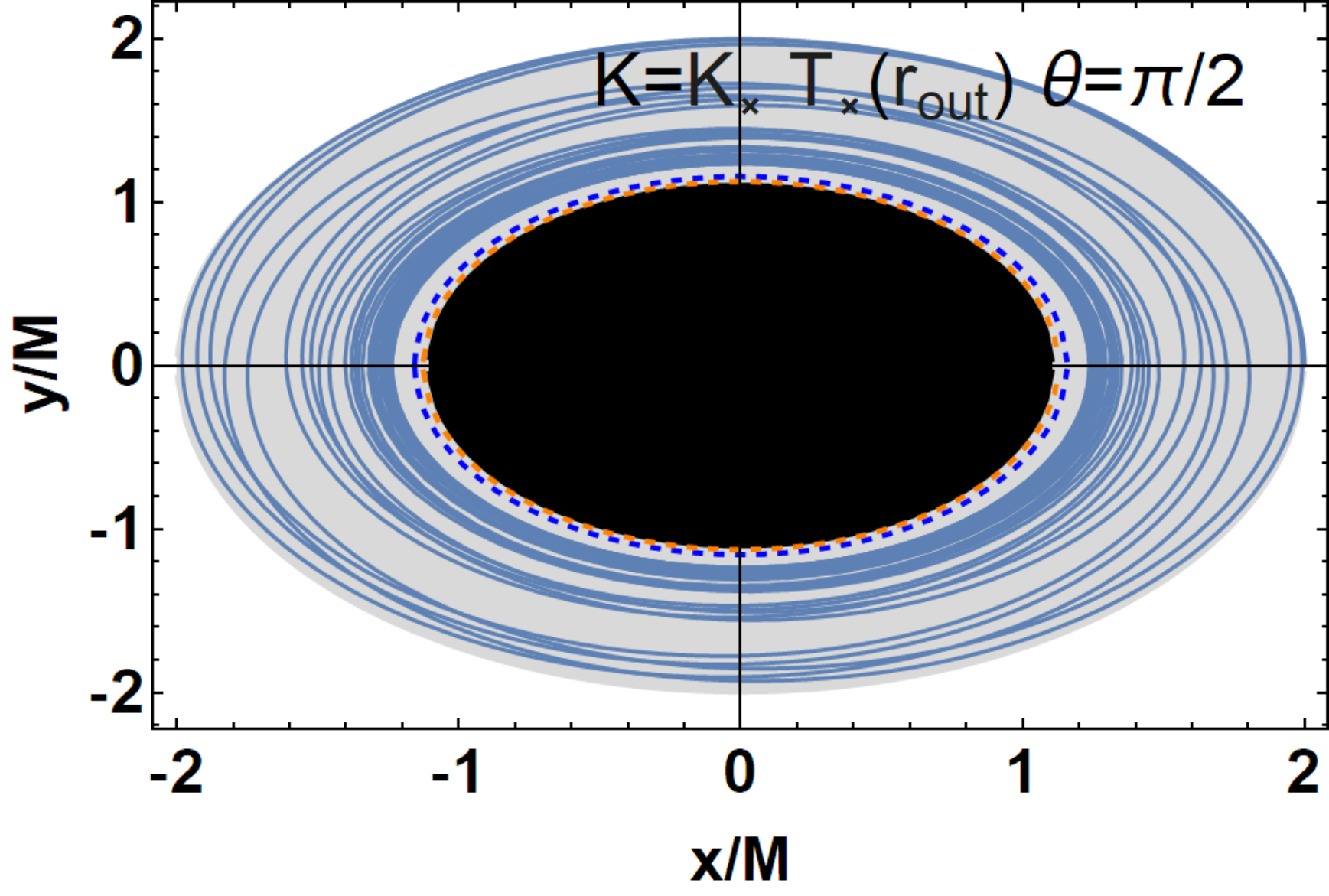}
  \includegraphics[width=5cm]{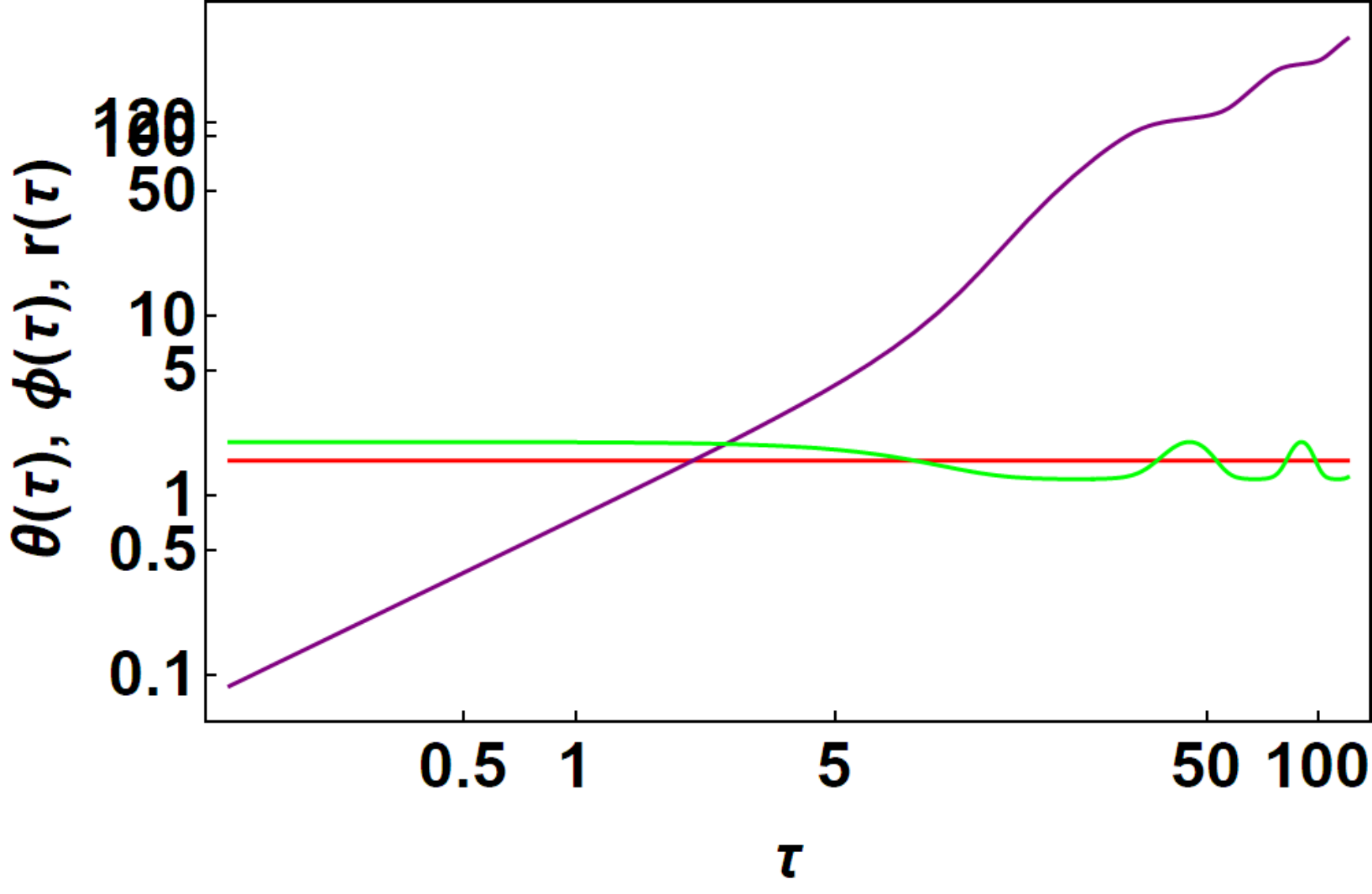}
  \includegraphics[width=5cm]{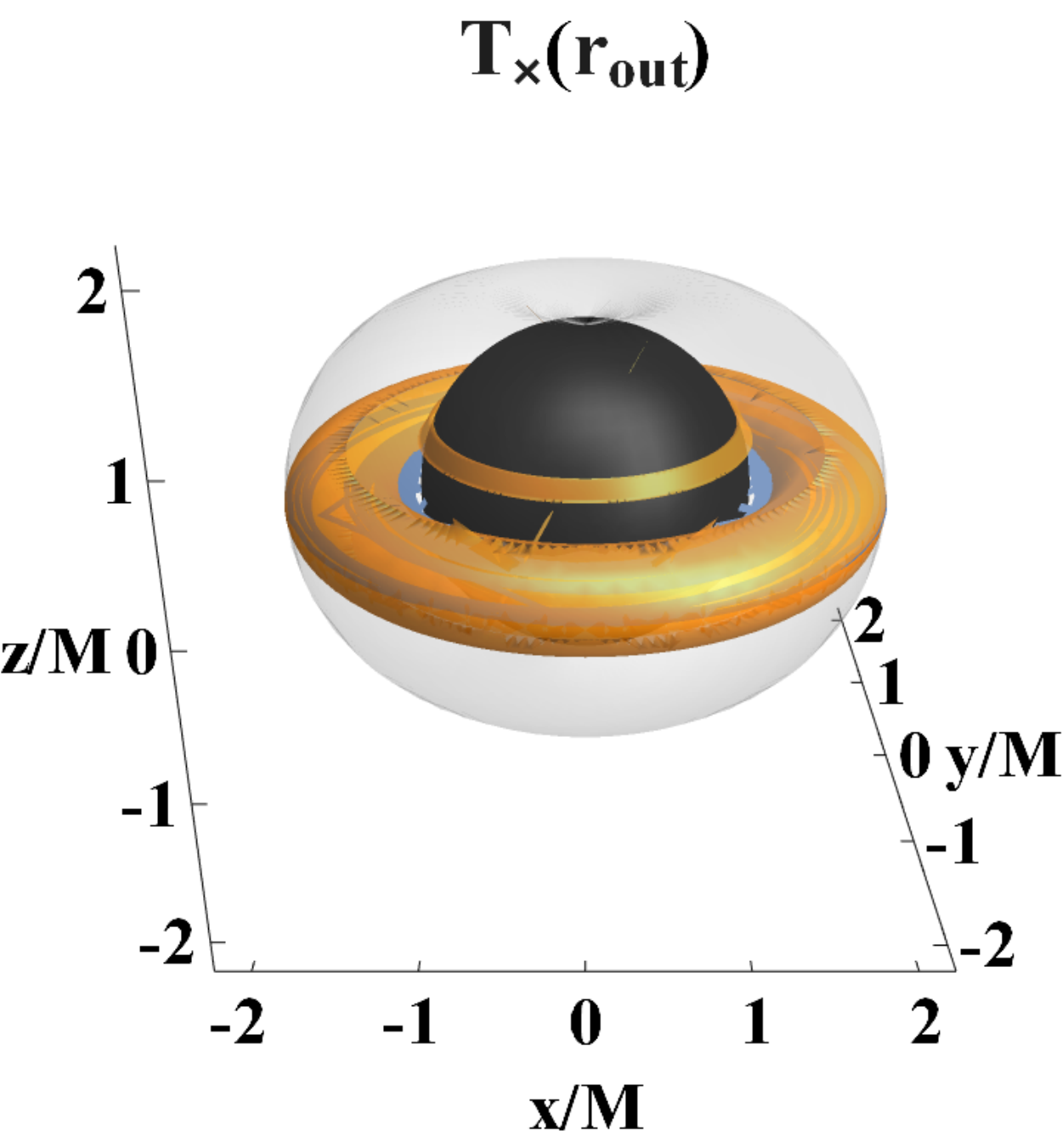}
  \includegraphics[width=5cm]{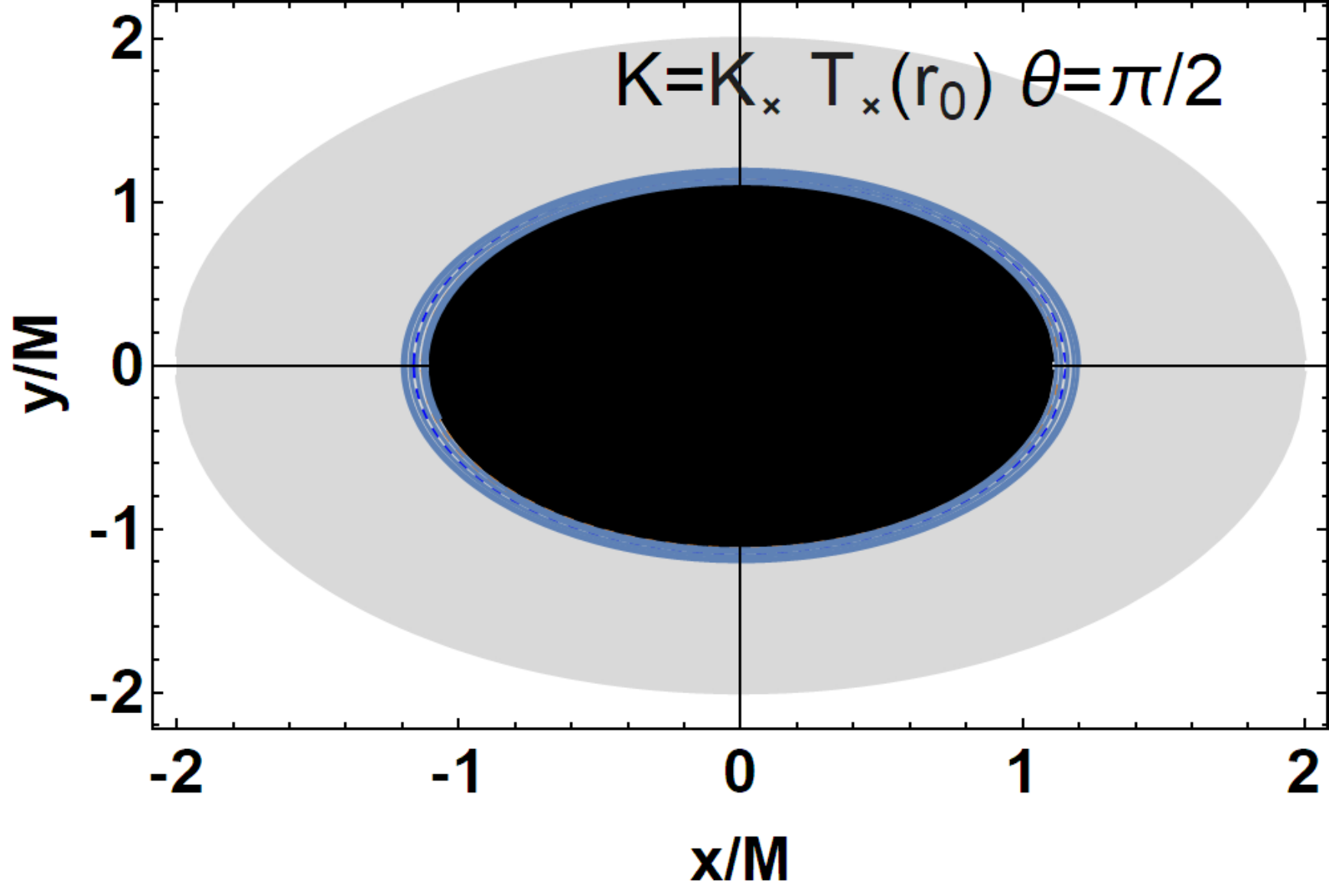}
  \includegraphics[width=5cm]{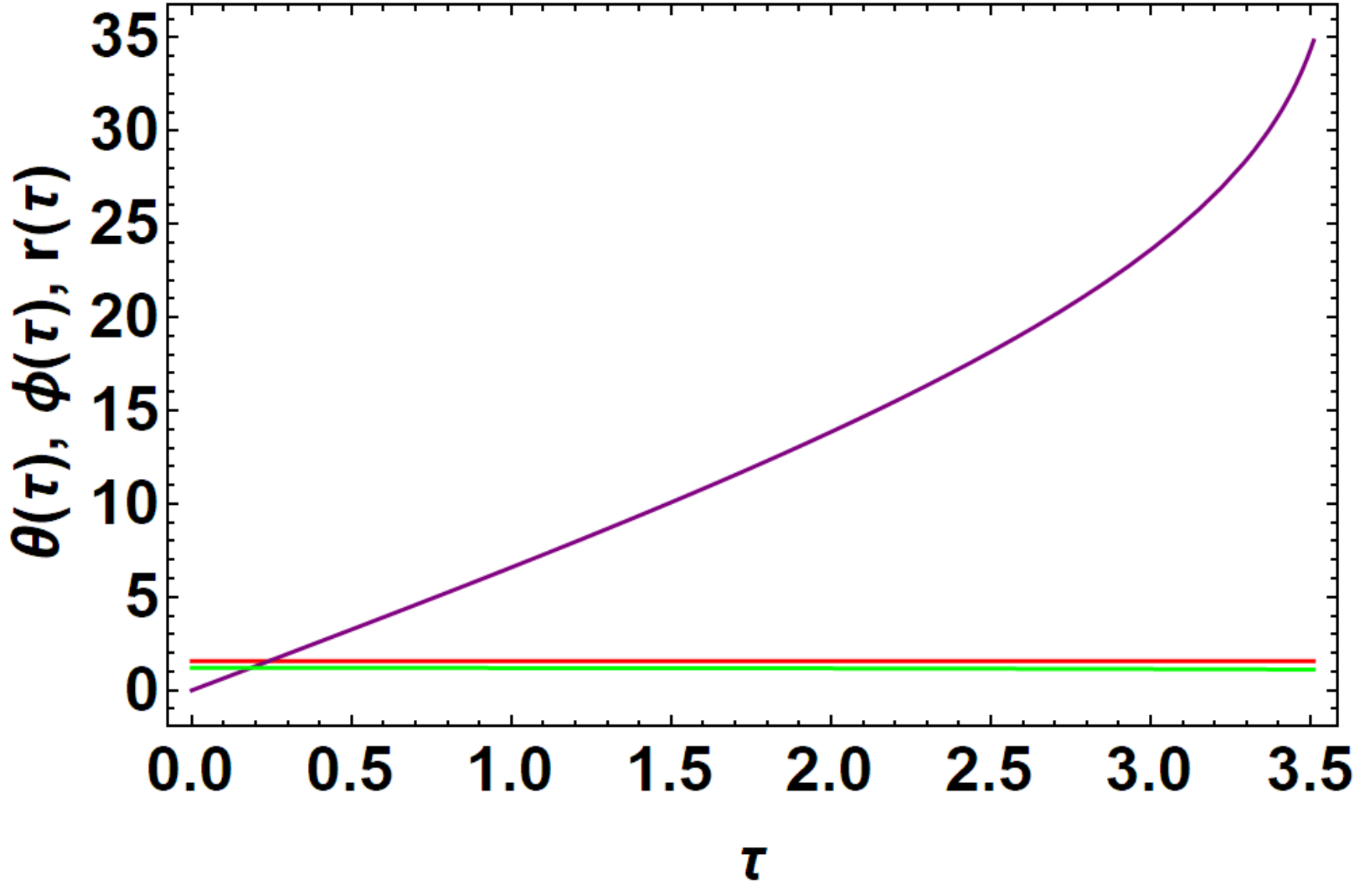}
  \includegraphics[width=5cm]{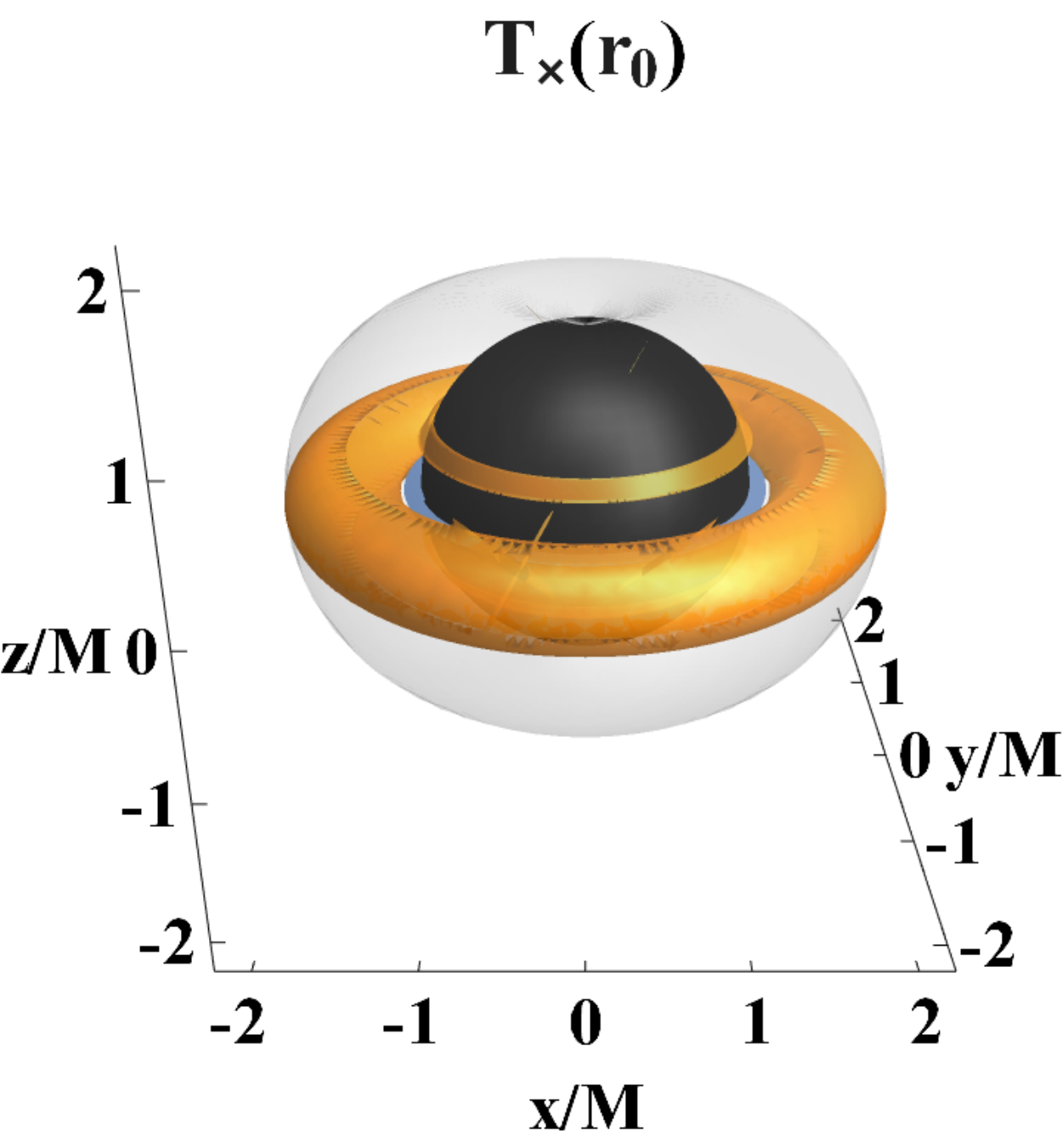}
     \caption{Timelike particles analysis of configurations of Figs\il(\ref{Fig:PlotTRial}) on the equatorial plane $\theta=\pi/2$.
     Torus  $T_\times$,   represented in Figs\il(\ref{Fig:PlotTRial}), and described in Sec.\il(\ref{Sec:mid-w-t}) is considered. There is $\{x=r \sin\theta \cos\phi,y=r \sin\theta \sin\phi,z=r \cos\theta\}$.
Black center is the \textbf{BH} $r<r_+$, gray region is the outer ergoregion $r\in ]r_+,r_{\epsilon}^+]$, where $r_+$ is the \textbf{BH} event horizon and $,r_{\epsilon}^+$ is the outer ergosurface.
 Yellow surface is the torus. The central  \textbf{BH}  has  spin $a_{mbo}^b$,  the  torus specific angular momentum $\ell$ is  signed in figure.
Complete  particles trajectories are shown. Considering Table\il(\ref{Table:Particle-models})  upper  line panels show details of  particles model \emph{\textbf{4.}}, central line panels show refer to particles model   \emph{\textbf{5.}}, the bottom line panels explore particles model \emph{\textbf{6.}}.
Left panels show motion on the \textbf{BH}   equatorial plane,  center panels show particles coordinates  angles $\theta$ (red curve), $\phi$ (purple curve) and radius  $r$ (green curve). }\label{Fig:Plotcredcatra}
    \end{figure}

\begin{figure}\centering
  % Requires \usepackage{graphicx}
  \includegraphics[width=5cm]{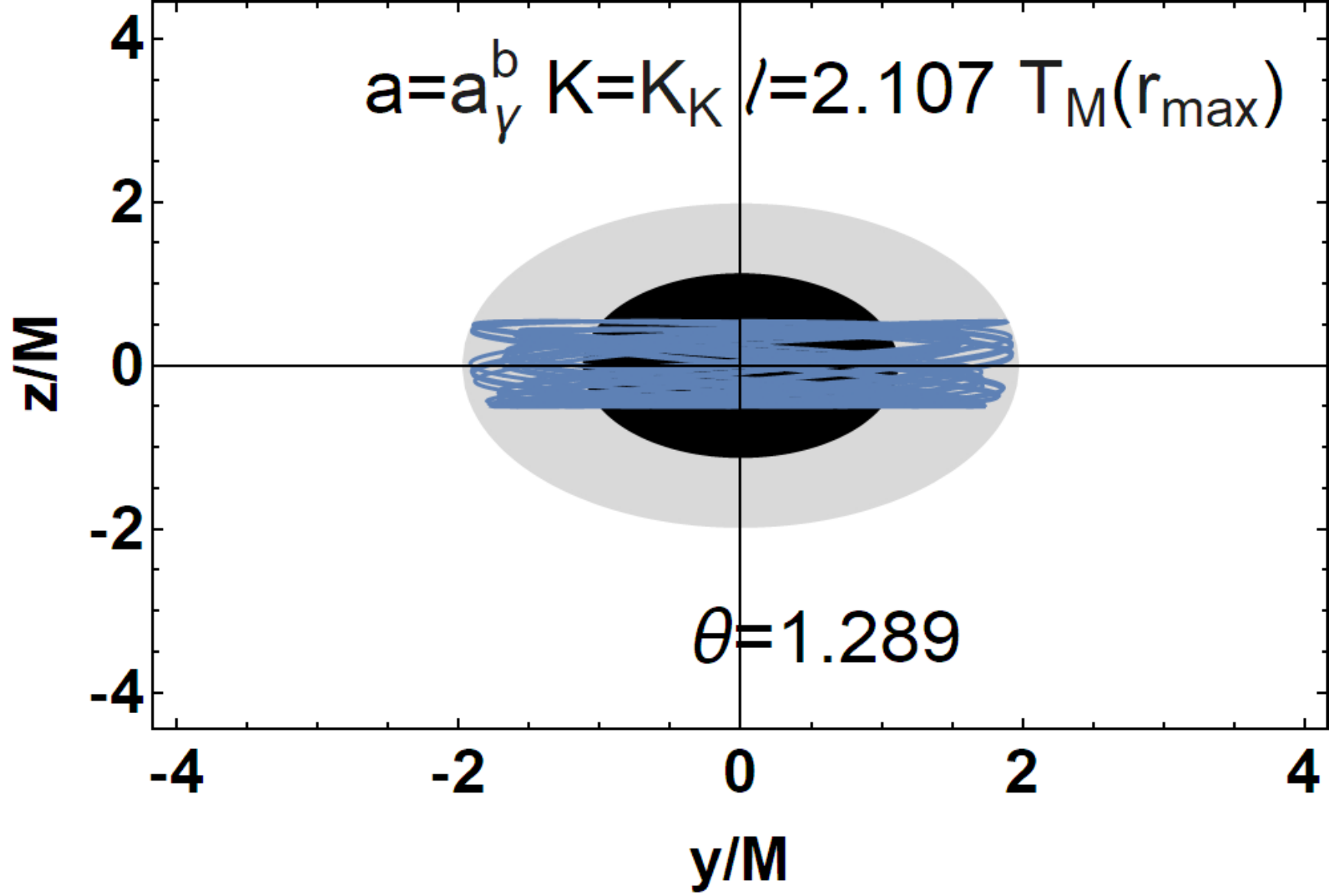}
  \includegraphics[width=5cm]{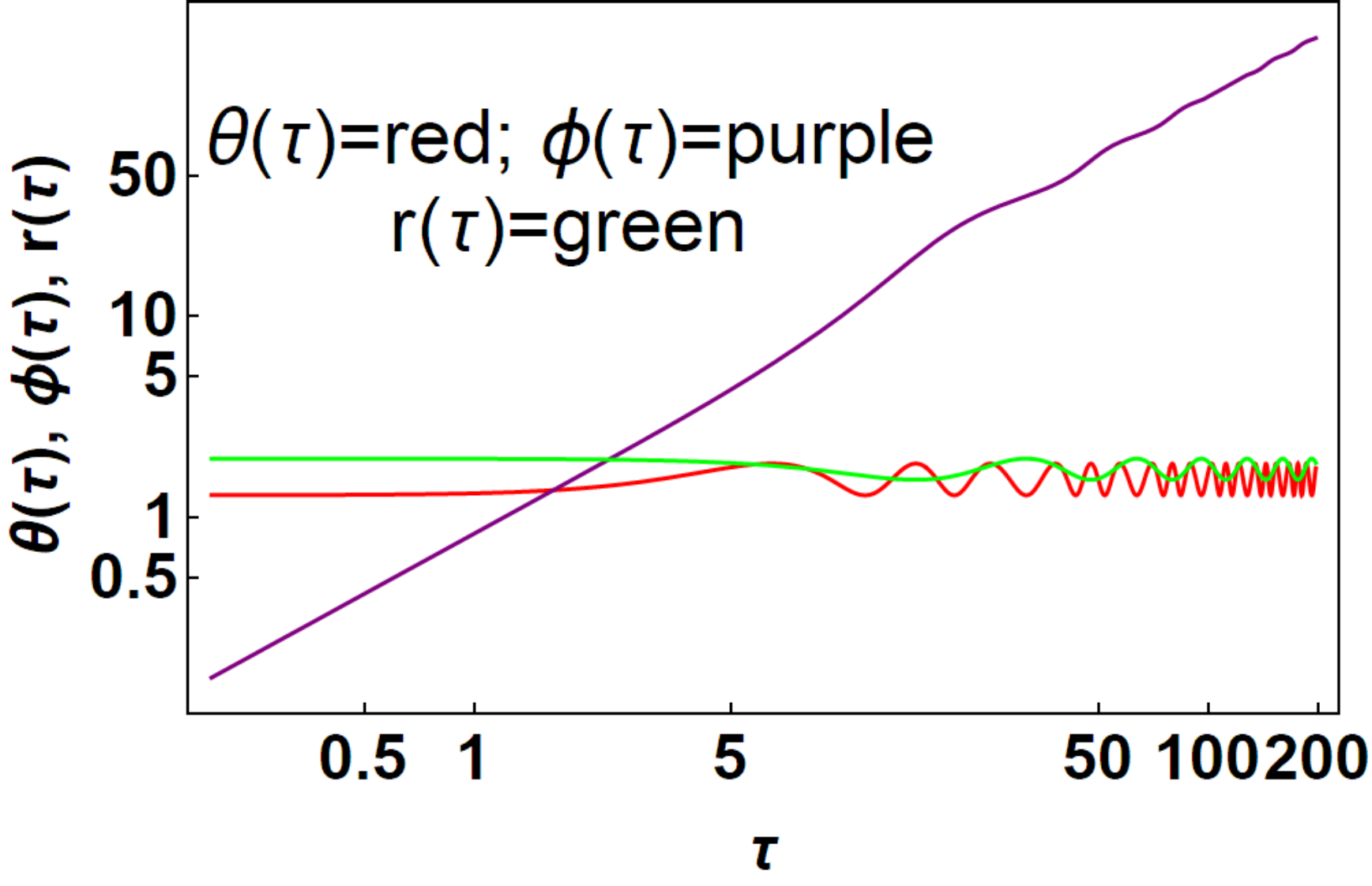}
  \includegraphics[width=5cm]{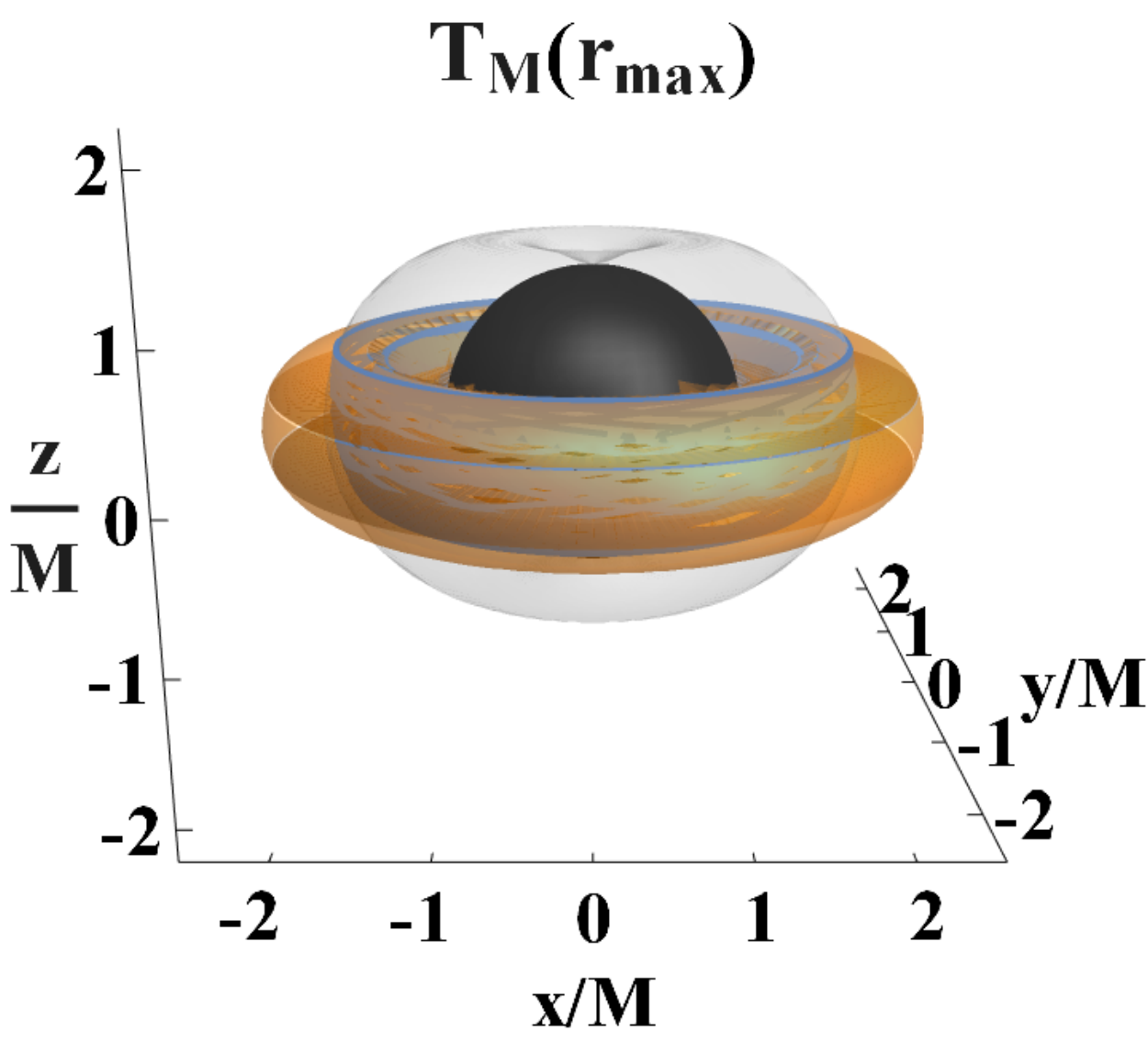}
  \includegraphics[width=5cm]{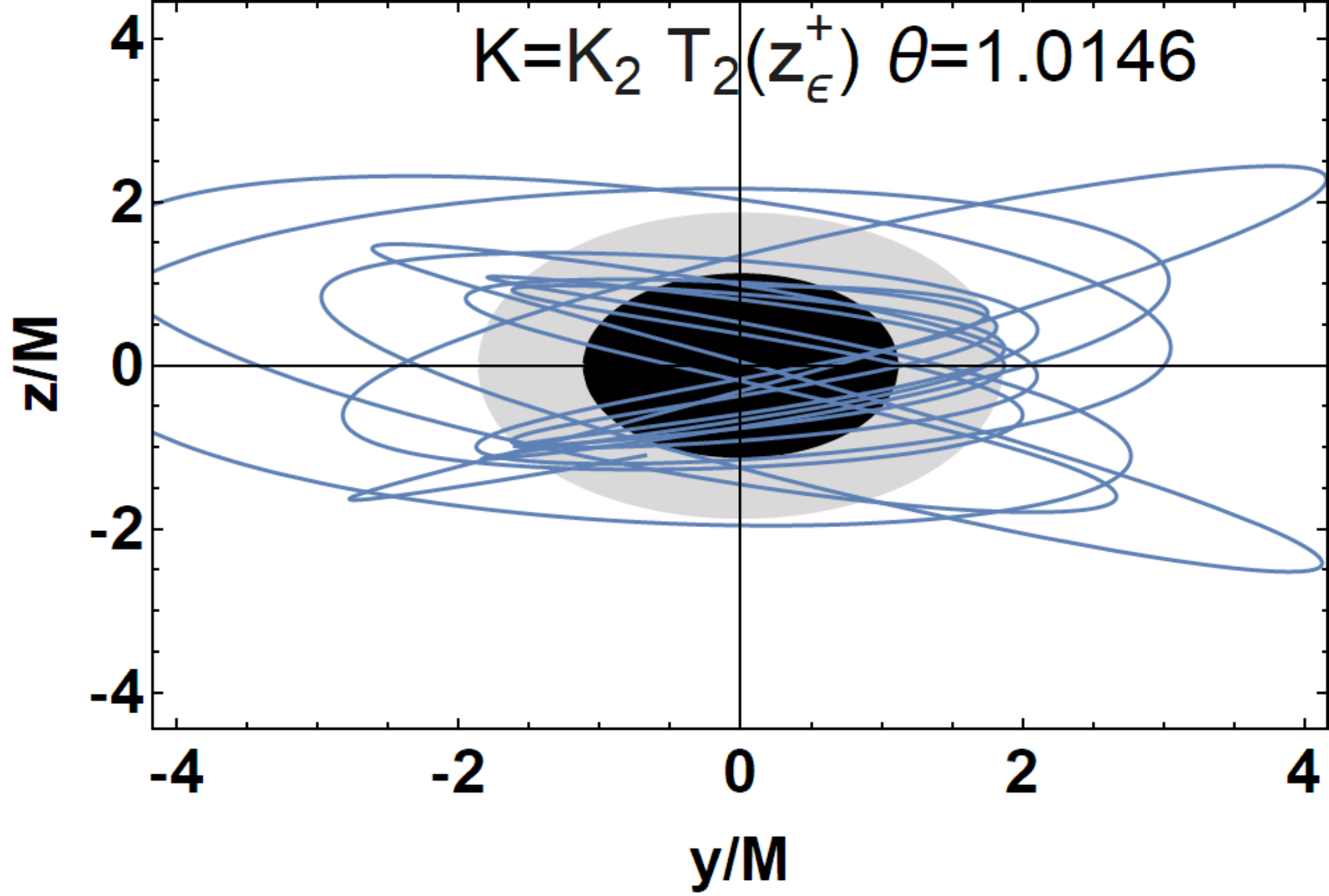}
  \includegraphics[width=5cm]{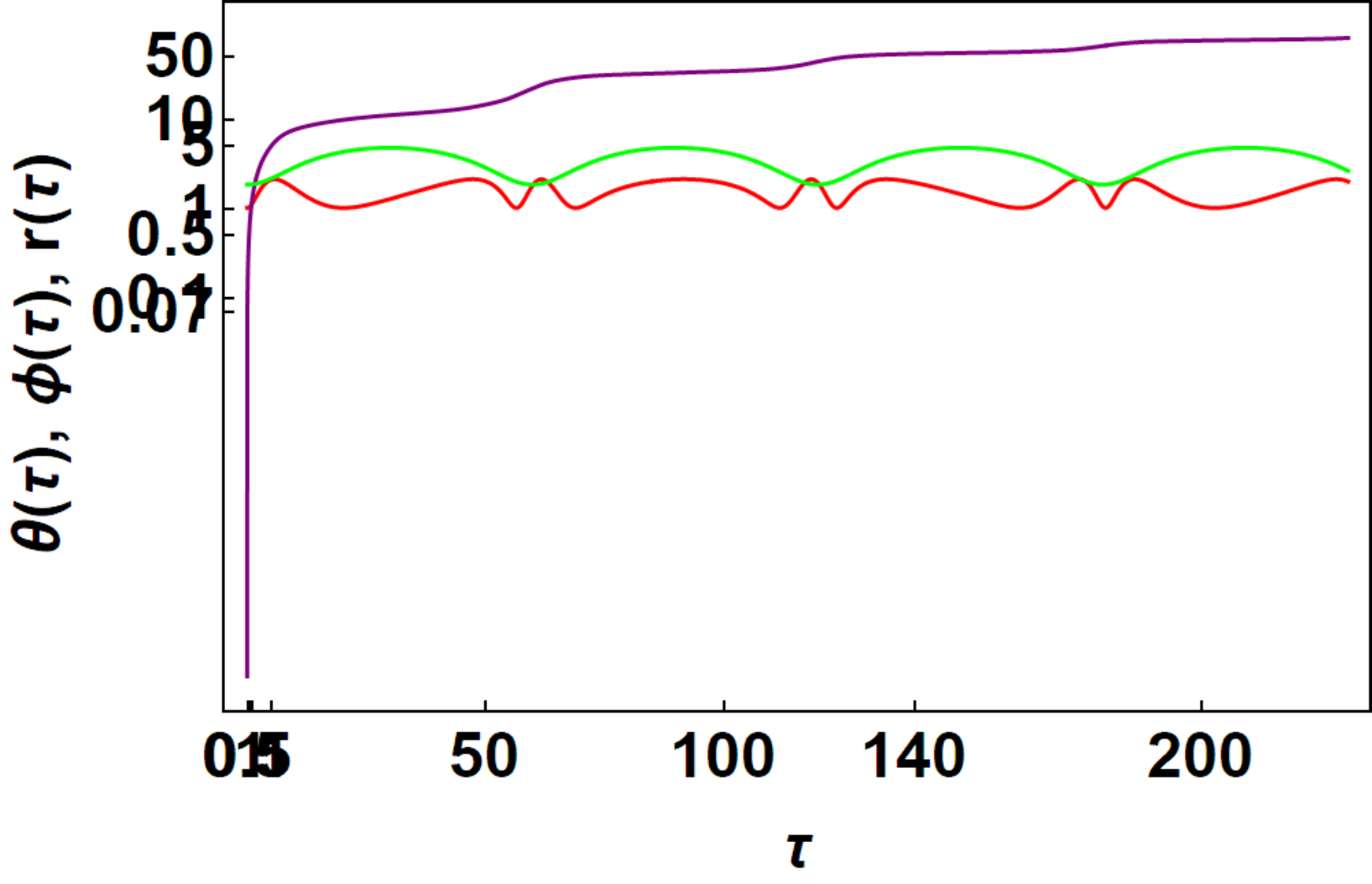}
  \includegraphics[width=5cm]{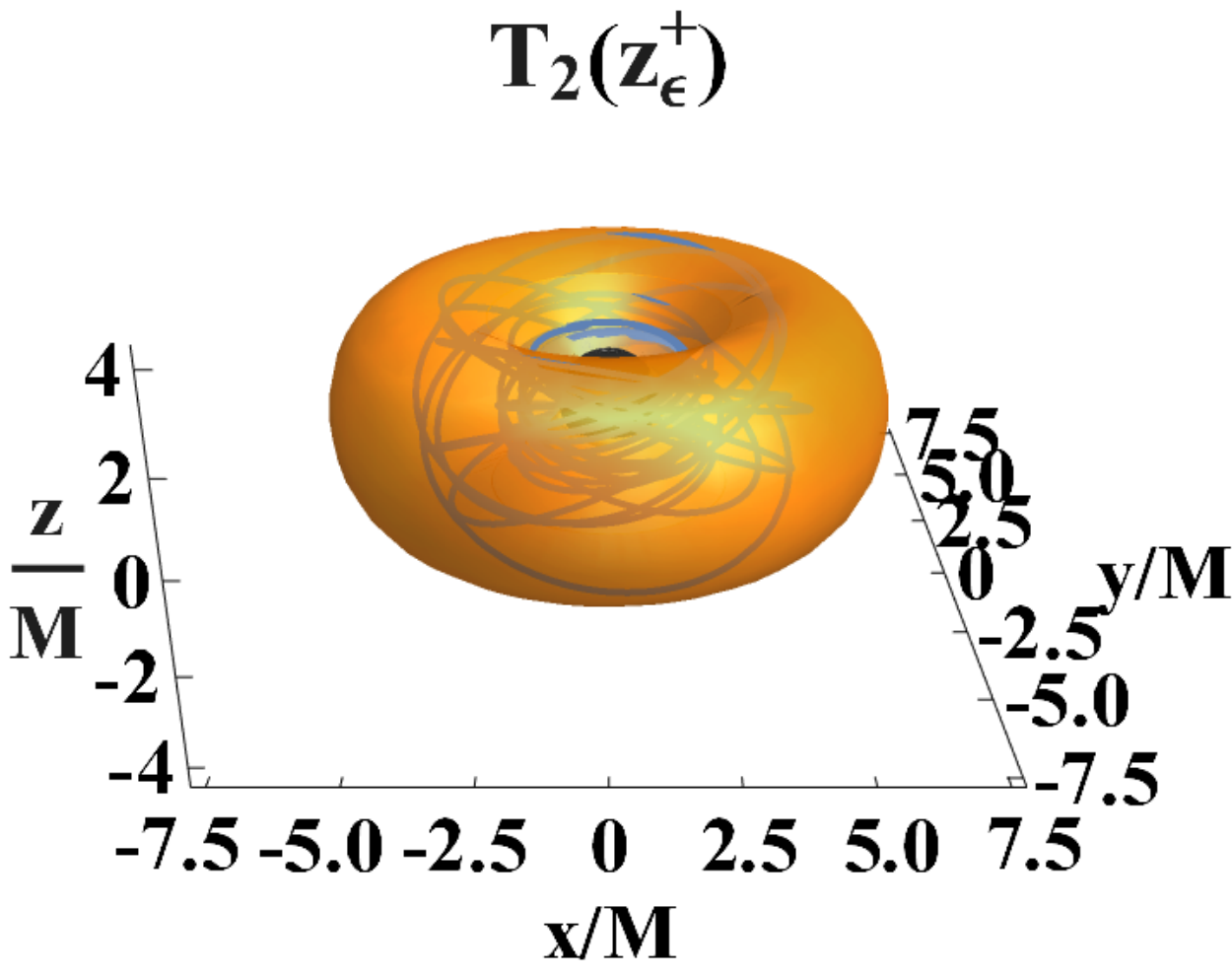}
  \includegraphics[width=5cm]{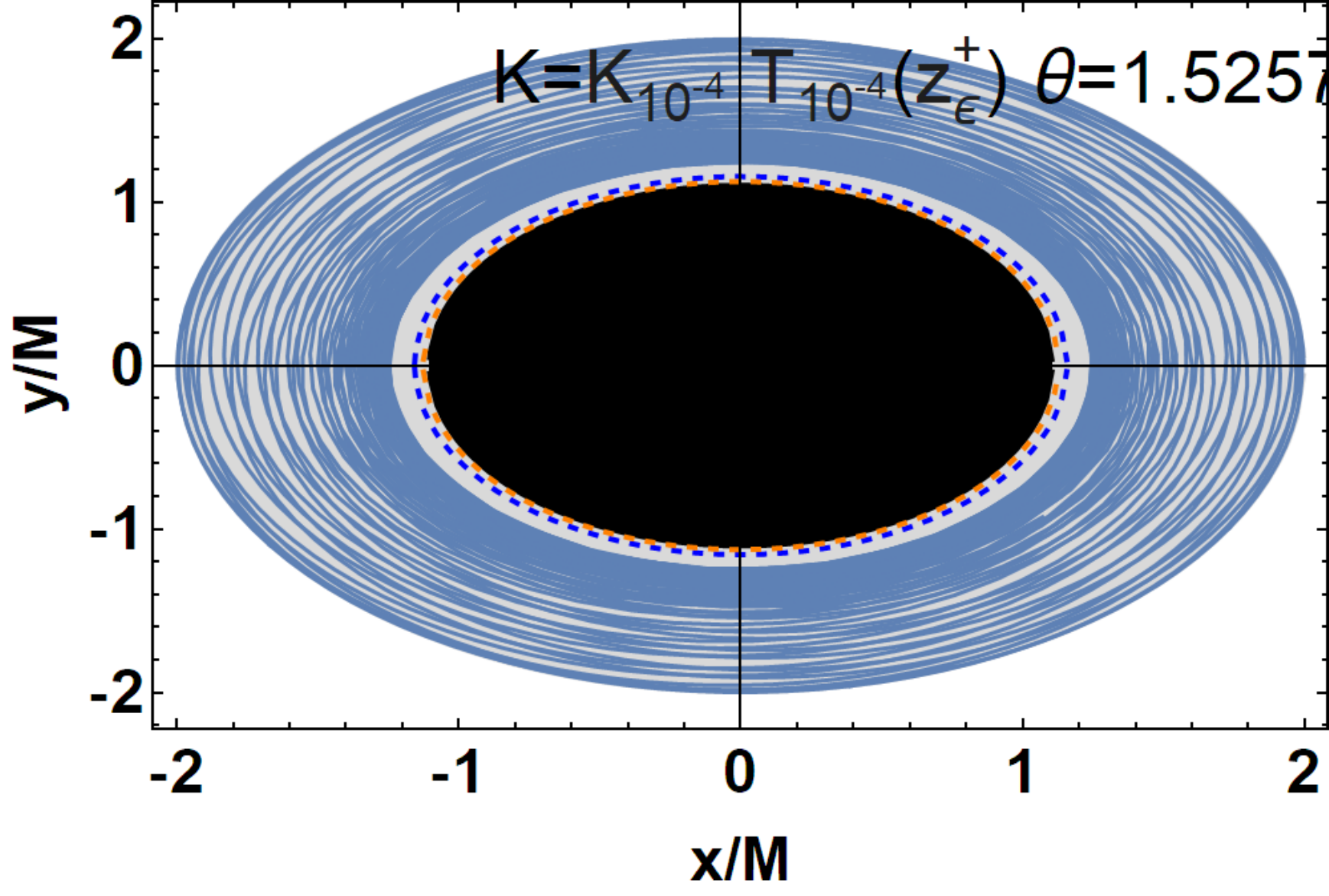}
  \includegraphics[width=5cm]{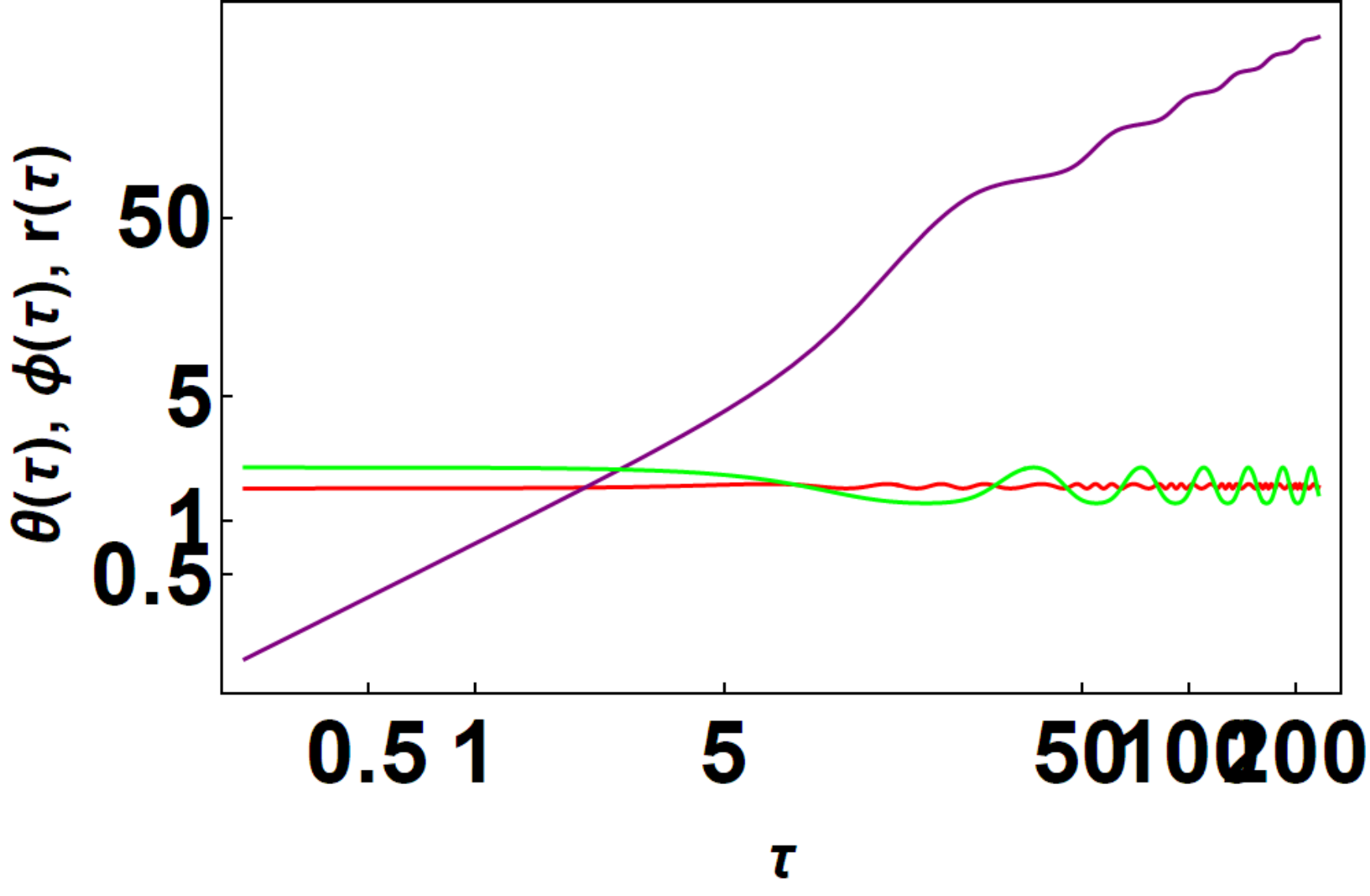}
  \includegraphics[width=5cm]{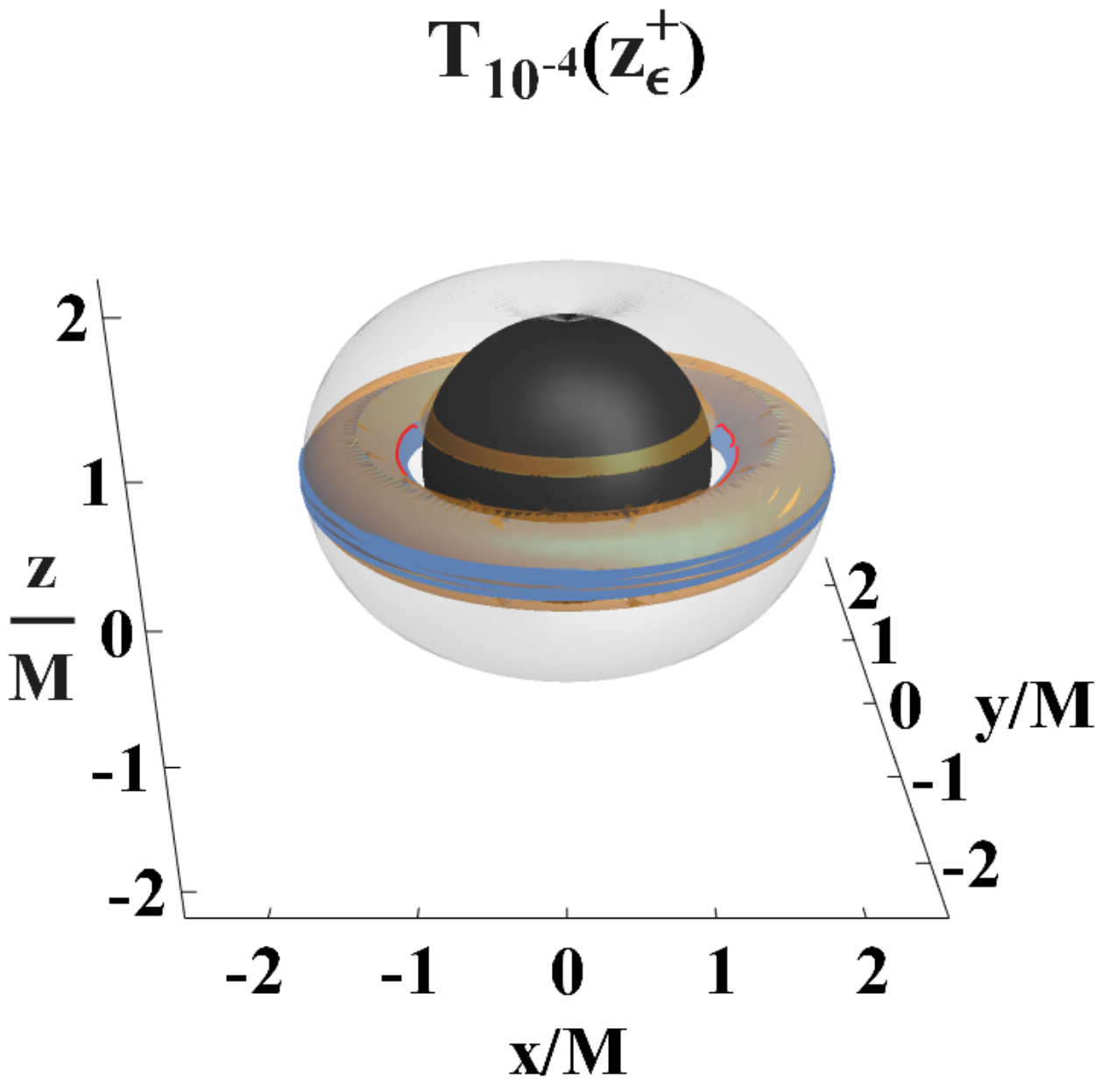}
    \caption{\textbf{BH} spacetime with spin $a_{mbo}^b$,  torus specific angular momentum $\ell$ is signed in figure.
    Figures show examples of timelike particles leaving the torus  from a point of the  equipressure surface.
Tori models  $T_M$, $T_2$ and $T_{10^{-4}}$ represented in Figs\il(\ref{Fig:PlotTRial})  and described in Sec.\il(\ref{Sec:mid-w-t}).
Particles leave a crossing point $z_{\epsilon}^+$ of the toroidal surface with the ergoregion (clearly for the problem symmetry there are four "equivalent" points at fixed radius $r_0$ and  four $\theta$s) on plane different from the equatorial plane, $r_{\max}$ is the geometrical maximum of the surface in the considered model coincident with a point of the outer ergosurface.  Complete  particles trajectories are shown. Considering Table\il(\ref{Table:Particle-models})  upper  line panels show details of  particles model \emph{\textbf{1.}}, central line panels show refer to particles model   \emph{\textbf{2.}}, the bottom line panels explore particles model \emph{\textbf{3.}}.
 Gray region is the outer ergosurface,
black region is the central \textbf{BH}
orange surface is the torus.
Right panel:  equatorial plane is at $z=0$,  center panel:  angles $\theta$ (red curve), $\phi$ (purple curve) and radius  $r$ (green curve). Note that in the bottom right panel  the projection on the equatorial plane of the motion is shown, as $\theta_0\approx\pi/2$. (There is $\{x=r \sin\theta \cos\phi,y=r \sin\theta \sin\phi,z=r \cos\theta\}$.).}\label{Fig:Plotcredcatr}
    \end{figure}

Therefore, according to the torus and the particle model, we fix
$\theta_0$, $r_0 $  in $\mathcal{R_{\emph{free}}}$, for simplicity we refer to these points as "emission points",
with  initial data $\Qa_0\equiv \Qa(0)$, where $\Qa_0$ is for any quantity evaluated at initial time $\tau=0$.
We consider the toroidal models of  Figs\il(\ref{Fig:PlotTRial}).
Using the Kerr geometry symmetries, there are two emission points obtainable in accordance with the reflection symmetries with respect to the  \textbf{BH} rotation axis (coincident  with the torus rotation axis) for the emission points on the equatorial plane, and  four positions for emission points on  planes  different from the equatorial, found considering  also the symmetry for reflection with respect to the disk equatorial plane  (coincident with the \textbf{BH} equatorial plane, note there is no need to fix $\phi_0$).
In the  orbits representations of  Figs\il(\ref{Fig:Plotcredcatra}) and (\ref{Fig:Plotcredcatr}) for convenience we used one or two emission points.
Furthermore we  considered the entire trajectory  in the internal area of the disk to enlighten the trajectories impacting the disk, on its inner edge, as this  back-reaction from the emission   to  the inner regions  of the tori resolved  eventually in photons and dust (or plasma) colliding with the toroid matter,   being  a characterizing factor of  partially contained  or  dragged toroids. Particle leaving  the surface can evolve in a swarm of particles created  in an  atmosphere wind around the  torus surfaces. Or eventually, they can part from the torus, for example from the outer edge of the outer region crossing  ergosurface  as a wind of particles and photons, or  part  of the emission could  impact on the  torus surface if emitted from the inner region (note  the model considered here is opaque, cooled with advection and  with super Eddington luminosity).

In this analysis we should take into account the  size of partially contained and  dragged tori, studied in Figs\il(\ref{Fig:raisePlot}), the  torus maximum   height  and the location of  the tori outer edge.

The torus  maximum dimensions are  constrained largely by the \textbf{BH}  dimensionless spin. The reduced dimensions of the tori mean that the dragged surfaces are  similar to globules where exfoliation  should take start.
Another important factor  in this process is the back-reaction effect through tori self gravity for the huge tori, combined with accretion and runaway instability.
The analysis of  PPI or MRI  for these tori should be also  taken into consideration \citep{Gammie}.

Runaway instability (RI), however, is an important process  for  thick disks orbiting \textbf{SMBHs},  particularly for partially contained and dragged surfaces. In this type of instability, the eventual rotational law $\ell\neq$constant,  defining different tori model, can be significant.
Neglecting  the torus self-gravity, and using
stationary models, RI can be studied with a  fully relativistic
hydrodynamics analysis.
 The RI features a sequence of
 varying mass \textbf{BH} phases  following the   mass transfer from the torus during the accretion and the consequent new status of the torus.  The torus can   become stable or unstable, depending  on several factors as   the mass ratio  of the  disk and the \textbf{BH}, if the \textbf{BH} is spinning, and most importantly the location of  torus   inner edge and cusp with respect to the central  \textbf{BH} at the initial stage of the process. We expect therefore the RI  be  significant for dragged surfaces considered here and especially the partially contained tori, having the cusp close to the \textbf{BH}, but  being large enough to   interfere with RI with the eventual mass-transfer slowing-down and the torus exfoliation characterizing the smallest tori in $\Sigma_{\epsilon}^+$.
  The RI  is generally a very fast process  enduring
on a dynamical timescale   of a few orbital periods.
In this process the geometry changes  consequently to the increase of the  \textbf{BH}  mass   for accretion from the disk. It is then clear that there can be  also a \textbf{BH} spin variation due to rotational energy extraction,  via the Blandford--Znajek mechanism, where
the amount of  extractable energy  depends on the disk mass  and can be ejected in jets or winds   observable for example in  GRBs. (Some considerations on the extraction of rotational energy are in Sec.\il(\ref{Sec:poly-altr})).  The accretion disk therefore  adapts to the new situation leading possibly to a series  of  steady states.
The  torus inner edge plays a key role in  this process, the  disk filling its Roche lobe  can transfer mass
through the cusp. Following the geometry change,  the  cusp  can stretch towards the horizon, slowing the mass transfer,
which can eventually stabilize the toroid, or  otherwise the cusp can move outwardly,
into the toroid interior, increasing the mass transfer,
 eventually the toroid could be "absorbed"   by the central \textbf{BH} in a very fast process.
The disk inner region, $[r_{inner},r_{center}]$, regulates the two cases, therefore this  analysis singles out the relevant tori .

\begin{figure}\centering
  % Requires \usepackage{graphicx}
  \includegraphics[width=8cm]{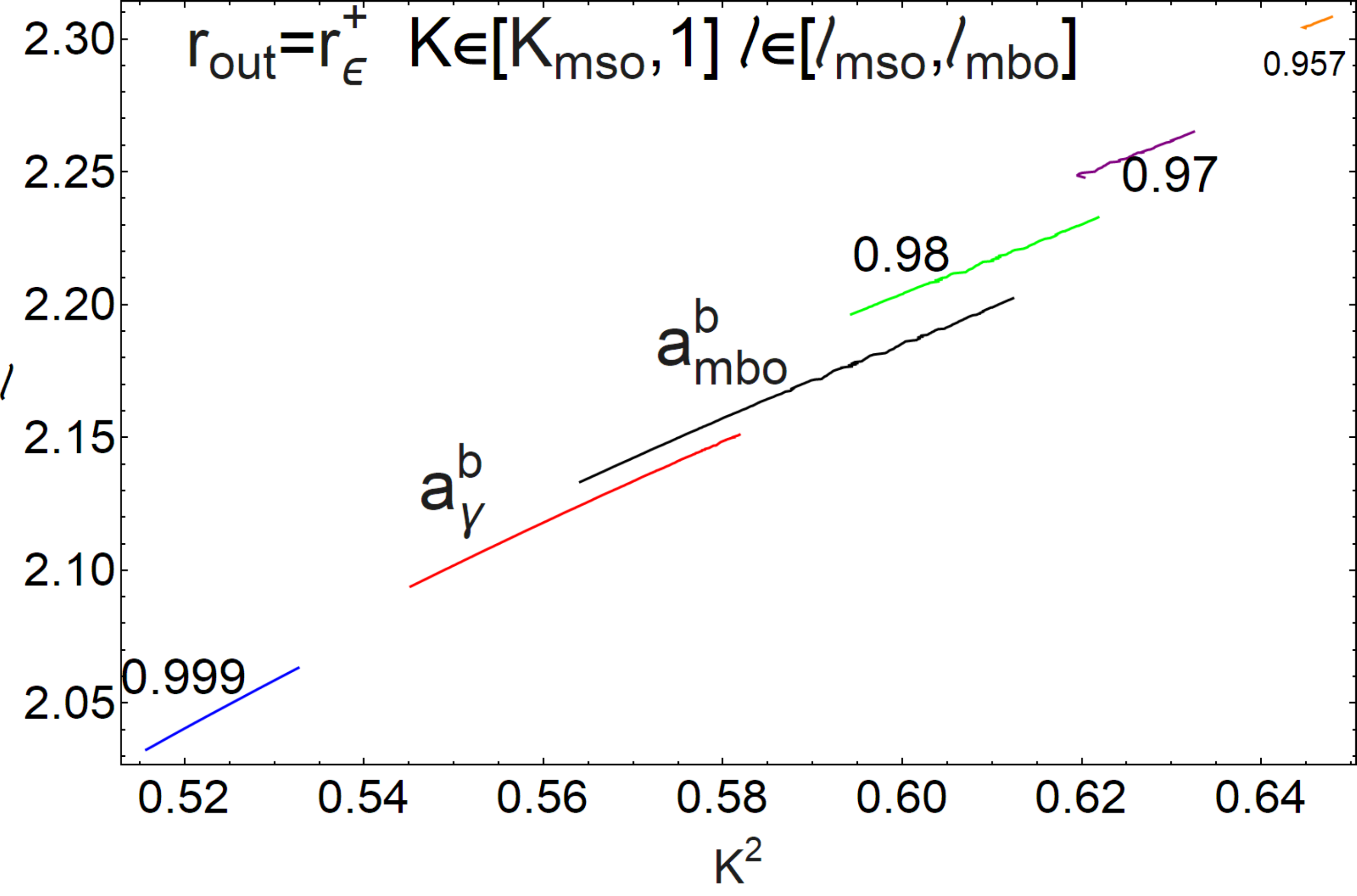}
  \includegraphics[width=8cm]{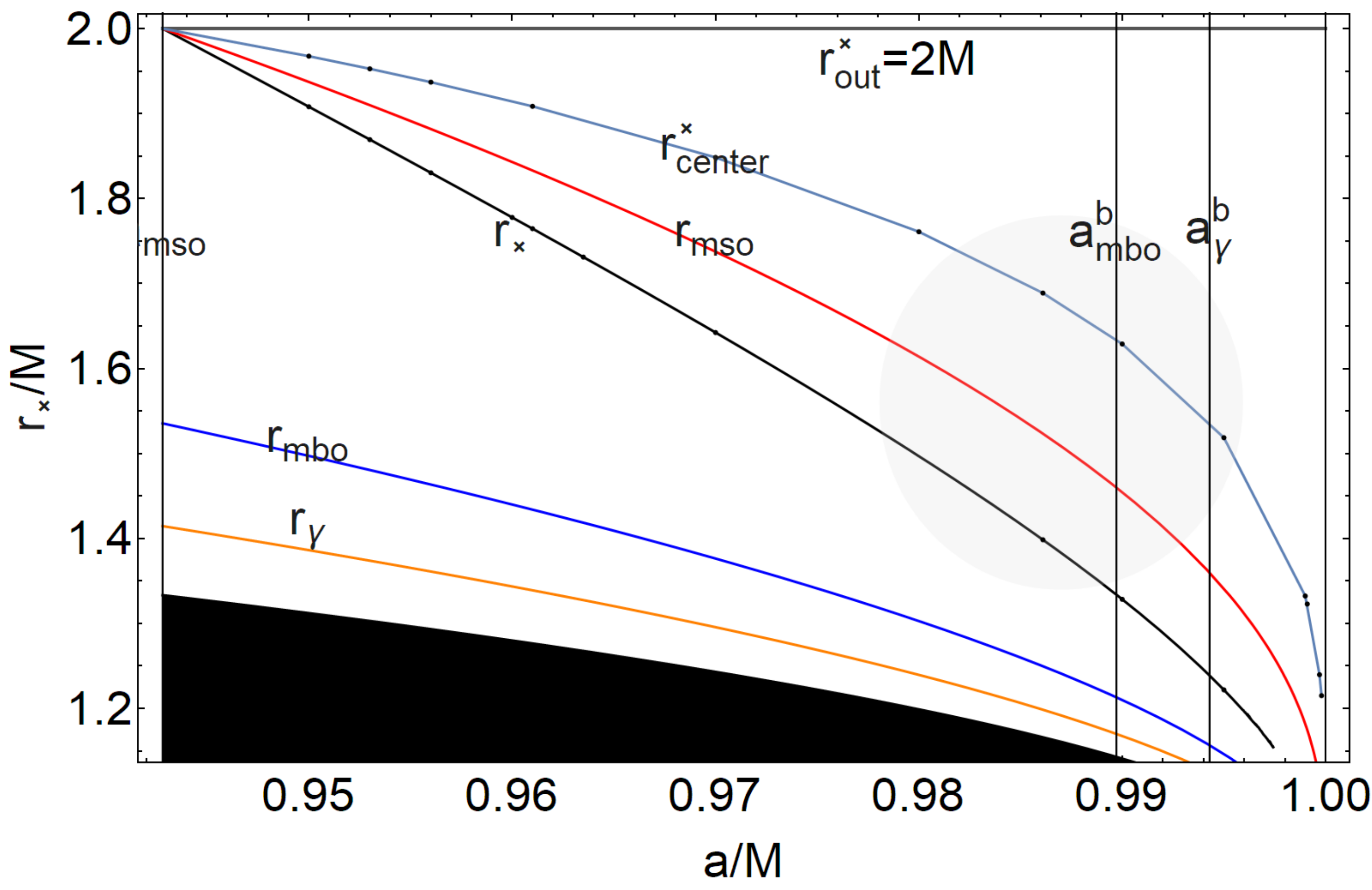}
    \caption{Analysis of  conditions for  the occurrence of the the torus outer  edge crossing   the outer ergosurface  $r_{outer}=r_{\epsilon}^+=2M$ for tori   with specific angular momentum $\ell\in \mathbf{L_1}$ and parameter $K\in ]K_{mso},1[$. Left panel:  quiescent and cusped tori are considered.  Curves show the relation $\ell(K)$ for the condition $r_{outer}=r_{\epsilon}^+$ for different \textbf{BH} dimensionless spins $a/M$ signed on the curves.  Right panel: black region is the \textbf{BH} $r<r_+$ where $r_+$ is the outer horizon, Black curve is the cusp $r_{\times}$ of the cusped tori with outer edge  $r^{\times}_{outer}=r_{\epsilon}^+=2M$, the center of the tori is the light blue curve. Red curve is the marginally stable orbit $r_{mso}$, blue curve is the marginally  bounded orbit $r_{mbo}$,  orange curve is the marginally circular orbit which is the  photon  orbit $r_{\gamma}$. Spin range $[a_{mso},M]$ is shown. Gray disk indicates   region $\mathrm{P}_{mbo}$ under analysis around $a\approx a_{mbo}^b$ and $r\approx 1.6M$.
    }\label{Fig:PlotJurY}
    \end{figure}

In $\Sigma_{\epsilon}^+$,  we consider quiescent tori with  $r_{inner}>r_{\times}$, cusped tori  $r_{cusp}$ and over critical tori  with an accretion  throat whose minima are given by the pressure extremes curve.
Naturally we can evaluate the force term to which the particles of the different points of the toroidal surface are subjected by the pressure gradients in Euler's law,   in points other than the cusp where the free force hypothesis is guaranteed--Figs\il(\ref{Fig:PlotJurY}).
It is expectable that the closer the inner edge is  to the \textbf{BH}   horizon, the more significant is  the RI, which is therefore significant particularly for  \textbf{BH} with high spins. In this case then, we note that tori with outer edge $r_{outer}=r_{\epsilon}^+$, represented in Figs\il(\ref{Fig:PlotJurY}), have inner edge closer to the $r_{mso}$ and therefore  far from the horizon
\citep{Abranobili,Abra1998,Korobkin:2012gj,Font:2002bi}.

Analysis of Figs\il(\ref{Fig:Plotcredcatra}) and Figs\il(\ref{Fig:Plotcredcatr}), considers  the \textbf{BH} geometry with spin  $a_{\gamma}^b$, studying  tori with specific   angular momentum
$\ell= \ell_v\equiv2.10728$, considering the four  toroidal configurations  defined by four  values of $K$ parameter $\{K_\times,K_M,K_2,K_{10^{-4}}\}$ of Figs\il(\ref{Fig:PlotTRial}). This first  study of the particles  can be implemented point by point on the surface.
In Figs\il(\ref{Fig:PlotTRial}) the blue line sets the center and  the geometrical maximum points of all the tori at $\ell=\ell_v$ and the minima of the accretion throats. Therefore, all the considered tori  have at least the inner part   of the disk contained in the ergoregion.
   Figs\il(\ref{Fig:PlotJurY})  show the results of the analysis  of the torus outer  edge crossing   the outer ergosurface  $r_{outer}=r_{\epsilon}^+=2M$ on the equatorial plane, for cusped tori  $\cc_1$,  with specific angular momentum $\ell\in \mathbf{L_1}$ and parameter $K\in ]K_{mso},1[$.
Tori having these characteristics can be observed  only orbiting \textbf{BHs} with spin  $a>a_{mso}$.
     Increasing the \textbf{BH} spin, the range of values of the parameters $\ell$ and $K^2$, for quiescent or cusped tori $\cc_1$, decreases according to the fact that the tori are corotating. Relation
    $\ell=\ell(K^2)$  can be considered  almost linear.
  The analysis shows a larger range of values for  $\ell$ and  $K^2$, for \textbf{BH}  with spin in a range centered on the value $a_{mbo}^b$.  The  peculiarity of the geometries of this range is also shown  in right panel,  with  the results for  cusped tori. Note that cusped tori have  equatorial elongation $\lambda_{\times}$  larger then the quiescent tori with equal specific angular momentum (a different case occurs for the {overcritical} tori having an accretion throat).
    The extended geodetic structure  of this geometry is studied in Figs\il(\ref{Fig:PlotVampb1}).
    Inner, outer edge  and center of the tori of  Figs\il(\ref{Fig:PlotVampa1}) are shown as functions of the specific angular momentum.
An alterative study of the   outer edge, center and cusps of the tori  $\cc_{\times}$  is  in  Figs\il(\ref{Fig:spessplhoke1}).
The analysis is repeated in Figs\il(\ref{Fig:PlotVamp1}) as function of the cusp radius  $r_{\times}$.
A prospect of the possible toroidal  surfaces orbiting in these geometries is in  Figs\il(\ref{Fig:weirplot}).
Analysis  of the disks verticality is in   Figs\il(\ref{Fig:gatplot8}),  Figs\il(\ref{Fig:gatplot17}) and
 Figs\il(\ref{Fig:Plotsoorr}).
Distance  cusps-centers is considered in    Figs\il(\ref{Fig:Plotbicilia}),
the von Zeipel surfaces are  Figs\il(\ref{Fig:collag}).

Tori $\cc_{\times}$ in Figs\il(\ref{Fig:PlotJurY})-right panel have outer edge at $r_{outer}^{\times}=r_{\epsilon}^+=2M$.
Curves show different characteristics of the cups and center of the tori in a  region  $\mathrm{P}_{mbo}$, defined by spins  $a_{mbo}^b$ and $a_{\gamma}^b$, around the radius  $r=1.6M$ and shown in Figs\il(\ref{Fig:PlotJurY}). The tori cusp  at variation of the \textbf{BH} spin remains  relatively close to the limiting radius $r_{mso}$, increasing the distance in  $\mathrm{P}_{mbo}$,  being small tori. Tori are smaller and  closer to the ergosurface   for spins close to  $a_{mso}$,   and are relatively small for the larger spins $a\approx M$ where the cusp approaches the radius $r_{mbo}$,  maximum distance is reached in   $\mathrm{P}_{mbo}$.
The cusp decreases increasing  the \textbf{BH} spin,  approaching $r_{mbo}$. Therefore increasing the \textbf{BH} spin  the tori elongation $\lambda_{\times}$ increases.

However, the tori  stability and tori  exfoliation depend on their inner region, whose radial  dimension is    $\lambda_{inner}\equiv r_{center}-r_{\times}$ (the outer region  is $\lambda_{outer}\equiv r_{outer}-r_{center}=2M -r_{center}=\lambda-\lambda_{inner}$).
We  observe that there is always  $\lambda_{inner}\ll\lambda_{outer}$, where  $\lambda_{inner}$ is maximal in  $\mathrm{P}_{mbo}$.

Tori studied   in Figs\il(\ref{Fig:PlotTRial}) orbit  around a \textbf{{BH}} with spin $a_{mbo}^b$ and they are defined by the following four models with fixed $\ell=$constant:
\begin{description}
\item[-]
\textbf{{{{(1)}}}}: Torus $\mathbf{T_2}$, yellow curve in Figs\il(\ref{Fig:PlotTRial}), is defined by parameter  $K =K_2\equiv K_{\times}+(1-K_{\times})/2>K_{\times}$ (the torus is characterized by the presence of  an accretion throat),  and it is  partially contained in $\Sigma_{\epsilon}^+$  (with  the torus  inner part).
\item[-]
\textbf{\textbf{{(2)}}}: Torus  $\mathbf{T_M}$, blue surface in Figs\il(\ref{Fig:PlotTRial}), is a   partially contained torus, characterized by an accretion throat defined by parameter$K_M= 0.765671>K_{\times}$. The torus  geometrical maximum crosses  the outer ergosurface.
\item[-]
\textbf{(3)}: Torus  $\mathbf{T_{10^{-4}} }$, green-dashed surface of Figs\il(\ref{Fig:PlotTRial}), is defined by $K=K_{10^{-4}}=K_{\times} + 5\times 10^{-4}>K_{\times}$, it is a  partially contained torus characterized by a small accretion throat, crossing the outer ergosurface on a plane $\theta\neq\pi/2$.
\item[-]
\textbf{(4)}: Torus  $T_{\times} $, black surface of  Figs\il(\ref{Fig:PlotTRial}), is the  cusped torus with $K=K_{\times}$, which is a  dragged surface, with  the  outer edge  coincident with the outer ergosurface $r_{\epsilon}^+=2M$ on the equatorial plane.
\end{description}
We set    the  initial data
$\{r_0, \theta_0,\phi'(0), E, K, r_0, y_0, x_0\}$  for the integration defining  six particles models\footnote{ Note in the formulation of toroidal models of    Figs\il(\ref{Fig:PlotTRial}) and initial data  in Table\il(\ref{Table:Particle-models}) it is not  necessary to integrate the equation for the  torus surface. We have numerically integrated    the condition $V_{eff}^2=K^2$ of Eq.\il(\ref{Eq:scond-d}), fixing the two tori parameters $(\ell,K)$, according to the  tori topology and using the rotational law $\ell(r)$ and the function $K(r)$ to fix the parameters values.
 On the other hand, the inner and outer edges of the cusped  tori, the torus center (maximum pressure point inside the disk), the geometrical maxima of the  torus surface  are in fact directly   obtainable, these radii are given  in \citet{pugtot}  and  for example in  \citet{Multy,letter}--see also \citep{abrafra}.}  listed in Table\il(\ref{Table:Particle-models})
\begin{table*}[ht!]
\centering
%\resizebox{.9\textwidth}{!}{%
\begin{tabular}{|l|}
\hline
\textbf{\emph{1.}} Torus $\mathbf{T_M}$, particles are initially located  on  the  geometric  maximum   $r_{\max}=r_{\epsilon}^+$ on a plane $\theta\neq\pi/2$.
 \\
 \hline
\textbf{Data: }$\{\phi'(0) = 0.83,% t_1'(0) =3.064,
 r(0) =1.96, \theta(0)=1.289,
  E= 0.766, y(0)=1.884, x(0)= 0.545\}$
\\
\hline \hline
\textbf{\emph{2.}} Torus   $\mathbf{T_2}$.  Particles leave the  crossing point of the torus   with the ergosurface.
\\
\hline
\textbf{Data: } $\{\phi'(0)=1.2157, % t_1'(0)= 3.7088,
 r(0)= 1.851,   \theta(0) =1.015, E=0.872, y(0)=1.5722, x(0)=0.9773\}$
 \\
\hline \hline
\textbf{\emph{3.}} From the torus    $\mathbf{T_{10^{-4}}}$. Particles leave  the crossing point of the torus surface  with the outer ergosurface.
\\
\hline
\textbf{Data:} $\{ \phi'(0)=0.750, %t'(0)=2.924,
 r(0)=1.99899, \theta(0)=1.52575, E=K_{10^{-4}}= 0.744353, y(0)=1.57217, x(0)= 0.0900212\}$
 \\
\hline \hline
\textbf{\emph{4.}} From the  torus  $\mathbf{T_{\times}}$.  Particles leave  a point close to the cusp $r_{\times}$ of the torus  $T_{\times}$.
\\
\hline
\textbf{Data:}  $\{\phi'(0)=5.0598,% t_1'(0)= 12.01,
 r(0)=r_{\times}=1.23872,  \theta(0)=\pi/2, E=K_{\times}= 0.744\}$.
 \\
\hline \hline
 \textbf{\emph{5.}} From the torus $\mathbf{T_{\times}}$. Particles are initially located on  the  torus outer edge,  coincident  with the outer ergosurface.
\\
\hline
\textbf{Data:}  $\{\phi'(0)=0.748, %t_1'(0)=2.9,
 r(0)=1.99992, \theta(0)=\pi/2, E=K_{\times}=0.743853\}$.
 \\
\hline \hline
  \textbf{\emph{6.}}  From the  torus  $\mathbf{T_{\times}}$, particles leave a point  $r_0$.
\\
\hline
\textbf{Data:}  $\{\phi'(0)=6.505,% t_1'(0)=15.063,
r(0)=1.2,  \theta(0)=\pi/2, E=K_{\times}=0.743853\}$
\\
\hline
\end{tabular}%}
\caption{Timelike particle models  initial data
$\{r_0, \theta_0,\phi'(0), E, K, y_0, x_0\}$ related to  tori models $\{T_\times,T_M,T_2,T_{10^{-4}}\}$ of Figs\il(\ref{Fig:PlotTRial}), dimensionless units have been used. $(')$ is intended a derivation  with respect to the proper time. There is $\Qa_0\equiv \Qa(0)$ for any quantity $\Qa$ evaluated at the initial time, where $r'(0)=0$ and $\theta'_0=0$.}
\label{Table:Particle-models}
\end{table*}
In Figs\il(\ref{Fig:PlotTRial}) we consider particular  toridal  surfaces at different $K$ from the  set of
     Figs\il(\ref{Fig:polodefin1})--(upper line central panel).
    The analysis is a first simple attempt to consider the  swarm of free particles and photons which  is  grounded  on the reduced  dimension of the tori in the ergoregion and the model of zero-pressure  fluid at the tori edges--see Figs\il\ref{Fig:raisePlot}).  We have used same approximation in Sec.\il(\ref{Sec:qpos})  for the analysis of  the oscillatory models of the torus using the characteristic Keplerian frequencies to discuss the adaptability of QPOs models on these structures.
   In general the test particles  hypothesis  is of  astrophysical significance, also in  the case of a tori  atmosphere.
    In this situation we can also combine in characteristics processes of these tori as the Poynting--Robertson effects  for partially included and dragged surfaces\footnote{The  radiation field carrying out energy and momentum interacts back with  the accretion disk plasma. The radiation field  induces  a radiation pressure which combines with a radiation drag force (which is the Poynting-Robertson effect), and  the mass transfer  during accretion. We should also note that the tori considered here are opaque and super-Eddington. The radiation drag can  act as  dissipative force  for the orbiting matter,
removing energy and angular momentum.
The Poynting-Robertson effect can  remove angular momentum and energy from small-sized test particles.
Test particle  radial motion  is therefore affected by the  Poynting-Robertson effect and radiation pressure and the  accreting matter can
lose angular momentum  combining to accretion.
(It should be noted that  Poynting flux (in presence of magnetic field)
carries away  from the \textbf{BH} energy and angular momentum, combining in the \textbf{BH} evolution  with the accretion.)
There can be therefore (radiatevely
and  thermally) outflows and inflow followed eventually also by a change into the disk structure, considering  outgoing
or ingoing radial photon flux.} \citep{Igumenshchev:2007bh,Bini:2014ooa,Bini:2014jra,Lee:2001by,Ballantyne:2005qp,Zanni,2007prpl.conf..277P,bakala} .

We show some results  considering light-like particles from the torus surface and with   fixed initial $\ell$ parameters, using the constraint and  the definition of $(E,L)$.
We used Eq.\il(\ref{Eq:uf})  for the photon to obtain initial data   $u^{\phi}(0)$,   determined by conditions on tori parameter and assuming the circular motion symmetries preserved, where there is  $u^r(0)=0$ and $u^{\theta}(0)=0$, determined by  parameter  $\ell$ and $E$,  the first fixed by the  torus,  while the second is fixed for the timelike particle by the  $K$-parameter of the torus. Eq.\il(\ref{Eq:uf})  does not differ for photon and timelike particles,  defined by $ E$ and  $L$ with the definition of  Killing fields. We considered $L=\ell E$, using for timelike particles parameters  $\ell$ and $E$.
For the photons  $\ell$,  the initial data $u^t(0)$  has been fixed  using the normalization condition  used also in the set of equation to be integrated.   We also considered the metric symmetries
 $(t,\phi)\rightarrow(-t,-\phi)$ for the two solutions in $u^t$.
 Details of  particles model \emph{\textbf{1.}},   \emph{\textbf{2.}}  and  \emph{\textbf{3.}} of Table\il(\ref{Table:Particle-models})   are shown in Figs\il(\ref{Fig:Plotcredcatr}).  Models
  \emph{\textbf{4.}},   \emph{\textbf{5.}}  and  \emph{\textbf{6.}}  considering particles on the torus and \textbf{BH} equatorial plane are shown Figs\il(\ref{Fig:Plotcredcatra}).

\begin{figure}\centering
  % Requires \usepackage{graphicx}
    \includegraphics[width=5cm]{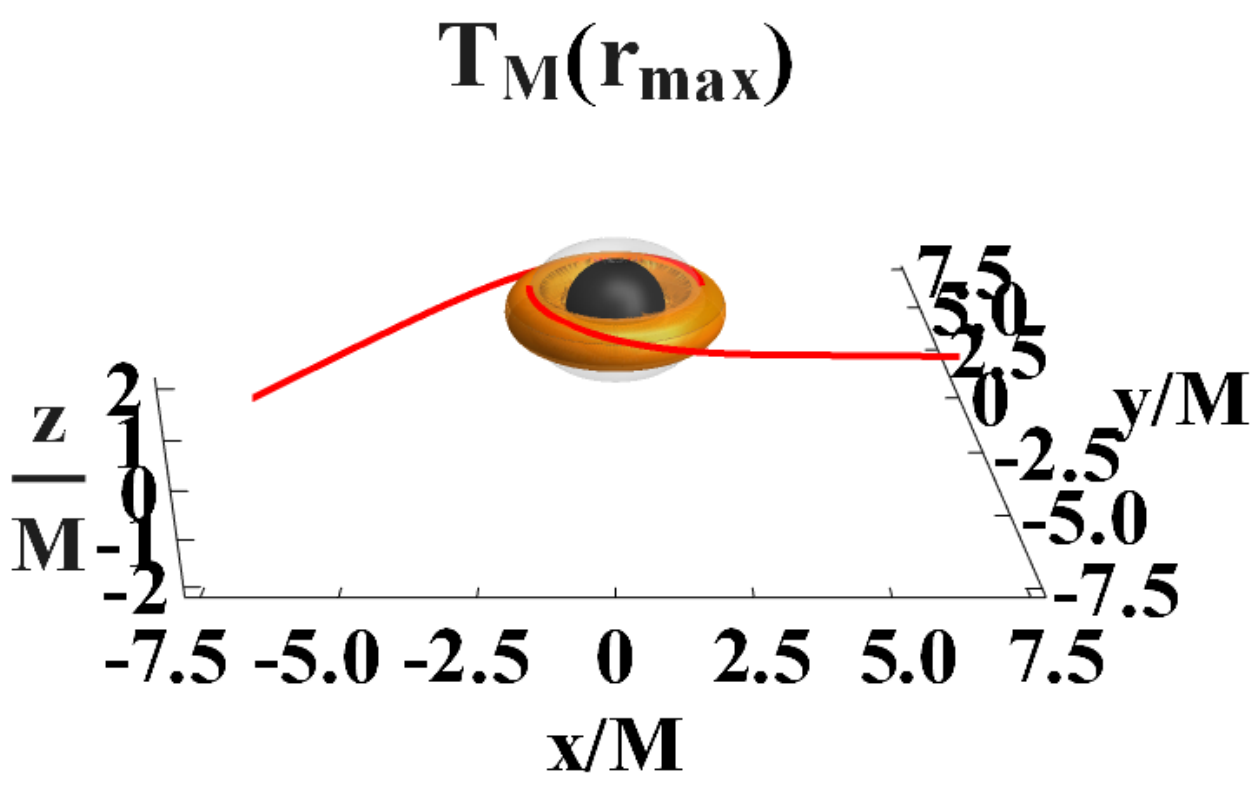}
  \includegraphics[width=5cm]{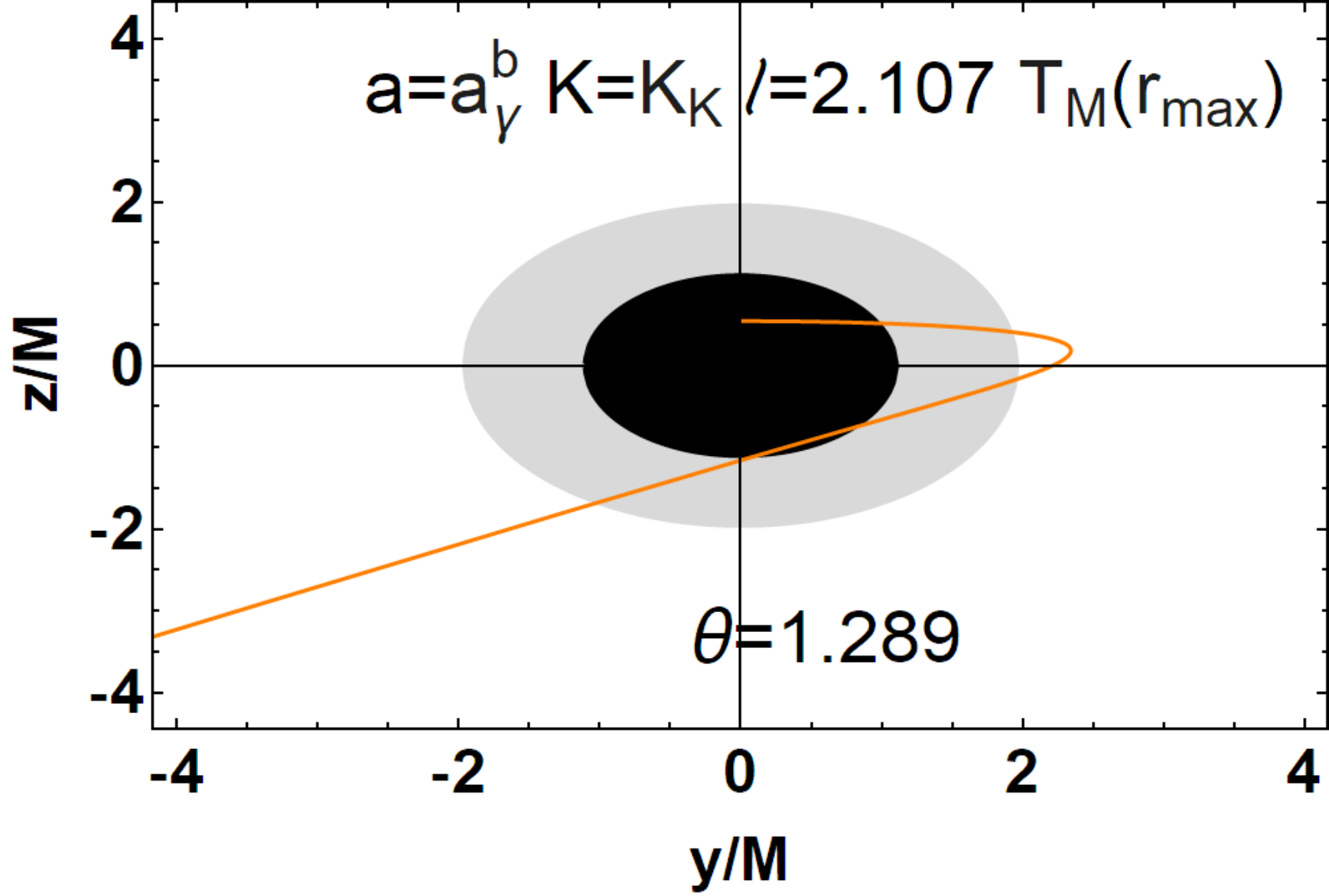}
  \includegraphics[width=5cm]{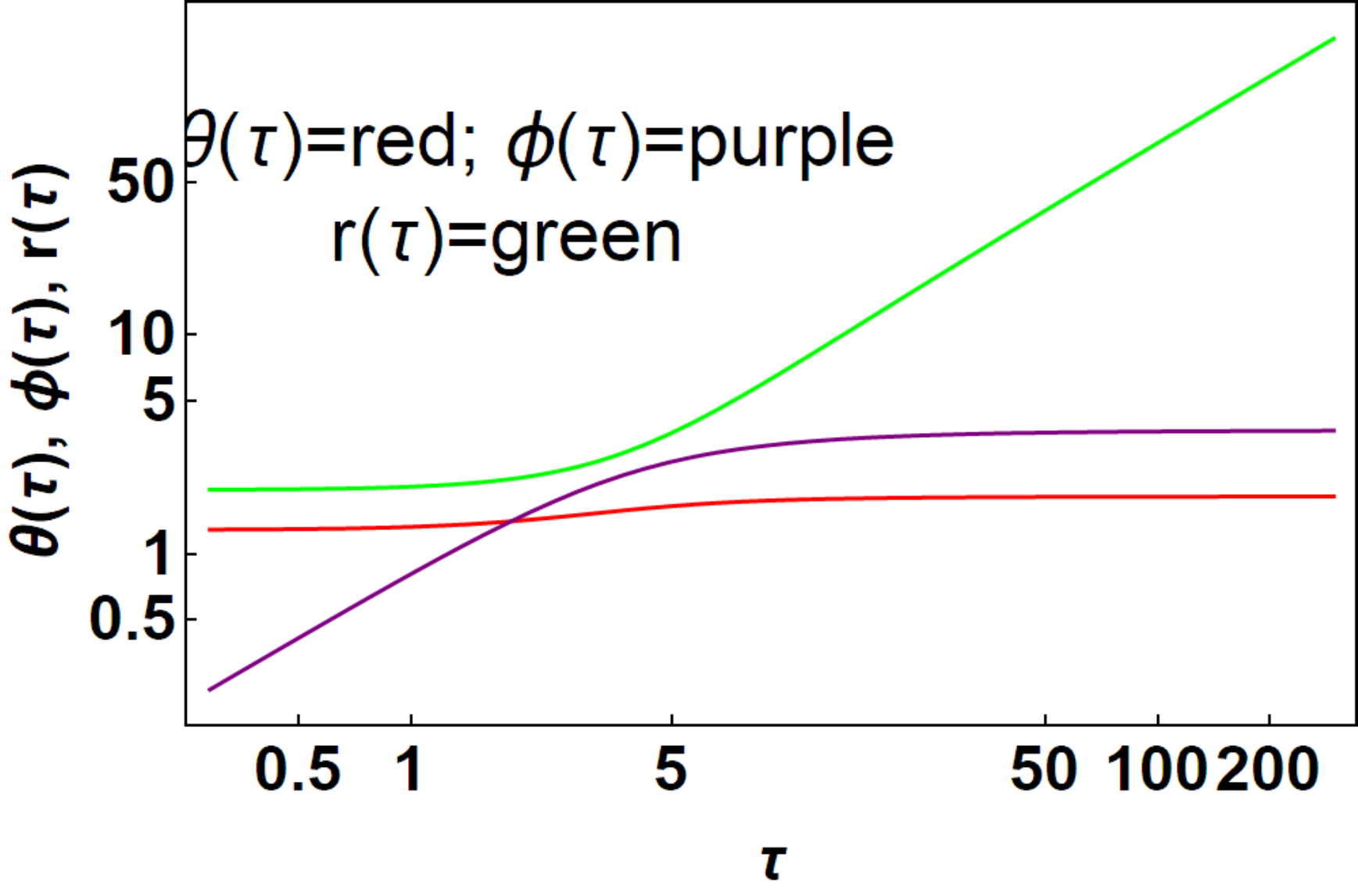}\\
    \includegraphics[width=5cm]{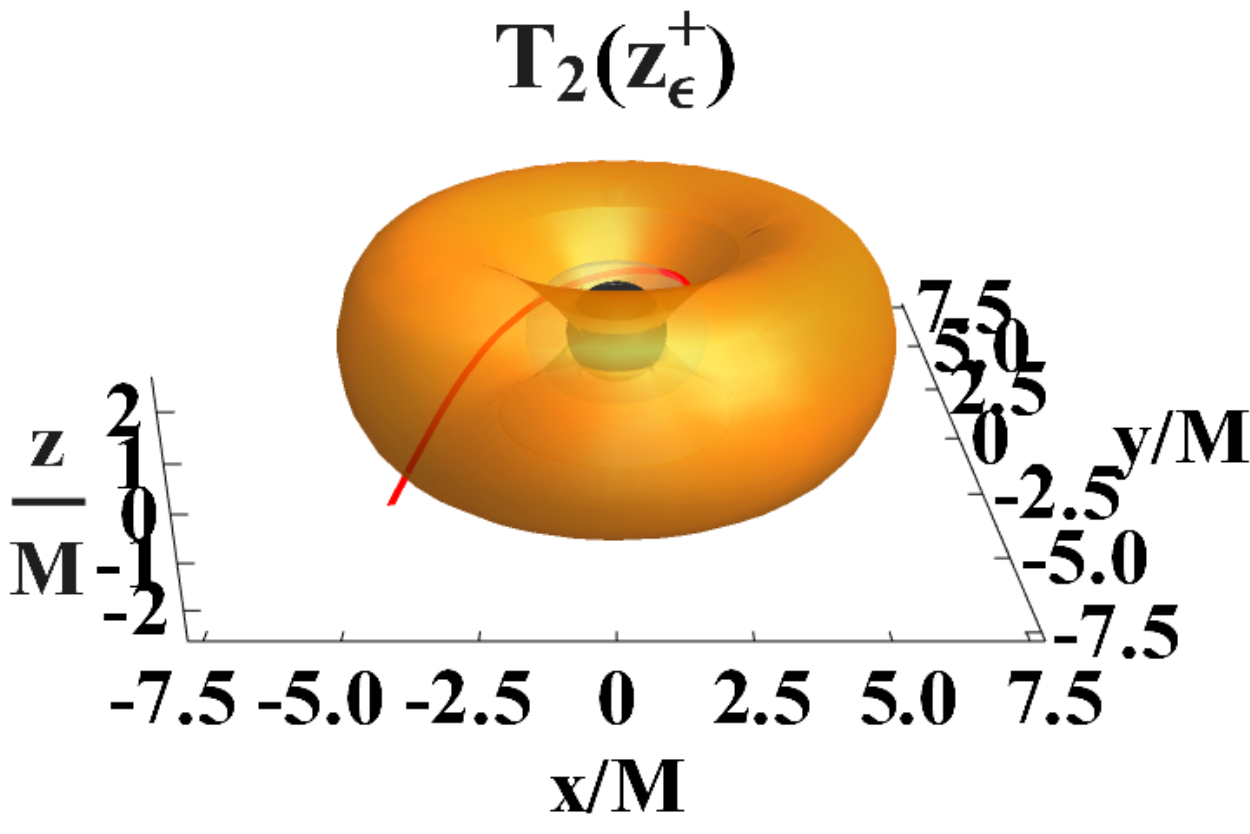}
      \includegraphics[width=5cm]{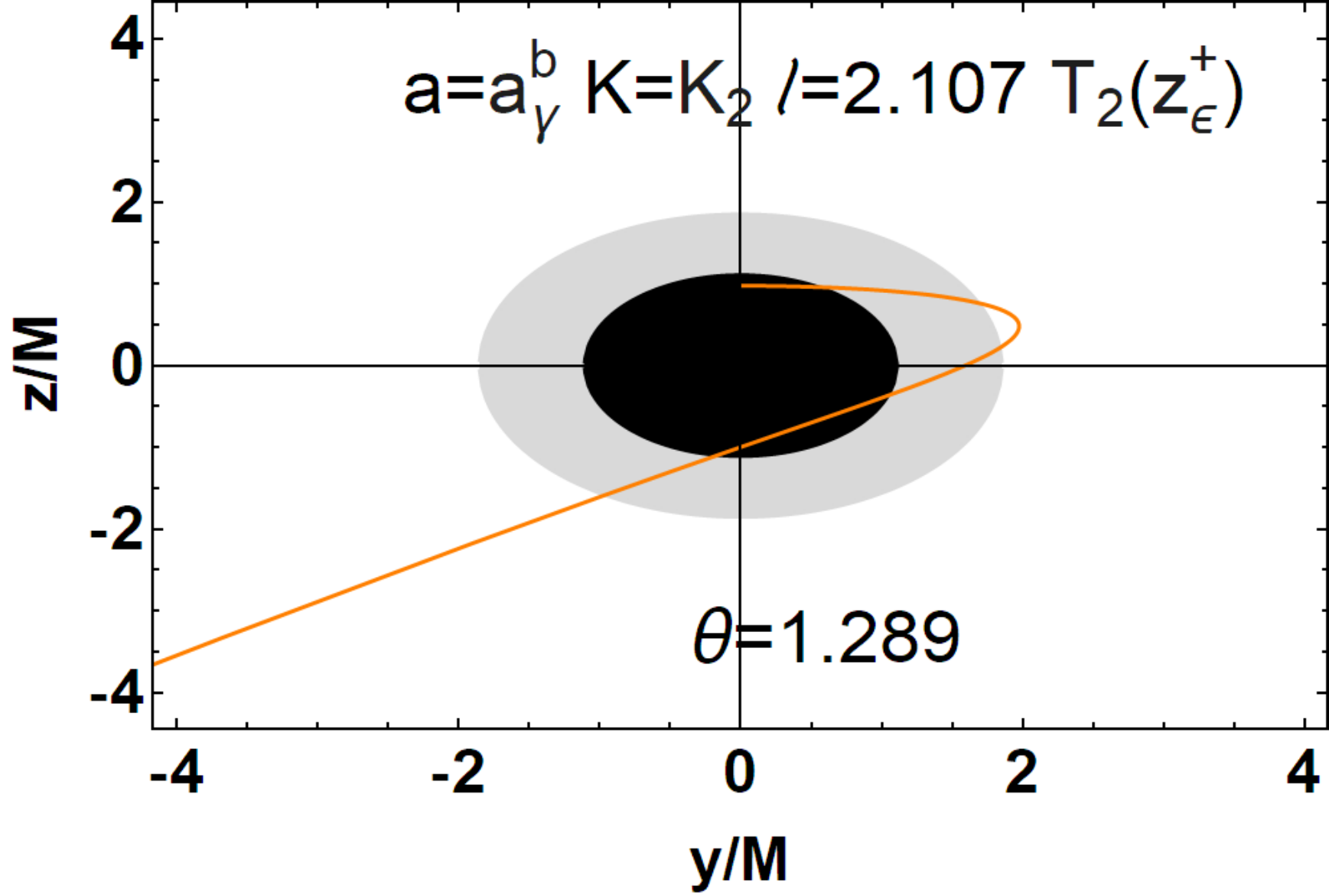}
    \includegraphics[width=5cm]{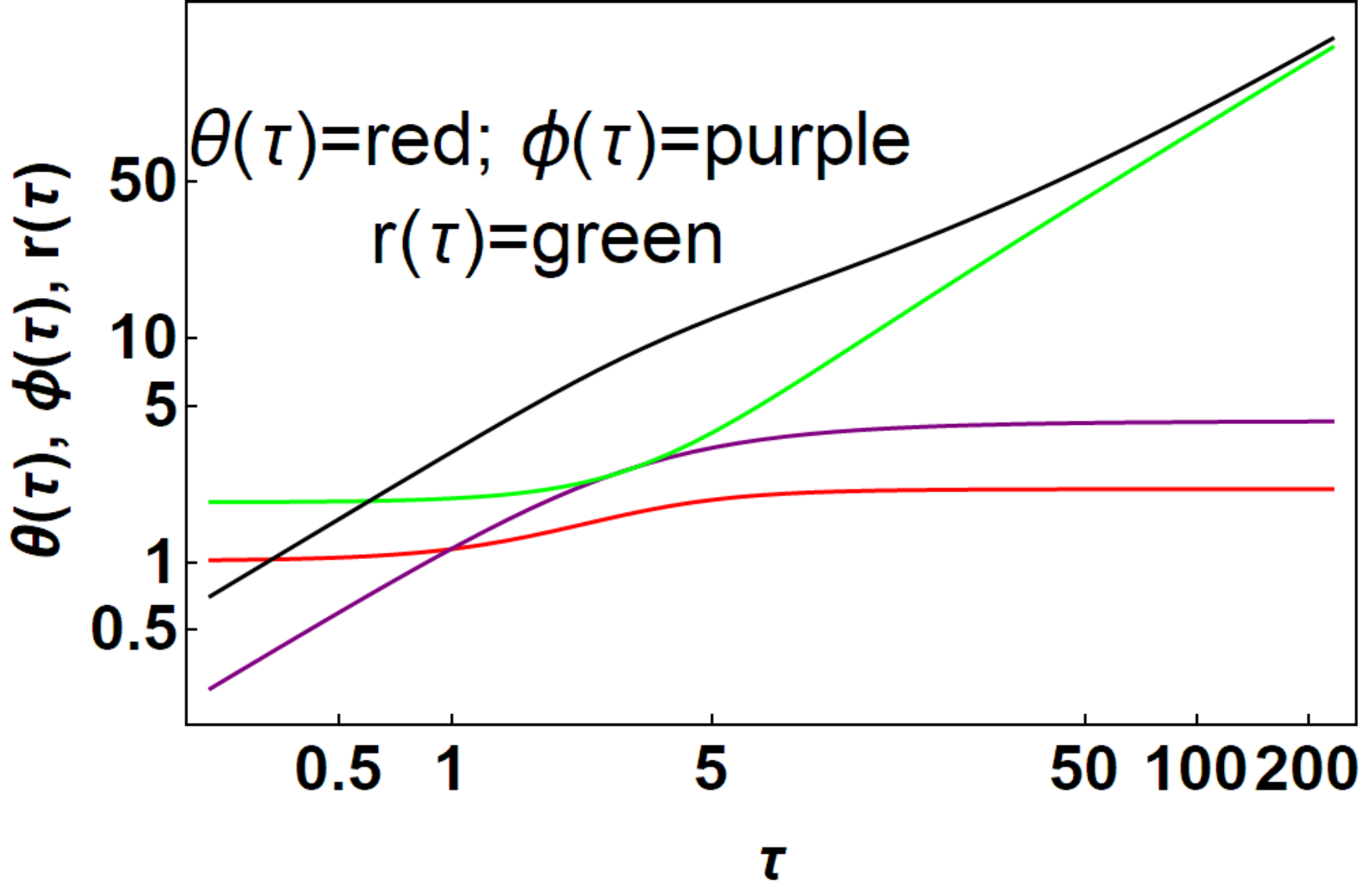}\\
      \includegraphics[width=5cm]{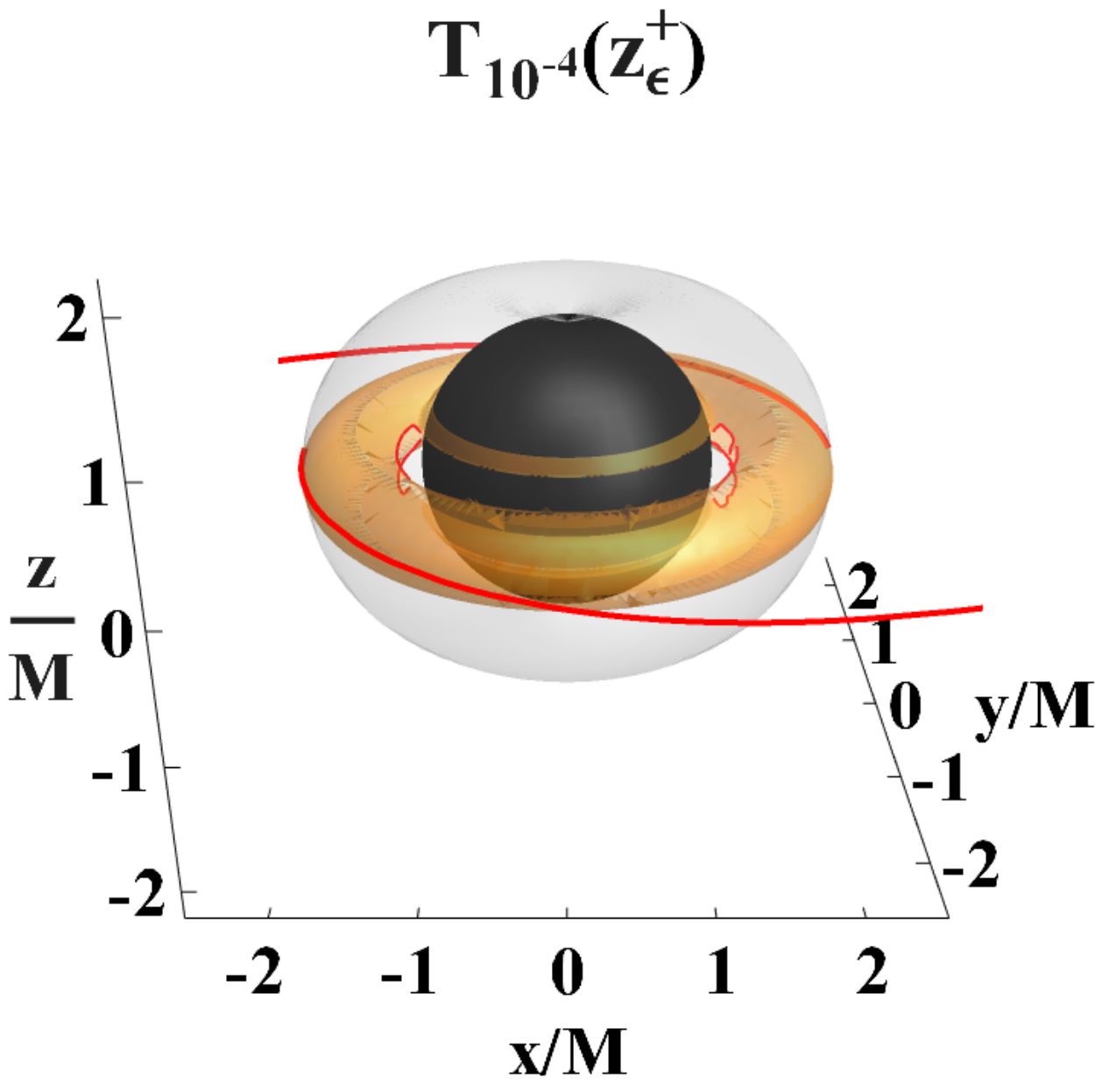}
        \includegraphics[width=5cm]{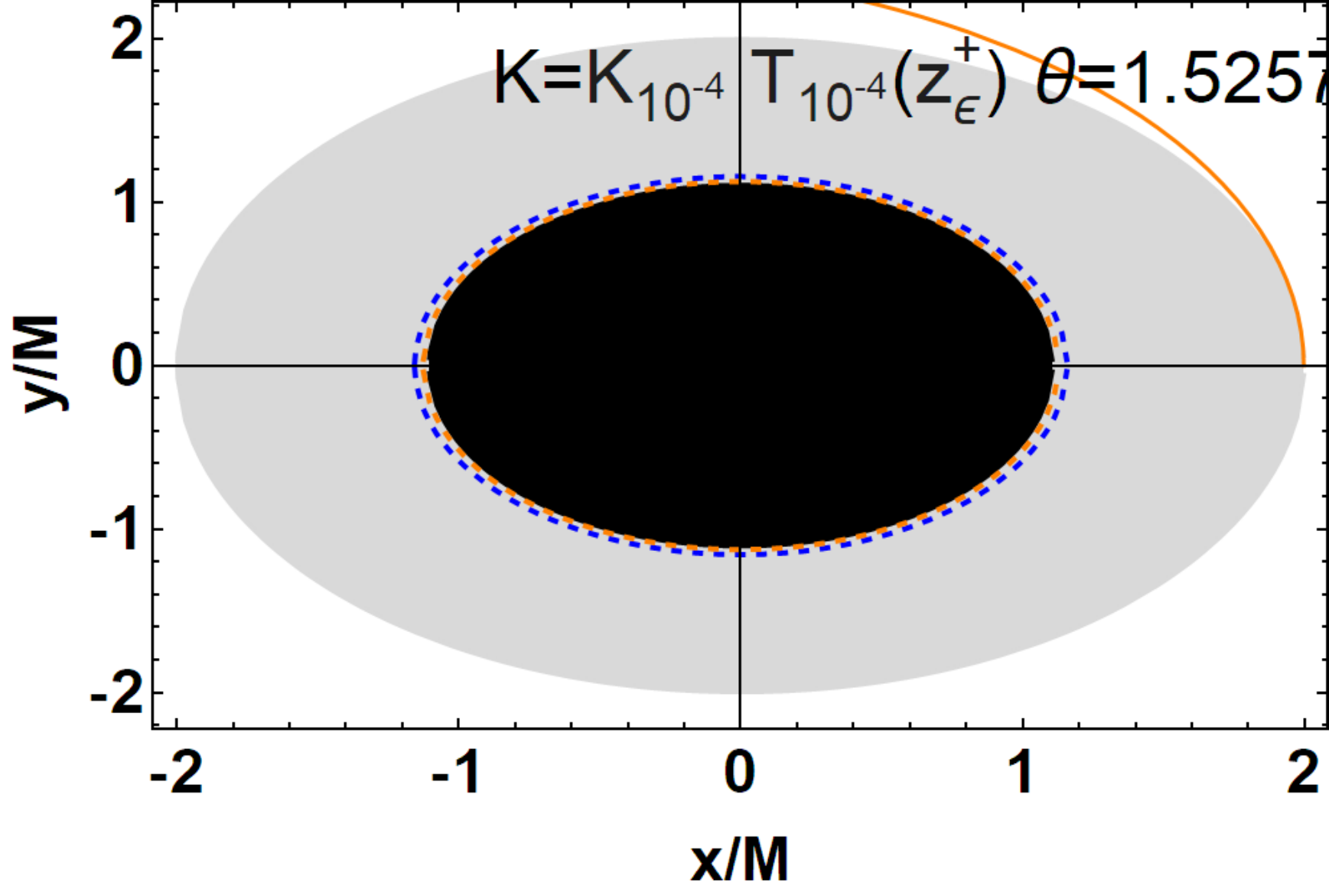}
    \includegraphics[width=5cm]{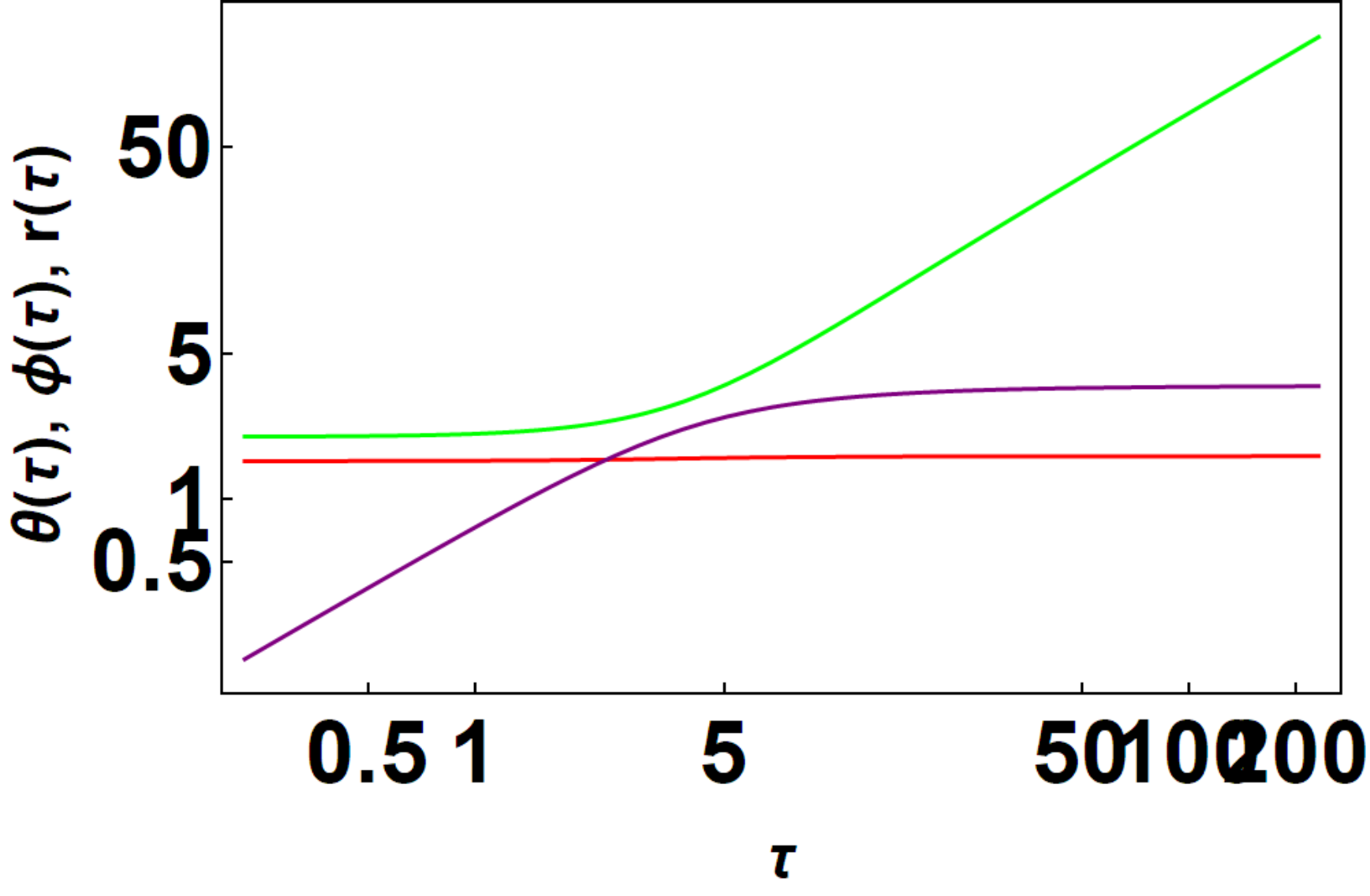}
    \caption{Light-like  particles analysis of configurations of Figs\il(\ref{Fig:PlotTRial}).
Black center is the \textbf{BH} $r<r_+$, gray region is the outer ergoregion $r\in ]r_+,r_{\epsilon}^+]$. (There is $\{x=r \sin\theta \cos\phi,y=r \sin\theta \sin\phi,z=r \cos\theta\}$.).
 Yellow surface is the torus.  \textbf{BH} spacetime has  spin $a_{mbo}^b$.  The torus specific angular momentum $\ell$ signed in figure.
 Upper line panels:  photons leave radius $r_0=r_{\max}=r_{\epsilon}^+$ (i.e. the  torus geometrical maximum,  coincident with the outer ergosurface $r_{\epsilon}^+$).
Second  and  third  line panels: the initial radius $r_0$ is the   the crossing points of the torus surface with the outer ergosurface.
Third line photon leaves the torus  outer edge coincident with $r_{\epsilon}^+=2M$.
Right panels show particles coordinates  $\theta$ (red curve), $\phi$ (purple curve) and radius  $r$ (green curve).}\label{Fig:zzPlotcredcatr}
    \end{figure}

\begin{figure}\centering
  % Requires \usepackage{graphicx}
  \includegraphics[width=4cm]{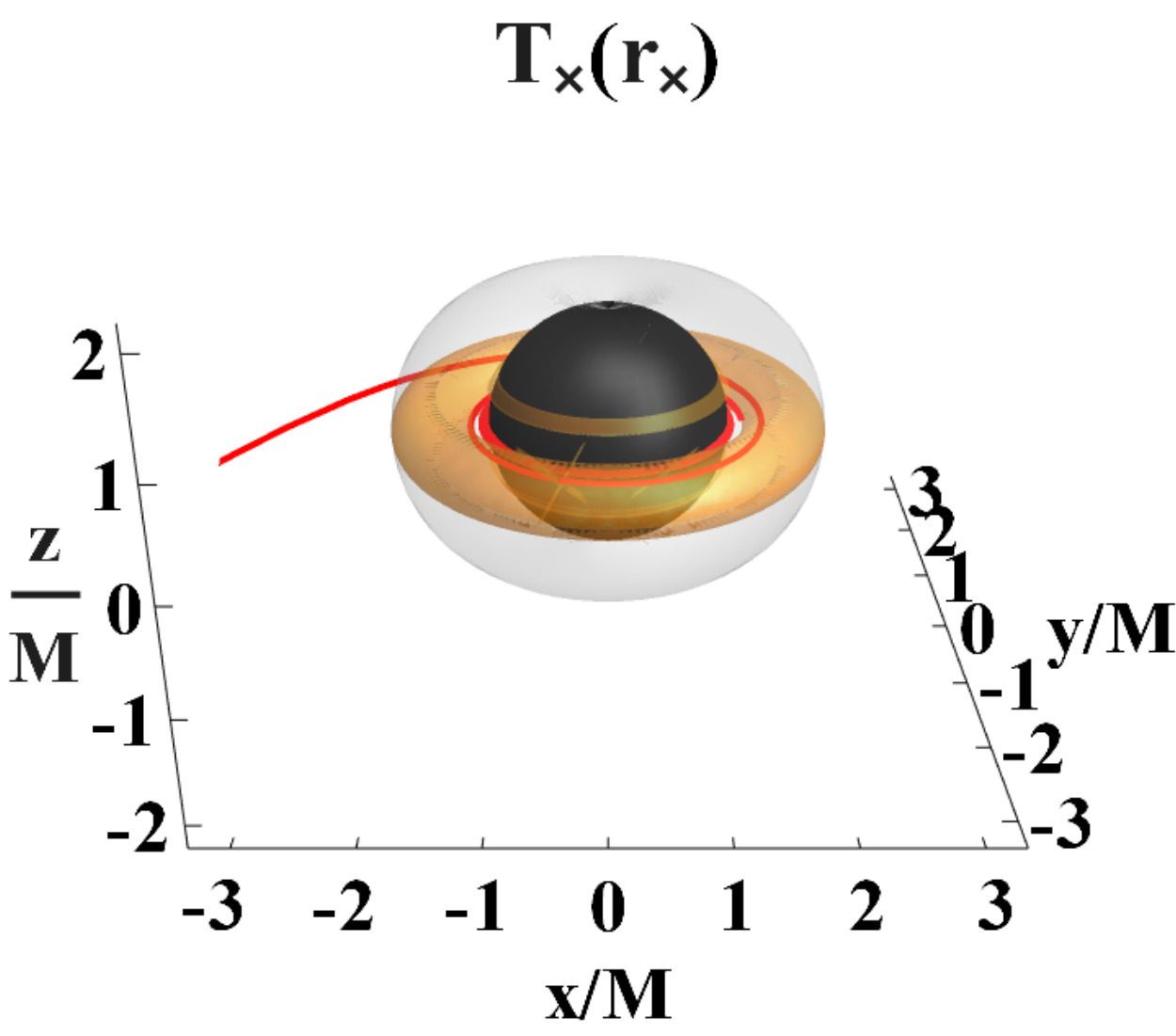}
  \includegraphics[width=4cm]{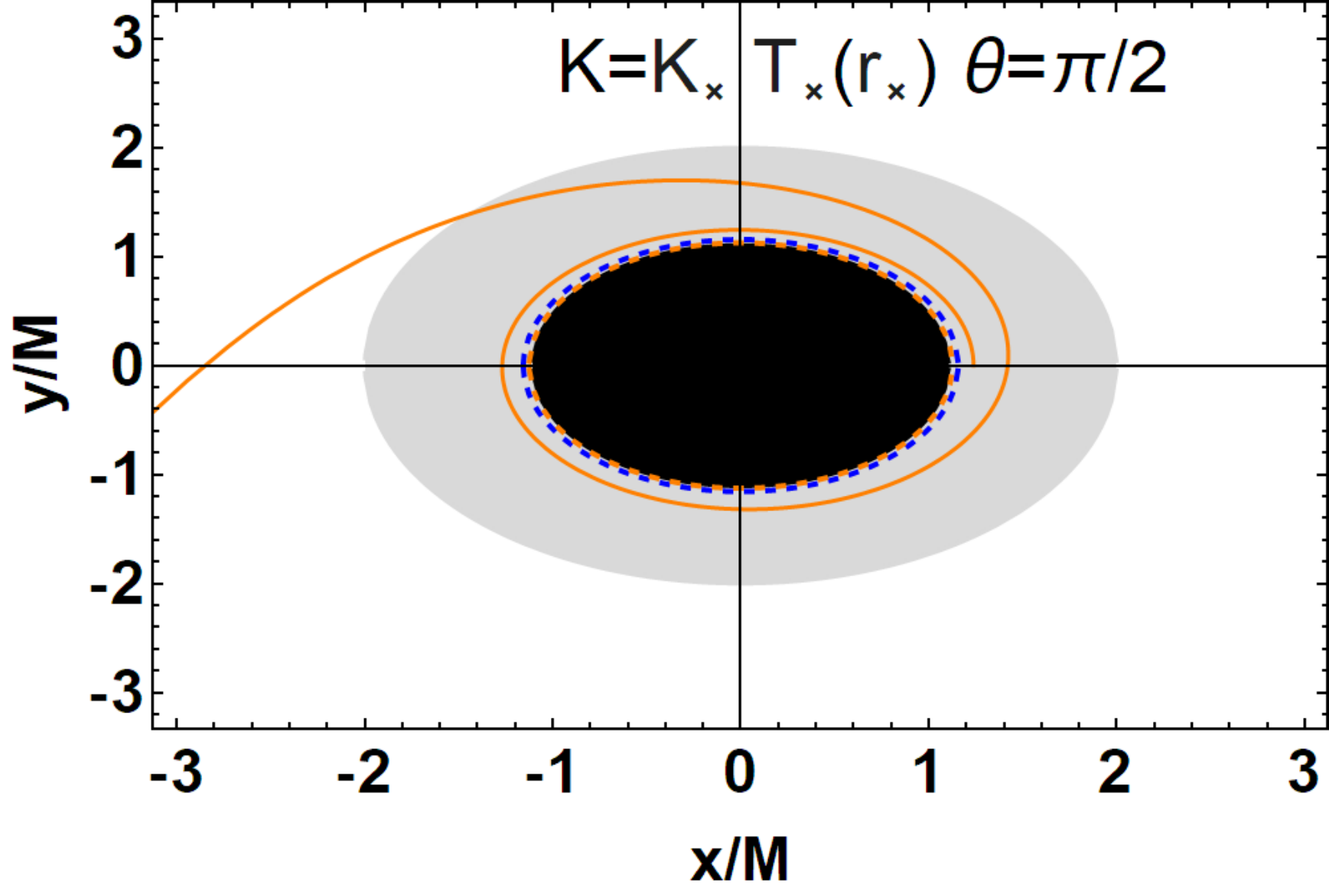}
   \includegraphics[width=4cm]{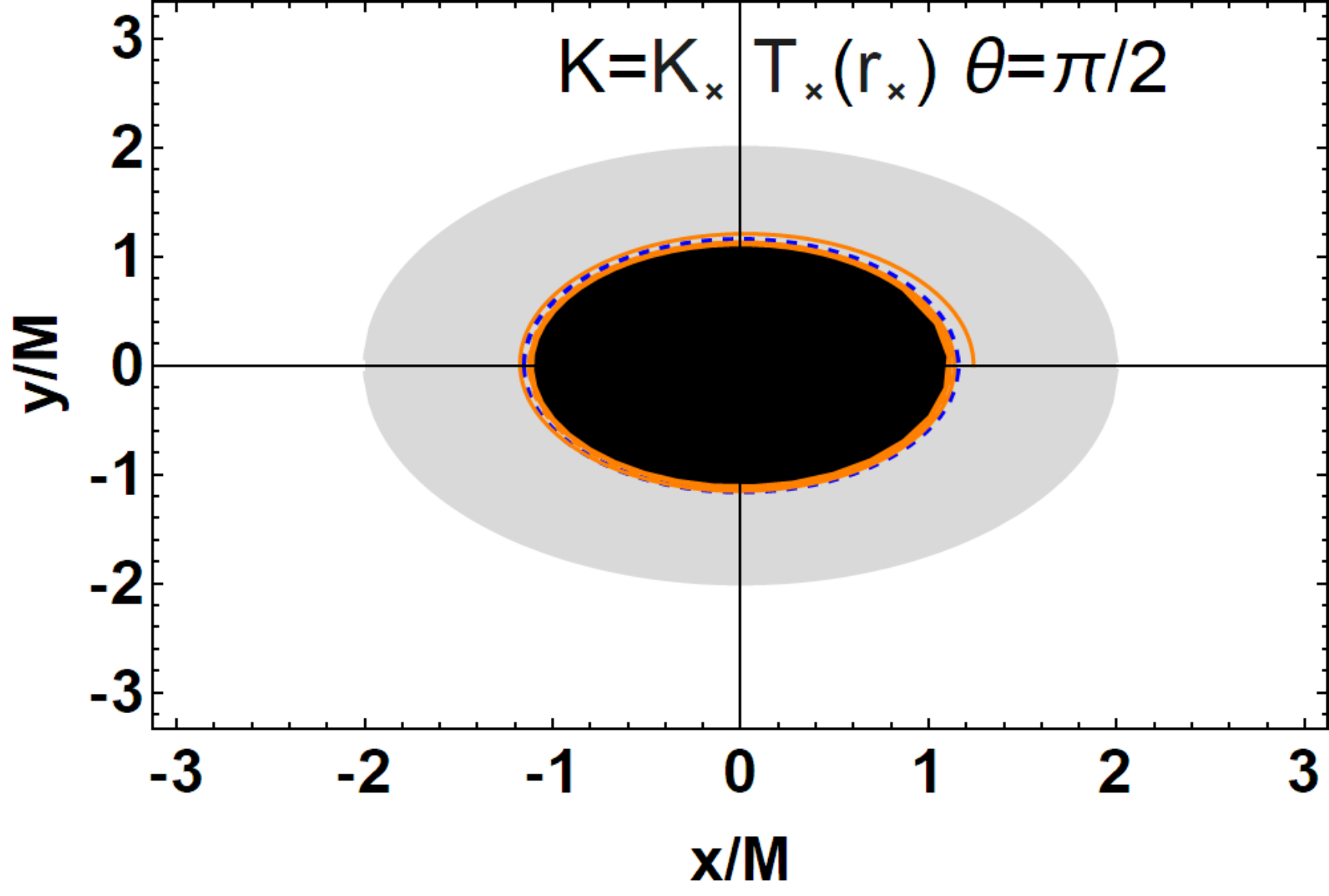}
     \includegraphics[width=4cm]{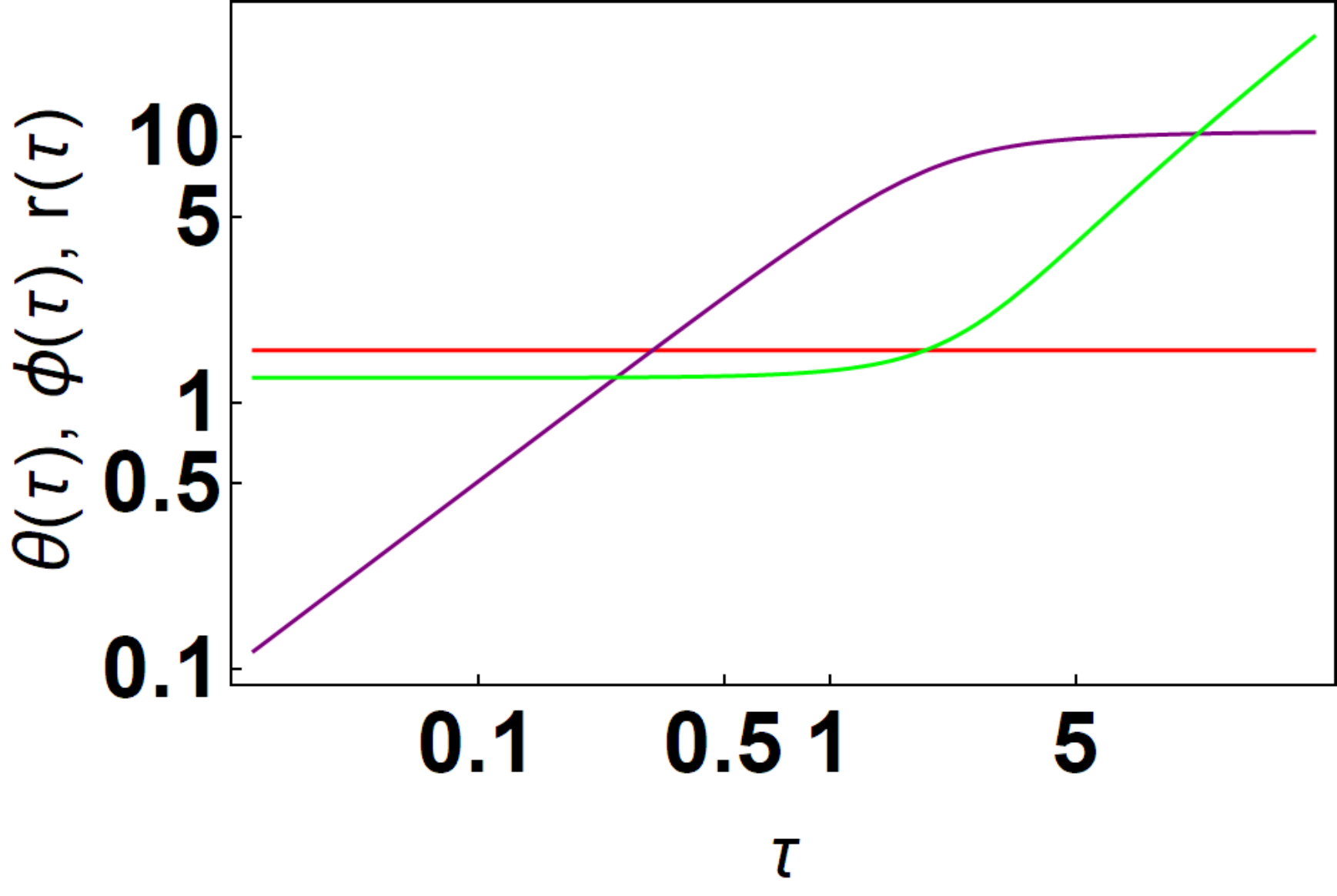}
   \includegraphics[width=4cm]{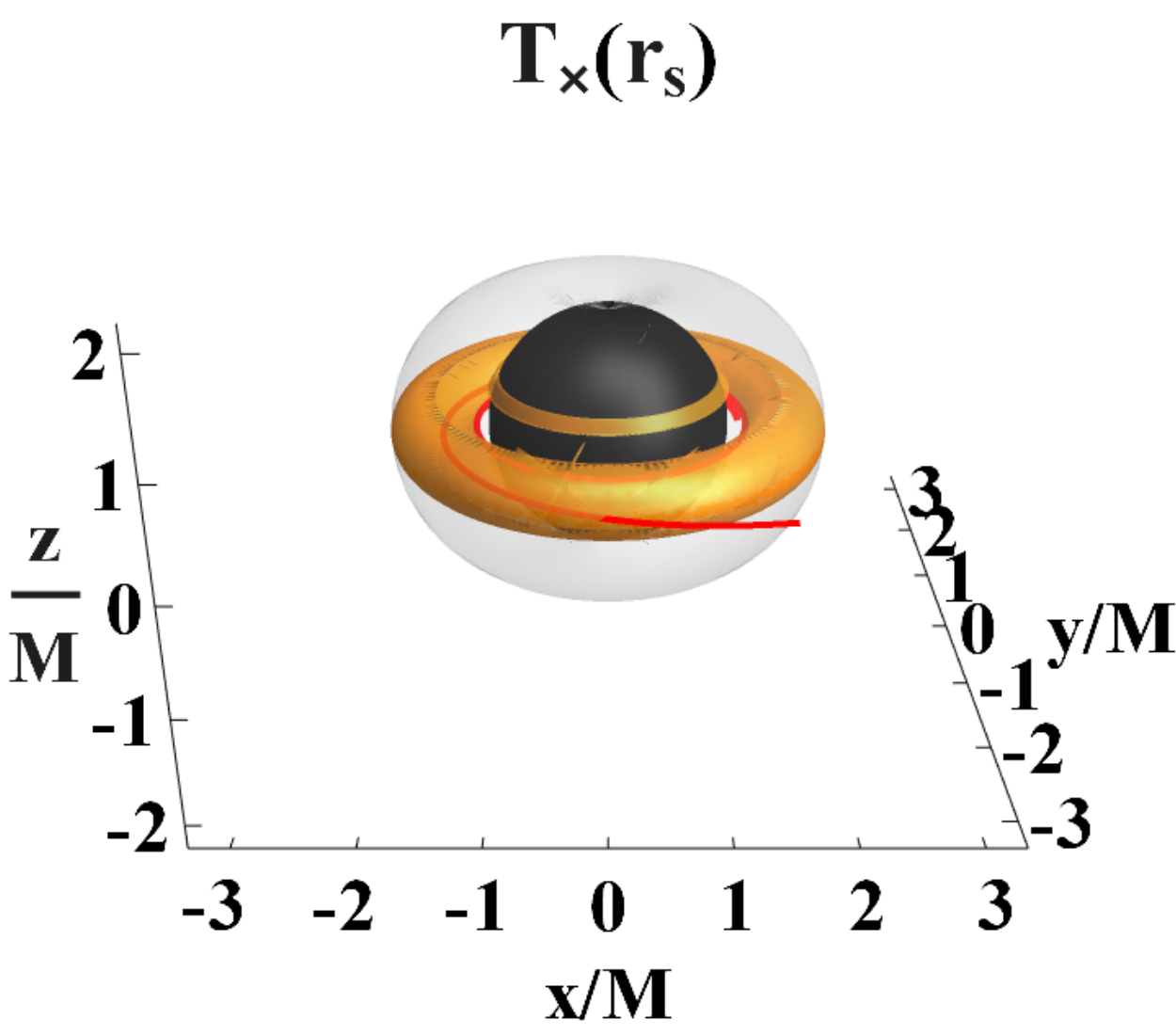}
     \includegraphics[width=4cm]{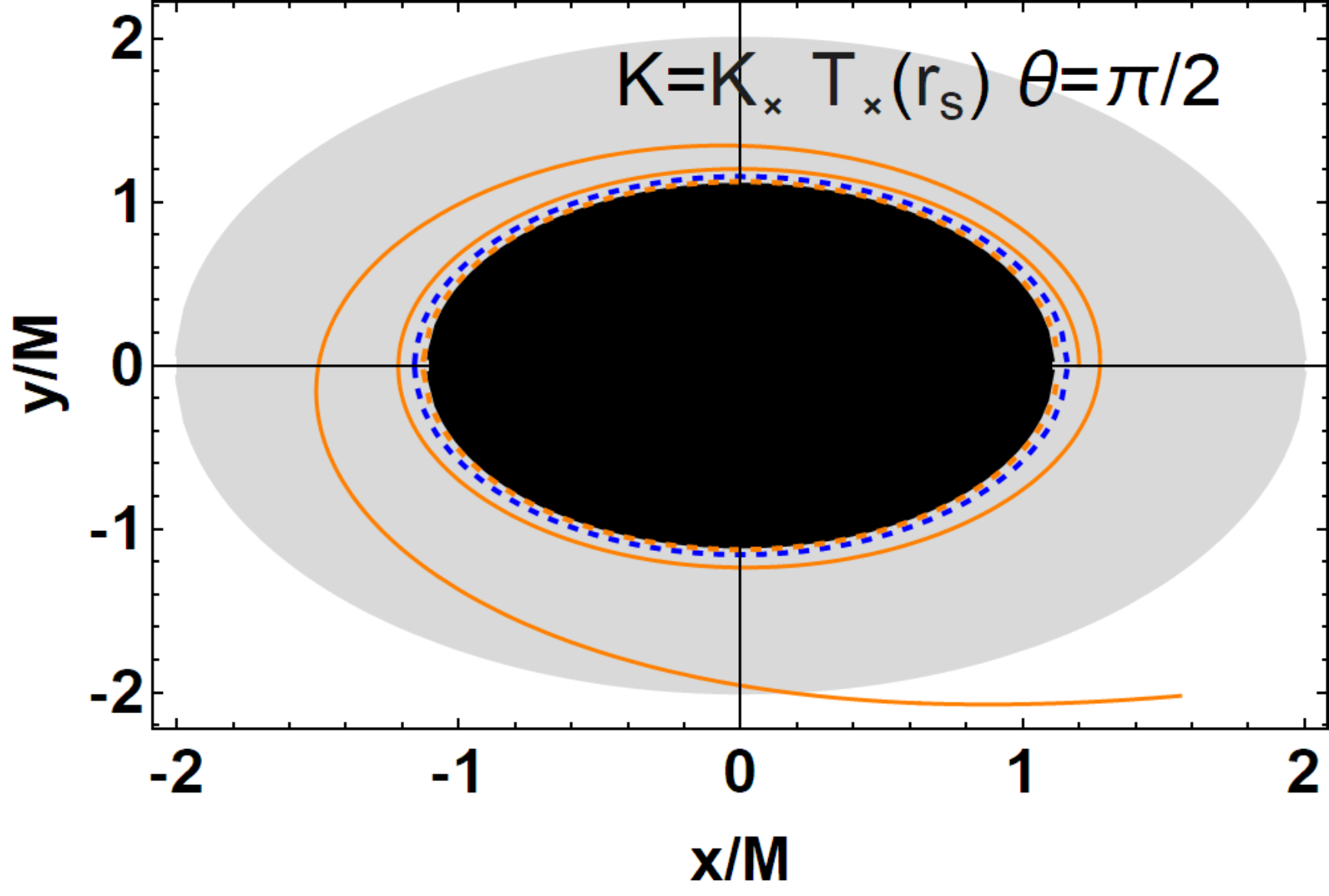}
         \includegraphics[width=4cm]{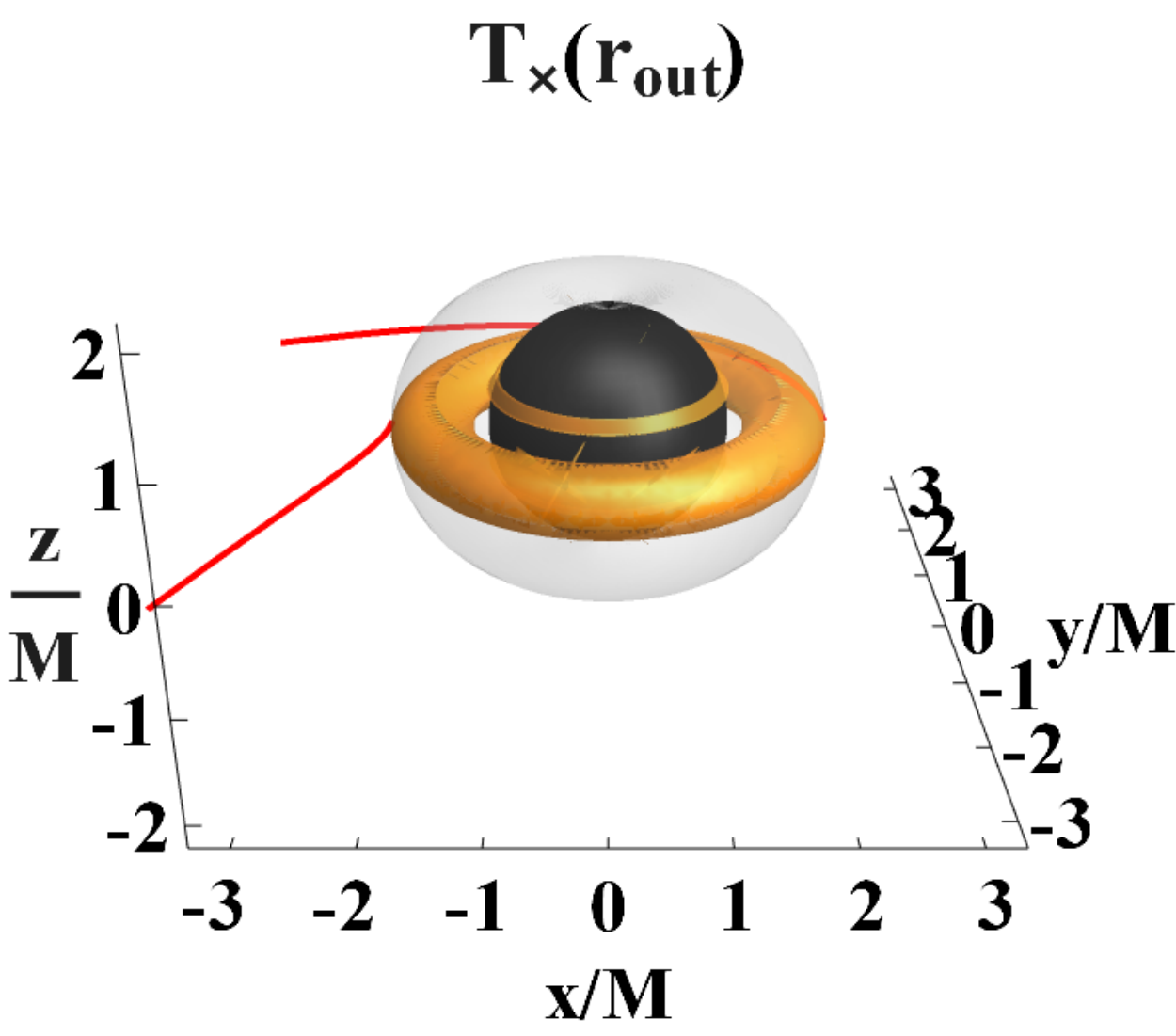}
    \caption{Light-like  particles analysis of configurations of Figs\il(\ref{Fig:PlotTRial}). Particles from cusped torus $T_{\times}$, with  cusps
$r_{\times}$. There is $\{x=r \sin\theta \cos\phi,y=r \sin\theta \sin\phi,z=r \cos\theta\}$.
Bottom line
left  and center panels:   particles  leave a point $r_s=1.2M$  in the inner region. Bottom right panel, particles leave the torus outer edge $r_{outer}=r_{\epsilon}^+$.
Upper right panel  shows particles coordinates  $\theta$ (red curve), $\phi$ (purple curve) and radius  $r$ (green curve).
Black center is the \textbf{BH} $r<r_+$  with spin $a_{mbo}^b$, gray region is the outer ergoregion $r\in ]r_+,r_{\epsilon}^+]$.
 Yellow surface is the torus.  Torus specific angular momentum is $\ell=2.107$.}\label{Fig:zzPlotcredcatr1}
    \end{figure}

  As exemplified by the cases studied in this example, particles follow three  classes of trajectories: particles   can leave the ergoregion; particles can remain  trapped in the  inner region (leading to collision with torus inner edge), or be absorbed by the  \textbf{BH}.
Photon motion is shown in Figs\il(\ref{Fig:zzPlotcredcatr}), (\ref{Fig:zzPlotcredcatr1}). Particle models of  Table\il(\ref{Table:Particle-models})  have been considered for the initial locations and the tori models.  The  disk exfoliation is based on  the free particles hypotheses, grounded on low pressure gradients, Lense-Thirring effect and small tori  sizes,  (a further open  question  concerns the process time scale for the total destruction of the  dragged torus). The example of   Figs\il(\ref{Fig:zzPlotcredcatr}), (\ref{Fig:zzPlotcredcatr1}) has therefore  to be understood as indicative, and the analysis  should  be framed  considering the  emission-reflection processes from the disk surface.
\subsubsection{Extreme configurations: multi-tori, proto-jets and  $\cc_2$ tori}\label{Sec:gir-c2}
\begin{figure}\centering
  % Requires \usepackage{graphicx}
   \includegraphics[width=15cm]{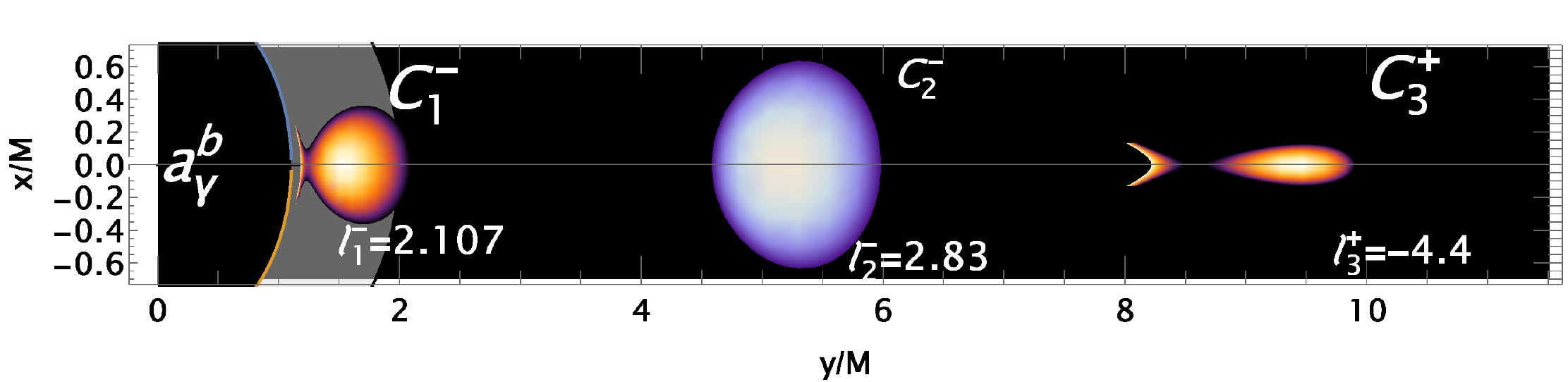}
  \caption{Closed equipotential surfaces  $C_i^*$ are shown in the  \textbf{BH} spacetime  with spin $a_{\gamma}^b$.  Gray region is the outer ergoregion.  $\ell^{*}$ is for the fluid specific angular momentum. Sign $*=\pm$ is for counterrotating/corotating configurations respectively.  The  middle corotating torus is  quiescent. The inner corotating cusped torus is  a dragged configuration.  See Figs\il(\ref{Fig:polodefin1}), (\ref{Fig:PlotVamp1}), (\ref{Fig:spessplhoke1}), (\ref{Fig:PlotVampa1})
(\ref{Fig:PlotVampb1}). }\label{Fig:Behav1}
\end{figure}

On the sidelines of this analysis we also consider the proto-jets existence  in $\Sigma_{\epsilon}^+$.
An  issue that can affect   stability  of tori orbiting in  $\Sigma_{\epsilon}^+$ is  the possibility that  several orbiting toroidal configurations  may be formed  in  aggregates of toroids  orbiting the central \textbf{BH}, occurring, for example, in composed systems of tori agglomerates as  in the  \textbf{eRAD}  framework \citep{ringed,dsystem}, developed  as aggregates of axisymmetric, corotating and counter-rotating  toroidal configurations, coplanar and centered on the equatorial plane of the central  Kerr attractor   in \textbf{AGNs}-Figs\il(\ref{Fig:Behav1}).
Figs\il(\ref{Fig:Behav1}) show for the \textbf{BH} spacetime with $a=a_{\gamma}^b$, the case of three closed configurations, the inner cusped torus contained in the ergoregion  as  in Figs\il(\ref{Fig:polodefin1}).
The outer critical torus is counterrotating, the middle of the triplet is quiescent and corotating. This is a case of double accretion onto the central \textbf{BH} with a middle screening torus.

\begin{figure}\centering
  % Requires \usepackage{graphicx}
  \includegraphics[width=8cm]{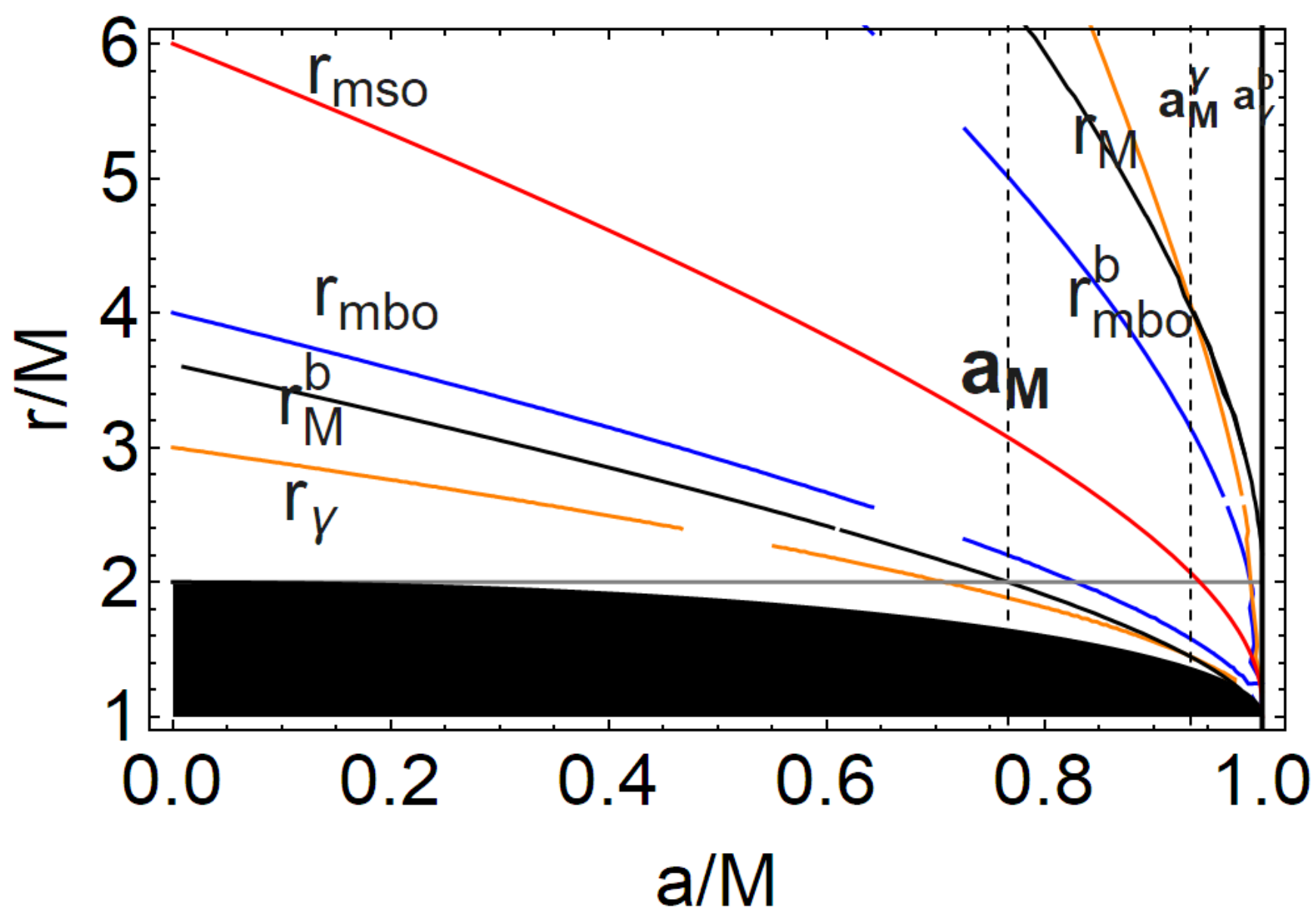}
  \includegraphics[width=8cm]{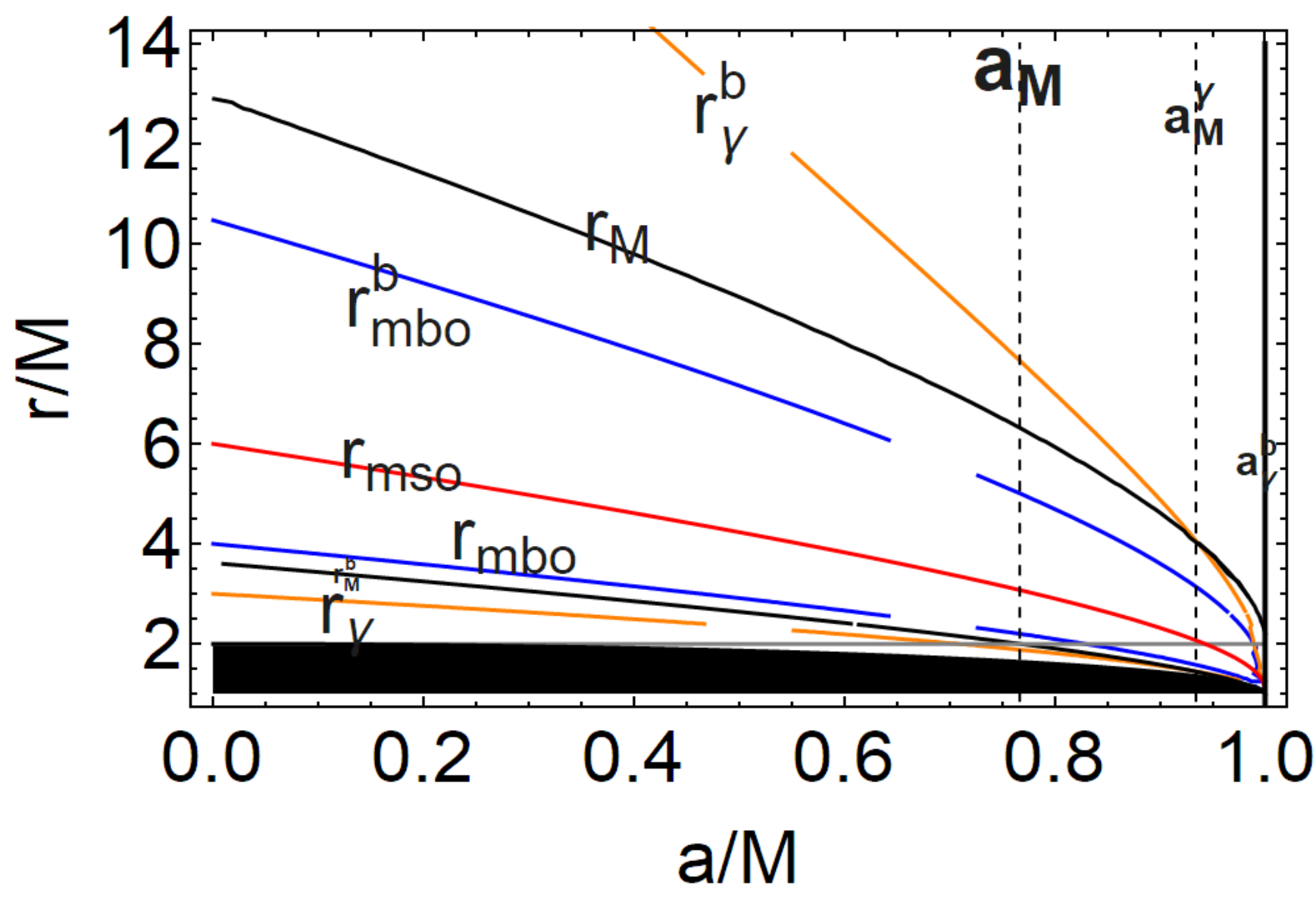}
  \caption{Analysis of the solutions $r_{\Mie}: \partial^2_r \ell=0$ and $r^b_{\Mie}: \ell(r_{\Mie})=\ell$ as function of the \textbf{BH} dimensionless spin $a/M$. Black region is the \textbf{BH}, region $r<r_+$, where $r_+$ is the outer horizon. $r_{mbo}$ is the marginally bounded orbit, $r_{mso}$ is the marginally stable orbit, $r_{\gamma}$ is the photon orbit, and marginally circular orbit. These are corotating orbits. Radii $r^b_{mbo}: \ell(r_{mbo})=\ell$ and
   $r^b_{\gamma}: \ell(r_{\gamma})=\ell$, set the center of the toroidal configurations. The static limit  on the equatorial plane  $r_{\epsilon}^+=2M$ is the gray line. Dashed lines are $a_{\Mie}=0.766745M: \ell(r^b_{\Mie})=\ell(r_{\epsilon}^+)$ and $a^{\gamma}_{\Mie}=0.934313M: \ell({r}_{\Mie})=\ell(r^b_{\gamma})$ and $\ell(r^b_{\Mie})=\ell({r}_{\gamma})$.}\label{Fig:vendiplot}
\end{figure}

\begin{figure}\centering
  % Requires \usepackage{graphicx}
  \includegraphics[width=8cm]{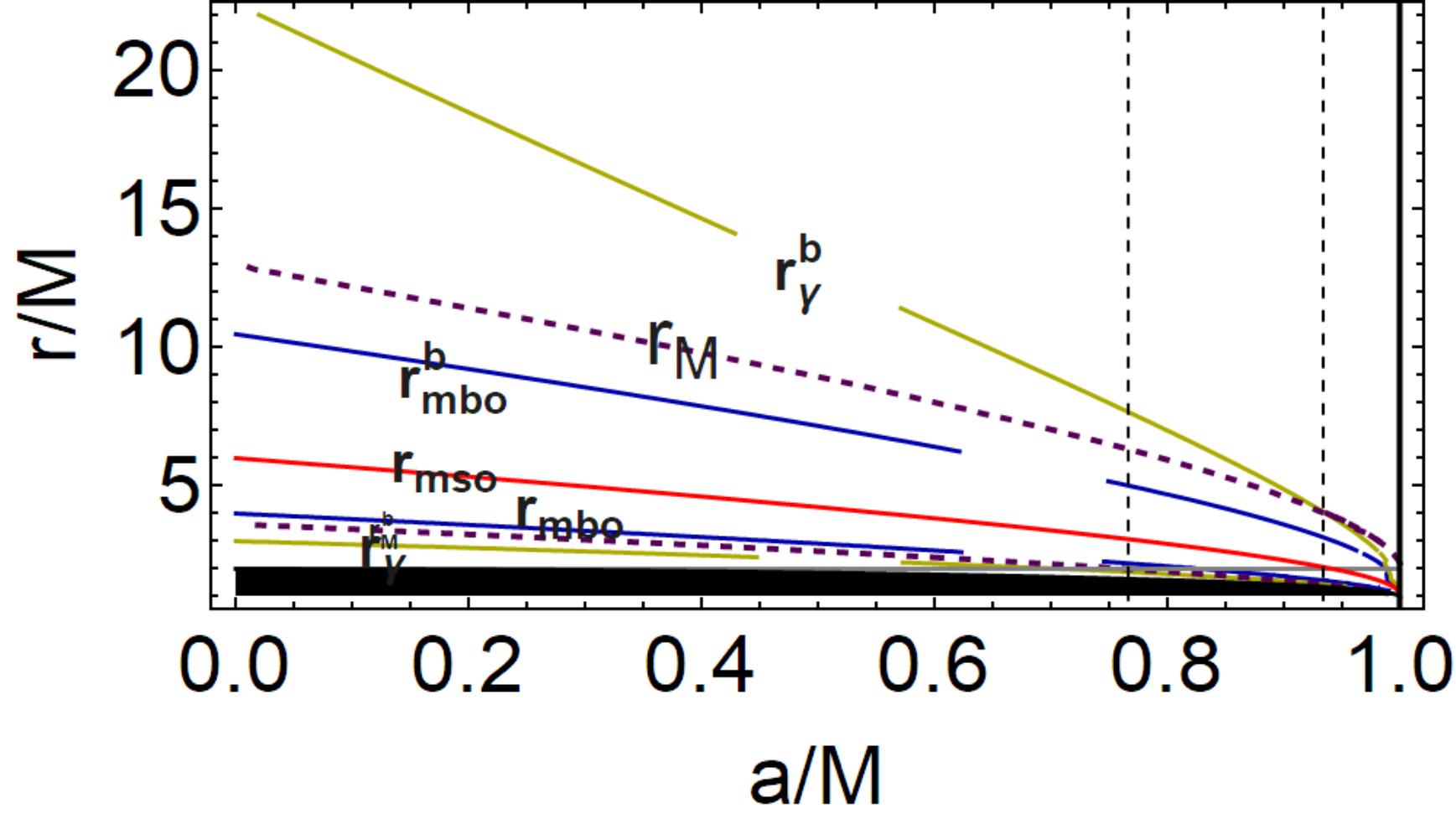}
  \includegraphics[width=8cm]{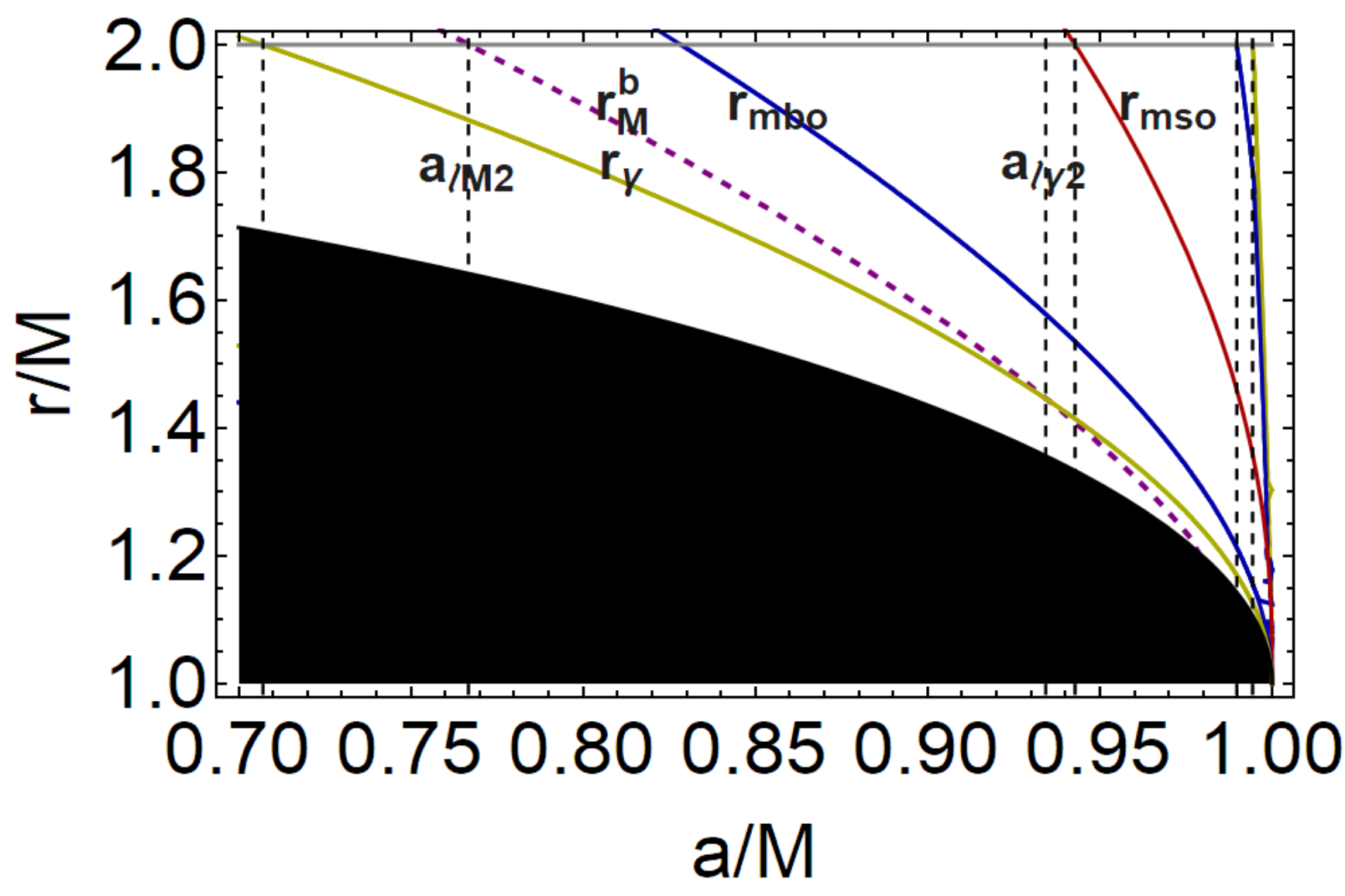}
  \caption{Analysis of  configurations  $\cc_2$ having specific angular momentum $\ell\in\mathbf{ L_2}$. The analysis is focused on  tori  inner edge and  the  proto-jets cusps configurations located  in the ergoregion and correspondent to the "accumulation" points of the pressure i.e. points $r_{\mathcal{M}}:\partial_r^2\ell=0$, which correspond to the  maximum pressures points in the disks.  There are, associated to tori with center  in $r_{\mathcal{M}}$ the  minima pressure points $r_{\mathcal{M}}^b:\ell(r_{\mathcal{M}})=\ell(r)$. Black region is the \textbf{BH}, $r<r_+$ where $r_+$ is the outer Killing horizon. The outer ergosurface on the equatorial plane  $r_{\epsilon}^+=2M$ is also shown (gray-line). Right panel is a zoom  on the ergoregion. Dashed lines are spins $\{a_{mbo},a_{mbo}^b,a_{\gamma},a_{\gamma}^b,a_{mso}\}$--see Figs\il(\ref{Fig:polodefin1}).        In the right panel spins $a_{\ell\gamma 2}=0.93431: r_{\Mie}^b=r_{\gamma} $
and $a_{\ell M2}=0.7667: r_{\Mie}^b=r_{\epsilon}^+$ are represented. $r_{mso}$ is the marginally stable orbit, $r_{mbo}$ is the marginally bounded orbit, $r_{\gamma}$ is the photon orbit and
$r_{\gamma}^b: \ell(r_{\gamma})=\ell(r)$, $r_{mbo}^b: \ell(r_{mbo}^b)=\ell(r)$. }\label{Fig:ProDRA2}
\end{figure}

It is clear that each toroidal component of the aggregate of partially contained or dragged surfaces is constrained according to the analysis of Secs\il(\ref{Sec:deta-hall}). Especially  tori orbiting around faster  rotating  \textbf{BH} can rise to collision.  These tori would be extremely small and at  close distance (i.e. small displacement $\bar{\lambda}_{(1,2)}\equiv r_{inner}^{\mathbf{(2)}}-r_{outer}^{\mathbf{\mathbf{\mathbf{(1)}}}}$ where  $r_{inner}^{\mathbf{(2)}}$ is the inner edge of the outer torus while $r_{outer}^{\mathbf{\mathbf{\mathbf{(1)}}}}$ is the outer edge of the inner torus). There  is    $r_{\Mie}>r_{\epsilon}^+$, and this property   suggests that the dragged tori  distribution (density of tori centers in a region around the central \textbf{BH}) has no extreme as function of $r/M$ for any spin $a/M$,  if the agglomerate  does not vary with the spin or radius. However, this may not be the case for  partially contained configurations because  of  the presence of the radius  $r_{\mathcal{M}}\in[r^b_{mbo},r^b_{\gamma}]$ in the stability range for the centers of $\cc_2$ tori.  Curve $r_{\mathcal{M}}(a)$ represents an "accumulation" point for the tori centers. Radius  $r_{\mathcal{M}}$  would be a  center for the configuration with specific angular momentum in the range  $\mathbf{L_2}$, where proto-jets are possible. Correspondingly, there is  the radius $r^b_{\Mie}\in[r_{\gamma},r_{mbo}]$  for the  proto-jets cusps in the geometries  $a\in[0,a_{\Mie}^{\gamma}[$, where
$a^{\gamma}_{\Mie}=0.934313M: \ell({r}_{\Mie})=\ell(r^b_{\gamma})$, and   $\ell(r^b_{\Mie})=\ell({r}_{\gamma})$;  in this spin range there is
$r^b_{\Mie}<r_{\gamma}$, and   ${r}_{\Mie}>r^b_{\gamma}$  for larger spin.
Radius  $r_{\Mie}$  is a center for tori in  $\mathbf{L_2}$ correspondent to unstable points  $r^b_{\Mie}$--
 see Figs\il(\ref{Fig:vendiplot}) and Figs\il(\ref{Fig:Plotsoorr})- Figs\il(\ref{Fig:PlotVampb1}).
On the other hand,
there is a \textbf{BH} geometry with $\ell_{\Mie}=\ell(r_{\epsilon}^+)$ where $\ell_{\Mie}\in[\ell_{mbo},\ell_{\gamma}]$,
in the range $[a_{\gamma},a_{mbo}]$.
 Then $r_{cusp}=r_j$ and  the   quiescent torus  $\cc_2$  inner edge  is  $r_{inner}>r_{j}\in]r_\gamma, r_{mbo}]$, which can be in ergoregion also for relatively small \textbf{BH} spin, opening an interesting possibility for jet emission and the role of the ergoregion, which  would be suppressed at large spin.
The possibility of existence of configurations $\mathbf{L_2}$ with the solutions of $\partial_r^2\ell=0$, points of {accumulation} of pressure $r_{\Mie}$, in the ergoregion, and therefore of inner edge of  $ \cc_2$, has been investigated   in Figs\il(\ref{Fig:ProDRA2}). It is clear that this possibility is regulated by the   spins $a_{\ell\gamma 2}=0.93431: r_{\Mie}^b=r_{\gamma} $
and $a_{\ell M2}=0.7667: r_{\Mie}^b=r_{\epsilon}^+$. In $a\in[a_{\ell M 2},a_{\ell\gamma 2}]$ the critical points $r_j$ lie in the ergoregion.
\subsection{Characteristic frequencies}\label{Sec:carac-fre}
\begin{figure}\centering
  % Requires \usepackage{graphicx}
  \includegraphics[width=5.6cm]{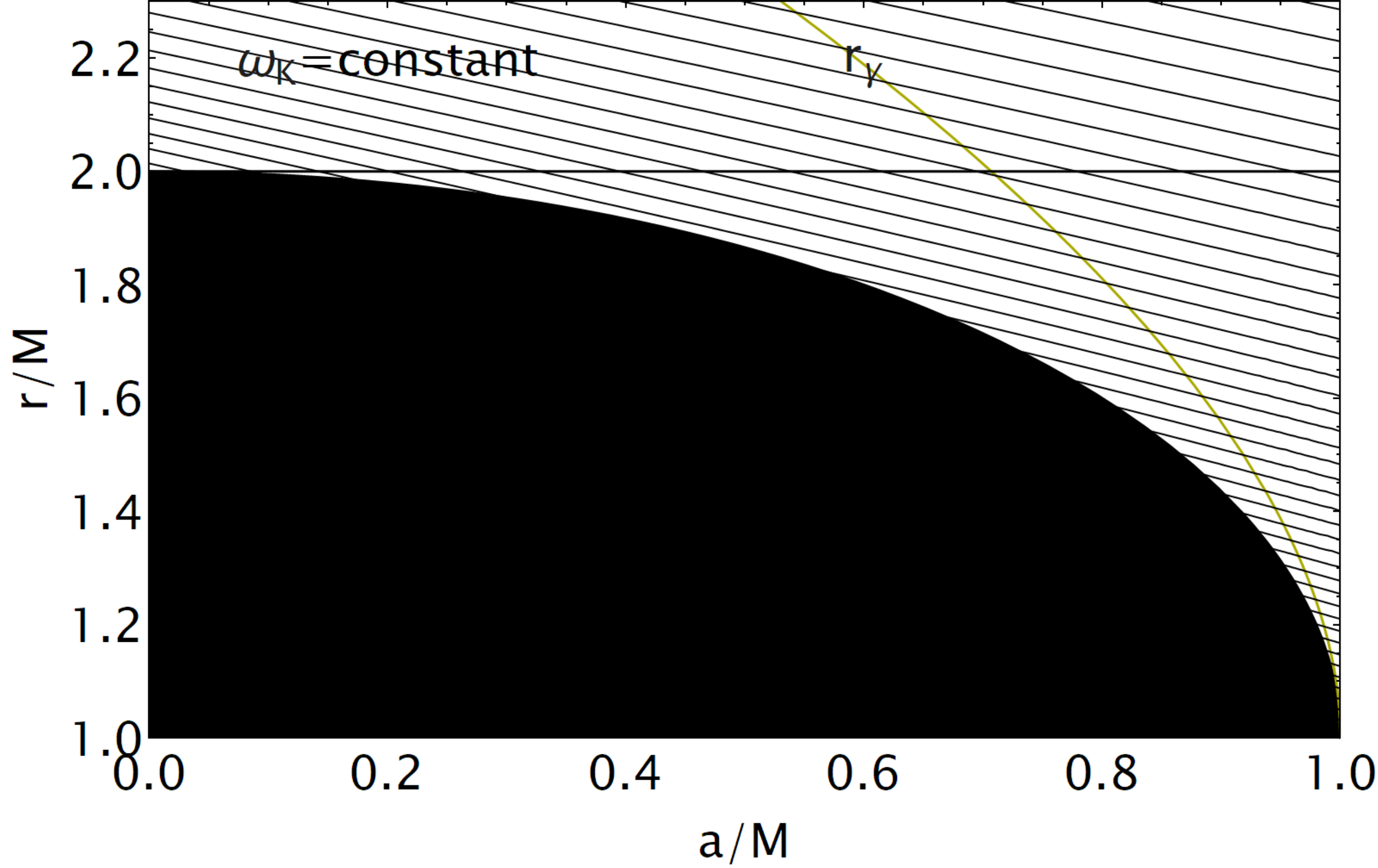}
  \includegraphics[width=5.6cm]{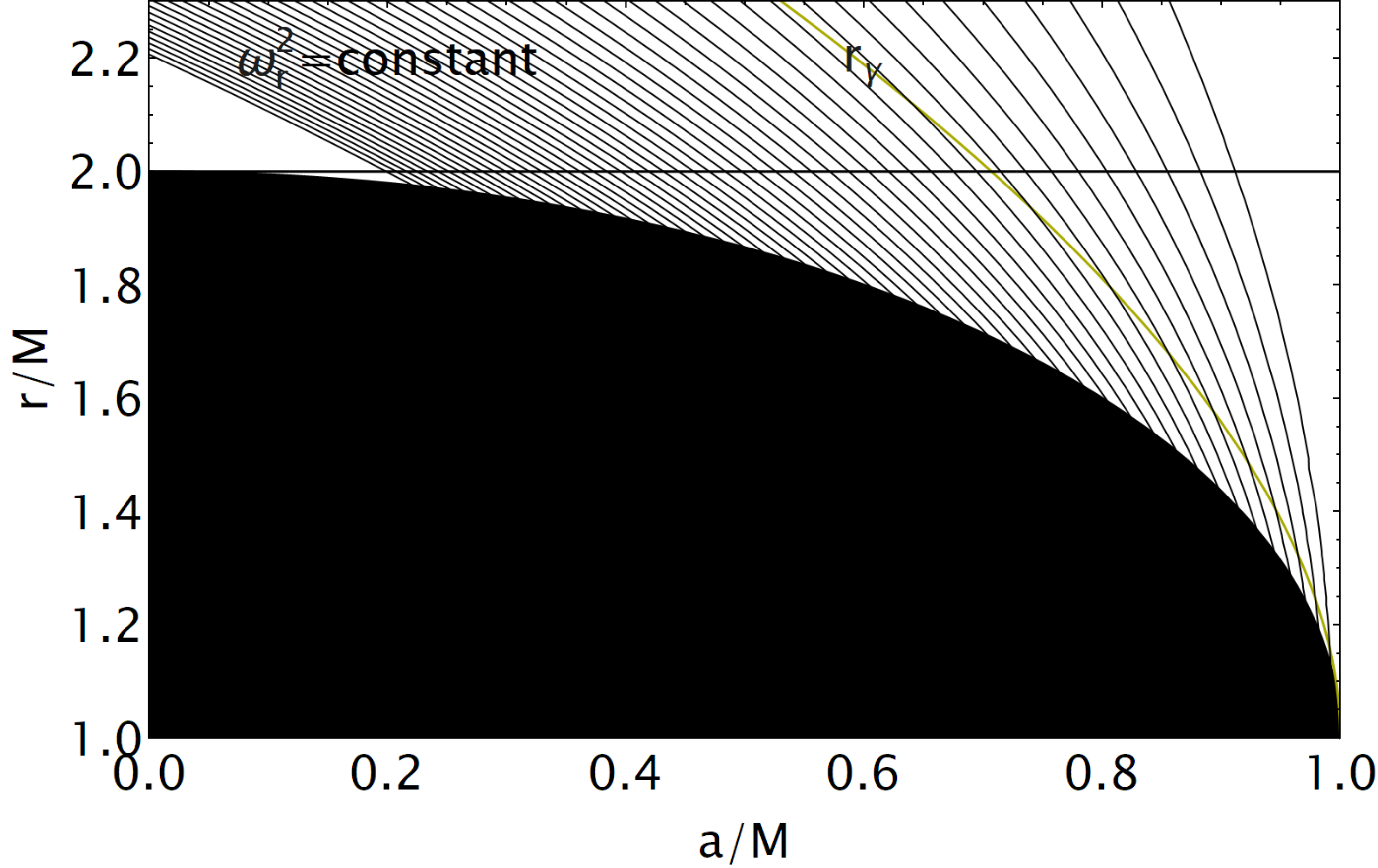}
  \includegraphics[width=5.6cm]{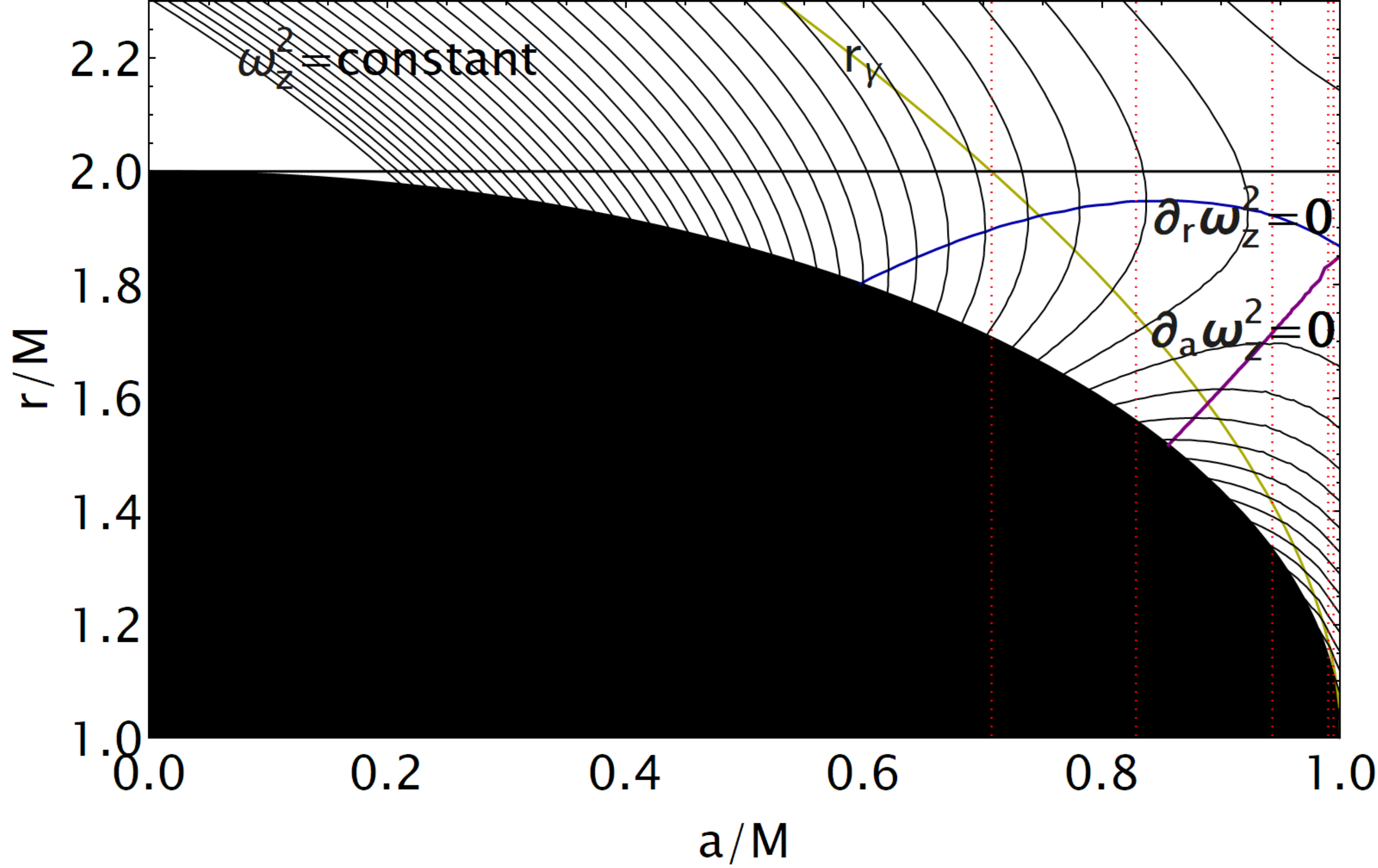}
  \caption{Frequencies  $\omega=$constant in the plane $r/M-a/M$ for  $\omega_k$ (relativistic angular frequency) see also Eqs\il(\ref{Eq:flo-adding}) the radial $\omega_r^2$ and vertical $\omega_z^2$ frequencies of    Eqs\il(\ref{Eq:ome}). Black region is the \textbf{BH} $r<r_+$ as function of the \textbf{BH} spin $a/M$.  The outer ergosurface on the equatorial plane $\sigma=1$ is the line $r_{\epsilon}^+=2M$. $r_{\gamma}$ is the corotating photon orbit.  Right panel solutions of $\partial_r \omega_z^2=0$ and $\partial_a\omega^2_z=0$ are shown. }\label{Fig:Behav}
\end{figure}
\begin{figure}\centering
  % Requires \usepackage{graphicx}
  \includegraphics[width=5.5cm]{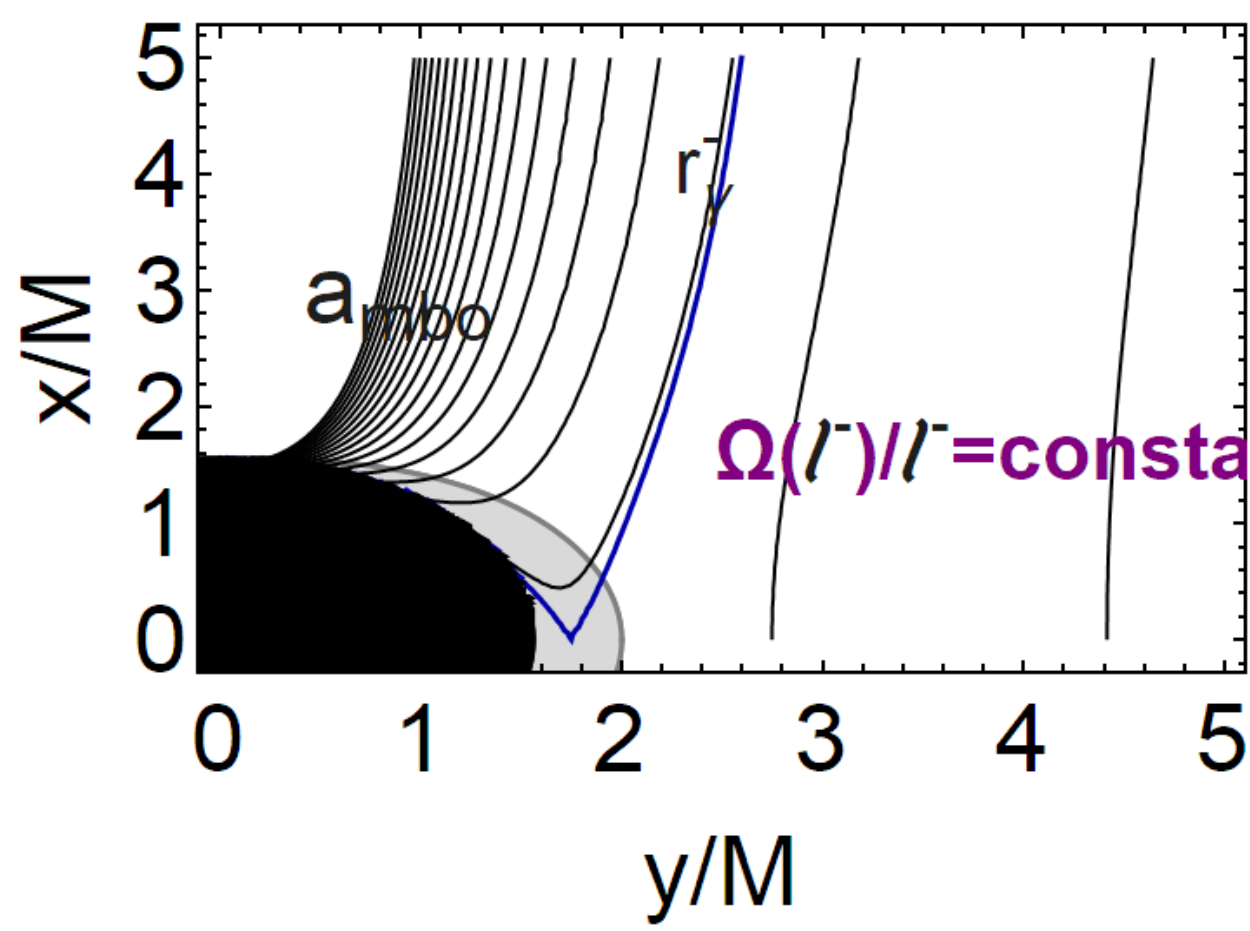}
  \includegraphics[width=5.5cm]{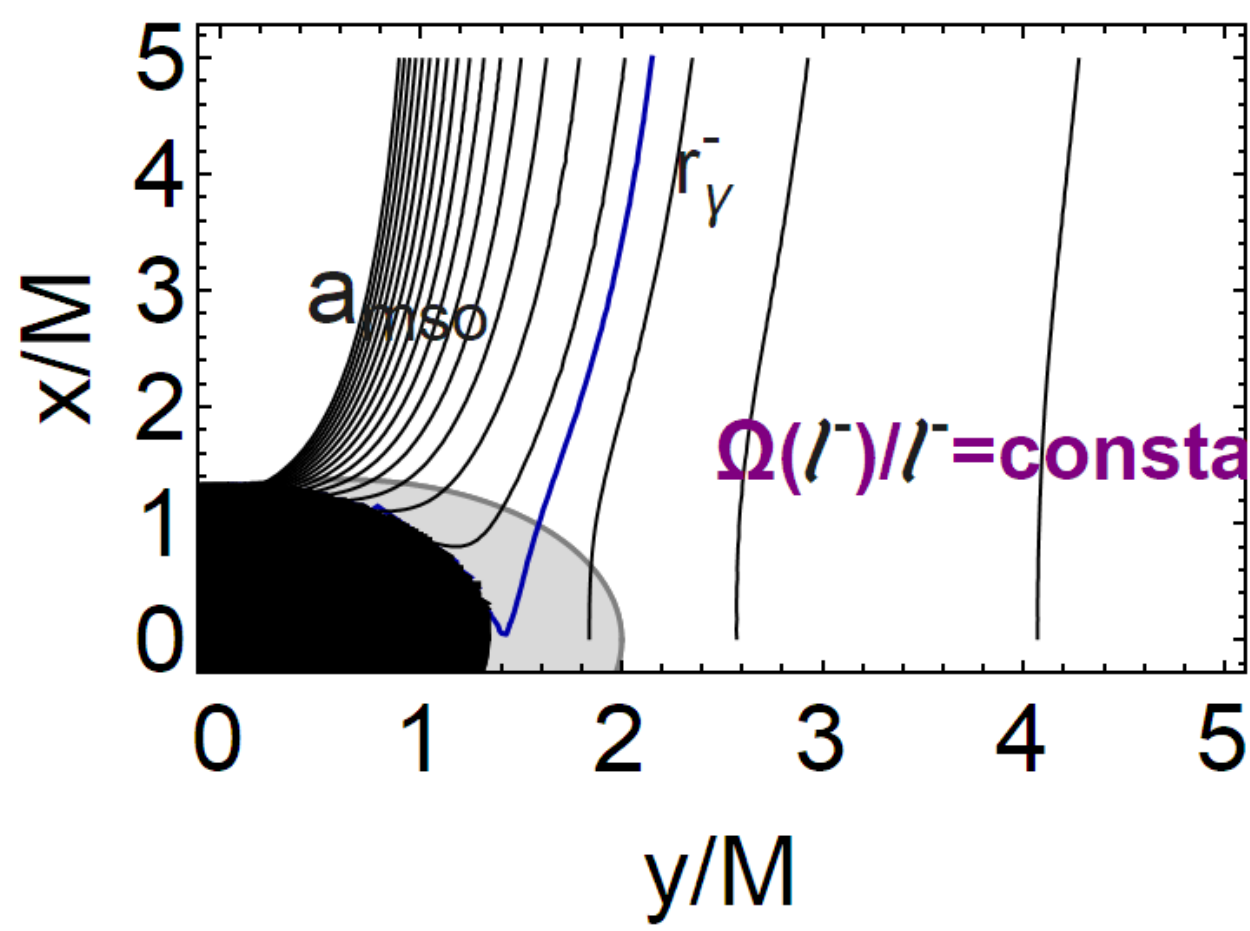}
    \includegraphics[width=5.5cm]{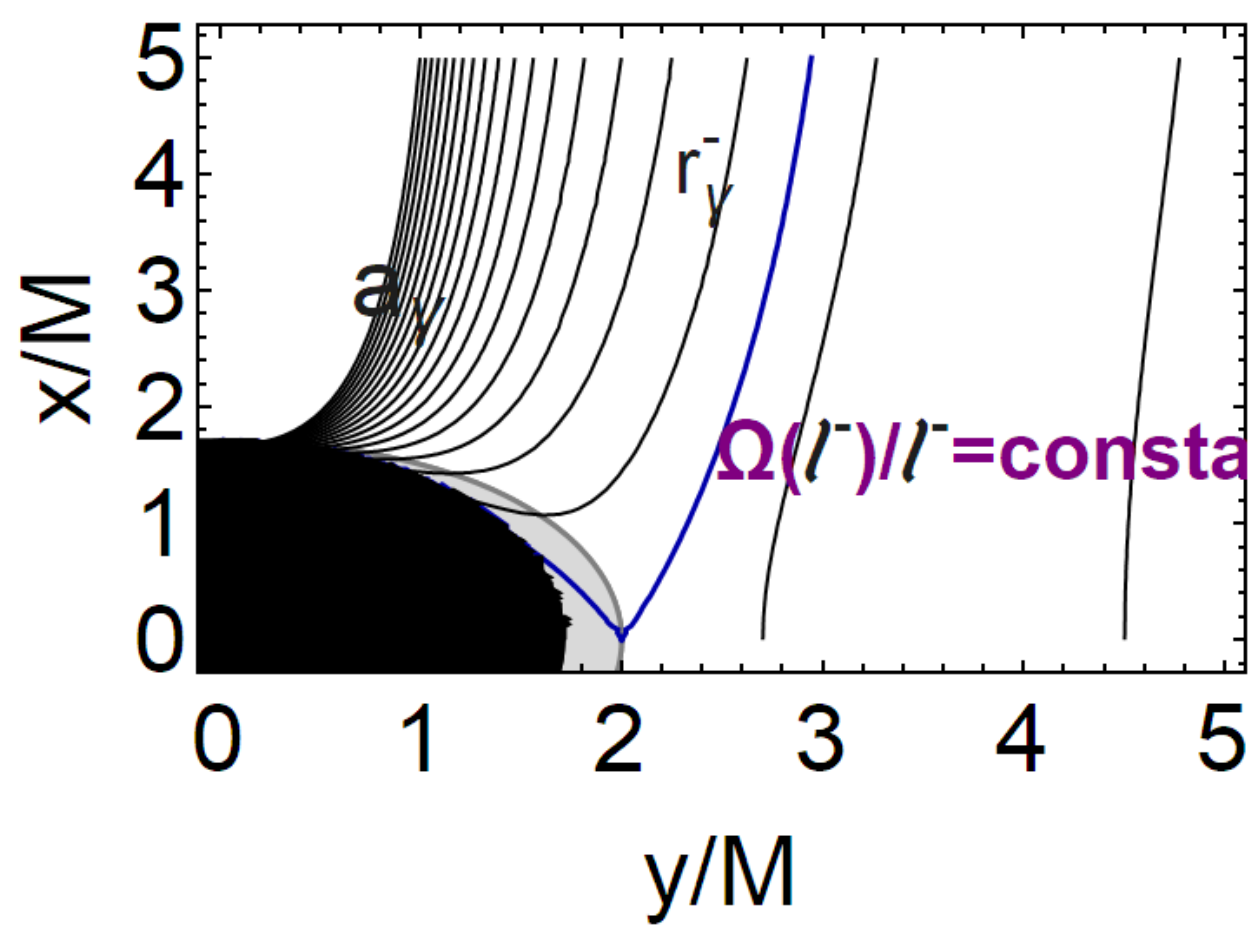}
  \includegraphics[width=5.5cm]{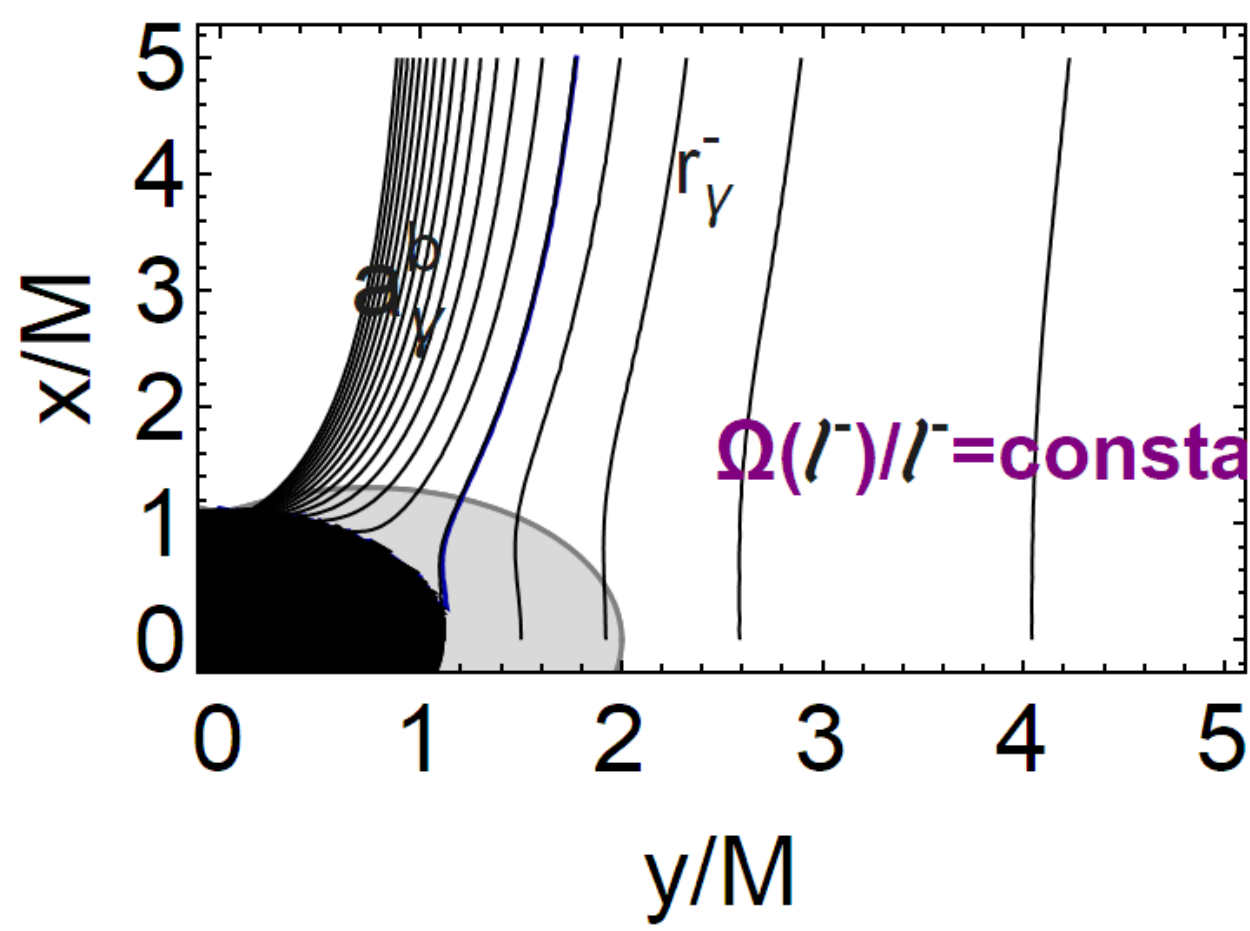}
    \includegraphics[width=5.5cm]{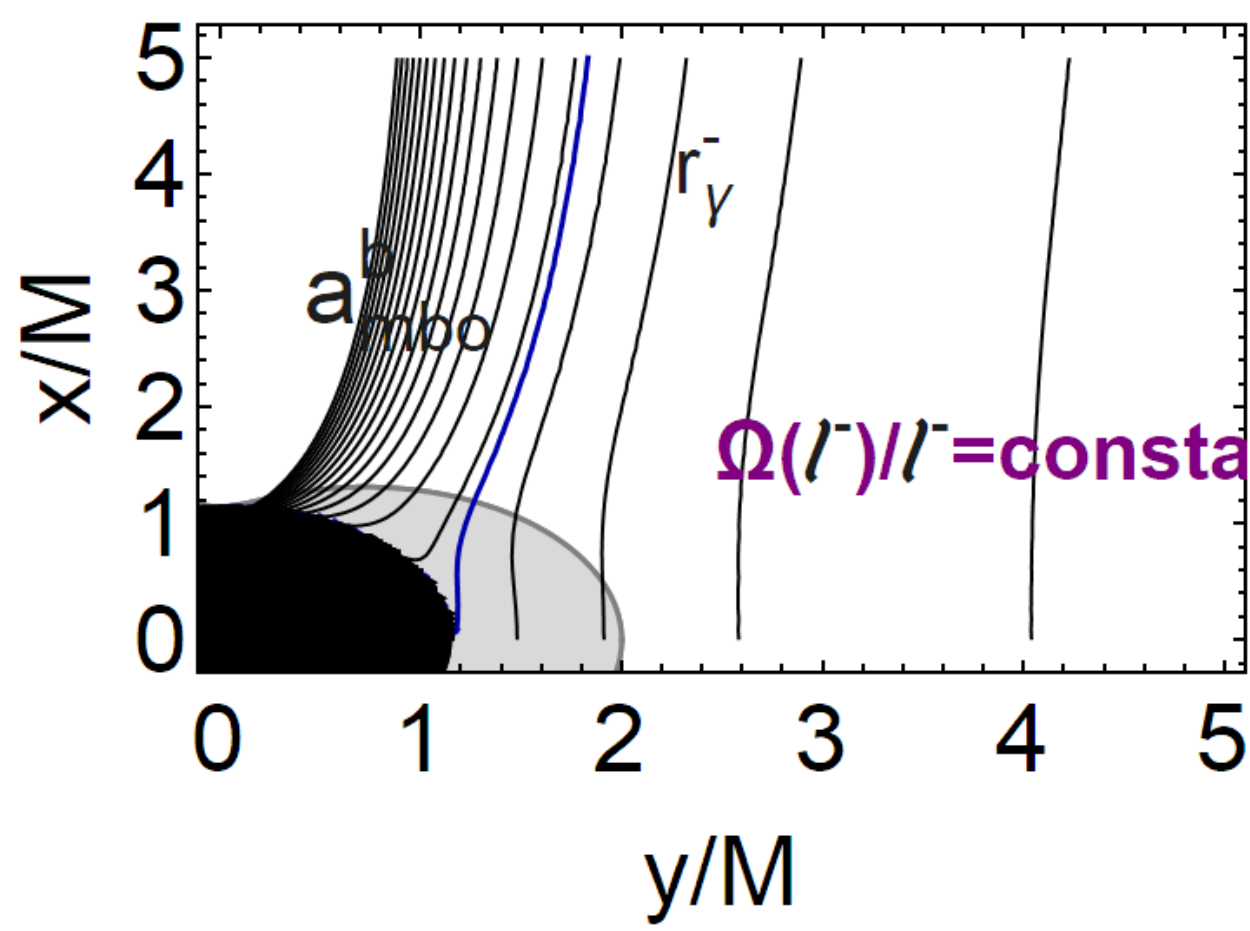}
  \includegraphics[width=5.5cm]{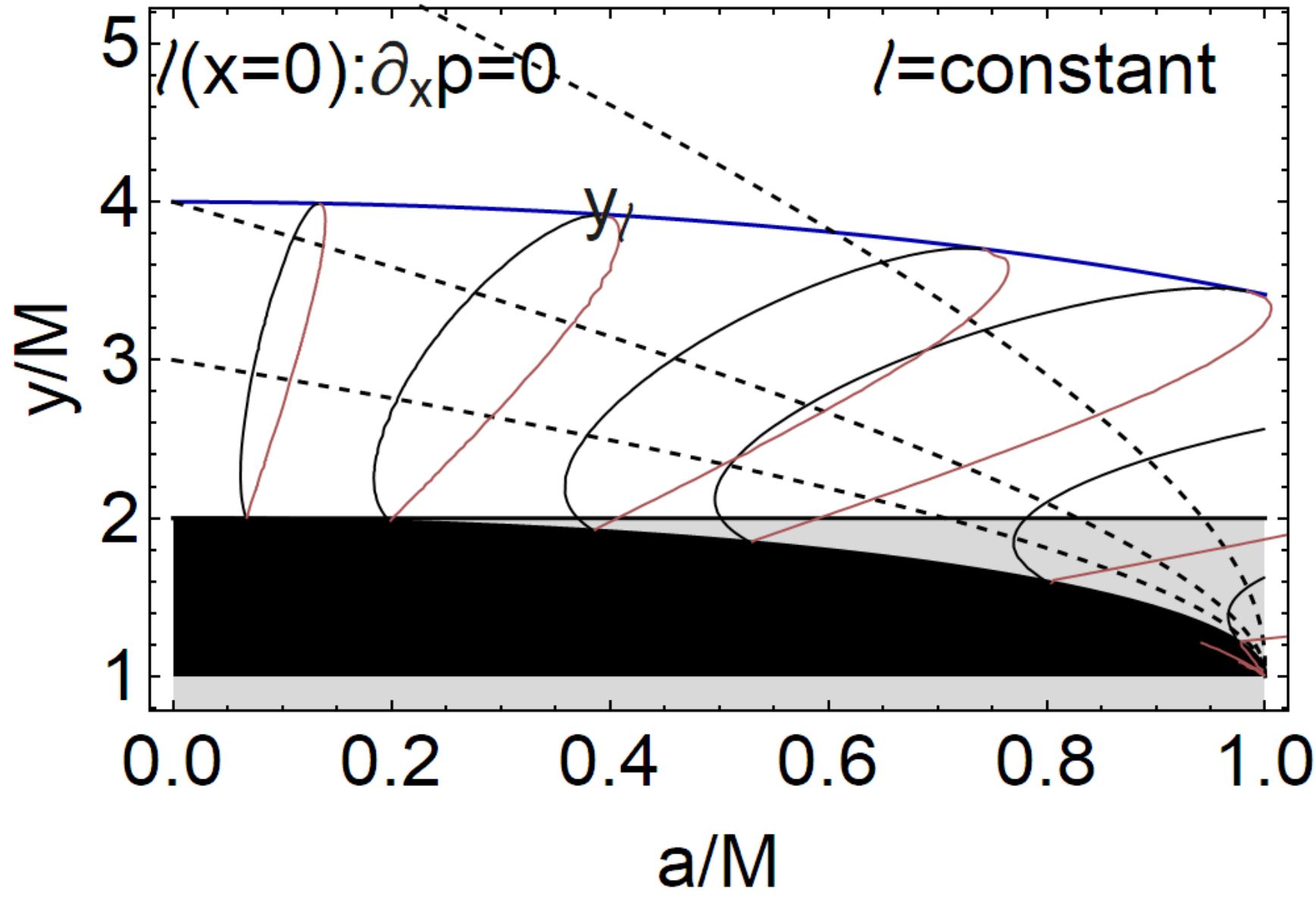}
  \caption{Von Zeipel surfaces  for \textbf{BHs} with spins $\mathbf{A}_{\epsilon}^+\equiv \{a_{mbo},a_{mbo}^b,a_{\gamma},a_{\gamma}^b,a_{mso}\}$--see Figs\il(\ref{Fig:polodefin1}). Black region is $r<r_+$ where $r_+$ is the outer Killing horizon. $r_\gamma$ is the photon orbit, sign $(-)$ stands for a corotating orbit. Gray region is the outer ergosurface.  $\Omega$ is the relativistic angular momentum of the fluids--see Eqs\il(\ref{Eq:flo-adding}), $\ell$ is the specific angular momentum. Right bottom panel: equatorial plane ($\sigma=1$ and $y=r$), study of the disks verticality, vertical gradients of the pressure, solutions $\ell=$constant for $\partial_x V_{eff}=0$ on the axes $x=0$ (equatorial plane, there is ($x=r\cos\theta, y=r\sin\theta)$, on the equatorial plane there is $y\equiv r$), see Figs\il(\ref{Fig:weirplot}). Each curve is for a value of $\ell=$constant, representing the zeros of the curves in   Figs\il(\ref{Fig:weirplot}). Note that the (closed) curves are bounded by $(a_\ell,y_\ell)$ of Eqs\il(\ref{Eq:quesalyl}). Dashed curves are $r_{\gamma}<r_{mbo}<r_{mso}$, photon orbit, marginally bounded orbit and marginally stable orbit.}\label{Fig:collag}
\end{figure}

The investigation of the disk verticality in the ergoregion includes  the exploration of the characteristic frequencies affecting the torus.
The analysis in Figs\il(\ref{Fig:Behav}) shows the vertical frequency $\omega_z$ in the ergoregion considering therefore  the frame dragging for very high spinning \textbf{BH}, as clear by the presence of extreme  of $\omega_z$ as function of the spin $a$ and as function of $r$.
Concerning more generally the tori  dynamical oscillations   it should be said that the spectrum modes are not completely studied.
The axisymmetric, incompressible modes related  to global oscillations  comprise the  epicyclic frequencies.
 The  Keplerian, radial and vertical angular   frequencies take the form \citep{2013A&A...552A..10S}
 \bea\label{Eq:ome}
\omega_k\equiv\frac{1}{a+r^{3/2}};\quad \omega_r^2\equiv\omega_k^2 \left(-\frac{3 a^2}{r^2}+\frac{8 a}{r^{3/2}}-\frac{6}{r}+1\right),\quad
\omega_z^2\equiv\omega_k^2 \left(\frac{3 a^2}{r^2}-\frac{4 a}{r^{3/2}}+1\right).
\eea
The frequencies are treated in various so called geodesic epicyclic models of HF QPOs
\penalty-10000\citep{Straub&Sramkova(2009),2011AA...525A..82S,2013A&A...552A..10S,2017AcA....67..181S,2015A&A...578A..90S,
2005PhRvD..71b4037S,2007A&A...470..401S,2016A&A...586A.130S,2011AA...525A..82S,
2007A&A...463..807S,2016MNRAS.457L..19T,2016ApJ...833..273T,2011AA...531A..59T,2008CQGra..25v5016K,
2017A&A...607A..69K,
2005A&A...437..775T}.  We apply then and the results
are presented in Figs\il(\ref{Fig:Behav})--(\ref{Fig:dedichinPlot}),
 frequencies are  given with respect to the Boyer-Lindquist time coordinate $t$, i.e.  with respect to  an observer at infinity. (Note
the frequencies in  the proper  time $\tau$ can be obtained by multiplying for $u^t\equiv dt/d\tau$.).
 Frequencies on the static limit are  shown in Figs\il(\ref{Fig:plotgeopo}) as function of the central \textbf{BH} dimensionless spin $a/M$.
  The $\Omega=\omega_K$  frequency is the relativistic orbital  angular frequency.  The last two frequencies derive from the first derivative in the radius and angle $\theta$ respectively, of the effective potential of the geodesic motion. Their properties were extensively  studied in \citet{2005A&A...437..775T}
In this model the entropy is constant along the flow. According to the von Zeipel condition, the surfaces of constant angular velocity $\Omega=\omega_z$ and of constant specific angular momentum $\ell$ coincide and  the rotation law $\ell=\ell(\Omega)$ is independent of the equation of state.
 This property of the rotational law is  linked to scale-times of the main physical processes involved in the disks, the accretion mechanism for transporting angular momentum in the disk,  and  the turbulence emergence dependent the magnetic fields and on the vertical structure of the torus \citep{M.A.Abramowicz,Chakrabarti0,
 Chakrabarti,Zanotti:2014haa,Lei:2008ui}.
The von Zeipel condition is also to be considered for the determination of the analytical models for the magnetic field in the magnetized  plasma  fluids\citep{Zanotti:2014haa}.
Von Zeipel surfaces are shown in Figs\il(\ref{Fig:collag}).
    Essentially the application of the so called Boyer condition within the conditions of the  von Zeipel results  reduces to an integrability condition on the Euler equations. In the case of a barotropic fluid,  the differential equation (\ref{E:1a0}) can be integrated reducing  to a gradient of a scalar,  which is possible if and only if
$\ell=\ell(\Omega)$. (Dragged tori are   far from the quasi-spherical conditions).
In general  in these  models of accretion disks the angular momentum of matter in the  disks  is considered to be sufficiently high   for  the centrifugal force to be  a predominant component of the four forces regulating the disks balance (centrifugal, gravitational, pressure and  magnetic forces, and eventually dissipative effects).  This holds  for  situations where the gravitational background is generated by a \textbf{SMBHs} shaping morphology and a great part of dynamics on (micro and macro-scopical scale of) the  disks.
In general    accretion disks, there must be an extended region where there is  $\mp\ell^{\pm}>\mp L^{\pm}$  in the same  orbital region (explicitly  including  counterrotating fluids on Kerr background). This limiting condition is assumed to hold for a general  accretion torus with a general   angular momentum distribution.   The  Bondi quasi-spherical  accretion constitutes a situation  when the condition $|\ell|>|L|$ is not fulfilled. In the  Bondi quasi spherical  accretion, the fluid angular momentum is  everywhere  smaller  than  the  Keplerian  one and
therefore   dynamically  unimportant.
However, the models under examinations here are based on   a full GR onset for each \textbf{RAD} toroid, where in fact there exists
an   extended region where the fluids angular momentum in the torus  is larger or equal  in magnitude than the Keplerian (test particle) angular momentum.
Frequencies $\{\omega_K,\omega_r,\omega_z\}$ of Eq.\il(\ref{Eq:ome}) are then combined the lower and higher frequency  peaks in the particles models of the
\textbf{QPOs} emission.
\subsubsection{On the origin of the QPOs emission   and  the ergoregion}\label{Sec:qpos}
We  discuss the possibility presented in different literature that the origin of the \textbf{QPOs} (Quasi-periodic oscillations) emission  could be  attributable to and immediately traceable to the oscillatory phenomena  reducible to the test particles, considering the case of particles in  the ergoregion\citep{2013A&A...552A..10S}.
This  hypotheses  is  plausible in these models of dragged and partially contained tori as they can be considered  small tori, relatively close to the maximum pressure radius constituting  the torus center. The small  geometric thickness of the disks is  investigated  in  Figs\il(\ref{Fig:raisePlot}),  although  the thickness parameter considered for the construction on   \textbf{QPOs} emission  model based on the  tori geometrical thickness  has been discussed in 	 \citet{2016MNRAS.457L..19T,Straub&Sramkova(2009)}.

\textbf{QPOs} are  observed in the X-ray brightness
  at low (Hz) and high (kHz) frequencies. Observed \textbf{QPO} frequencies are
 associated in many models  with frequencies of orbital and oscillatory motion of the particles from accretion disk matter.
The peaks
of high frequencies  turn to be near the orbital frequency of the
marginally stable orbit extreme, limit for  the inner edge
of thin  disks.
The upper   $\omega_U$ and lower $\omega_L$ frequencies of the peak are in the particles geodesic models expressed  in terms of the radial and vertical oscillations and  the azimuthal  frequency.
Particular attention is given to recognize the emergence of the twin HF \textbf{QPOs} with resonant frequency ratios  $\omega_U/\omega_L={{3:2}}$. The observed ratio of the twin peak frequencies $\approx \mathrm{3:2}$
and  other  resonances  characterized by
ratios  as  $(\mathrm{2:1}, \mathrm{3:1}, \mathrm{5:2})$.
could  explain observed \textbf{QPOs} frequencies (with same
$3:2$ ratio), considering the combinational frequencies,
occurring in
 the inner parts of accretion flow around a black hole \citet{2013A&A...552A..10S}. Therefore we shall consider the resonant frequency ratios {\footnotesize{$\mathbf{R1}=2:1$, $\mathbf{R2}=3:1$, $\mathbf{R3}={3}:{2}$, $\mathbf{R4}={4}:{3}$, $\mathbf{R5}={5}:{4}$}}.

\begin{figure}\centering
  % Requires \usepackage{graphicx}
  \includegraphics[width=8cm]{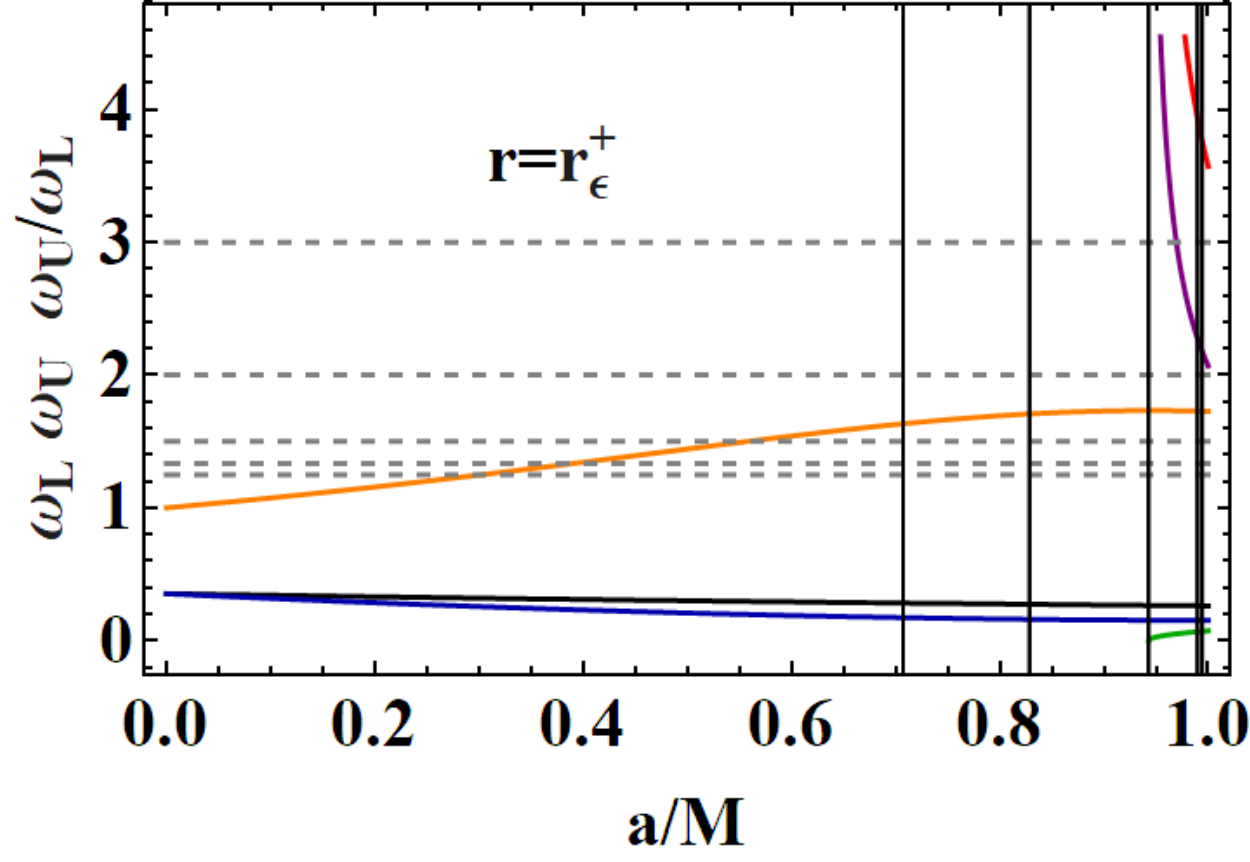}
  \includegraphics[width=8cm]{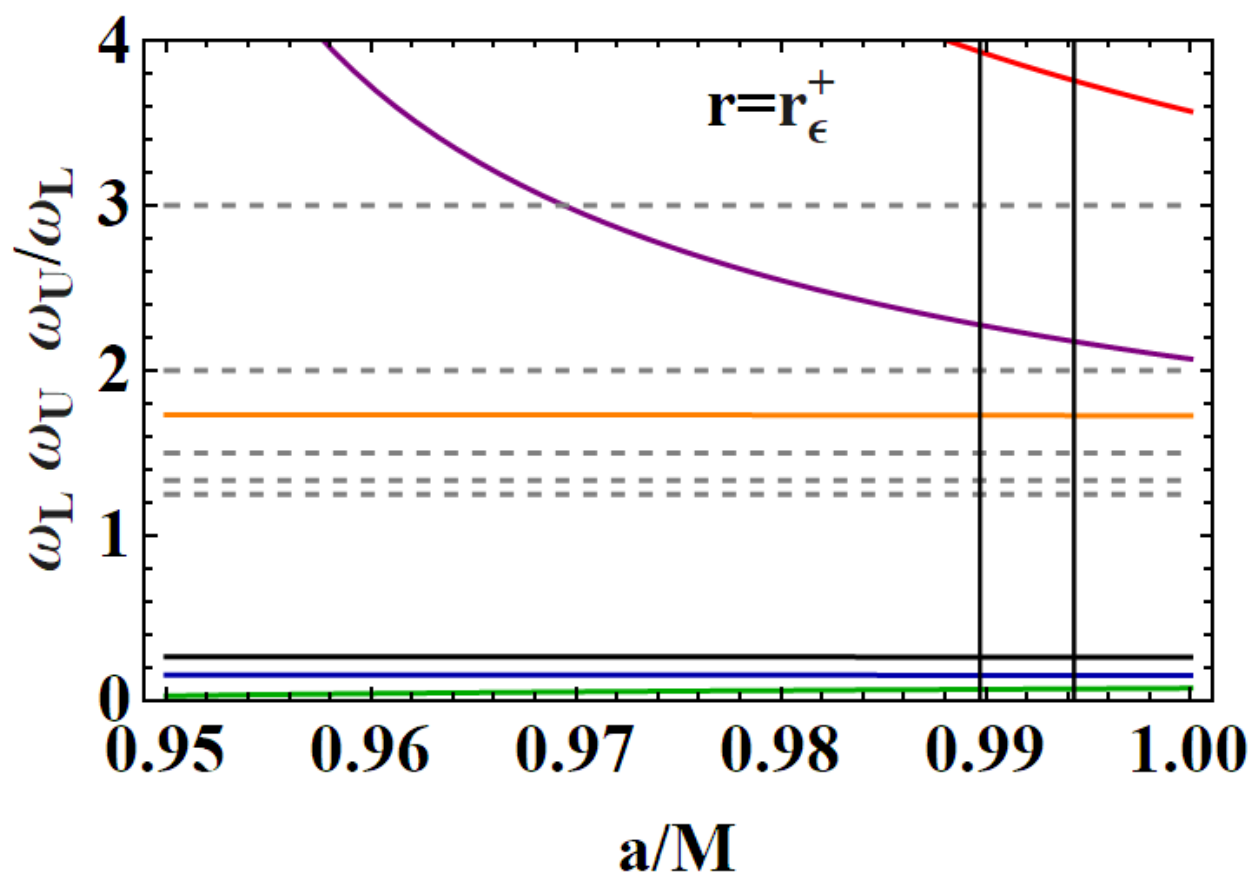}
  \caption{The  \textbf{(RE)} and \textbf{(Kepl)} QPO models frequencies of Eqs\il(\ref{Eq:posittru})  evaluated on the static limit  $r_{\epsilon}^+=2M$ on the equatorial plane  are plotted   as functions of the \textbf{BH} spin $a/M$.  Black lines are spins $\mathbf{A}_{\epsilon}^+\equiv\{a_{mbo},a_{mbo}^b,a_{\gamma},a_{\gamma}^b,a_{mso}\}$-- see also  Figs\il(\ref{Fig:polodefin1}).   Dashed lines are resonant frequency ratios  {{$\mathbf{R1}=2:1$, $\mathbf{R2}=3:1$, $\mathbf{R3}={3}:{2}$, $\mathbf{R4}={4}:{3}$, $\mathbf{R5}={5}:{4}$}}.  Plot of frequencies  $\omega_U$, $\omega_L$  and the ratio $\omega_U/\omega_L$. Right panel is a zoom in the region $a/M\in [0.95,1]$. Frequencies ratios are $\omega_k/\omega_r$  (red curve)   \textbf{(Kepl)} QPO model, $\omega_z/\omega_r$ (purple curve)  \textbf{(RE)} model and $\omega_k/\omega_z$ (orange curve). Frequencies  are $\omega_K$  (black), $\omega_{r}$ (green) and $\omega_z$ (blue).}\label{Fig:plotgeopo}
\end{figure}

\begin{figure}\centering
  % Requires \usepackage{graphicx}
  \includegraphics[width=5.6cm]{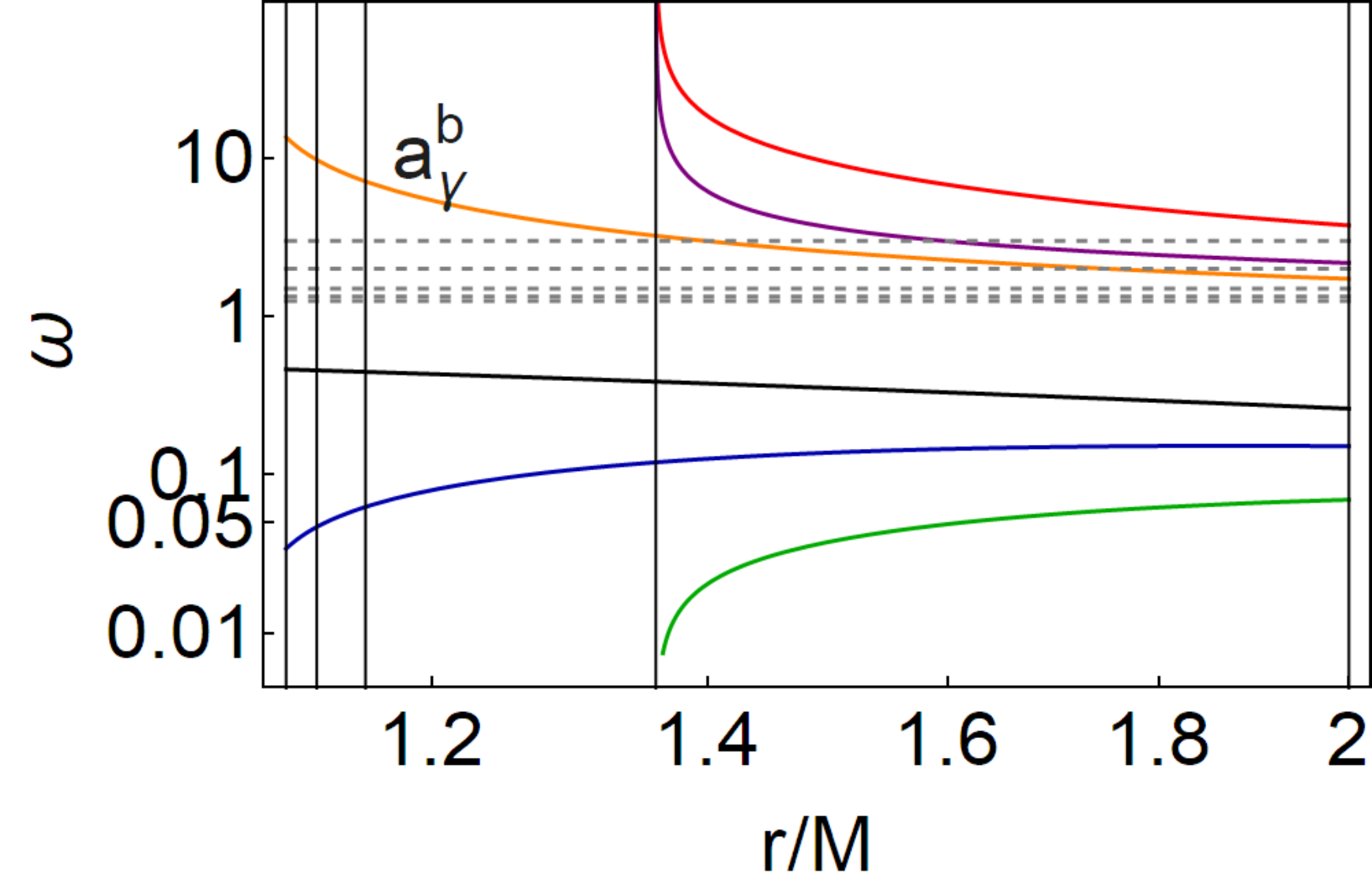}
  \includegraphics[width=5.6cm]{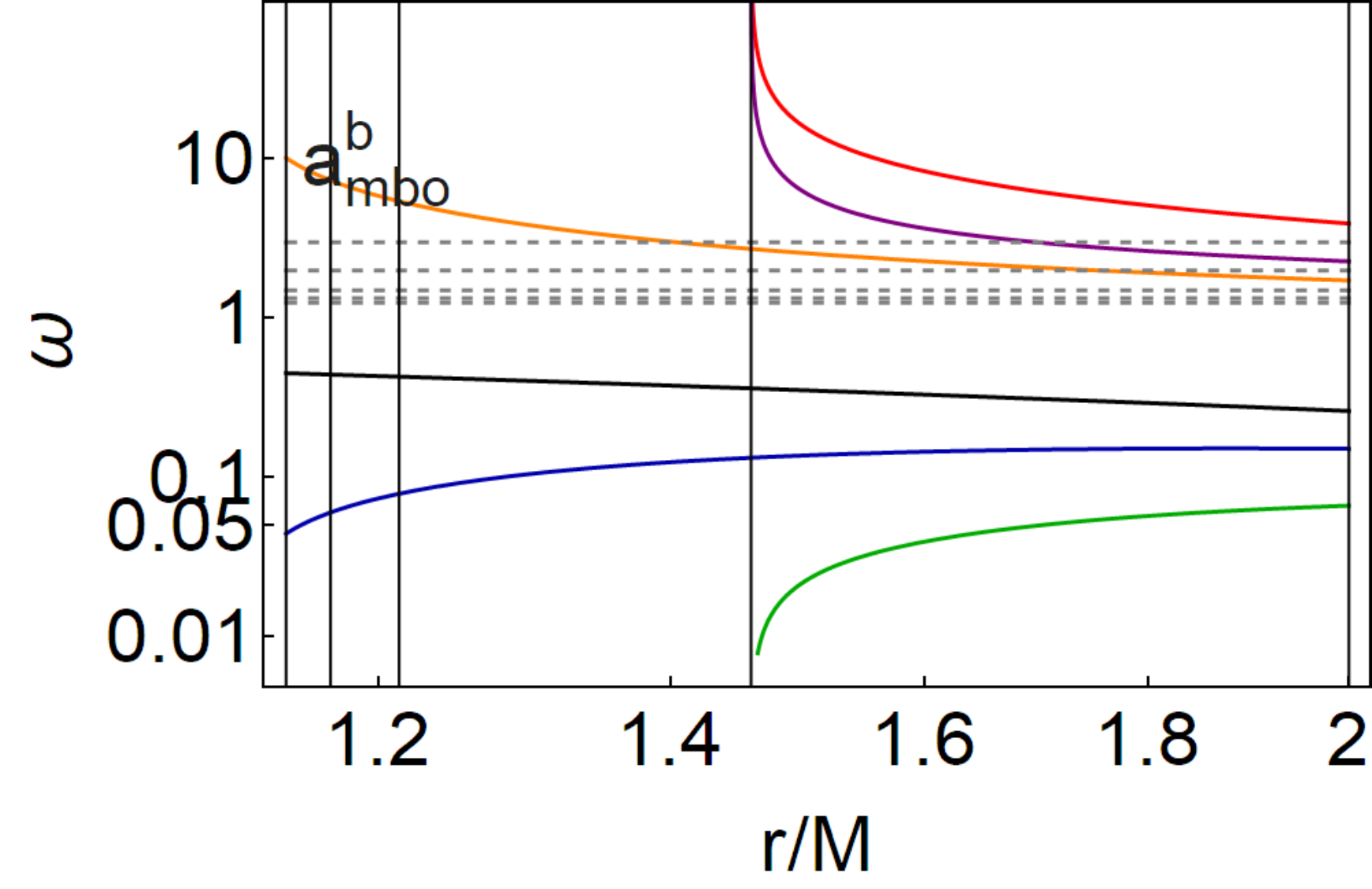}
  \includegraphics[width=5.6cm]{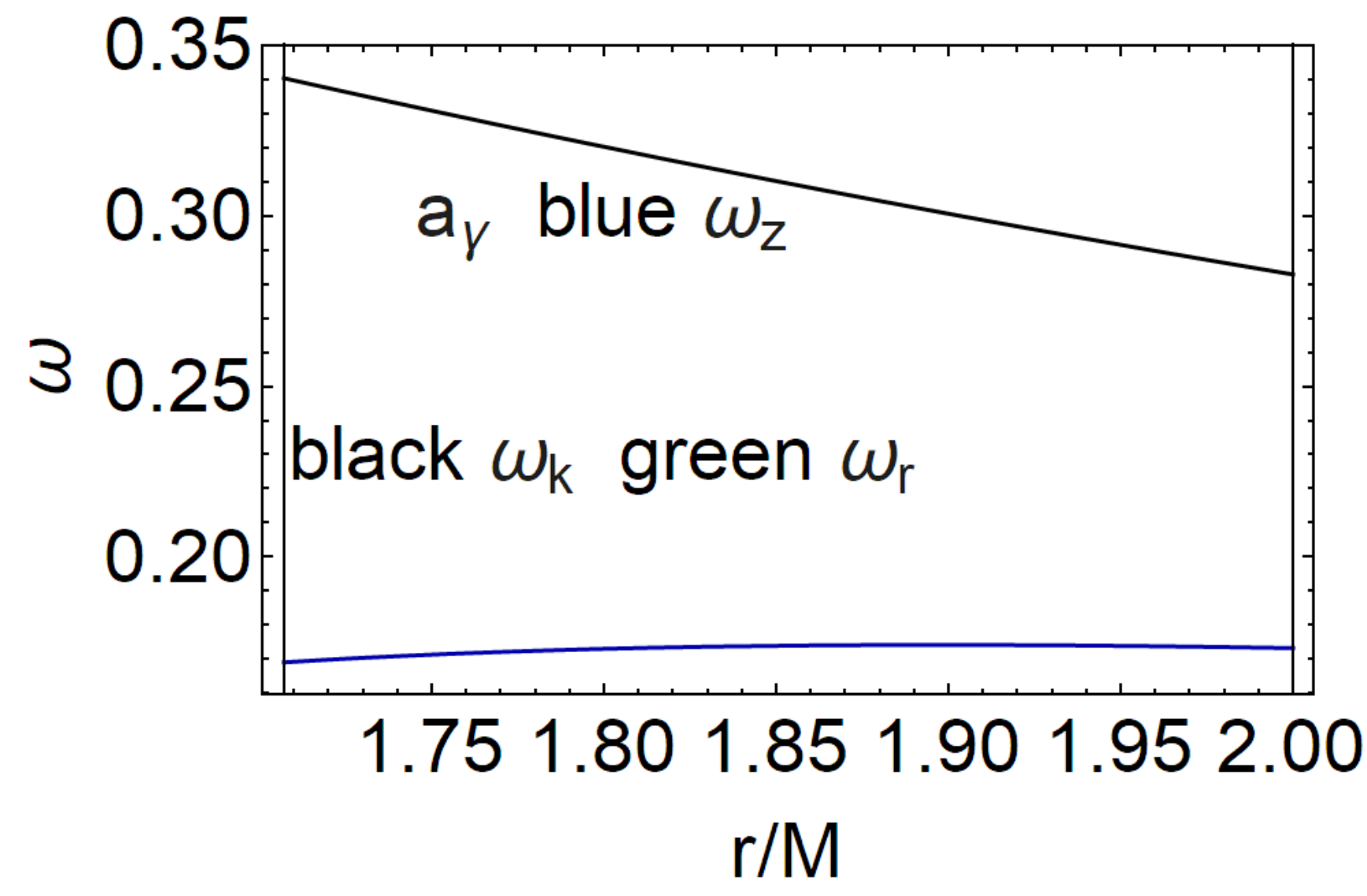}
  \includegraphics[width=5.6cm]{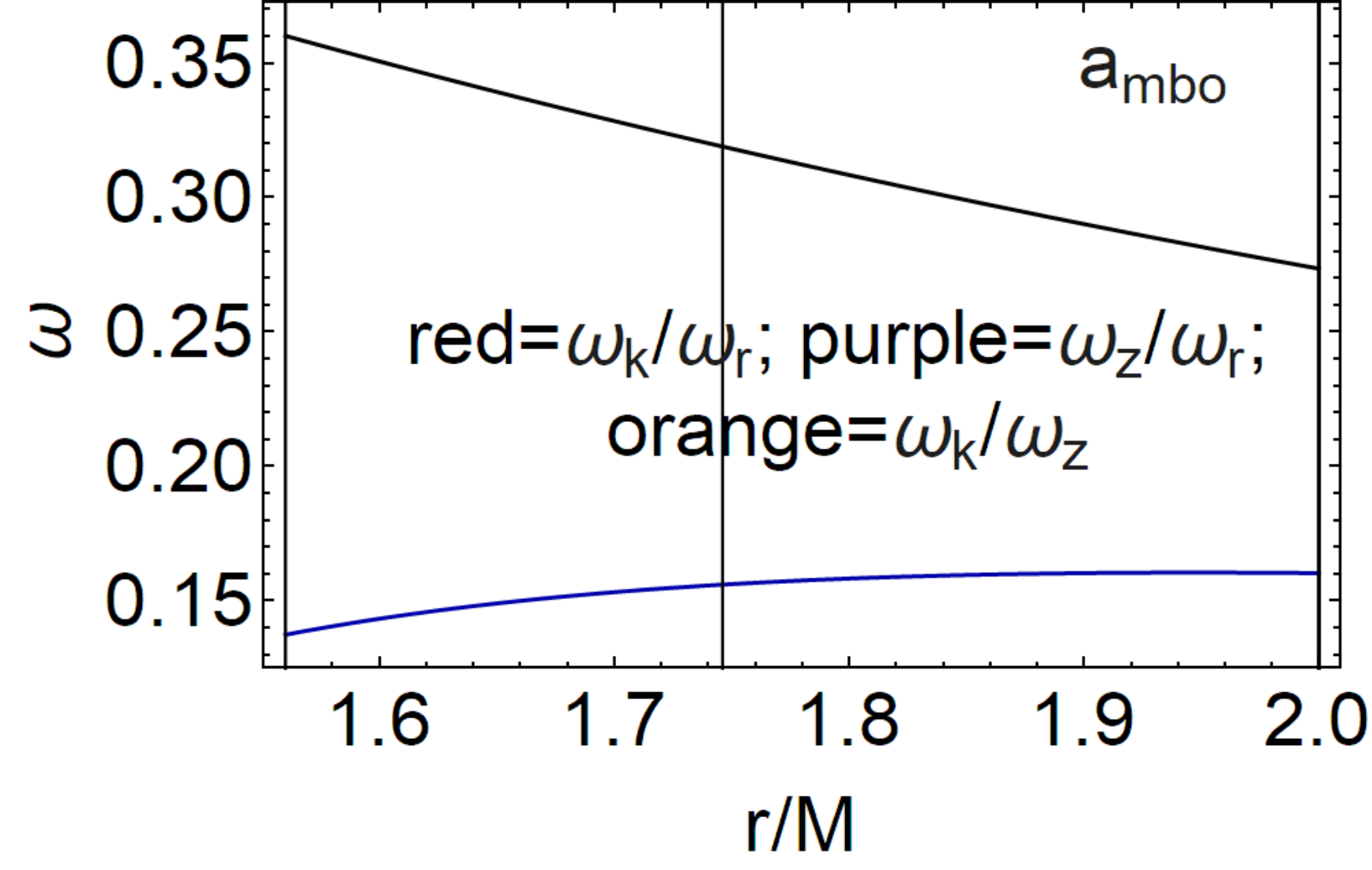}
  \includegraphics[width=5.6cm]{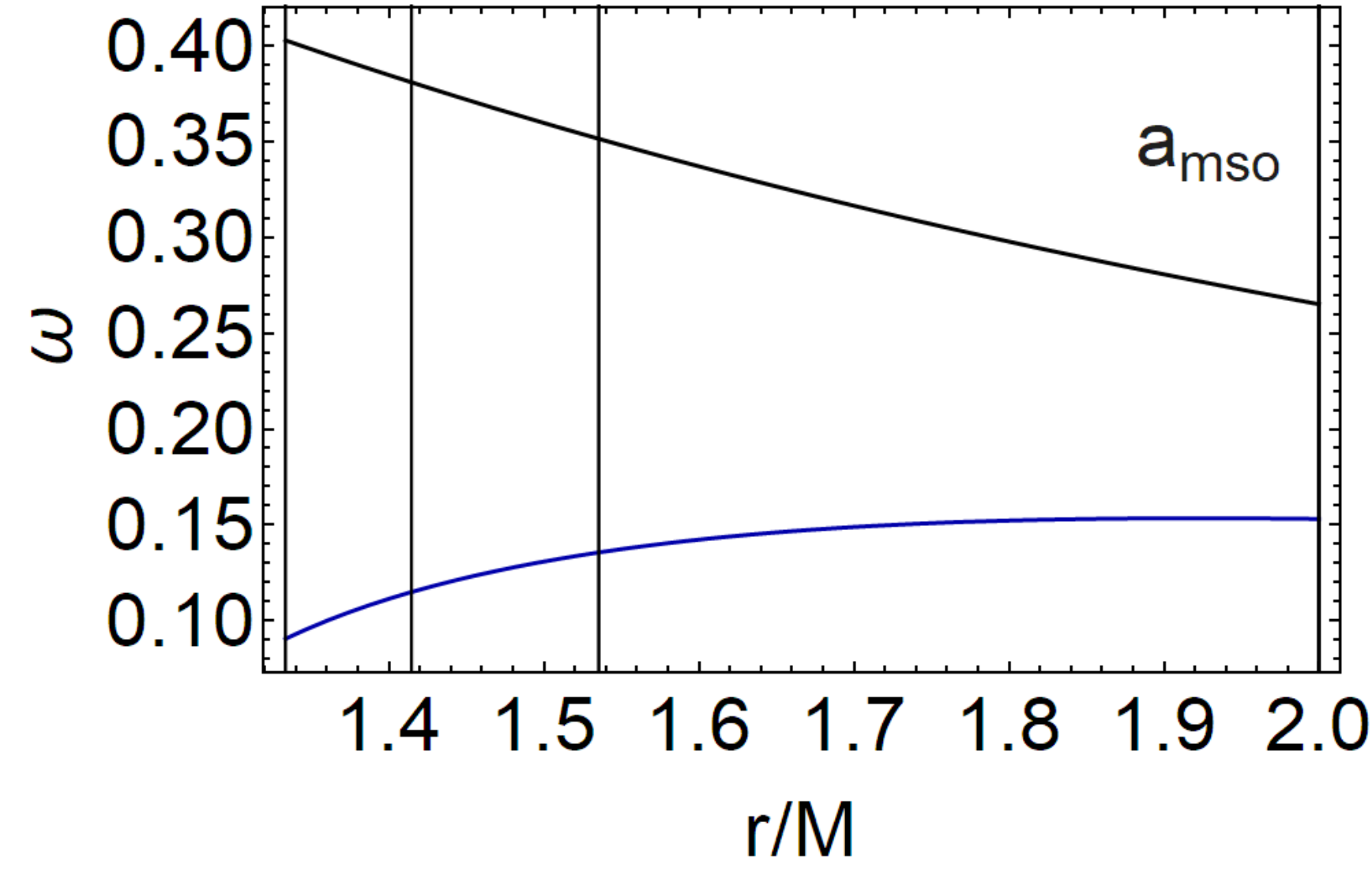}
  \caption{Plot of frequencies  $\omega_U$ $\omega_L$  and the ratio $\omega_U/\omega_L$ as functions of the radius $r/M\in[r_+,r_{\epsilon}^+]$ in the ergoregion, where the outer ergosurface on the equatorial plane is $r_{\epsilon}^+=2M$. Resonant frequency ratios  {{$\mathbf{R1}=2:1$, $\mathbf{R2}=3:1$, $\mathbf{R3}={3}:{2}$, $\mathbf{R4}={4}:{3}$, $\mathbf{R5}={5}:{4}$}} (dashed lines) are also shown. Black lines are  $r_+<r_{\gamma}<r_{mbo}<r_{mso}$, where $r_{\gamma}$ is the marginally circular orbit, $r_{mbo}$ is the marginally bounded orbit, $r_{mso}$  is the marginally stable orbit. Spins $\mathbf{A_{\epsilon}^+}\equiv\{a_{mbo},a_{mbo}^b,a_{\gamma},a_{\gamma}^b,a_{mso}\}$ are represented--see also  Figs\il(\ref{Fig:polodefin1}).   Frequencies  are $\omega_K$  (black), $\omega_{r}$ (green) and $\omega_z$ (blue)  of Eqs\il(\ref{Eq:ome}) as functions of $r/M$--see also Figs\il(\ref{Fig:Behav}). Frequencies ratios are $\omega_k/\omega_r$  (red curve)   \textbf{(Kepl)} QPO model, $\omega_z/\omega_r$ (purple curve)  \textbf{(RE)} model and $\omega_k/\omega_z$ (orange curve). } \label{Fig:dedichinPlot}
\end{figure}

The cusped tori are associated with the cusp  point of  the minimum null-pressure.
In this respect the torus   cusp could be possibly connected to \textbf{QPOs} emission.
\begin{figure}\centering
  % Requires \usepackage{graphicx}
  \includegraphics[width=5.6cm]{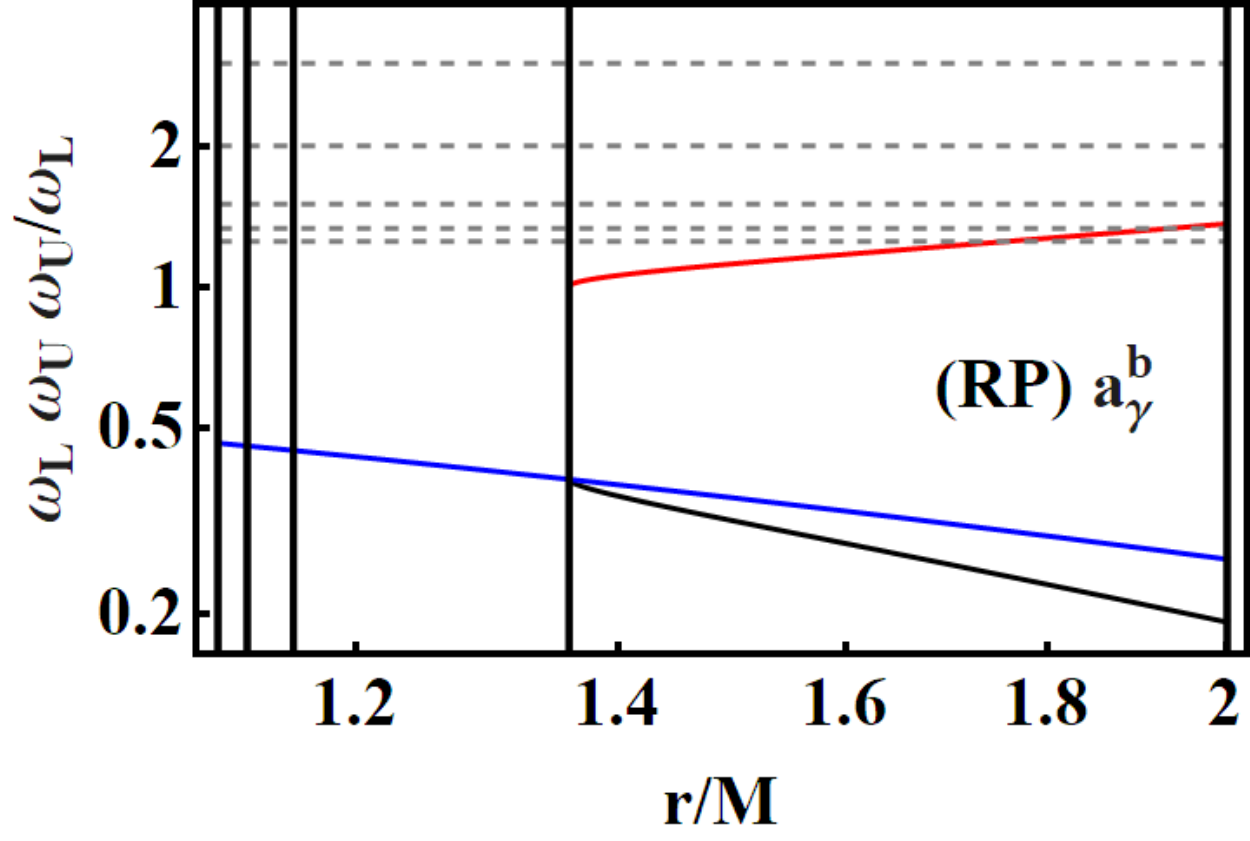}
  \includegraphics[width=5.6cm]{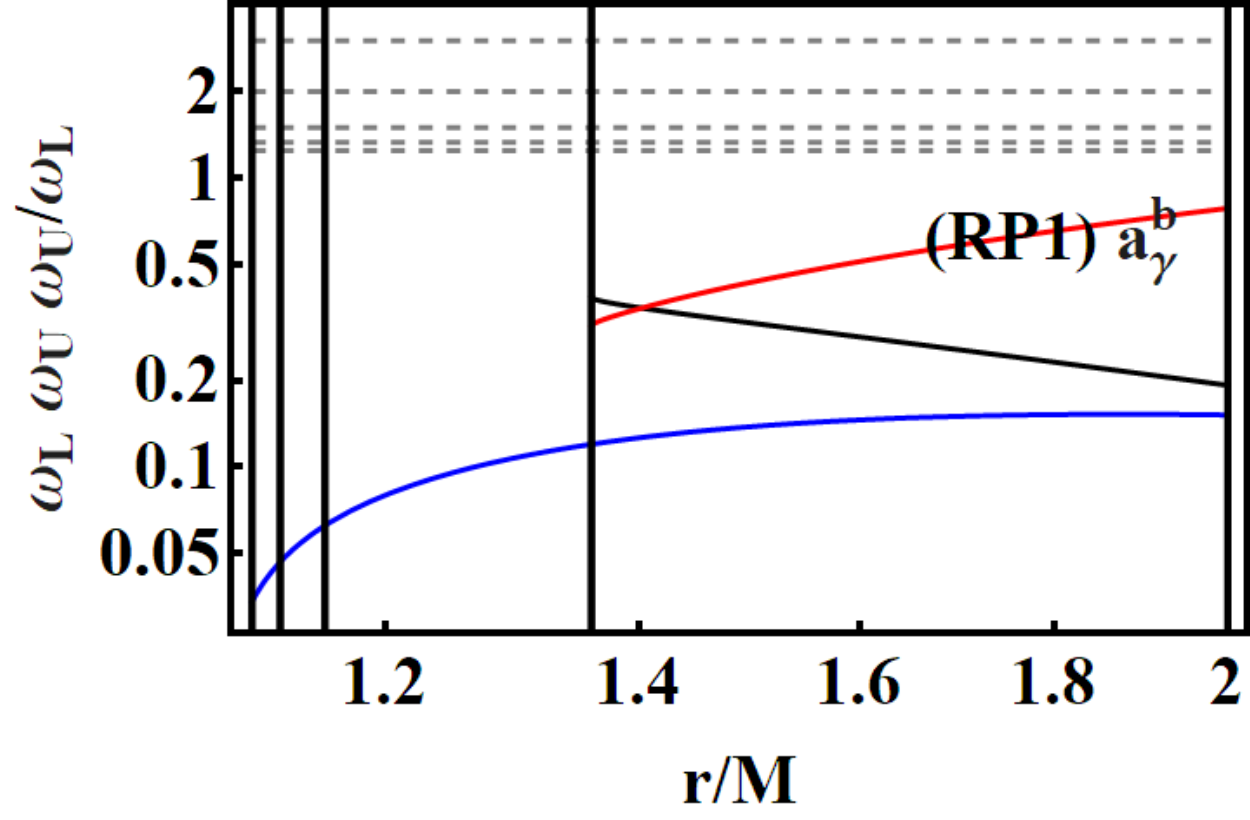}
  \includegraphics[width=5.6cm]{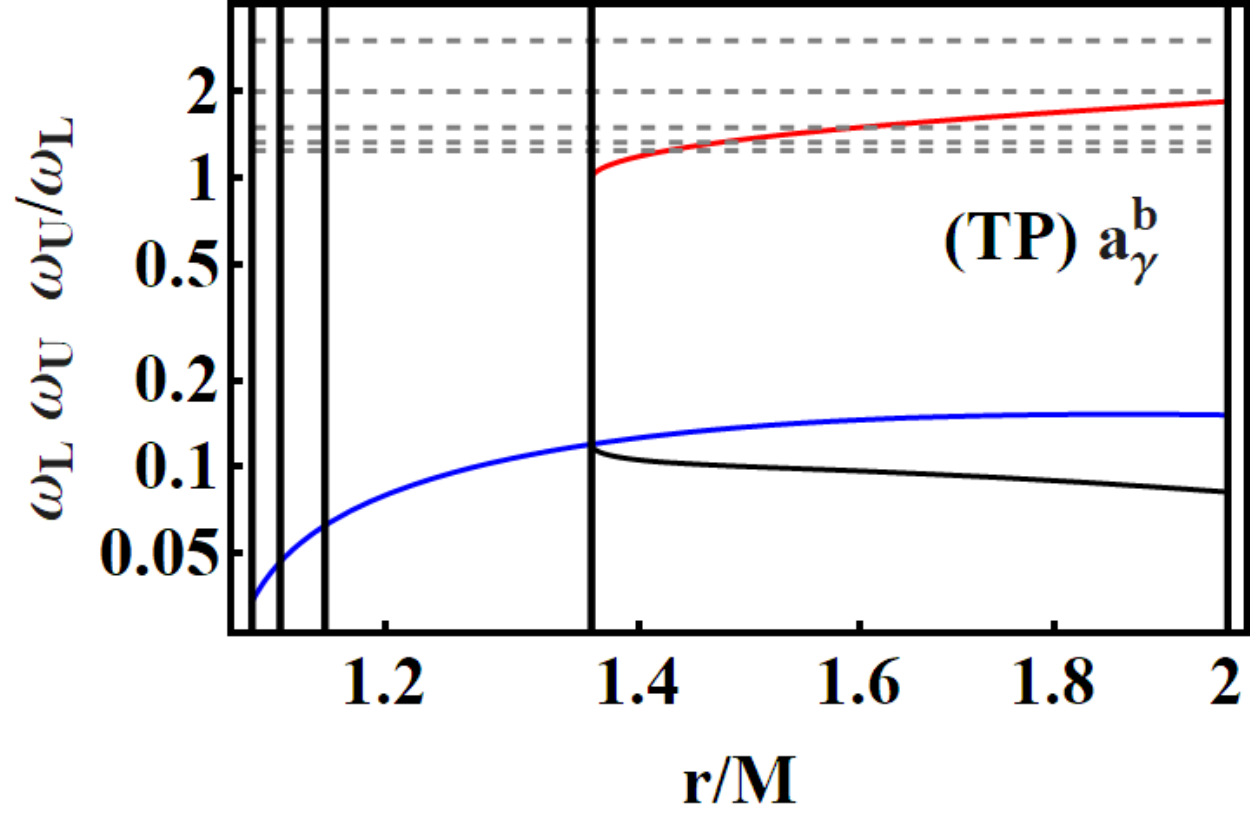}
  \includegraphics[width=5.6cm]{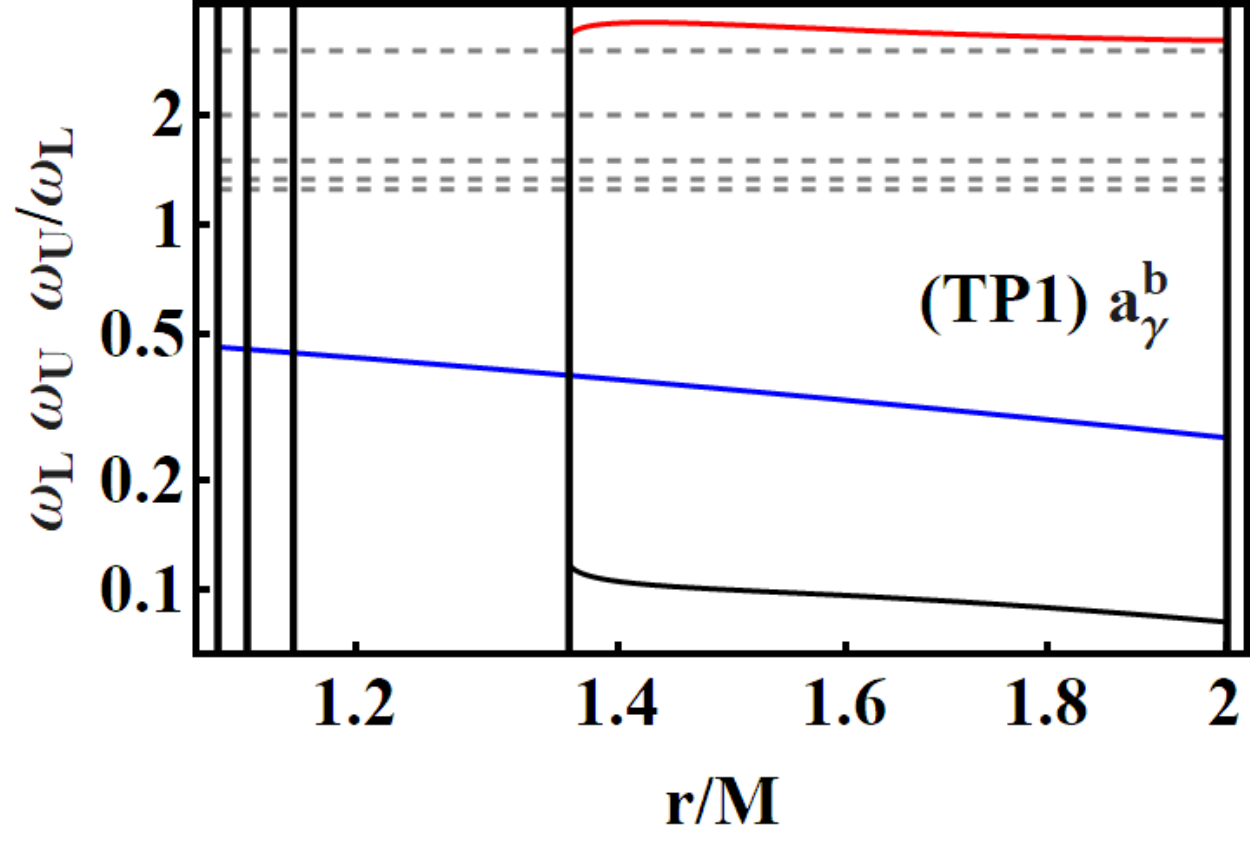}
    \includegraphics[width=5.6cm]{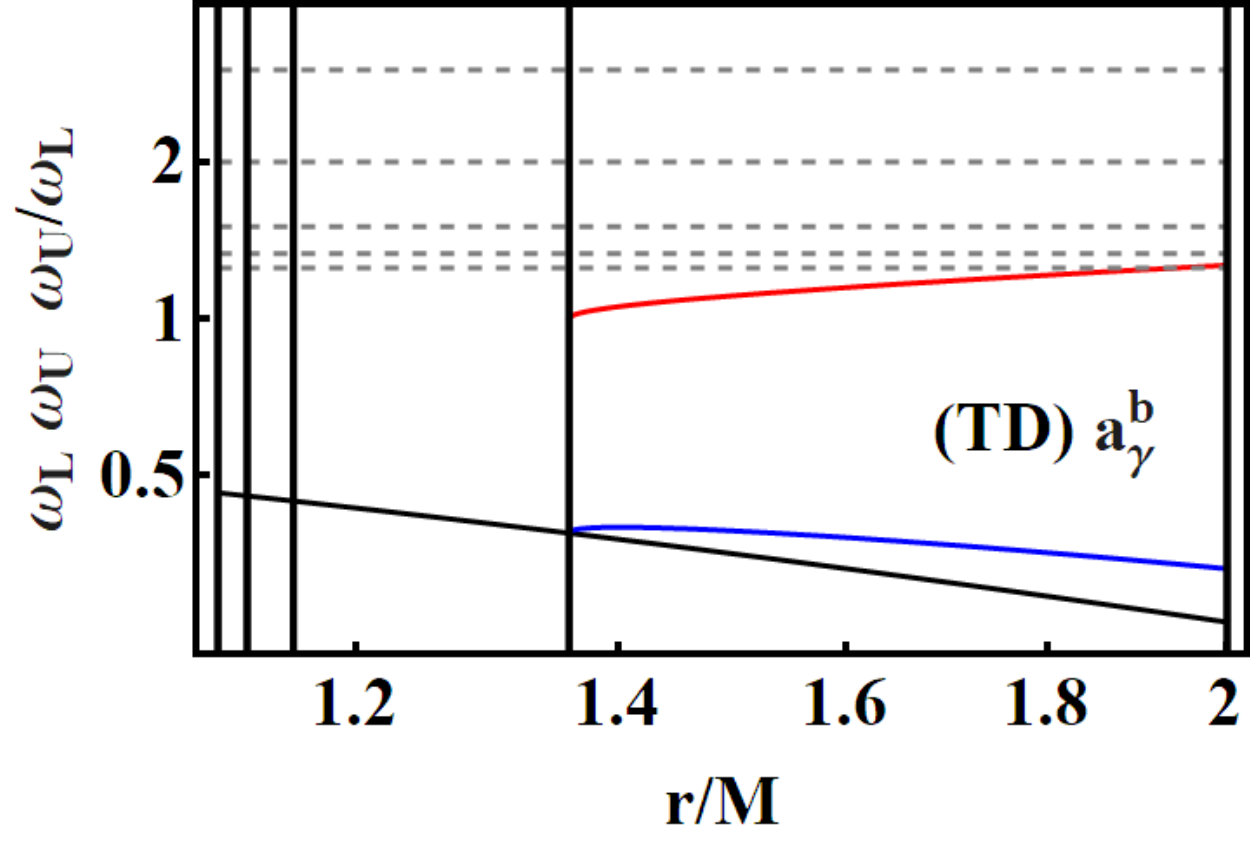}
  \includegraphics[width=5.6cm]{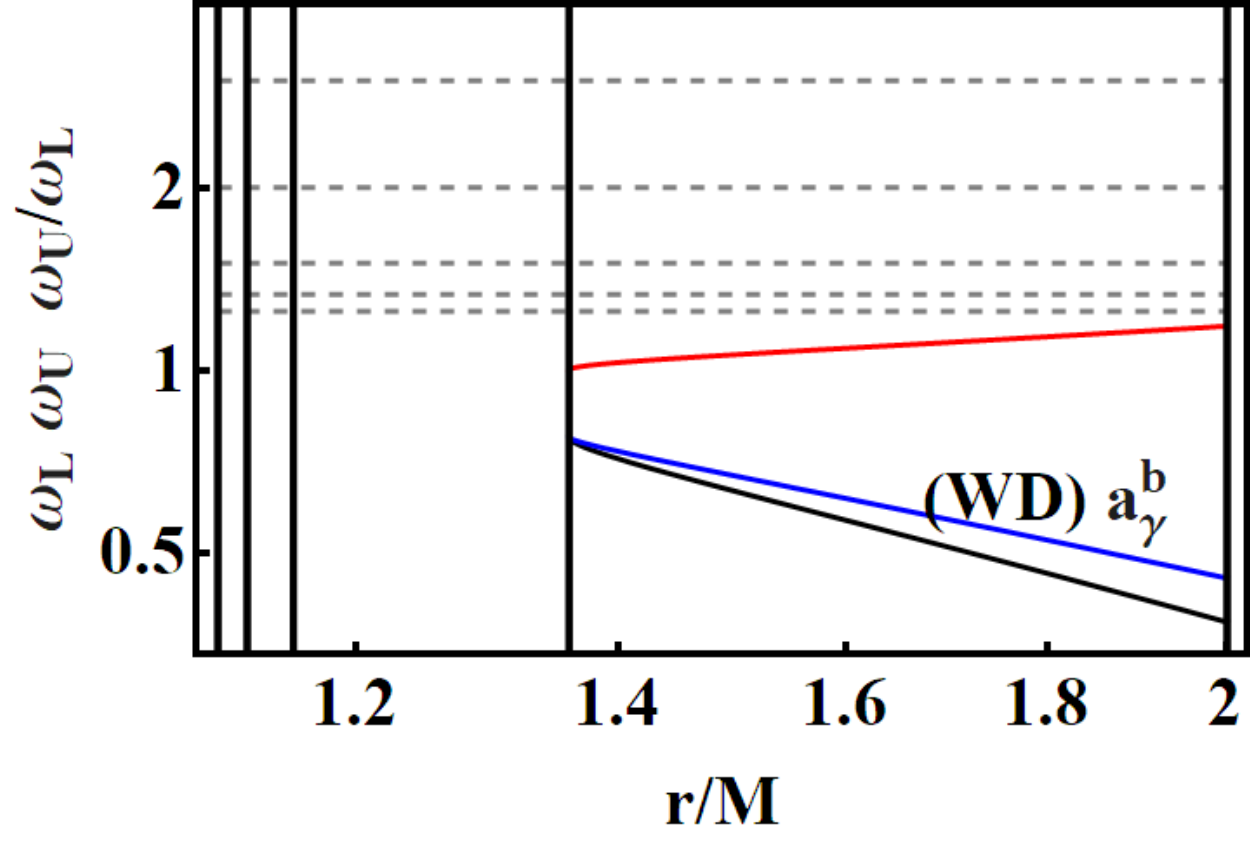}
   \includegraphics[width=5.6cm]{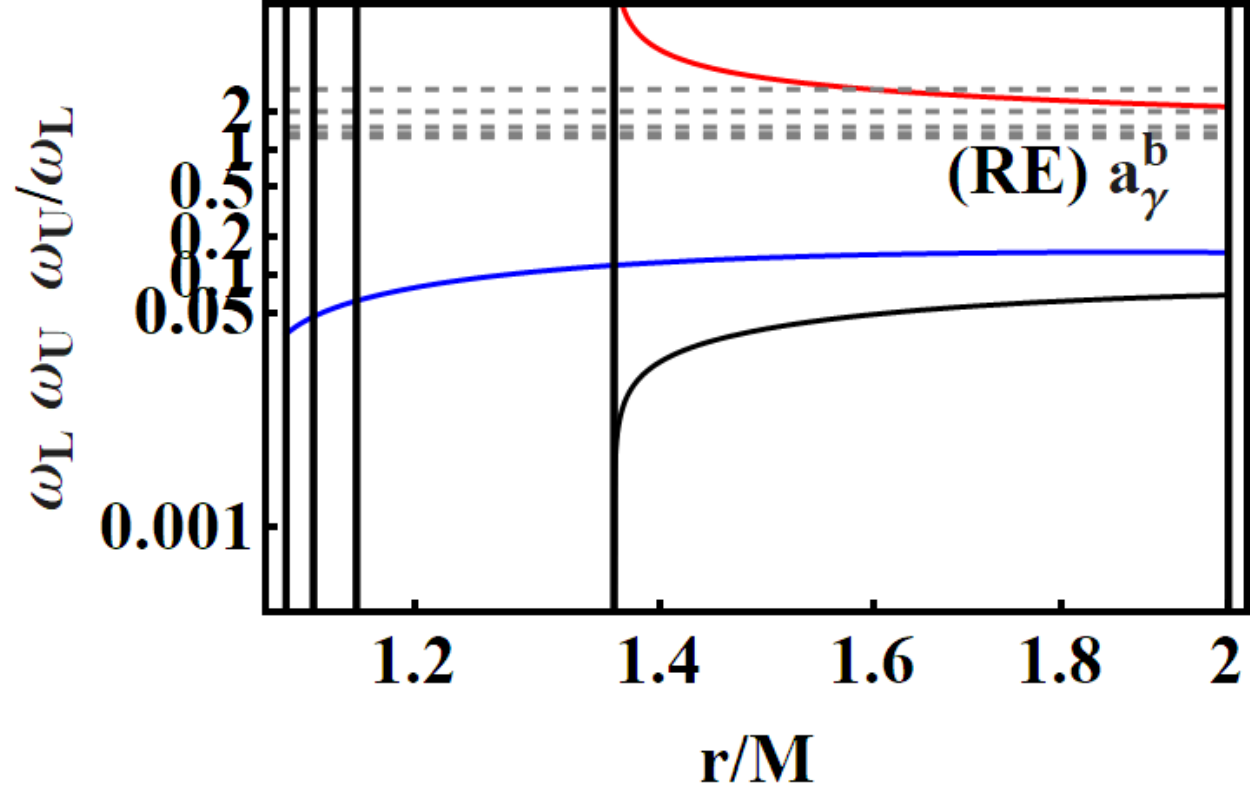}
  \includegraphics[width=5.6cm]{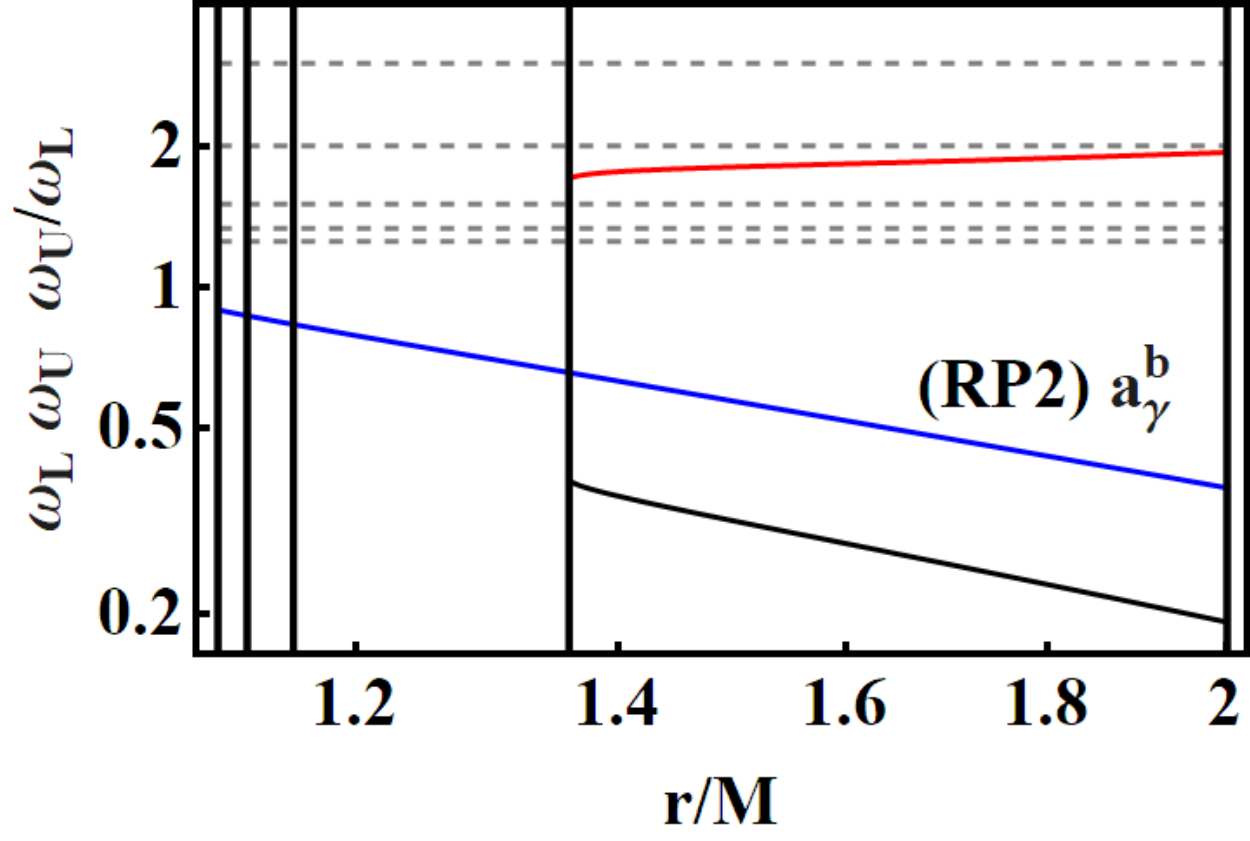}
  \caption{
  Plot of frequencies  $\omega_U$ (blue) $\omega_L$ (black)  and the ratio $\omega_U/\omega_L$ (red curve) as functions of the radius $r/M\in[r_+,r_{\epsilon}^+]$ in the ergoregion, in the \textbf{BH} spacetime with spin $a_{\gamma}^b$--see also  Figs\il(\ref{Fig:polodefin1}). Where the outer ergosurface on the equatorial plane is $r_{\epsilon}^+=2M$.
   Oscillation models  { {$\{ ( \mathbf{WD}),( \mathbf{TD}),(\mathbf{TP}),( \mathbf{TP1}),( \mathbf{RP}),( \mathbf{RP1}),( \mathbf{RP2}),
 (\mathbf{RE}) \}$} }
 of Eqs\il(\ref{Eq:posittru}) are studied. Resonant frequency ratios {  $\mathbf{R1}=2:1$, $\mathbf{R2}=3:1$, $\mathbf{R3}=3:2$, $\mathbf{R4}={4}:{3}$, $\mathbf{R5}={5}:{4}$ } (dashed lines) are also shown. Black lines are  $r_+<r_{\gamma}<r_{mbo}<r_{mso}$, where $r_{\gamma}$ is the marginally circular orbit, $r_{mbo}$ is the marginally bounded orbit, $r_{mso} $  is the marginally stable orbit.
   }\label{Fig:apopof4}
\end{figure}
The   individual \textbf{QPO} models frequencies considered here are functions of three fundamental frequencies of
perturbed circular geodesic motion and are:
\bea&&\label{Eq:posittru}
\textbf{RE:}\quad\left(\omega_L=\omega_r ,\quad
\omega_U=\omega_z\right),\quad \textbf{Kepl:}\quad\left(\omega_L=\omega_r,\quad
\omega_U=\omega_K\right),\\&&\nonumber
\textbf{RP:}\quad \left(\omega_L= \omega_K-\omega_r,\quad
 \omega_U= \omega_K\right),\quad
\textbf{RP1:}\quad \left(\omega_L= \omega_K-\omega_r,\quad
 \omega_U= \omega_z\right)
 \\
 &&\textbf{RP2:}\quad \left(\omega_L= \omega_K-\omega_r,\quad \omega_U=2 \omega_K-\omega_z\right);
 \\&&\nonumber
 \textbf{TP:}\quad  \left(\omega_L= \omega_z-\omega_r,\quad
 \omega_U= \omega_z\right)
,\quad
 \textbf{TP1:}\quad \left(\omega_L= \omega_z-\omega_r,\quad
 \omega_U= \omega_K\right)
 \\&&\nonumber
 \textbf{TD:}\quad \left(
 \omega_U=\omega_K ,\quad
 \omega_U=\omega_K+\omega_r\right)
,\quad
\textbf{WD:}\quad  \left(\omega_L=2(\omega_K-\omega_r),\quad
 \omega_U=2\omega_K-\omega_r\right)
\eea
\textbf{(RE) } and \textbf{(Kepl)}  models are investigated in Figs\il(\ref{Fig:dedichinPlot}).
Frequencies on the static limit are in Figs\il(\ref{Fig:plotgeopo}).
The radial frequency is well defined,  $\omega_r^2 >0$,  for  $r >
 r_ {mso}$, where it vanishes, determining  orbits stability for radial oscillations
 around the point of the perturbed particle.
  We  consider this problem and  various oscillation models
as in Eqs\il(\ref{Eq:posittru}) in the ergoregion, where $r_{mbo}\in[r_+,  r_ {\epsilon}^+] $  for $a\geq a_ {mso} $, including geometry $a_{\gamma}^b$ and $a_{mbo}^b$.
\begin{figure}\centering
  % Requires \usepackage{graphicx}
  \includegraphics[width=5.6cm]{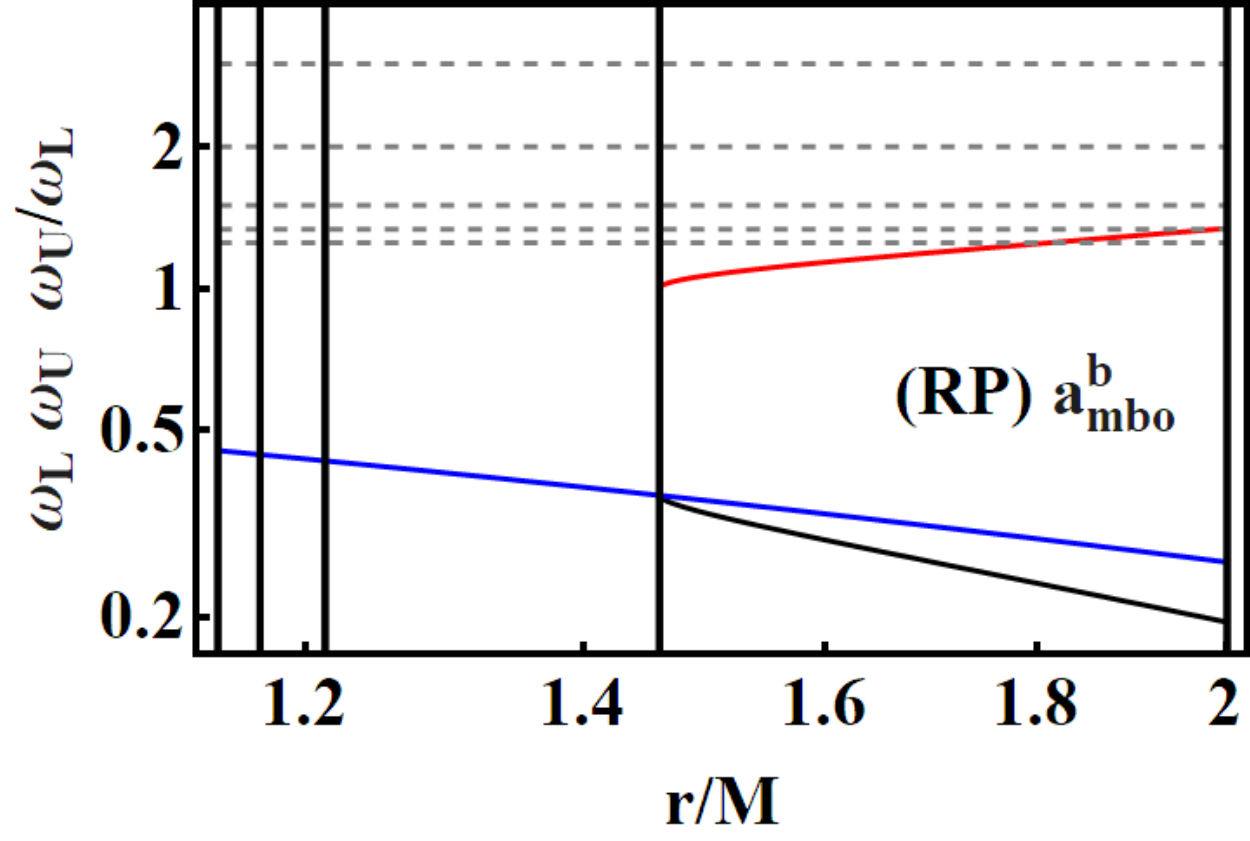}
  \includegraphics[width=5.6cm]{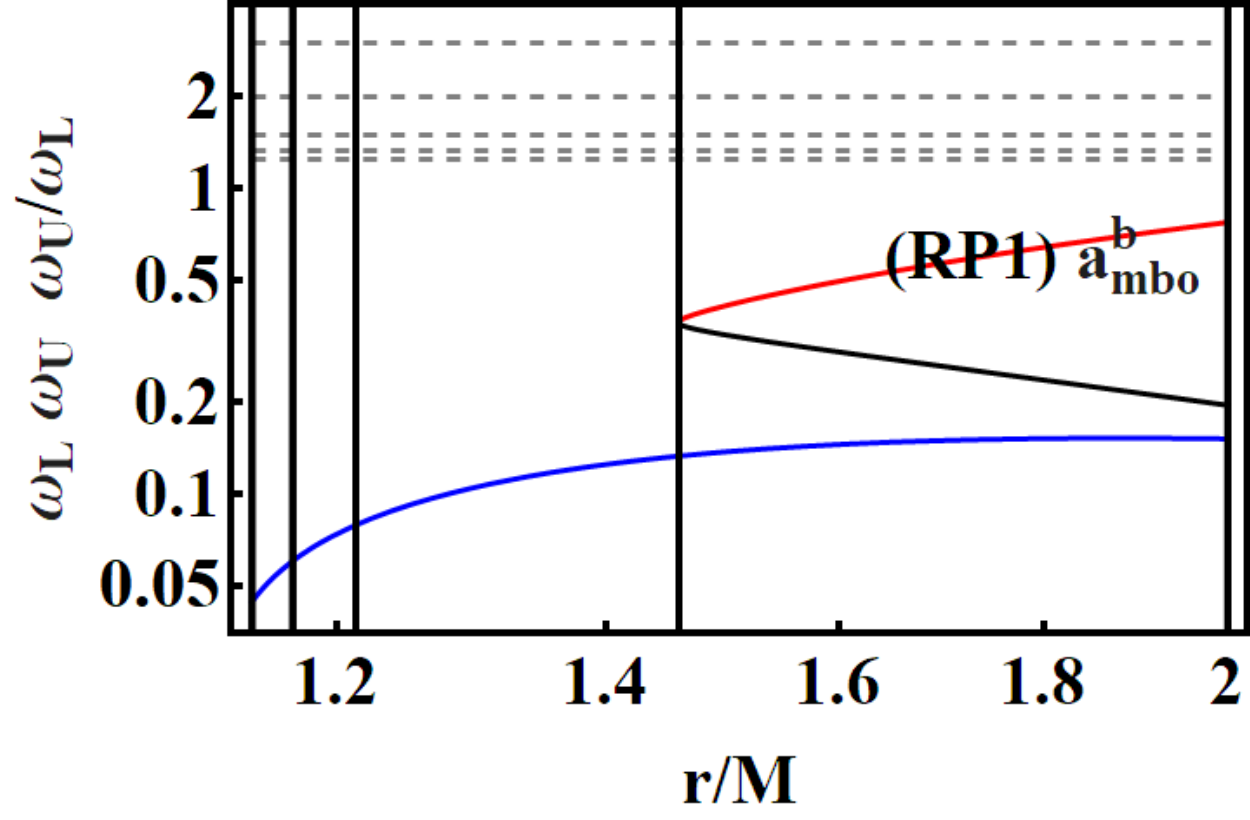}
  \includegraphics[width=5.6cm]{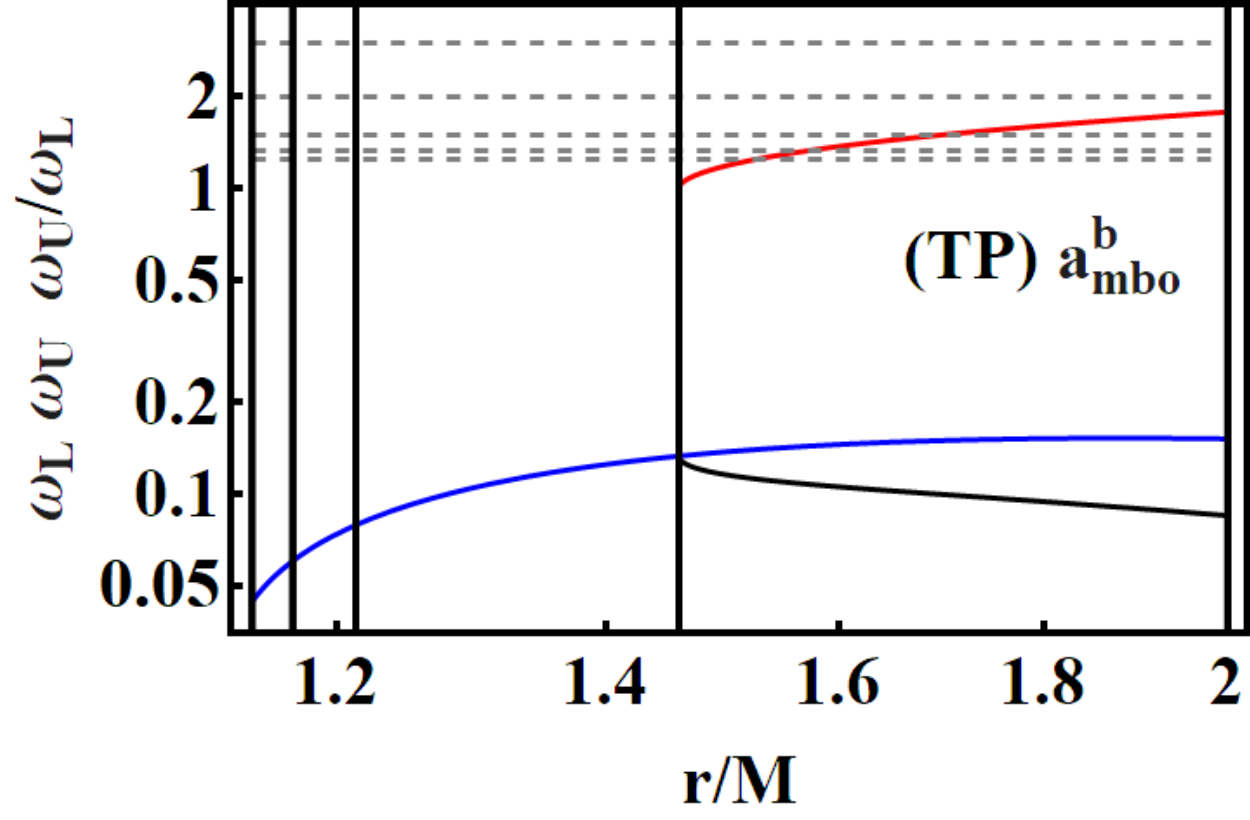}
  \includegraphics[width=5.6cm]{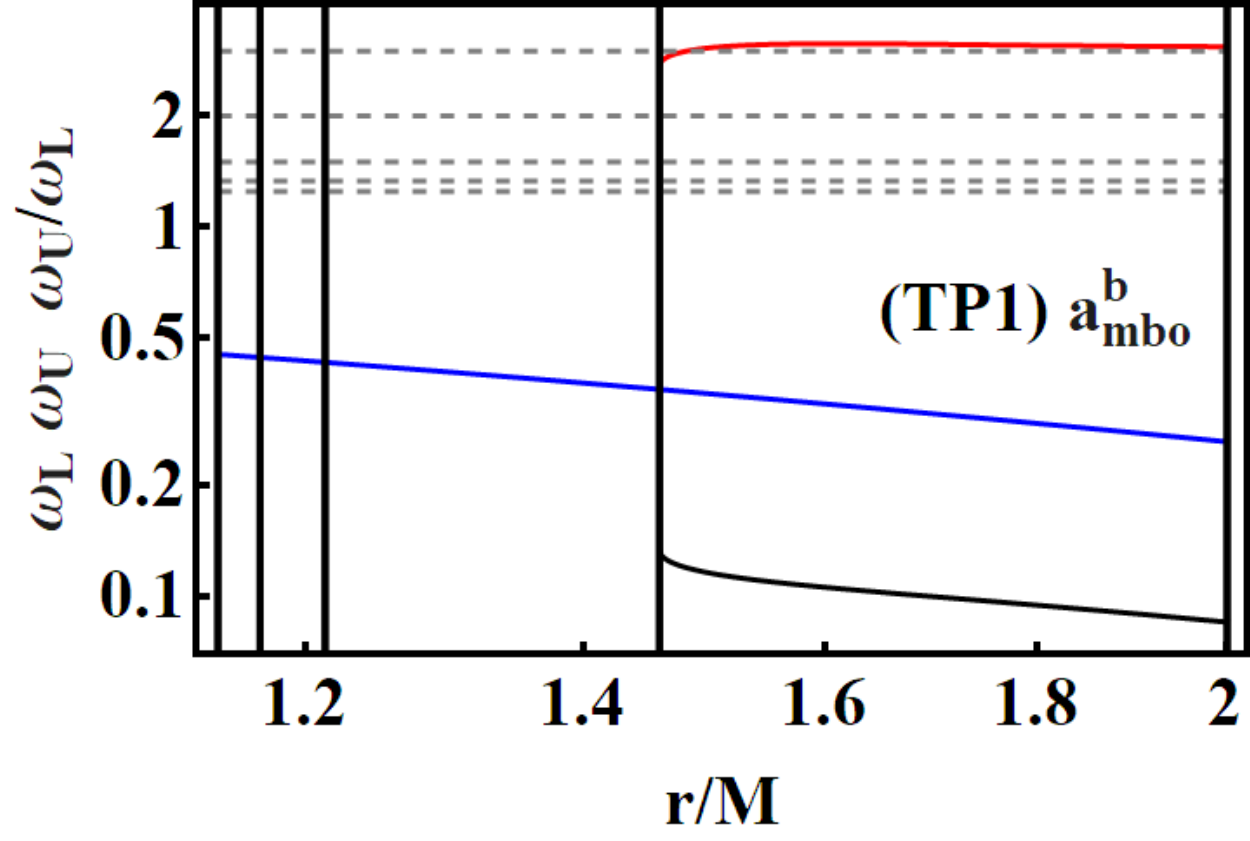}
    \includegraphics[width=5.6cm]{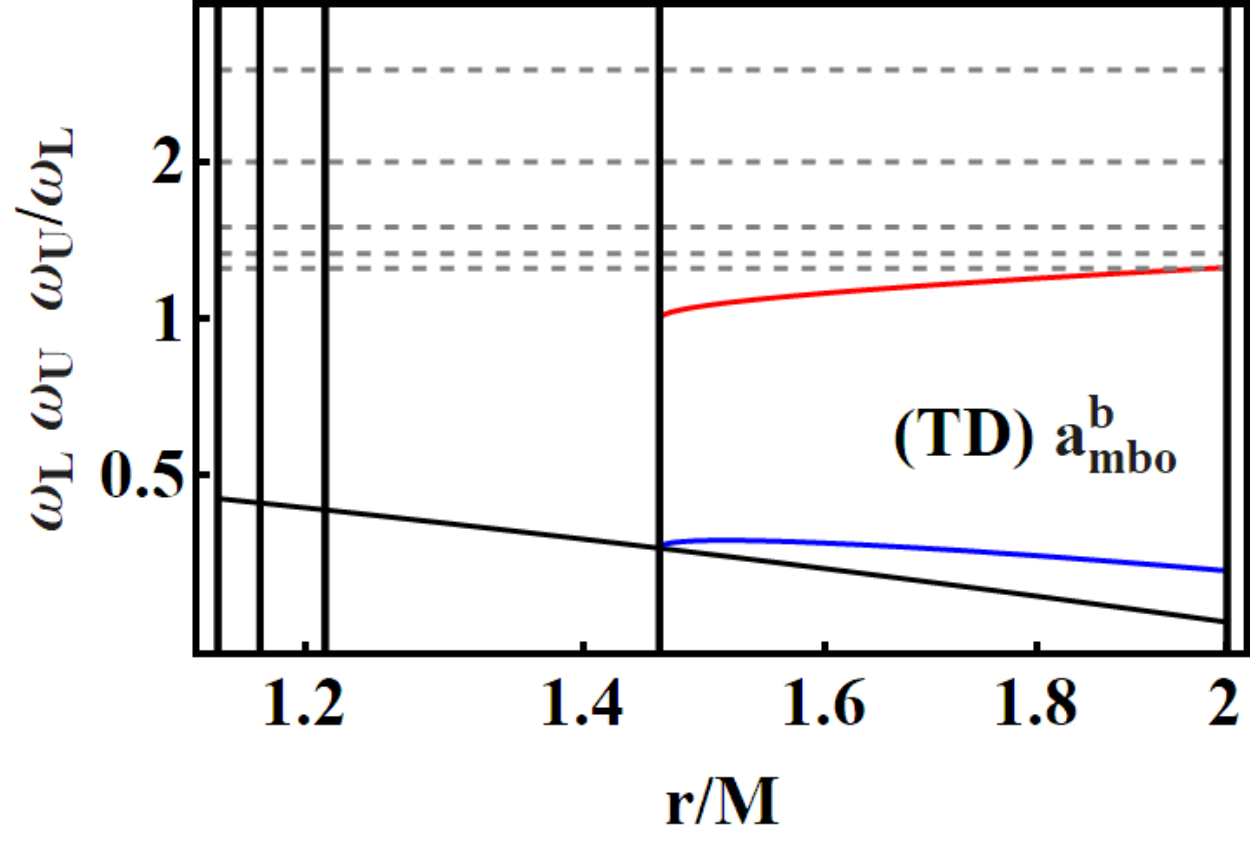}
  \includegraphics[width=5.6cm]{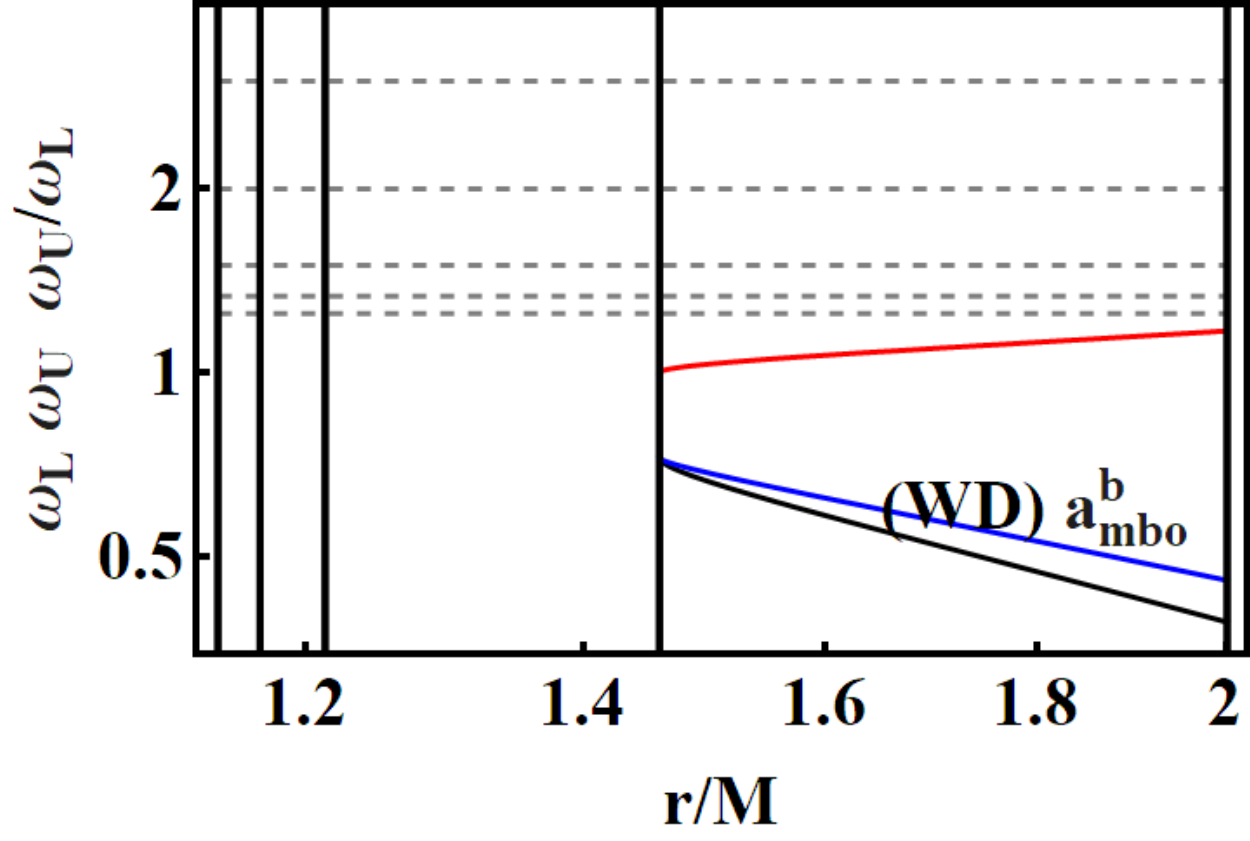}
   \includegraphics[width=5.6cm]{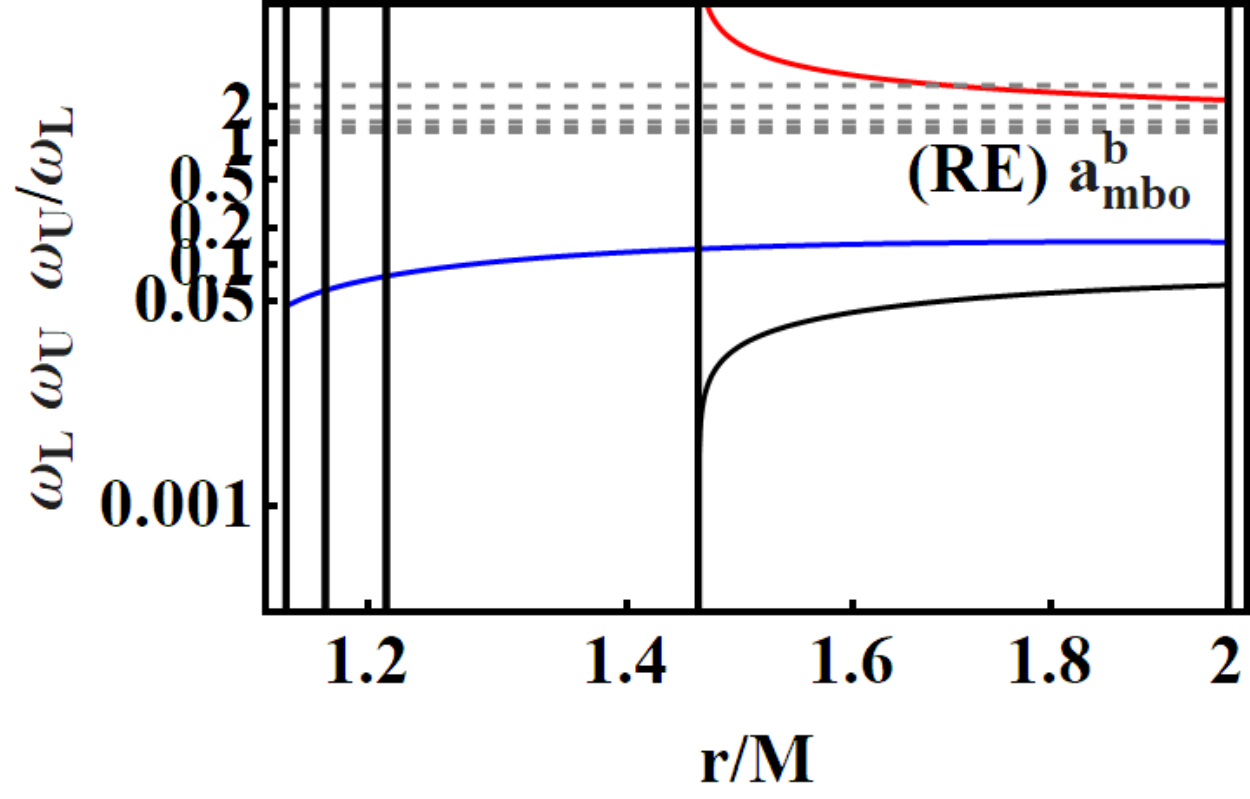}
    \includegraphics[width=5.6cm]{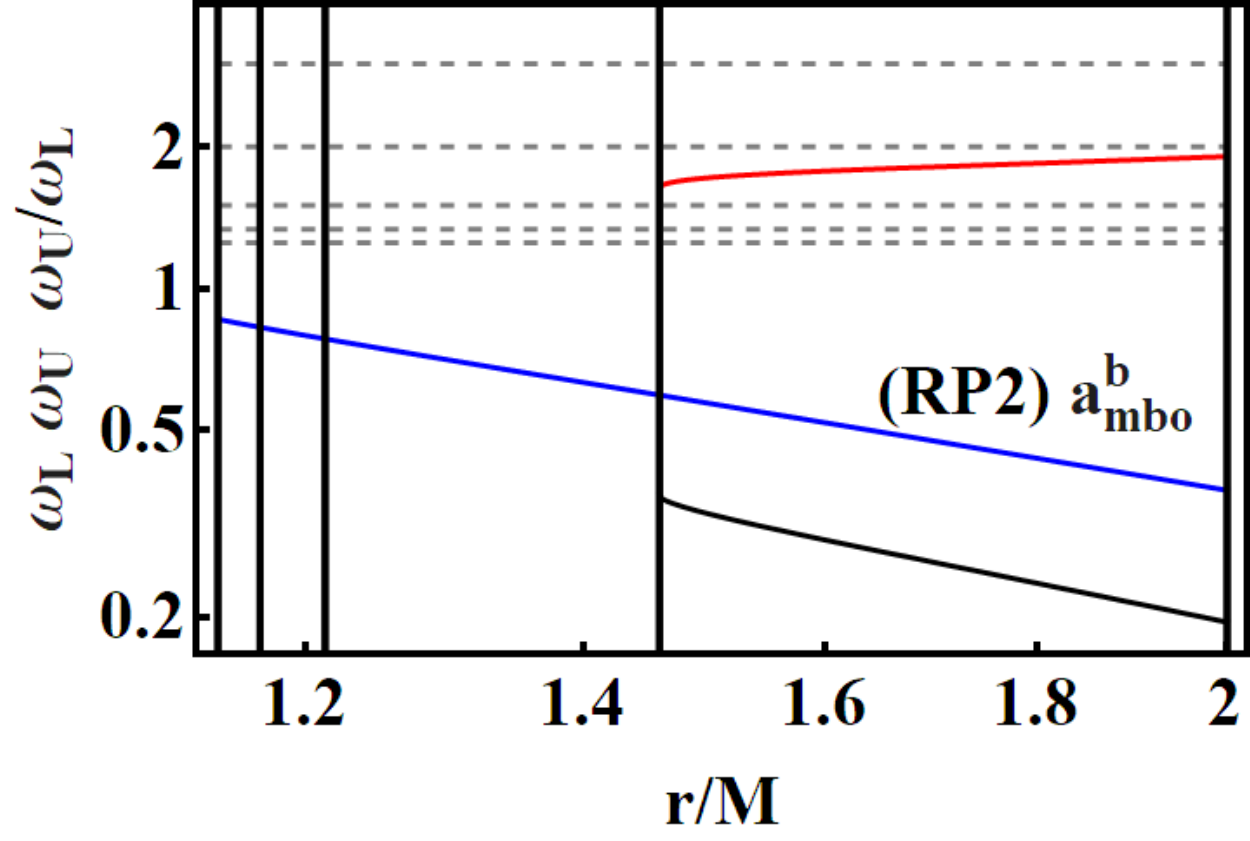}
  \caption{Plot of frequencies  $\omega_U$ (blue curve) $\omega_L$ (black curve)  and the ratio $\omega_U/\omega_L$ (red curve) as functions of the radius $r/M\in[r_+,r_{\epsilon}^+]$ in the ergoregion, where the outer ergosurface on the equatorial plane is $r_{\epsilon}^+=2M$.  The \textbf{BH} spacetime has spin   $a_{mbo}^b$--see also  Figs\il(\ref{Fig:polodefin1}).  Oscillation models  {{ $\{(\textbf{WD}),(\textbf{TD}),(\textbf{TP}),(\textbf{TP1}),(\textbf{RP}),(\textbf{RP1}),(\textbf{RP2}),(\textbf{RE})\}$}} of Eqs\il(\ref{Eq:posittru}). Resonant frequency ratios  {{$\mathbf{R1}=2:1$, $\mathbf{R2}=3:1$, $\mathbf{R3}={3}:{2}$, $\mathbf{R4}={4}:{3}$, $\mathbf{R5}={5}:{4}$}} (dashed lines) are also shown. Black lines are  $r_+<r_{\gamma}<r_{mbo}<r_{mso}$, where $r_{\gamma}$ is the marginally circular orbit, $r_{mbo}$ is the marginally bounded orbit, $r_{mso}$  is the marginally stable orbit.}\label{Fig:bapopof4}
\end{figure}
In Figs\il(\ref{Fig:bapopof4}) we show the models in the \textbf{BH} geometry $a_{mbo}^b$ in comparison with different resonant ratios, in Figs\il(\ref{Fig:apopof4})  we show the models in the \textbf{BH} geometry $a_{\gamma}^b$.
  Each  frequency model  we study  is borrowed  from a specific context from which they are derived including  slender tori and  hot spot models (assuming radiating spots in thin accretion disks).
Therefore frequencies used in the fit of the models   {\footnotesize{$\{(\textbf{WD}),(\textbf{TD}),(\textbf{TP}),(\textbf{TP1}),(\textbf{RP}),(\textbf{RP1}),(\textbf{RP2}),(\textbf{RE}),(\textbf{Kepl})\}$}} are therefore expressions of particle  frequencies that can  demonstrate Keplerian and
epicyclic frequencies:
 \textbf{(WD)} is for   "warped disk" model introduced for
oscillatory modes in a warped accretion disk.
  $\mathbf{(TD)}$  is for tidal disruption model
(related to  tidal disruption of  accreting inhomogeneities) \citep{Infras}.
  \textbf{TP}  and \textbf{TP1} are for  total precession models.
$\mathbf{(RP)}$, $\mathbf{(RP1)}$  an $\mathbf{(RP2)}$  are for
relativistic-precession models,
 attributing  the HF QPOs to modes of blobs  in the inner parts of
the accretion disk.
\textbf{(RE)} is for  "epicyclic resonance"
model dealing with radial and vertical epicyclic oscillations, and
the  \textbf{(Kepl) } for "Keplerian resonance"  model assuming a resonance
between the orbital Keplerian motion and the radial epicyclic
oscillations \citep{2017AcA....67..181S,2016A&A...586A.130S,2007A&A...463..807S,2005A&A...437..775T,2017A&A...607A..69K,2016ApJ...833..273T,2015A&A...578A..90S,2013A&A...552A..10S,2011AA...531A..59T,2011AA...525A..82S,2008CQGra..25v5016K,2007A&A...470..401S}
\subsection{Tori from polytropic fluids and the ergoregion}\label{Sec:poly-altr}
In this section we specify the  polytropic equation of state (EoS) for dragged  tori.
An  interesting issue  to be investigated   is  whether  the frame dragging of the Kerr spacetime may  differentiate  dragged and partially contained  tori  or configurations close to the outer ergosurface  with different equations of state, for example if the  polytropics  can affect differently   dragged tori.
We consider  a polytropic equation of state, assuming the pressure $p$ be a function of the matter density $\varrho$: $p=\tilde{k} \varrho^{\gamma}$, where  $\tilde{k}>0$  and $\gamma$ is the polytropic index.
In \citet{pugtot}  different polytropic EoS  for  tori were studied,  distinguishing classes in relation with the tori parameters ranges and the locations, here we re-focus on this concept in relations  with tori in the ergoregion.
It has been shown that  there is a specific classification of eligible geometric polytropes  and a specific class of polytropes is characterized by a discrete  range of values for the index $\gamma$, \citep{Fi-Ringed,Raine,pugtot,2009CQGra..26u5013S,2020Univ....6...26S}.
For the matter density $\rho$  there is
\bea\label{peter}
\bar{\varrho}_{\gamma}\equiv\left[\frac{1}{\tilde{k}}\left(V_{eff}^{-\frac{\gamma -1}{\gamma
}}-1\right)\right]^{\frac{1}{(\gamma-1)}}\quad\mbox{for}\quad\gamma\neq1,
\quad%\label{peter1}
\varrho_{\tilde{k}}\equiv V_{eff}^{-\frac{1+\tilde{k}}{ \tilde{k}}}\frac{1}{1+\tilde{k}},\quad\mbox{for}\quad \gamma=1,
\eea
where  $\varrho_{\gamma}\equiv
\tilde{k}^{1/(\gamma-1)}\bar{\varrho}_{\gamma}$,  which is independent from $\tilde{k}$. %The following
We  consider  here $\gamma\neq1$. %it is
%\(
%
The following two cases occur:
\begin{description}
\item[\textbf{\emph{1.}}] There is  $\mathcal{C}>0$,  which is for $\gamma\neq 1$, where   $V_{eff}^2 \geq 1$ for
$ \gamma\in]0,1[$ and   $V_{eff}^2 \in ]0,1]$ for $\gamma> 1$.
\item[\textbf{\emph{2.}}] There is  $\mathcal{C}< 0$, which is for   $\gamma\neq1$, within the condition $V_{eff}^2\in]0,1[ $ for $\gamma\in]0,1[$, and the condition   $
	   V_{eff}^2 >1$ for $\gamma>1$. The polytropic  index satisfies the condition
  $\gamma = {1}/({2n}) +1$ where  $ n\in\mathbb{Z}$ and
    $n\geq 1$, and $n\leq -1$.
    \end{description}
(Note, the case \textbf{\emph{1.}} includes polytropes with index $\gamma=4/3$.).
\subsubsection{Polytropic fluids and tori energetics}\label{Sec:ener}
\begin{figure}\centering
  % Requires \usepackage{graphicx}
  \includegraphics[width=8cm]{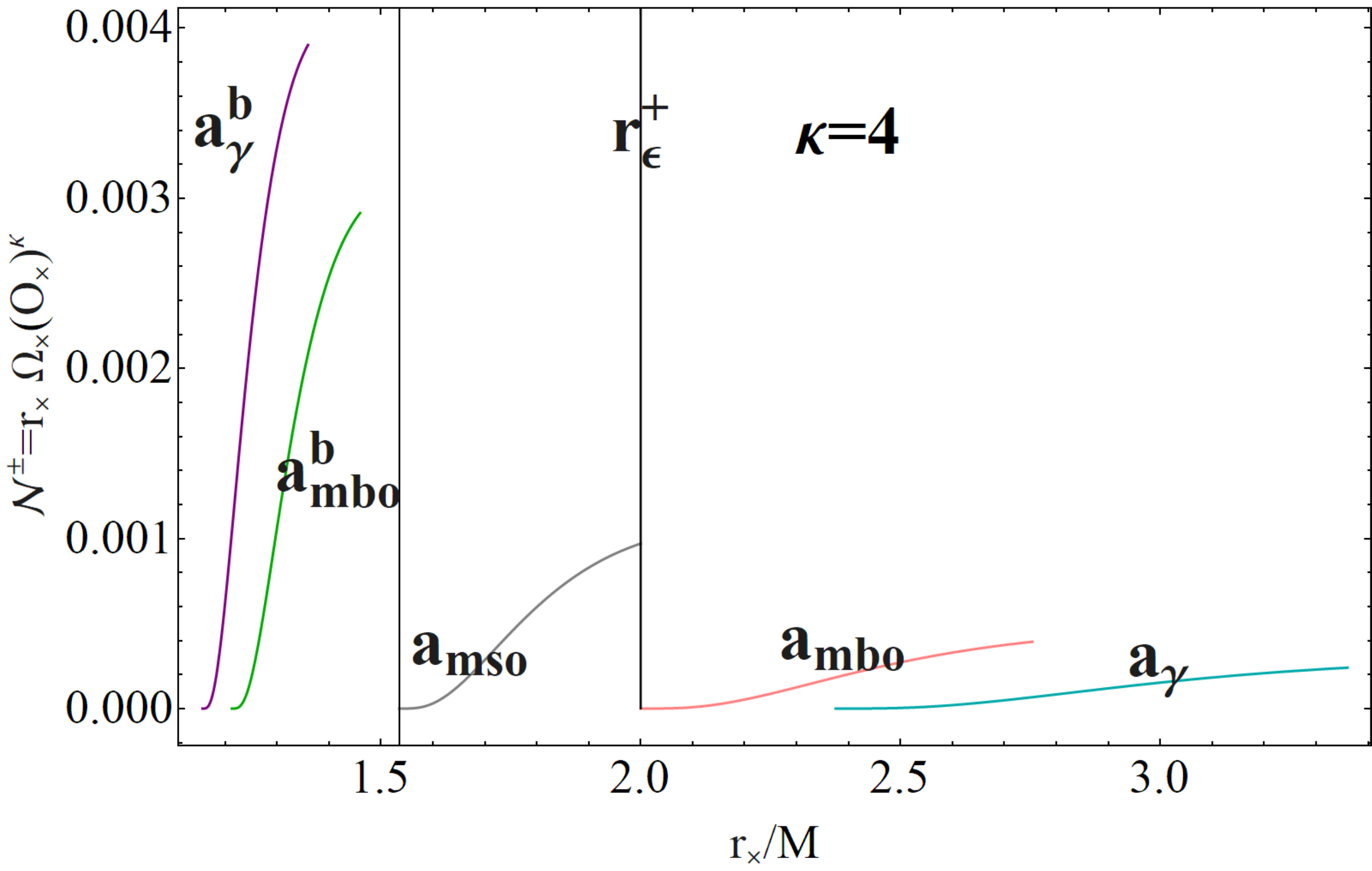}
    \includegraphics[width=8cm]{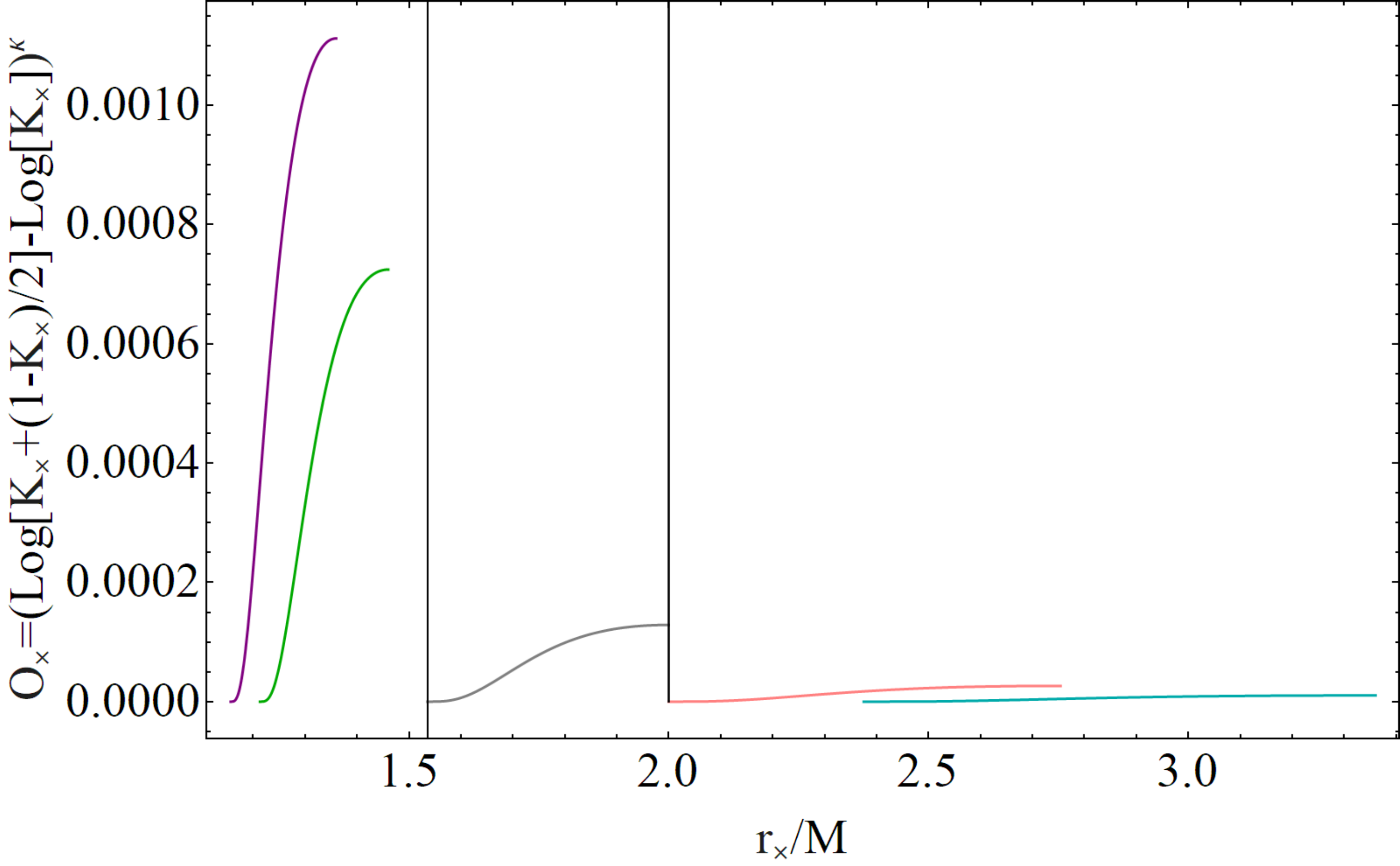}
  \caption{Evaluation of $\mathcal{N}$- and $\mathcal{O}$-quantities of Table\il(\ref{Table:Q-POs})  for corotating  tori, versus the cusp location $r_{\times}/M$,  for the spins $\mathbf{A_{\epsilon}^+}\equiv\{a_{mbo},a_{mbo}^b,a_{\gamma},a_{\gamma}^b,a_{mso}\}$--see Figs\il(\ref{Fig:polodefin1}).        The quantities at  $r_{s}<r_{\times} $ are evaluated according to $K_*\equiv K(r_s)=  K_{\times}+(1-K_{\times})/\epsilon$ where  $\kappa=n+1=4$, with  $\gamma=1/n+1$ is the polytropic index and $\epsilon=2$. $r_{\epsilon}^*=2M$ is the outer ergosurface.}\label{Fig:Twomodestate}
\end{figure}

We consider  tori   with  polytropic fluids:  $p=\tilde{\kappa} \varrho^{1+1/n}$,  providing an estimation of the   mass-flux,  enthalpy-flux (evaluating also the temperature parameter),
and  the flux thickness based on geometric consideration only on the thick disks in the ergoregion or close to ergoregion--see \citet{abrafra,Japan}. In details, these quantities  are listed in Table\il(\ref{Table:Q-POs}).
\begin{table*}[ht!]
\begin{center}
%\resizebox{.9\textwidth}{!}{%
\begin{tabular}{|l|l|}
\hline \hline
%{\textwidth}{@{\extracolsep{\fill}}lrcl@{}}%
 %\{\textwidth}{@{\extracolsep{\fill}}lrcl@{}}
 \mbox{\textbf{Quantities}}$\quad  \mathcal{O}(r_\times,r_s,n)\equiv q(n,\tilde{\kappa})(W_s-W_{\times})^{d(n)}$ &   $\mbox{\textbf{Quantities}}\quad  \mathcal{P}\equiv \frac{\mathcal{O}(r_{\times},r_s,n) r_{\times}}{\omega_K(r_{\times})}$\\\hline\hline
$\mathrm{\mathbf{Enthalpy-flux}}=\mathcal{D}(n,\tilde{\kappa} ) (W_s-W)^{n+3/2},$&  $\mathbf{torus-accretion-rate}\quad  \dot{m}= \frac{\dot{M}}{\dot{M}_{Edd}}$  \\
 $\mathrm{\mathbf{Mass-Flux}}= \mathcal{C}(n,\tilde{\kappa}) (W_s-W)^{n+1/2}$& $\textbf{Mass-accretion-rates }\quad
\dot{M}_{\times}=\mathcal{\widetilde{A}}(n,\tilde{\kappa}) r_{\times} \frac{(W_s-W_{\times})^{n+1}}{\omega_K(r_{\times})}$
 \\
 $\frac{\mathcal{\widetilde{L}}_{\times}}{\mathcal{\widetilde{L}}}= \frac{\mathcal{B}}{\mathcal{\widetilde{A}}} \frac{W_s-W_{\times}}{\eta c^2}$&     $\textbf{Cusp-luminosity}\quad  \mathcal{\widetilde{L}}_{\times}=\mathcal{B}(n,\tilde{\kappa}, r_{\times}) \frac{(W_s-W_{\times})^{n+2}}{{\omega_K(r_{\times})}}$
 \\
\hline\hline
{\textbf{[$\mathcal{O}$-quantities]}:} $ \mathcal{R}_{*}^{\pm}\equiv(W^{\pm}(r_{s})-W^{\pm}_{*})^\kappa$
&\label{Eq:soft-chogra}
{\textbf{[$\mathcal{N}$-quantities]}:  }     $\mathcal{N}_{*}^{\pm}\equiv\frac{{r_*} (W^{\pm}(r_{s})-W^{\pm}_{*})^\kappa}{\omega_K(r^{\pm}_*)}$
\\
\hline \hline
\end{tabular}%}
\end{center}
\caption{Quantities $\mathcal{O}$ and $\mathcal{N}$.  $\mathcal{\widetilde{L}}_{\times}/\mathcal{\widetilde{L}}$ stands for  the  fraction of energy produced inside the flow and not radiated through the surface but swallowed by central \textbf{BH}. %\rtb{These quantities are evaluated for polytropic indices $\gamma=4/3$ in Figs\il(\ref{Fig:Manyother}).}
Efficiency
$\eta\equiv \mathcal{\widetilde{L}}/\dot{M}c^2$,    $\mathcal{L}$ representing the total luminosity, $\dot{M}$ the total accretion rate where, for a stationary flow, $\dot{M}=\dot{M}_{\times}$,  $W=\ln V_{eff}$ is the potential  of Eq.\il(\ref{Eq:scond-d}), $\omega_K$ is the Keplerian (relativistic)  angular frequency,
$W_s\geq W_{\times}$ is the value of the equipotential surface, which is taken with respect to the asymptotic value, $ W_{\times}=\ln K_{\max}$  is the  function at the cusp (inner edge of accreting torus), $\mathcal{D}(n,\tilde{\kappa}), \mathcal{C}(n,\tilde{\kappa}), \mathcal{\widetilde{A}}, \mathcal{B}$ are functions of the polytropic index and the polytropic constant.
}
\label{Table:Q-POs}
\end{table*}
Parameters  $(\tilde{\kappa},n)$ within the constraints  $q(n,\tilde{\kappa})=\bar{q}=$constant,  fix  a  polytropic-family.
Then $\kappa=n+1$, with  $\gamma=1/n+1$  being the polytropic index. We use the fact that the %
 pressure forces   are vanishing at the edges of the accretion torus.
These quantities  can be express in  the  general  form  $\mathcal{O}(r_\times,r_s,n)$.
The   $\mathcal{O}(r_\times,r_s,n)$ depends on the location of the inner edge of the accreting torus, being  determined only by the angular momentum, and on  the radius   $r_s$ which is  related to the thickness of the matter flow:
$K_{\times}$ corresponds to the accreting point $r_{\times}$, while $r_s$ is directly associated to the accreting flux thickness, where  $K_s\in[K_{\times},1[$. 
In  $\mathcal{O}(r_\times,r_s,n)$,  $q(n,\tilde{\kappa})$ and $d(n)$ are different functions of the polytropic index  $\gamma=1+1/n$ and of the  polytropic constant $\tilde{\kappa}$.  The mass flow rate through the cusp (mass loss, accretion rates)  $\dot{M}_{\times}$,  and the cusp luminosity $\mathcal{L}_{\times}$ (and the accretion efficiency $\eta$),
 measuring the
rate of the thermal-energy    carried at the  cusp could be expressed as    $\mathcal{P}$   quantities, or alternatevley $\mathcal{N}$-quantities, as in Table\il(\ref{Table:Q-POs}).  $\mathcal{N}$- and $\mathcal{O}$-quantities of Table\il(\ref{Table:Q-POs})  are evaluated for corotating  tori in Figs\il(\ref{Fig:Twomodestate})
 providing a simple way to evaluate the  trends of these functions of the \textbf{BH} spin and  the cusp location.
The quantities at  $r_{s}<r_{\times} $ are evaluated according to $K_*\equiv K(r_s)=  K_{\times}+(1-K_{\times})/\epsilon$, where $\epsilon=2$ in  Figs\il(\ref{Fig:Twomodestate}). The behaviour of the curves at different spins as functions of $r_{crit}$ (therefore as functions of the specific fluid momentum) is qualitatively independent from variation of $\epsilon$ (an increase of $\epsilon$ correspondent to a lowering of $K$ parameter  over the critical maximum value  and lowering of the matter level towards the cusp  is associated to similar qualitative  behaviour of the curves and  decrease in magnitude). We note that clearly the choice of $K_*$ corresponds to the choice of a point on the curves $\partial_y V_{eff}=0$ at $y<r_{crit}=r_{\times}$
 as in Figs\il(\ref{Fig:polodefin1}).
 An analysis  of the $K_{crit}$ and location of the critical points for different \textbf{BH} spins in relation  to the possibility of multi-orbiting tori around the central \textbf{BH}, consisting  by both corotating or counterotating fluids, can be found in Figs\il(\ref{Fig:slowtoga}).
We can note that there is an increase of  the spin  corresponding to a decreases of the range $r_{\times}$; on the contrary, the curves increase due to  increasing the  distance from the \textbf{BH} of the disk cusp which corresponds  to the increase of the specific angular momentum (at fixed \textbf{BH} spin $a/M$).
 Despite the reduced dimensions of the dragged configurations it is clear that $\mathcal{N}$ and $\mathcal{O}$ are quite large inside the ergoregion $r_{\times}\in ]r_+.r_{\epsilon}^+]$.
 Considering Figs\il(\ref{Fig:PlotVampb1}) we can see that the analysis of cusped configurations in  Figs\il(\ref{Fig:Twomodestate}) is in agreement with the existence of the cusp in the ergoregion at
 $a_{\gamma}^b$, $a_{mbo}^b$ and $a_{mso}$.
 This analysis, however, does not take into account both the deviations present from the perfectly stationary situation, and the fate of the flow in free falling after the cusp. We can analyze this case, taking into account that that region corresponds in fact to the situation of minimum hydrostatic pressure.

\begin{figure}\centering
  % Requires \usepackage{graphicx}
  \includegraphics[width=8cm]{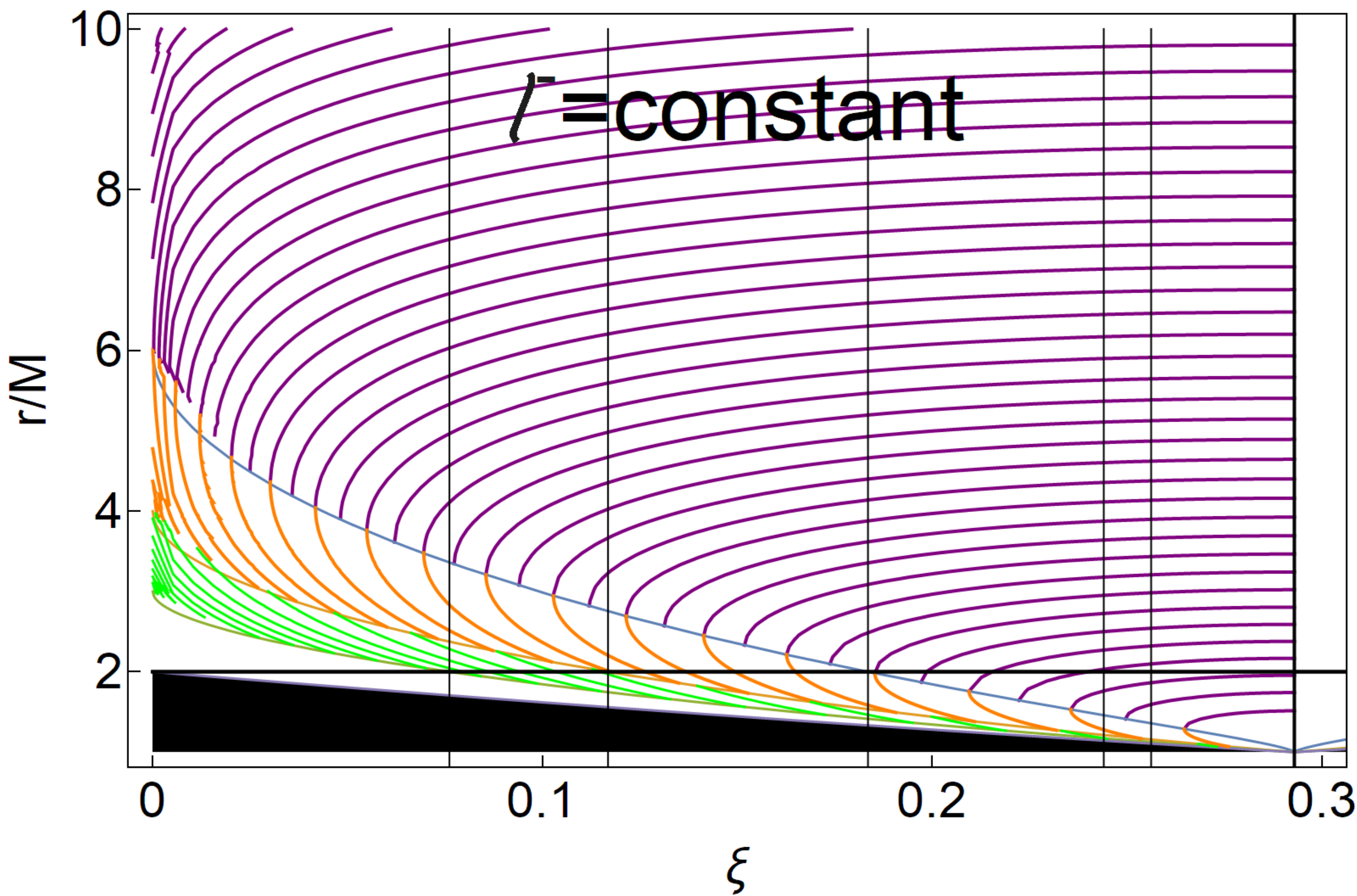}
  \includegraphics[width=8cm]{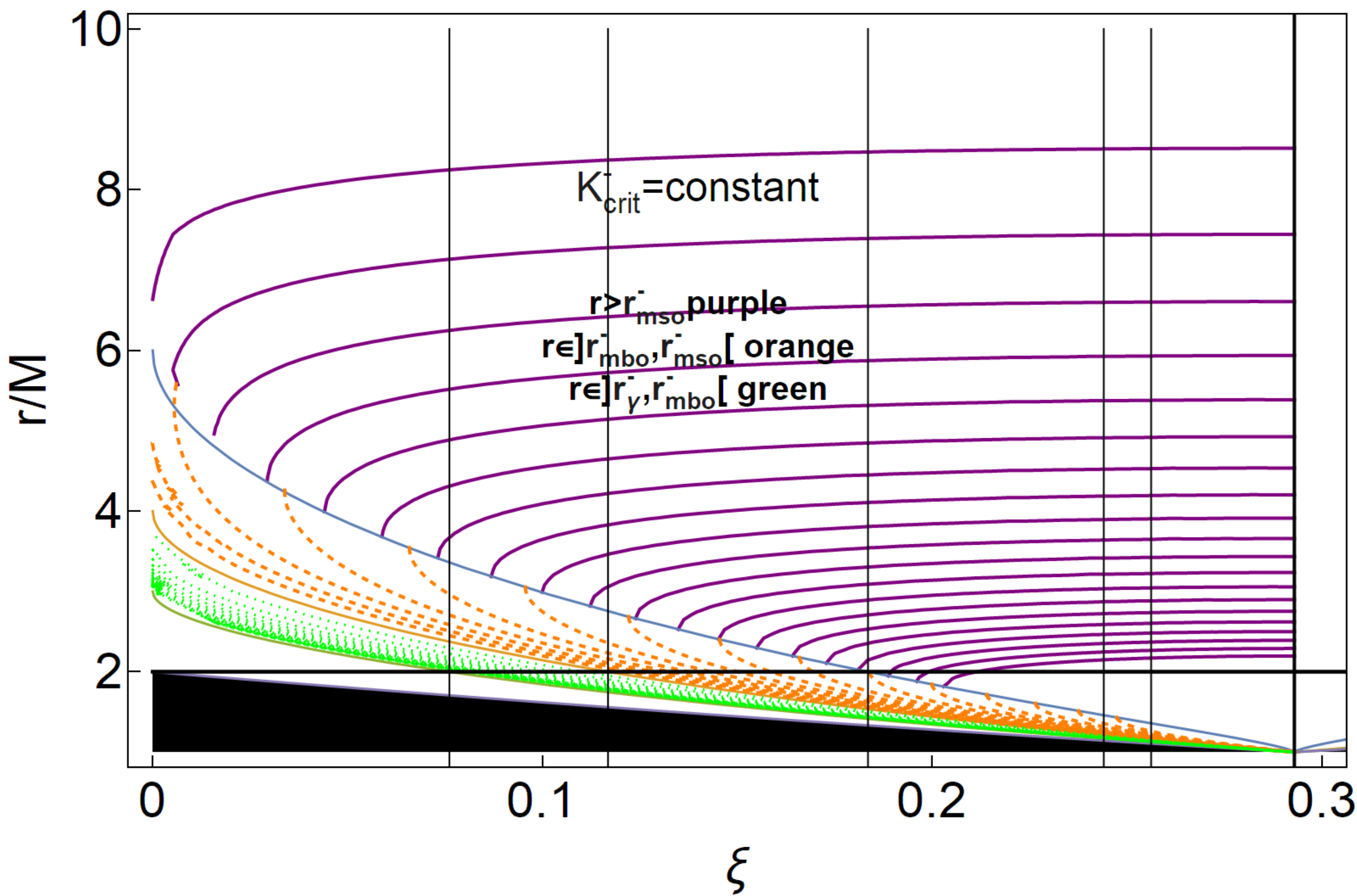}
  \caption{Plane $r/M$ and $\xi$, where $\xi$ is the maximum  extractable  rotational energy parameter.
Gray curves  are the marginally bounded orbit, $r_{mbo}$, marginally stable orbit $r_{mso}$ and $r_{mco}=r_{\gamma}$ marginally circular orbit.  Left panel: Curves   $\ell(\xi;r)=$constant (on a vertical line of the plane each curve describes  one torus).
 Right panel: curves
 $K_{crit}=$constant    on the equatorial plane.  Spins $\mathbf{A_{\epsilon}^+}\equiv \{a_{mbo},a_{mbo}^b,a_{\gamma},a_{\gamma}^b,a_{mso}\}$ are plotted as black lines ---see Figs\il(\ref{Fig:polodefin1}).  Black region is the \textbf{BH} at $r<r_+$, where  $r_+$ is the outer Killing horizon. The outer ergosurface $r_{\epsilon}^+=2M$ is represented as a black  horizontal line. Quantities  refer to the status of the \textbf{BH}-accretion disk system at its stationary state (0) prior the energy extraction.}\label{Fig:ManyoBoF}
\end{figure}

\medskip

\textbf{The  extractable rotational  energy  $\xi$.}

In Figs\il(\ref{Fig:ManyoBoF})  the analysis has been developed   showing explicitly  the relations between the  dragged toroidal surfaces characteristics  in the  ergoregion and the maximum extractable rotational  energy  $\xi$ from the \textbf{BH} horizon, through the analysis  of the  leading function $\ell(r)$ and the energy function $K(r)$.
The plots show the situation one might expect  in a phase prior the total extraction of the energy from the \textbf{BH}.
We start by considering the  spin function $a(\xi)$ \citep{PS21}:
\bea\label{Eq:exi-the-esse.xit}
a(\xi)\equiv 2 \sqrt{-(\xi -2) (\xi -1)^2 \xi },
\eea
 relating the  dimensionless \textbf{BH} spin $a/M$ to   the dimensionless ratio $\xi$, (total released rotational  energy   versus \textbf{BH}  mass measured by an observer at infinity). We are  assuming a  process ending with  the  \textit{total} extraction of  the    rotational energy of the central Kerr \textbf{BH}. Considering  $M(0)$ and  $ J(0)$ the mass and  angular
momentum of the initial state of the \textbf{BH},   the upper limit for  of the energy extraction from a stationary process bringing the \textbf{BH} at the state $\mathbf{\mathbf{\mathbf{(1)}}}$ is $M(0) - M_{irr}(M(0), J(0))$. The \textbf{BH} angular momentum in the new state  $\mathbf{\mathbf{\mathbf{(1)}}}$ is zero.  All the quantities therefore are evaluated at the state $(0)$ prior the process, which implies that  all the quantities evaluated here inform on the status of the \textbf{BH}-accretion disk system at its stationary state $(0)$.
We use the  approach introduced in \citet{Daly0,PS21}, which is ,  based on
 the definition of  \textbf{BH} irreducible mass and   rotational energy, and it is    independent from the
 energy extraction mechanism,  using
 the  \textbf{BH} classical thermodynamical law.

 Considering the  dimensionless rotational energy as $\xi=1-M_{irr}/M$ ($M_{irr}$ is the irreducible mass), with quantities evaluated at the stationary state prior the process, we obtain  the restricted range  $\xi\in[0,\xi_{\ell}]$, where $\xi_{\ell}\equiv \frac{1}{2} \left(2-\sqrt{2}\right)$ limiting therefore the energy extracted to a superior of  $\approx 29\%$ of the mass $M$, where at the state $(0)$ (prior the extraction) there is an extreme Kerr \textbf{BH} spacetime. The larger extractable rotational energy is related to the \textbf{BH}  with the {smallest  dragged tori}, including the dragged or partially included tori in the ergoregion. Plots show also details on the location of the inner edge and center of maximum pressure, in the possible configurations around the \textbf{BH}.
 The small dimension of tori is intended in the following sense:  it is clear  from  Figs\il(\ref{Fig:raisePlot}) that
 the thickness of tori is larger for larger \textbf{BHs} spin for  the parameter $\epsilon\geq 1.18$,
  the limit being associates to cusps   very close to the $r_{mbo}$,
  therefore to very large $K_{\times}$ .
  For $\epsilon\leq 1.18$, we find tori with  greater thickness   in geometries of smaller \textbf{BH}  dimensionless spins.
   From Figs\il(\ref{Fig:PlotJurY})  and from  the analysis of conditions for  the outer edge  $r_{outer}^{\times}=r_{\epsilon}^+$  of the cusped tori to be coincident with the static limit $r_{\epsilon}^+$, it is clear that there is  a total disk elongation $\lambda$  larger for the larger \textbf{BH} spins,  while the inner elongation $\lambda_{inner}=r_{center}-r_{\times}$  decreases with the spin (with respect to the observers at infinity)  $a>a_{mbo}^b$
\footnote{In relations to  \textbf{BHs} of the set $\mathbf{A}_{\epsilon}^+$,  there is $ \xi^b_{mbo}=0.244$, $\xi_{mbo}=0.117$, $\xi_{\gamma}^b=0.256$, $\xi_{\gamma}=0.0761$,
$\xi_{mso}=0.183503$, $
\xi_{\max} =\frac{1}{2} \left(2-\sqrt{2}\right)$, where $\xi_{\max}$ is the maximum extractable rotational  energy, occurring for an extreme Kerr \textbf{BH}.}.
\begin{figure}\centering
  % Requires \usepackage{graphicx}
  \includegraphics[width=5.6cm]{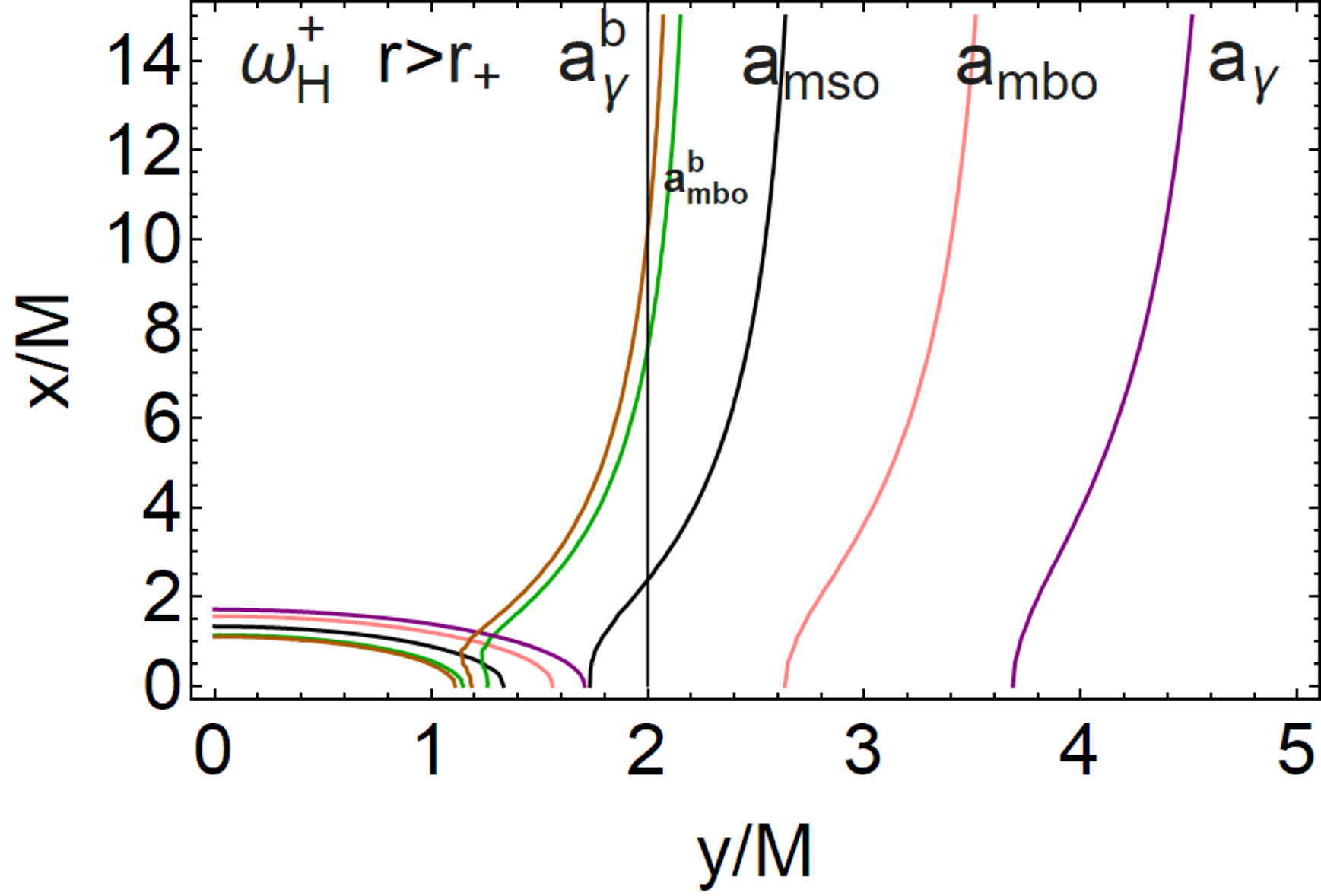}
  \includegraphics[width=5.6cm]{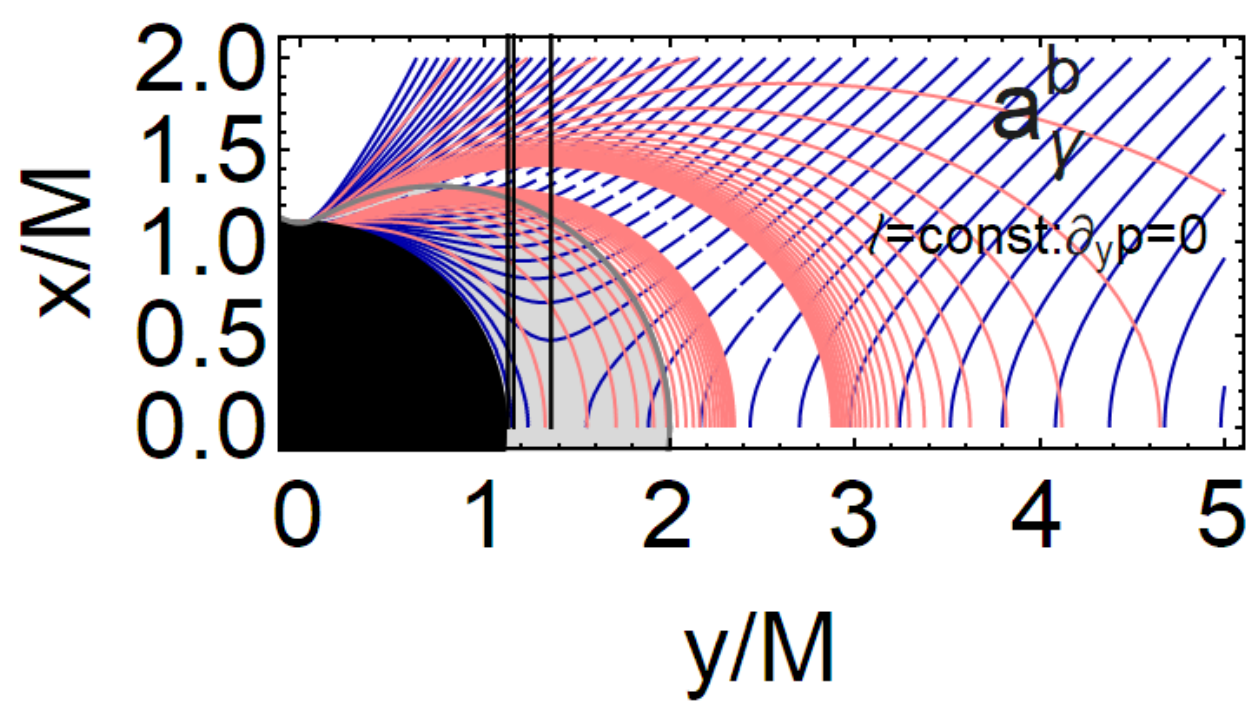}
  \includegraphics[width=5.6cm]{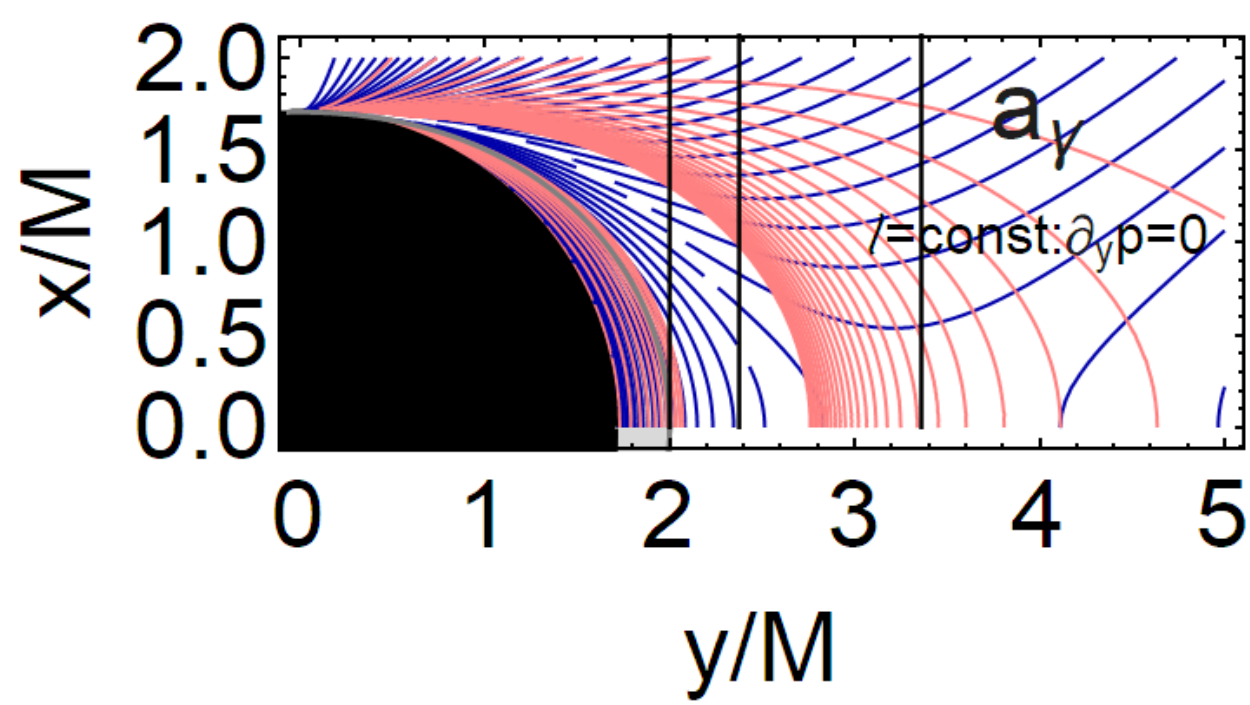}
  \includegraphics[width=5.6cm]{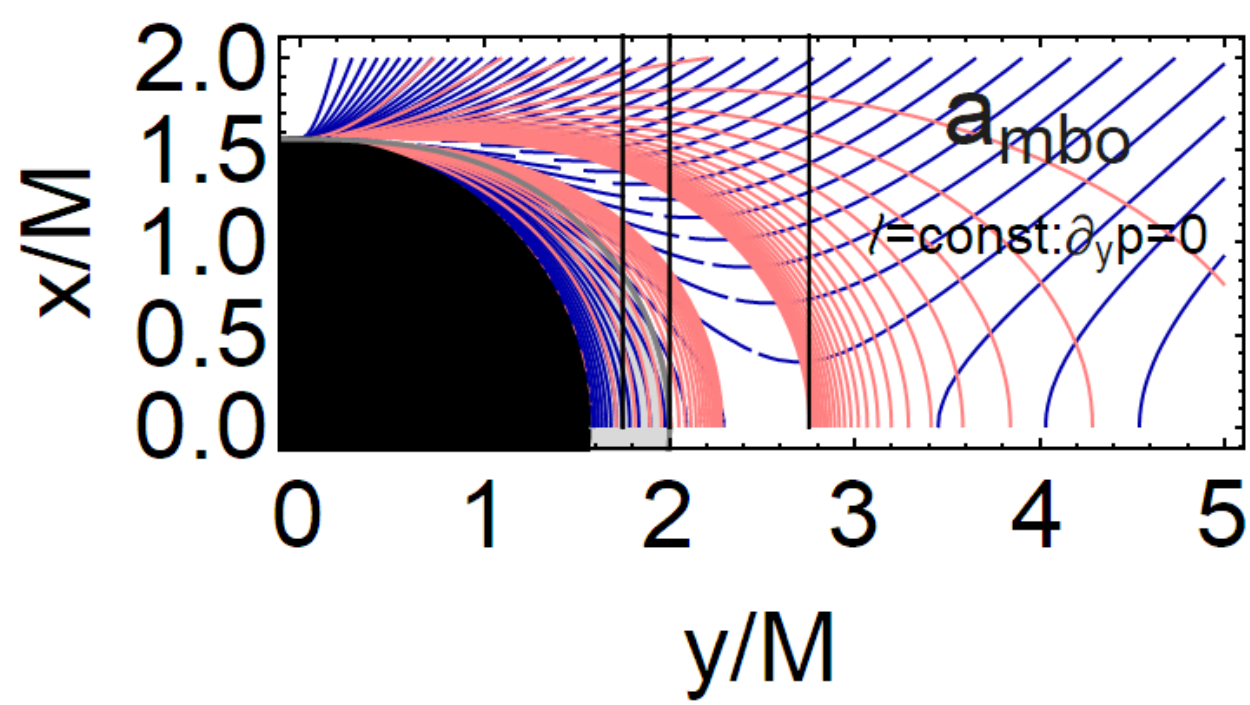}
    \includegraphics[width=5.6cm]{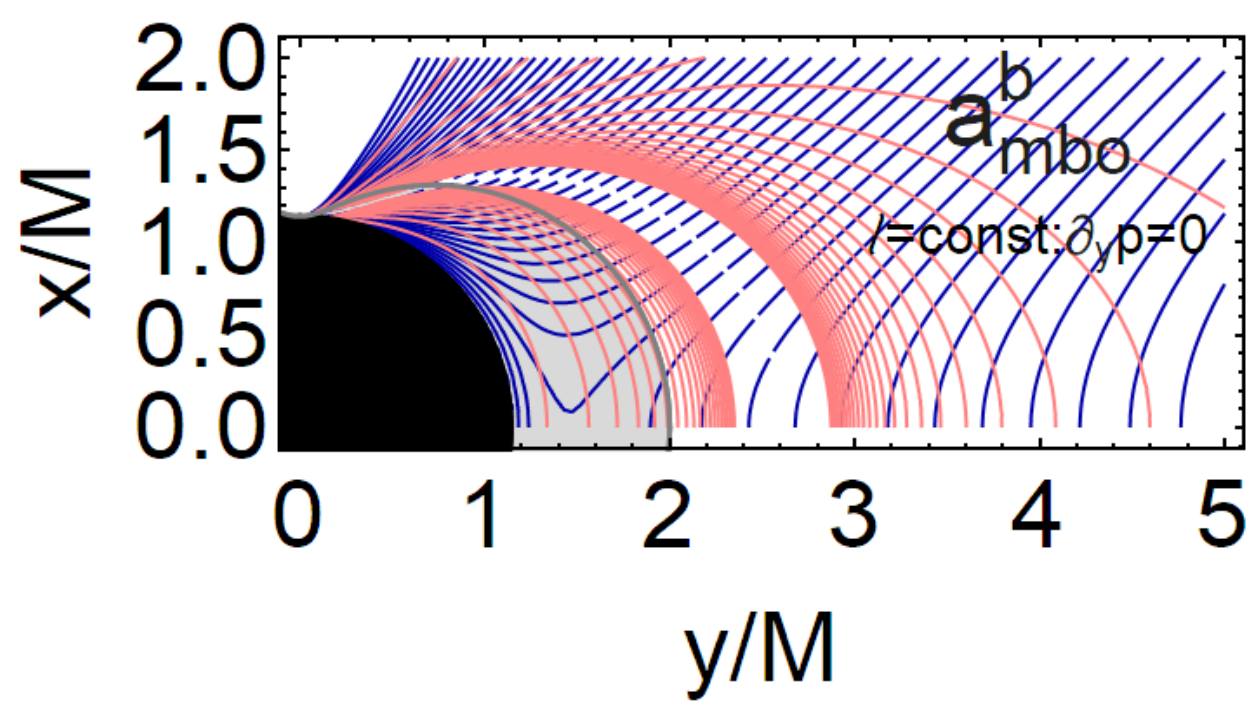}
  \includegraphics[width=5.6cm]{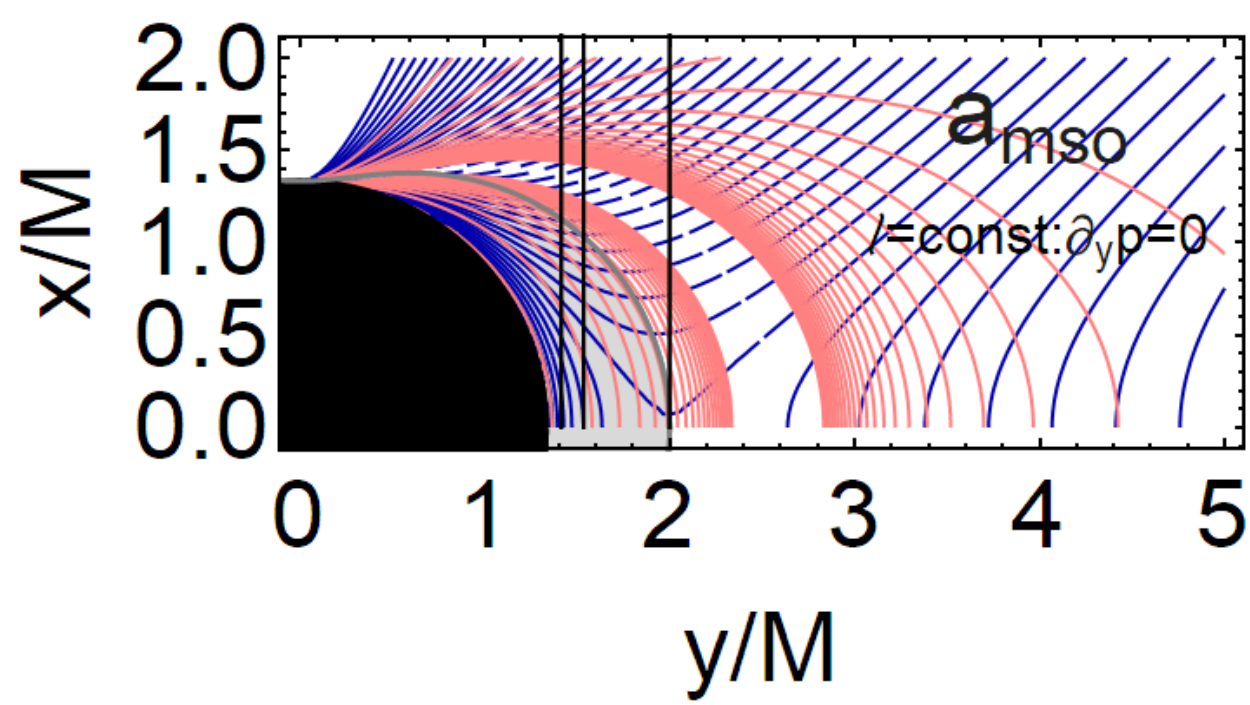}
  \caption{Upper left panel: plot of the light surfaces, defined by $\mathcal{L}\cdot\mathcal{L}=0$, (where $\mathcal{L}\equiv \xi_t+\omega\xi_\phi$) for angular velocity  $\omega=\omega_H^+$ of the outer horizon. at $r>r_+$.  Center and left upper line panel and below panels show the curves $\ell=$constant $\ell$ specific angular momentum solution of $\partial_y V_{eff}(x,y)=0 $ (there is ($x=r\cos\theta, y=r\sin\theta)$, on the equatorial plane there is $y\equiv r$), obtaining two solutions plotted as pink and blue curves. Spins $\mathbf{{A_{\epsilon}}^+}\equiv\{a_{mbo},a_{mbo}^b,a_{\gamma},a_{\gamma}^b,a_{mso}\}$ are represented--see also  Figs\il(\ref{Fig:polodefin1}). Black region is $r<r_+$, where $r_+$is the outer horizon. Gray region is the outer ergosurface. On the equatorial plane $(x=0)$ the outer ergosurface is $y=r_{\epsilon}^+=2M$.  }\label{Fig:Plotexale}
\end{figure}
Figs\il(\ref{Fig:Plotexale})  show the solution of the zeros of radial gradients of pressure, which is related to the definition of $\ell(r)$. The analysis considers the surfaces $\ell(x,y)=$constant, where the effective potential is  function of Cartesian coordinates $(x,y)$ and  we can note the role of the  marginally stable orbits--see also  Figs\il(\ref{Fig:collag}).
In Sec.\il(\ref{Sec:magn}) we include some notes on the stationary  magnetized tori in the ergoregion, considering the GRHD tori with a  toroidal magnetic field according to the Komissarov solution \citep{Komissarov:2006nz}.
\section{Conclusions}\label{Sec:conclusion}
In this work we investigated     orbiting  extended  matter configurations  in the ergoregion of  a spinning black hole, introducing  in  Sec.\il(\ref{Sec:deta-hall}) the concept of dragged  tori  and partially contended tori.
To this end,  we considered
  corotating perfect fluids   tori, constituting    geometrically  thick  accretion   disks,  using the disk   model detailed in Sec.\il(\ref{Sec:model}) for tori orbiting a central Kerr \textbf{BH}.

It is  argued that the  smaller
  dragged tori could be  subjected to a characteristic instability, disk exfoliation, consequence of the  geometry  frame-dragging. We discussed  this hypothesis assessing  the tori features   for the occurrence of this phenomena, the possible outcomes as  particle and photon emission, or in an  enhanced accretion. This  process   is expected   to lead to the  destruction of the torus  in combination with accretion, emission and other  processes typical  of  the regions very close to the black hole horizon as, for example, the Runway Instability.
  An  important parameter for the tori exfoliation is
the torus verticality, given by   the evaluation of the dragged torus  geometrical thickness.  In Sec.\il(\ref{Sec:dv})  we face this aspect,  discussing also tori   possible formation, and  providing  in Sec.\il(\ref{Sec:influ-ergosra}) an throughout investigation of the
dragged disks thickness  and   exploring the    influence of the dragging frame on the disk thickness.
Results of the analysis of the outer  edge location with the respect to the static limit are in Figs\il(\ref{Fig:PlotVampa1},\ref{Fig:spessplhoke1},\ref{Fig:PlotVamp1},\ref{Fig:PlotJurY}).
The results of the investigation of  the inner edge location with the respect to the static limit are in Figs\il(\ref{Fig:PlotBlakPurp7}) and (\ref{Fig:Plotssot}).

The analysis eventually  provides indications on the attractors where dragged tori can be observed, and tori  characteristics, according to the central attractor dimensionless spin.
 We individuated  five classes of geometries, as in    Figs\il(\ref{Fig:PlotVampb1}),   defined  according to the  \textbf{BHs} spin--mass ratios, bounded by the spins $\mathbf{A}_{\epsilon}^+\equiv \{a_{\gamma},a_{mbo},a_{mso},a_{mbo}^b,a_{\gamma}^b\}$, and regulating the constraints on the location of the extremes of the  pressure  in the tori--Figs\il(\ref{Fig:polodefin1}),(\ref{Fig:principmill}) and  (\ref{Fig:Plotsoorr}).
 We studied the toroids in each of  these classes.
Configurations orbiting in the geometries defined in $\mathbf{A}_{\epsilon}^+$ are shown in  Figs\il(\ref{Fig:polodefin1})
The conditions for the tori outer edge $r_{outer}$  to be in $\Sigma_{\epsilon}^+$, in  the different  geometries of the set  $\mathbf{A}_{\epsilon}^+$,   are shown in Figs\il(\ref{Fig:PlotVampa1}), for    cusped tori are  in  Figs\il(\ref{Fig:spessplhoke1}), and   in Figs\il(\ref{Fig:PlotVamp1})  where the conditions are expressed in functions of the disks inner edge, $r_{inner}$, location.
We provided in Figs\il(\ref{Fig:spessplhoke1},\ref{Fig:PlotVamp1},\ref{Fig:Plotssot},\ref{Fig:PlotBlakPurp7})  indications on the  torus   inner region elongation, $\lambda_{inner}\equiv r_{center}-r_{inner}$, the most active and significant part of the tori for this analysis  (large tori  inner  elongation are expected  for small  \textbf{BH} spin).
While  results on   the   torus outer edge coincidence with the static limit  are in Figs\il(\ref{Fig:spessplhoke1},\ref{Fig:PlotVamp1},\ref{Fig:PlotJurY}).

 On the other hand, the pressure  gradients are also  the main factor to be evaluated for  the  exfoliation and the consequent  formation of  a possible atmosphere of  free-particle  swarm.
In order to asses the reliability of an eventual process of disk exfoliation we   studied  in Sec.\il(\ref{Sec:dv}) the pressure gradients in the disks, results are in  Figs\il(\ref{Fig:titlePlot}), (\ref{Fig:polodefin1}),(\ref{Fig:weirplot}),(\ref{Fig:collag}),(\ref{Fig:gatplot8}),(\ref{Fig:gatplot17}),(\ref{Fig:gatplot5}),(\ref{Fig:spessplhoke1}).
    In   Figs\il(\ref{Fig:weirplot}) and(\ref{Fig:Plotexale})   we show the results of the analysis of the  disk verticality in terms of the polar gradients of the pressure,  considering the lines of extremes  of the HD  pressure,   providing also the  surfaces of geometric maximum,  obtaining  therefore a clear indication of the maximum vertical extension of the torus in the ergoreigon.
   Figs\il(\ref{Fig:raisePlot}) summarize the  results on the dragged tori  geometrical thickness, distinguishing faster spinning  and slower spinning  black holes, and the situations where the inner Roche lobe of the tori is thicker then the outer Roche lobe for the dragged cusped tori.
In Sec.\il(\ref{Sec:mid-w-t}) we  explored the  process of
tori exfoliation considering five tori models,  six particle models from   four regions of the configurations crossing the static limit.
We considered  quiescent (inert) tori (topologically regularly surfaces) and  cusped tori,  characterized  by the  HD instability driven by a Paczy\'nski mechanism. The accretion occurs  through the torus cusp (from the outer Roche lobe of the disk ).
On the sidelines of this analysis we also considered in Sec.\il(\ref{Sec:gir-c2})  the proto-jets existence  in $\Sigma_{\epsilon}^+$, and extreme configurations as  agglomerates of toroids orbiting the ergoregion  constituting aggregates of corotating tori which can collide .

Dragged and partially contained tori  have been also studied in  relation to  the  \textbf{QPOs} emissions from the tori  inner edges.
More precisely, the  characteristic frequencies of the toroids are studied in Sec.\il(\ref{Sec:carac-fre}) and  in Sec.\il(\ref{Sec:qpos}) we discuss the
origin of the \textbf{QPOs} emission from  toroids in  the ergoregion.
 We focused on the resonant frequency ratios {\footnotesize $\mathbf{R1}=\mathrm{2:1}$, $\mathbf{R2}=3:1$, $\mathbf{R3}={3}:{2}$, $\mathbf{R4}={4}:{3}$, $\mathbf{R5}={5}:{4}$} and
  frequencies used to the fit of the models  {\footnotesize  $\{(\textbf{WD}),(\textbf{TD}),(\textbf{TP}),(\textbf{TP1}),(\textbf{RP}),(\textbf{RP1})
  ,(\textbf{RP2}),(\textbf{RE}),(\textbf{Kepl})\}$} are  expressions of particle geodesic   frequencies of Eqs\il(\ref{Eq:posittru}) related to dragged or partially contained tori (with   torus inner region contained in the ergoregion).
   Some results are in  Figs\il(\ref{Fig:bapopof4})-- (\ref{Fig:dedichinPlot}).

In   Sec.\il(\ref{Sec:poly-altr}) we specified   the  polytropic equation of state for dragged  and partially contained tori, investigating how  the frame dragging of the Kerr spacetime  could   differentiate  dragged and partially contained  tori  with different values of the  polytropic index.
Discussion on some aspects of  tori energetics for these tori  is in Sec.\il(\ref{Sec:ener}). We provided  an estimation of the    flux thickness, mass-flux and   enthalpy-flux
  basing  on geometric considerations  for dragged and partially contained   thick  disks.
These toroids  can be observed  orbiting   black holes  with  dimensionless spin $a>0.9897M$.

In Sec.\il(\ref{Sec:ener}),
  we also discussed  the relations between the  dragged tori  in the  ergoregion and the maximum extractable rotational  energy  $\xi$ from the \textbf{BH} horizon,  showing  the situation of  the state prior the total extraction of the energy from the \textbf{BH}, considering the  approach introduced in \citet{Daly0}
 focused  on the definition of  \textbf{SMBH} (irreducible) mass function and  the  definition of rotational energy.
Finally, in  Appendix\il(\ref{Sec:magn}), we included in the GRHD model developed in  Sec.\il(\ref{Sec:model}) a toroidal magnetic field a la' Komissarov \citep{Komissarov:2006nz},  developing the analysis on
 the  stationary  magnetized tori in the ergoregion.
As mentioned in Sec.\il(\ref{Sec:mid-w-t}),  disk exfoliation could  combine with others tori  characteristics processes  as the Poynting--Robertson effects  for partially included and dragged surfaces. Runaway instability  is  another  major   process  for  dragged  thick disks orbiting \textbf{SMBHs}.
We  intend to deepen the relation with    the Bardeen--Petterson  effect,  for  tori having  a slight inclination  with respect to the equatorial plane,  in regards to the possible origin on an inner corotating dragged or partially contained tori.

\section*{Data Availability}
No new data were generated or analysed in support of this research.

\appendix

\section{Notes on stationary  magnetized tori in the ergoregion: the toroidal magnetic field}\label{Sec:magn}
The study of the magnetized surfaces with toroidal magnetic field  is an interesting  issue  of  the toroidal  configurations  we consider here.
The presence of a magnetic field with a relevant toroidal component can be   related  to the disk differential rotation, viewed as a generating  mechanism of the magnetic field \citep{Komissarov:2006nz,Montero:2007tc},
The  Komissarov solution   is a well known and widely used magnetic field solution  based on the construction of a toroidal magnetic field considering the set  of results known as von Zeipel theorem- \citep{Komissarov:2006nz}. Here we consider the possibility  that such field could be defined in the ergoregion.
We re-consider briefly the onset developed in \citet{EPL,Fi-Ringed}.
We also refer to   \citet{EPL,Hamersky:2013cza,Karas:2014rka,Fi-Ringed,adamek} where this solution is dealt  in detail in the context of accretion disks.
 We consider  an infinitely conductive plasma, with    $F_{ab}u^a=0$ where   $F_{ab}$ is the Faraday tensor and $u^a B_a=0$, where $u^a$ is the fluid four-velocity and $B^a$ is the magnetic field with   $\partial_{\phi}B^a=0$ and $B^r=B^{\theta}=0$ .

The Euler equation for this system can  be exactly integrated for the background spacetime of Schwarzschild and Kerr \textbf{BHs}
   with    a magnetic field $B=B^{\phi}$ and magnetic pressure $p_B$
\bea&&\label{RSC}
B^{\phi }=\sqrt{\frac{2 p_B}{\mathcal{A}}}\quad \mbox{where}\quad p_B=\mathcal{M}\mathcal{A}_{\mathrm{T}}^{q-1}\varpi^q \quad \mbox{or alternatively}\\&&\nonumber   B^{\phi }=\sqrt{{2 \mathcal{M} \varpi^q}} \mathcal{A}_{\mathrm{T}}^{(q-2)/2} V_{eff}(\ell),
\\
&&
\mbox{where}\quad\mathcal{A}\equiv\ell ^2 g_{tt}+2 \ell  g_{t\phi}+g_{\phi \phi }, \quad \mathcal{A}_{\mathrm{T}}\equiv g_{t \phi }^2-g_{{tt}}g_{\phi \phi},
\eea
where $\varpi$ is the fluid enthalpy, $q$  and $\Mie$ are constant-\citet{Komissarov:2006nz,Montero:2007tc}. (A  barotropic equation of state is assumed).  Eq.\il(\ref{Eq:scond-d})  has been used in
 second term of  equation\il\ref{RSC}.
According to our set-up we  introduce a deformed (magnetized) \emph{Paczy{\'n}ski potential function}  $\tilde{W}$ and the
  Euler  equation  (\ref{Eq:scond-d})  becomes:
\bea\label{Eq:Kerr-case}
&&\partial_{\mu}\tilde{W}=\partial_{\mu}\left[\ln V_{eff}+ \mathcal{G}\right]\, \mbox{where}\quad \tilde{W}\equiv \mathcal{G}(r,\theta)+\ln(V_{eff})=K,
\\
&&
\mbox{where for} \quad a\neq0:\quad \mathcal{G}(r,\theta)=\Sie \left(\mathcal{A} V_{eff}^2\right)^{q-1}=\Sie\mathcal{A}_{\mathrm{T}}^{q-1}, \quad\mbox{and}\quad \Sie\equiv\frac{q \mathcal{M} \omega ^{q-1}}{q-1}.
\eea
We therefore consider  the equation for the function $\tilde{W}$ and we introduce  the function
\bea
&&\label{Eq:goood-da}
 \widetilde{V}_{eff}^2\equiv V_{eff}^2 e^{2 \Sie \left(\mathcal{A} V_{eff}^2\right){}^{q-1}}.
\eea
%%s
 The toroidal surfaces  are obtained from the equipotential surfaces, for the potential
 $\widetilde{V}_{eff}^2$ which, for $\Sa=0$, reduces to  the effective potential ${V}_{eff}^2$ for the non-magnetized case in
Eq.\il(\ref{Eq:scond-d}).
The tori are regulated by the modified rotational law   $\widetilde{\ell}^{\pm}(r):\, \partial_r\tilde{V}_{eff}=0$, for counterrotating and corotating magnetized  fluids  respectively-
{where there is } $\lim_{\mathcal{\Sie}\rightarrow0}\widetilde{\ell}^{\mp}=\lim_{q\rightarrow 1}\widetilde{\ell}^{\mp}=\ell^{\pm}$.
%\end{strip}
%
We  rephrase  the problem  introducing  an adapted function $\mathcal{\Sie}_{crit}(r;\ell,q)$,
whose  values $\mathcal{\Sie}_{crit}(r;\ell,q)=$constant,
 provide  the parameter  $\Sa$ defining  the  torus.
Re-considering the equation for the  hydrostatic pressure critical points, solution $\mathcal{\Sie}_{crit}(r;\ell,q)$  represents the values of $\Sa$ as a function of $r$, for  which    critical points of the function $\widetilde{V}_{eff}$ exist:
{{
\bea\nonumber
\mathcal{\Sie}_{crit}\equiv-\Sie_{\Qa}\frac{a^2 (a-\ell)^2+2 r^2 (a-\ell) (a-2 \ell)-4 r (a-\ell)^2-\ell^2 r^3+r^4}{2  r  (r-1)\left[r (a^2-\ell^2)+2 (a-\ell)^2+r^3\right]},\quad \left(\mbox{where}\quad\Sie_{\Qa}\equiv \frac{\Delta^{-\Qa}}{\Qa}\right)
\\
\label{Eq:Sie-crit}
\eea}}
(with $r\rightarrow r/M$ and $a\rightarrow a/M$ and $\Qa=q-1$),  giving the tori  centers    and, eventually the tori cusps.
(A negative solution for $\Sa_{crit}<0$ may appear   for $q>1$.).
\begin{figure}\centering
  % Requires \usepackage{graphicx}
  \includegraphics[width=8cm]{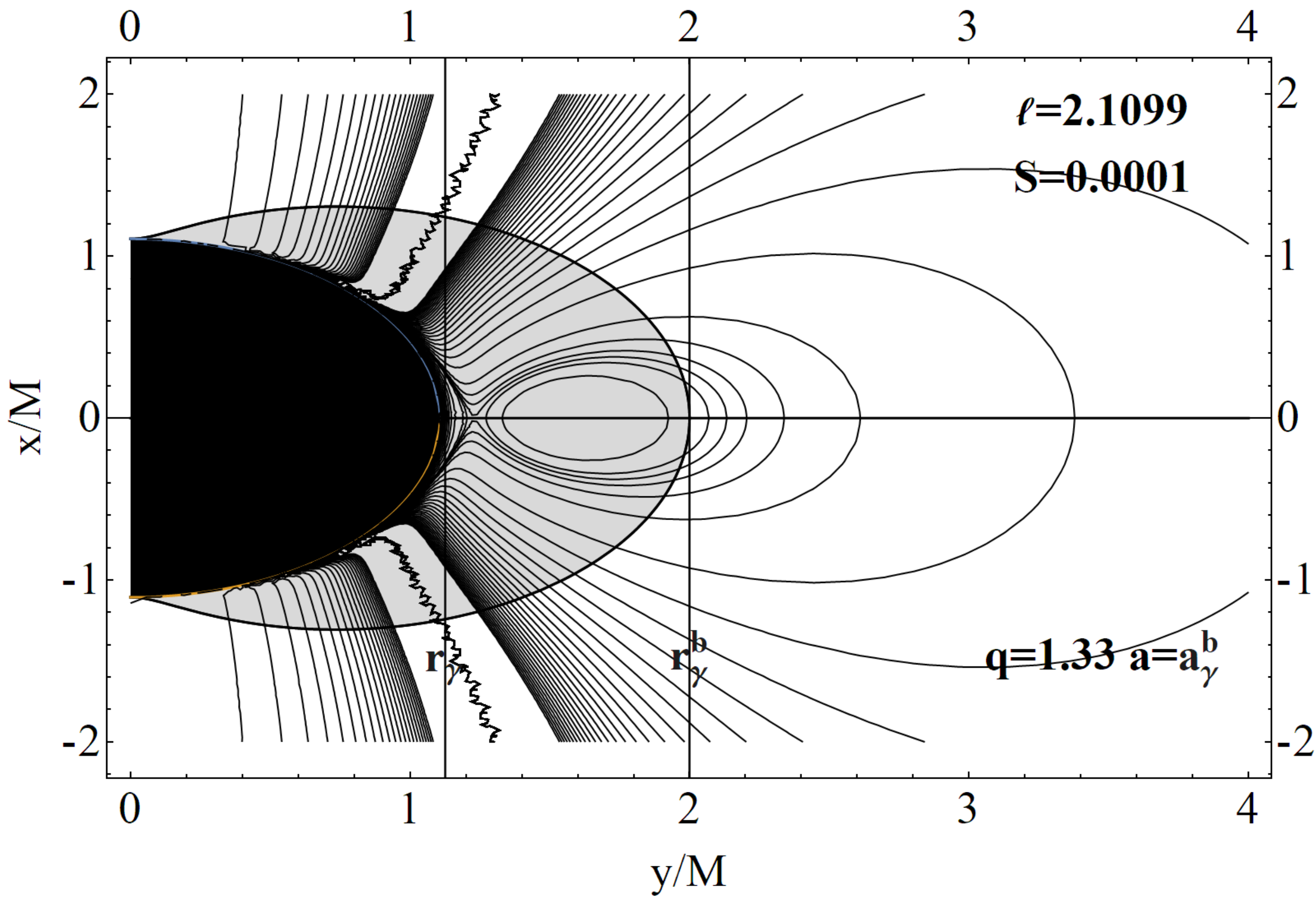}
  \includegraphics[width=8cm]{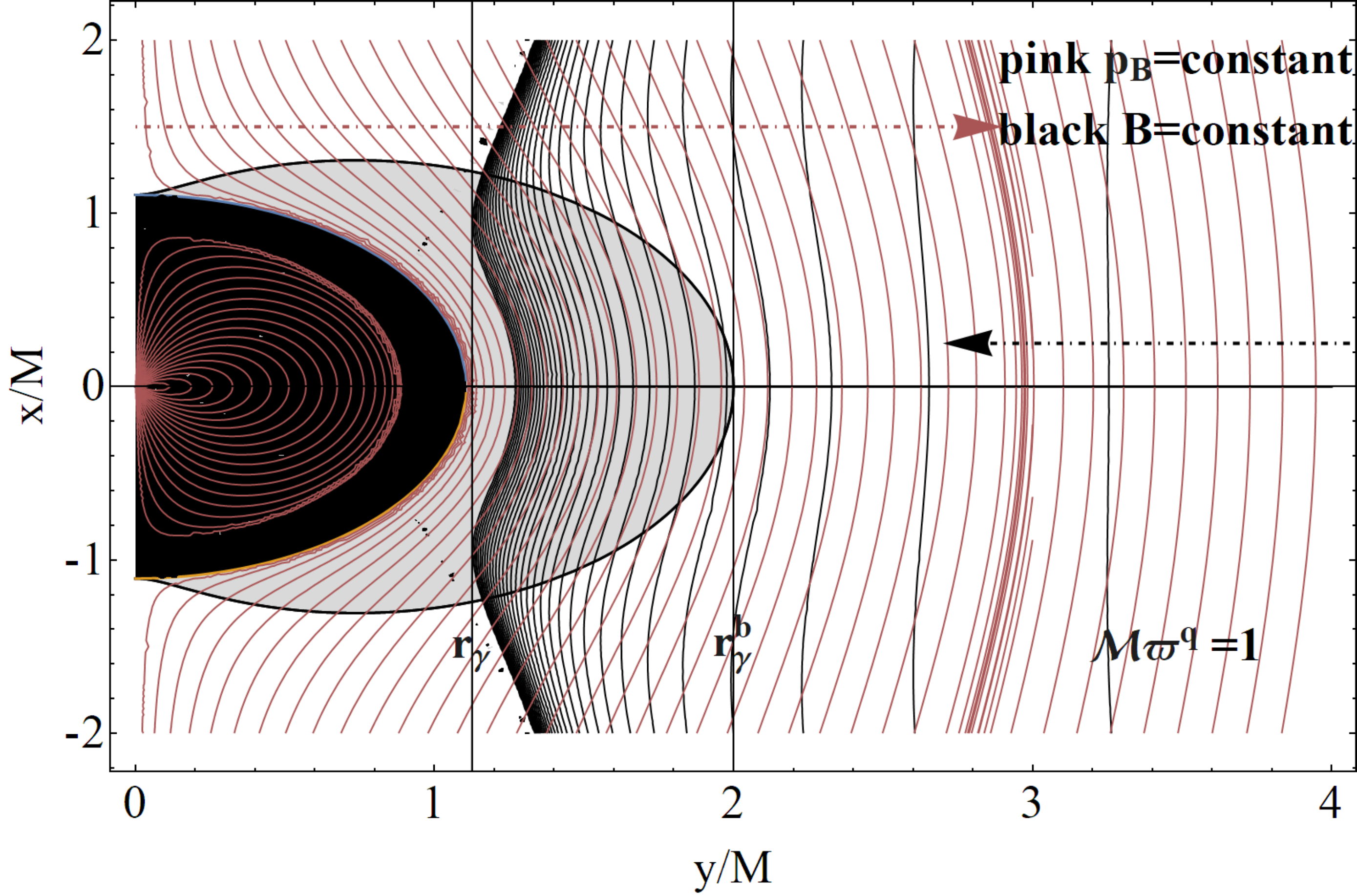}
  \includegraphics[width=8cm]{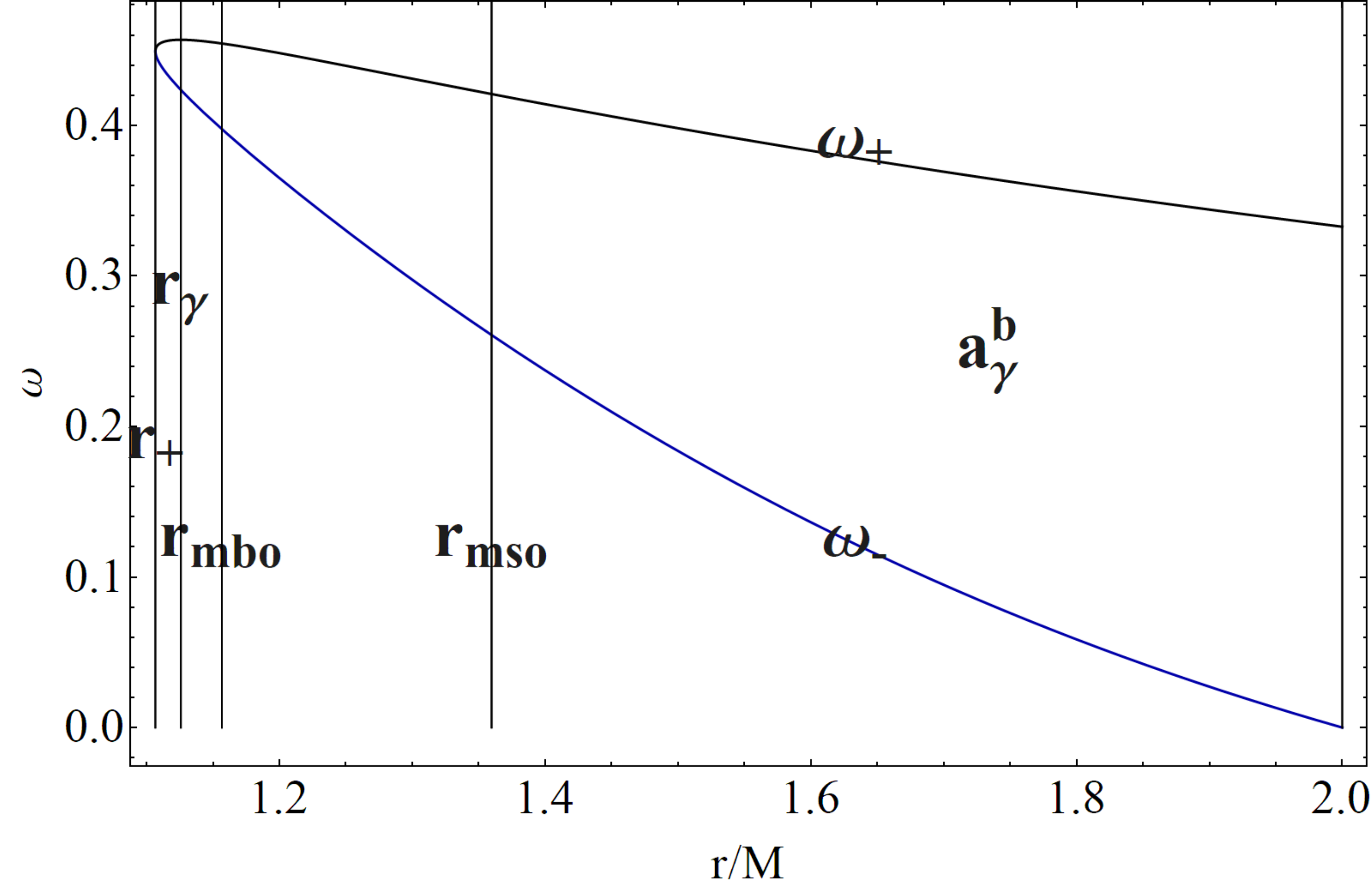}
   \includegraphics[width=8cm]{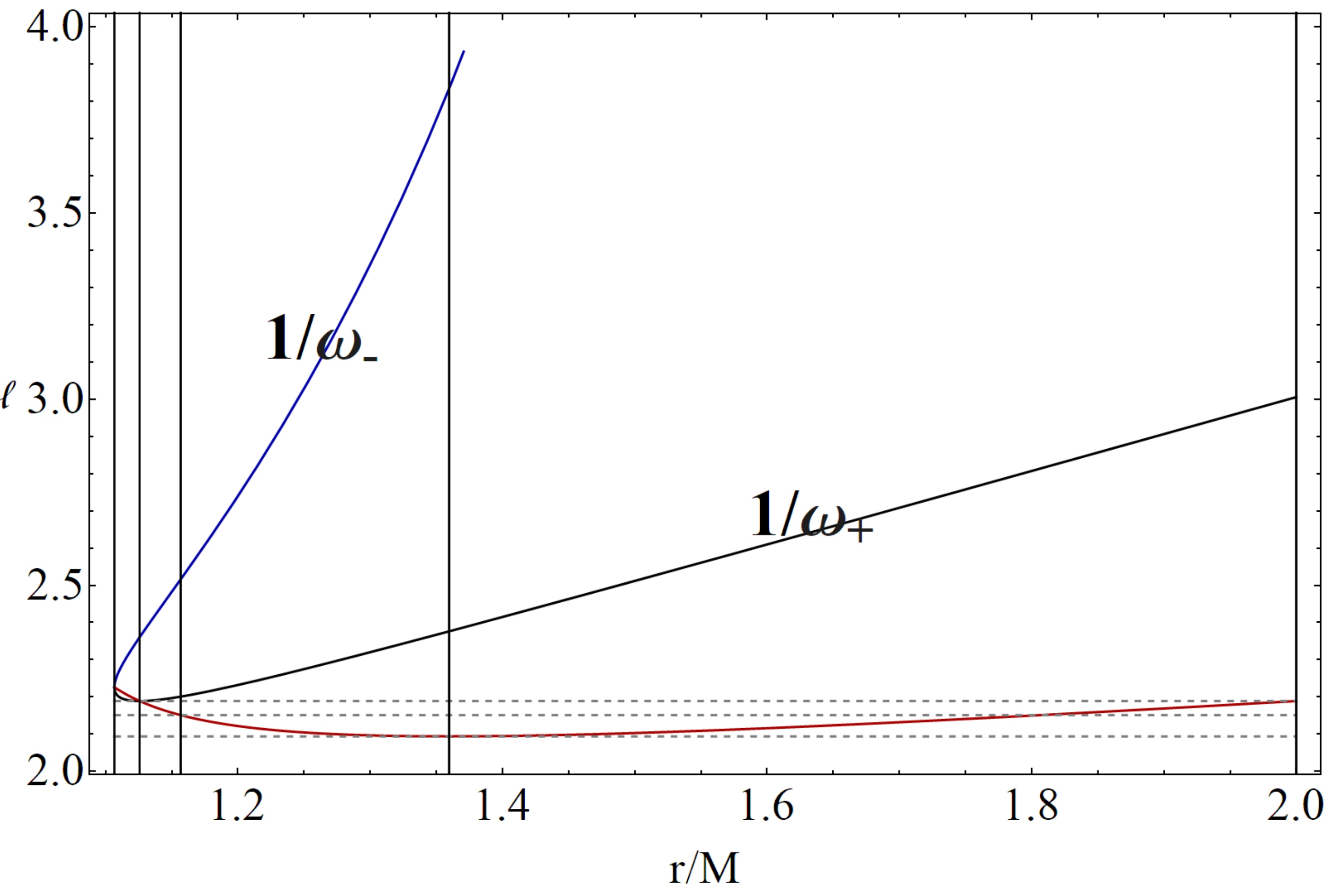}
  \caption{Magnetized tori in the ergoregion. Upper panel: Black region is the \textbf{BH} $r<r_+$, gray region is the outer ergosurface. For spacetime $a=a_{\gamma}^b$ right panel shows the tori surfaces at constant $K$  for a fixed value of the specific momentum $\ell$, the magnetic parameters $\Sie$  and $q$. $r_{\gamma}$ is the photon orbits and $r_{\gamma}^b: \ell(r_{\gamma})=\ell(r)$.  Right panel shows the magnetic pressure $p_B=$constant (pink) and the magnetic field $B=$constant  (black) of Eqs\il(\ref{RSC}), for parameters $\Mie \varpi^1=1$ ($\varpi$ is  the enthalpy).  Dotted-dashed arrows indicate  the increasing values  of the magnetic pressure (pink) and  of the magnetic field (black). For the analysis of the  spacetime  $a_{\gamma}^b$  properties see Figs\il(\ref{Fig:PlotVampb1})-(\ref{Fig:PlotVampa1})-(\ref{Fig:spessplhoke1})-(\ref{Fig:PlotVamp1})-(\ref{Fig:weirplot})-(\ref{Fig:collag})-(\ref{Fig:Plotexale}). Below  left panel:   photon orbital frequencies $\omega_{\pm}$ versus $r/M$ for the spacetime $a=a_{\gamma}^b$, stationary observers exist for $\omega\in]\omega_-,\omega_+[$ where $\omega_+=\omega_-$ on the horizon $r_+$.  $r_{mso}$ is the marginally stable orbit, $r_{mbo}$ is the marginally bounded orbit. Right panel: \textbf{BH} spacetime with spin $a_{\gamma}^b$, limits on the specific angular momentum $\ell<1/\omega_+$  and $\ell>1/\omega_-$ for the magnetic field in the ergoregion. Dashed horizontal lines are $\ell(r_{\gamma})>\ell(r_{mbo})>\ell(r_{mso})$. }\label{Fig:Sospende}
\end{figure}
  We are interested to  the conditions for  the magnetic field, defined in Eq.\il(\ref{RSC}), is   well defined in the ergoregion at any plane  $\theta=$constant.
  It is simple to see that  the magnetic pressure,  within the condition assumed on the enthalpy, is always positive.
  We note that, although the magnetic field is independent from the effective potential function  $V_{eff}$, it depends implicitly on the  $K$ parameter.
In  Figs\il(\ref{Fig:Sospende}) we considered the field $B$ in relation to tori present in the ergoregion.
The definition of the magnetic field is delineated by the light-surfaces, solutions of $\mathcal{L}\cdot\mathcal{L}=0$,   which provide solutions  in the ergoregion--Figs\il(\ref{Fig:Plotexale}).  (The  photon circular  orbit on the equatorial plane $r_{\gamma}$, is a particular relevant surface.). The field is not well defined approaching the horizon $r_+$.
There is   $\mathcal{A}=\ell^2 g_{tt}+ 2\ell g_ {t\phi} +
  g_ {\phi\phi} =\ell^2(\mathcal{L}\cdot\mathcal{L})= \ell^2\left (g_{tt}+
     2\omega_{\ell} g_ {t\phi} +\omega_{\ell}^2 g_ {\phi\phi} \right)$ where
$\omega_\ell\equiv 1/\ell$, and the problem is reduced to find  the solutions   of the equations for the frequencies $\omega_{\ell}$ of the stationary observers. Light surfaces  have frequencies  $\omega_{\pm}$. Spacelike solutions are given  for
  $\omega_\ell < \omega_-$ and $\omega_\ell > \omega_+$ (on all the planes $\theta$). On the ergosurface the condition
is  $\omega_\ell > \omega_+$  as $\omega_ -=
 0$.

%%%%%%%%%%%%%%%%%%%%%%%%\input{Fi-biblio}

%\bsp	% typesetting comment
%\label{lastpage}
\end{document}